\documentclass[siggraph]{acmart}

\usepackage{booktabs} 

\setcitestyle{square}

\usepackage{subfigure}
\usepackage{graphicx}
\usepackage{multicol}
\usepackage{amsmath}
\usepackage{calc}

\usepackage{amssymb}
\usepackage{hyperref}
\usepackage{url}
\usepackage{diagbox}
\usepackage{adjustbox}

\def\semichecked{\checkmark\!\!\!\raisebox{0.4 em}{\tiny$\smallsetminus$}}
\usepackage[ruled]{algorithm2e} 

\SetAlFnt{\small}
\SetAlCapFnt{\small}
\SetAlCapNameFnt{\small}
\SetAlCapHSkip{0pt}
\IncMargin{-\parindent}

\setcopyright{none}

\renewcommand\footnotetextcopyrightpermission[1]{} 
\newcommand{\tabincell}[2]{\begin{tabular}{@{}#1@{}}#2\end{tabular}}

\newlength{\tempheight}
\newcommand{\rowname}[1]
{\rotatebox{90}{\parbox{\tempheight}{\centering #1}}}

\hypersetup{draft}
\begin{document}
\title{Pointfilter: Point Cloud Filtering via Encoder-Decoder Modeling}

\author{Dongbo Zhang}
\affiliation{%
  \institution{Beihang University}
  \city{Beijing}
  \country{China}}
\email{zhangdongbo9212@163.com}
\author{Xuequan Lu}
\affiliation{%
  \institution{Deakin University}
  \country{Australia}
}
\email{xuequan.lu@deakin.edu.au}
\author{Hong Qin}
\affiliation{%
 \institution{Stony Brook University}
 \state{New York}
 \country{USA}}
\email{qin@cs.stonybrook.edu}
\author{Ying He}
\affiliation{%
  \institution{Nanyang Technological University}
  \country{Singapore}}
\email{YHe@ntu.edu.sg}

\begin{abstract}
Point cloud filtering is a fundamental problem in geometry modeling and processing. Despite of significant advancement in recent years, the existing methods still suffer from two issues: 1) they are either designed without preserving sharp features or less robust in feature preservation; and 2) they usually have many parameters and require tedious parameter tuning. In this paper, we propose a novel deep learning approach that automatically and robustly filters point clouds by removing noise and preserving their sharp features. Our point-wise learning architecture consists of an encoder and a decoder. The encoder directly takes points (a point and its neighbors) as input, and learns a latent representation vector which goes through the decoder to relate the ground-truth position with a displacement vector. The trained neural network can automatically generate a set of clean points from a noisy input. Extensive experiments show that our approach outperforms the state-of-the-art deep learning techniques in terms of both visual quality and quantitative error metrics. The source code and dataset can be found at \url{https://github.com/dongbo-BUAA-VR/Pointfilter}.
\end{abstract}

%
%


%
%


\maketitle

\section{Introduction}
\label{sec:introduction}
As the output of 3D scanning processes, point clouds are widely used in geometry processing, autonomous driving and robotics. Many point clouds, especially the ones obtained from consumer-level depth sensors, are corrupted with noise. Therefore, point cloud filtering is a necessary preprocessing step for the downstream applications. 

The last two decades have witnessed significant progress in point cloud filtering. Many elegant algorithms have been proposed. The LOP (Locally Optimal Projection) methods including LOP \cite{Lipman2007TOG}, WLOP (weighted LOP) \cite{Huang2009TOG} and CLOP (continuous LOP) \cite{Preiner2014TOG}), are robust to noise and outliers. RIMLS (robust implicit moving least squares) \cite{Ztireli2009CGF} and GPF (GMM-inspired Feature-preserving Point Set Filtering) \cite{Lu2018TVCG} can preserve sharp features. Nevertheless, these techniques still suffer from either feature smearing or less robustness in filtering. For example, the LOP family \cite{Lipman2007TOG, Huang2009TOG, Preiner2014TOG} are not designed for preserving sharp features, because of their inherent isotropic nature. RIMLS \cite{Ztireli2009CGF} and GPF \cite{Lu2018TVCG} depend heavily on normal filters and are not robust for large noise or non-uniform sampling. Furthermore, it is not easy for GPF \cite{Lu2018TVCG} to find a proper radius that balances noise removal and gaps near edges, and it is also slow due to high computational cost of Expectation-Maximization (EM) optimization. Finally, they all need careful trial-and-error parameter tuning which is tedious and time-consuming. 

Deep learning has proven highly effective in processing regular data, such as images and videos~\cite{He2015, vgg}. However, it is non-trivial to apply deep learning techniques to irregular data, such as point clouds. To our knowledge, point cloud filtering with resorting to deep learning has been rarely studied so far, such as EC-Net~\cite{Yu2018ECCV}, PCN \cite{Rakotosaona2019CGF} and TotalDenoising \cite{hermosilla2019ICCVl}. These data-driven approaches avoid the tedious parameter tuning, however, they still generate limited results with either smoothing out sharp features or poor generalization, which diminishes their usages in real-world applications. 
\begin{table*}[]
    \centering
    \setlength\tabcolsep {2pt}
    \caption{Comparison of the main characteristics (CHs) between the state-of-the-art point set filtering techniques and our method. For the normal independence, we only consider the inference stage for learning-based techniques.     $\checkmark$ and $\times$ denote YES and NO, respectively. $\semichecked$ indicates an ``intermediate'' level between YES and NO.}\label{table:featurechart}
    \begin{tabular}{|l|c|c|c|c|c|c|}
    \hline
     \diagbox[width=6em]{\textbf{Methods}}{\textbf{CHs}} & \tabincell{c}{normal\\independence} 
     & \tabincell{c}{noise} & \tabincell{c}{outliers}
     & \tabincell{c}{Feature\\aware} & \tabincell{c}{Parameters} & \tabincell{c}{Speed}
     \\ \hline
     \tabincell{l}{LOP~\cite{Lipman2007TOG} \&\\WLOP~\cite{Huang2009TOG}} & $\checkmark$ & $\checkmark$    & $\semichecked$  & $\times$       & $\times$  & $\semichecked$
     \\ \hline
     \tabincell{l}{CLOP~\cite{Preiner2014TOG}}                            & $\checkmark$ & $\checkmark$    & $\semichecked$  & $\times$       & $\times$  & $\checkmark$
     \\ \hline
     \tabincell{l}{RIMLS~\cite{Ztireli2009CGF}}                           & $\times$     & $\checkmark$    & $\times$        & $\checkmark$   & $\times$  & $\semichecked$
     \\ \hline
     \tabincell{l}{GPF~\cite{Lu2018TVCG}}                                 & $\times$     & $\checkmark$    & $\semichecked$  & $\checkmark$   & $\times$  & $\times$
     \\ \hline
     \tabincell{l}{EC-Net~\cite{Yu2018ECCV}}                              & $\checkmark$ & $\checkmark$    & $\times$        & $\checkmark$   & $\checkmark$ & $\checkmark$
     \\ \hline
     \tabincell{l}{PCN~\cite{Rakotosaona2019CGF}}                         & $\checkmark$ & $\checkmark$    & $\checkmark$    & $\times$       & $\checkmark$ & $\semichecked$ 
     \\ \hline
     \tabincell{l}{TotalDenoising~\cite{hermosilla2019ICCVl}}             & $\checkmark$ & $\checkmark$    & $\times$        & $\times$       & $\checkmark$ & $\checkmark$
     \\ \hline
     \tabincell{l}{Ours}                                                  & $\checkmark$ & $\checkmark$    & $\semichecked$  & $\checkmark$   & $\checkmark$ & $\checkmark$
     \\ \hline
    \end{tabular}
\end{table*}
In this paper, we propose a novel deep learning approach to overcome the above issues. Motivated by the successes of AutoEncoder and PointNet \cite{Schmidhuber2015NN, Qi2017CVPR}, we design an end-to-end neural network, called Pointfilter, for point cloud filtering.
Our network is an encoder-decoder based architecture which straightforwardly takes the raw neighboring points of each noisy point as input, and regresses a displacement vector to push this noisy point back to its ground truth position. In designing the loss function, we also take sharp features into account so that sharp features can be preserved by this network. Given a noisy point cloud as input, our trained model can  automatically and robustly predict a corresponding clean point cloud, by removing noise and preserving sharp features. Various experiments demonstrate that our method achieves better performance than the state-of-the-art techniques (or comparable to optimization based methods like RIMLS and GPF which require decent normals and trial-and-error parameter tuning), in terms of visual quality and error metrics. Our method is fast and avoids manual parameter tuning. We show the main features of the existing techniques and our method in Table \ref{table:featurechart}.
The main contributions of this paper are:
\begin{itemize}
\item a novel end-to-end neural network that achieves point cloud filtering by encoder-decoder modeling;
\item an effective loss function that takes sharp features into account.
\end{itemize}
We will make the source code and dataset publicly available.

\begin{figure*}[htp]
  \centering
  \includegraphics[width=0.8\textwidth]{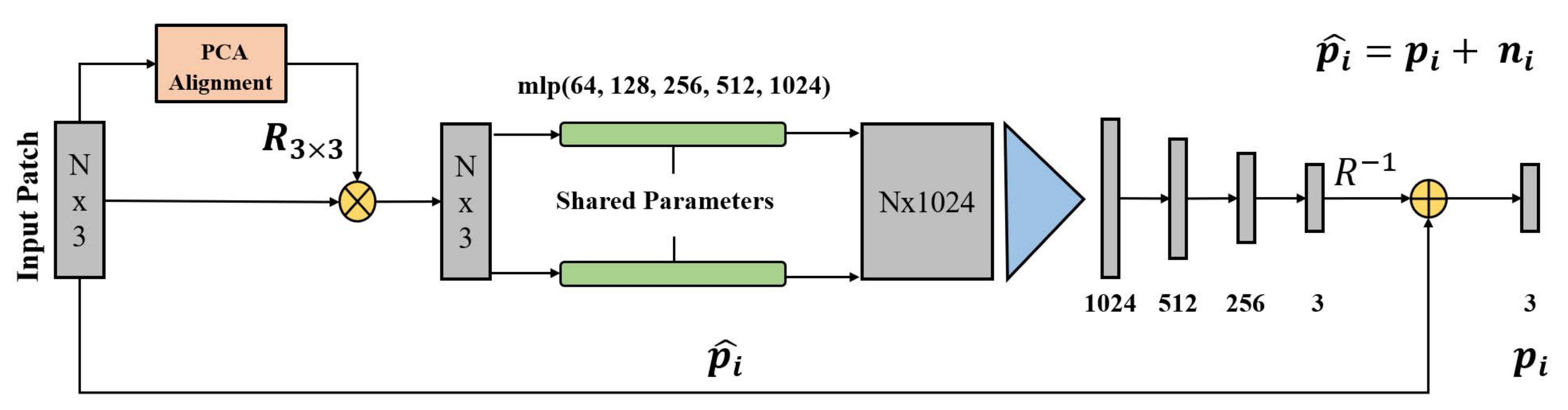}
  \caption{Pointfilter architecture: given a noisy patch with $N$ points (generated by the preprocessing step), we use PCA for alignment and feed the aligned patch into the neural network. Normalized input passes through the shared multi-layer perceptrons (MLPs) to extract local features and then aggregates each point feature by a max pooling layer. The MLPs consist of $5$ hidden layers with neuron sizes $64, 128, 256, 512$, and $1024$, respectively. Following the aggregated features, three fully connected layers, with neuron sizes $512, 256, 3$, are used to regress a displacement vector between the noisy point and the underlying surface. All layers except the last one adopt BatchNorm and ReLU, whereas the last layer only uses the activation function $tanh$ to constrain the displacement vector space. }
  \label{fig:overview}
\end{figure*}

\section{Related Work}
\label{sec:relatedwork}
We first review the methods for point cloud filtering, and then discuss the existing deep learning techniques for point clouds. 

\subsection{ Point Cloud Filtering}
\label{sec:traditionalPCfiltering}
Point cloud filtering can be generally classified into two types: two-step based techniques and projection-based methods. 

\textbf{Two-step based Methods.} The two-step based framework consists of at least two steps: normal smoothing and point position update under the guidance of the filtered normals. To preserve sharp features, Avron et al. \cite{Avron2010TOG} and Sun et al. \cite{Sun2015CAGD} introduced $L_1$ and $L_0$ optimization for point set filtering, respectively. Recently, with sharp feature skeletons, a point cloud smoothing technique \cite{Zheng2017TVC} was presented based on the guided filter and was extended to point clouds via a multi-normal strategy. Zheng et al. \cite{Zheng2018CAGD} extended the rolling guidance filter to point set filtering and designed a new point position updating strategy to overcome sharp edge shrinkage. Lu et al. \cite{Lu2018ArXiv} proposed a two-step geometry filtering approach for both meshes and point clouds. Although these two-step methods achieve promising results in point cloud filtering, they rely highly on normal estimation which is typically sensitive to heavy noise and non-uniform sampling. Most point set filtering methods achieve filtered results through projecting the input point set onto the underlying point set surface. 

\textbf{Projection-based Methods.} One popular category of this type is moving least squares and its variants \cite{Levin1998MoC,Levin2004GMfSV,Alexa2001Vis,Alexa2003TVCG,Amenta2004TOG,Fleishman2005TOG,Ztireli2009CGF}. The moving least squares (MLS) has been seminally formulated by Levin \cite{Levin1998MoC,Levin2004GMfSV}. Some works defined moving least squares (MLS) and extremal surfaces \cite{Alexa2001Vis,Alexa2003TVCG,Amenta2004TOG}. Later, two different variants have been presented for projection: statistics-based and robust implicit moving least squares (RIMLS) \cite{Fleishman2005TOG,Ztireli2009CGF}. Although RIMLS can preserve share features to some extent, it is not easy to find a proper support radius, which makes it difficult to noisy input. Lange et al. \cite{Lange2005CAGD} developed a method for anisotropic fairing of a point sampled surface using an anisotropic geometric mean curvature flow. Recently, the LOP (locally optimal projection) based methods have become increasingly popular. For example, Lipman et al. \cite{Lipman2007TOG} proposed the locally optimal projection operator (LOP) which is parameterization free. Later, Huang et al. \cite{Huang2009TOG} presented a weighted LOP (WLOP), which enhances the uniform distribution of the input points. A kernel LOP has also been proposed to speed up the computation of LOP \cite{Liao2013CAD}. More recently, a continuous LOP (CLOP) has been presented to reformulate the data term to be a continuous representation of the input point set and arrives at a fast speed \cite{Preiner2014TOG}. The LOP-based methods cannot preserve sharp features/edges, due to lack of consideration of sharp information. Note that a few projection-based methods utilize smoothed normals as prior to preserve sharp features, such as EAR \cite{Huang2013TOG} and GPF \cite{Lu2018TVCG}. Since these methods contain a number of parameters, careful trial-and-error parameter tuning is required to obtain decent results, especially for complex models.

\subsection{Deep Learning on Point Clouds}
\label{sec:relatedworkdponpointclouds}
\textbf{Point-based Network Architecture.} Qi et al. \cite{Qi2017CVPR} proposed the pioneering network architecture, named PointNet, which can consume raw points without voxelization or rendering. In contrast to the conventional CNNs that rely on convolution operators, PointNet adopts the multi-layer perceptrons for feature extraction, hereby working well for irregular domains. PointNet is simple and elegant, and provides a unified framework for shape classification and segmentation. However, PointNet processes the points individually, and cannot characterize local structures which are crucial for high-level semantic understanding. To overcome this limitation, Qi et al. developed an improved version, PointNet++, which aggregates local structures in a hierarchical way~\cite{Qi2017NIPS}. Following PointNet and PointNet++, lots of network architectures applied on raw point clouds emerged. For instance, based on dynamic local neighborhood graph structure, Wang et al. \cite{Wang2019TOG} designed an EdgeConv block to capture the relationships both in spatial and feature space. At the same time, an alternative convolutional framework, SpiderCNN \cite{Xu2018ECCV}, was proposed to aggregate neighboring features by a special family of parameterized weighted functions instead of MLPs. Leveraging the spatial-locally correlation, a novel block, named $\mathcal{X}$-Conv \cite{yangyan2018arXiv}, is proposed to tackle the degradation of shape information and variance to point ordering caused by directly applying convolution kernels on point clouds. Inspired by the Scale Invariance Feature Transform \cite{Lowe2004IJCV} (SIFT) which is a robust 2D representation, the SIFT-like module \cite{Jiang2018ArXiv} was developed to encode information of different orientations and scales and could be flexibly incorporated into PointNet-style networks. Besides shape classification and segmentation tasks, there are few point input network architectures applied on upsampling \cite{Yu2018CVPR,Yifan2019CVPR}, local shape properties estimation \cite{Guerrero2018CGF, lu2020deep} and so on. 

\textbf{Deep Learning on Point Cloud Filtering.} As for point cloud filtering, Roveri et al. \cite{Roveri2018CGF} proposed a filtering network, PointProNet, designed for consolidating raw point clouds corrupted with noise. Benefiting from powerful 2D convolution, PointProNet transfers 3D point clouds consolidation into 2D height map filtering. To preserve sharp edges while filtering, Yu et al. \cite{Yu2018ECCV} introduced a novel edge-aware network architecture, called EC-Net, by incorporating a joint loss function. It works well for models with sharp features, but training EC-Net requires manually labelling sharp edges. Combining with \cite{Guerrero2018CGF}, a two-stage network architecture, PointCleanNet (PCN) \cite{Rakotosaona2019CGF}, was developed for removing outliers and denoisnig separately. Recently, Hermosilla et al. \cite{hermosilla2019ICCVl} proposed an unsupervised method to filter noisy point clouds. By imposing priors, the method can directly train on noisy data without needing ground truth examples or even noisy pairs. However, this method cannot preserve share features due to lack of sharp feature information during the training stage. Duan et al.  \cite{duan2019ICASSP} presented a neural-network-based framework, named NPD, to smooth 3D noisy point clouds via projecting noisy points onto the reference planes. Since it handles points individually, NPD cannot exploit the local information which is important for projection. Also, the global $L_2$ loss is not effective for feature preservation.

\begin{figure}[hbt!]
  \centering
  \includegraphics[width=0.45\textwidth]{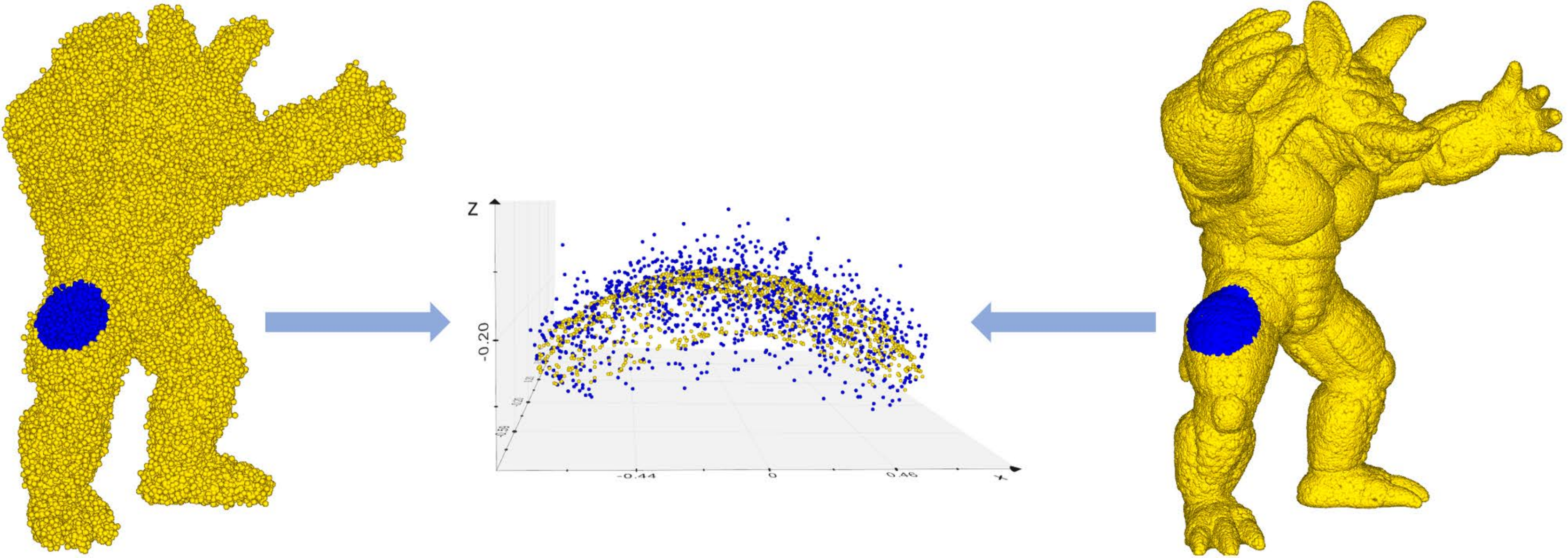}
  \caption{Pre-processing of Pointfilter: given a noisy point cloud (left) and a clean point cloud (right), we create a pair of noisy and clean patches (colored in blue). The pre-processing results are shown in the middle (clean patch is colored in yellow).}
  \label{fig:pacalignment}
\end{figure}

\section{Method}
\label{sec:method}
\subsection{Overview} Given a noisy point cloud, we aim to restore its clean version by our Pointfilter in a manner of supervised learning. Before introducing details of our Pointfilter framework, we first formulate a noisy point cloud as follows:
\begin{equation}\label{eq:noiseformulation}
\hat{\mathbf{P}} = \mathbf{P} + \mathbf{N},
\end{equation}
where $\hat{\mathbf{P}} = \{\hat{\mathbf{p}}_{1},\,...\,, \, \hat{\mathbf{p}}_{n} \, |\  \hat{\mathbf{p}}_{i} \in R^{3}, i = 1\,...\,n\}$ is an observed point cloud corrupted with noise, $\mathbf{P}$ is the corresponding clean point cloud (underlying surface) and $\mathbf{N}$ is the additive noise. In this work, we address the filtering problem in a local way, which means the filtered result of a noisy point only depends on its neighboring structure. As we know, point cloud filtering is an ill-posed problem and it is difficult to straightforwardly regress the additive noise for each noisy point like image filtering. As an alternative, we handle point cloud filtering by projecting each noisy point onto the underlying surface. More specifically, we treat the additive noise $\mathbf{N}$ as displacement vectors between the noisy point cloud $\hat{\mathbf{P}}$ and the clean point cloud $\mathbf{P}$, and learn the displacement vector for each noisy point. To achieve this, we propose an encoder-decoder architecture network, named Pointfilter, to regress the additive noise $\mathbf{N}$, shown in Fig. \ref{fig:overview}. We briefly introduce a pre-processing step for the input data in Section \ref{sec:preprocessing}, and then show how to model our Pointfilter in Section \ref{sec:pointfilterframework}. We finally explain how we train our network in Section \ref{sec:networktraining} and how we make inference with the trained network in Section \ref{sec:networkinference}. 

\subsection{Preprocessing}\label{sec:preprocessing}
Given a pair of point clouds $\mathbf{P}$ and $\hat{\mathbf{P}}$, the noisy patch $\hat{\mathbf{\mathcal{P}}}_{i}$ and its corresponding ground truth patch $\mathbf{\mathcal{P}}_{i}$ are defined as follows
\begin{equation}\label{equation:patchdefination}
    \hat{\mathbf{\mathcal{P}}}_{i} = \{\hat{\mathbf{p}}_{j}\ | \ \|\hat{\mathbf{p}}_{j} - \hat{\mathbf{p}}_{i}\| < r\}, \, \, \mathbf{\mathcal{P}}_{i} = \{\mathbf{p}_{j}\ |\ \|\mathbf{p}_{j} - \hat{\mathbf{p}}_{i}\| < r \}, 
\end{equation}
where $\hat{\mathbf{p}}_{i}, \hat{\mathbf{p}}_{j} \in \hat{\mathbf{P}}$,  $\mathbf{p}_{j} \in \mathbf{P}$ and $r$ is the patch radius. Once patches are generated, two issues need to be addressed in point cloud filtering: (1) how to avoid unnecessary degrees of freedom from observed space? (2) how to guarantee our Pointfilter is insensitive to certain geometric transformations (e.g. rigid transformations)? For the first issue, an immediate remedy is to translate patches into origin and then scale them into unit length, i.e., $\hat{\mathbf{\mathcal{P}}}_{i} = (\hat{\mathbf{\mathcal{P}}}_{i} - \hat{\mathbf{p}}_{i}) / r$. Similarly, the ground truth patch $\mathbf{\mathcal{P}}_{i}$ does the same thing, i.e., $\mathbf{\mathcal{P}}_{i} = (\mathbf{\mathcal{P}}_{i} - \hat{\mathbf{p}}_{i}) / r$. To be invariant to rigid transformations (e.g., rotations), a few methods \cite{Qi2017CVPR, Qi2017NIPS} attempted to predict rotation matrix $\mathbf{R} \in SO(3)$ via an additive spatial transformer network, while it has been proven to be fragile to rotations without massive data augmentation \cite{you2018ArXiv}. We align the input patches by aligning their principle axes of the PCA with the Cartesian space. Specifically, we first align the last principle axis with the z-axis, and then align the second principle axis with the x-axis. The alignment process is illustrated in Fig. \ref{fig:pacalignment}. To effectively tune network parameters with batches, the number of points in each input patch should be the same. In our experiments, we empirically set the default value $|\hat{\mathbf{\mathcal{P}}_{i}}| = 500$. We pad the origin for patches with insufficient points ($< 500$) and do random downsampling for patches with sufficient points ($> 500$). We set the patch radius $r$ to $5\%$ of the model's bounding box diagonal length. 

\subsection{The Pointfilter Framework}\label{sec:pointfilterframework}
The architecture of our point cloud filtering framework is demonstrated in Fig. \ref{fig:overview}. The key idea of our Pointfilter is to project each noisy point onto the underlying surface according to its neighboring structure. To achieve this, we design our Pointfilter network as an encoder-decoder network. Specifically, the encoder consists of two main parts: (1) feature extractors (i.e., MLPs) that are used to extract different scales of features; (2) a collector that is used to aggregate the features ($N\times 1024$) as a latent vector $\mathbf{z} \in R^{1024}$. The encoder module attempts to obtain a compact representation for an input patch. In the decoder module, a regressor is employed to evaluate the displacement vectors with the latent representation vector $\mathbf{z}$ as input. In this work, we adopt the recent PointNet \cite{Qi2017CVPR} as the backbone in our Pointfilter. In practice, the extractors and collector are realised by the shared MLPs and max pooling layer, respectively, and the regressor is constructed by three fully connected layers. Details of our Pointfilter network are shown in Fig. \ref{fig:overview}. At the beginning of our Pointfiler, a PCA-induced rotation matrix $\mathbf{R}$ is applied to transform the input patch to a canonical space. Therefore, at the end of our Pointfiler, an inverse matrix $\mathbf{R}^{-1}$ should be multiplied by the evaluated displacement vector to get the final displacement vector. 

\begin{figure*}
  \centering
   \subfigure[Noisy input]
    {
        \includegraphics[width=0.22\textwidth]{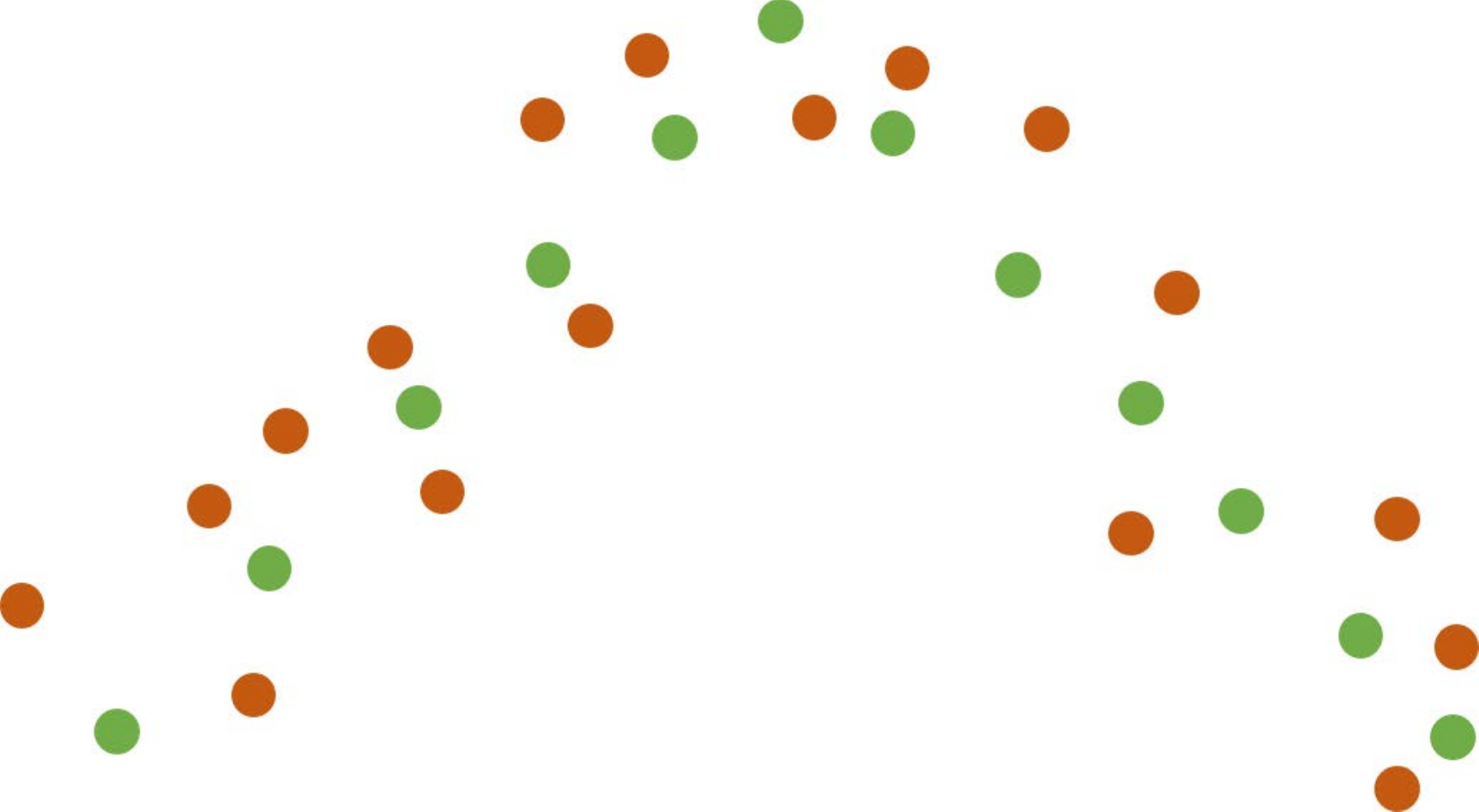}
    }
    \subfigure[$L_{2}$]
    {
        \includegraphics[width=0.22\textwidth]{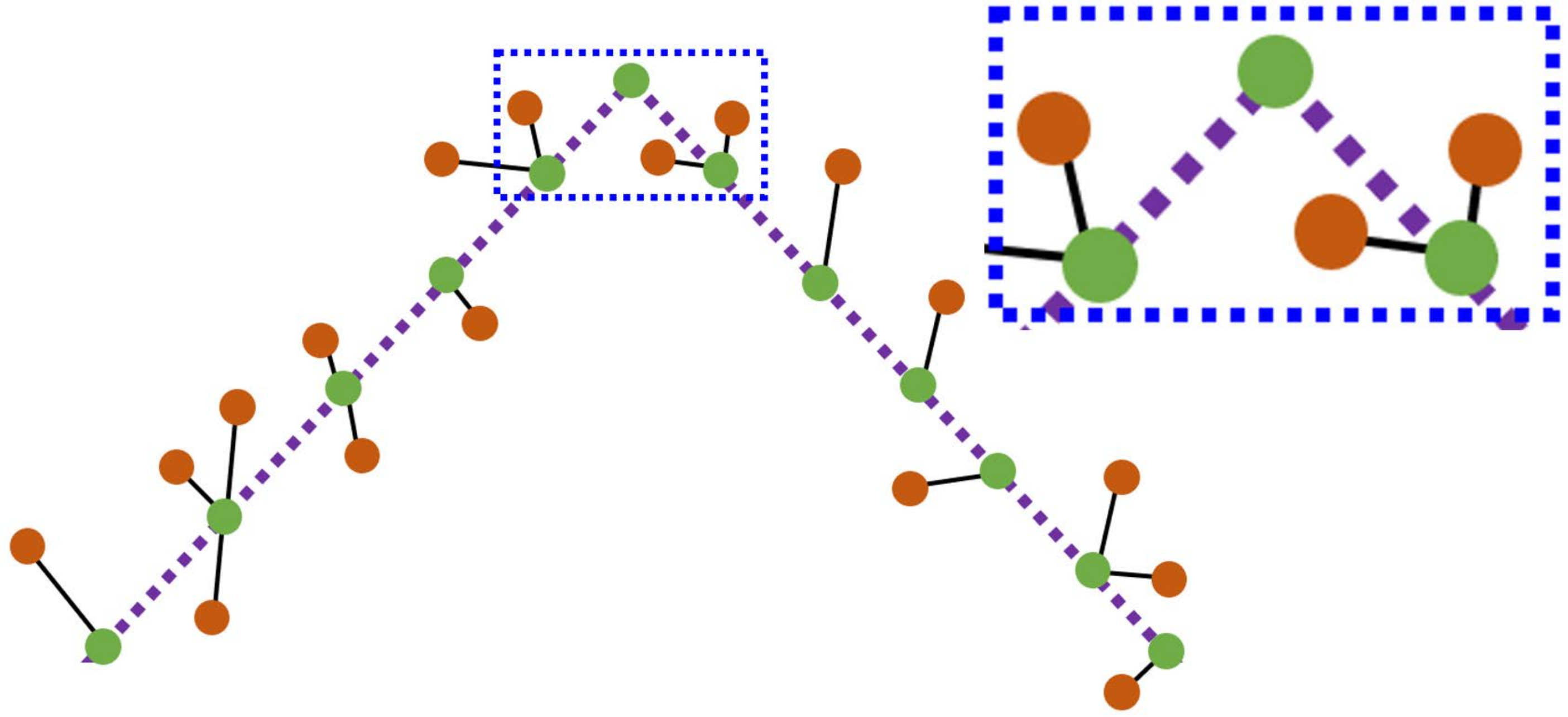}
    }
    \subfigure[$L_{proj}^{a}$]
    {
        \includegraphics[width=0.22\textwidth]{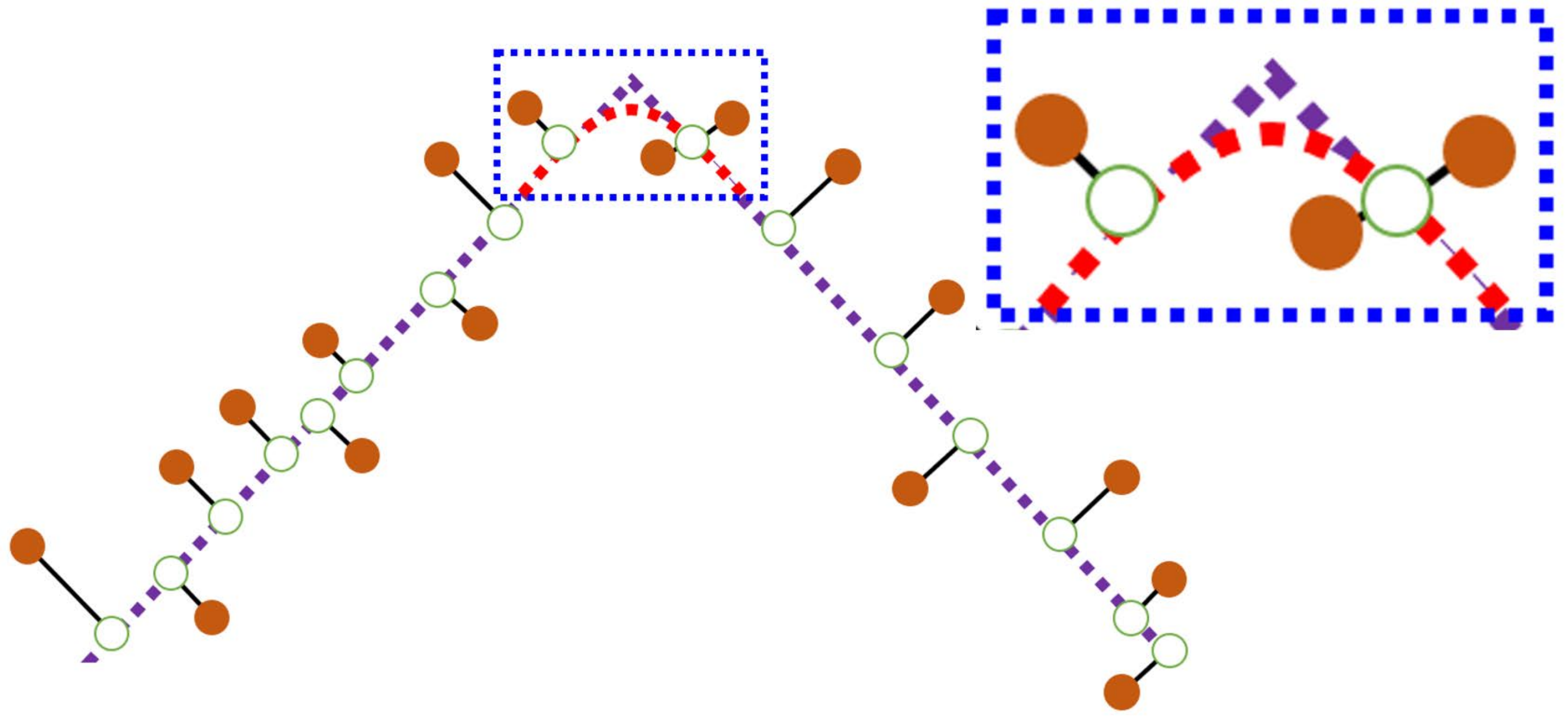}
    }
    \subfigure[$L_{proj}^{b}$]
    {
        \includegraphics[width=0.22\textwidth]{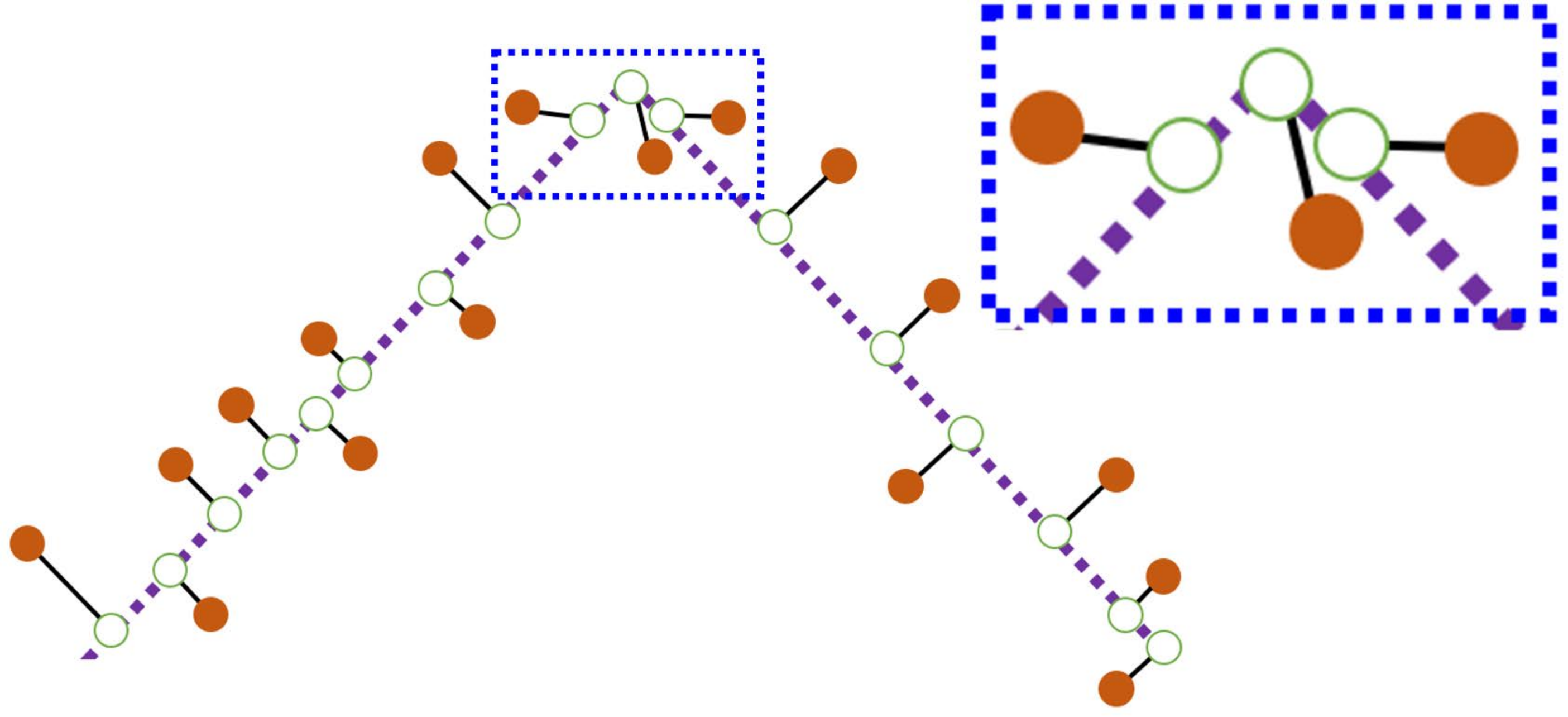}
    }
  \caption{Illustrating different loss functions on a toy model which simulates the side view of two planes. (a) We draw the ground-truth points in green and the noisy points in red, and the underlying surface by dashed purple lines. (b) The $L_2$ loss function (i.e., closest point here) is sampling dependent since it directly maps noisy points back to the closest sampled points. (c) The $L_{proj}^{a}$ loss would blur sharp edges (dashed curved red lines), though it could project noisy points onto the surface (hollow  circles). (d) Our proposed $L_{proj}^{b}$ projects each noisy point onto the underlying surface (hollow  circles) with considering normal information, hereby is more effective in preserving sharp features. See the close-up windows for the differences.}
  \label{fig:LossAlternative}
\end{figure*}

\begin{figure}[htb!]
  \centering
  \subfigure[Noisy]
  {
  \includegraphics[width=0.1\textwidth]{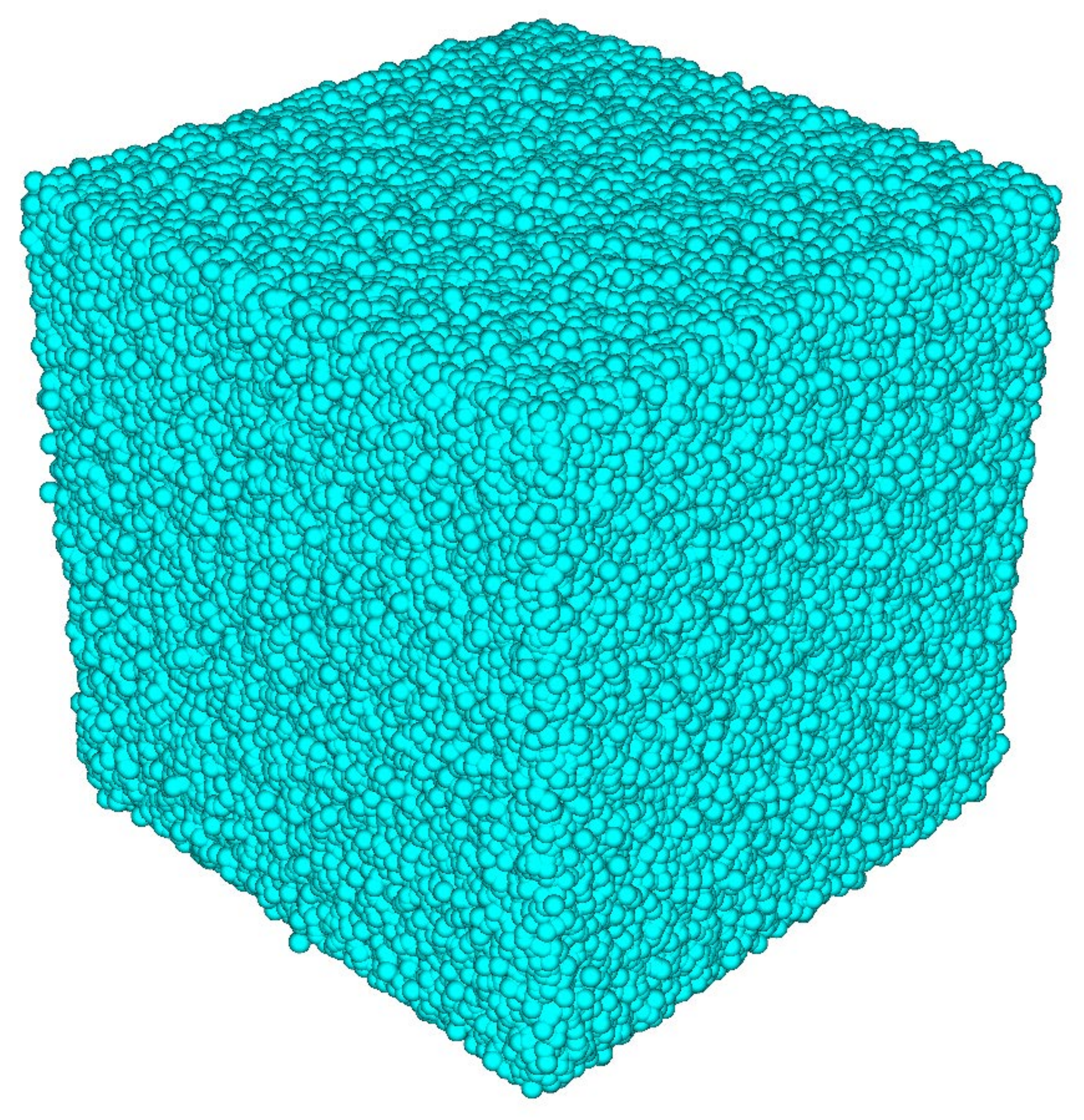}
  }
  \subfigure[$L_{proj}^{a}$]
  {
  \includegraphics[width=0.1\textwidth]{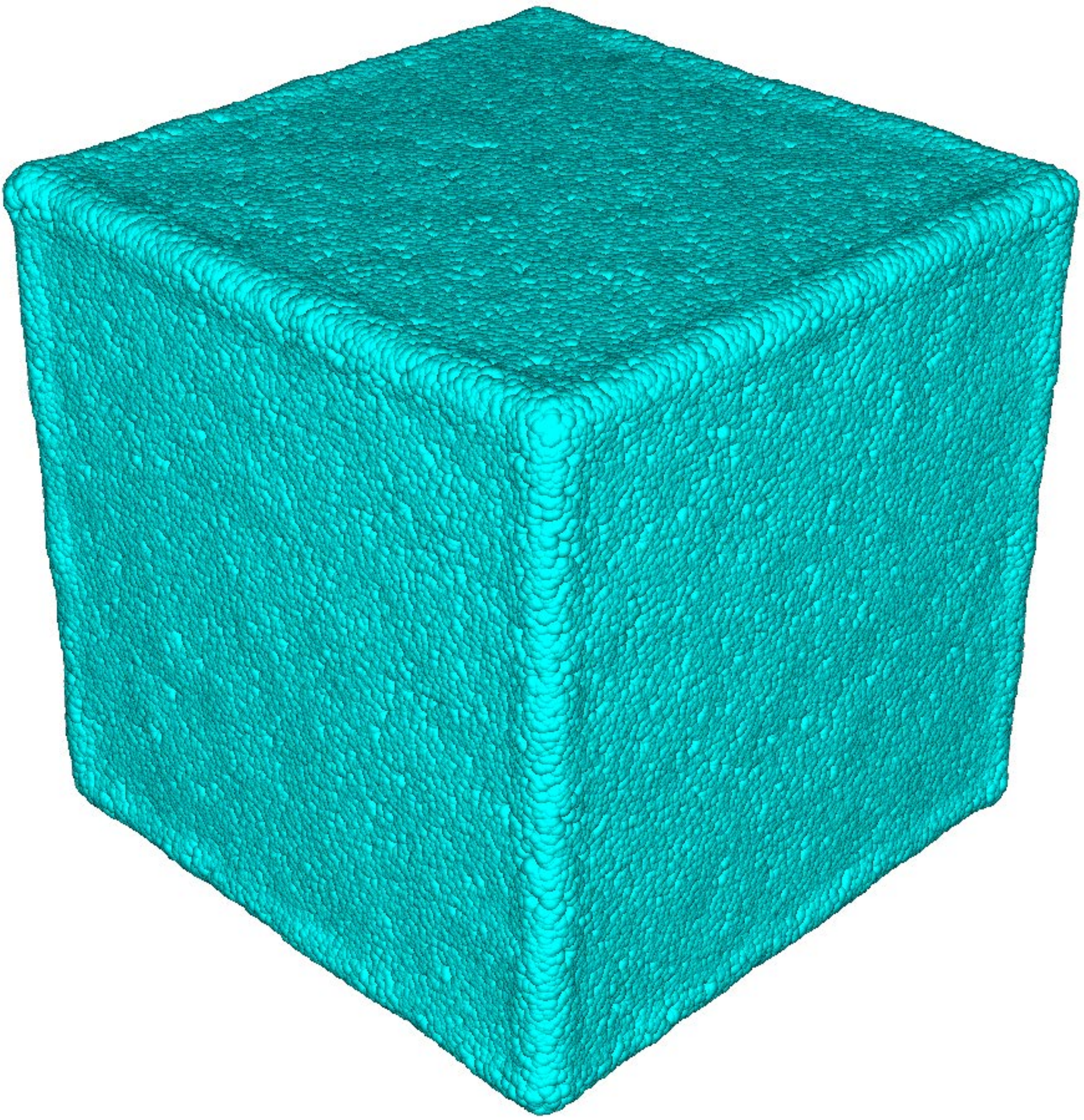}
  }
  \subfigure[$L_{2}$]
  {
  \includegraphics[width=0.1\textwidth]{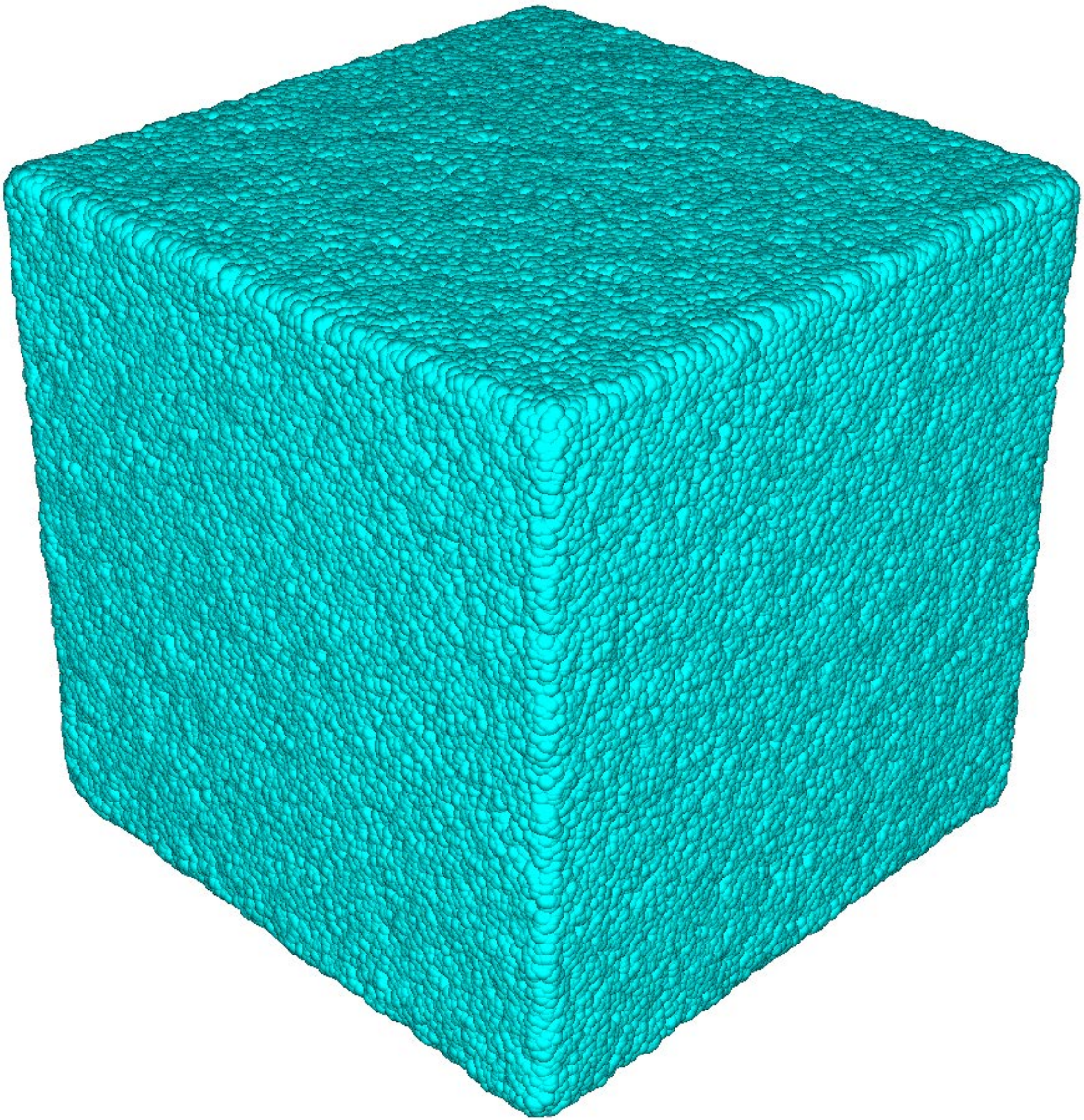}
  }
  \subfigure[$L_{proj}^{b}$]
  {
  \includegraphics[width=0.1\textwidth]{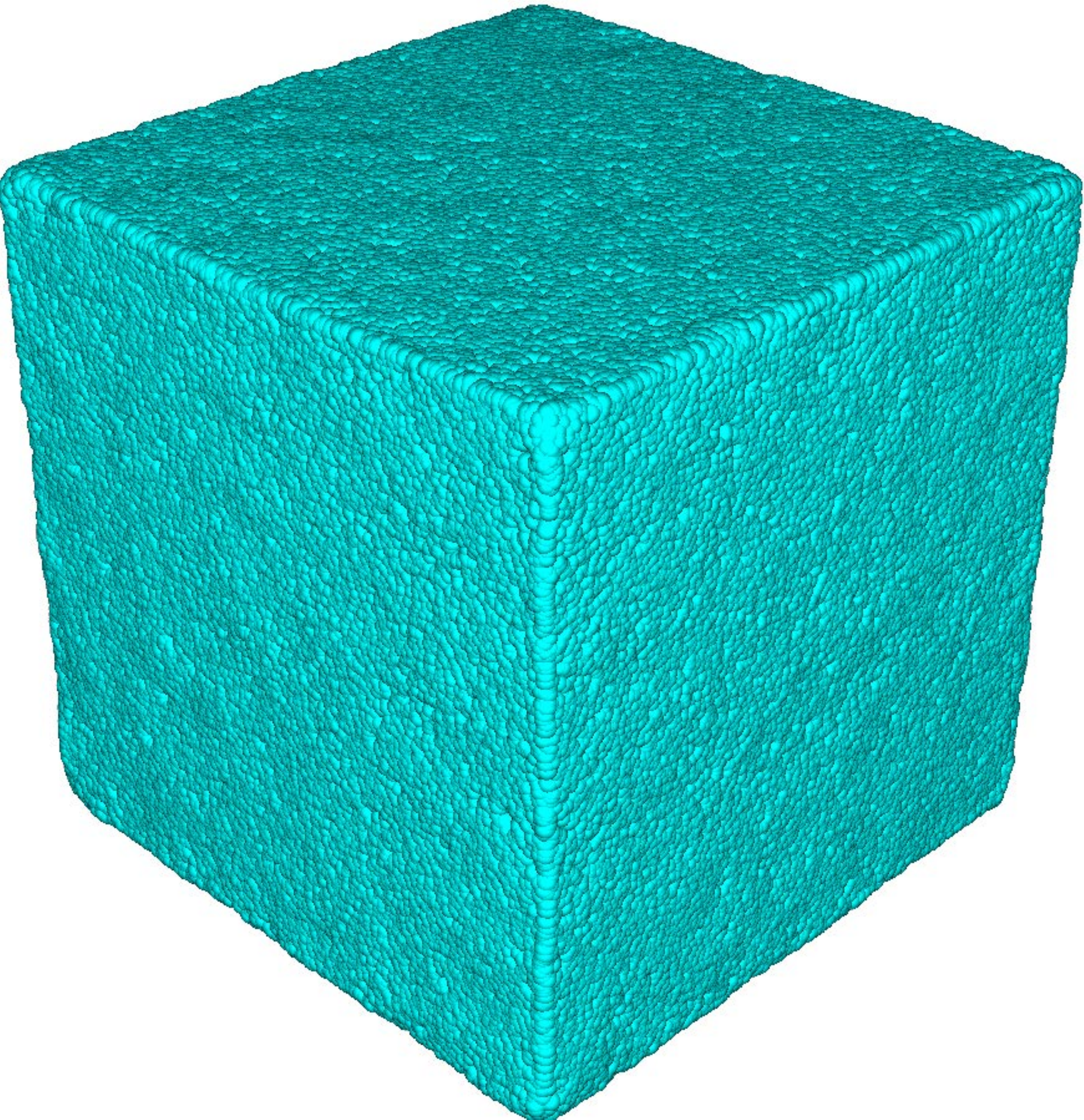}
  }
  \caption{Comparison of different loss functions ($L_{proj}^{a}$, $L_{2}$ and $L_{proj}^{b}$). $L_{proj}^{b}$ can better preserve sharp features than other loss functions since normal information is considered. Note that $L_{rep}$ is respectively added to three loss functions for fair comparisons.}
  \label{fig:AblationLoss}
\end{figure}

\textbf{Loss function.} To enable the filtered point cloud approximating the underlying surface while preserving sharp features, the loss function should be elaborately defined. A simple option for measuring the filtered point cloud would be the $L_2$ distance, which has been used in \cite{Rakotosaona2019CGF}. As shown in Fig. \ref{fig:LossAlternative}, compared with the $L_2$-based distance (4 (b)) which is sampling dependent, a more general alternative is to project noisy points onto the underlying surface (4 (c), (d)). Moreover, the $L_2$-based distance does not specifically consider feature information, since it simply finds the closest points on the ground truth, regardless of feature points or non-feature points. It leads to less sharp feature results (see Fig. \ref{fig:AblationLoss} (c)). Thus, the loss should be capable of measuring the projection distance. Inspired by \cite{Kolluri2008TOA}, our projection loss is defined as
\begin{equation}\label{eq:IMLS}
\begin{aligned}
L_{proj}^{a} = \dfrac{\sum_{\mathbf{p}_{j} \in \mathbf{\mathcal{P}}_{i}}|(\bar{\mathbf{p}}_{i} - \mathbf{p}_{j}) \cdot \mathbf{n}_{p_{j}}^{T}| \cdot \phi(\|\bar{\mathbf{p}}_{i} - \mathbf{p}_{j}\|)}{\sum_{\mathbf{p}_{j} \in \mathbf{\mathcal{P}}_{i}} \phi(\|\bar{\mathbf{p}}_{i} - \mathbf{p}_{j}\|)},
\end{aligned}
\end{equation}
where $\bar{\mathbf{p}}_{i}$ is the filtered point of the noisy point $\hat{\mathbf{p}}_{i}$, and $\mathbf{n}_{p_{j}}$ is the ground-truth normal of the point $\mathbf{p}_j$, and $\phi(\|\bar{\mathbf{p}}_{i} - \mathbf{p}_{j}\|)$ is a Gaussian function giving larger weights to the points near $\bar{\mathbf{p}}_{i}$, defined as
\begin{equation}\label{eq:SpatialGaussianWeight}
\begin{aligned}
\phi(\|\bar{\mathbf{p}}_{i} - \mathbf{p}_{j}\|) = \exp \left(-\dfrac{\|\bar{\mathbf{p}}_{i} - \mathbf{p}_{j}\|^2}{{\sigma_{p}}^2} \right).
\end{aligned}
\end{equation}
The kernel size $\sigma_{p}$ is defined as $\sigma_{p} = 4\sqrt{diag/m}$, where $diag$ is the length of the diagonal of the bounding box of patch $\mathbf{\mathcal{P}}_{i}$ and $m = |\hat{\mathbf{\mathcal{P}}}_{i}|$ \cite{Huang2009TOG}. Besides approximating the underlying surface, we also expect the filtered points are distributed uniformly. To achieve this goal, we adopt a repulsion term that penalizes point aggregation. Overall, we formulate the loss function as
\begin{equation}\label{eq:Initial_Loss}
\begin{aligned}
L = \eta L_{proj}^{a} + (1 - \eta)\,L_{rep}, \, \, \, \, \, L_{rep} = \max_{\mathbf{p}_{j} \in \mathbf{\mathcal{P}}_{i}}|\bar{\mathbf{p}}_{i} - \mathbf{p}_{j}|,
\end{aligned}
\end{equation}
where $\eta$ is a trade-off parameter to control the repulsion force in the filtering process, and we empirically set $\eta = 0.97$ in our training stage. 

We found that the projection loss $L_{proj}^{a}$  tends to blur sharp features (see Fig. \ref{fig:AblationLoss} (b)). We address this issue by considering normal similarity in our loss function, in which we introduce a bilateral mechanism to construct the projection distance formula (Eq. \eqref{eq:IMLS}). Specifically, the function is defined as the normal similarity between the current point and its neighboring points in the patch. For simplicity, the function is $\theta (\mathbf{n}_{\bar{p_{i}}}, \mathbf{n}_{p_j}) = \exp\left(-\dfrac{1 - \mathbf{n}_{\bar{p_{i}}}^{T}\mathbf{n}_{p_j}}{1 - \cos(\sigma_{n})}\right)$ \cite{Huang2013TOG}, where $\mathbf{n}_{\bar{p}_{i}}$ is the normal of the filtered point $\bar{\mathbf{p}}_{i}$ and $\sigma_{n}$ is the support angle (default to $15^{\circ}$). 
Therefore, our final projection function is defined as
\begin{equation}\label{eq:lossfunction}
\begin{aligned}
L_{proj}^{b} =\dfrac{\sum_{\mathbf{p}_{j} \in \mathbf{\mathcal{P}}_{i}}|(\bar{\mathbf{p}}_{i} - \mathbf{p}_{j}) \cdot \mathbf{n}_{p_{j}}^{T}| \cdot \phi(\|\bar{\mathbf{p}}_{i} - \mathbf{p}_{j}\|) \theta (\mathbf{n}_{\bar{p_{i}}}, \mathbf{n}_{p_j})}{\sum_{\mathbf{p}_{j} \in \mathbf{\mathcal{P}}_{i}} \phi(\|\bar{\mathbf{p}}_{i} - \mathbf{p}_{j}\|) \theta (\mathbf{n}_{\bar{p_{i}}}, \mathbf{n}_{p_j})}.
\end{aligned}
\end{equation}
For efficiency and simplicity, the normal of the filtered point $\mathbf{n}_{\bar{p_{i}}}$ is assigned by the normal of the ground truth point which is nearest to the filtered point. 
\textit{It is worth noting that Pointfilter only requires ground-truth point normals in the training stage. }

We chose the encoder-decoder structure because: (1) it is stable and mature; (2) it can learn complex and compact representations of a point cloud; (3) the learned latent representations are helpful to regress the displacement vector according to the input noisy patch. 

\subsection{Training}\label{sec:networktraining}
We implementing Pointfilter in PyTorch on a desktop PC with an Intel Core I7-8750H CPU (2.20 GHz, 16GB memory) and a GeForce GTX 1060 GPU (6GB memory, CUDA 9.0). We set $50$ training epochs and a mini-batch size of $64$. We adopted SGD as our optimizer and decreased the learning rate from 1e-4 to 1e-8 with increasing epochs. We also adopted batch-normalization \cite{Ioffe2015ArXiv}, ReLU \cite{Nair2010ICML} and $tanh$ in Pointfilter.

\subsection{Network Inference}\label{sec:networkinference}
Given a trained Pointfilter, our approach filters a noisy point cloud in a point-wise way. Firstly, we build a patch  for each noisy point and transform it to a canonical space as described in Section \ref{sec:preprocessing}. Secondly, we feed each aligned patch into Pointfilter and obtain a displacement vector. Finally, we map the displacement vector back to the underlying space. The inference is formulated as follows:
\begin{equation}\label{eq:inferece}
\begin{aligned}
\bar{\mathbf{p}}_{i} = r \mathbf{R}^{-1}    f(\mathbf{R} (\hat{\mathbf{\mathcal{P}}_{i}} - \hat{\mathbf{p}_{i}}) / r) + \hat{\mathbf{p}}_{i},
\end{aligned}
\end{equation}
where $\bar{\mathbf{p}}_{i}$ and $\hat{\mathbf{p}}_{i}$ are the filtered point and noisy point, respectively. $f$ represents our Pointfilter. $\mathbf{R}\in SO(3)$ is the PCA-induced rotation matrix, and $r$ is the patch radius. To get better filtered results, we adopt multiple iterations of inference to progressively filter the noisy point cloud, especially for point clouds corrupted with larger noise. 

\begin{figure}[htp!]
\centering
\begin{tabular}{ll}
\rowname{Training Set} & \includegraphics[width=0.43\textwidth]{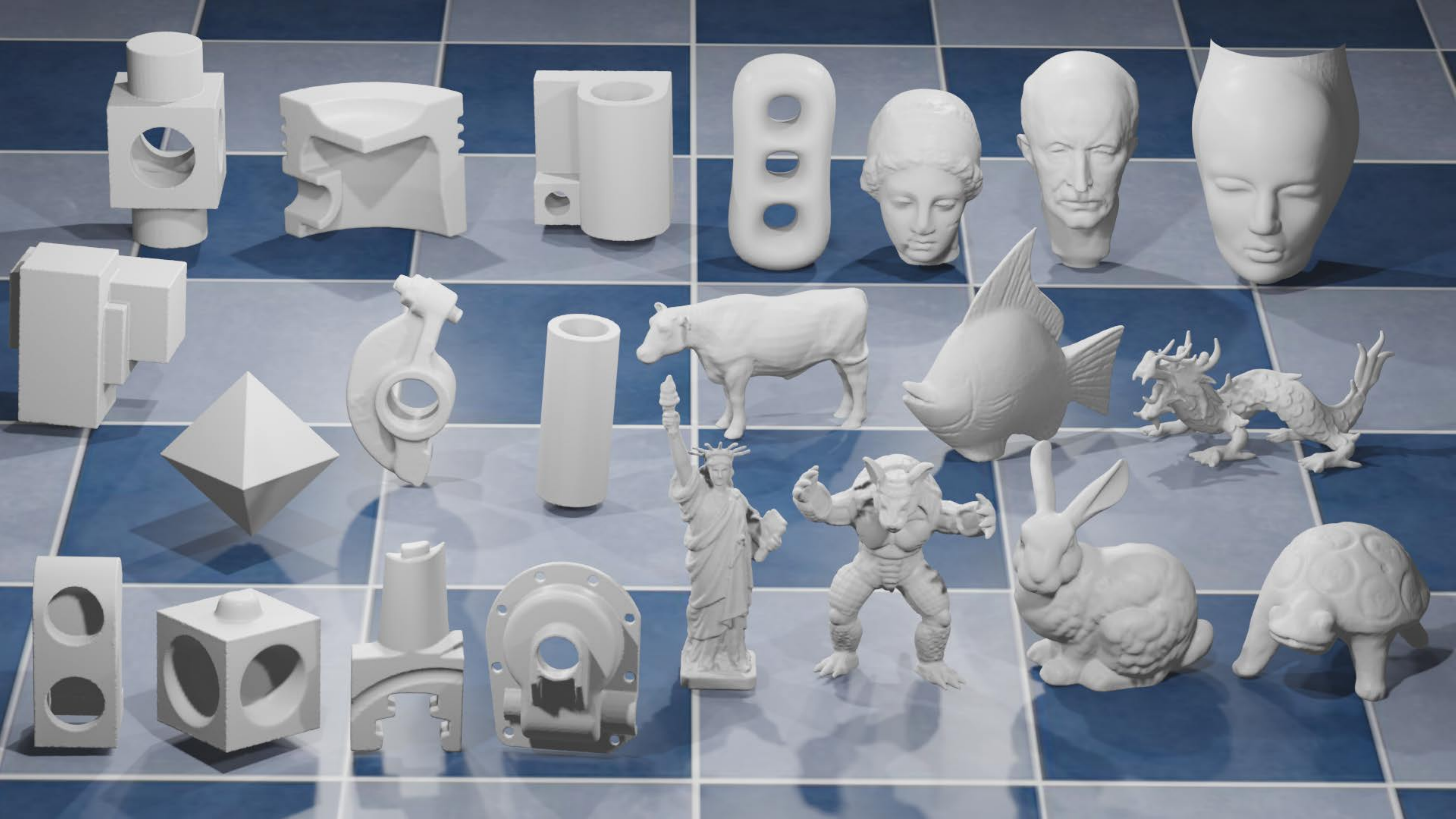}\\
\rowname{Test Set (Synthetic)} & \includegraphics[width=0.43\textwidth]{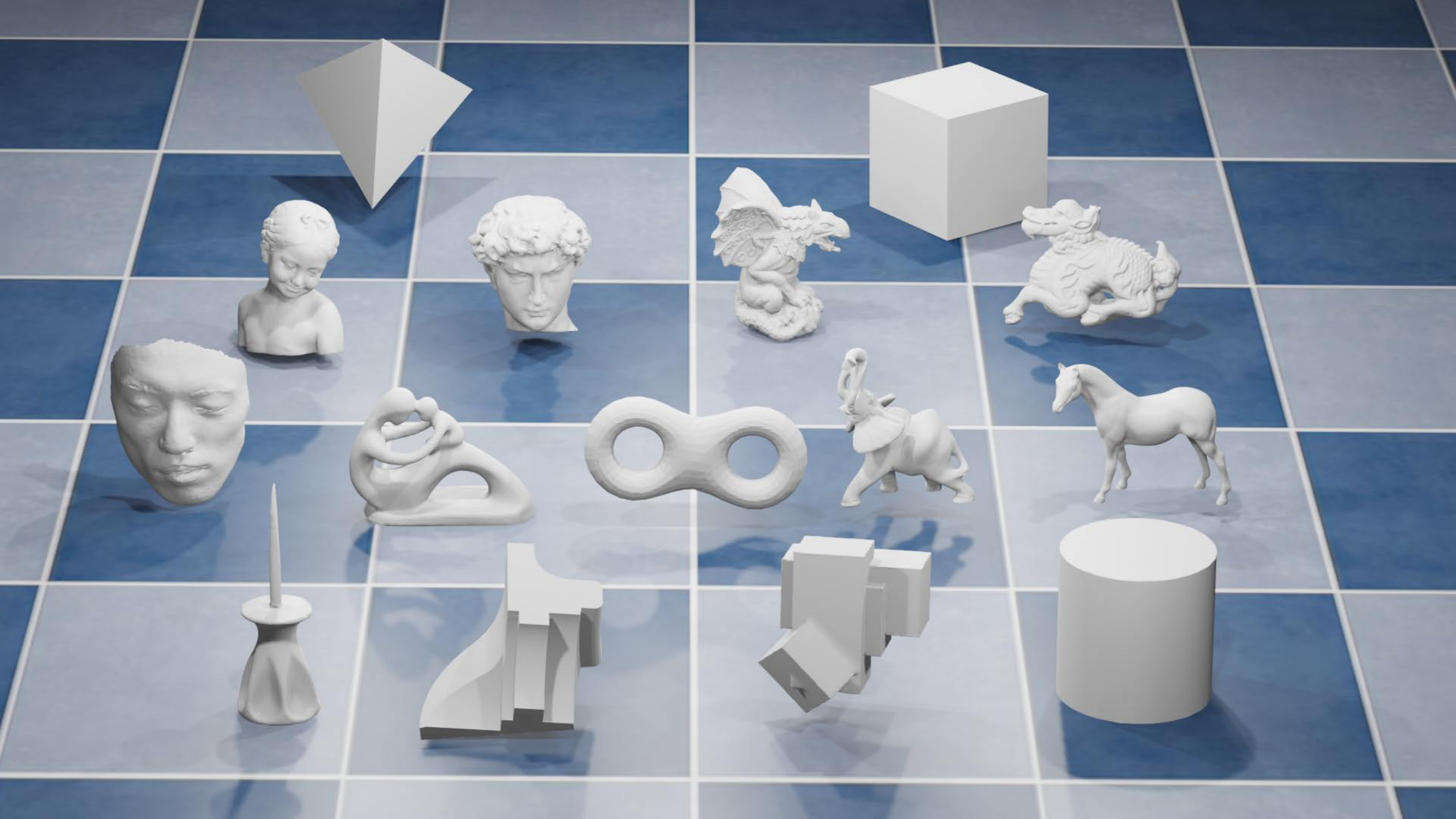}
\end{tabular}
\caption{Our dataset. Note that 7 scanned models are tested in Section \ref{sec:results}.}
\label{fig:dataset}
\end{figure}
\section{Experimental Results}\label{sec:results}
\subsection{Dataset}
As a supervised learning method, we prepare a training dataset consisting of $22$ 3D clean models ($11$ CAD models and $11$ non-CAD models) which are shown in Fig. \ref{fig:dataset}. Each model is generated by randomly sampling $100$k points from its original surface. Given a clean model, its corresponding noisy models are synthesized by adding Gaussian noise with the standard deviations of $0.0\%$, $0.25\%$, $0.5\%$, $1\%$, $1.5\%$ and $2.5\%$ of the clean model's bounding box diagonal length. In short, our training dataset contains $132$ ($22\times 6$) models. Specifically, for each epoch in the training stage, we randomly sample $8,000$ patches from each model as training samples. \textit{Notice that these models are our final training dataset, and we do not augment any data on-the-fly during training}. Besides, the normal information for clean models are required for training, as indicated in Eq. \eqref{eq:lossfunction}.

To demonstrate the generalization of the proposed Pointfilter, our test dataset includes both synthesized noisy models and raw-scan models, which will be explained in the following experiments (Section \ref{sec:visualcomparison} and \ref{sec:quantitativecomparison}). For synthesized data, $15$ 3D clean models and their corresponding noisy models are synthesized by adding Gaussian noise with the standard deviations of  $0.5\%$, $1\%$, $1.5\%$ and $2.5\%$ of the clean model's bounding box diagonal length. Each clean model contains $100$k points randomly sampled from its original surface. In addition, we also test 7 raw scanned point clouds including both shapes and scenes.

\begin{figure*}[htp]
    \centering
    \subfigure[Noisy]
    {
        \begin{minipage}[b]{0.105\textwidth} 
        \includegraphics[width=1\textwidth]{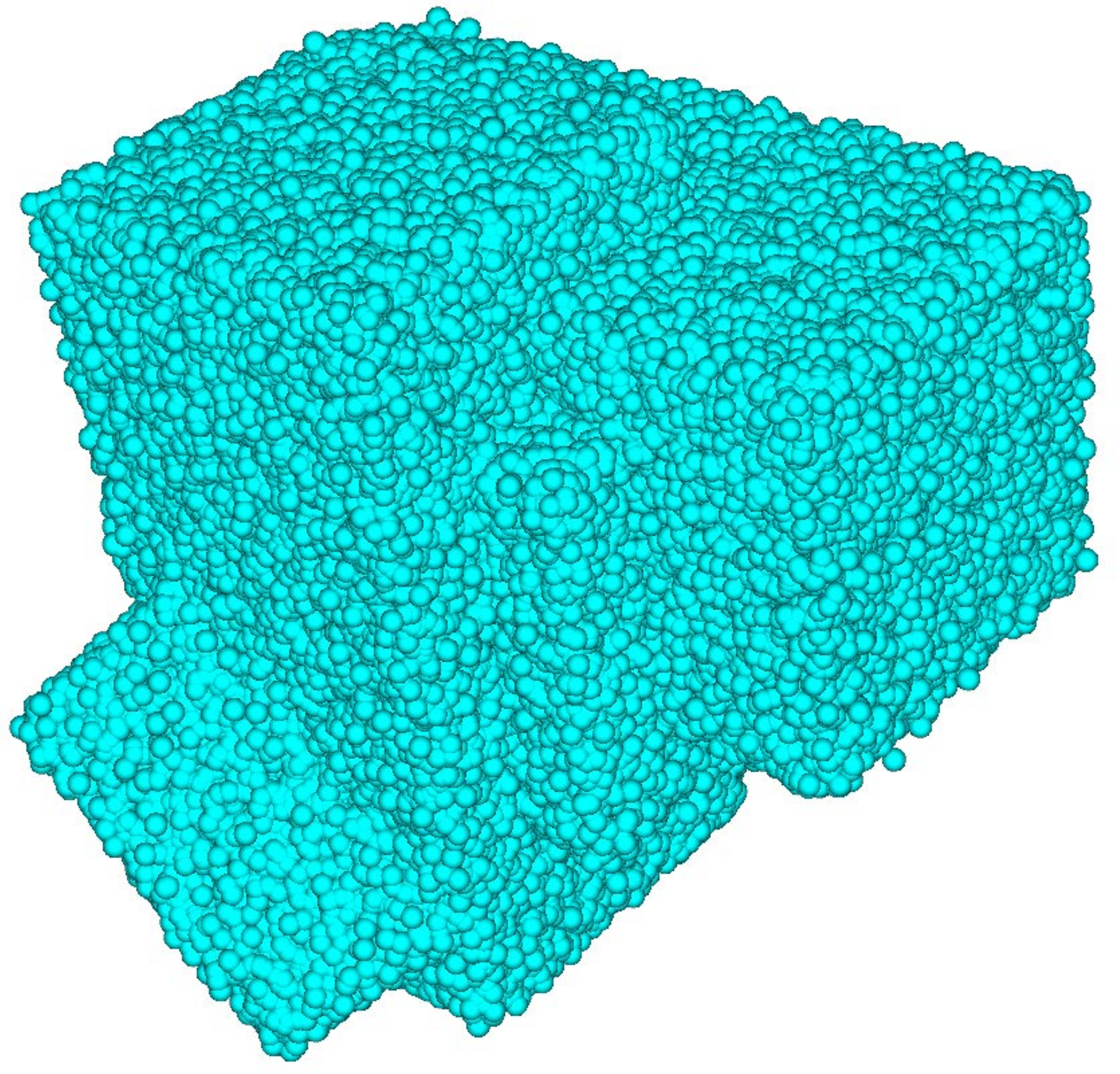}\\ 
        \includegraphics[width=1\textwidth]{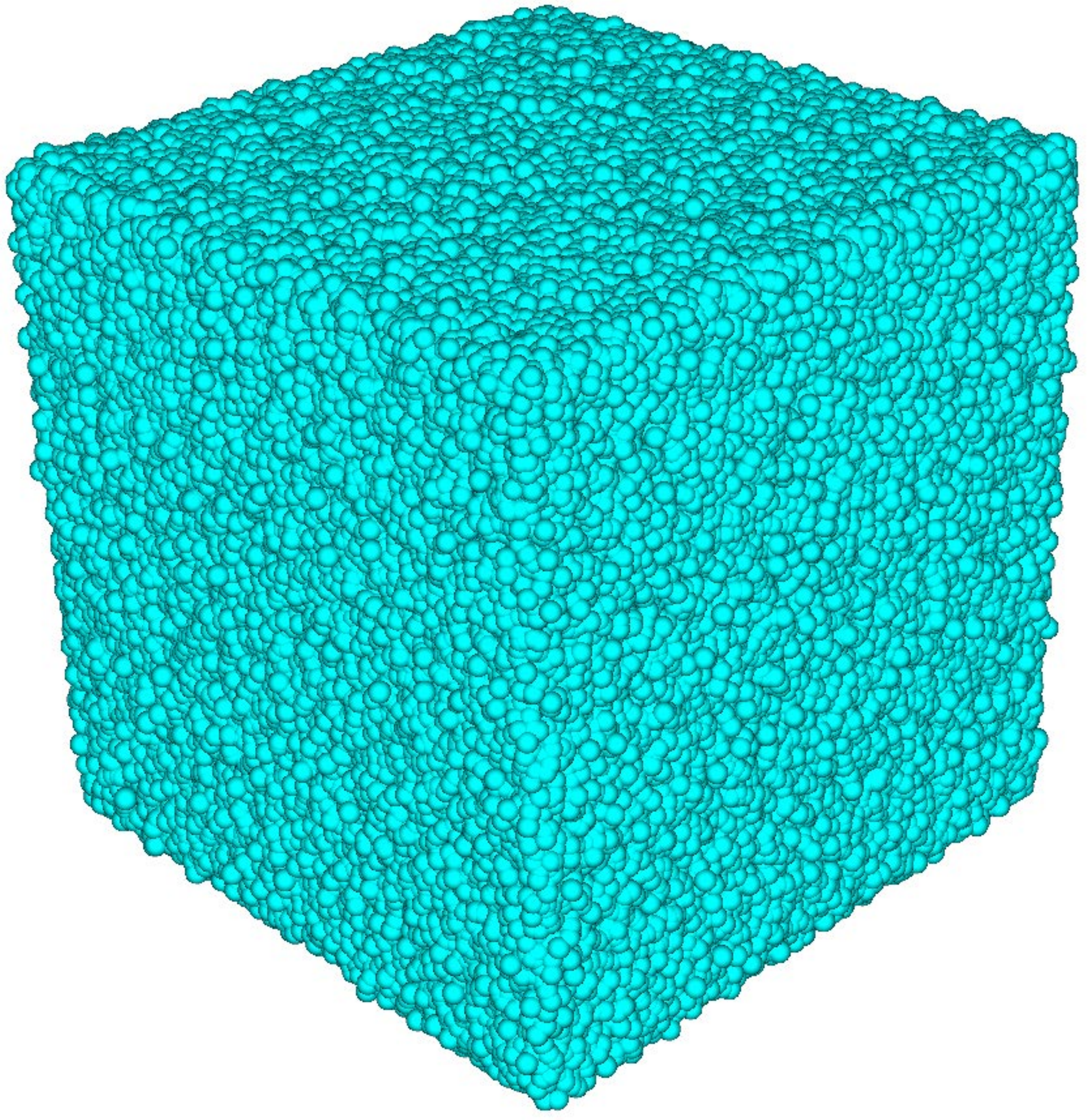}\\ 
        \includegraphics[width=1\textwidth]{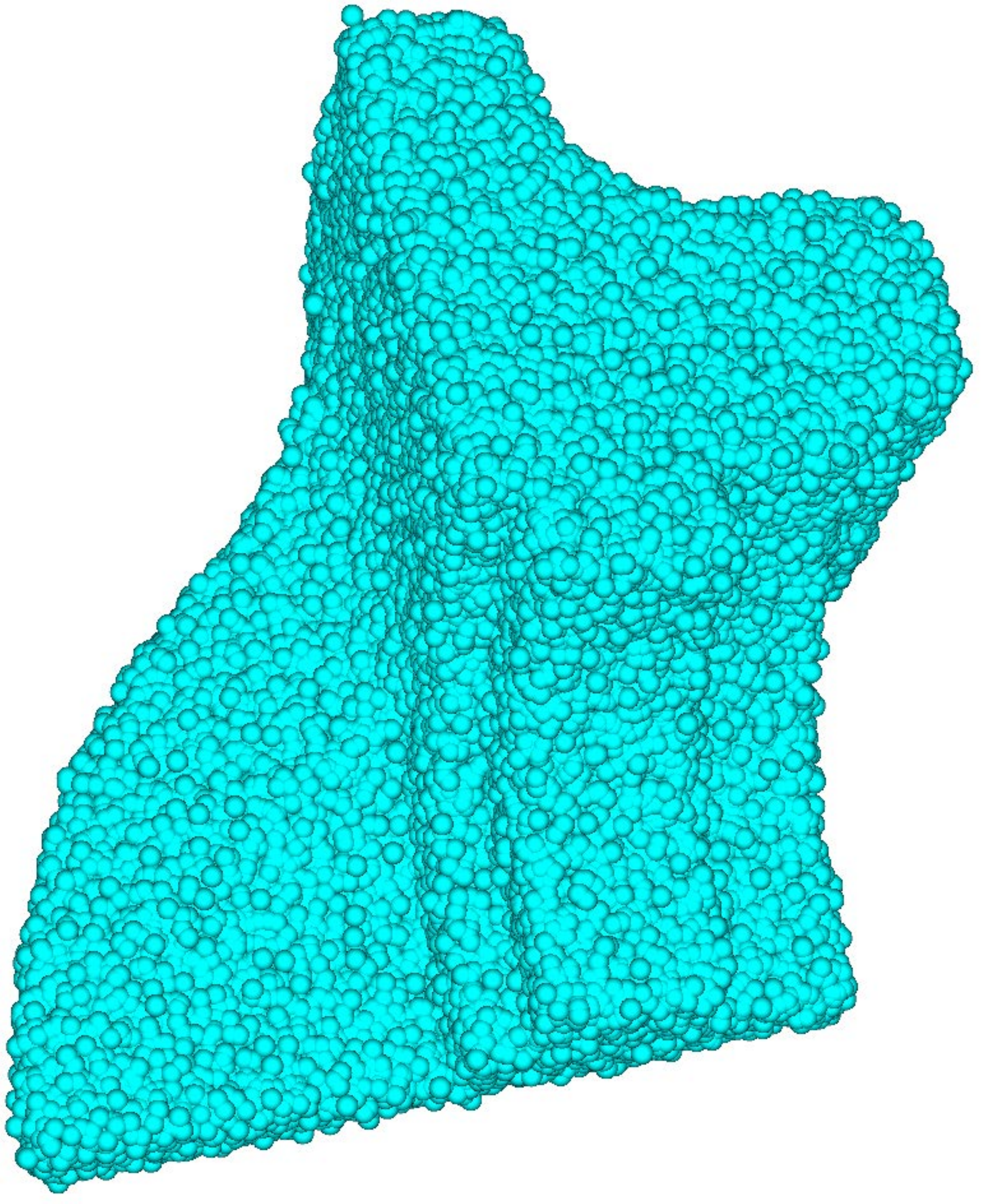}\\
        \includegraphics[width=1\textwidth]{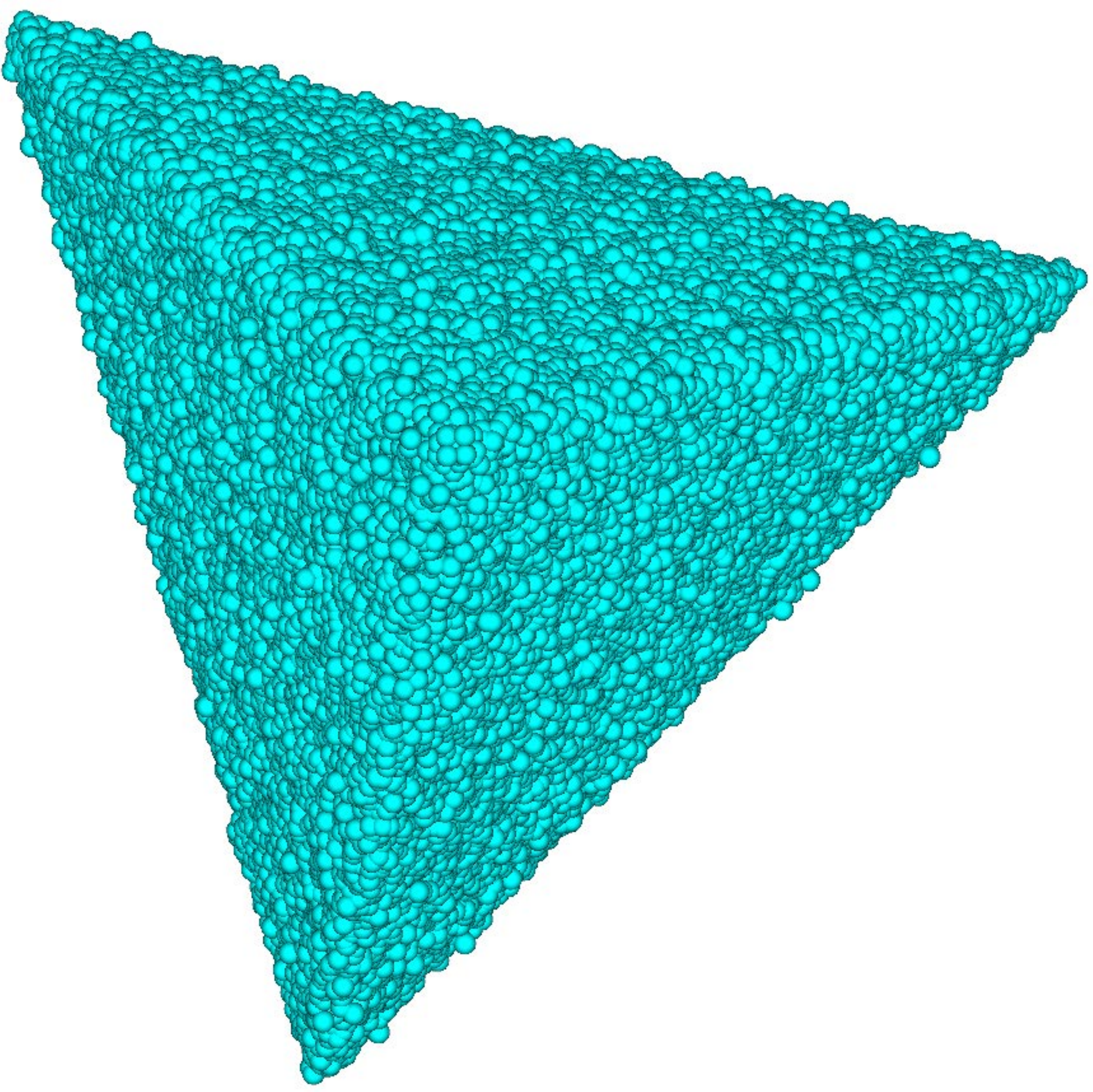}
        \end{minipage}
    }
    \subfigure[RIMLS]
    {
        \begin{minipage}[b]{0.105\textwidth} 
        \includegraphics[width=1\textwidth]{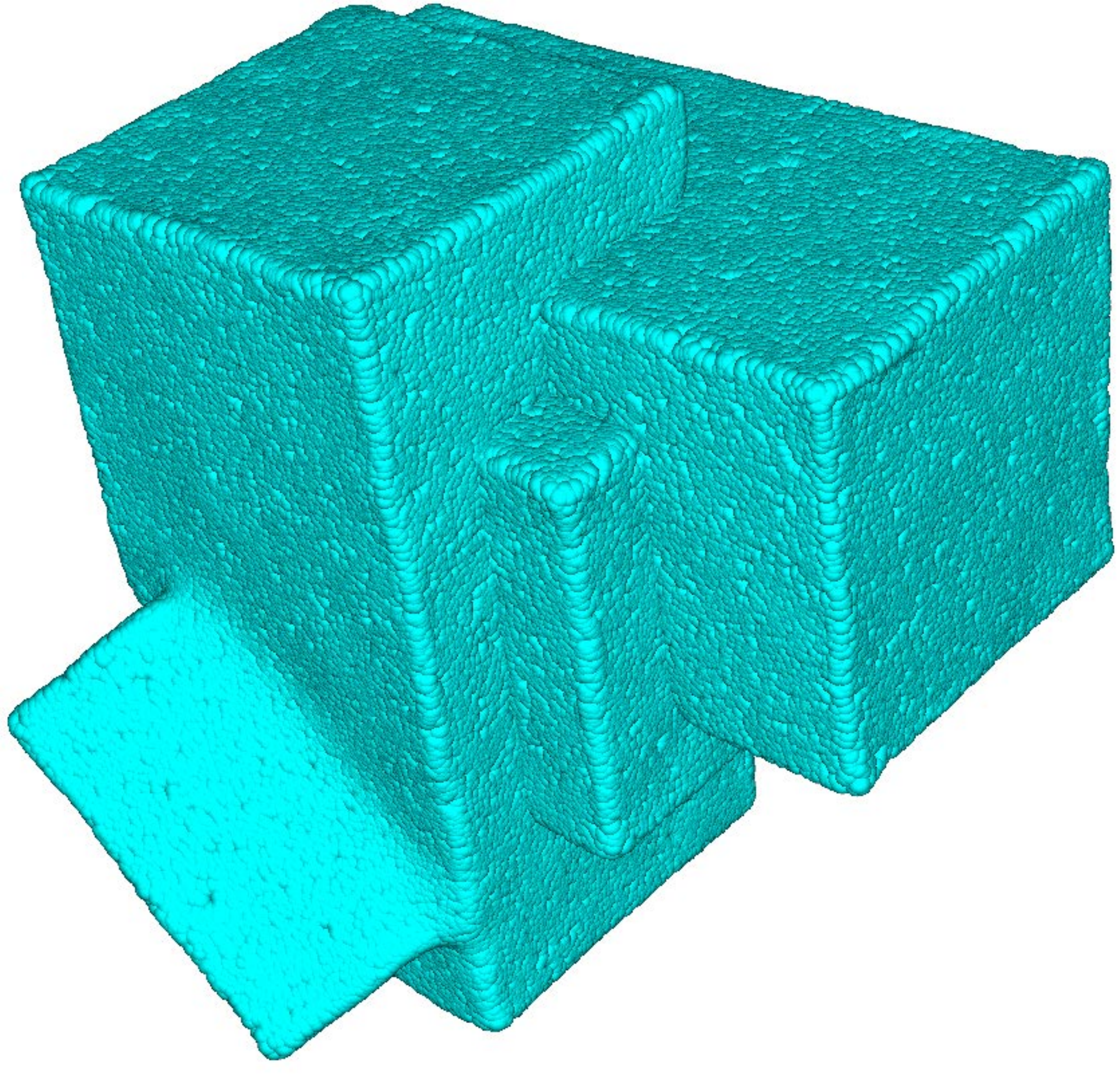}\\ 
        \includegraphics[width=1\textwidth]{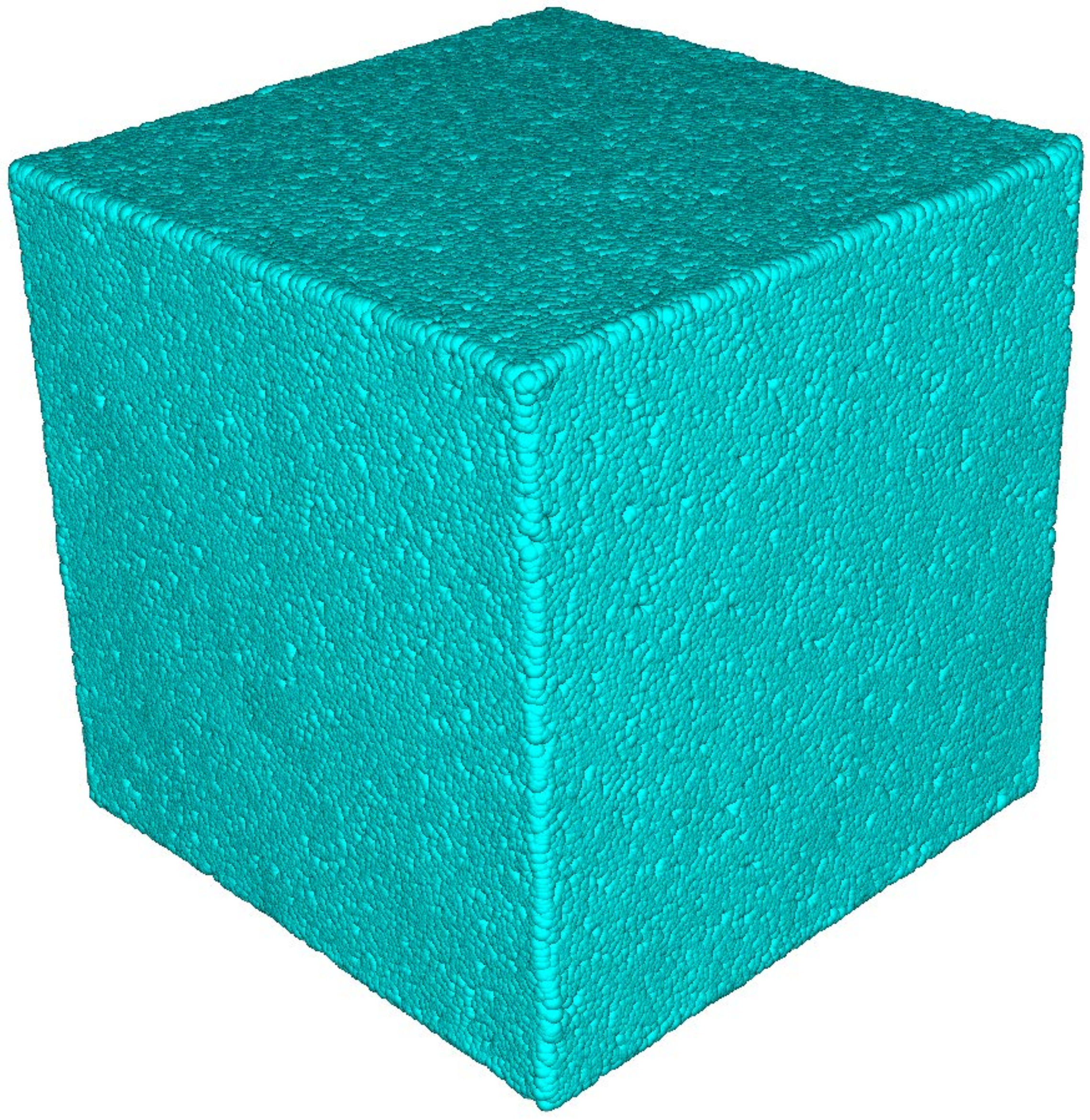}\\ 
        \includegraphics[width=1\textwidth]{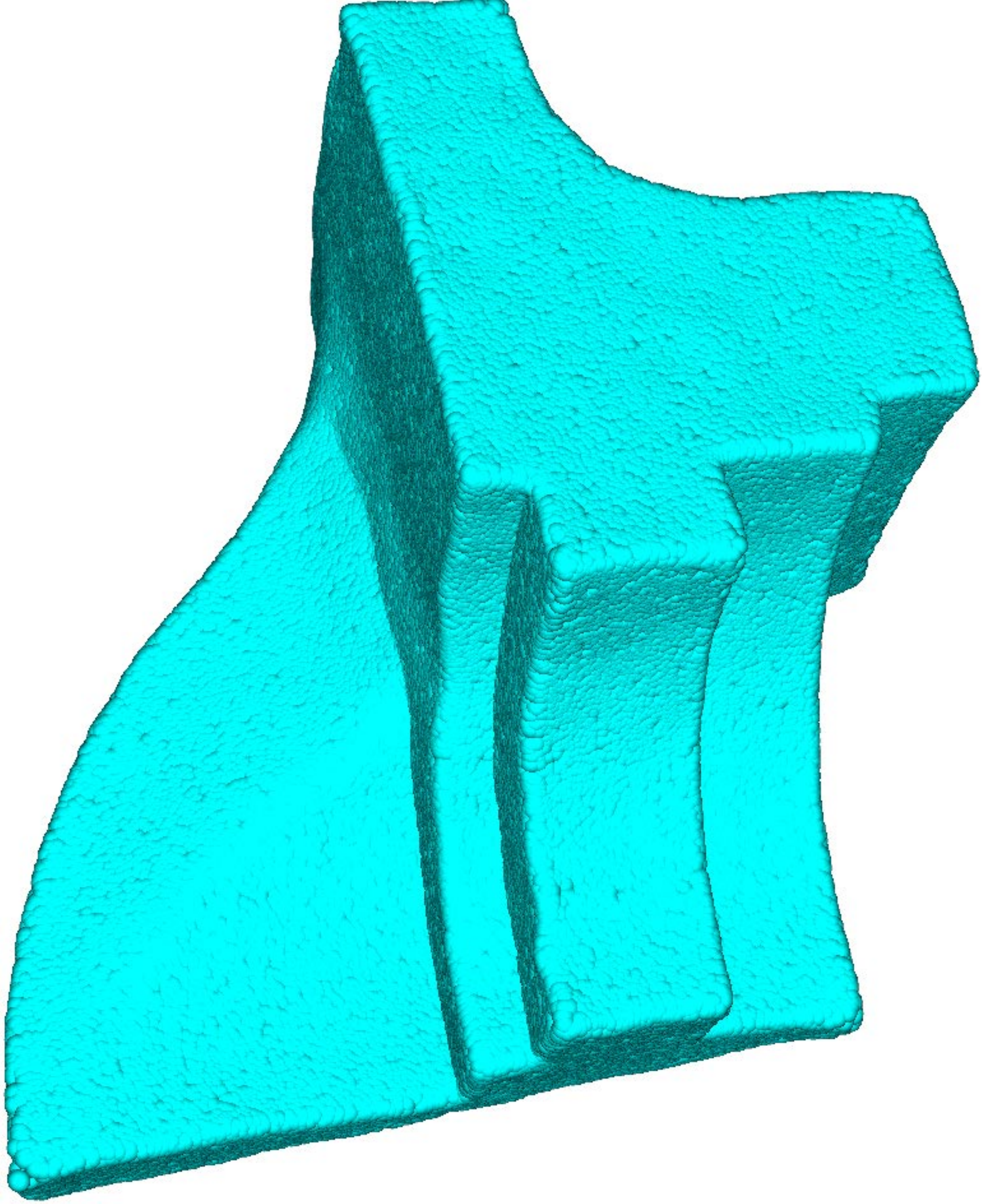}\\
        \includegraphics[width=1\textwidth]{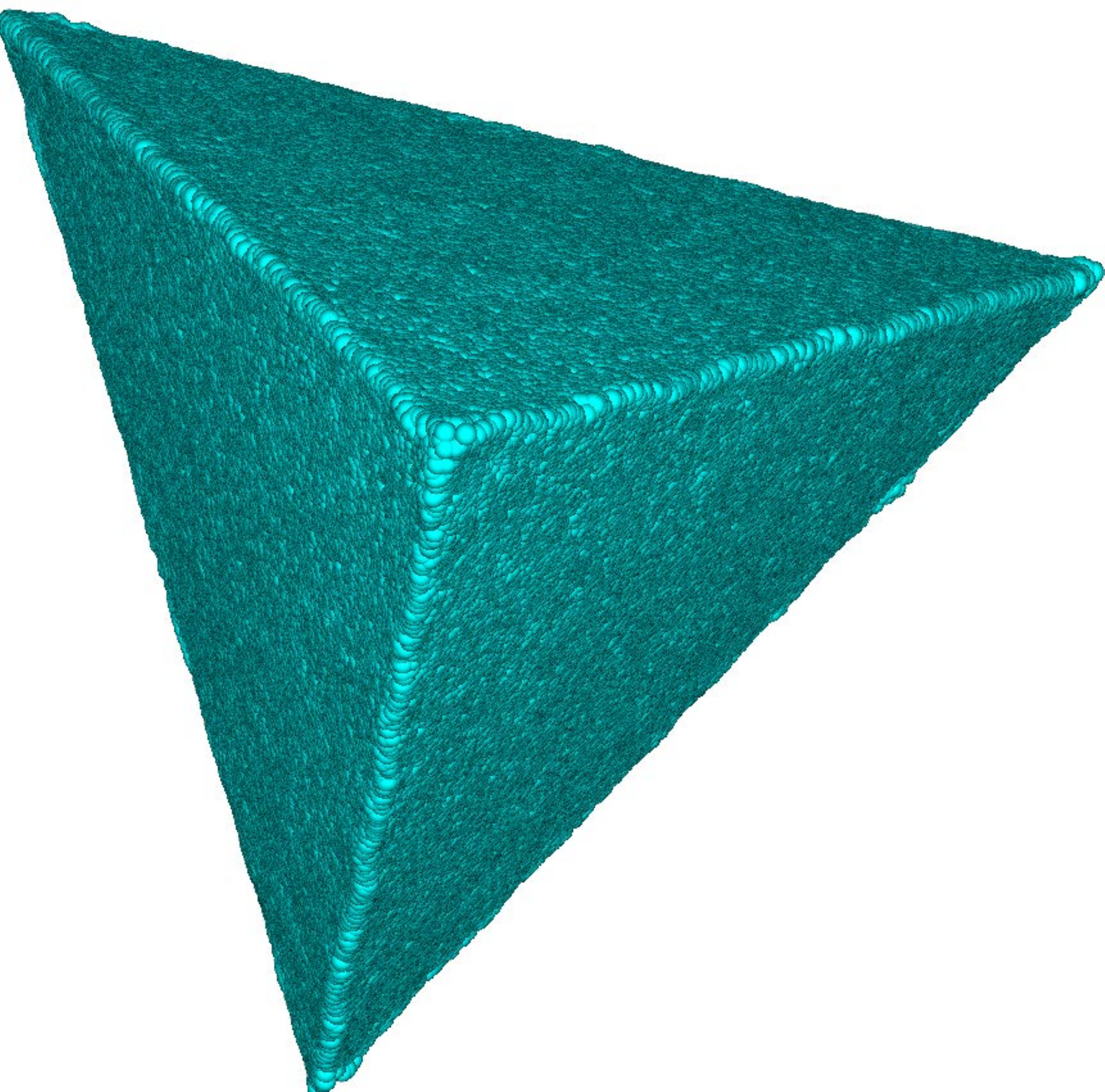}
        \end{minipage}
    }
    \subfigure[GPF]
    {
        \begin{minipage}[b]{0.105\textwidth} 
        \includegraphics[width=1\textwidth]{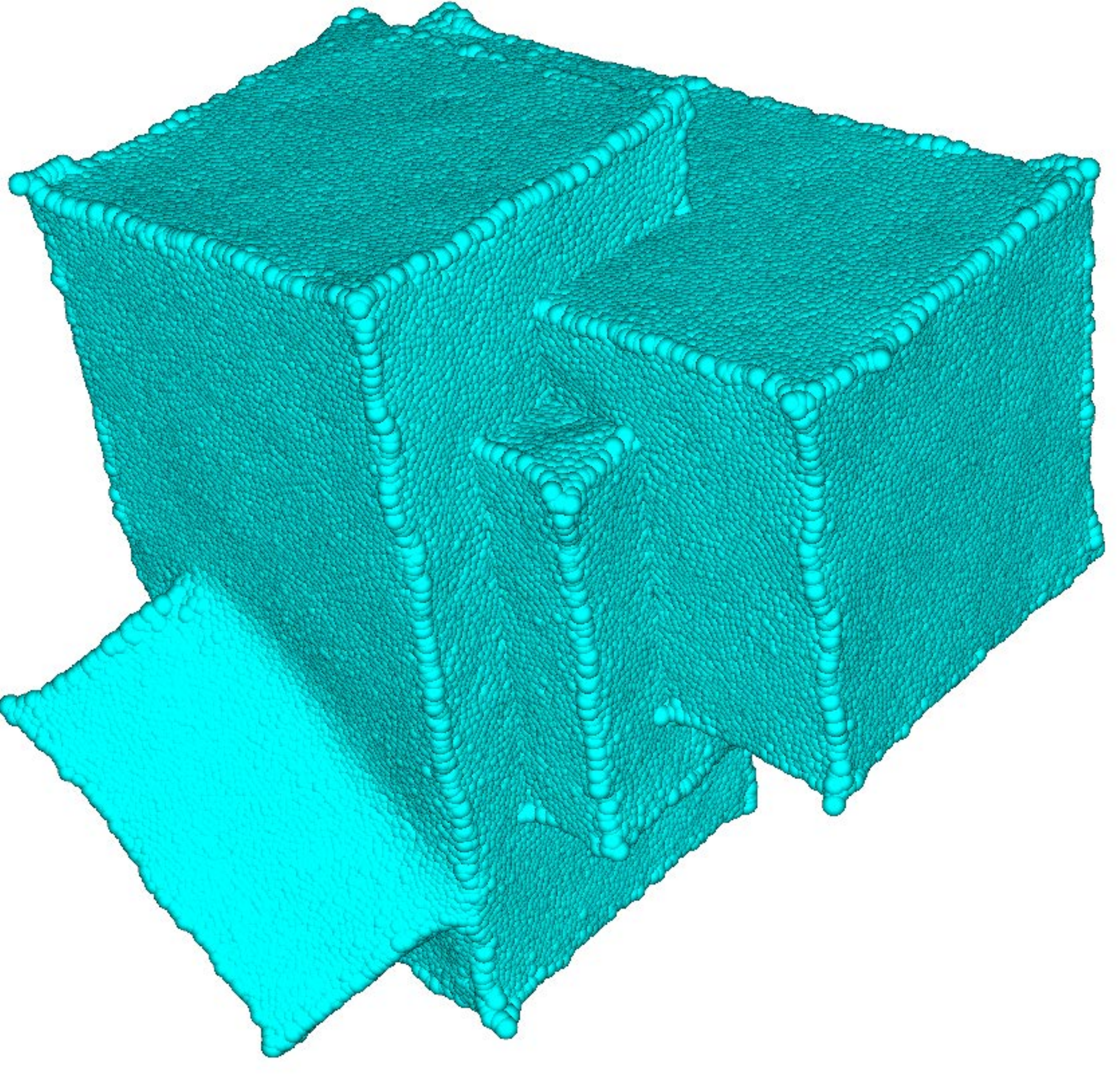}\\ 
        \includegraphics[width=1\textwidth]{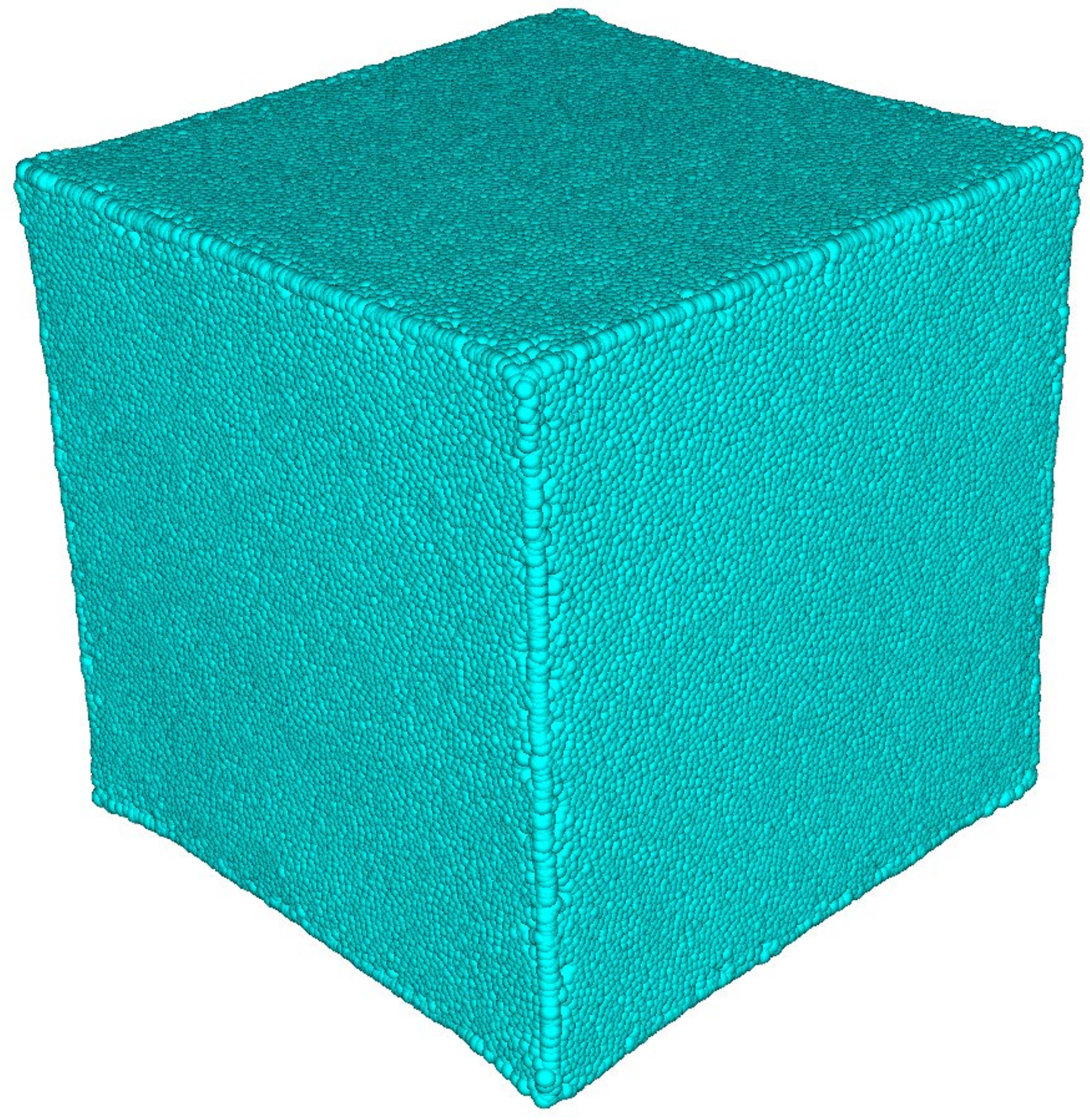}\\ 
        \includegraphics[width=1\textwidth]{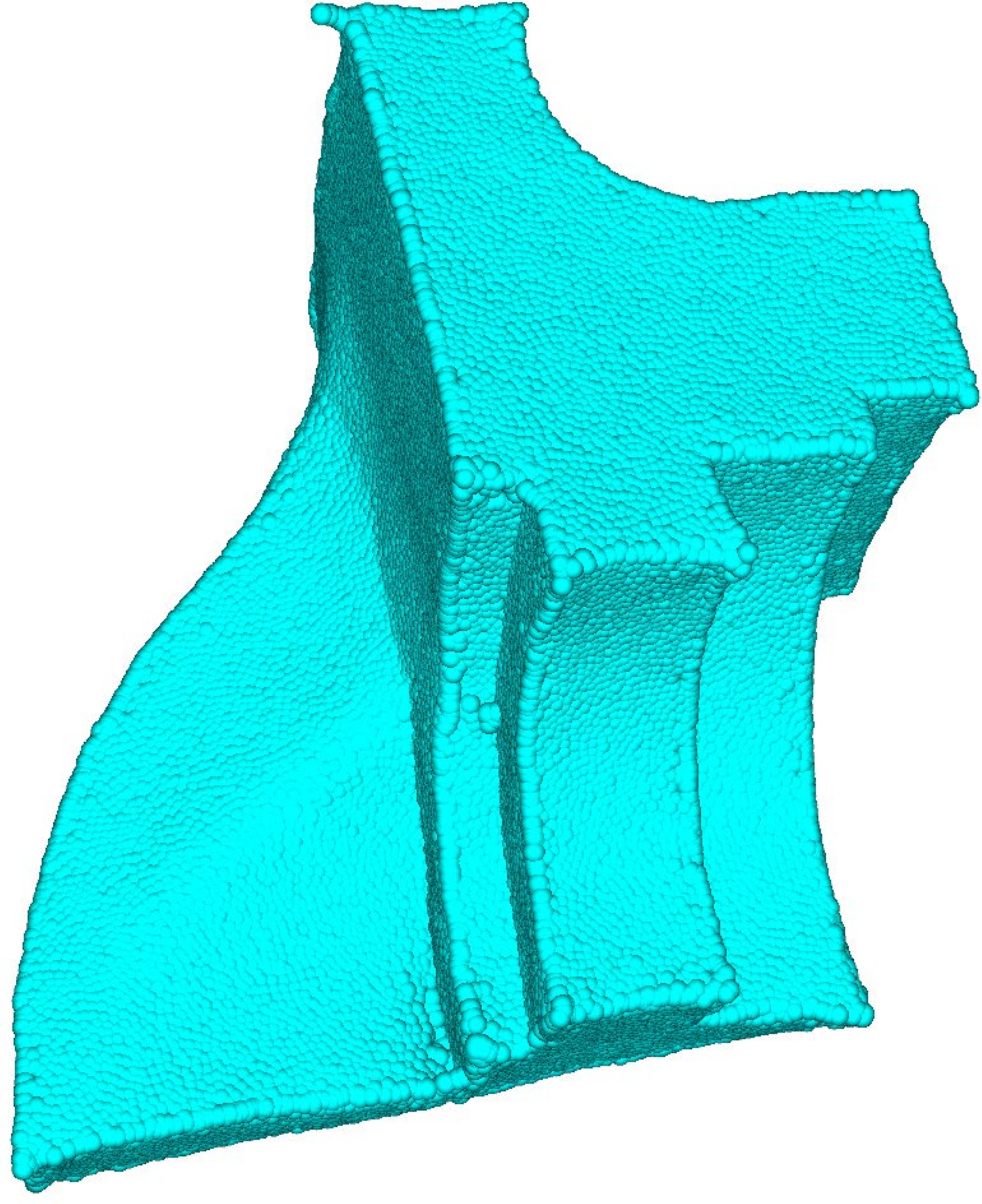}\\
        \includegraphics[width=1\textwidth]{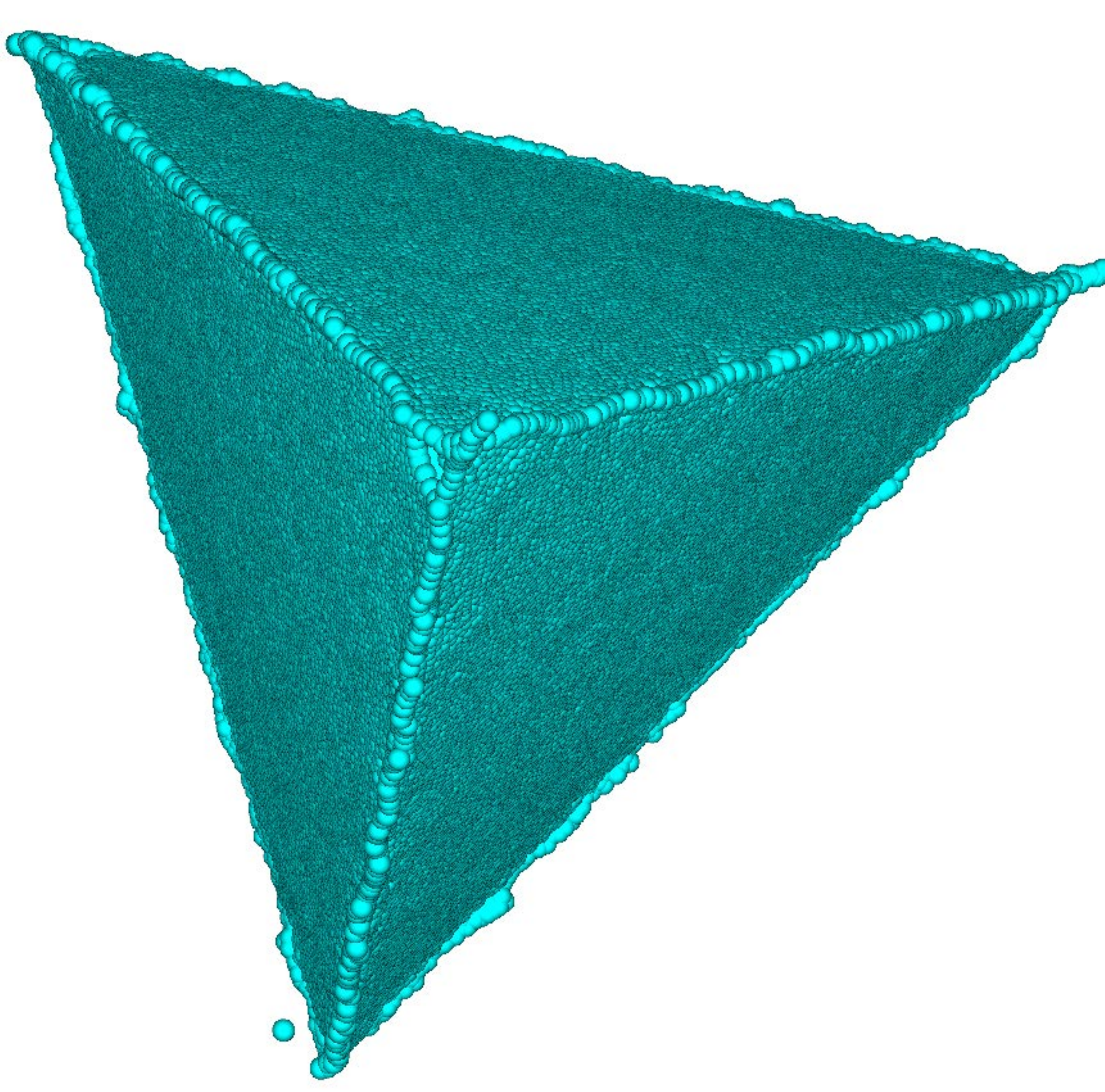}
        \end{minipage}
    }
    \subfigure[WLOP]
    {
        \begin{minipage}[b]{0.105\textwidth} 
        \includegraphics[width=1\textwidth]{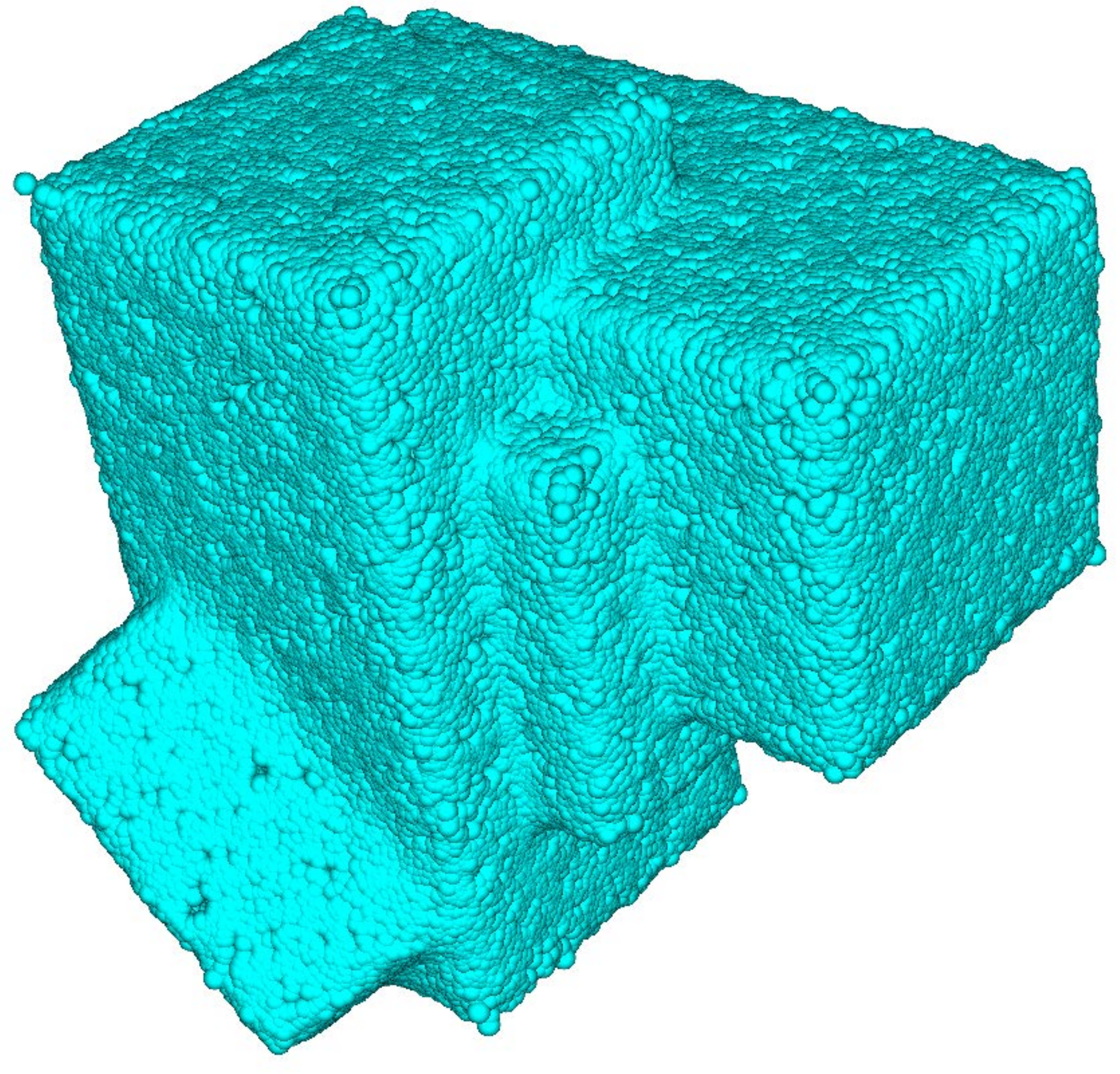}\\ 
        \includegraphics[width=1\textwidth]{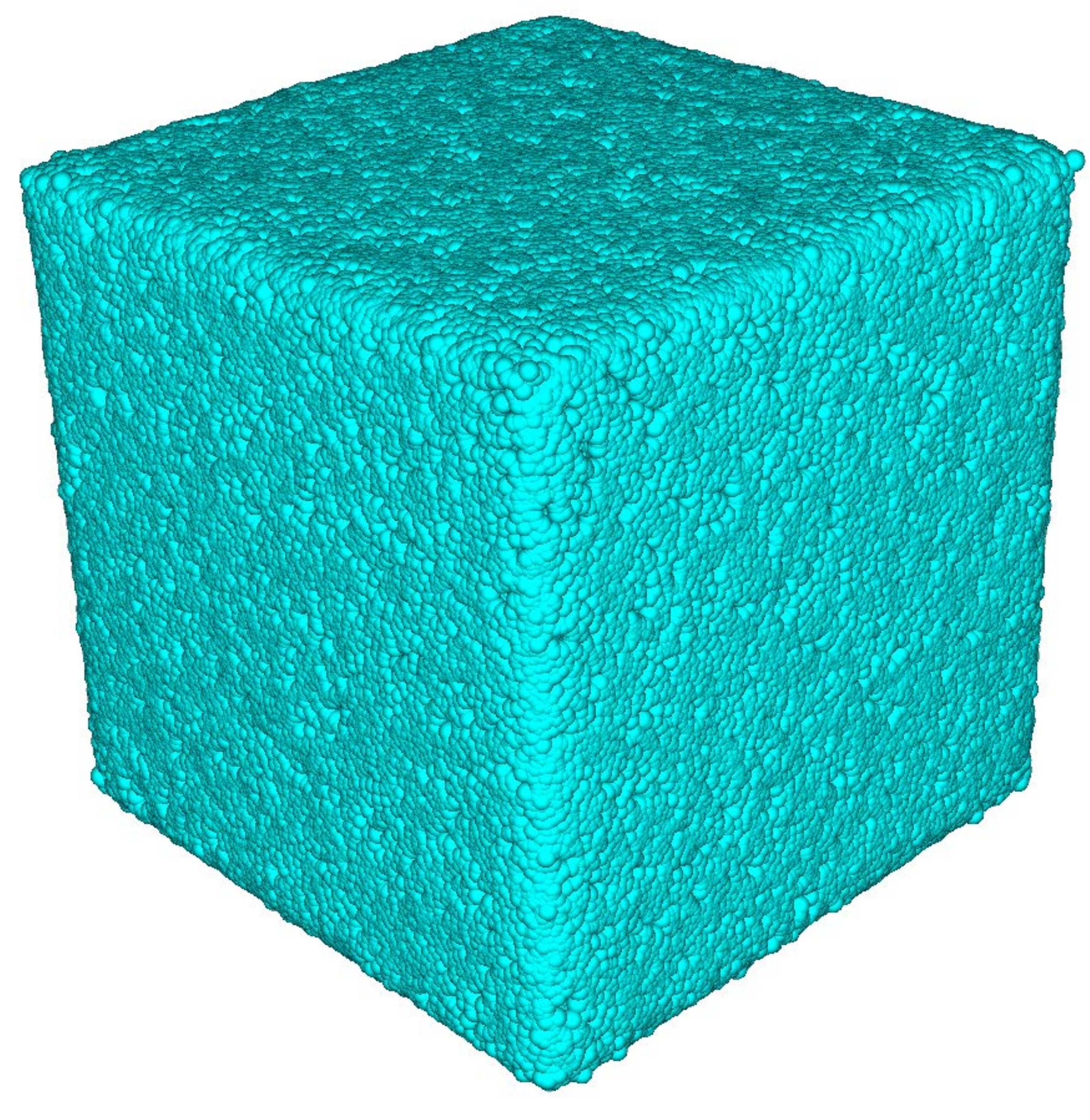}\\ 
        \includegraphics[width=1\textwidth]{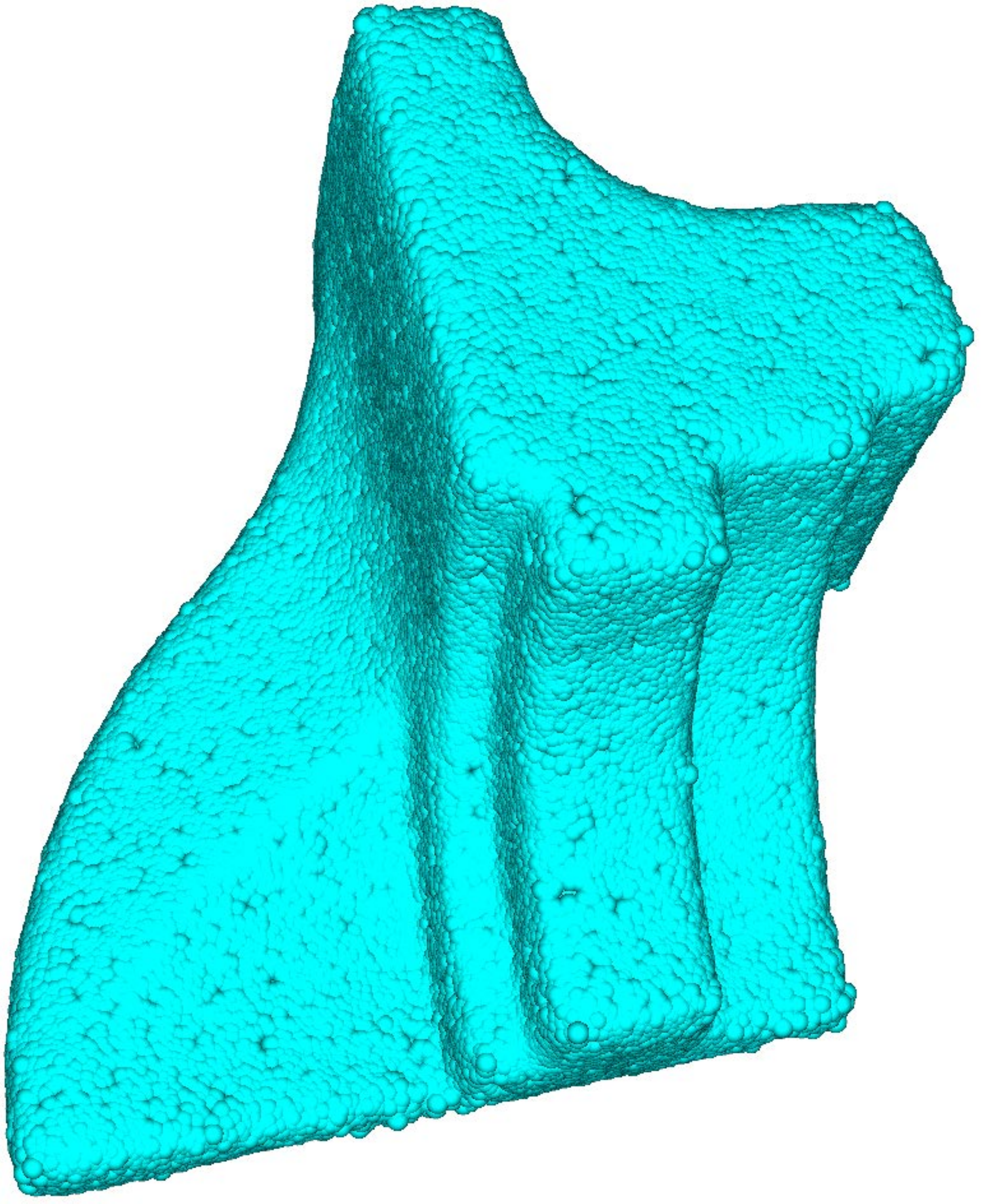}\\
        \includegraphics[width=1\textwidth]{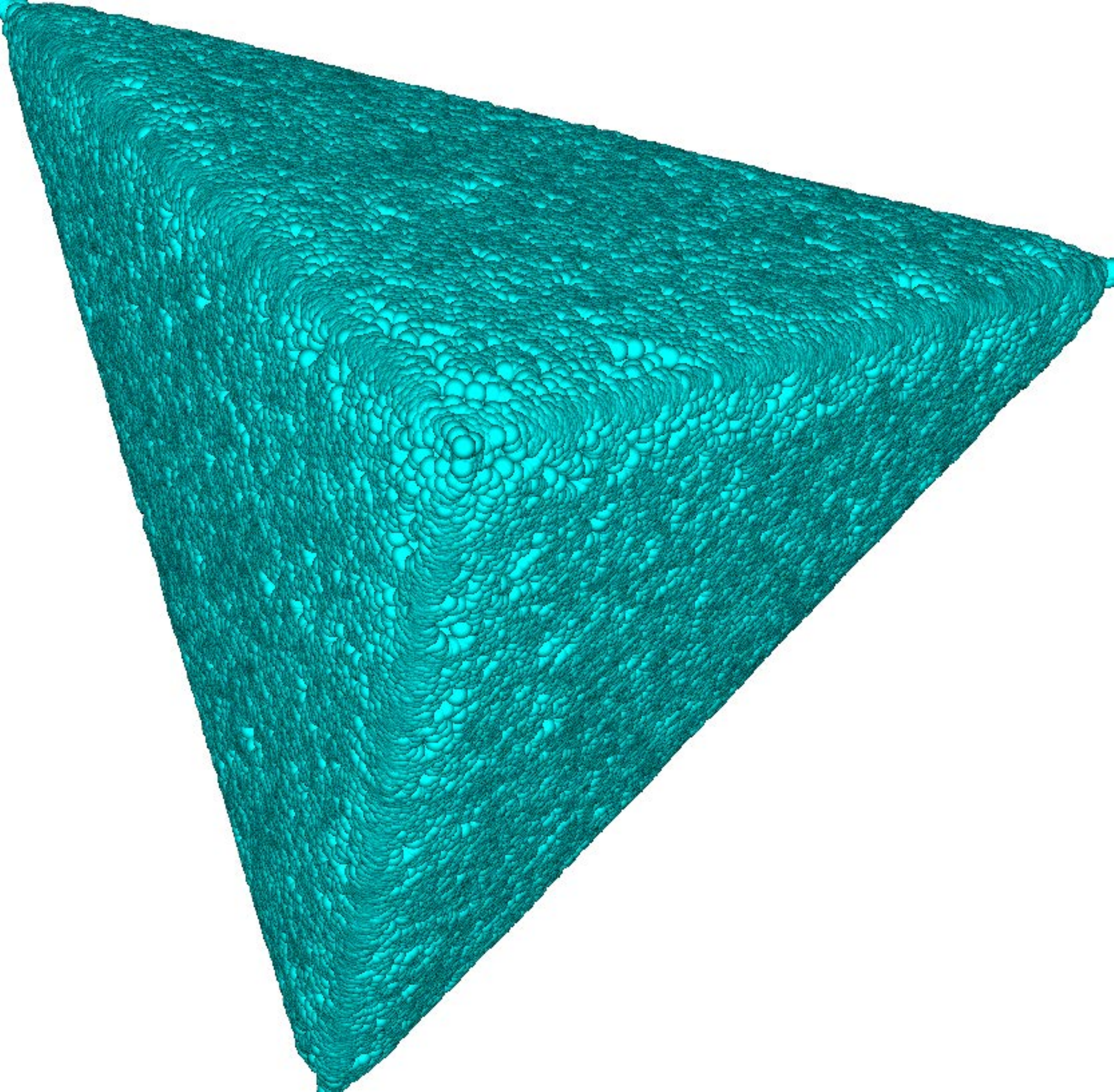}
        \end{minipage}
    }
    \subfigure[CLOP]
    {
        \begin{minipage}[b]{0.105\textwidth} 
        \includegraphics[width=1\textwidth]{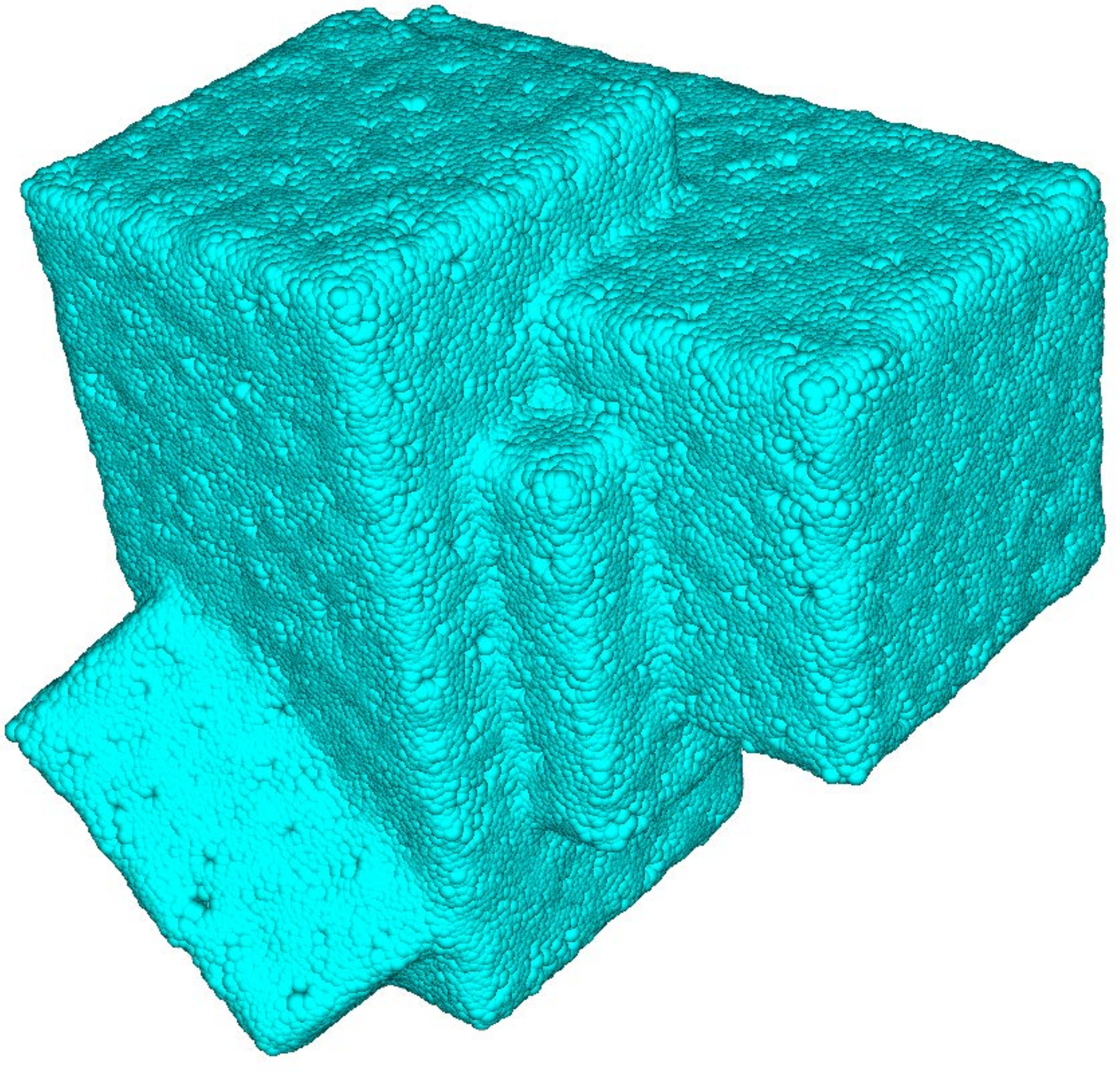}\\ 
        \includegraphics[width=1\textwidth]{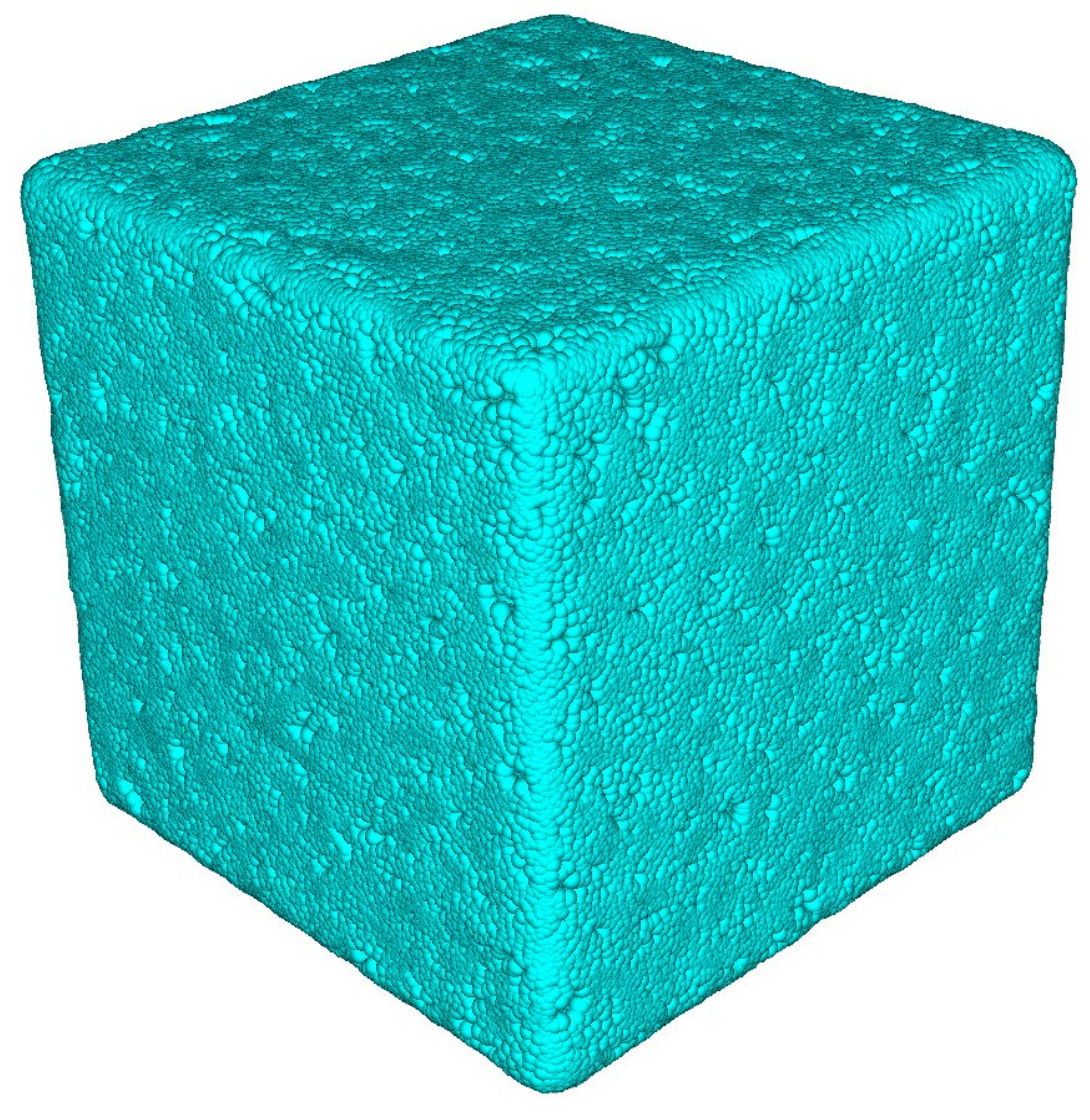}\\ 
        \includegraphics[width=1\textwidth]{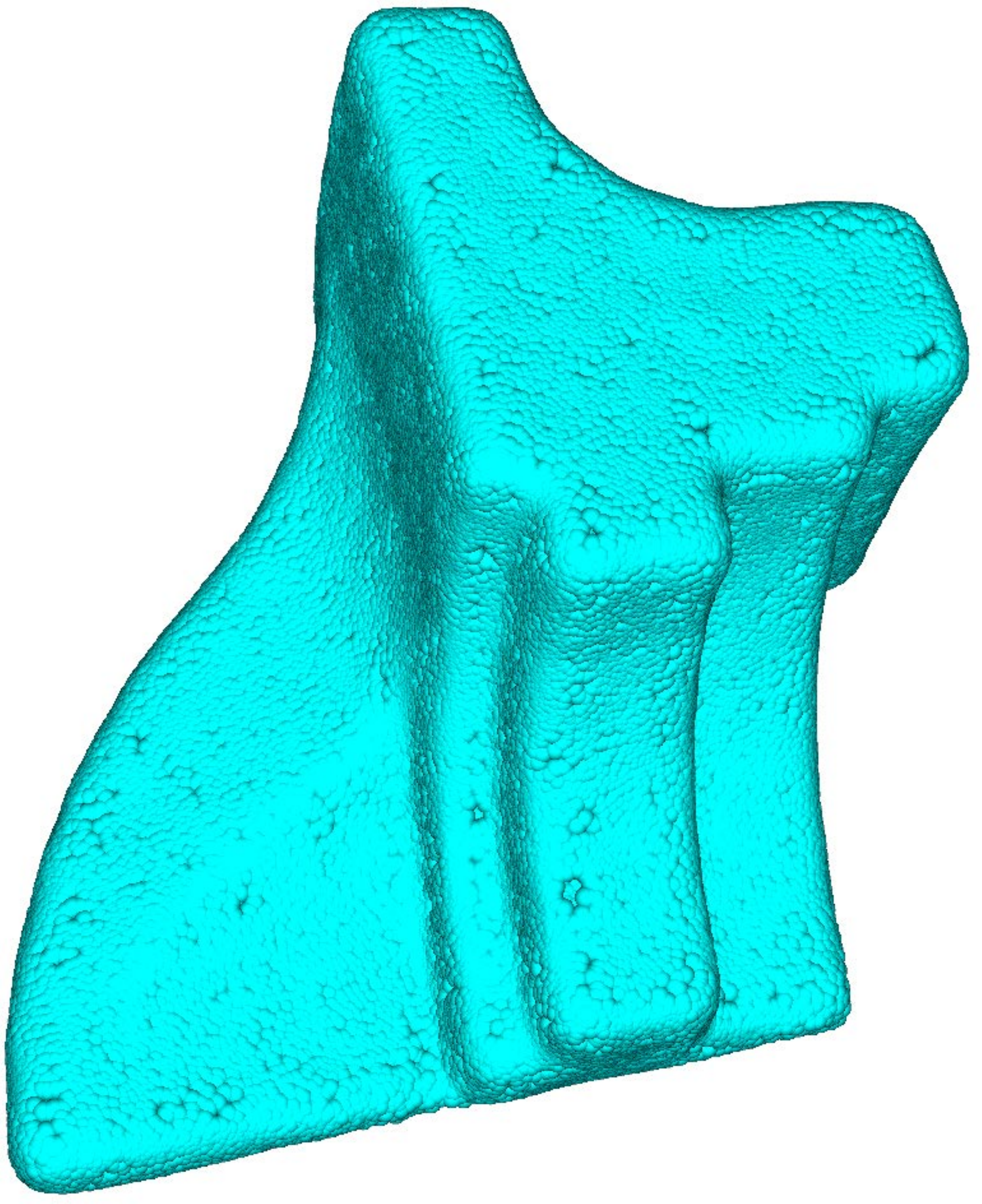}\\
        \includegraphics[width=1\textwidth]{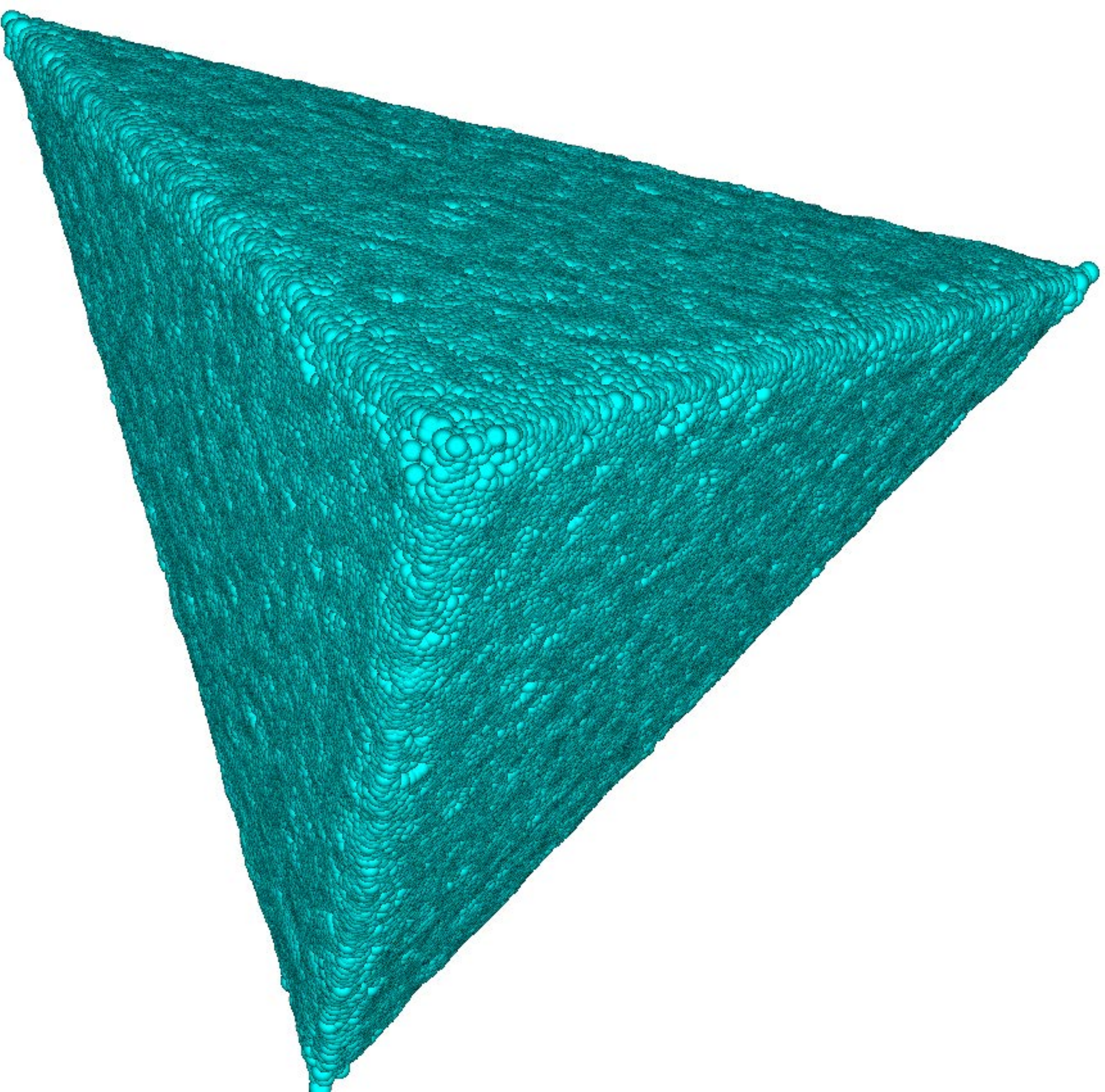}
        \end{minipage}
    }
    \subfigure[PCN]
    {
        \begin{minipage}[b]{0.105\textwidth} 
        \includegraphics[width=1\textwidth]{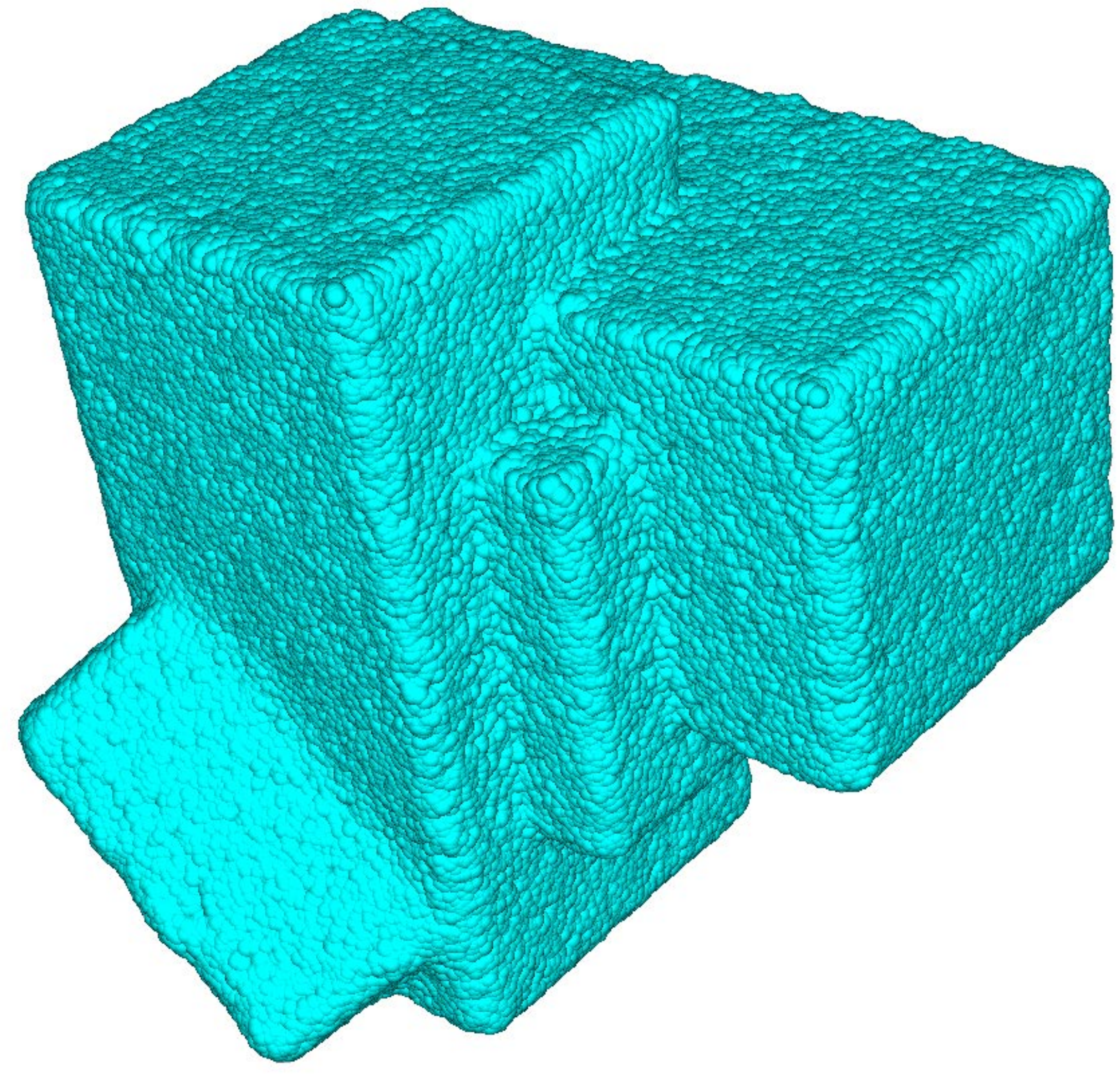}\\ 
        \includegraphics[width=1\textwidth]{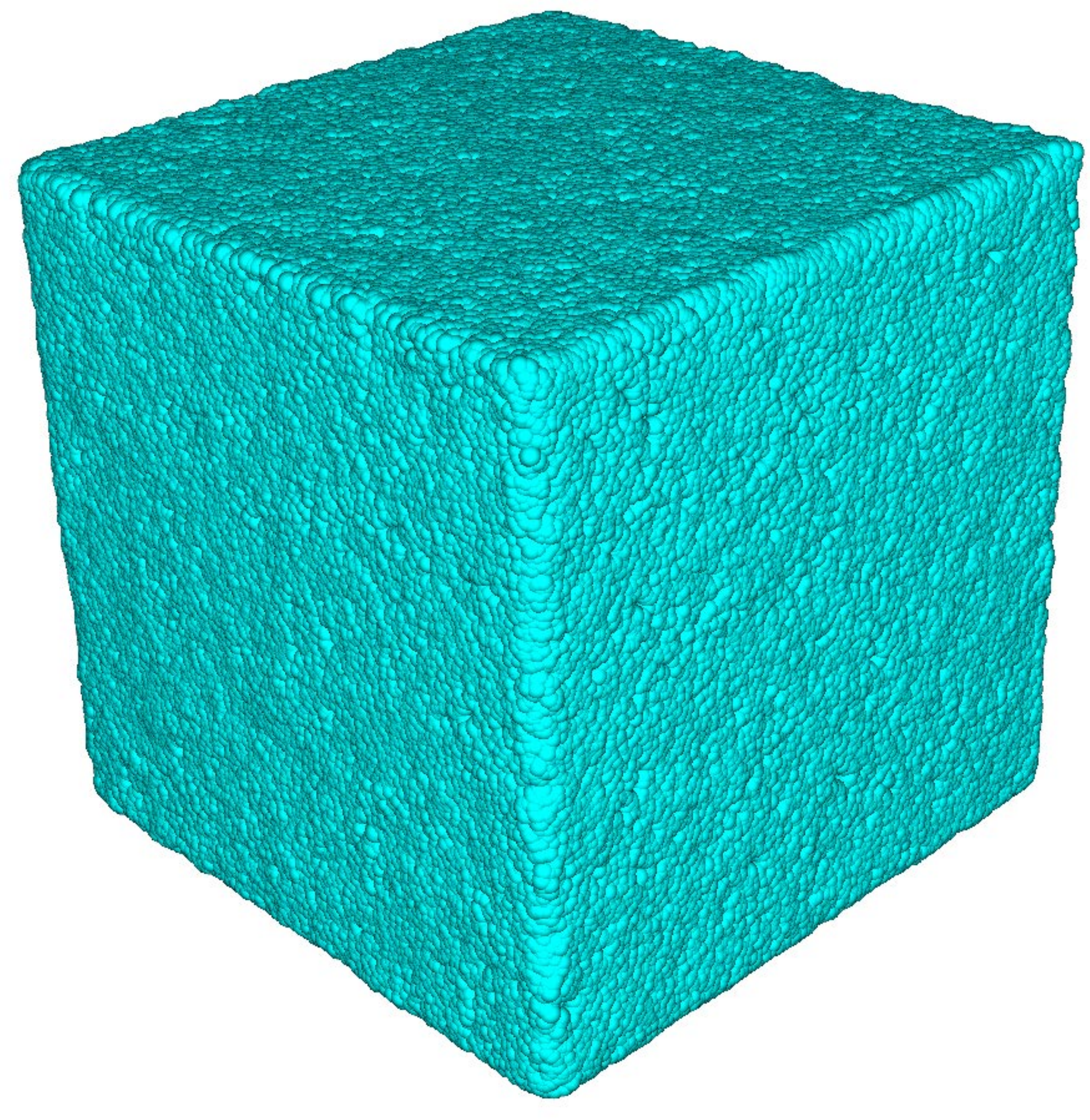}\\ 
        \includegraphics[width=1\textwidth]{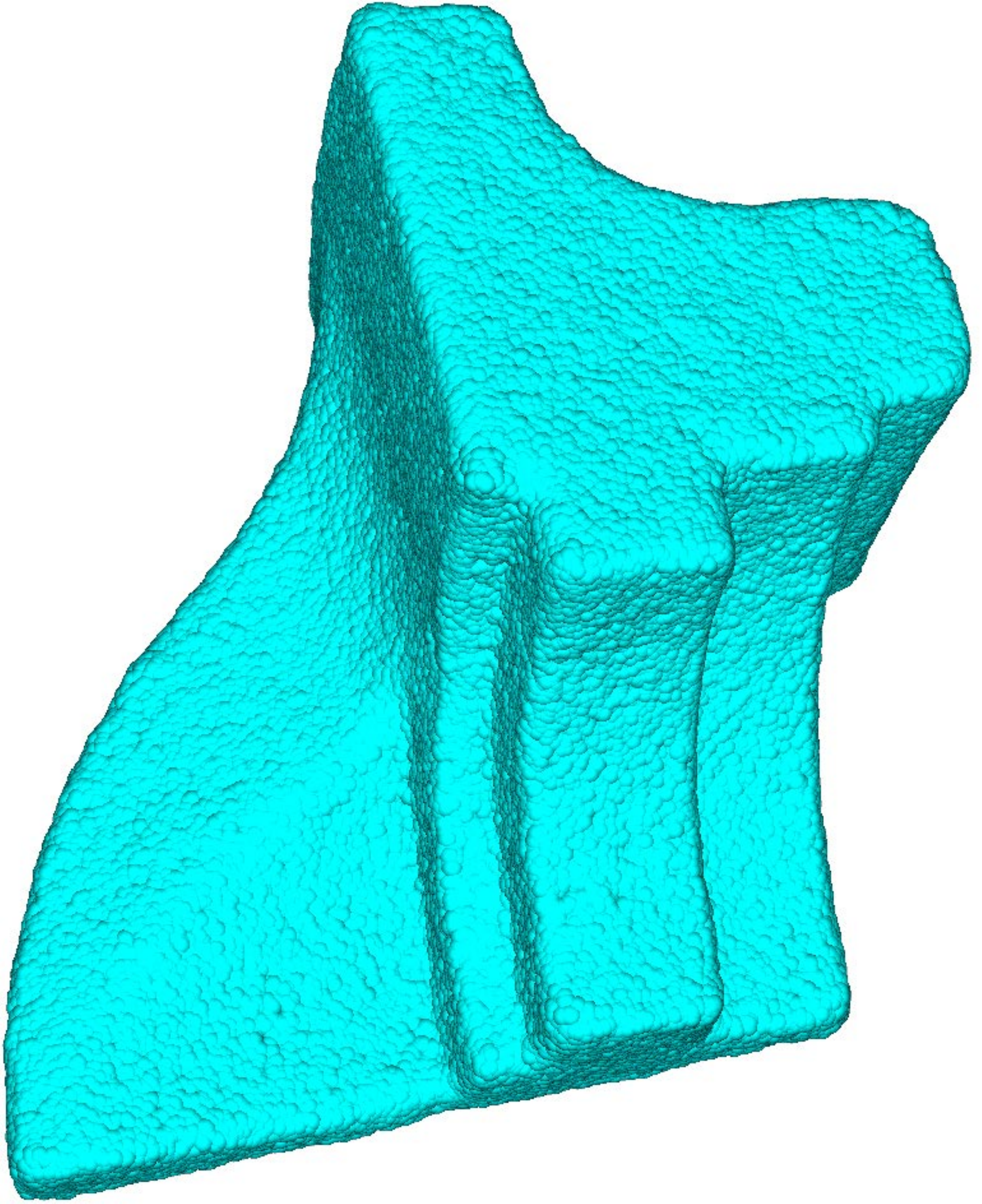}\\
        \includegraphics[width=1\textwidth]{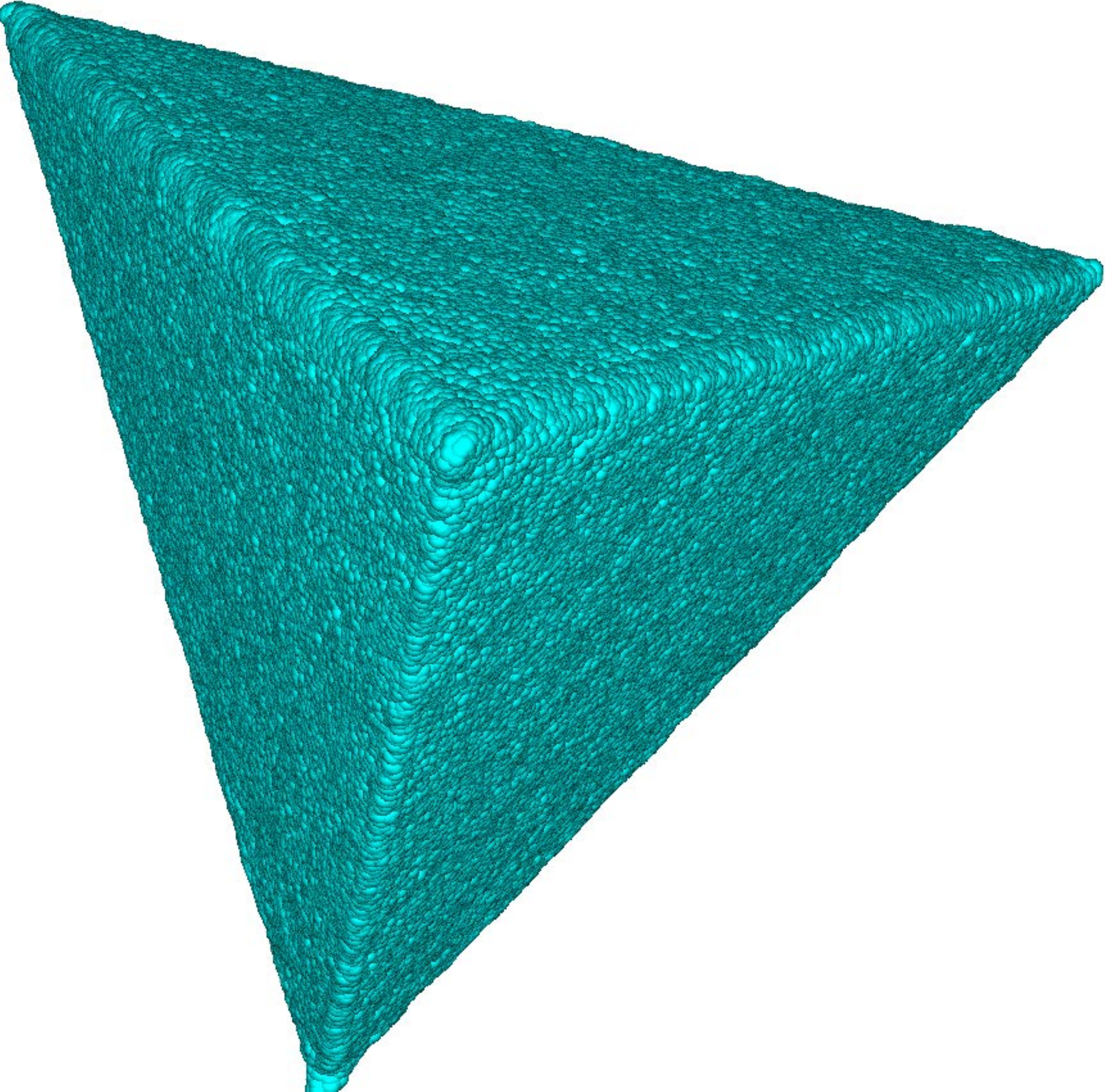}
        \end{minipage}
    }
    \subfigure[TD]
    {
        \begin{minipage}[b]{0.105\textwidth} 
        \includegraphics[width=1\textwidth]{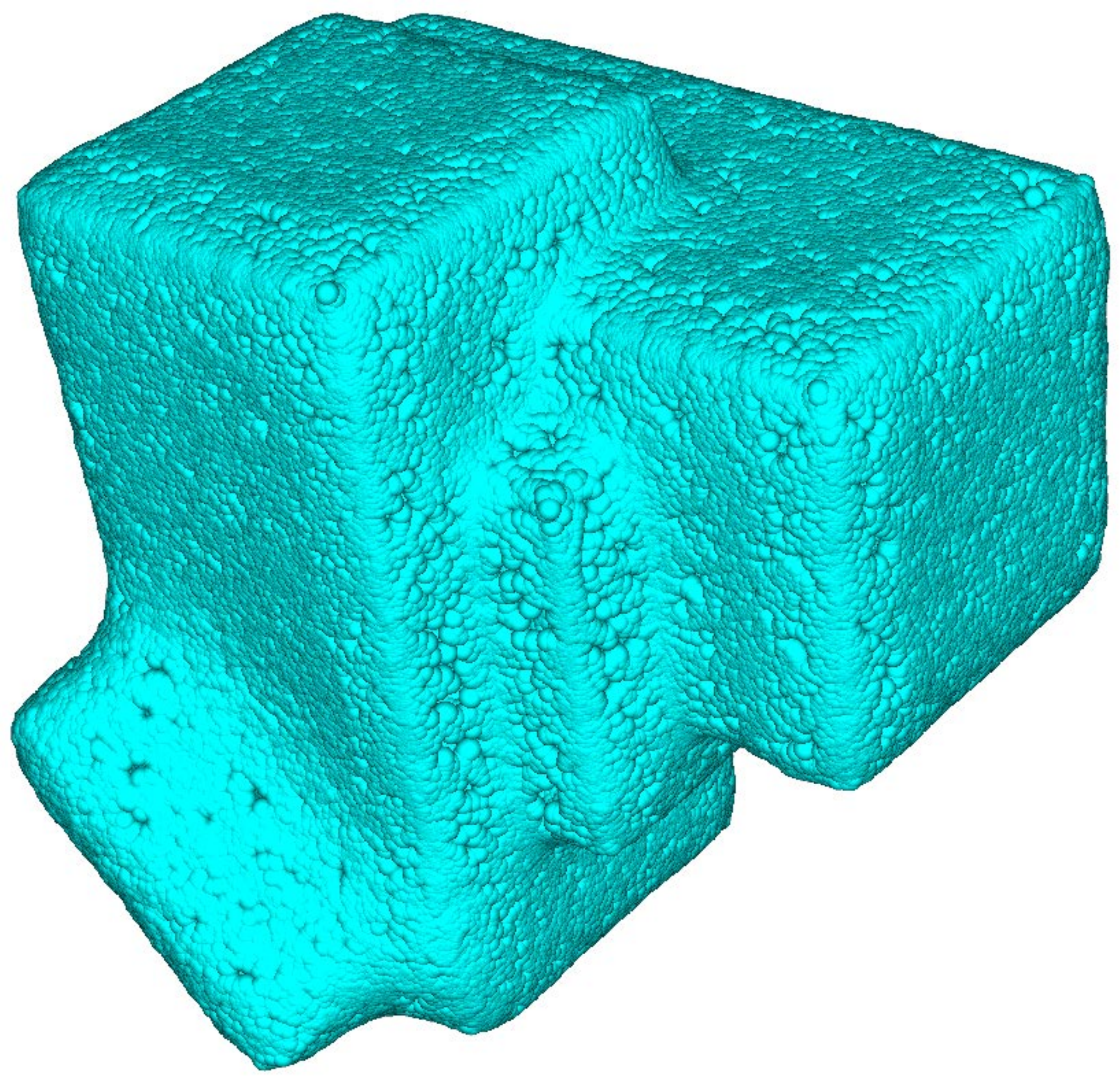}\\ 
        \includegraphics[width=1\textwidth]{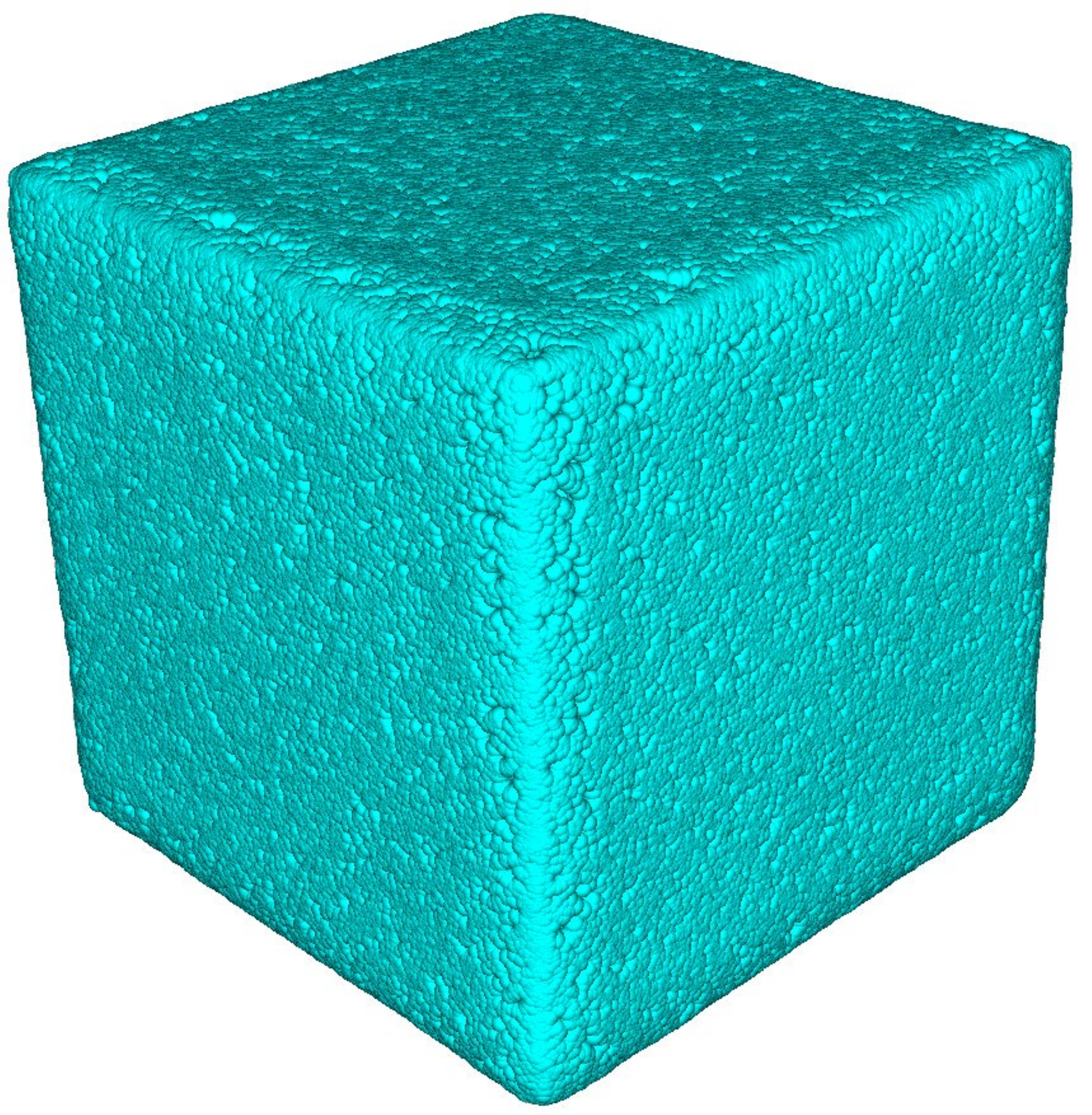}\\ 
        \includegraphics[width=1\textwidth]{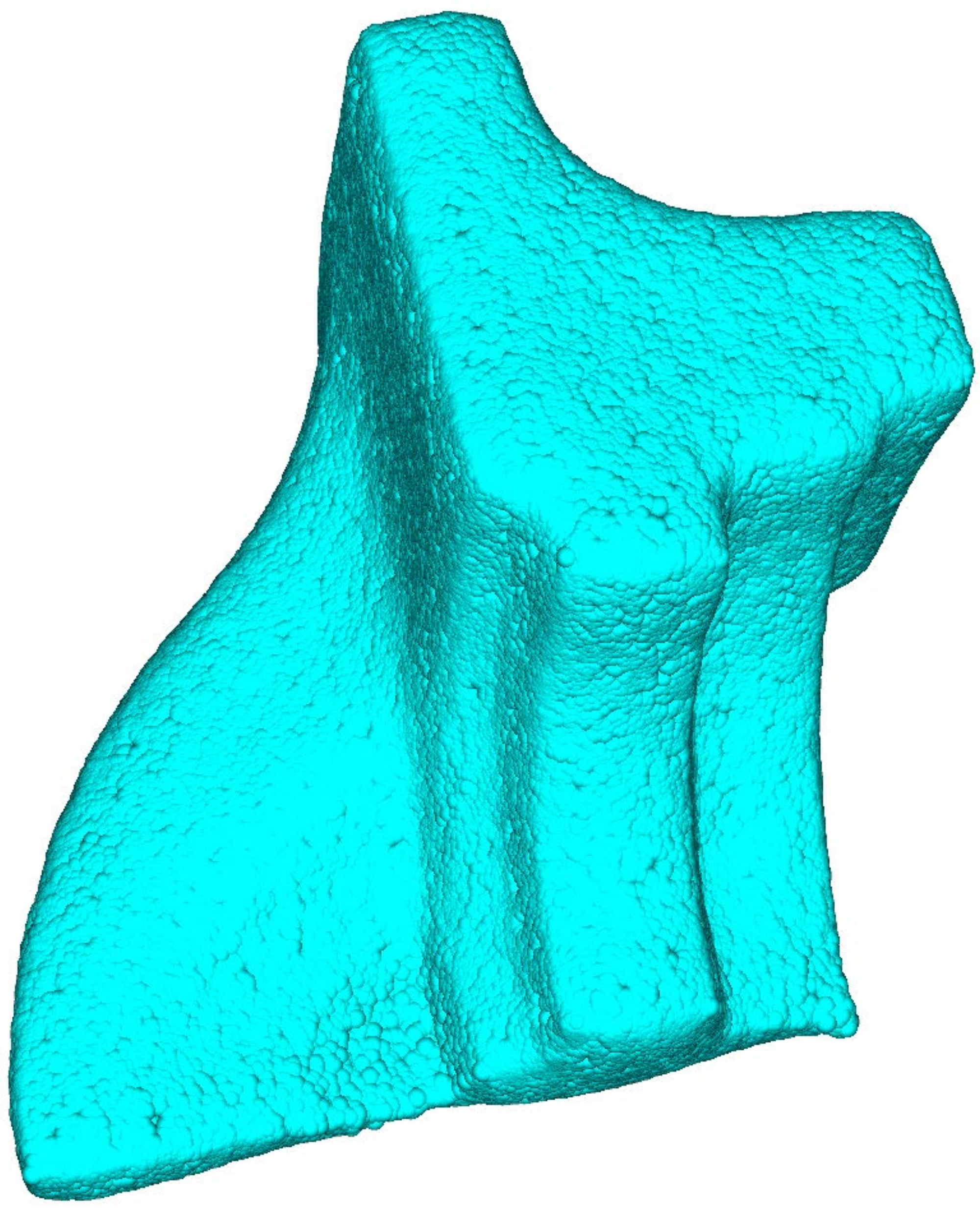}\\
        \includegraphics[width=1\textwidth]{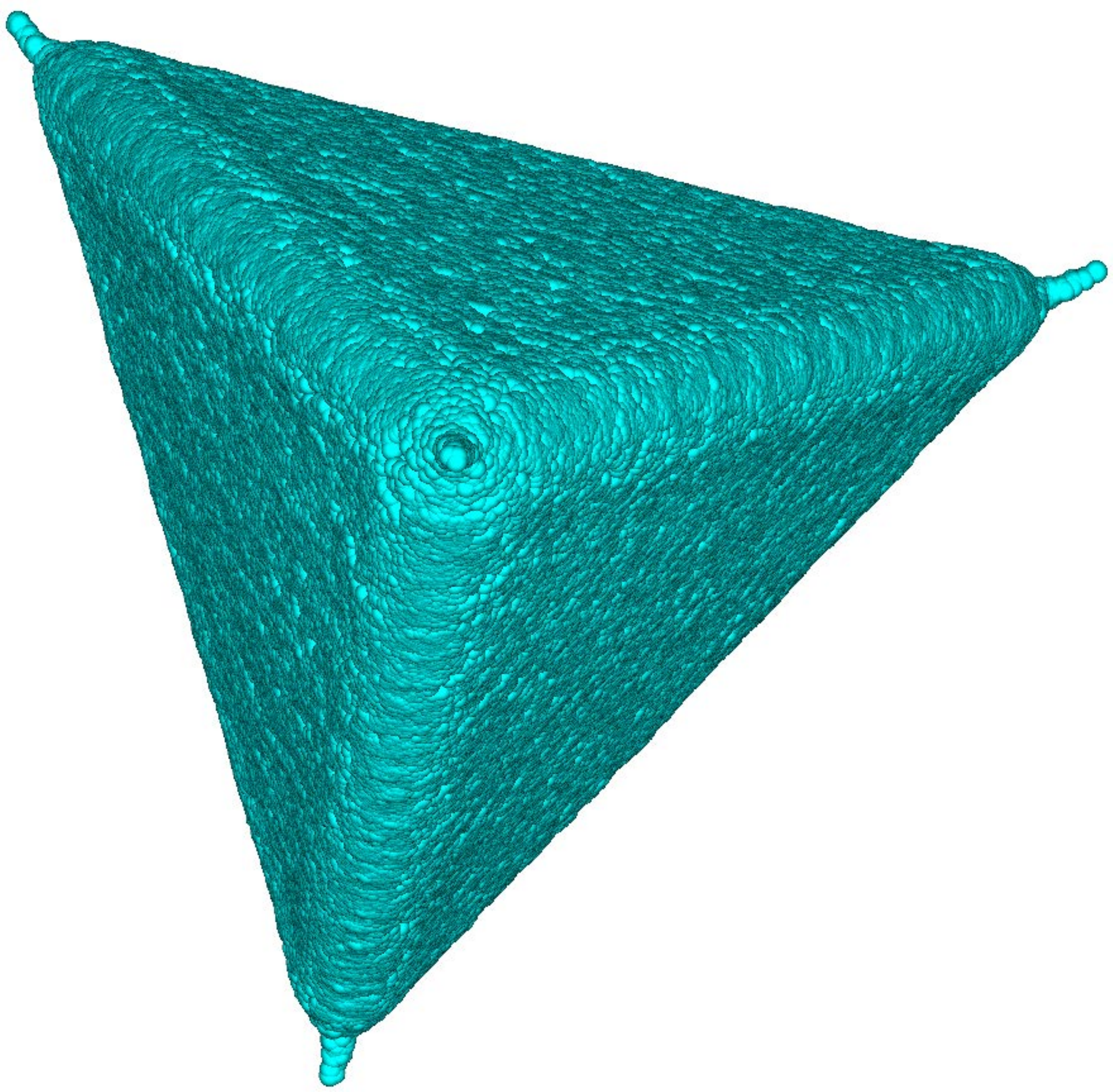}
        \end{minipage}
    }
    \subfigure[Ours]
    {
        \begin{minipage}[b]{0.105\textwidth} 
        \includegraphics[width=1\textwidth]{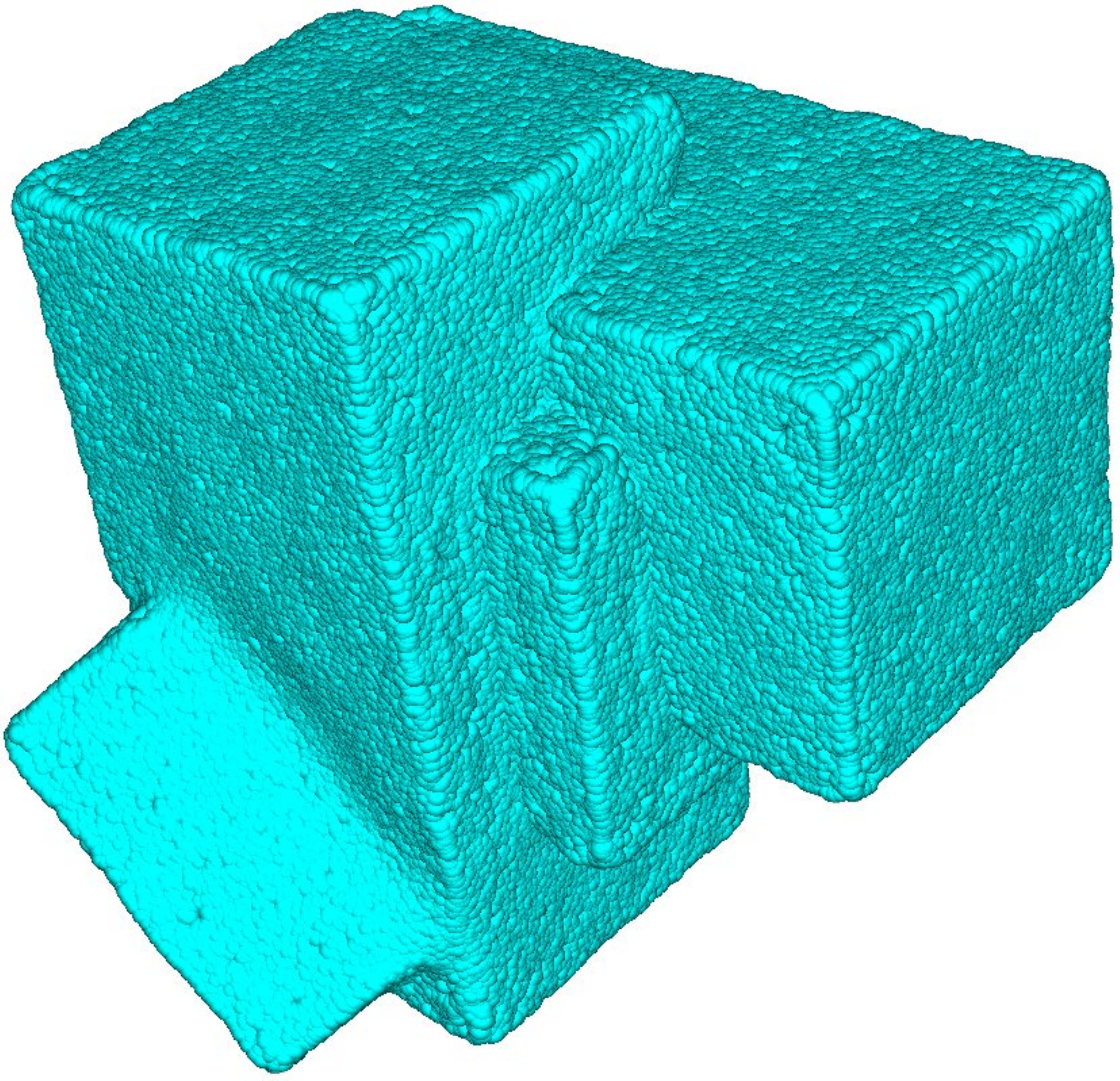}\\ 
        \includegraphics[width=1\textwidth]{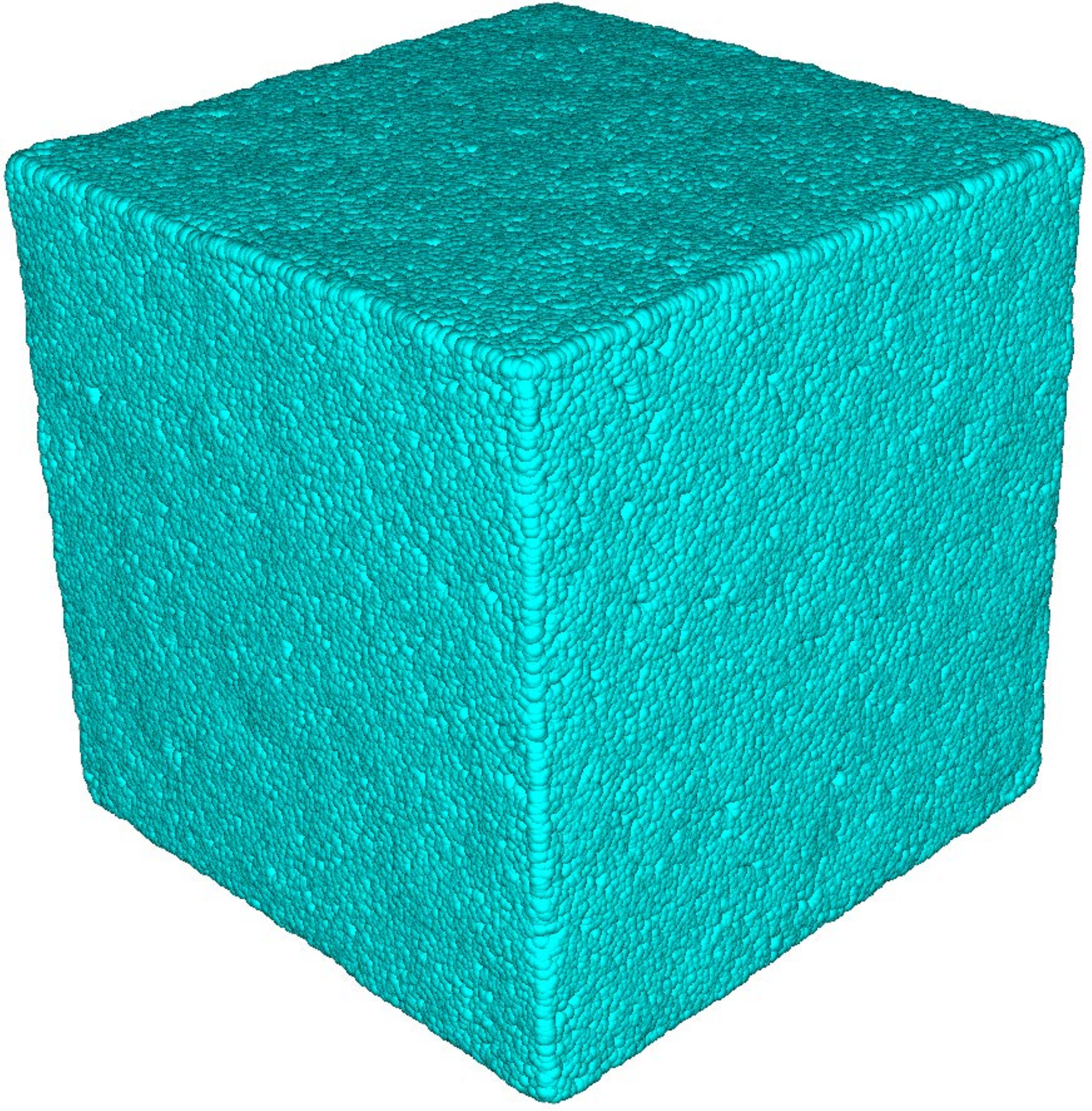}\\ 
        \includegraphics[width=1\textwidth]{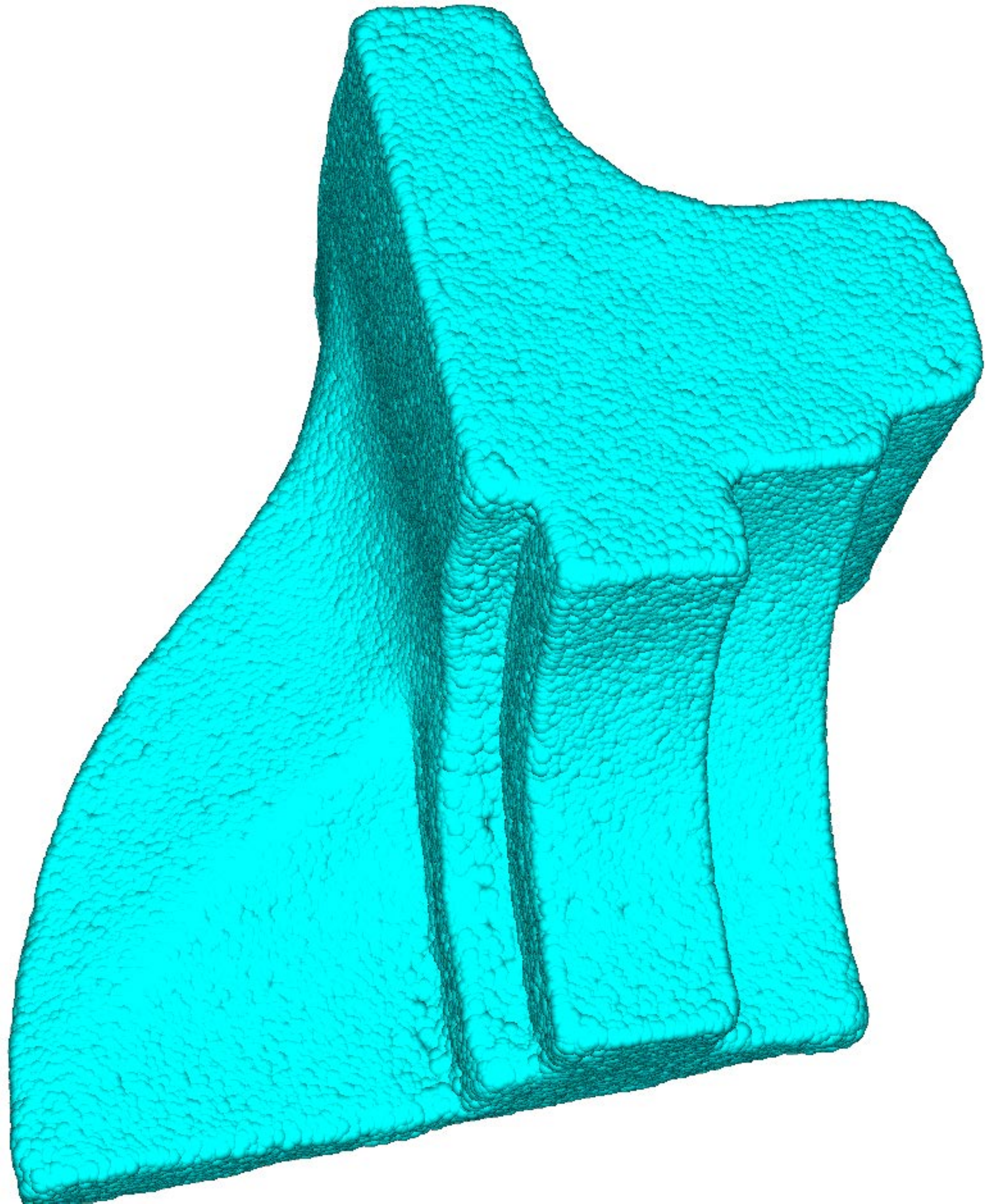}\\
        \includegraphics[width=1\textwidth]{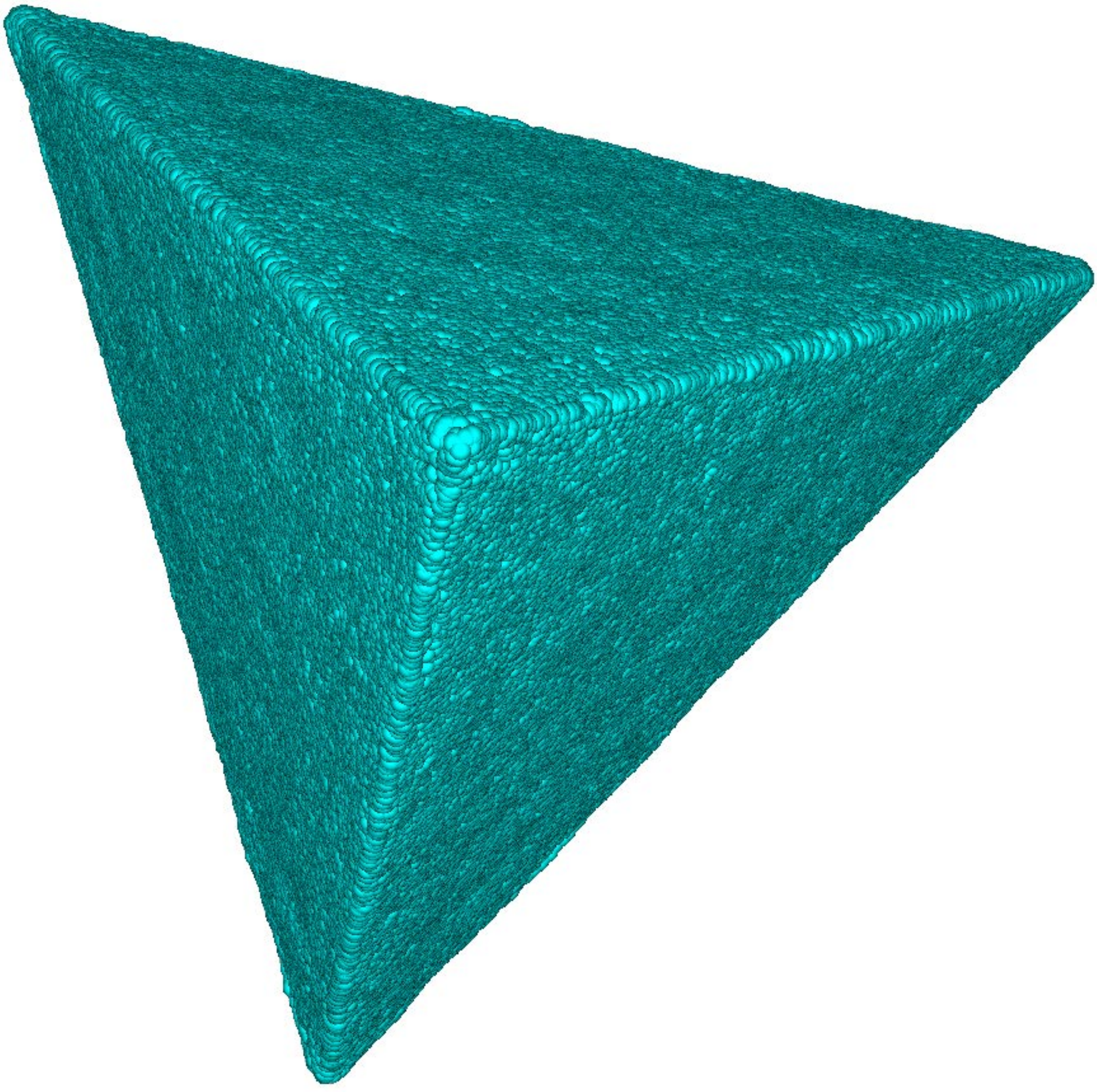}
        \end{minipage}
    }
    \caption{Visual comparison of point clouds with $0.5\%$ noise. Our method preserves sharp features better than the existing methods.}
    \label{fig:cadvisualcomparison}
\end{figure*}
\begin{figure*}[htb]
    \centering
    \subfigure[Noisy]
    {
        \begin{minipage}[b]{0.105\textwidth} 
        \includegraphics[width=1\textwidth  ]{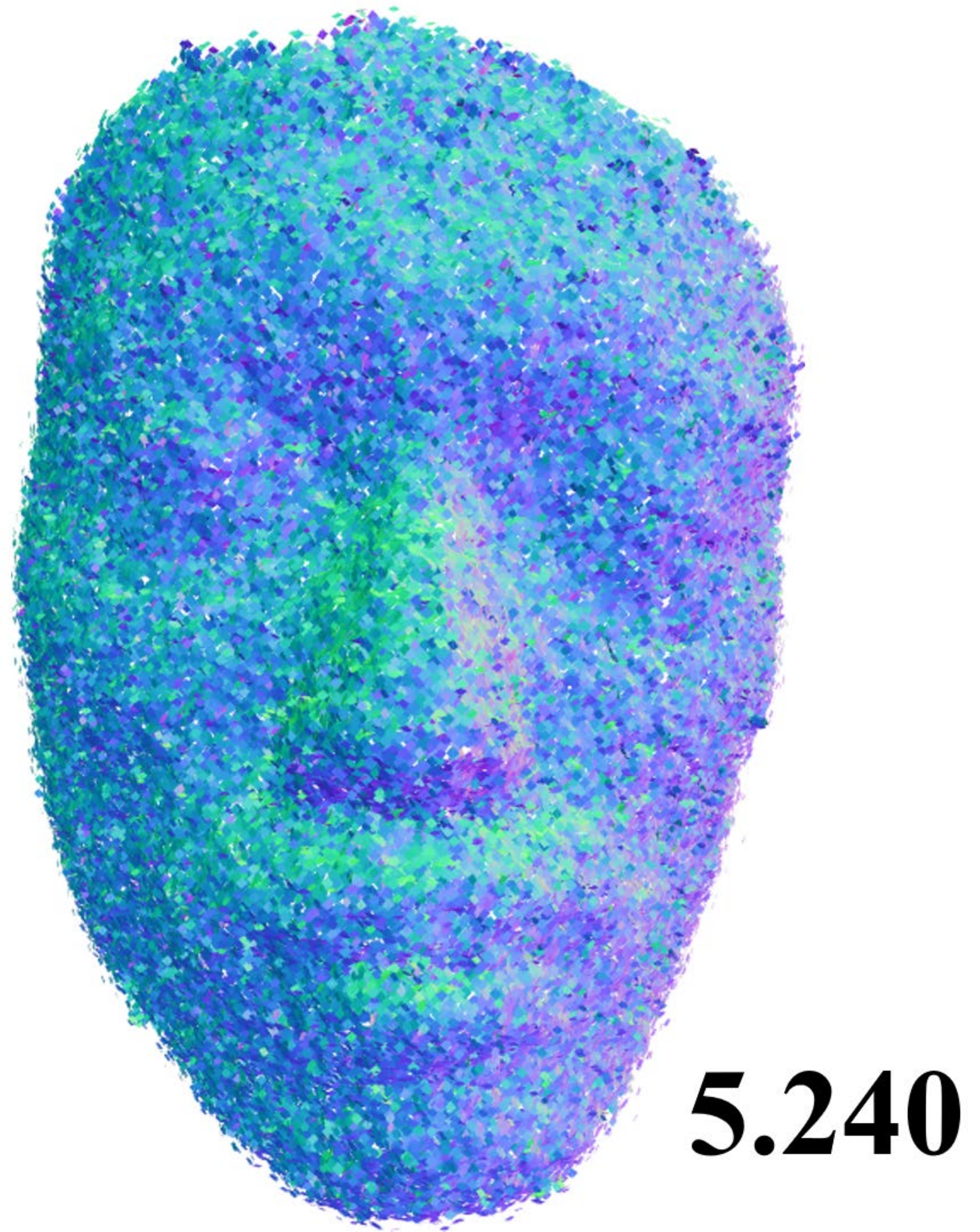}\\ 
        \includegraphics[width=1\textwidth  ]{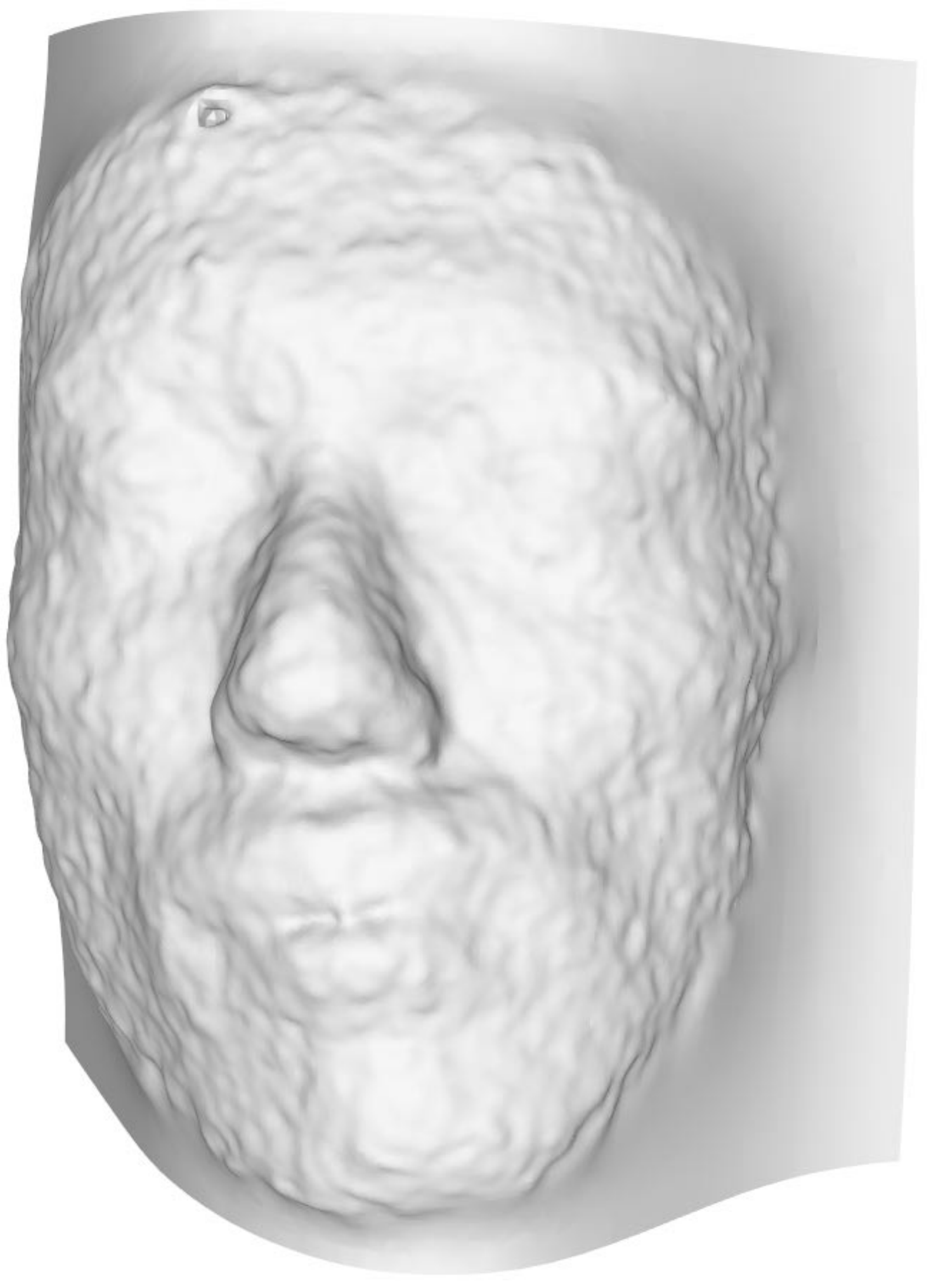}
        \end{minipage}
    }
    \subfigure[RIMLS]
    {
        \begin{minipage}[b]{0.105\textwidth}
        \includegraphics[width=1\textwidth  ]{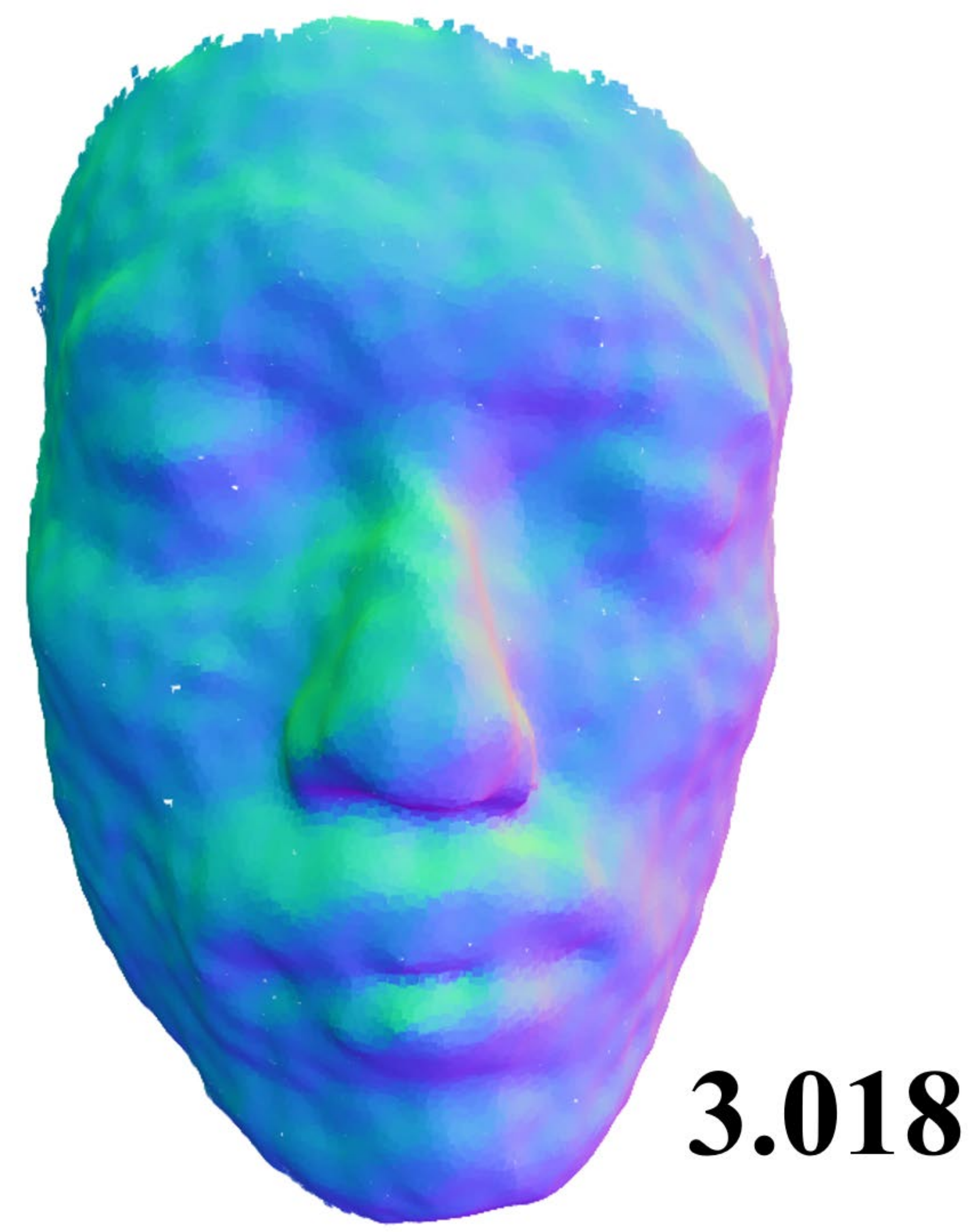}\\ 
        \includegraphics[width=1\textwidth  ]{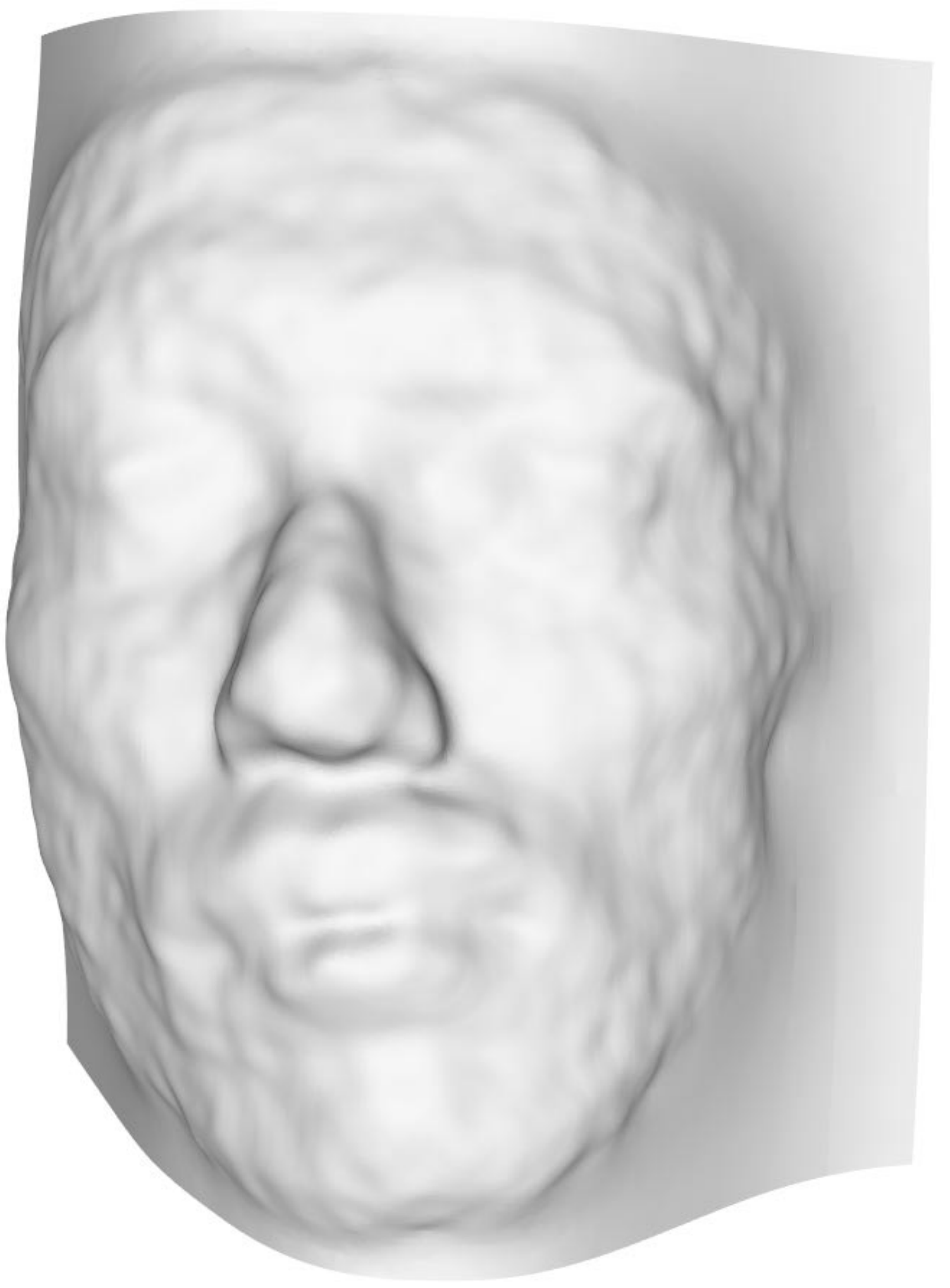}
        \end{minipage}
    }
    \subfigure[GPF]
    {
        \begin{minipage}[b]{0.105\textwidth}
        \includegraphics[width=1\textwidth  ]{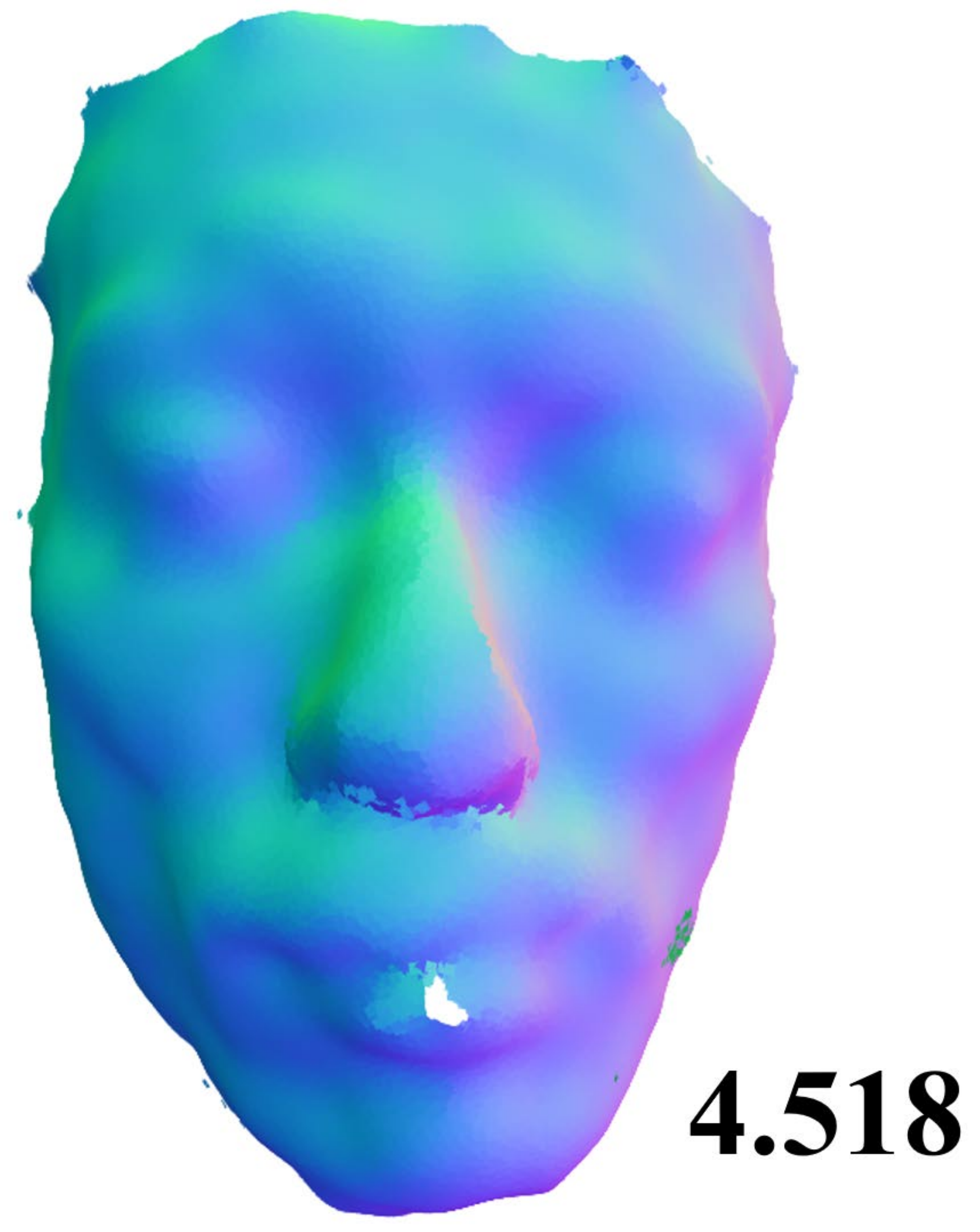}\\ 
        \includegraphics[width=1\textwidth  ]{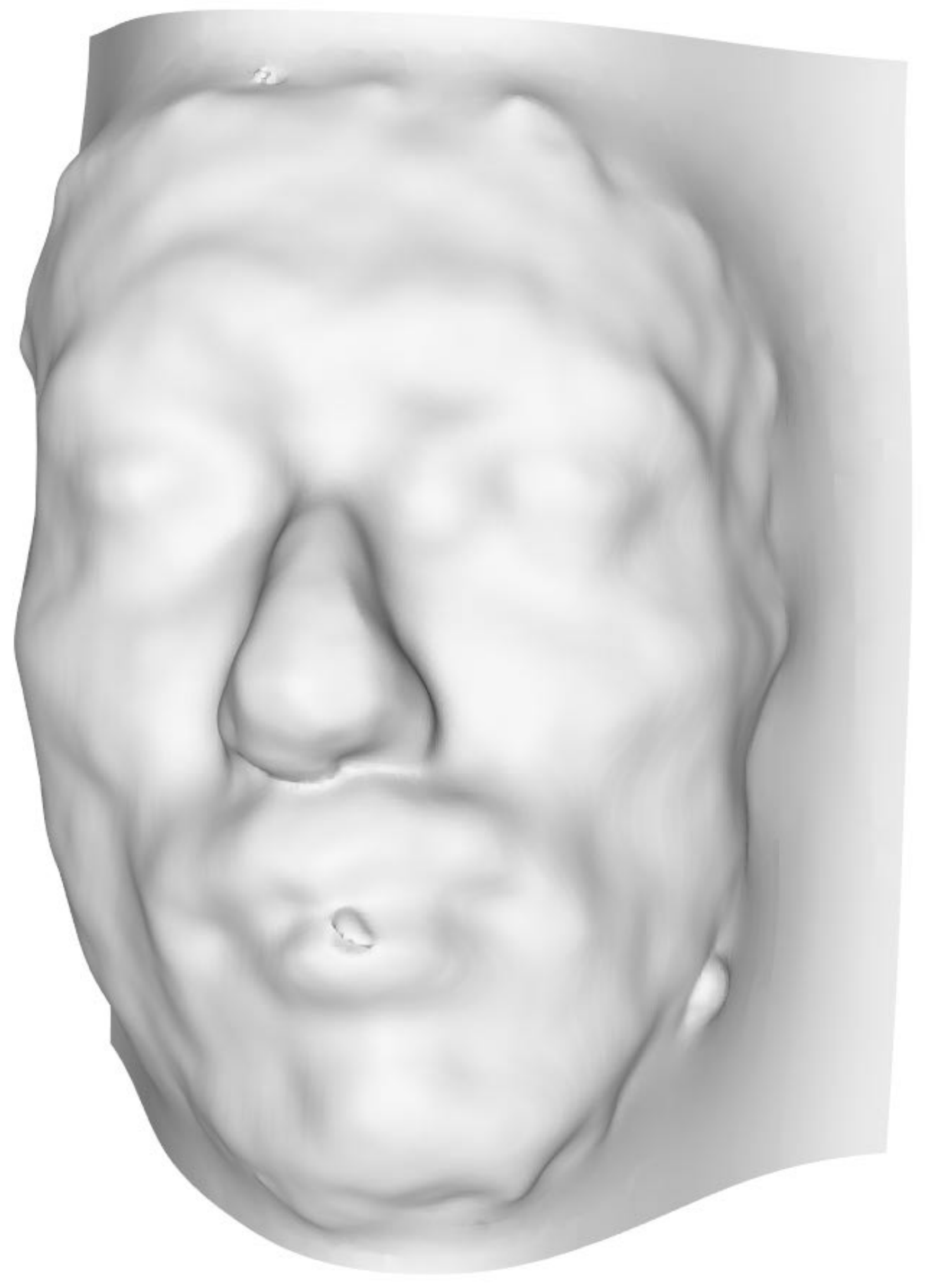}
        \end{minipage}
    }
    \subfigure[WLOP]
    {
      \begin{minipage}[b]{0.105\textwidth} 
        \includegraphics[width=1\textwidth  ]{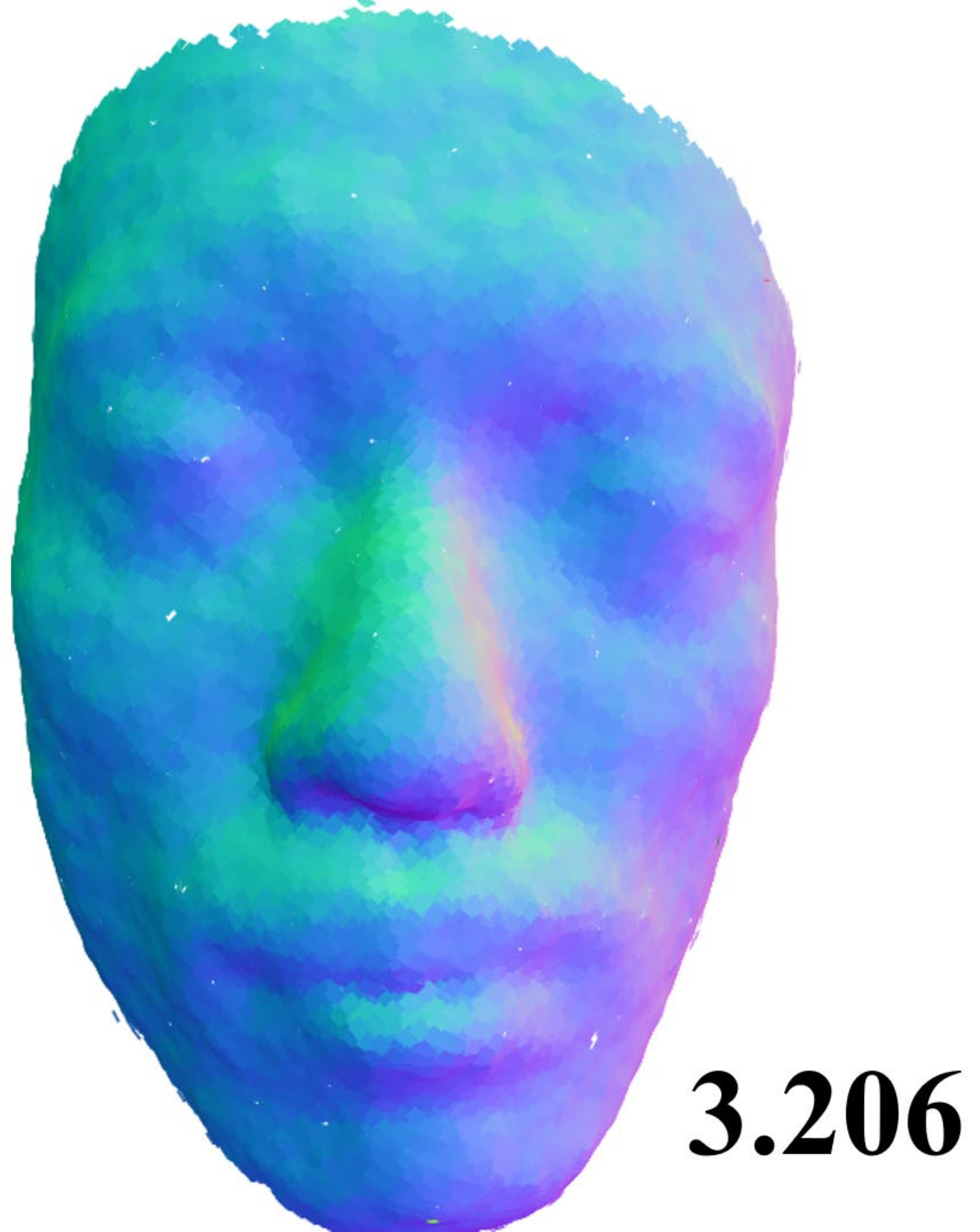}\\
        \includegraphics[width=1\textwidth  ]{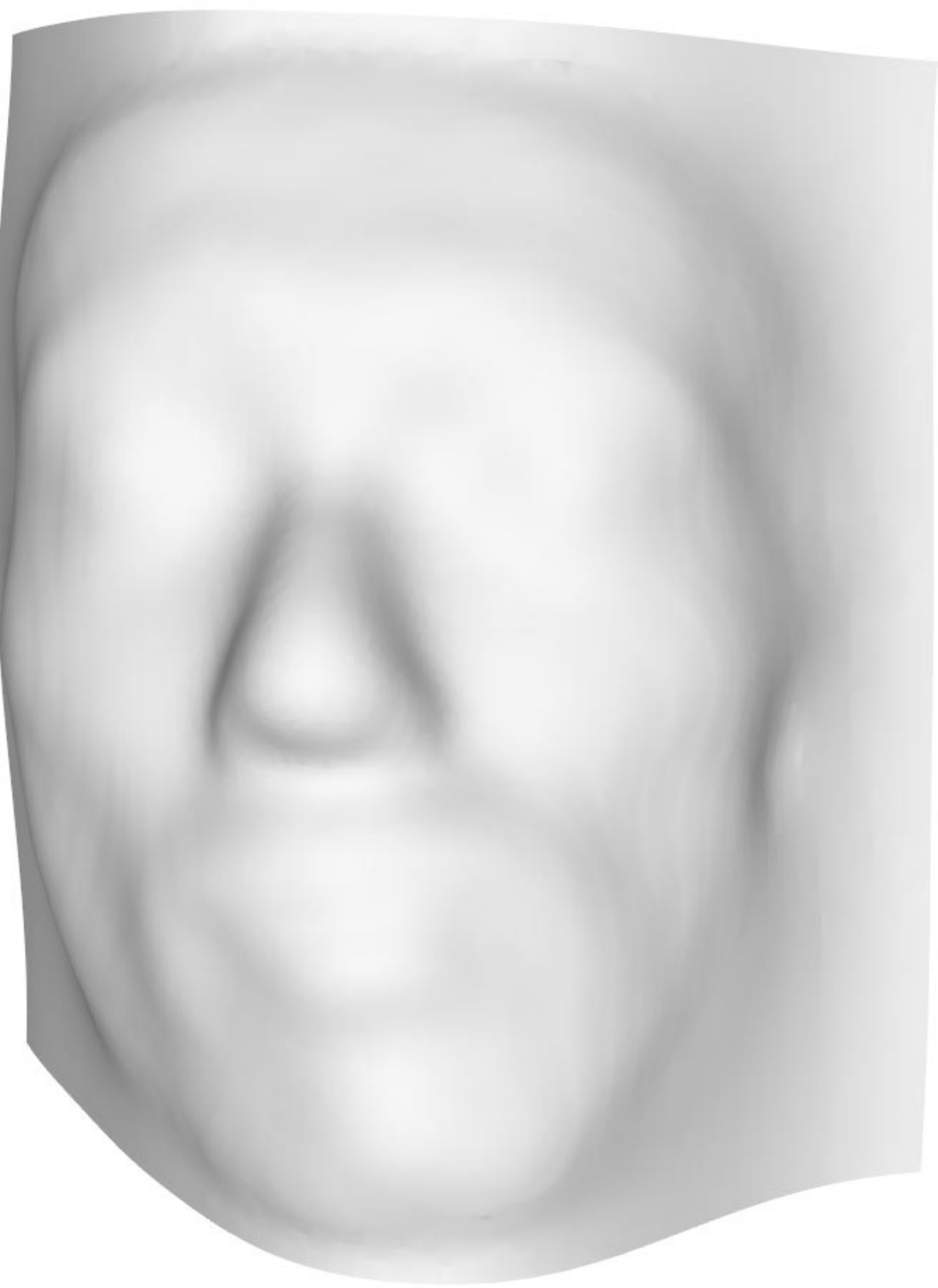}
        \end{minipage}
    }
    \subfigure[CLOP]
    {
        \begin{minipage}[b]{0.105\textwidth}
        \includegraphics[width=1\textwidth  ]{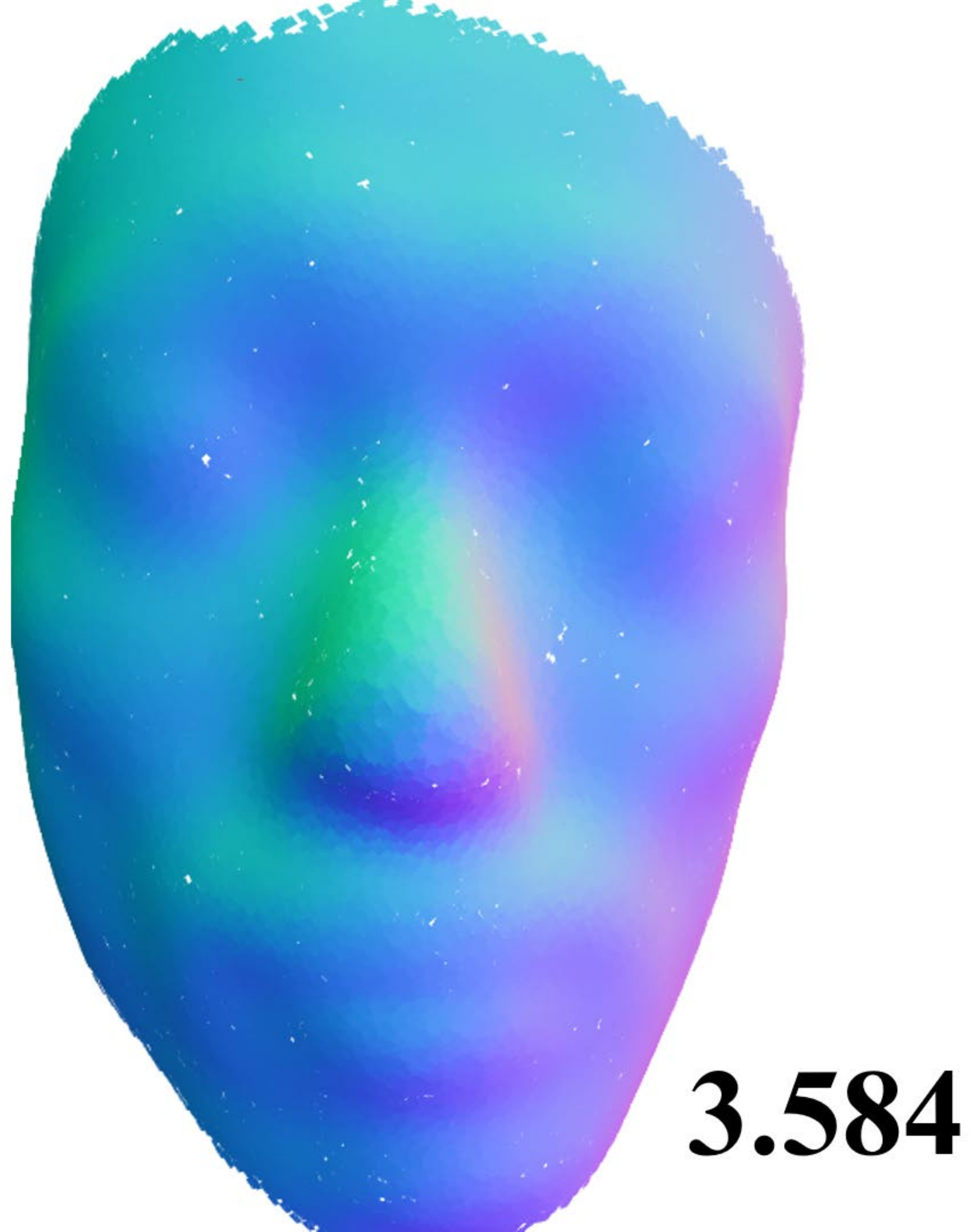}\\
        \includegraphics[width=1\textwidth  ]{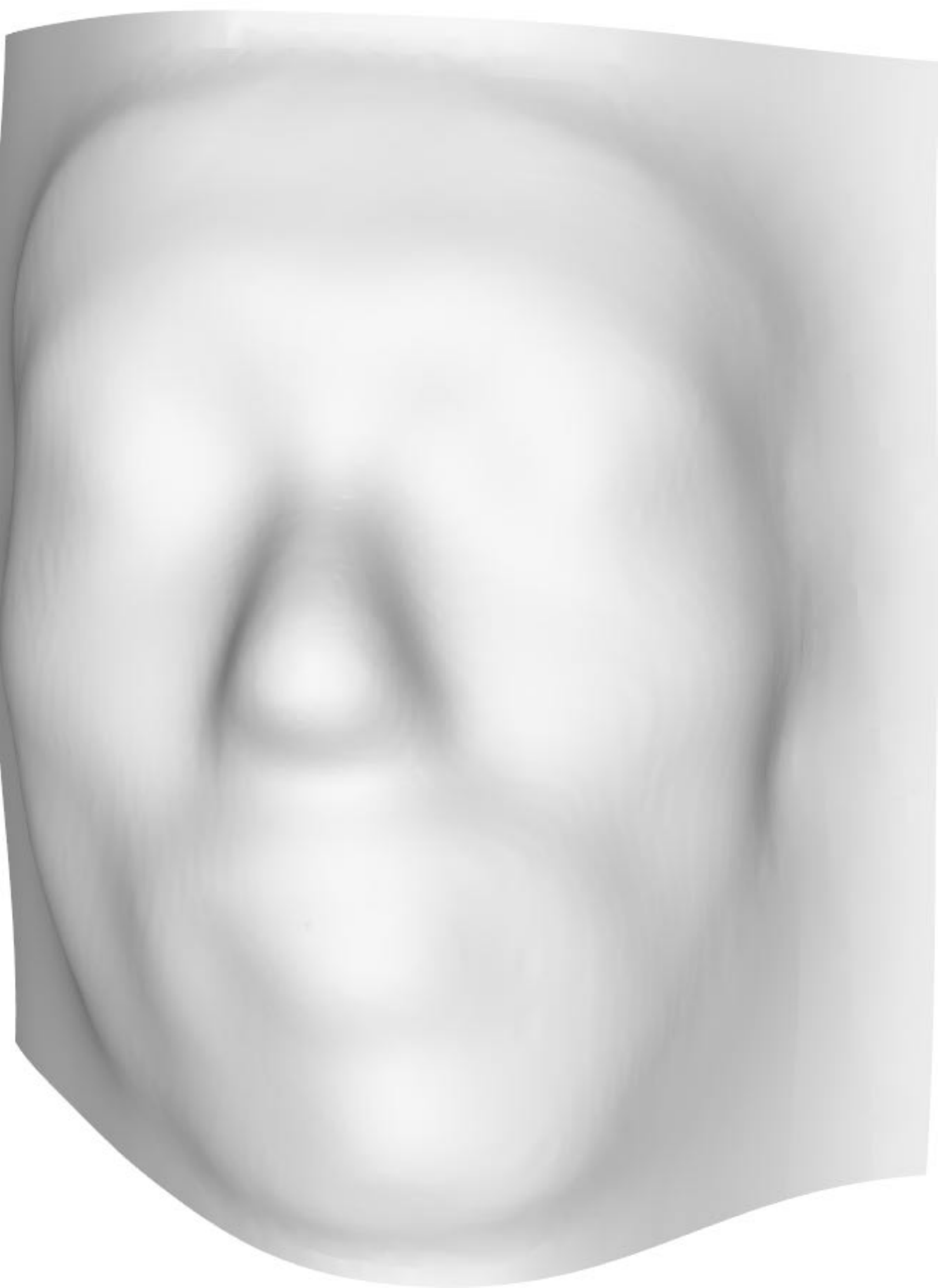}
        \end{minipage}
    }
    \subfigure[PCN]
    {
      \begin{minipage}[b]{0.105\textwidth} 
        \includegraphics[width=1\textwidth  ]{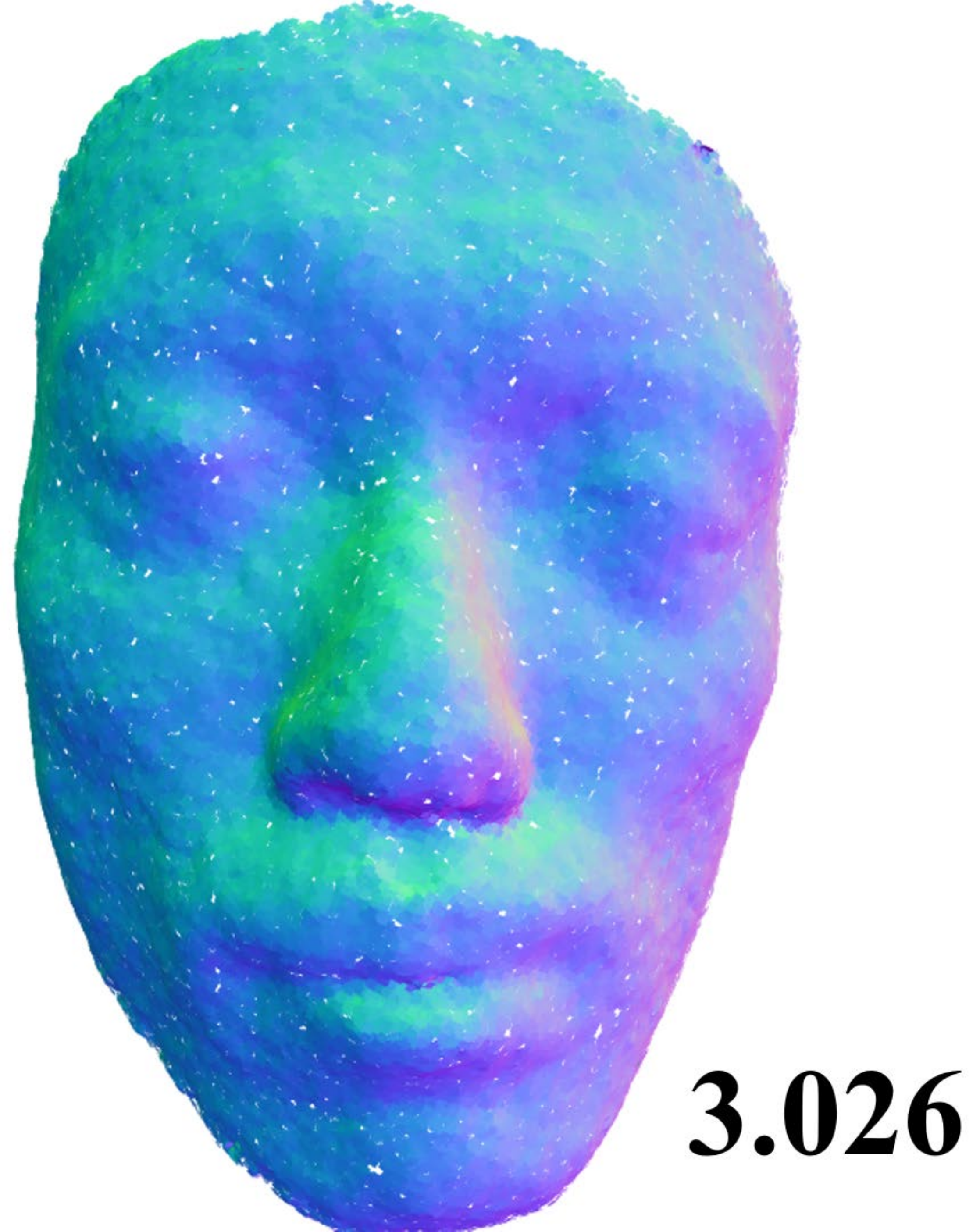}\\
        \includegraphics[width=1\textwidth  ]{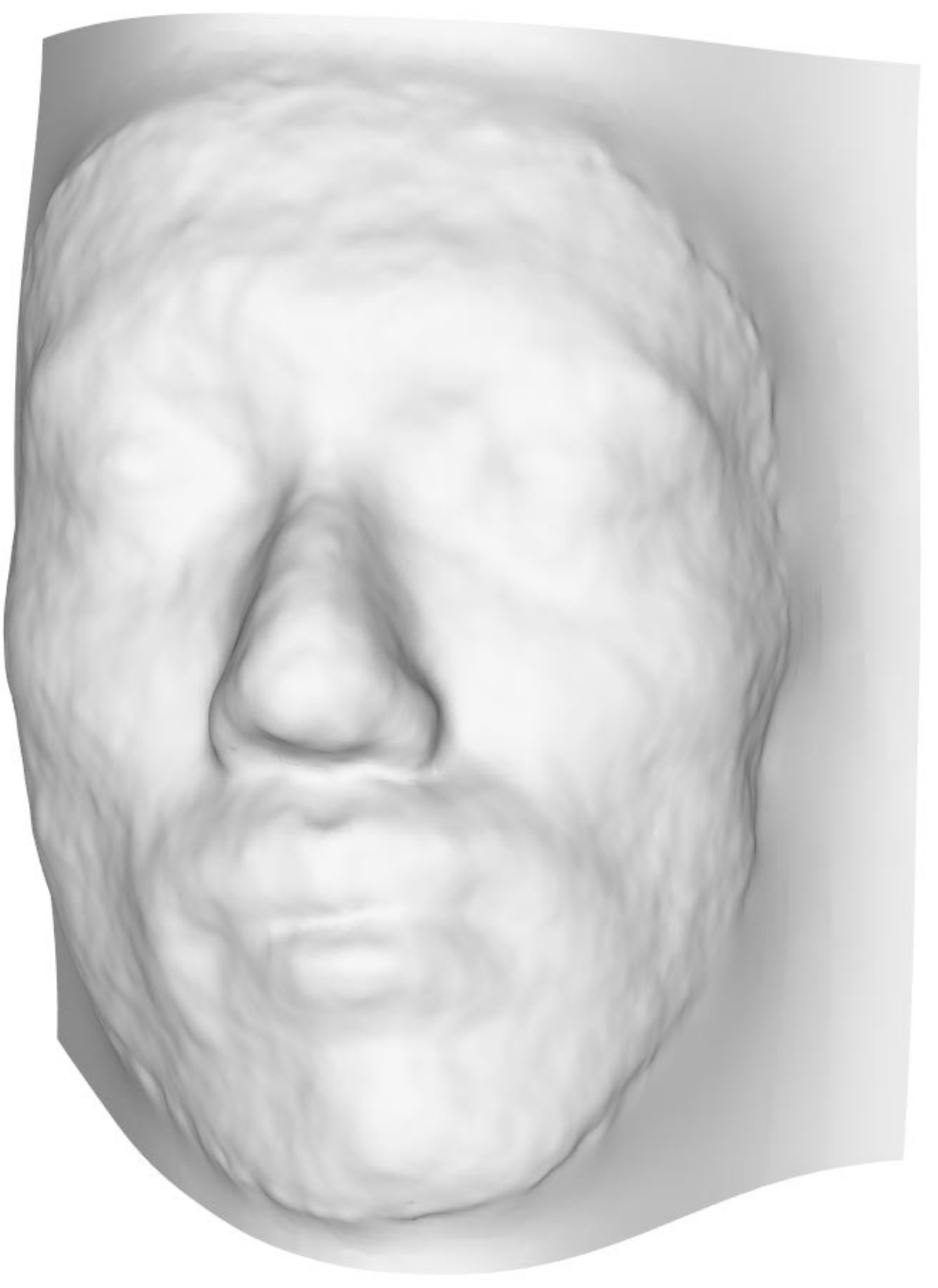}
        \end{minipage}
    }
    \subfigure[TD]
    {
      \begin{minipage}[b]{0.105\textwidth} 
        \includegraphics[width=1\textwidth  ]{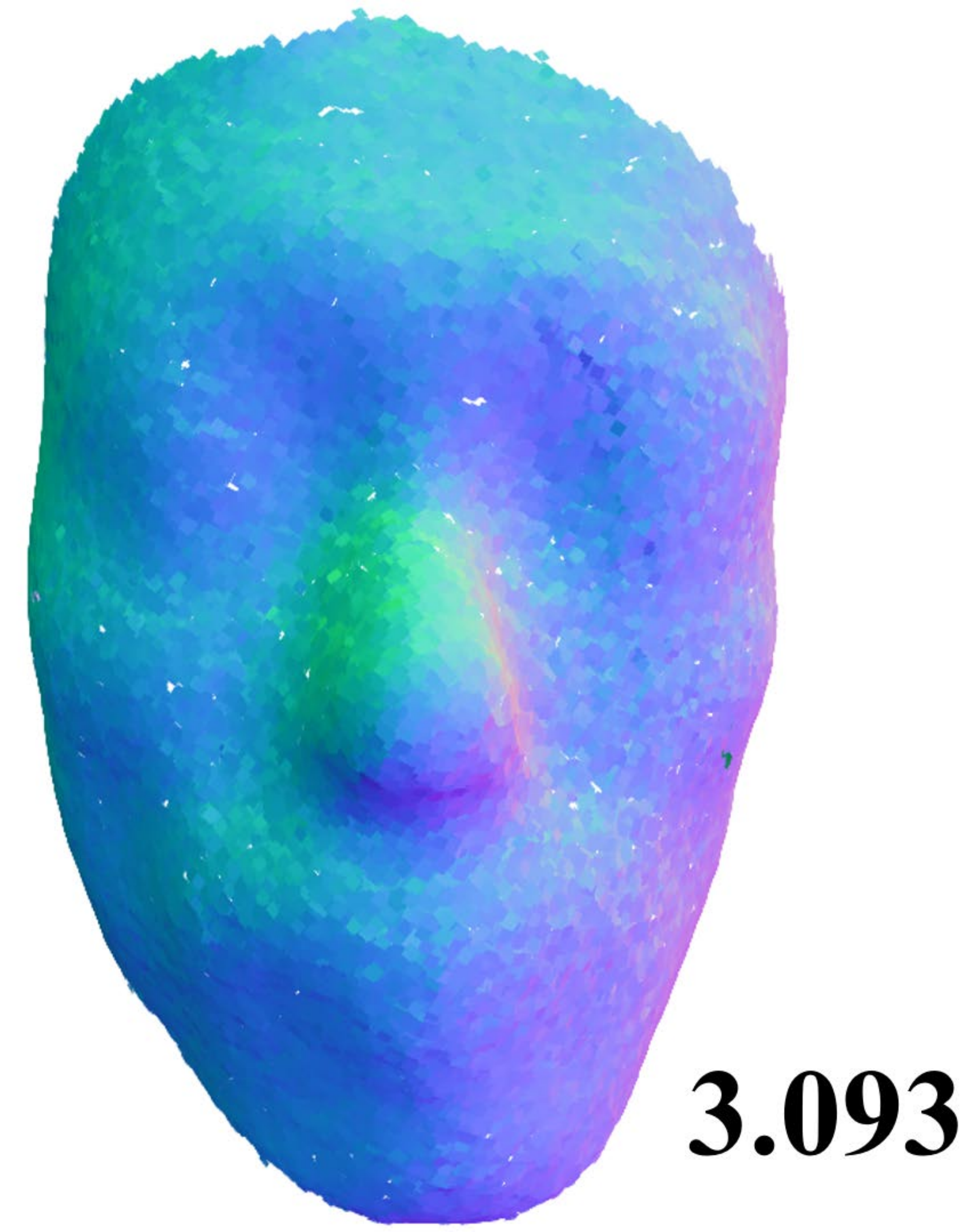}\\
        \includegraphics[width=1\textwidth  ]{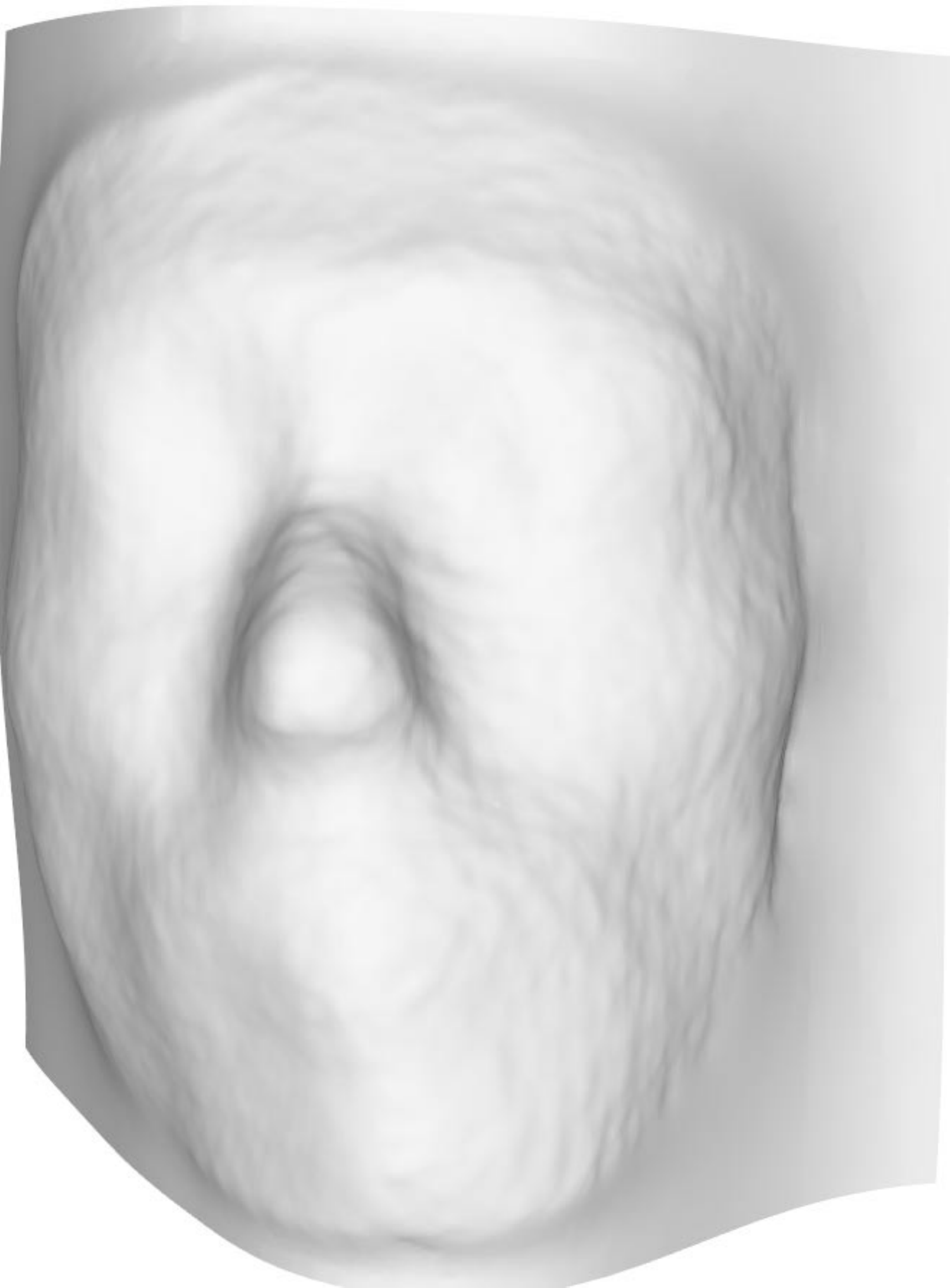}
        \end{minipage}
    }
    \subfigure[Ours]
    {
        \begin{minipage}[b]{0.105\textwidth}
        \includegraphics[width=1\textwidth  ]{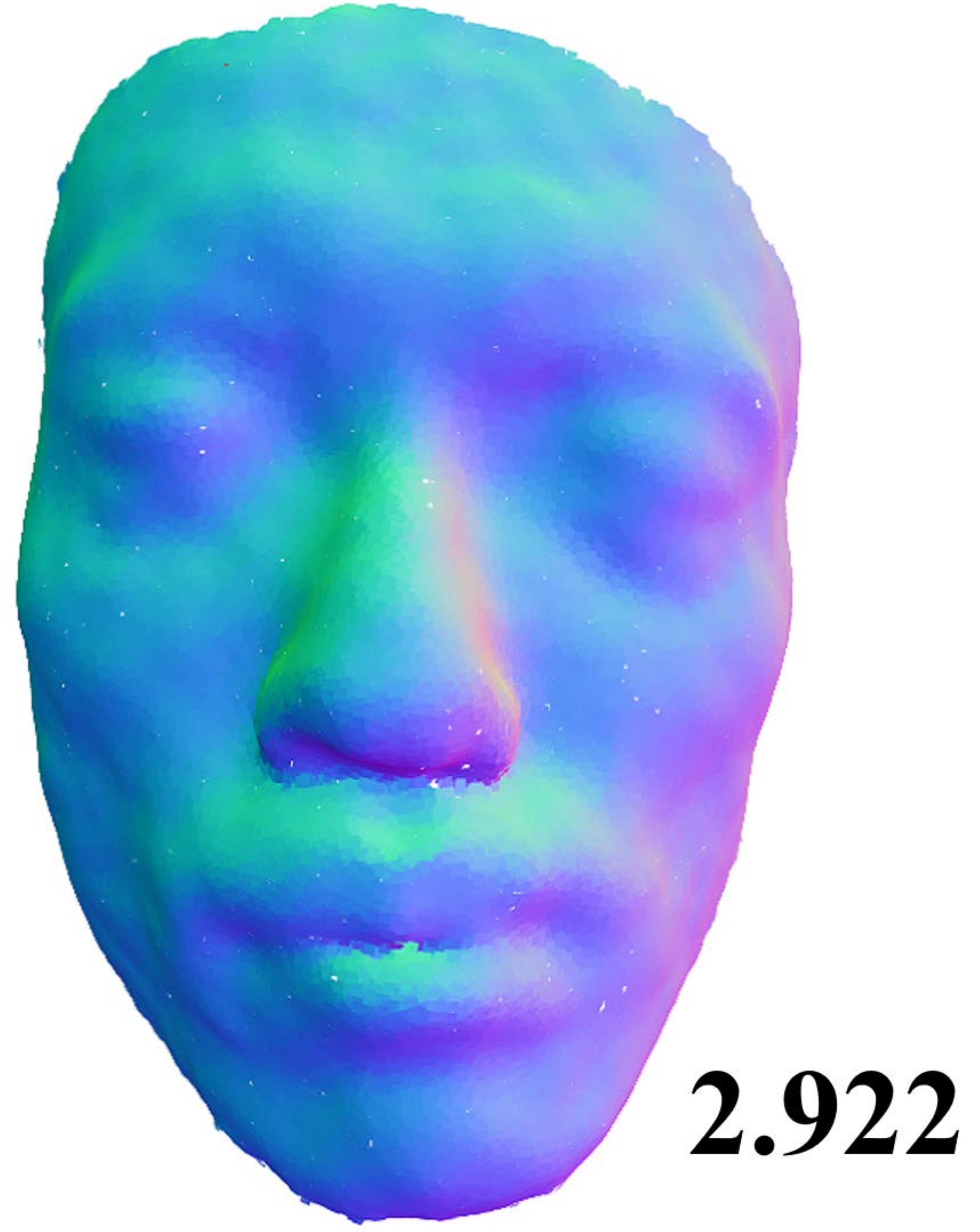}\\
        \includegraphics[width=1\textwidth  ]{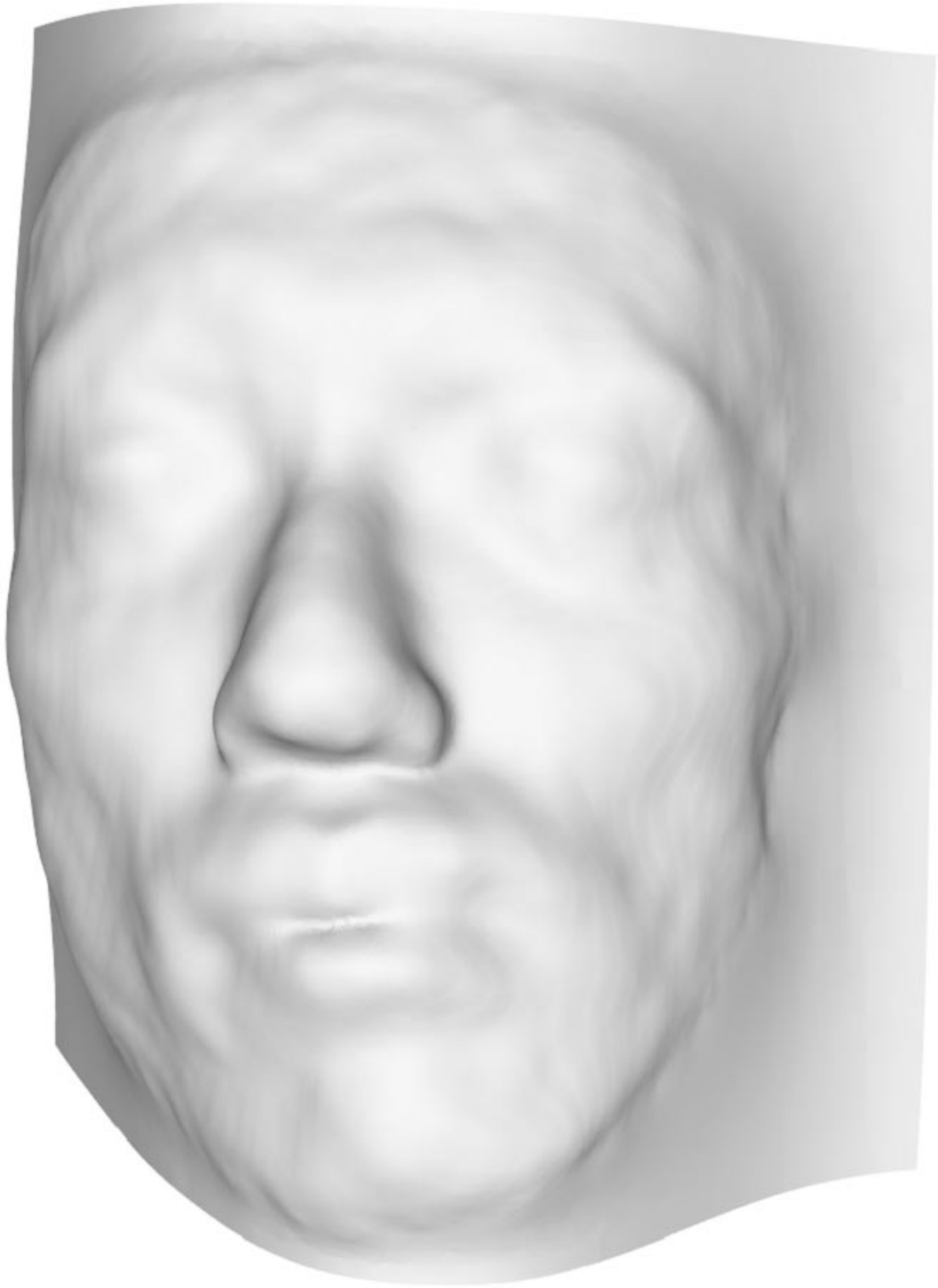}
        \end{minipage}
    }
    \caption{Visual comparison of point cloud filtering with $0.5\%$ synthetic noise. The overall MSEs ($\times 10^{-3}$) for different methods are shown in the figure.}
    \label{fig:facesurfacereconstruction}
\end{figure*}
\begin{figure*}[htb!]
    \centering
    \subfigure[Noisy]
    {
        \begin{minipage}[b]{0.105\textwidth} 
        \includegraphics[width=1\textwidth  ]{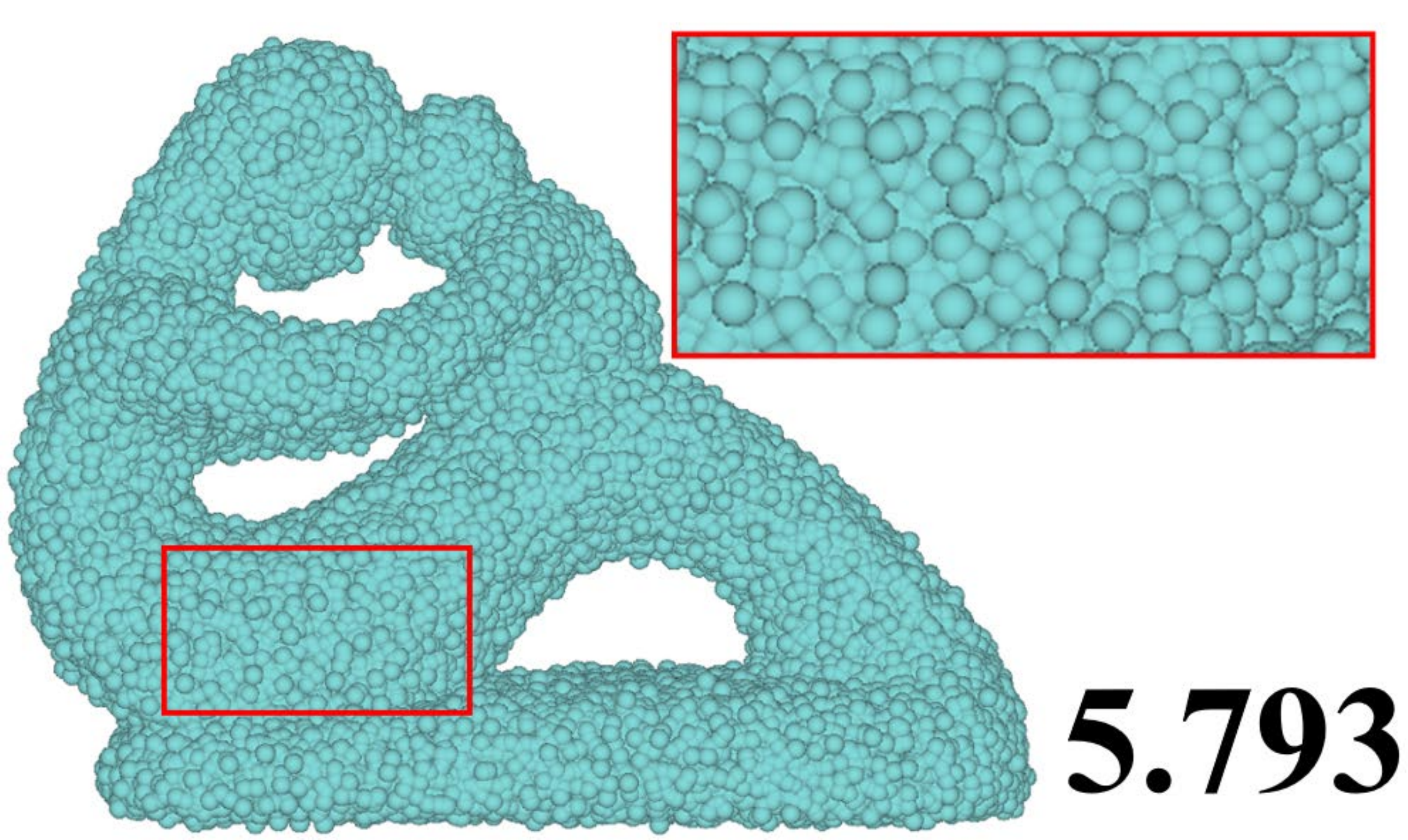}\\
        \includegraphics[width=1\textwidth  ]{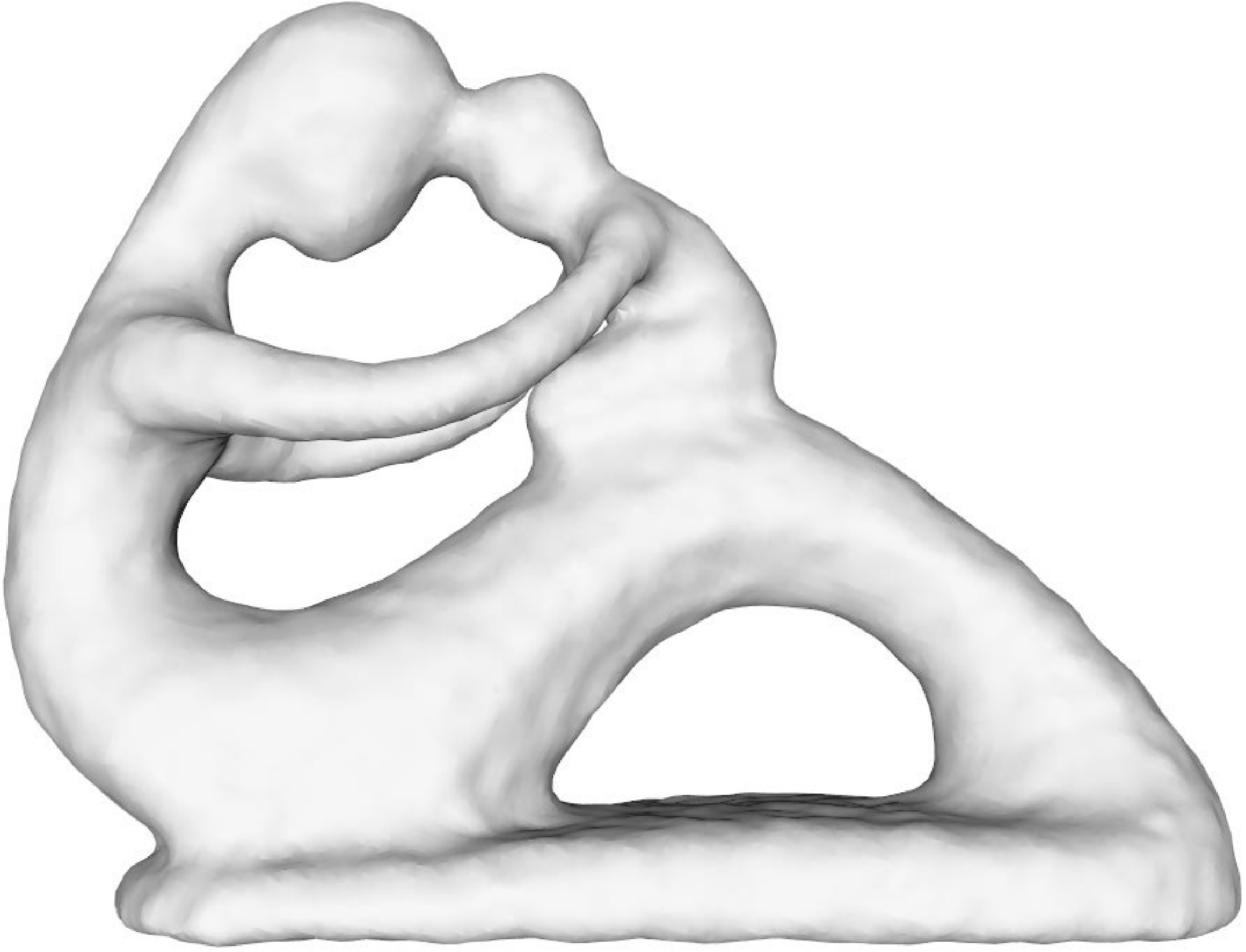}\\ 
        \includegraphics[width=1\textwidth  ]{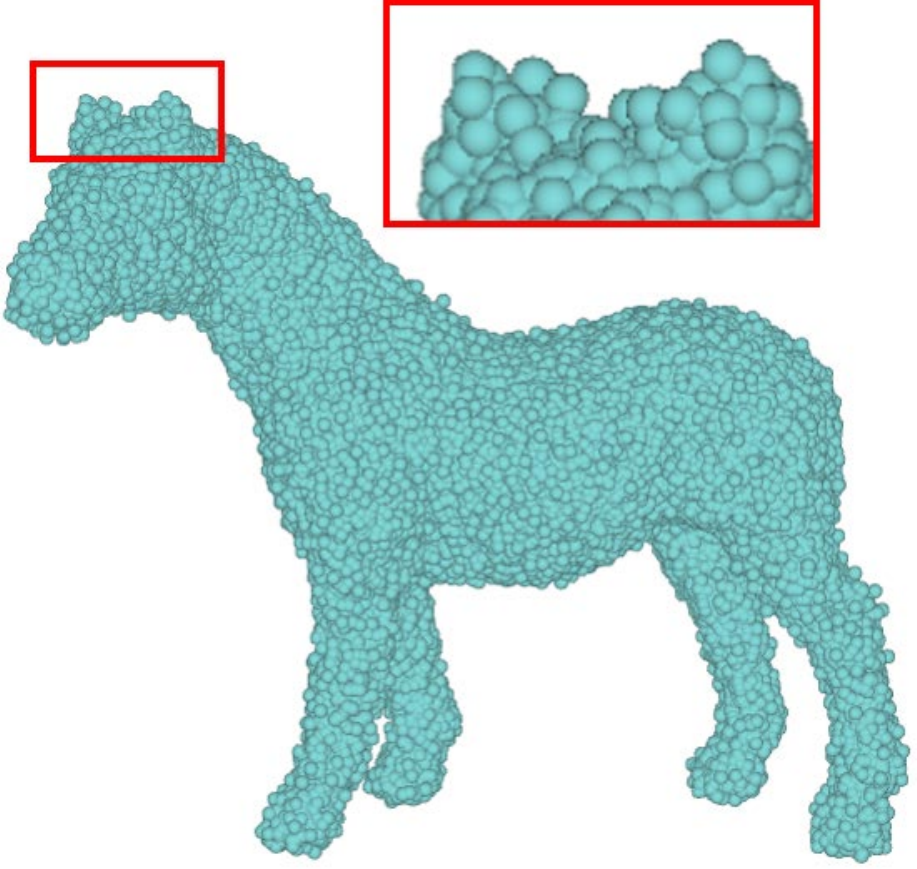}\\
        \includegraphics[width=1\textwidth  ]{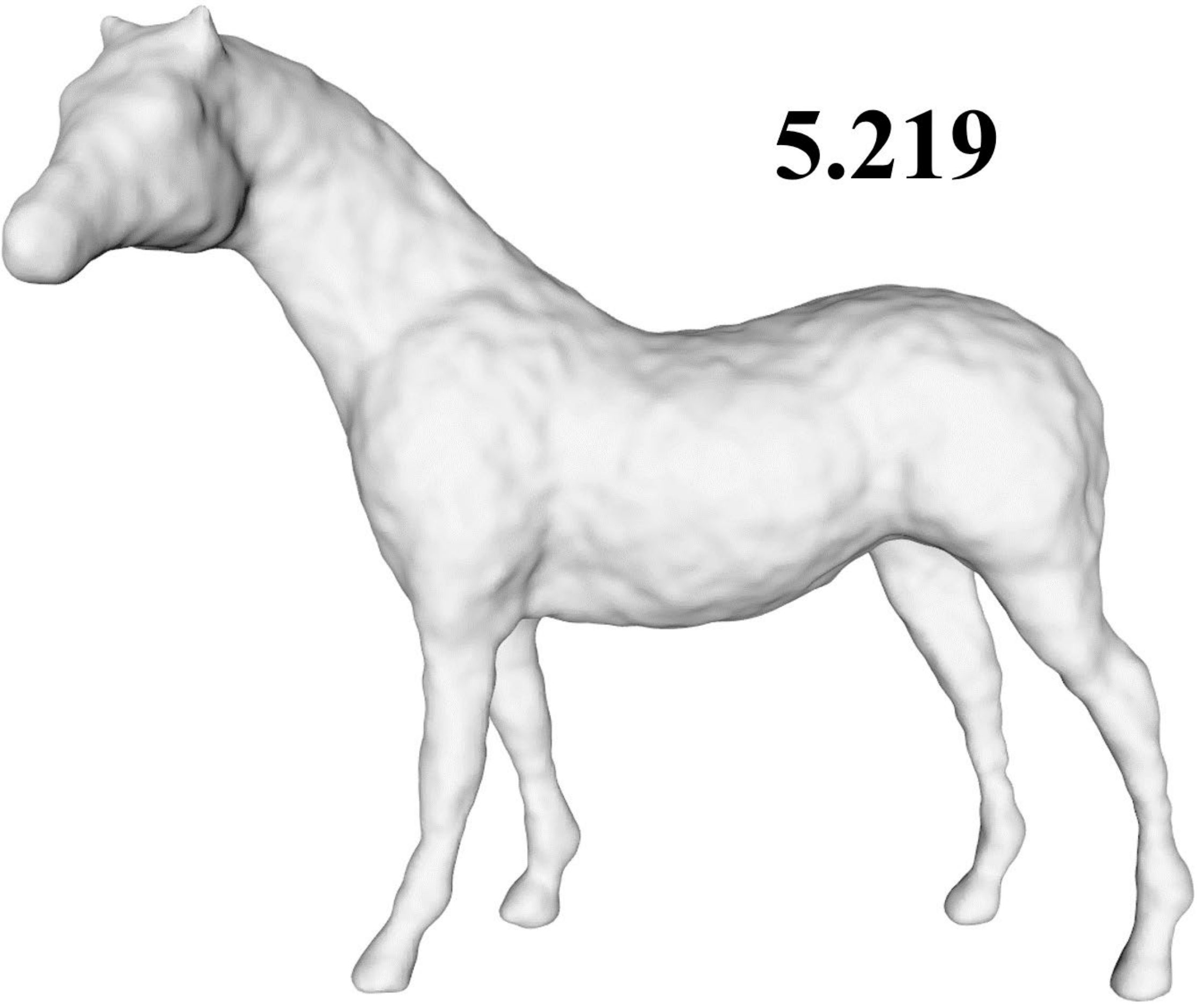}
        \end{minipage}
    }
    \subfigure[RIMLS]
    {
        \begin{minipage}[b]{0.105\textwidth}
        \includegraphics[width=1\textwidth  ]{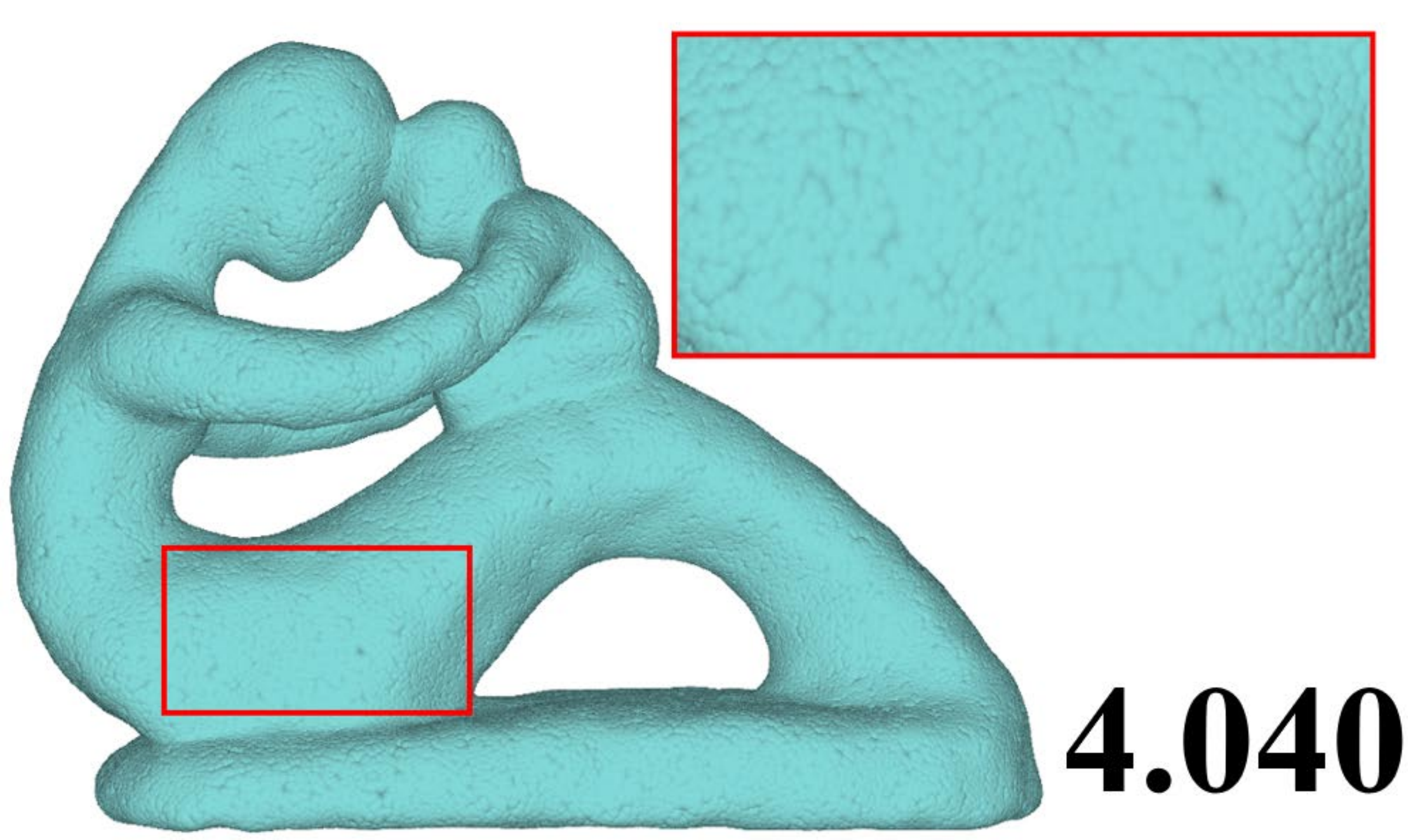}\\
        \includegraphics[width=1\textwidth  ]{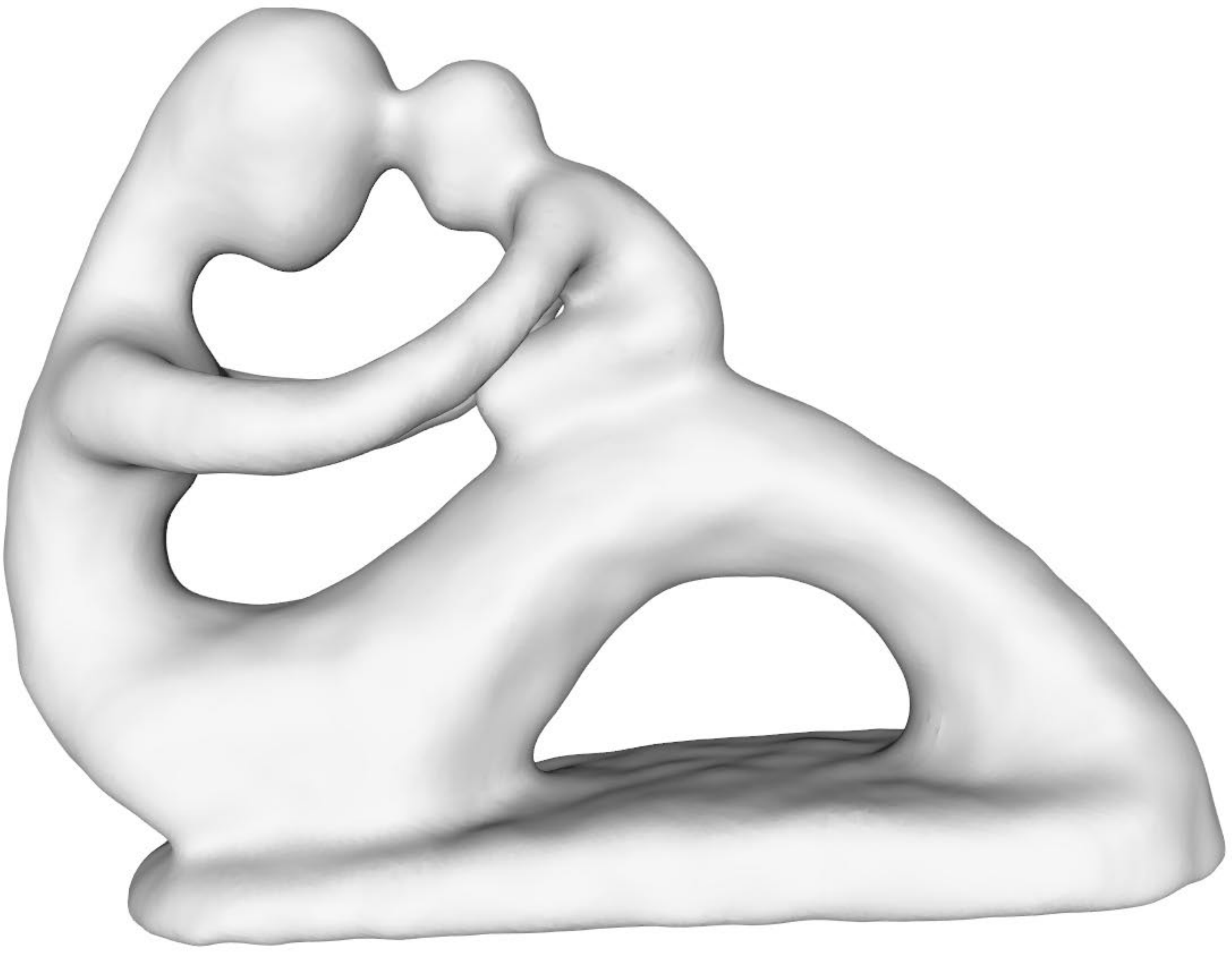}\\ 
        \includegraphics[width=1\textwidth  ]{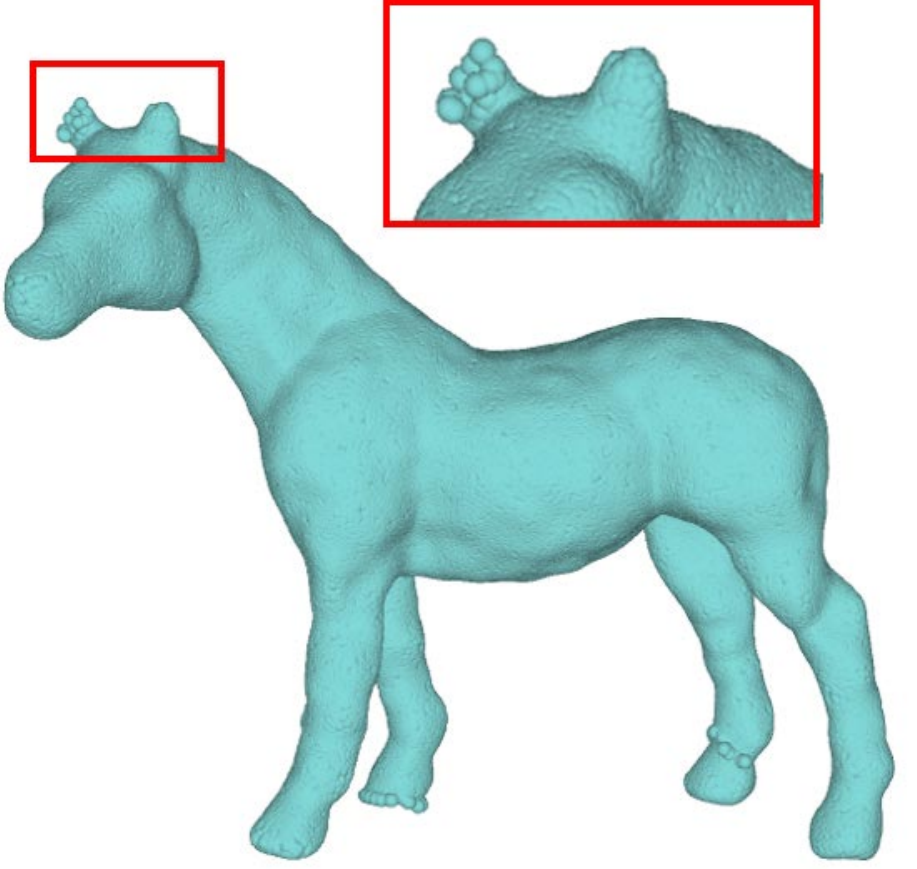}\\
        \includegraphics[width=1\textwidth  ]{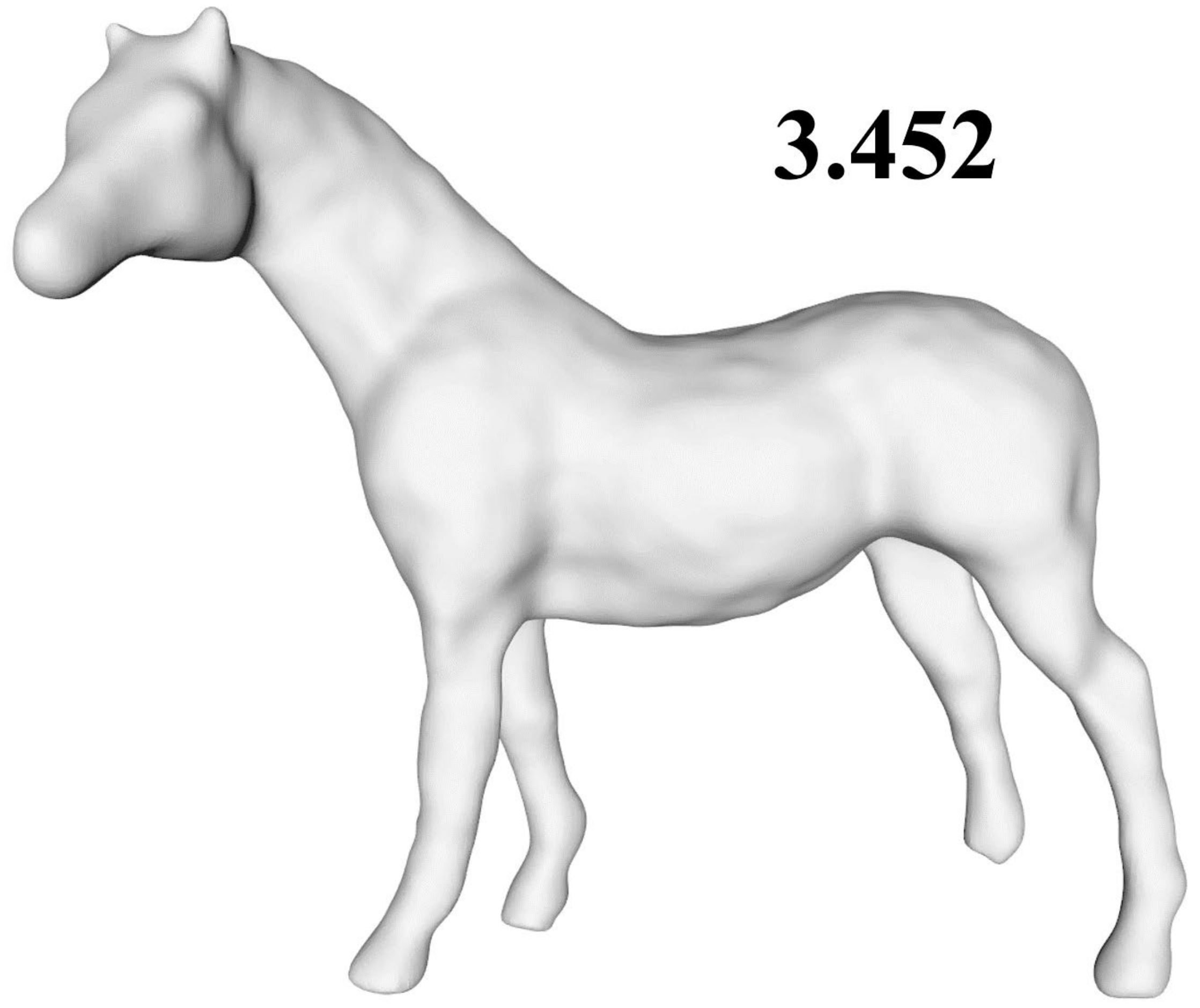}
        \end{minipage}
    }
    \subfigure[GPF]
    {
        \begin{minipage}[b]{0.105\textwidth}
        \includegraphics[width=1\textwidth  ]{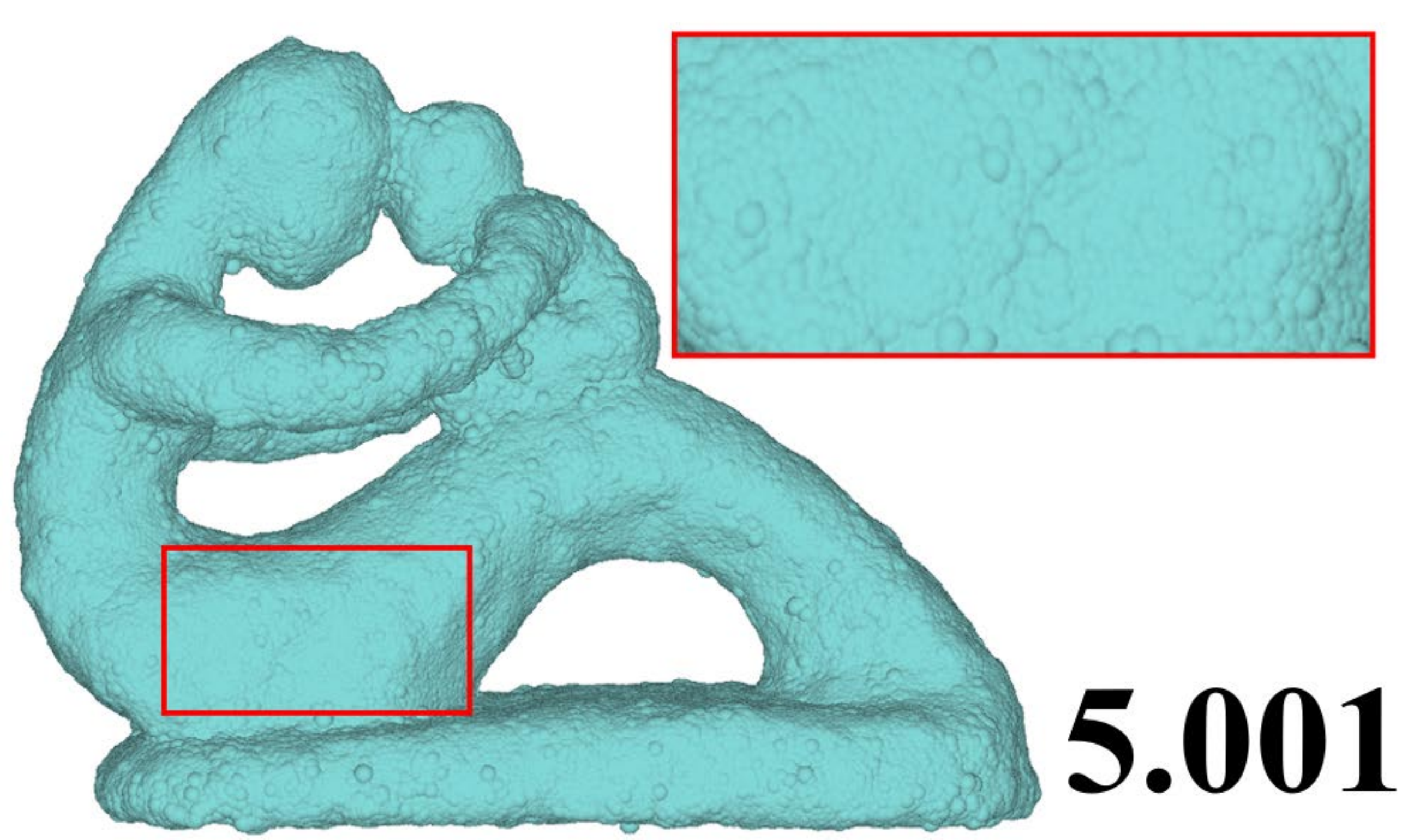}\\
        \includegraphics[width=1\textwidth  ]{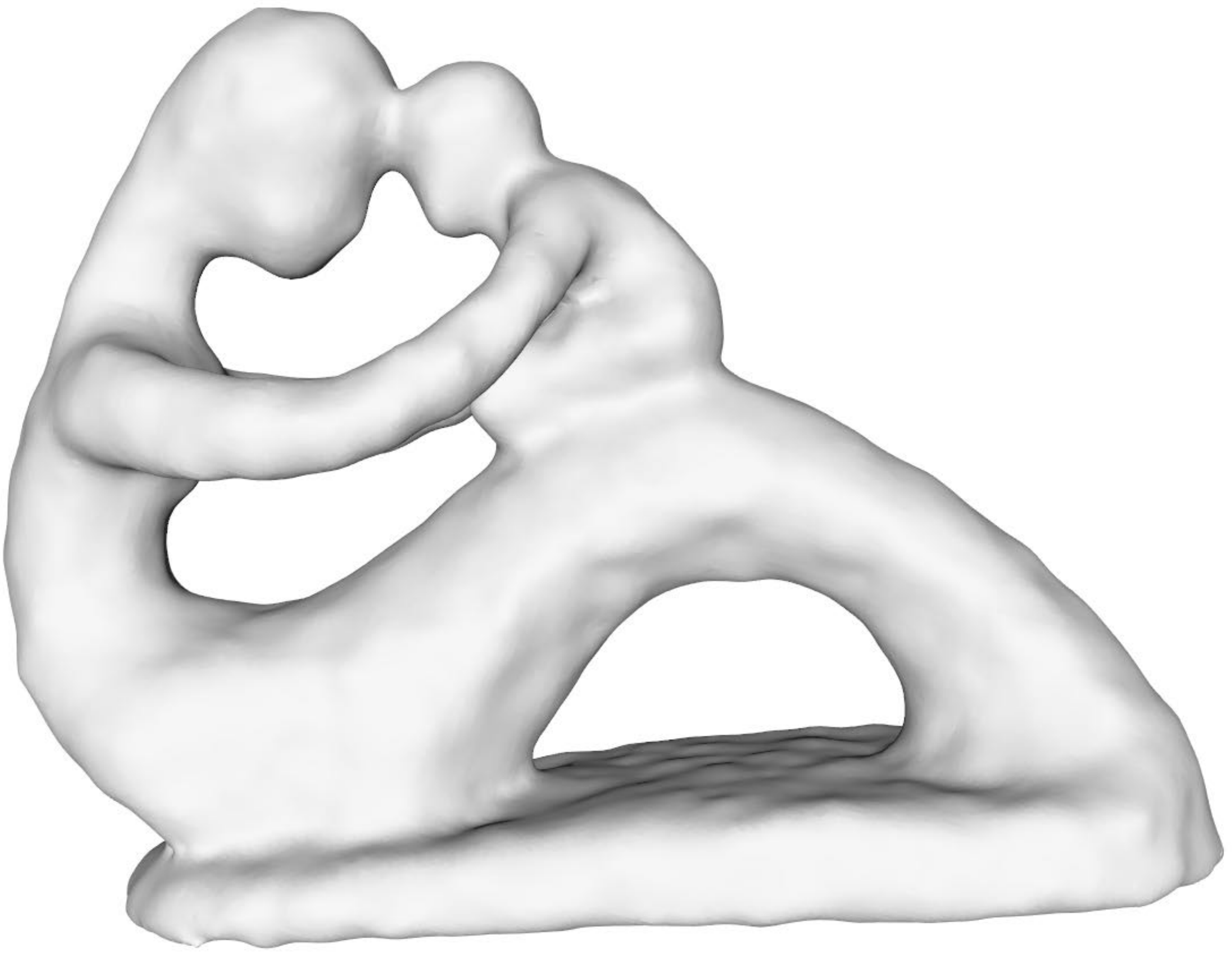}\\ 
        \includegraphics[width=1\textwidth  ]{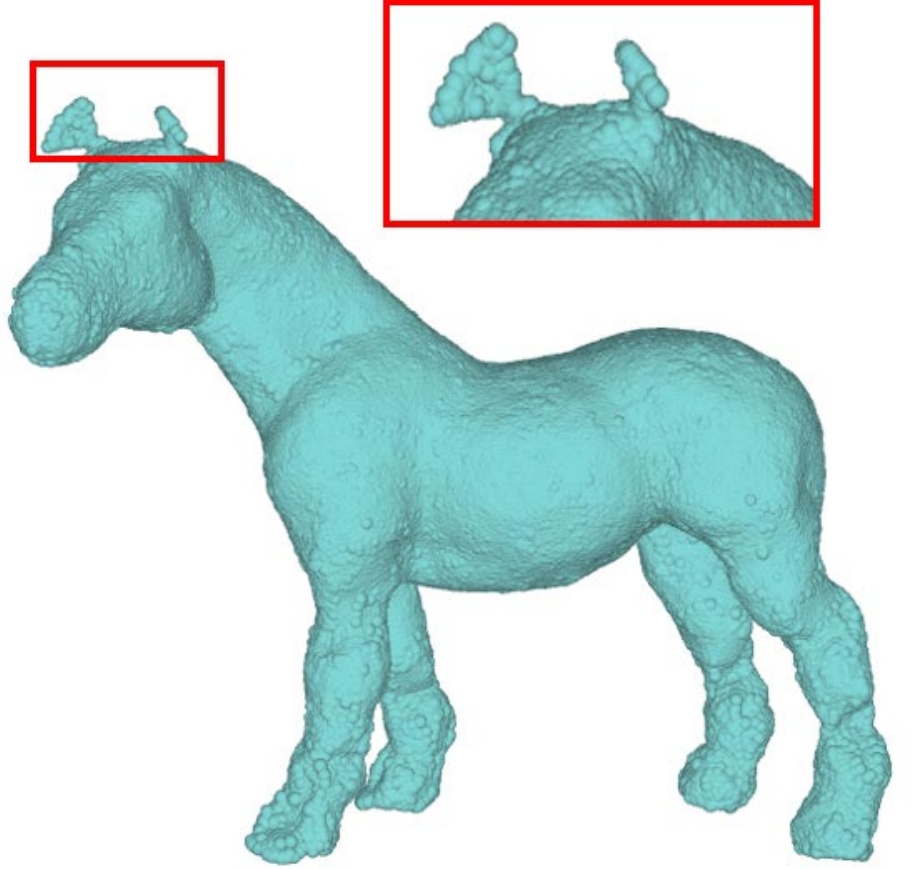}\\
        \includegraphics[width=1\textwidth  ]{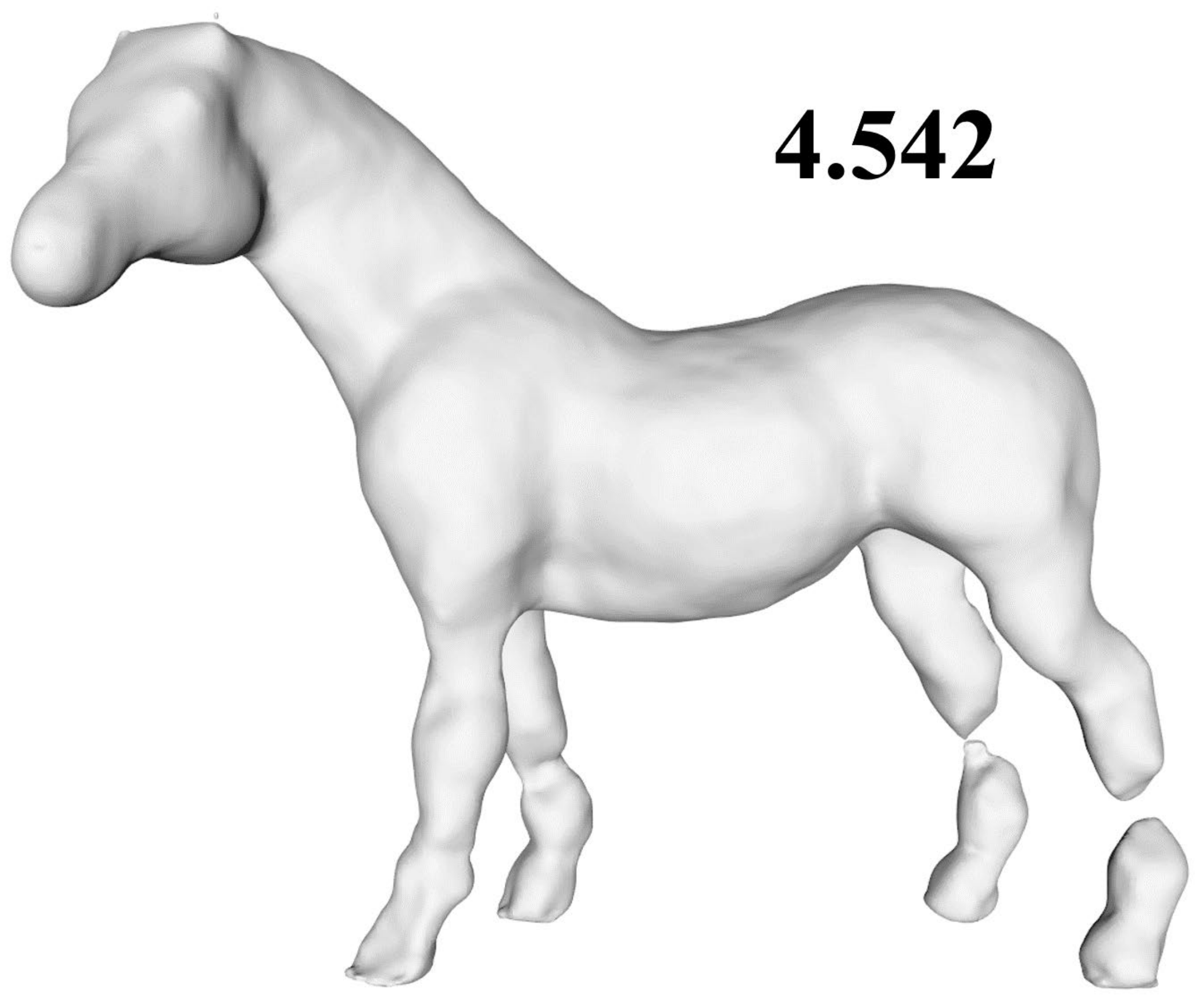}
        \end{minipage}
    }
    \subfigure[WLOP]
    {
        \begin{minipage}[b]{0.105\textwidth}
        \includegraphics[width=1\textwidth  ]{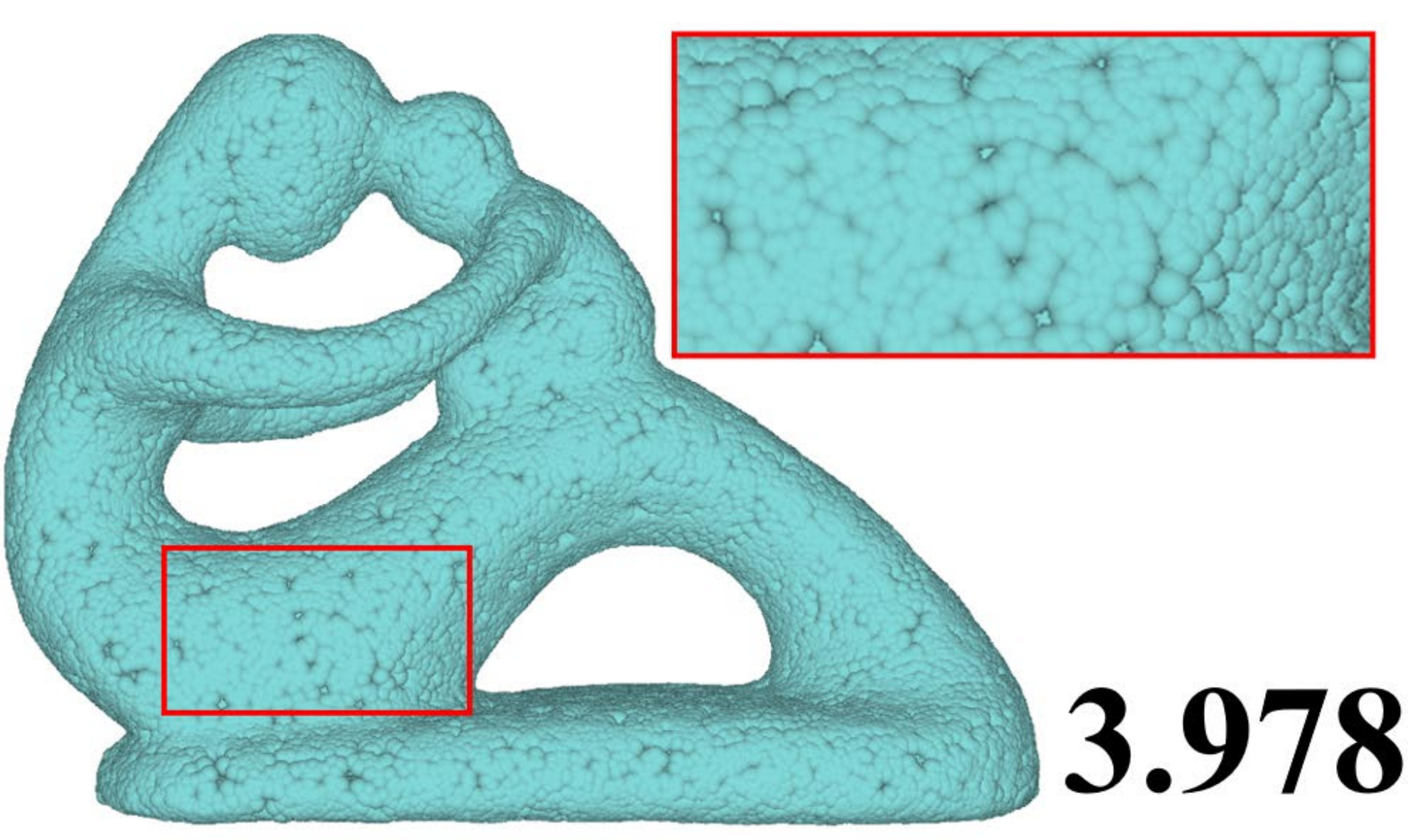}\\
        \includegraphics[width=1\textwidth  ]{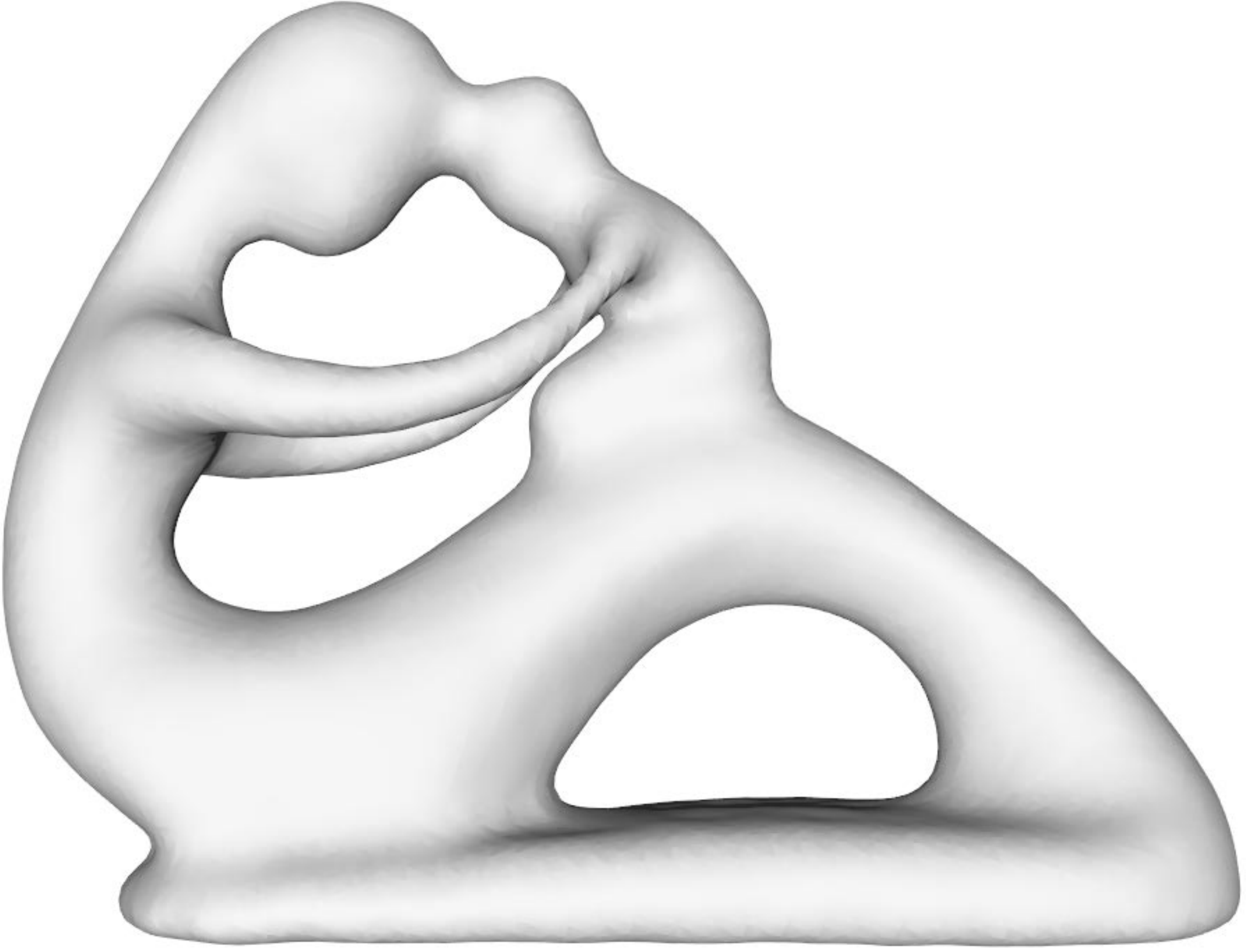}\\ 
        \includegraphics[width=1\textwidth  ]{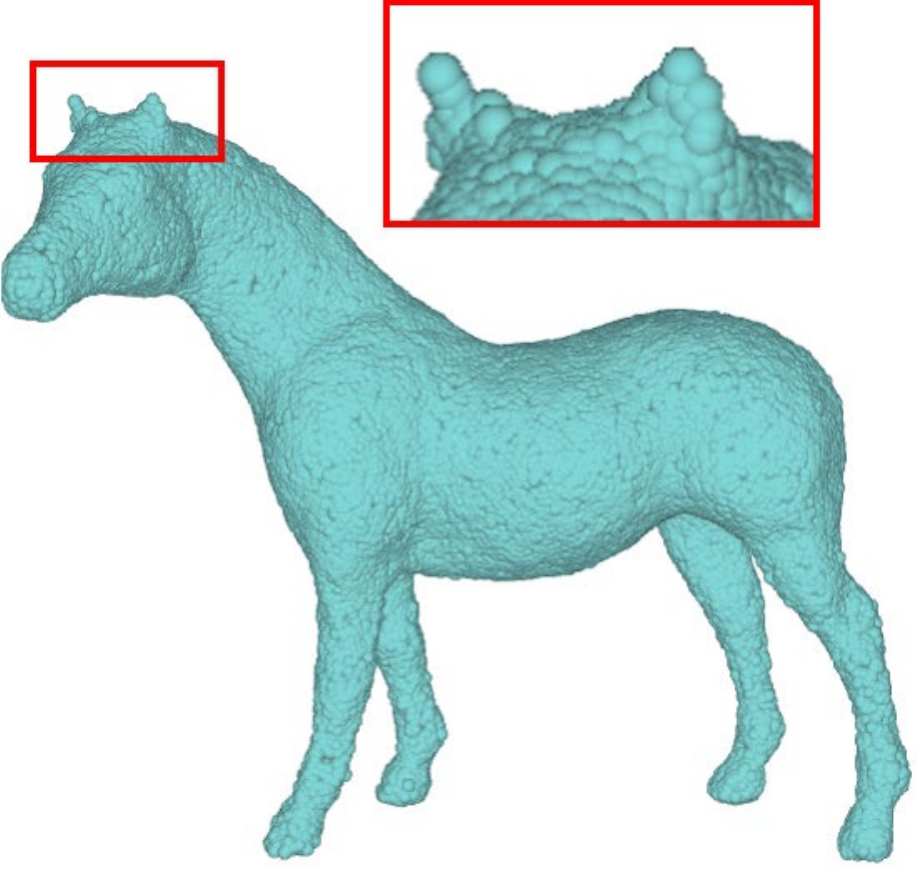}\\
        \includegraphics[width=1\textwidth  ]{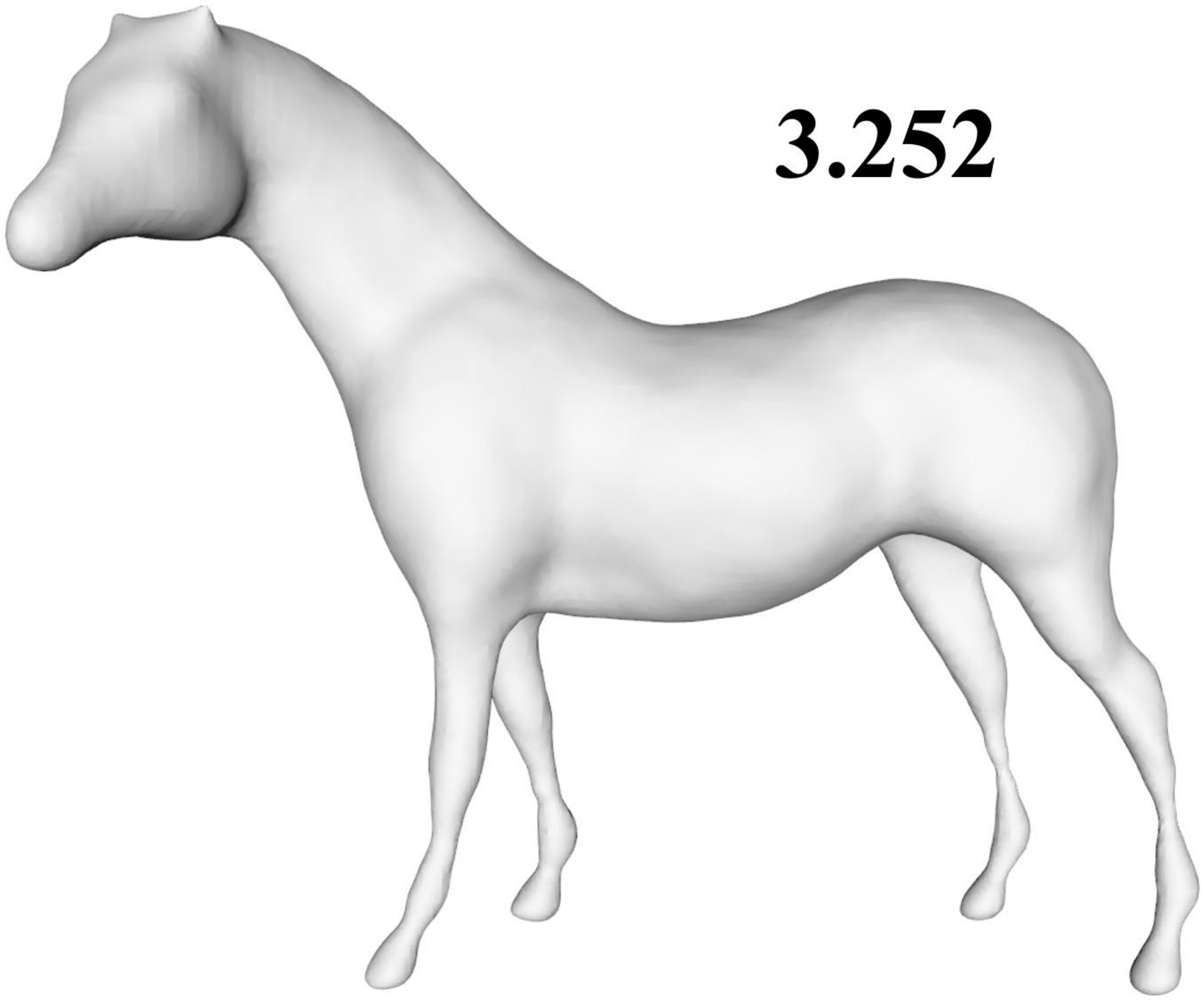}
        \end{minipage}
    }
    \subfigure[CLOP]
    {
        \begin{minipage}[b]{0.105\textwidth}
        \includegraphics[width=1\textwidth  ]{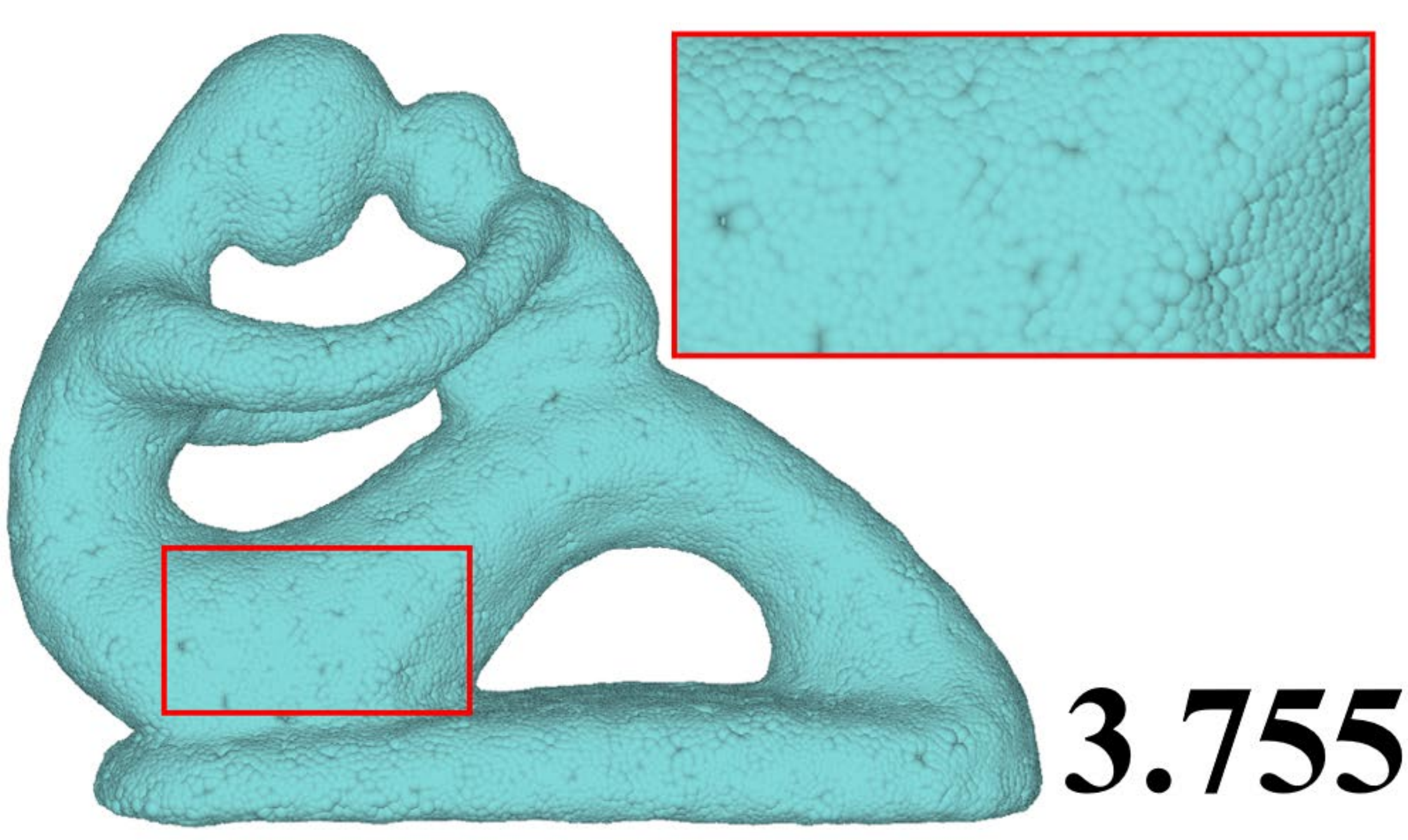}\\
        \includegraphics[width=1\textwidth  ]{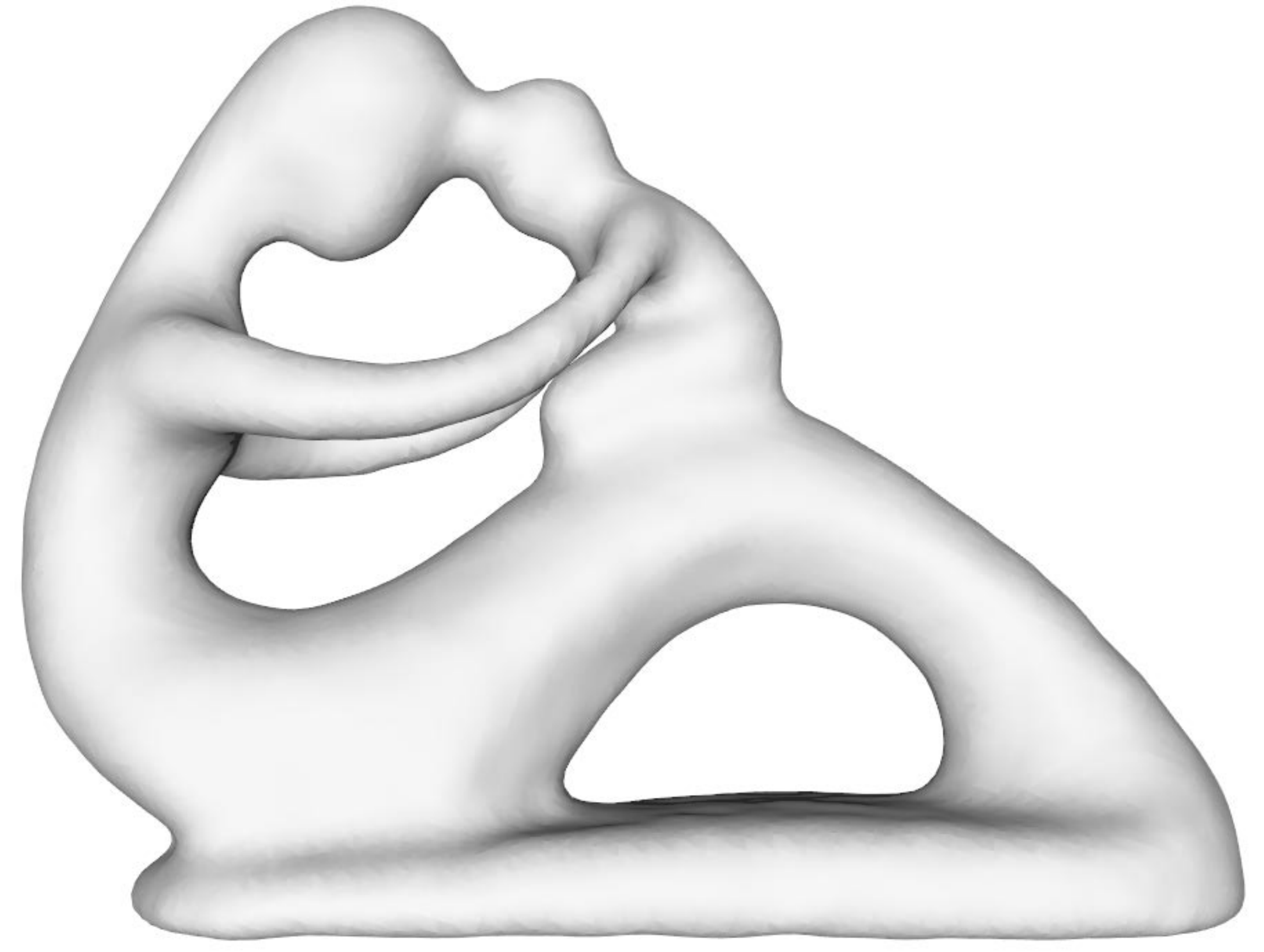}\\ 
        \includegraphics[width=1\textwidth  ]{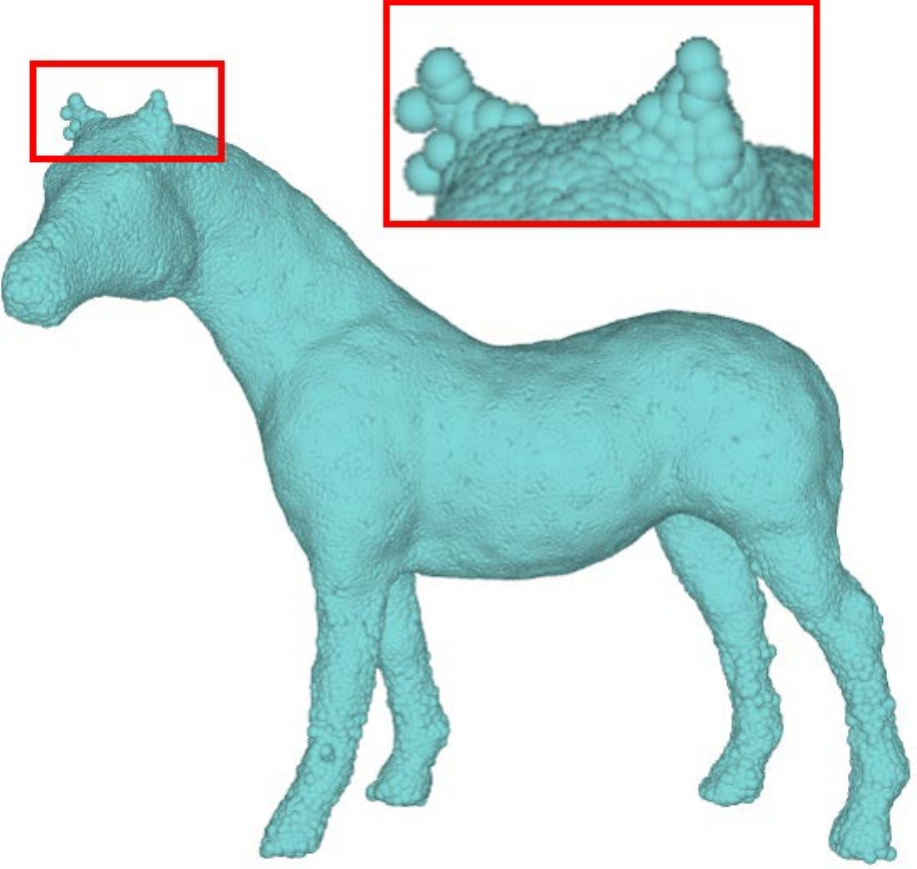}\\
        \includegraphics[width=1\textwidth  ]{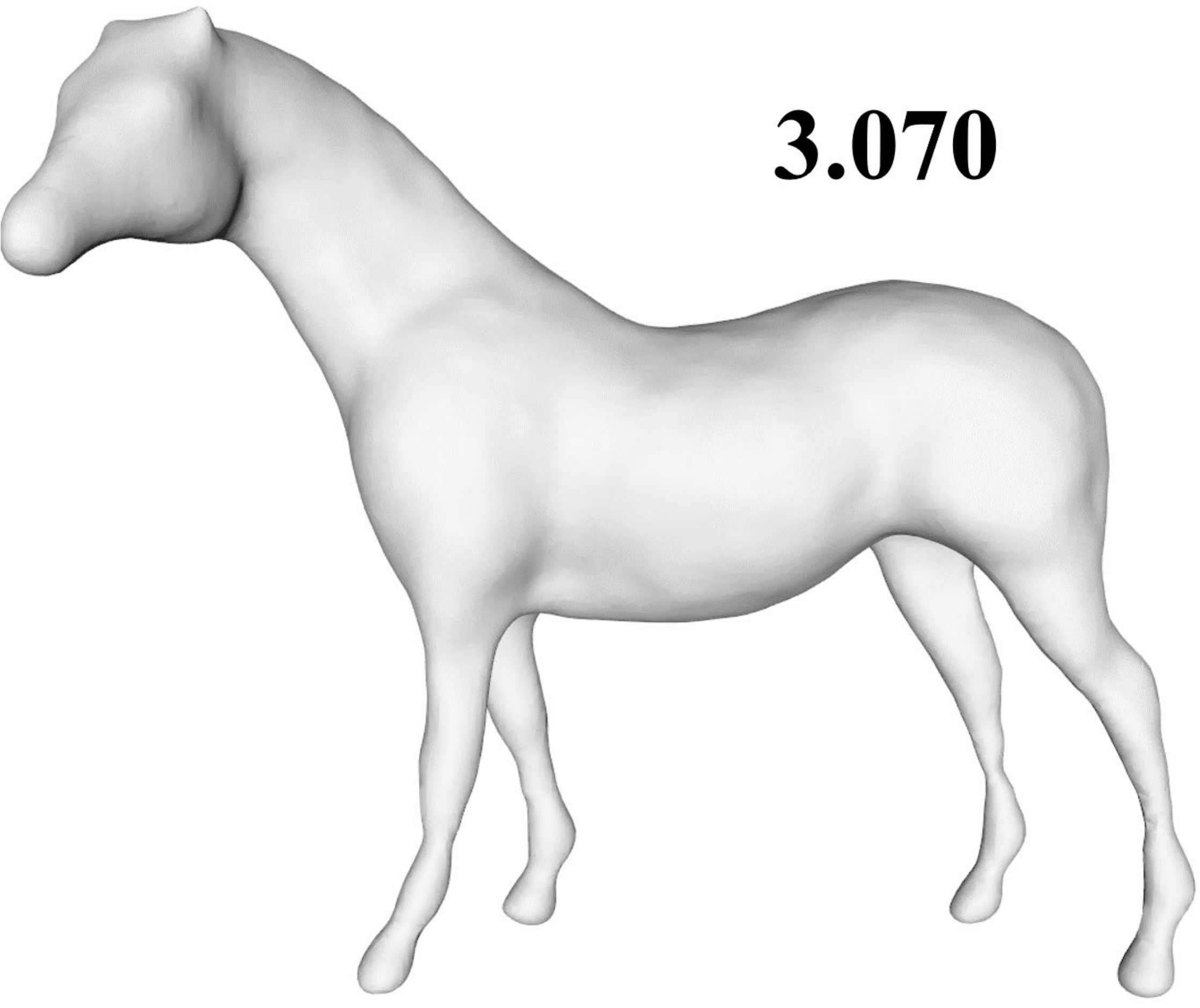}
        \end{minipage}
    }
    \subfigure[PCN]
    {
        \begin{minipage}[b]{0.105\textwidth}
        \includegraphics[width=1\textwidth  ]{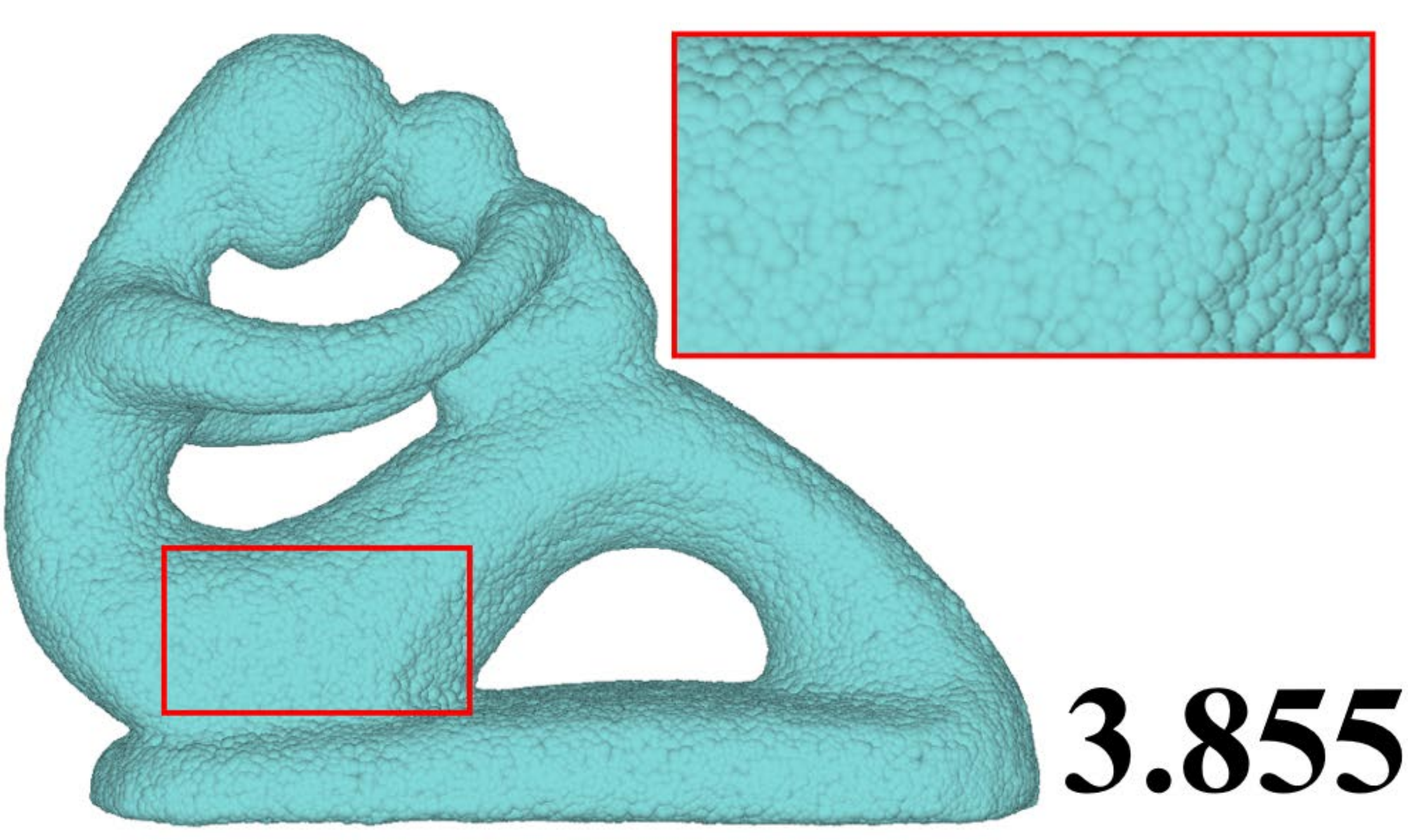}\\
        \includegraphics[width=1\textwidth  ]{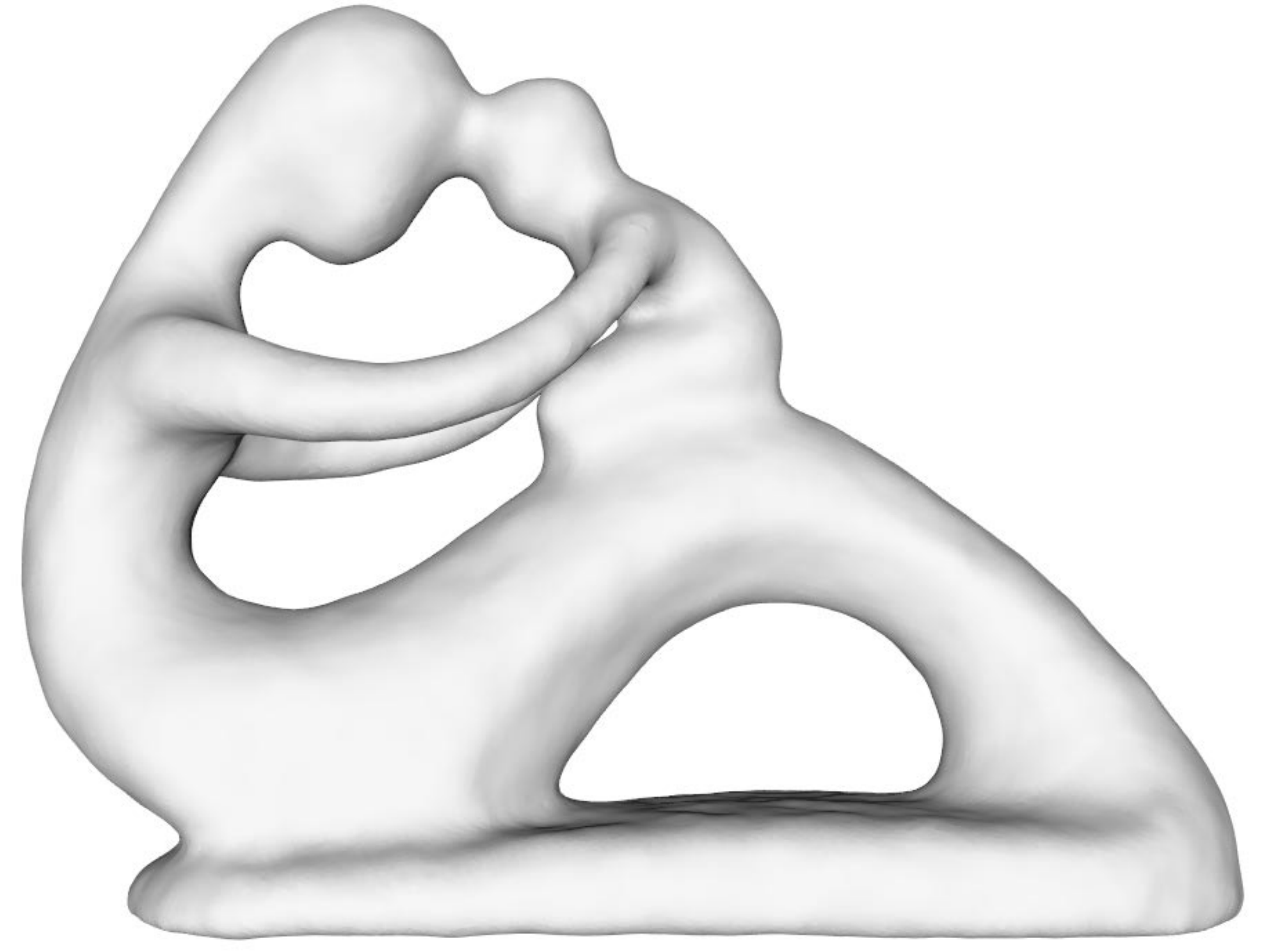}\\ 
        \includegraphics[width=1\textwidth  ]{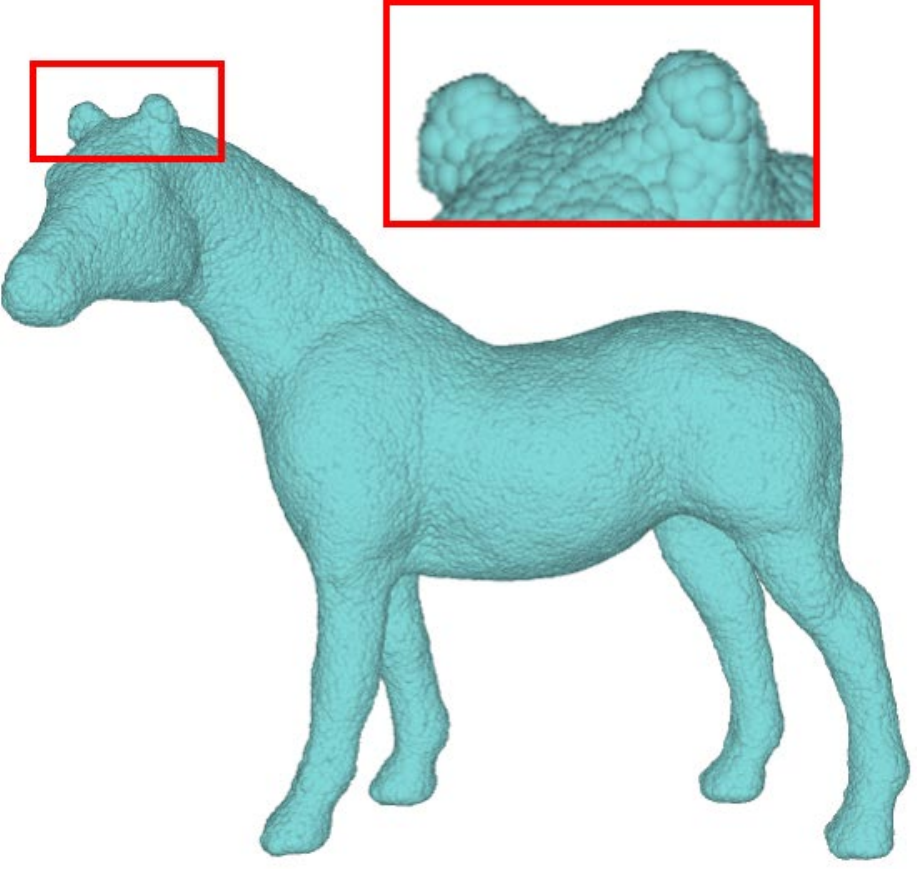}\\
        \includegraphics[width=1\textwidth  ]{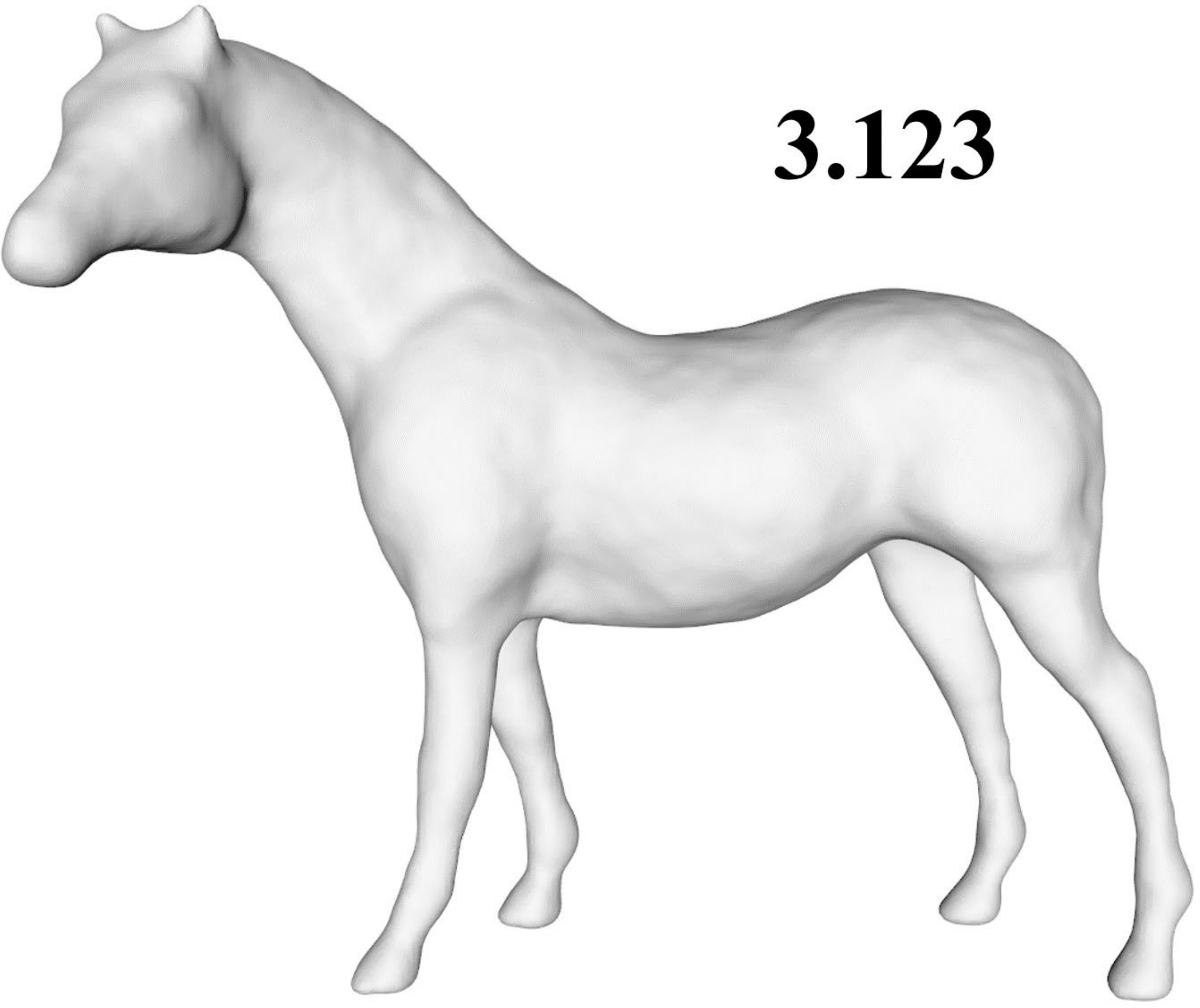}
        \end{minipage}
    }
    \subfigure[TD]
    {
        \begin{minipage}[b]{0.105\textwidth}
        \includegraphics[width=1\textwidth  ]{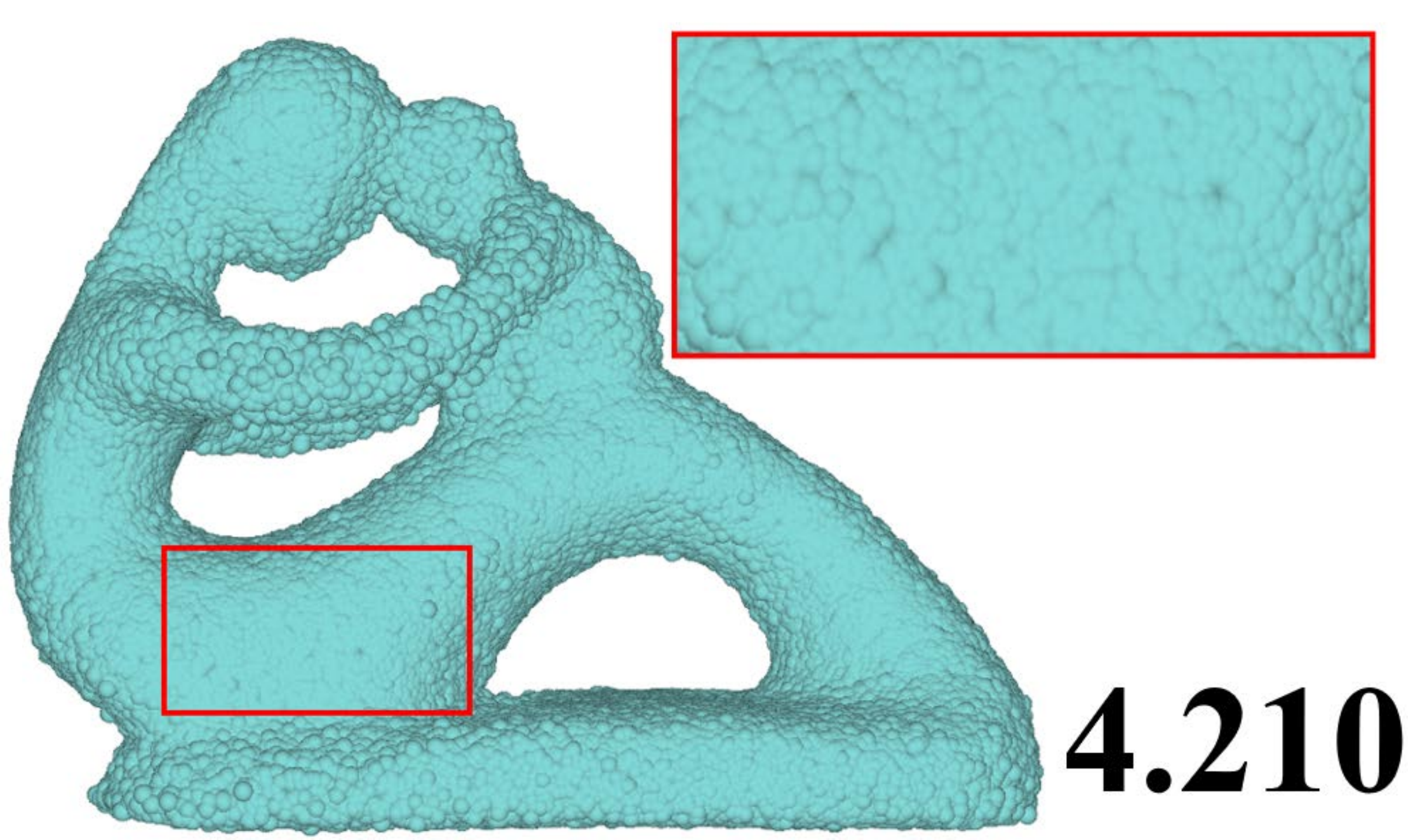}\\
        \includegraphics[width=1\textwidth  ]{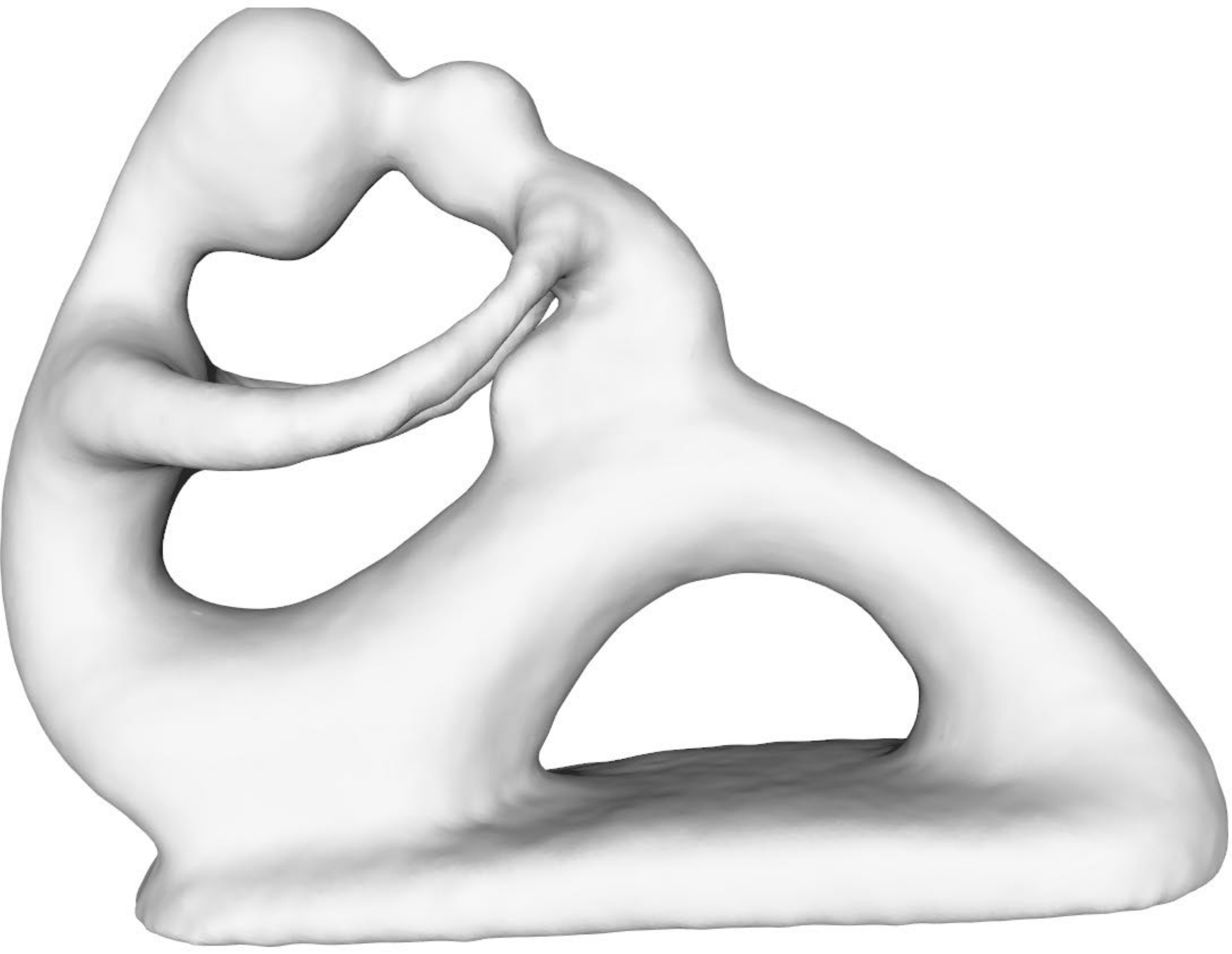}\\ 
        \includegraphics[width=1\textwidth  ]{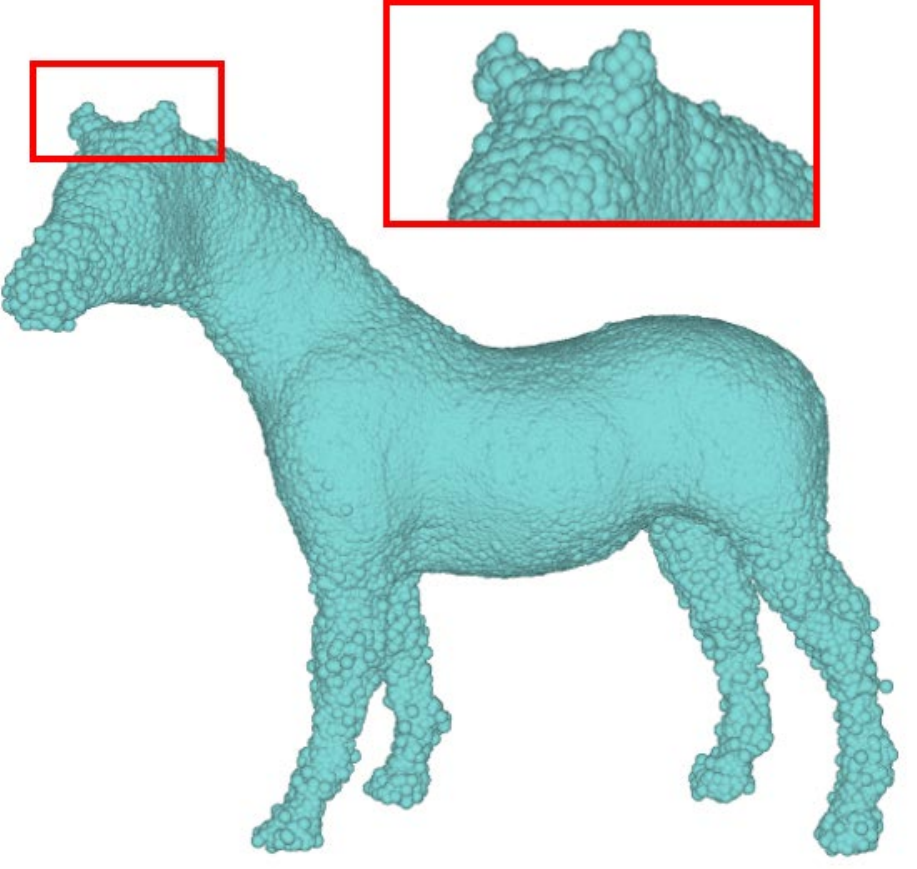}\\
        \includegraphics[width=1\textwidth  ]{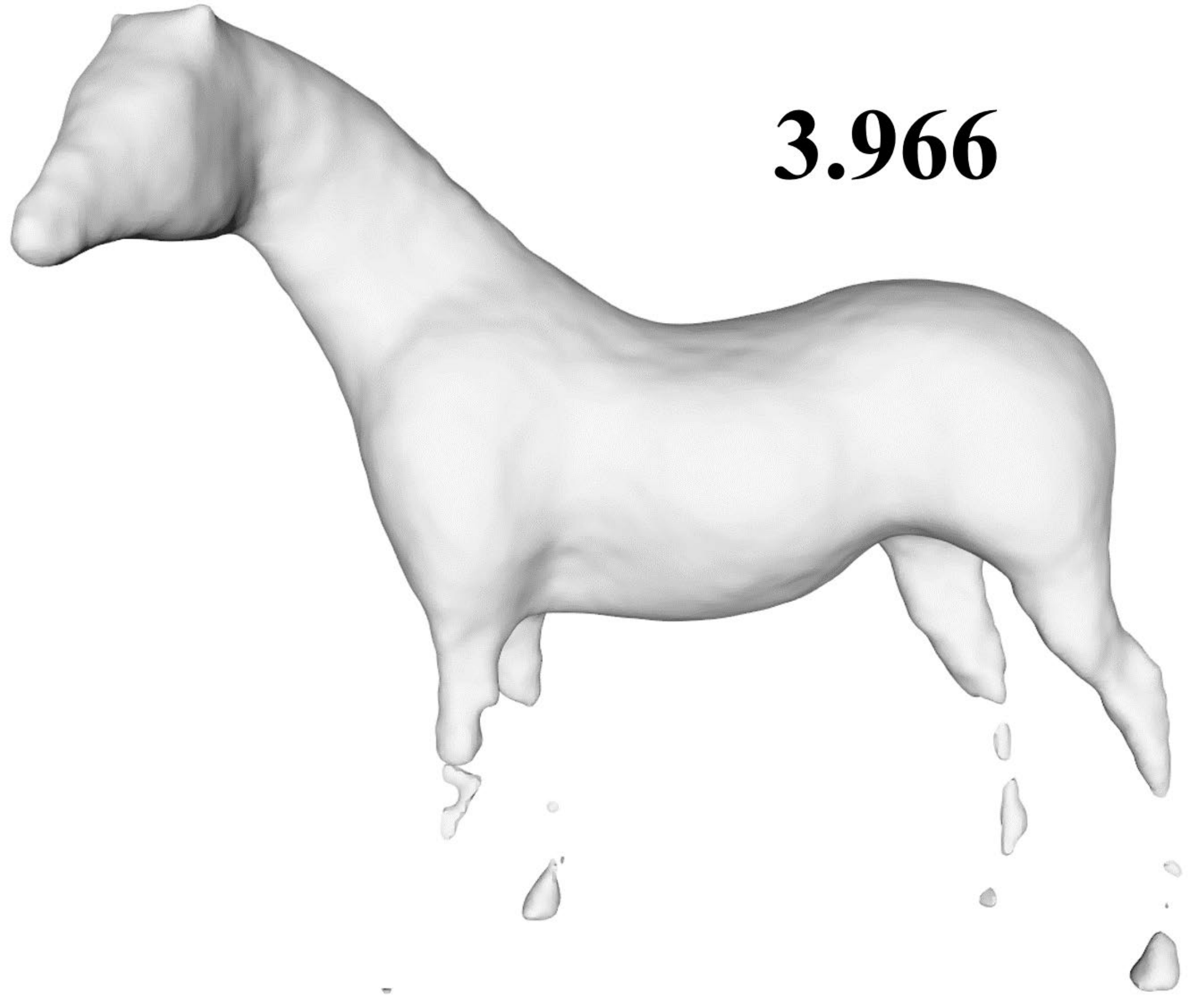}
        \end{minipage}
    }
    \subfigure[Ours]
    {
        \begin{minipage}[b]{0.105\textwidth}
        \includegraphics[width=1\textwidth  ]{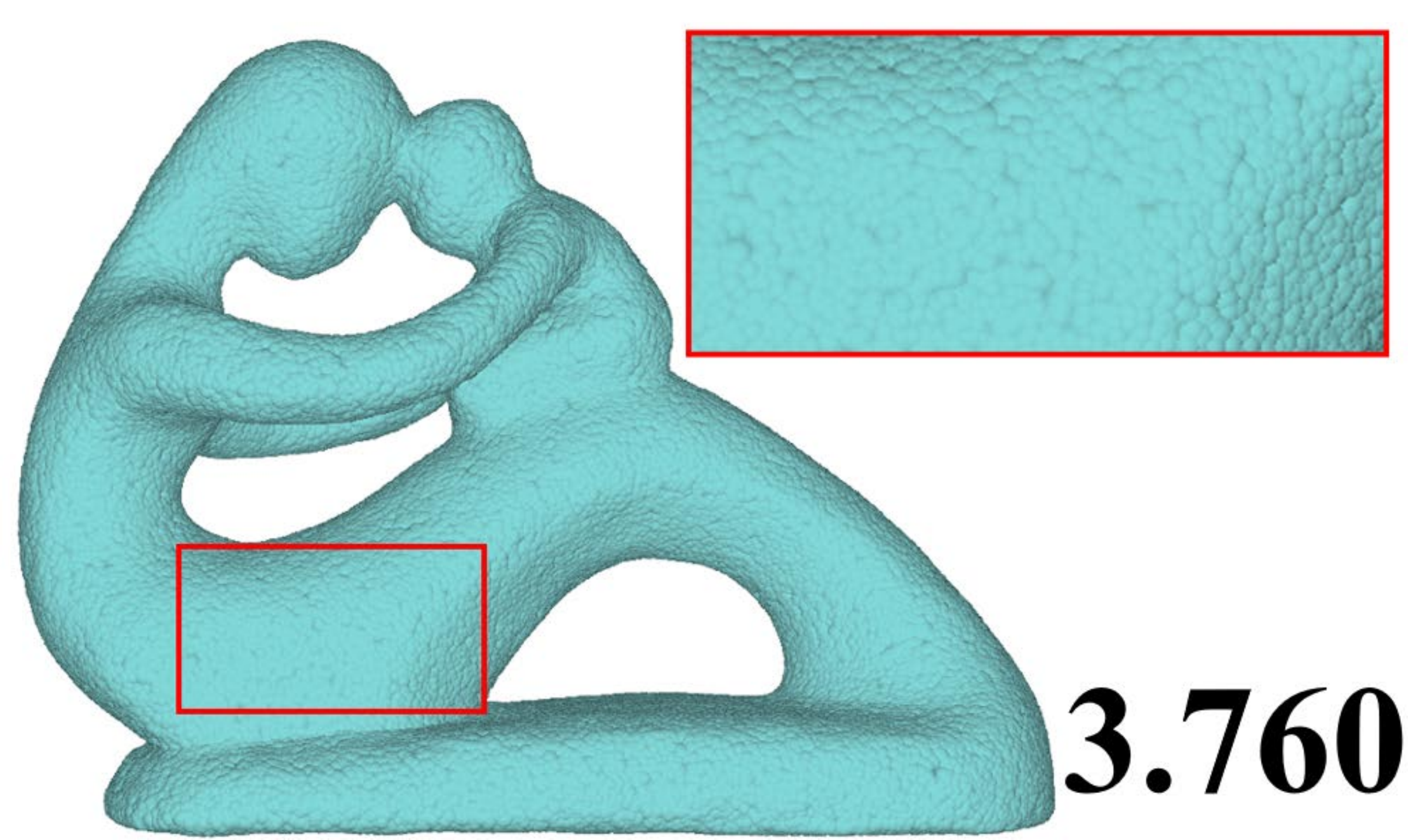}\\
        \includegraphics[width=1\textwidth  ]{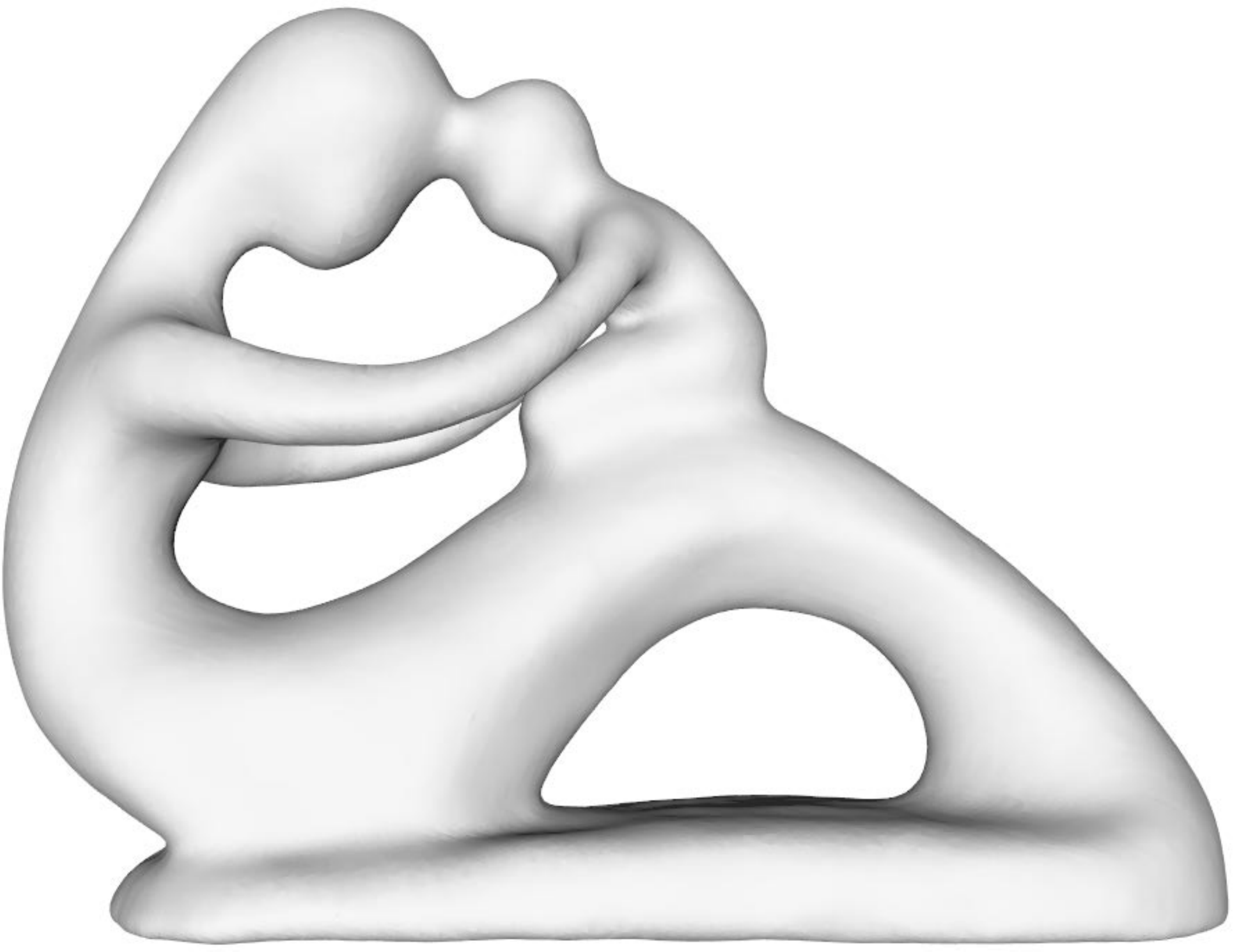}\\
        \includegraphics[width=1\textwidth  ]{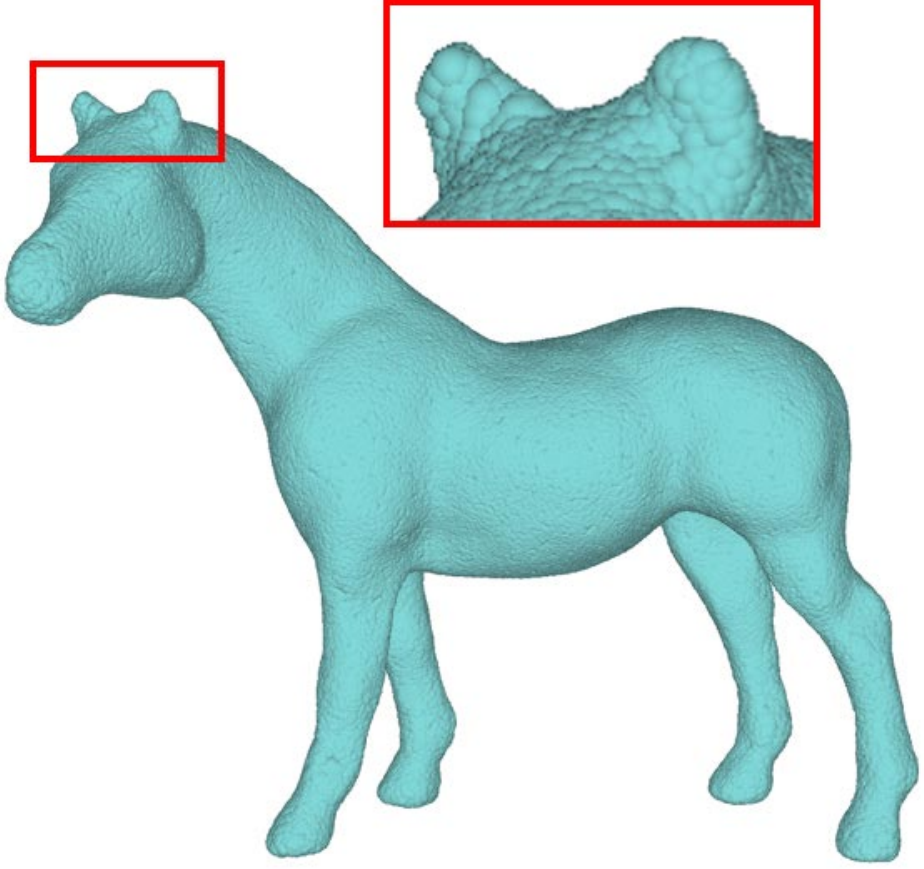}\\
        \includegraphics[width=1\textwidth  ]{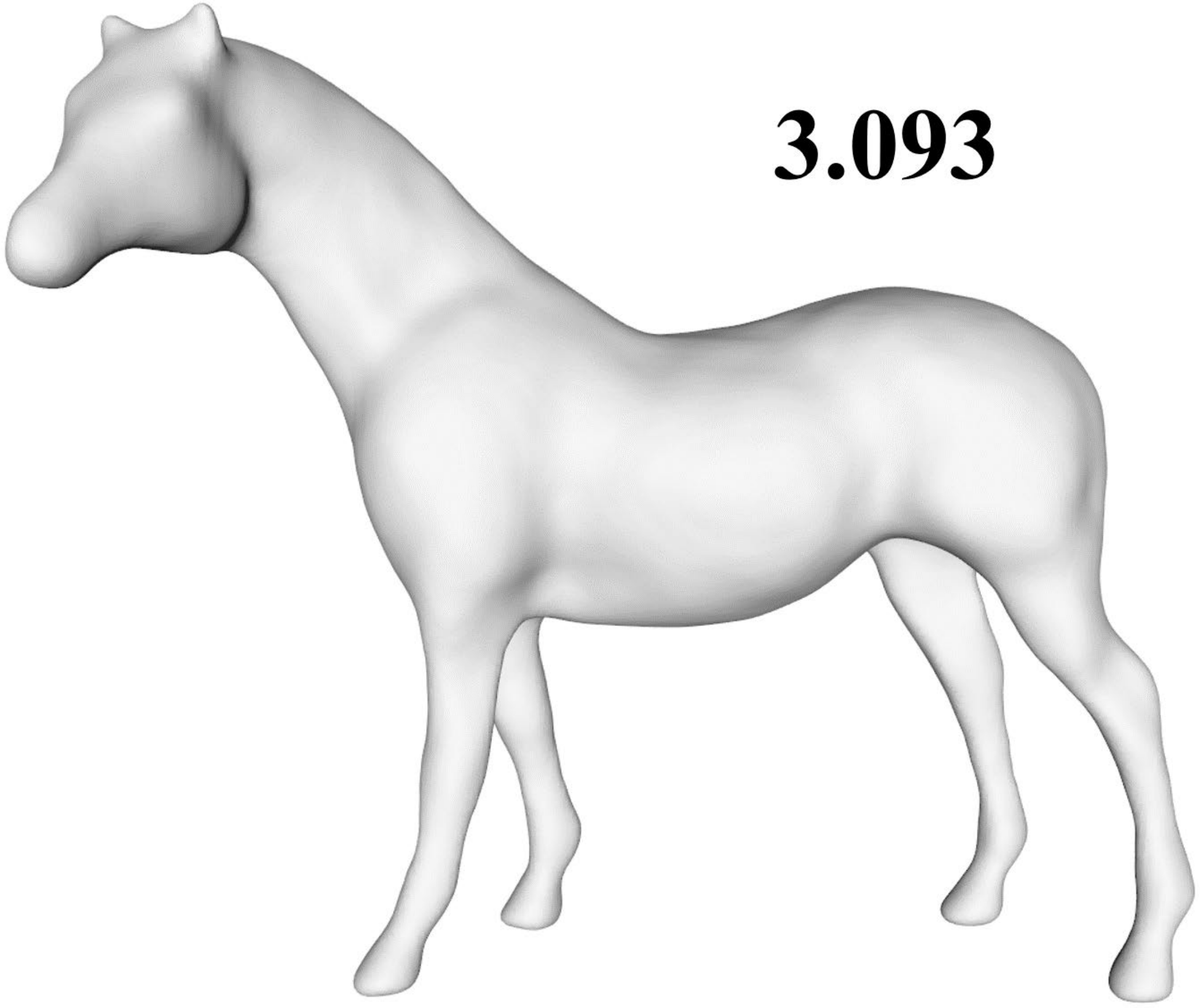}
        \end{minipage}
    }
    \caption{Visual comparison of point clouds filtering with $0.5\%$ synthetic noise. The overall MSEs ($\times 10^{-3}$) for different methods are shown in the figure.}
    \label{fig:surfacereconstruction}
\end{figure*}

\subsection{Compared Techniques}
\label{sec:comparedtechniques}
We compare our approach with the state-of-the-art point cloud filtering techniques, namely WLOP \cite{Huang2009TOG}, CLOP \cite{Preiner2014TOG}, RIMLS \cite{Ztireli2009CGF}, GPF \cite{Lu2018TVCG}, EC-Net \cite{Yu2018ECCV}, PointCleanNet (PCN) \cite{Rakotosaona2019CGF} and TotalDenoising (TD) \cite{hermosilla2019ICCVl}. Specifically, RIMLS and GPF are designed to preserve sharp features by incorporating smooth normals which are achieved by the bilateral smoothing \cite{Huang2013TOG}. For fair comparisons and visualization purposes, we (i) tune the main parameters of each state-of-the-art technique to achieve as good visual results as possible (EC-Net, PCN and our method have fixed parameters); (ii) employ the same surface reconstruction parameters for the same model. Notice that surface reconstruction is straightforwardly applied to the filtered point sets. As for PCN and TotalDenoising, we use the source codes released by the authors to train a new model over our training dataset. Since EC-Net requires manually labelling polylines of sharp edges for training, we simply utilize the trained model released by the authors instead. 
We compare our method with these methods, in terms of both visual quality and quantity (if ground truth is available). 
\begin{figure*}[htb!]
    \centering
    \subfigure[RIMLS]
    {
        \begin{minipage}[b]{0.11\textwidth} 
        \includegraphics[width=1\textwidth  ]{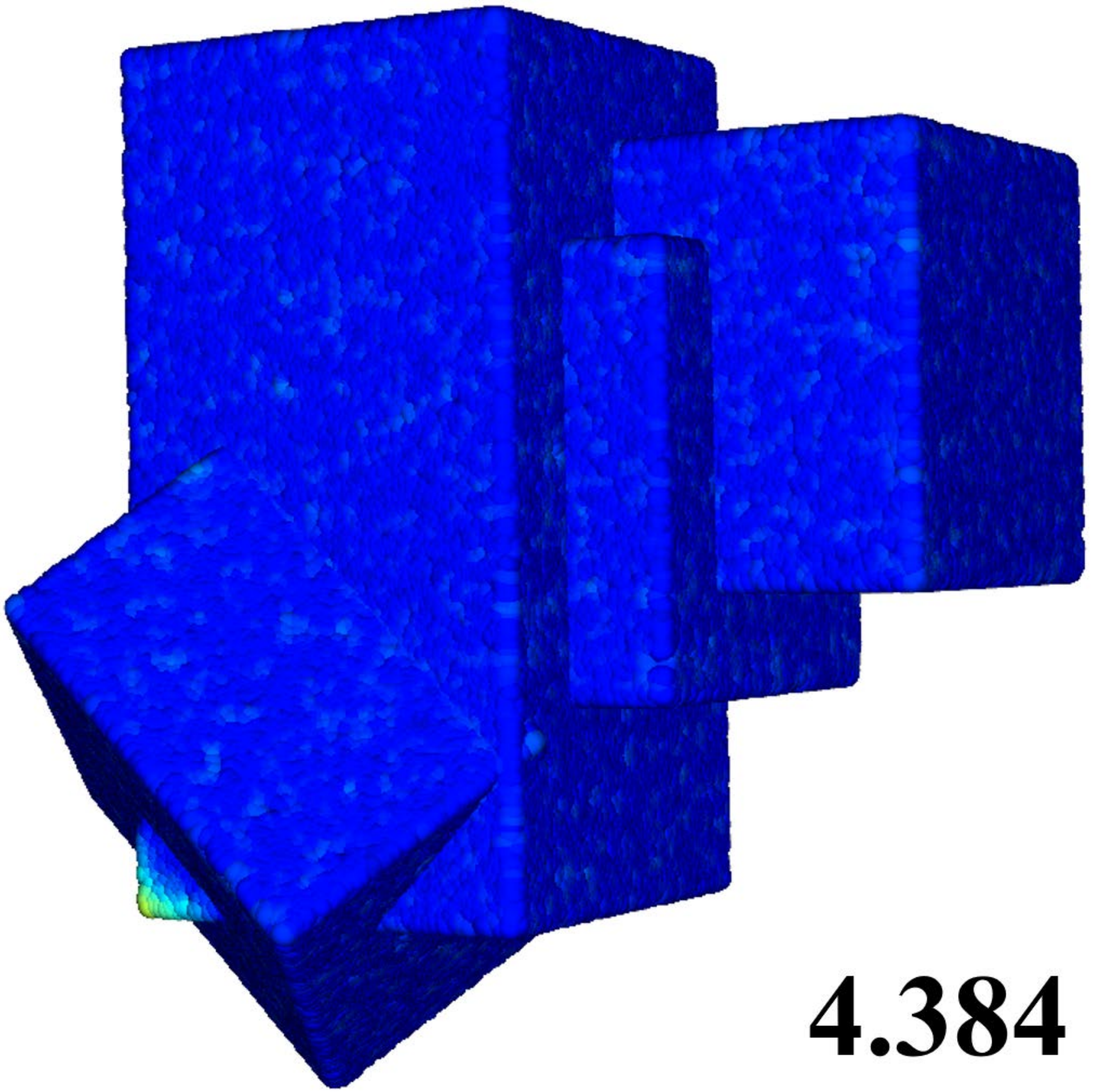}\\ 
        \includegraphics[width=1\textwidth  ]{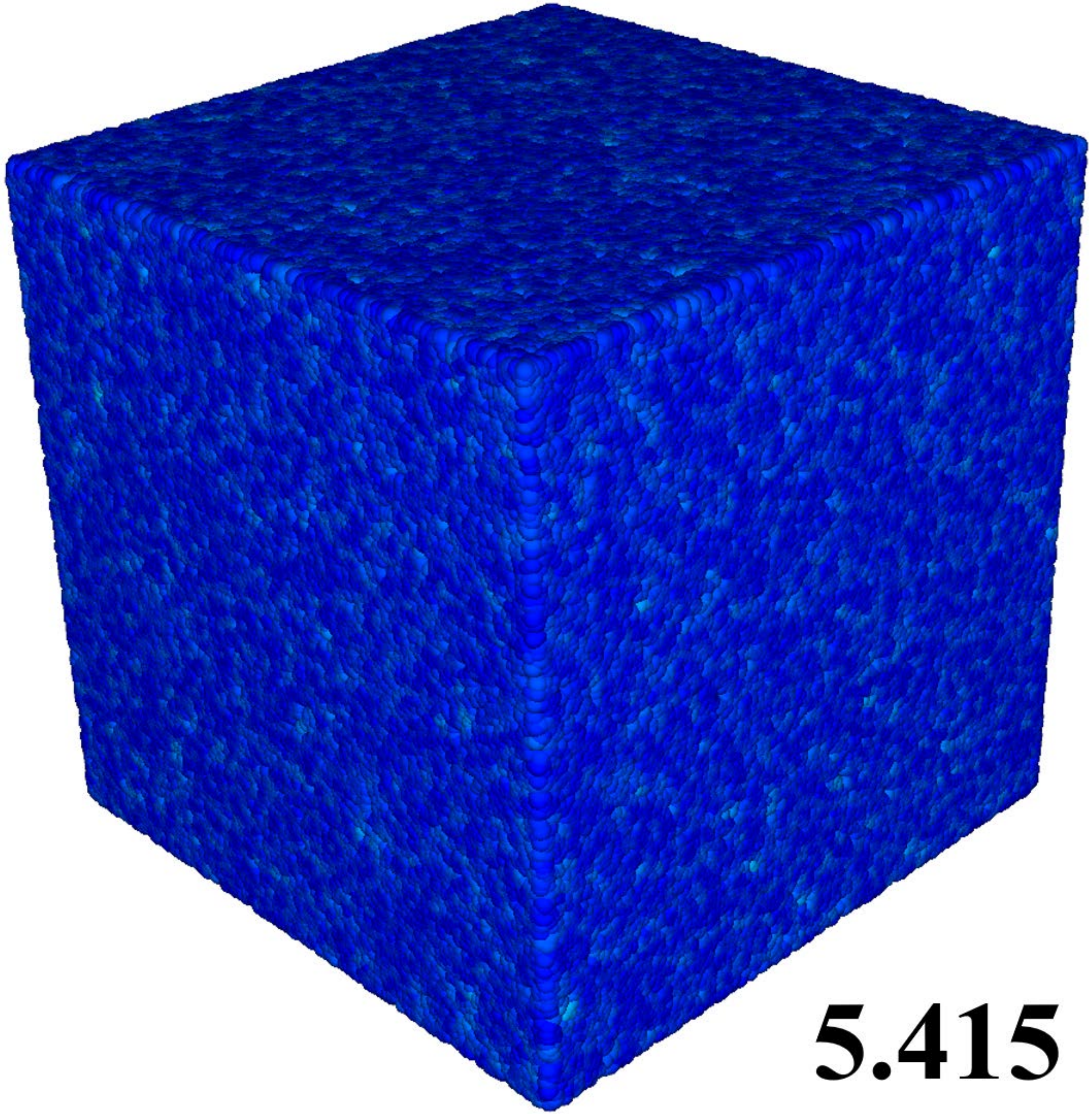}\\ 
        \includegraphics[width=1\textwidth  ]{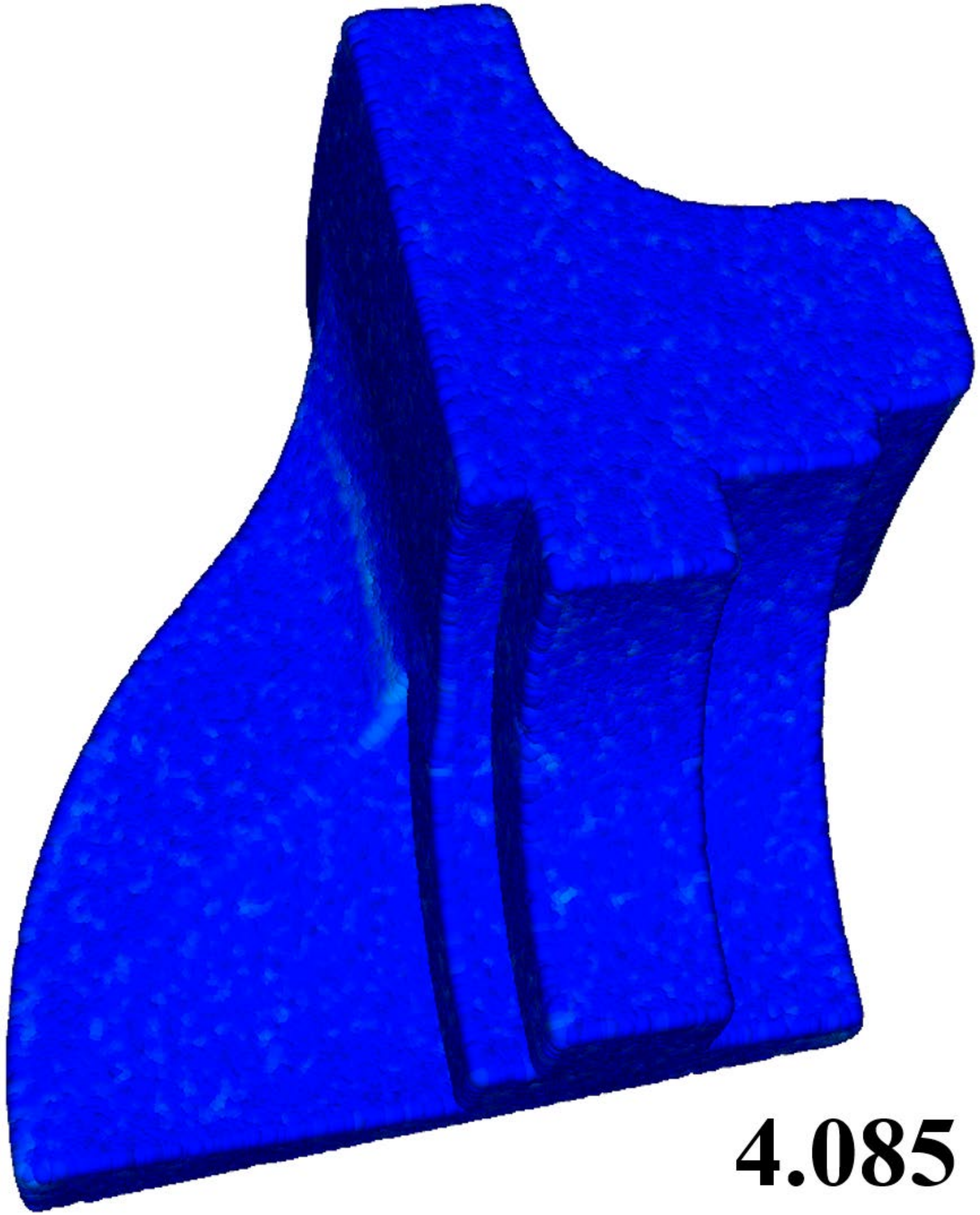}\\
        \includegraphics[width=1\textwidth  ]{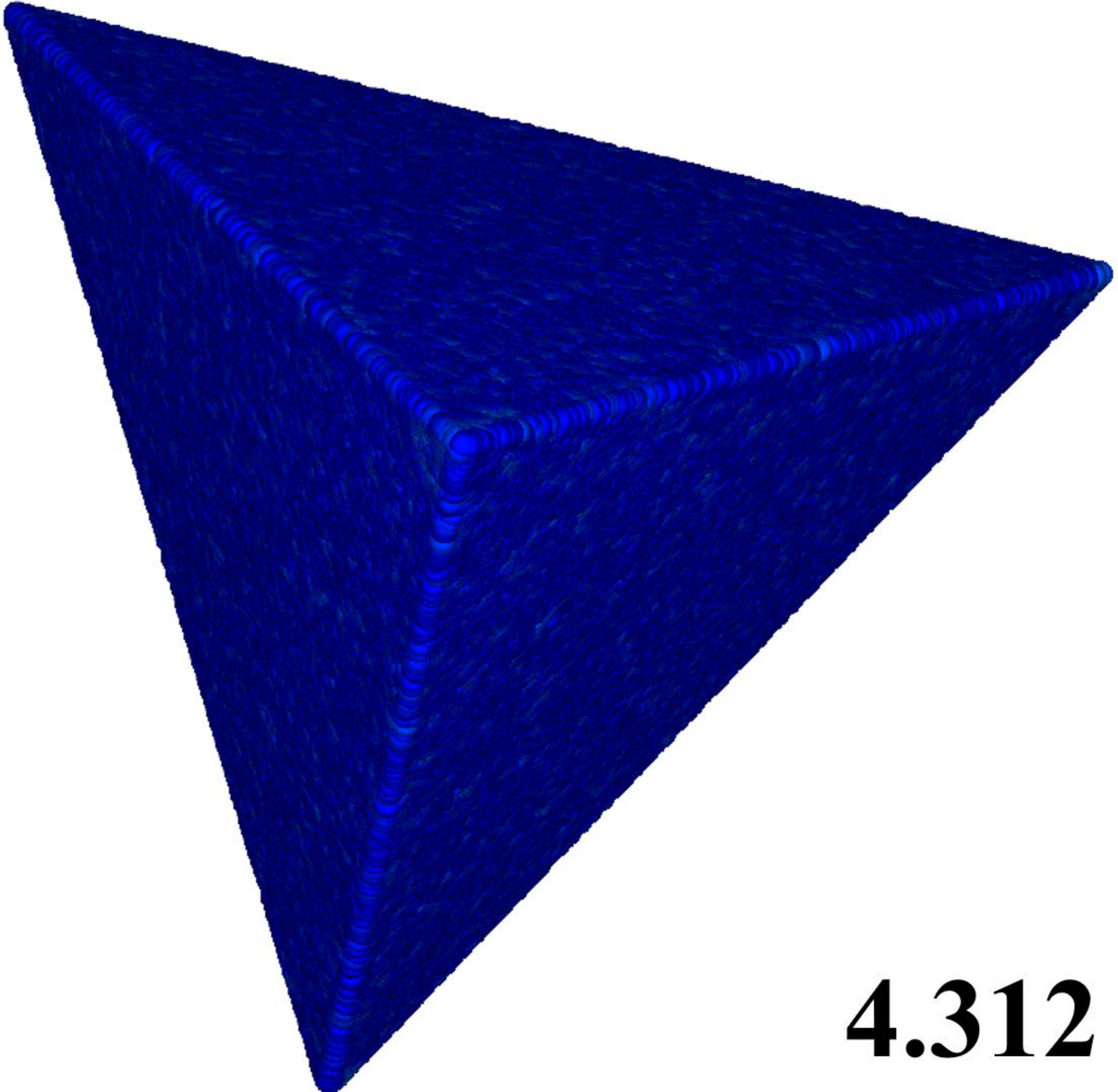}
        \end{minipage}
    }
    \subfigure[GPF]
    {
        \begin{minipage}[b]{0.11\textwidth} 
        \includegraphics[width=1\textwidth  ]{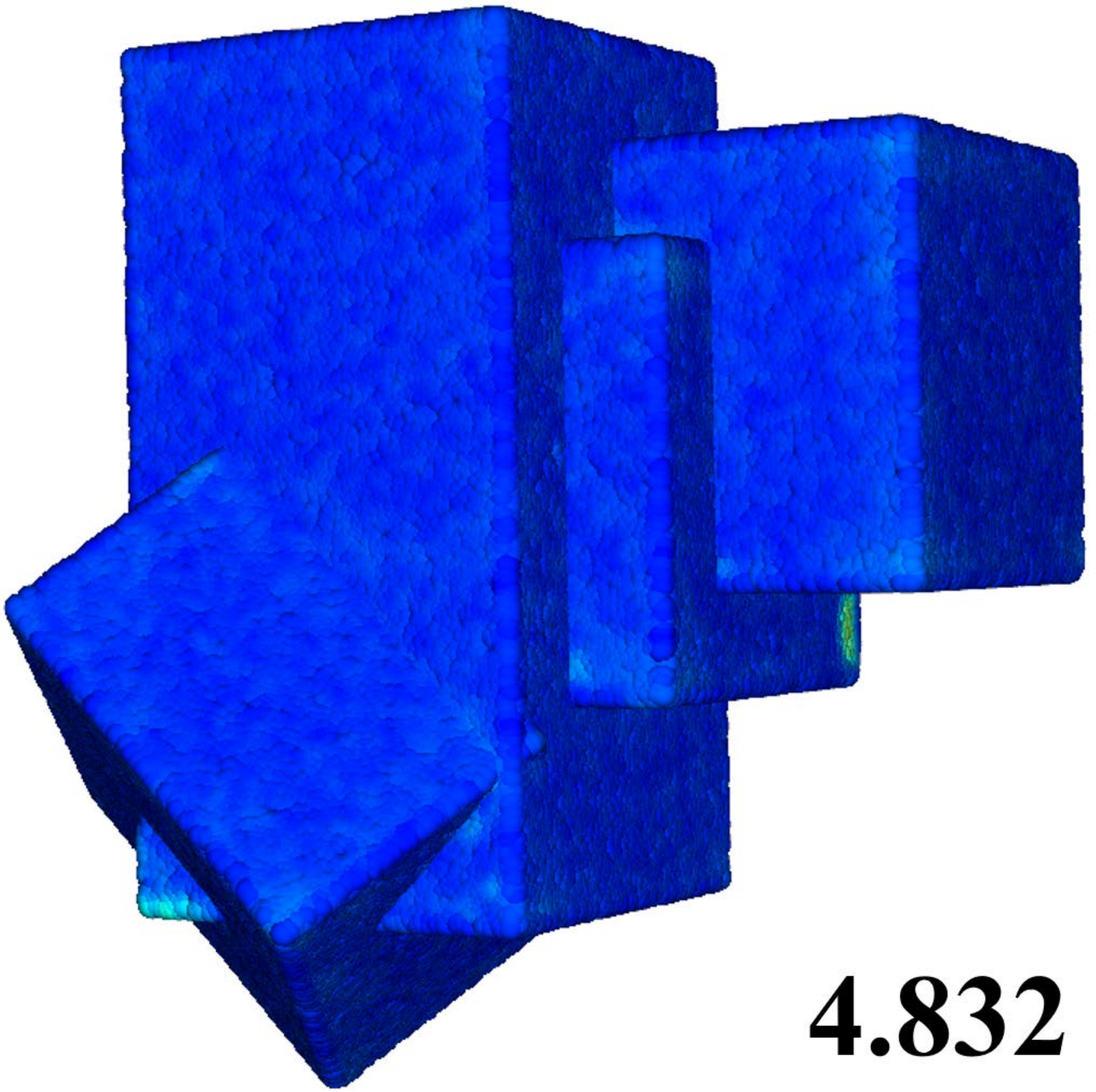}\\ 
        \includegraphics[width=1\textwidth  ]{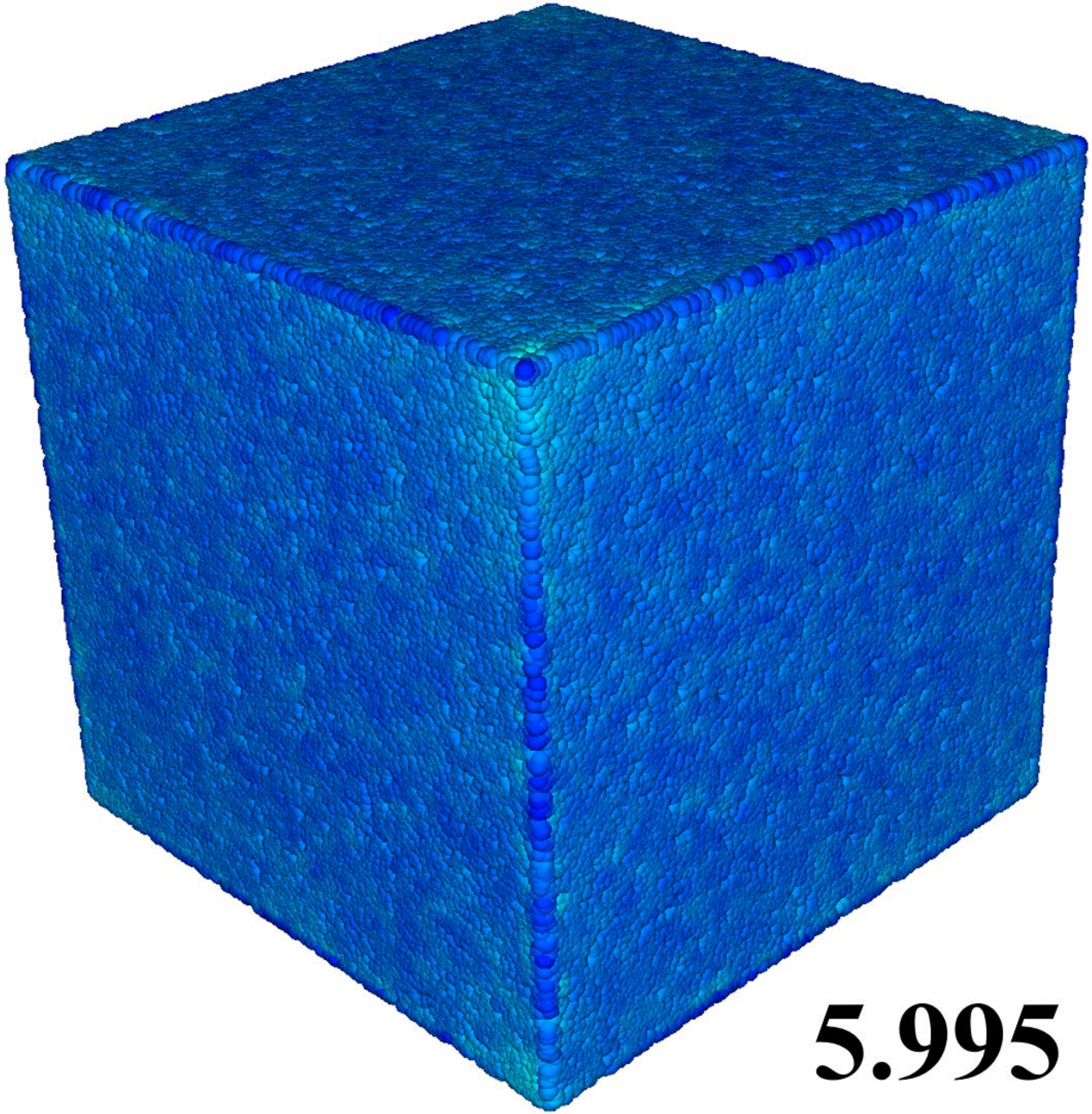}\\ 
        \includegraphics[width=1\textwidth  ]{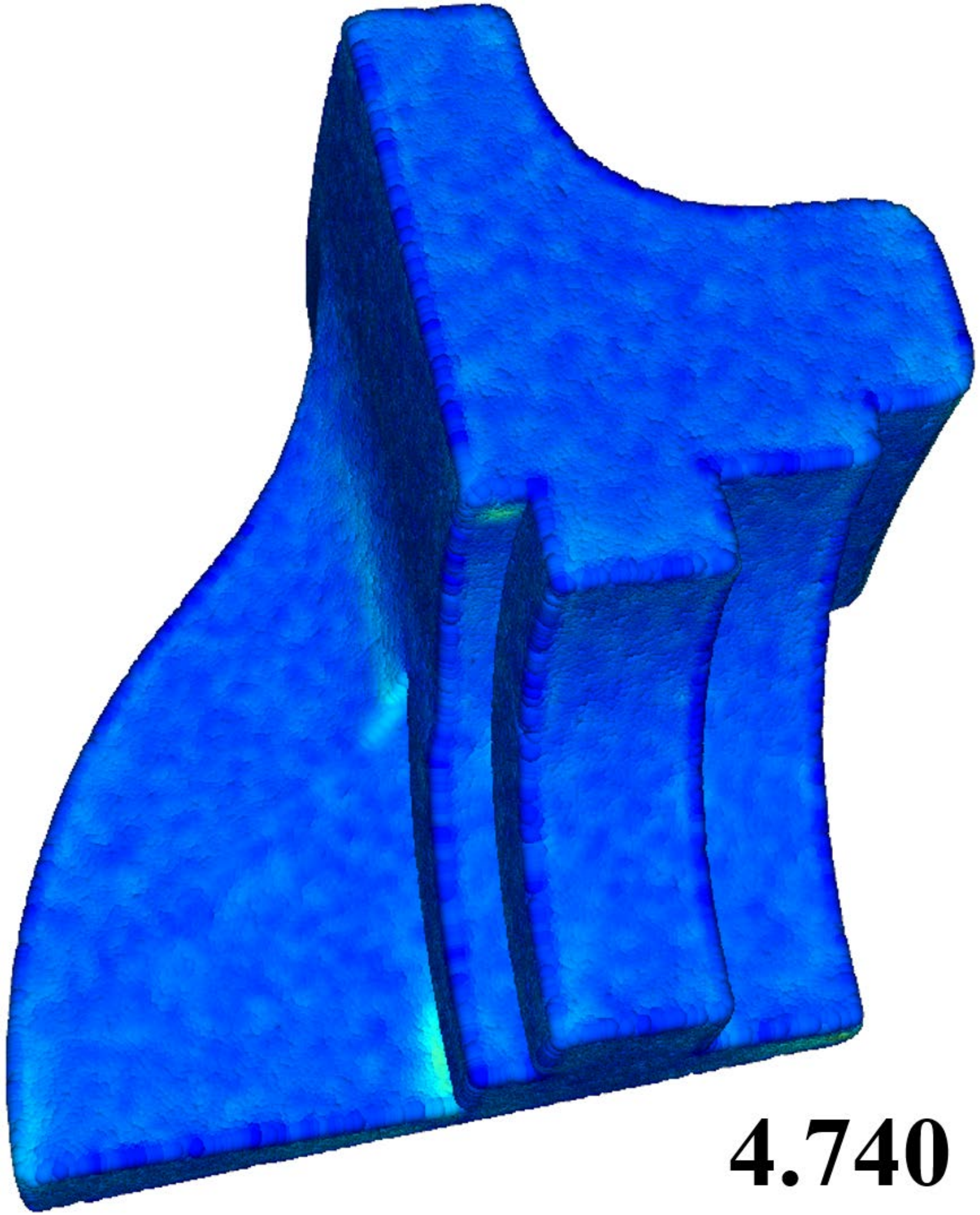}\\
        \includegraphics[width=1\textwidth  ]{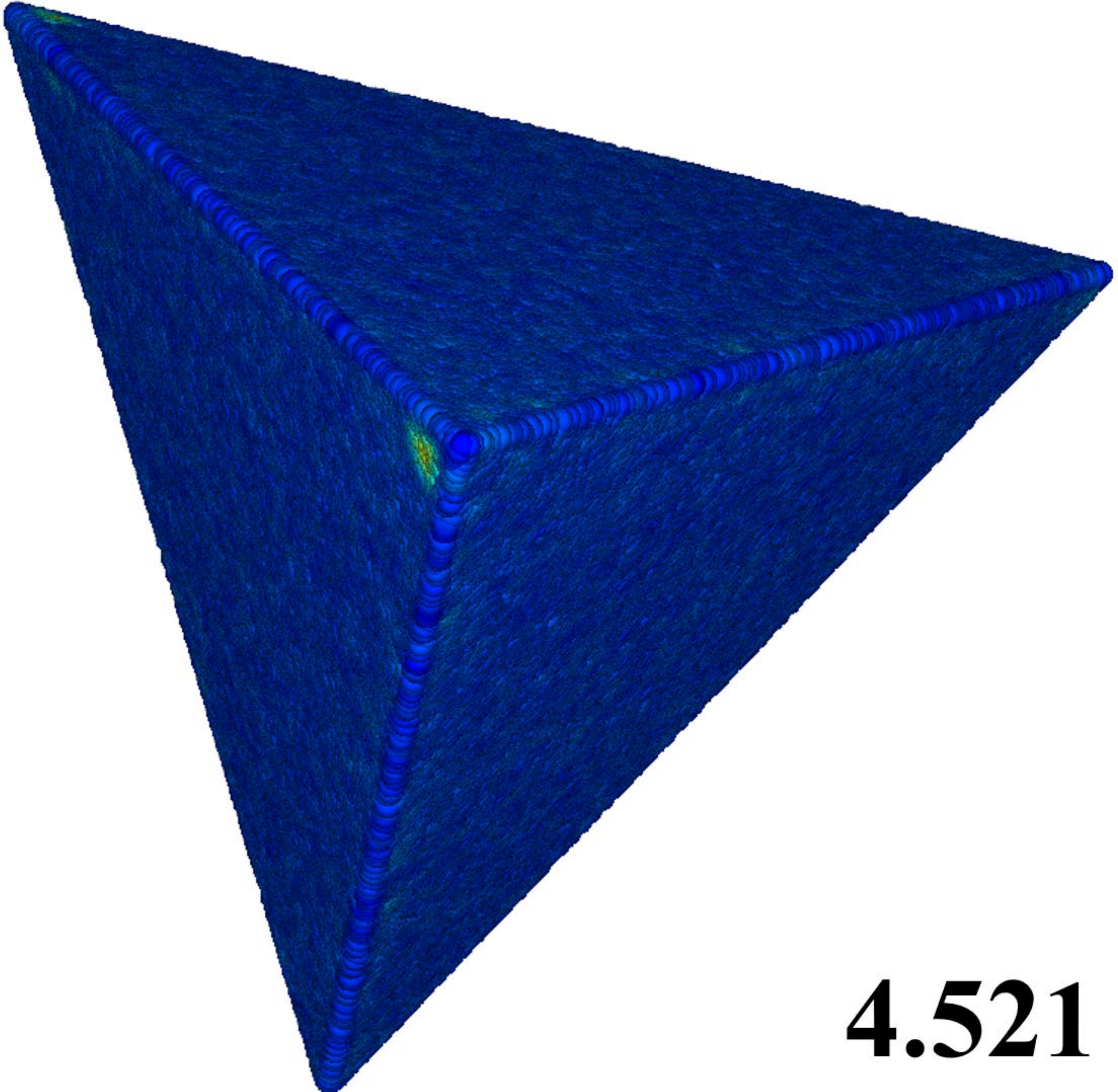}
        \end{minipage}
    }
    \subfigure[WLOP]
    {
        \begin{minipage}[b]{0.11\textwidth} 
        \includegraphics[width=1\textwidth  ]{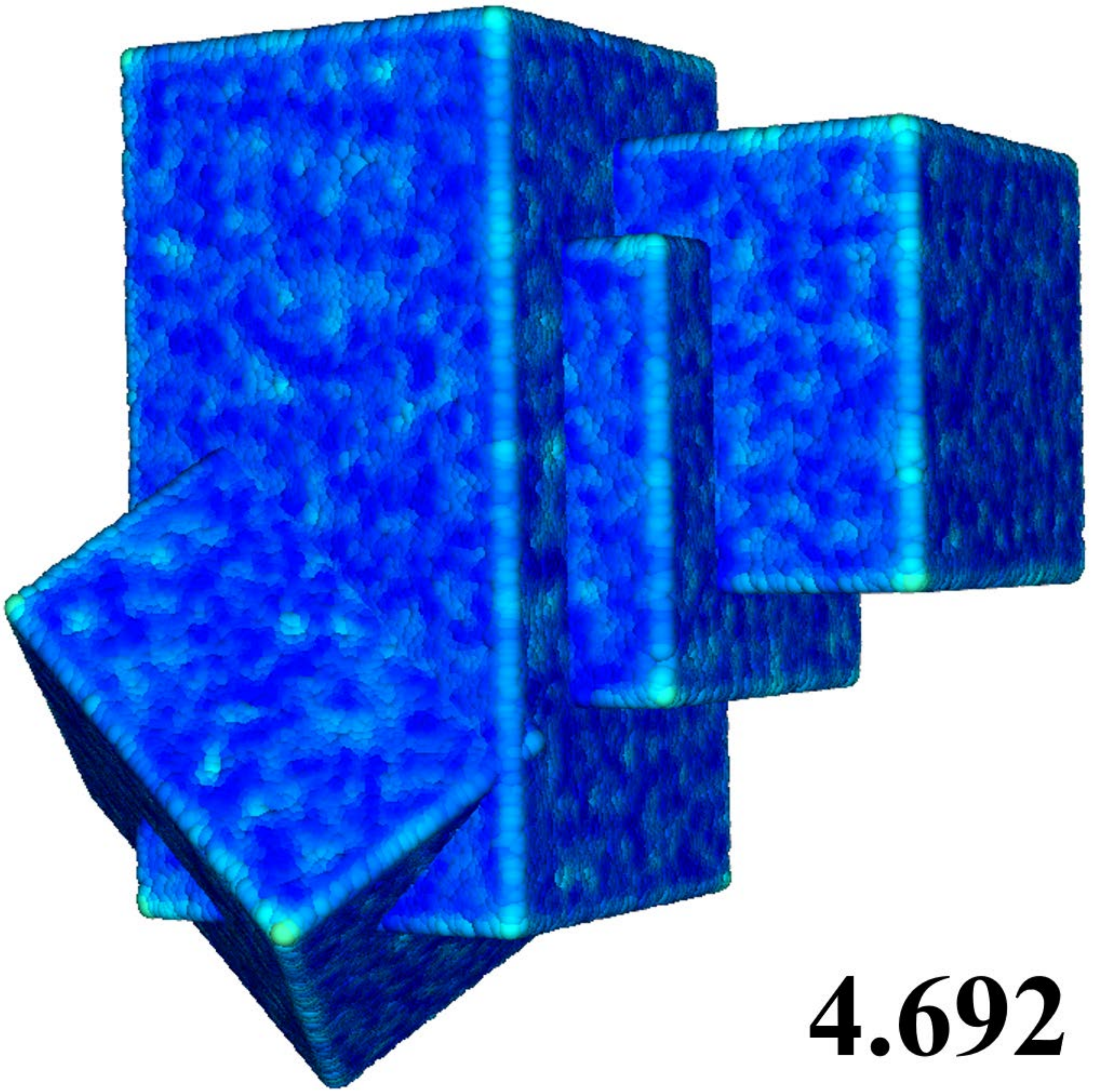}\\ 
        \includegraphics[width=1\textwidth  ]{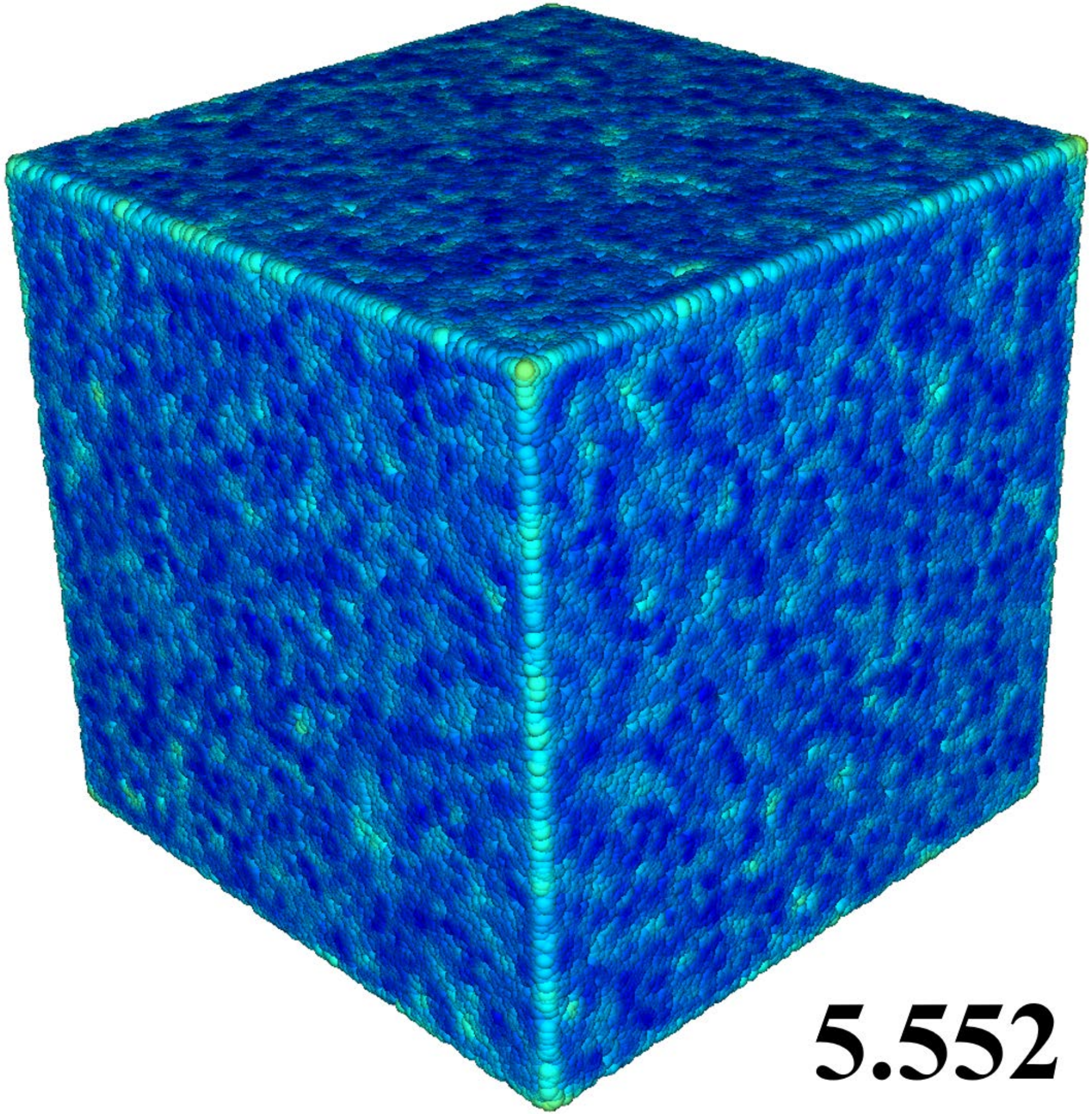}\\ 
        \includegraphics[width=1\textwidth  ]{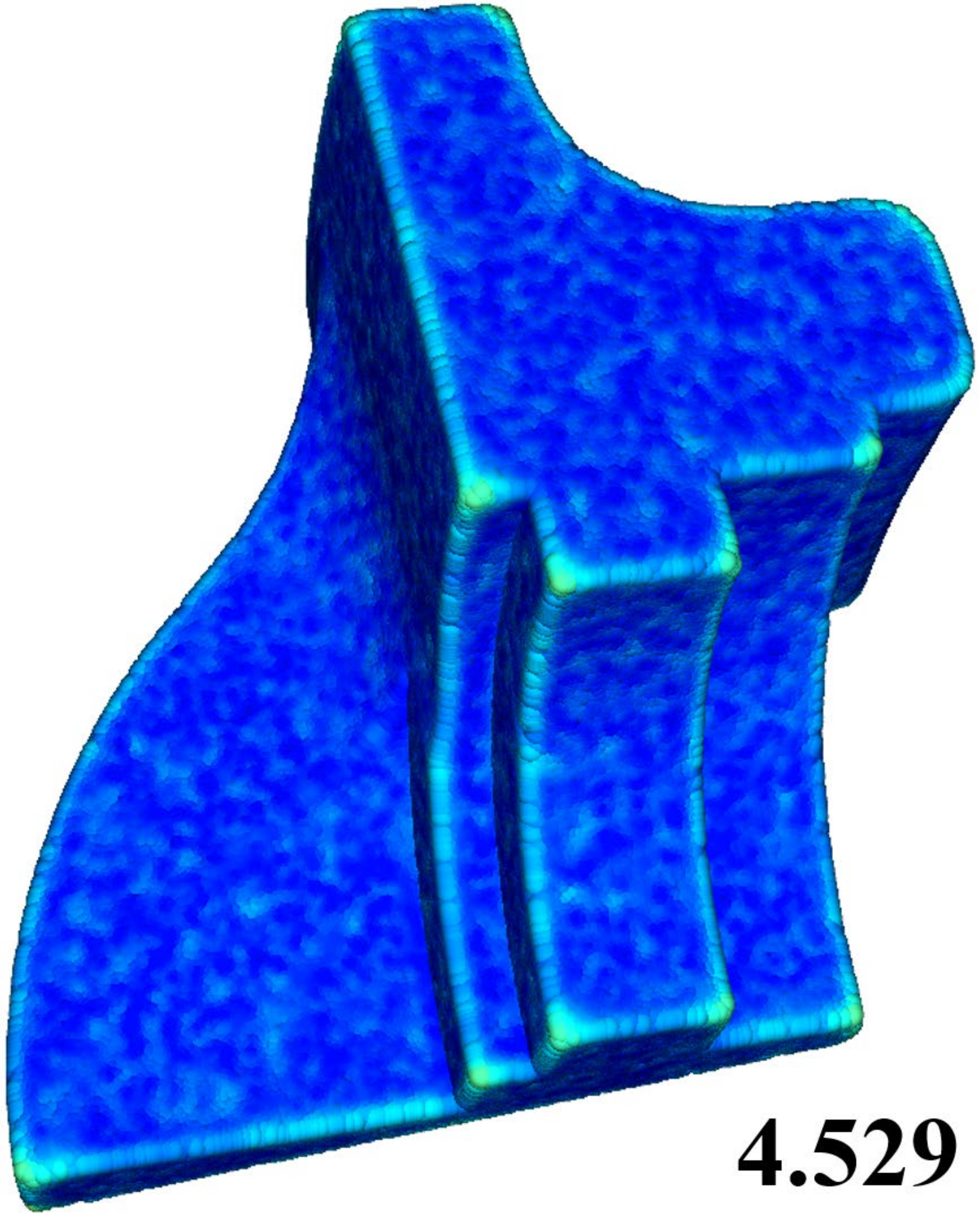}\\
        \includegraphics[width=1\textwidth  ]{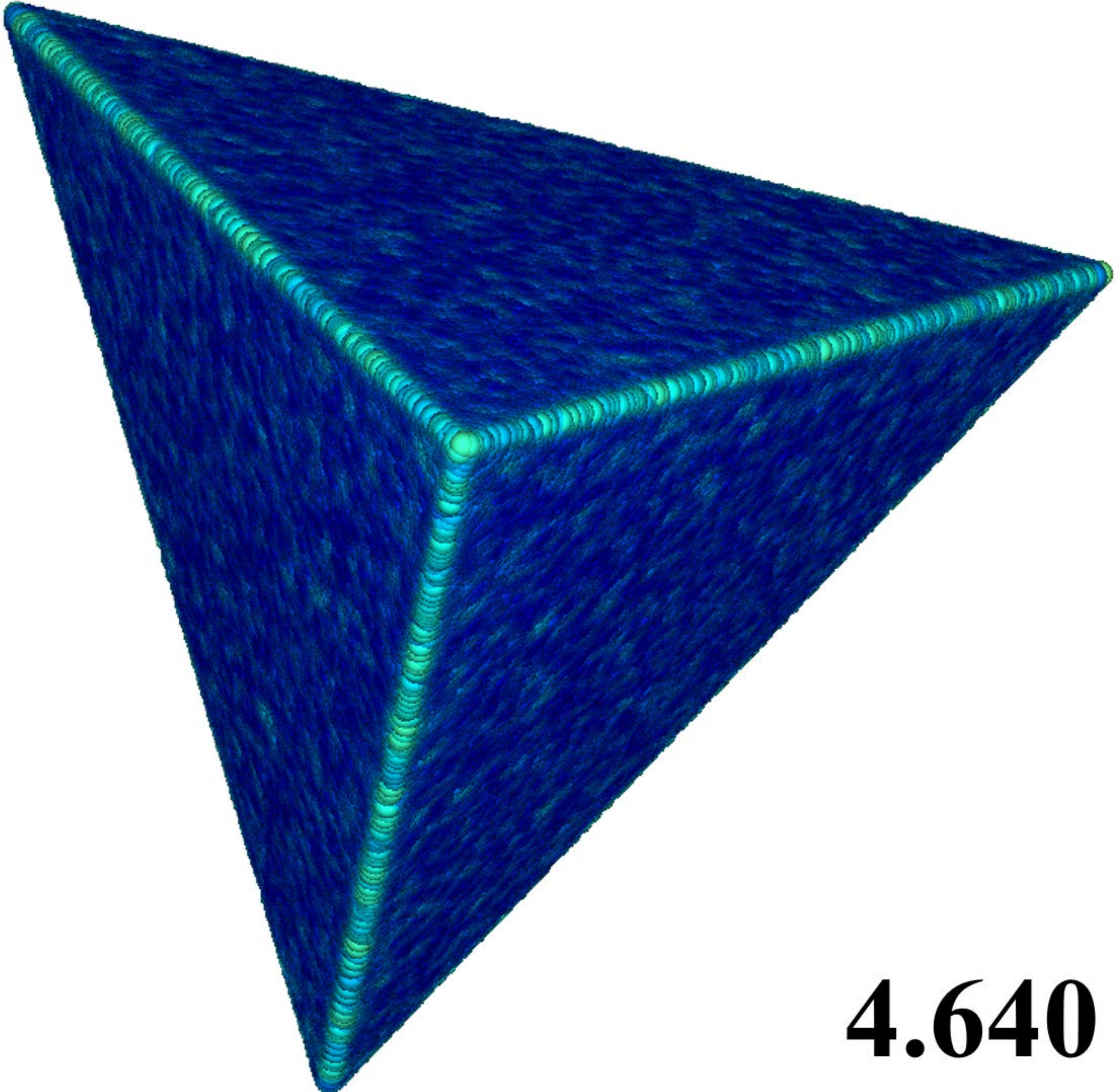}
        \end{minipage}
    }
    \subfigure[CLOP]
    {
        \begin{minipage}[b]{0.11\textwidth} 
        \includegraphics[width=1\textwidth  ]{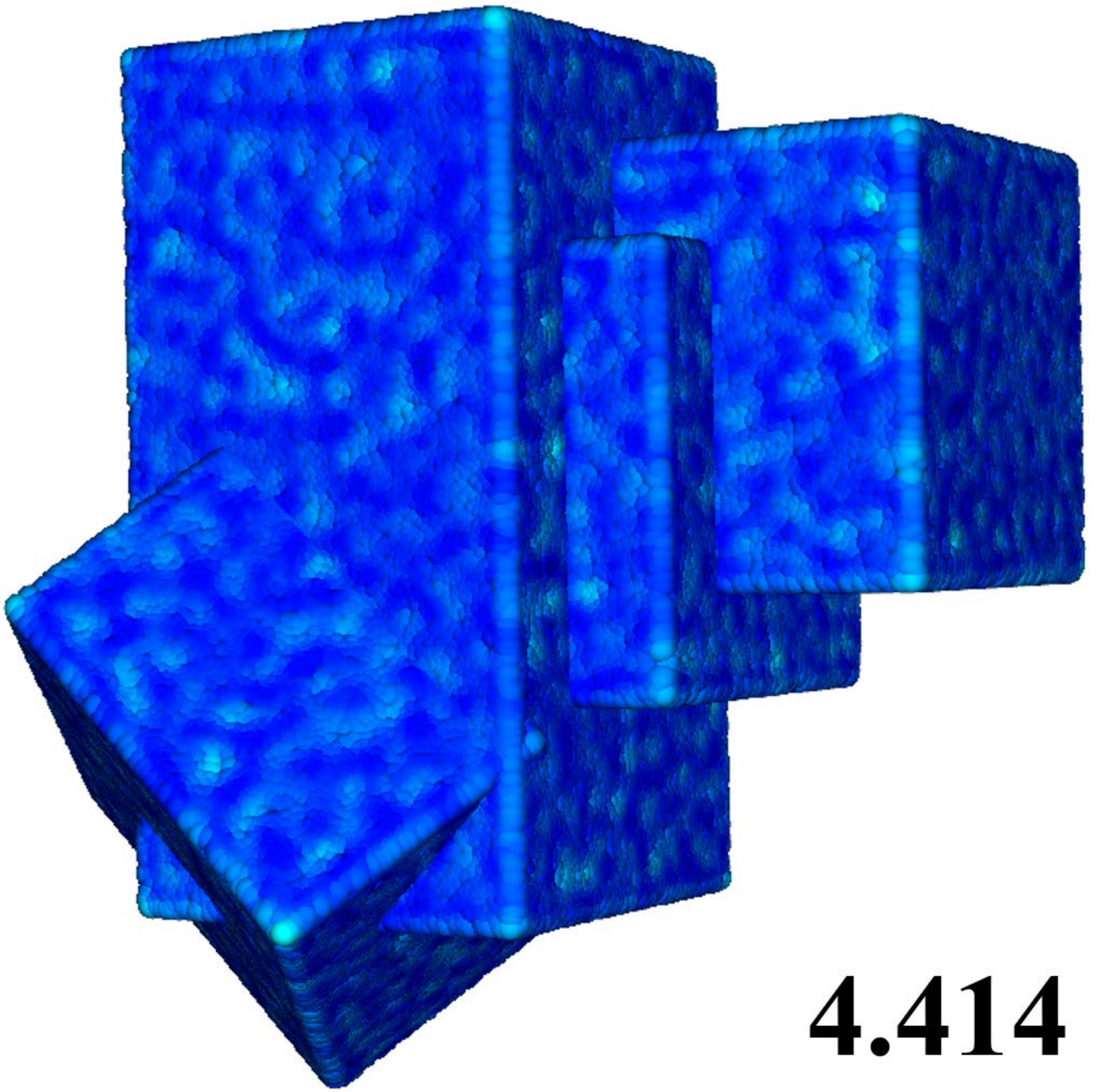}\\ 
        \includegraphics[width=1\textwidth  ]{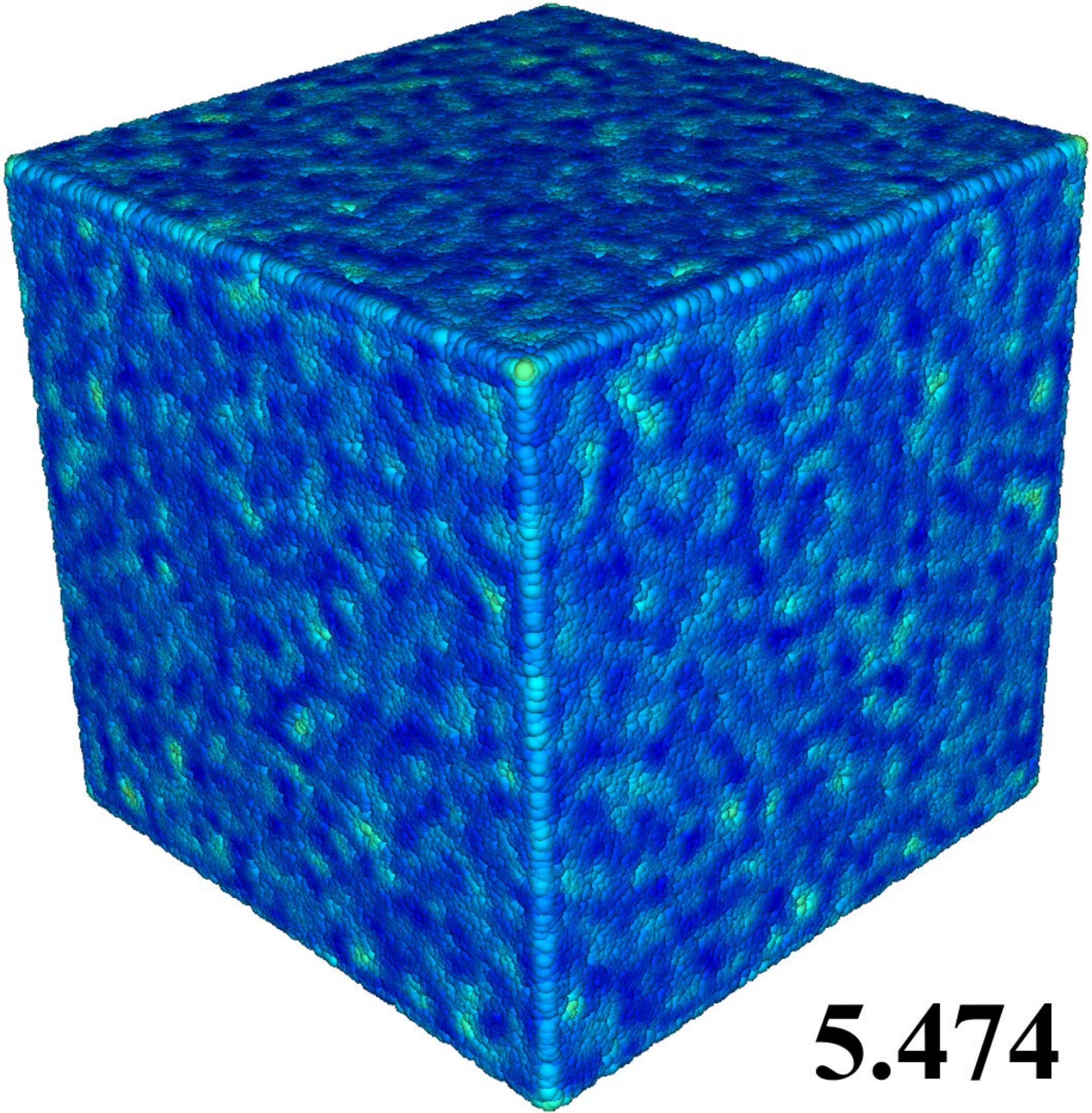}\\ 
        \includegraphics[width=1\textwidth  ]{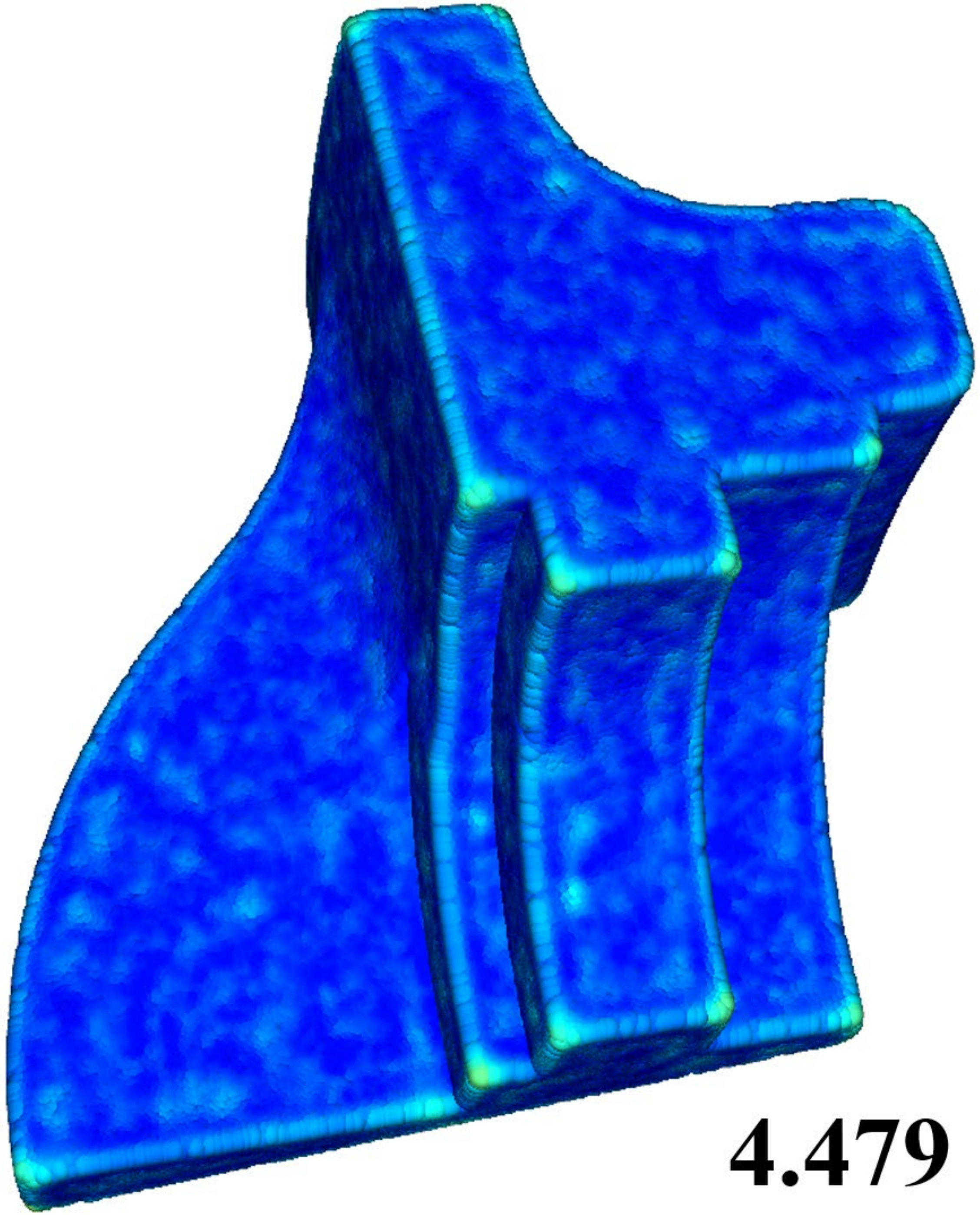}\\
        \includegraphics[width=1\textwidth  ]{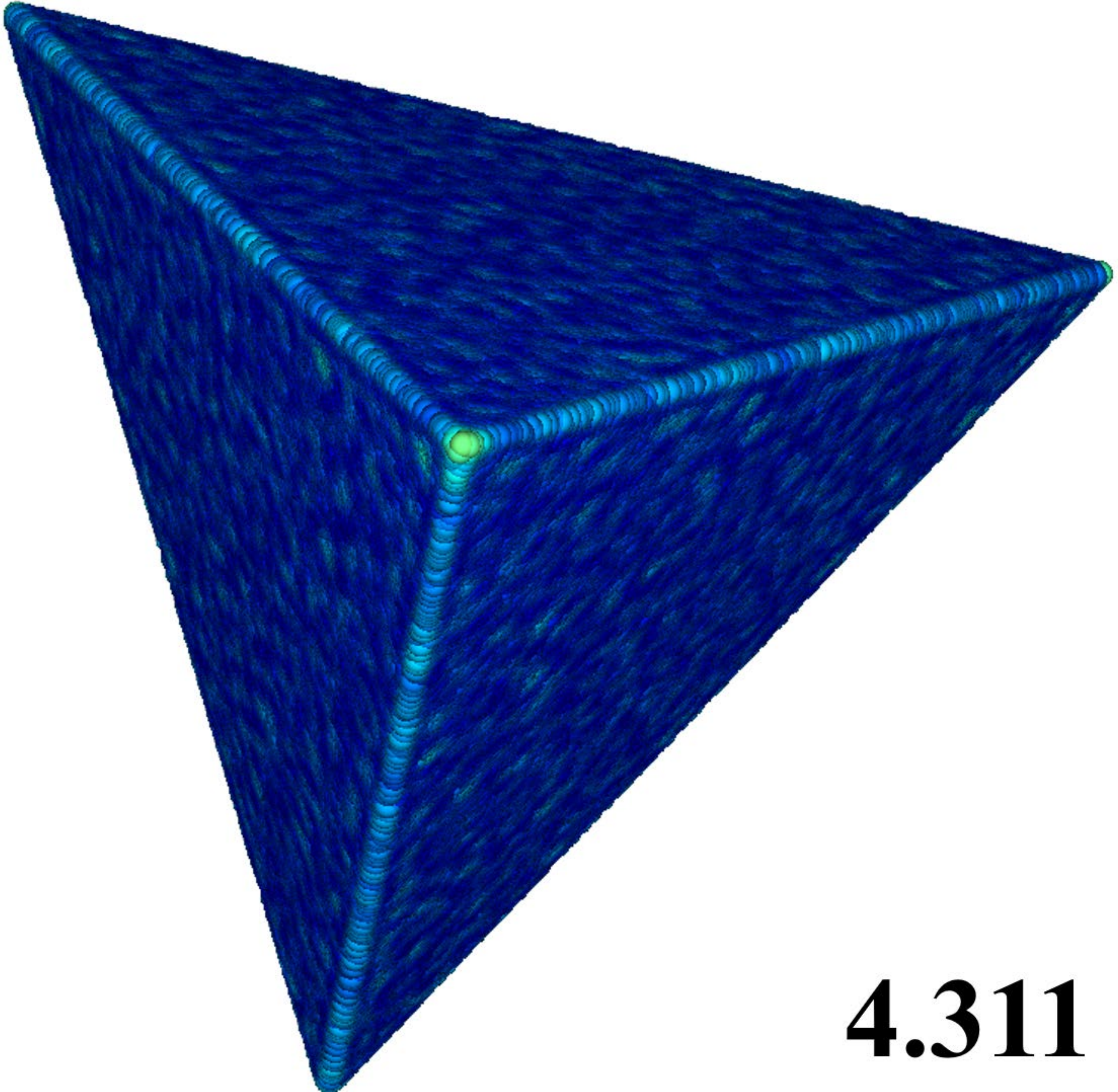}
        \end{minipage}
    }
    \subfigure[PCN]
    {
        \begin{minipage}[b]{0.11\textwidth} 
        \includegraphics[width=1\textwidth  ]{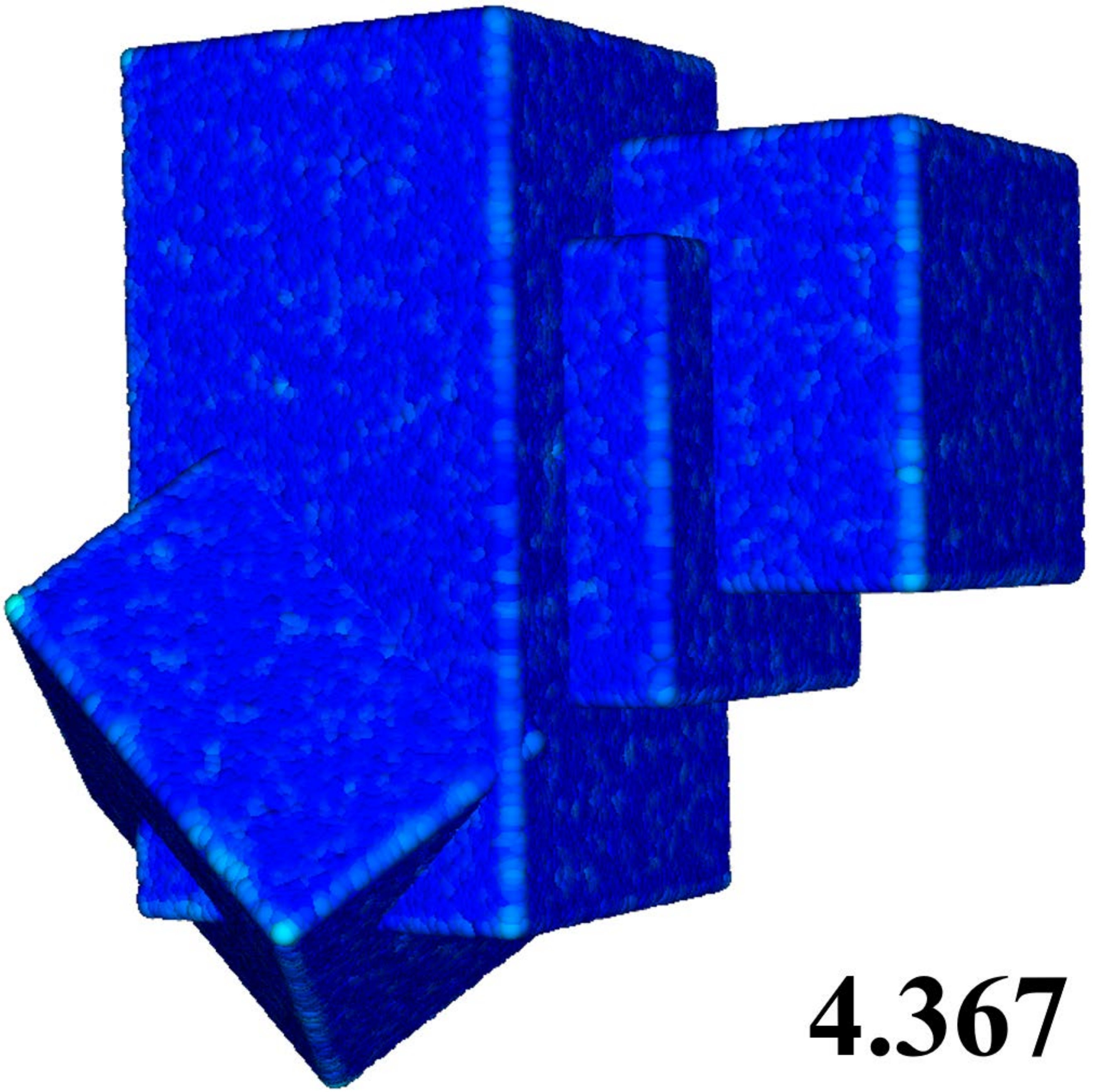}\\ 
        \includegraphics[width=1\textwidth  ]{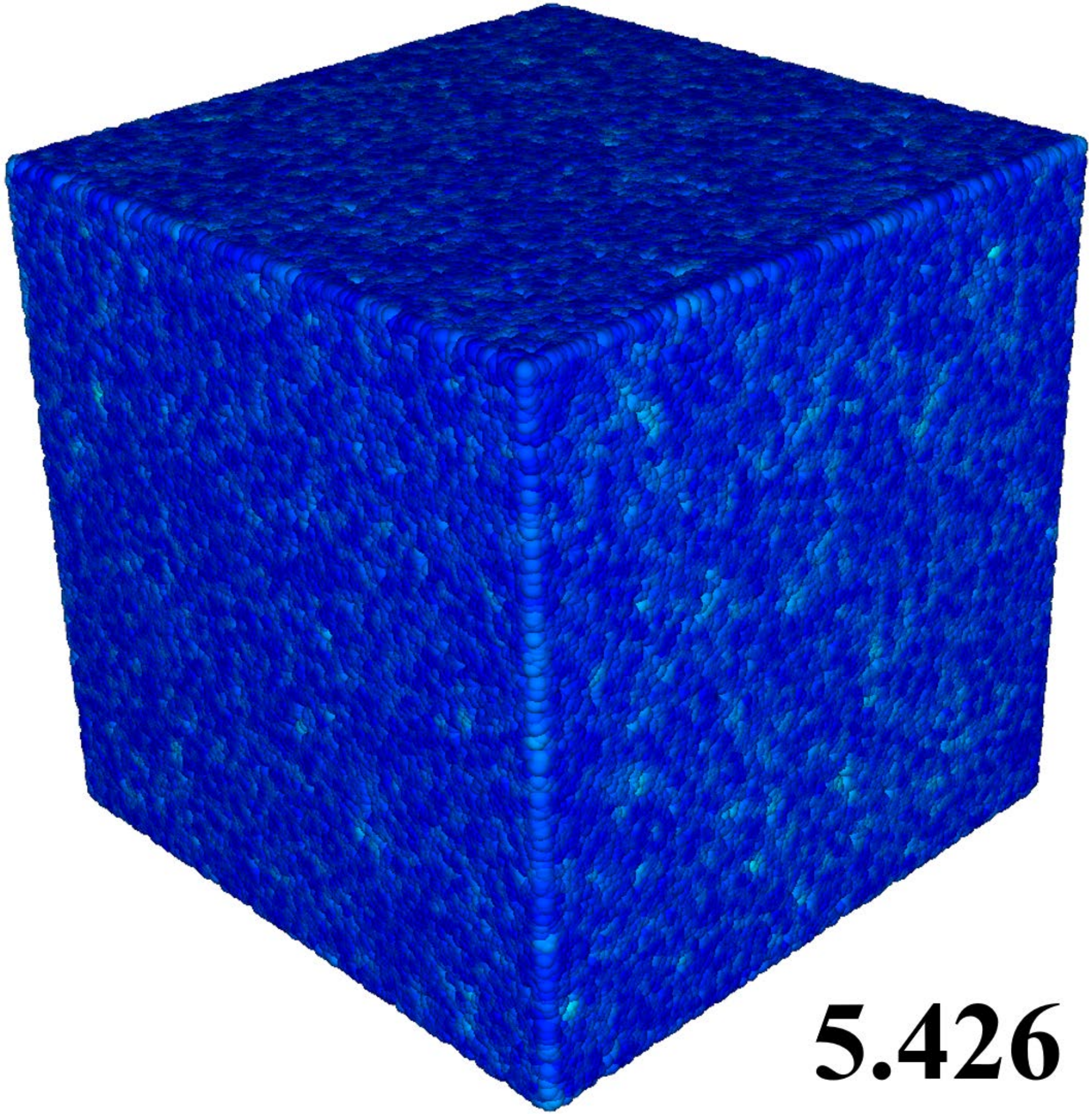}\\ 
        \includegraphics[width=1\textwidth  ]{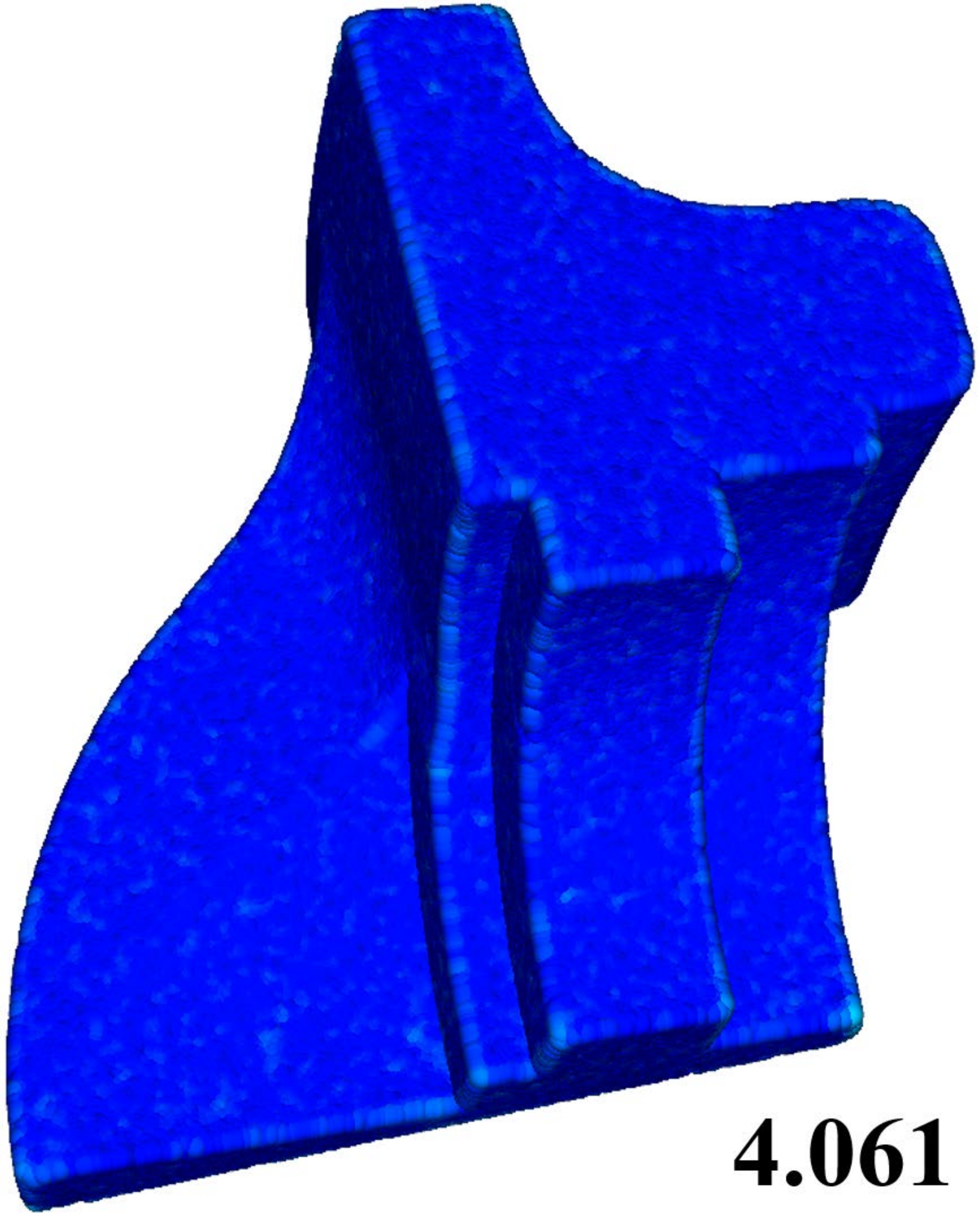}\\
        \includegraphics[width=1\textwidth  ]{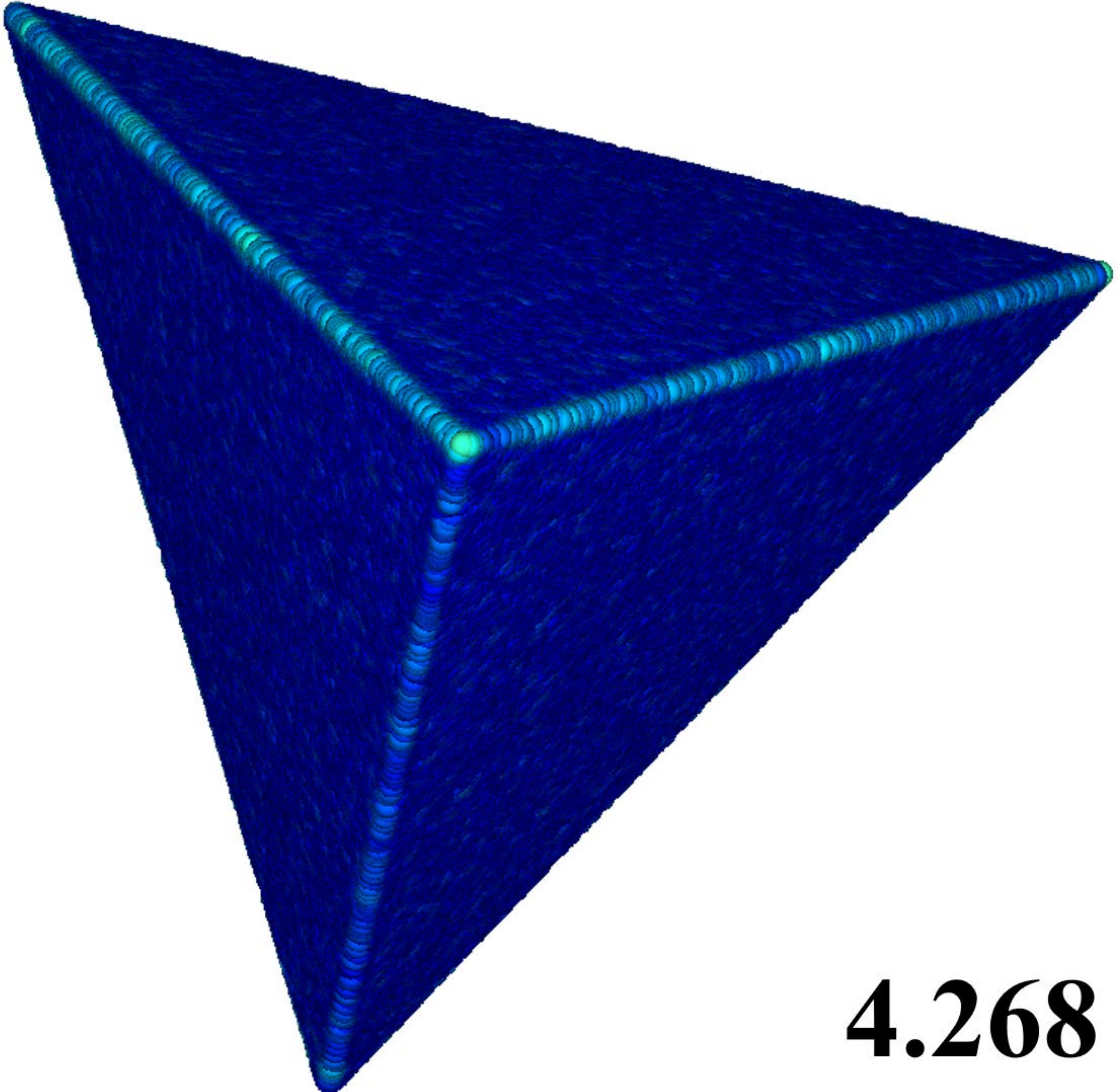}
        \end{minipage}
    }
    \subfigure[TD]
    {
        \begin{minipage}[b]{0.11\textwidth} 
        \includegraphics[width=1\textwidth  ]{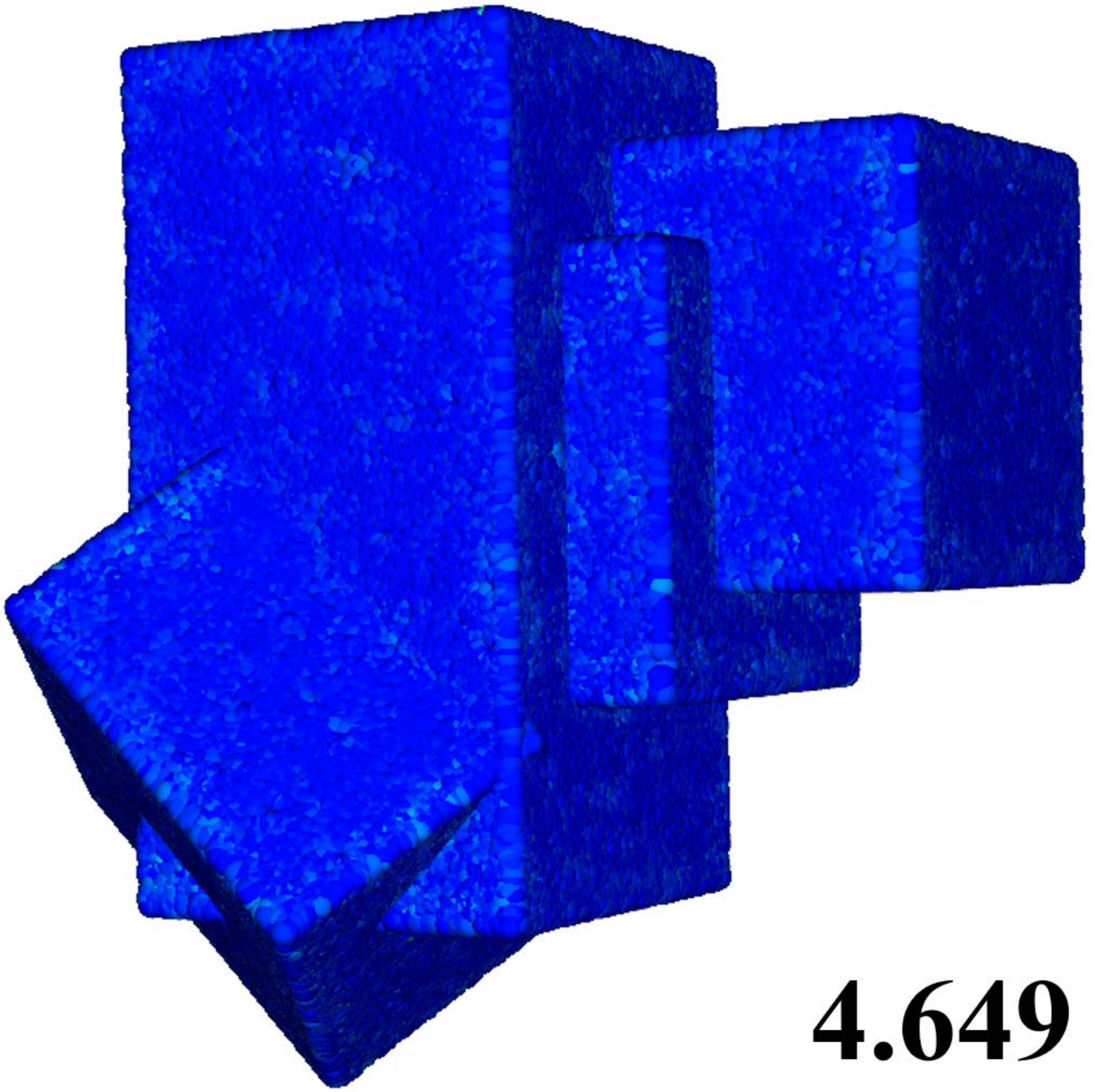}\\ 
        \includegraphics[width=1\textwidth  ]{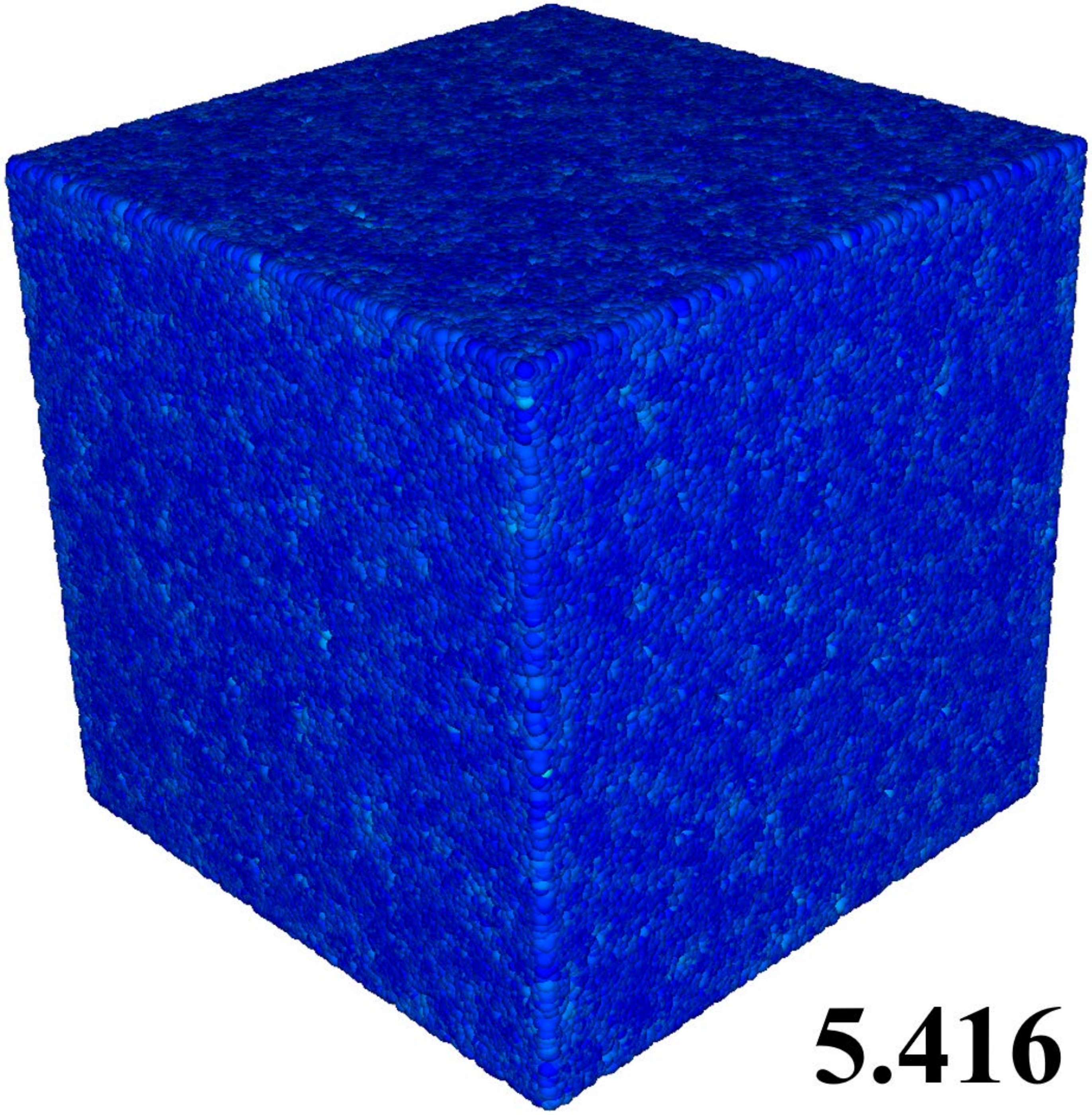}\\ 
        \includegraphics[width=1\textwidth  ]{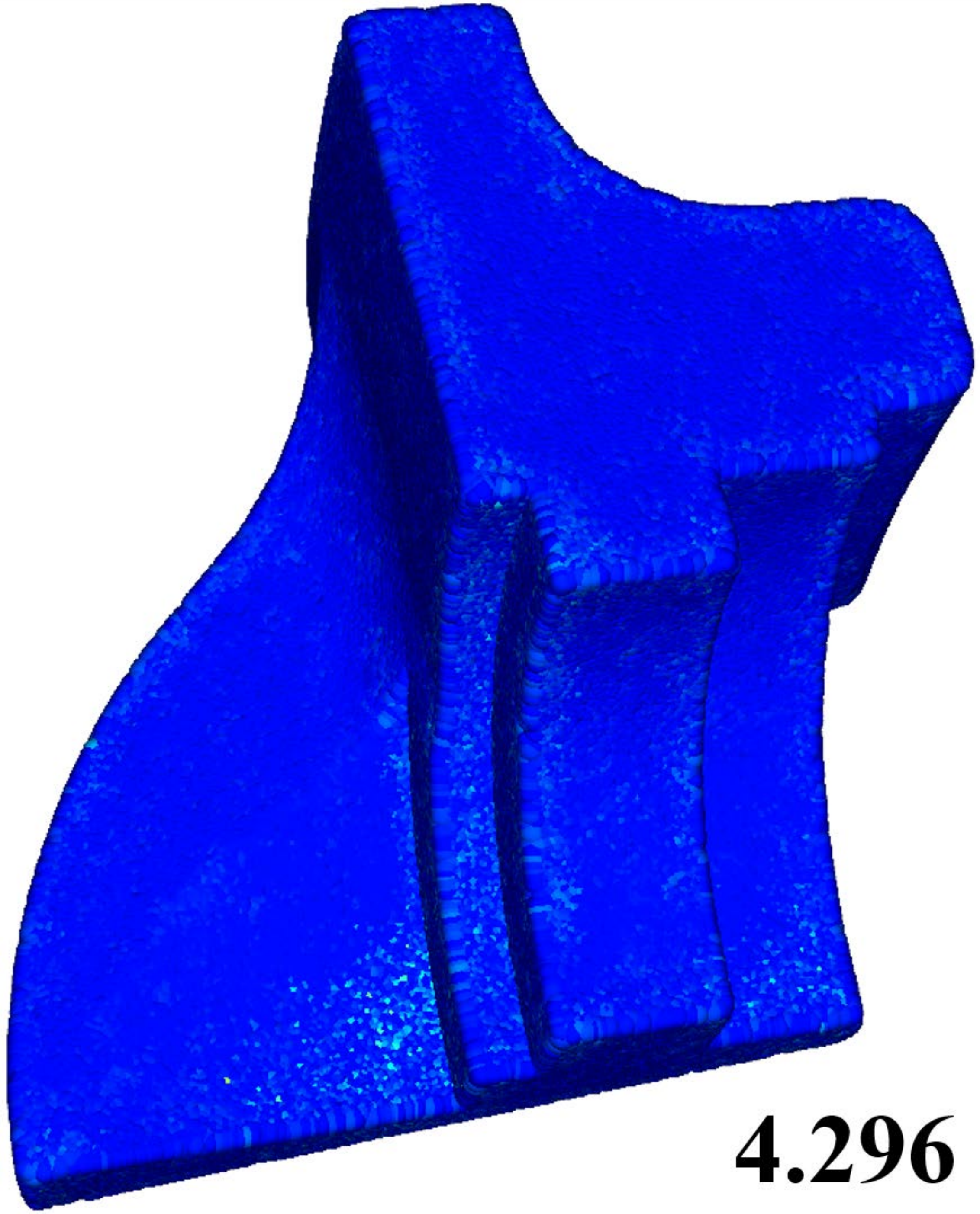}\\
        \includegraphics[width=1\textwidth  ]{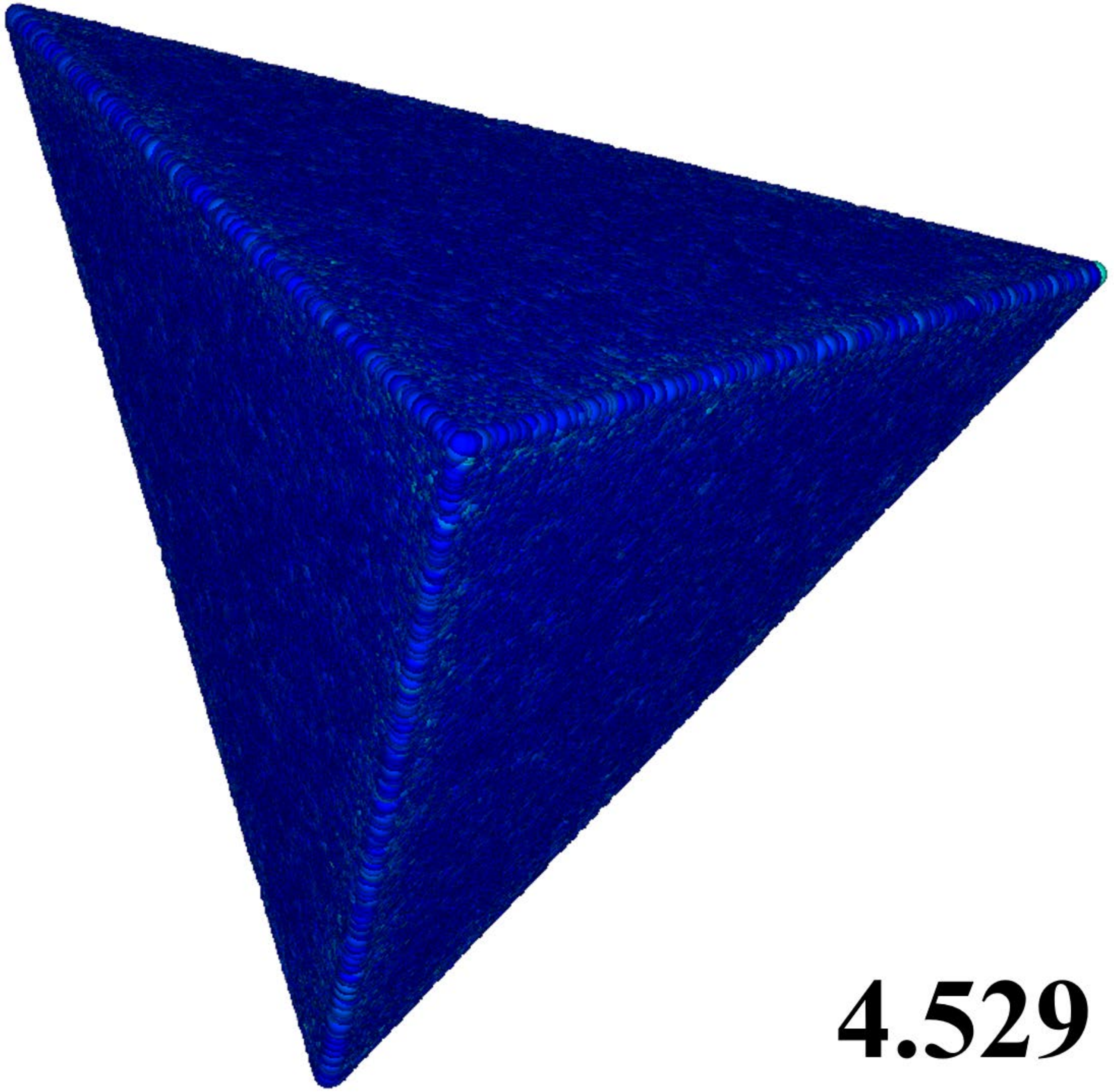}
        \end{minipage}
    }
    \subfigure[Ours]
    {
        \begin{minipage}[b]{0.11\textwidth} 
        \includegraphics[width=1\textwidth  ]{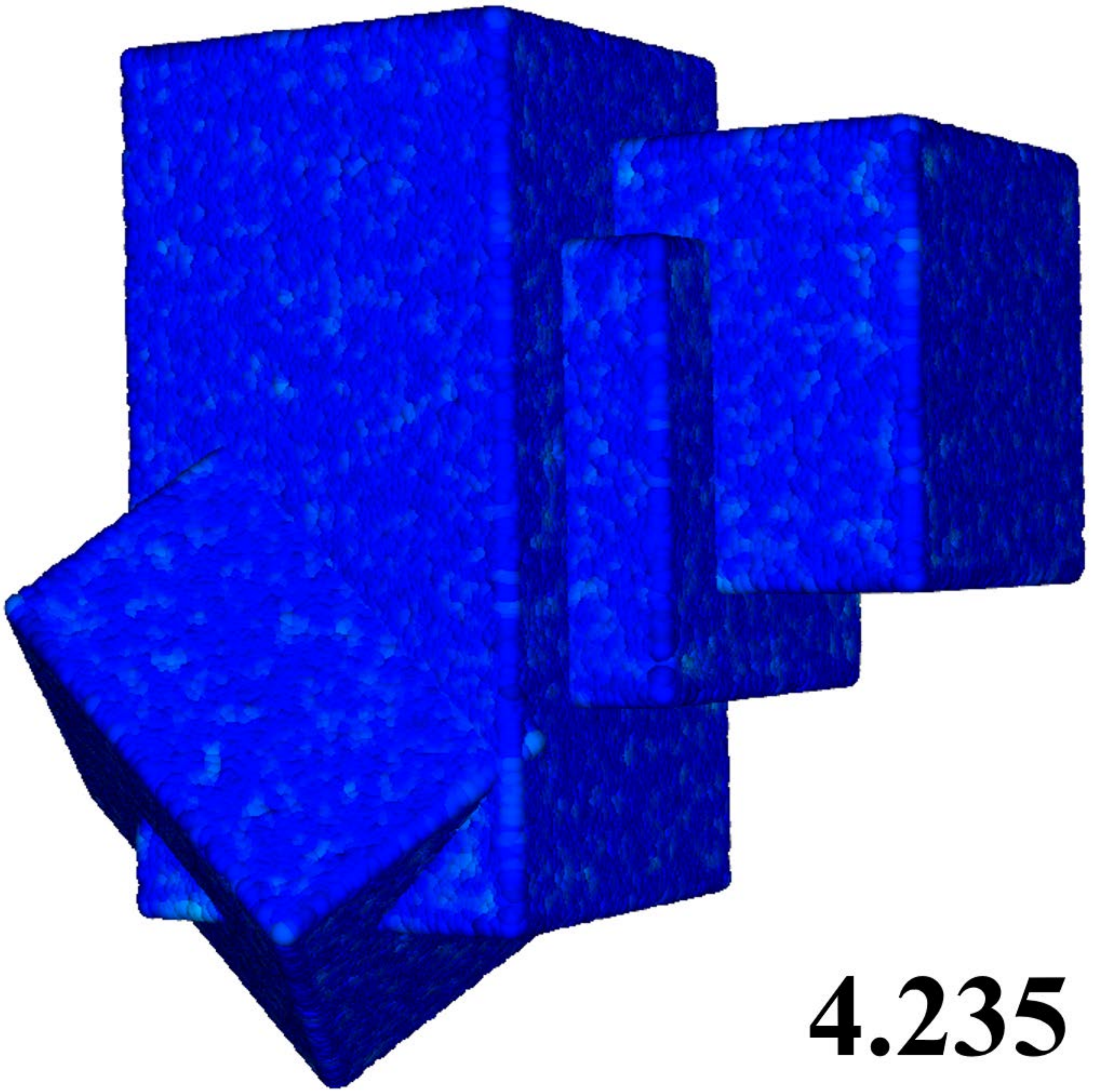}\\ 
        \includegraphics[width=1\textwidth  ]{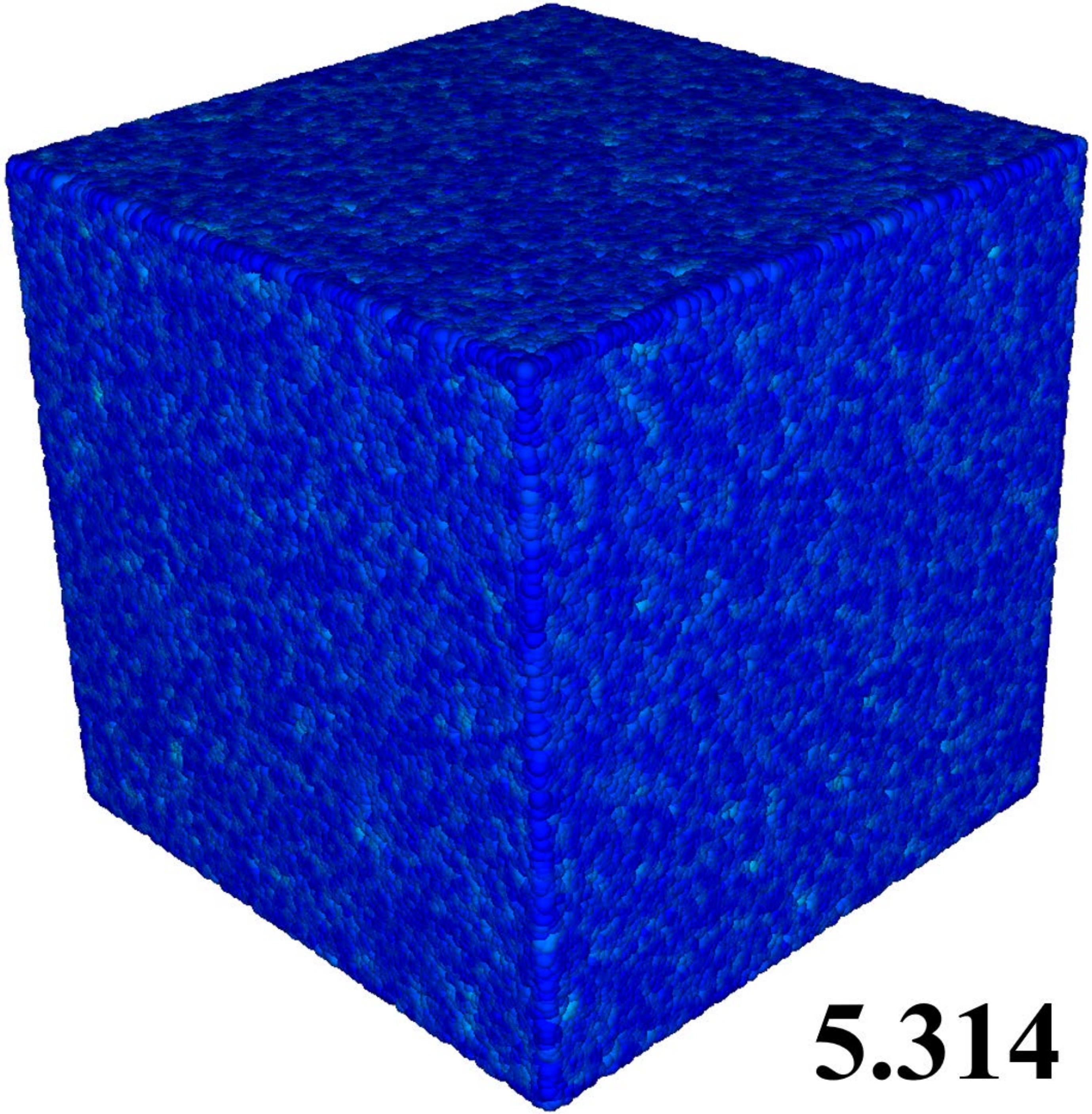}\\ 
        \includegraphics[width=1\textwidth  ]{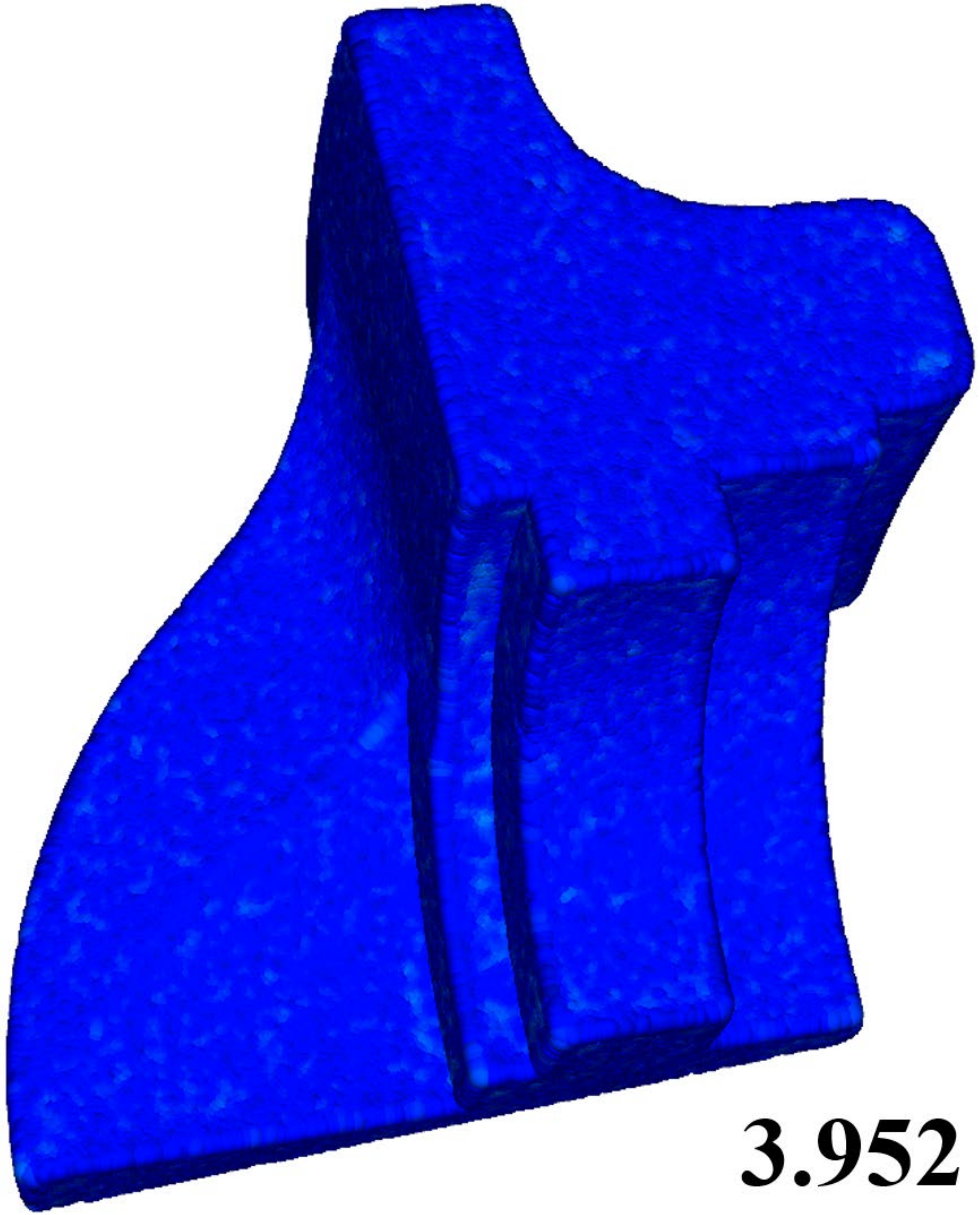}\\
        \includegraphics[width=1\textwidth  ]{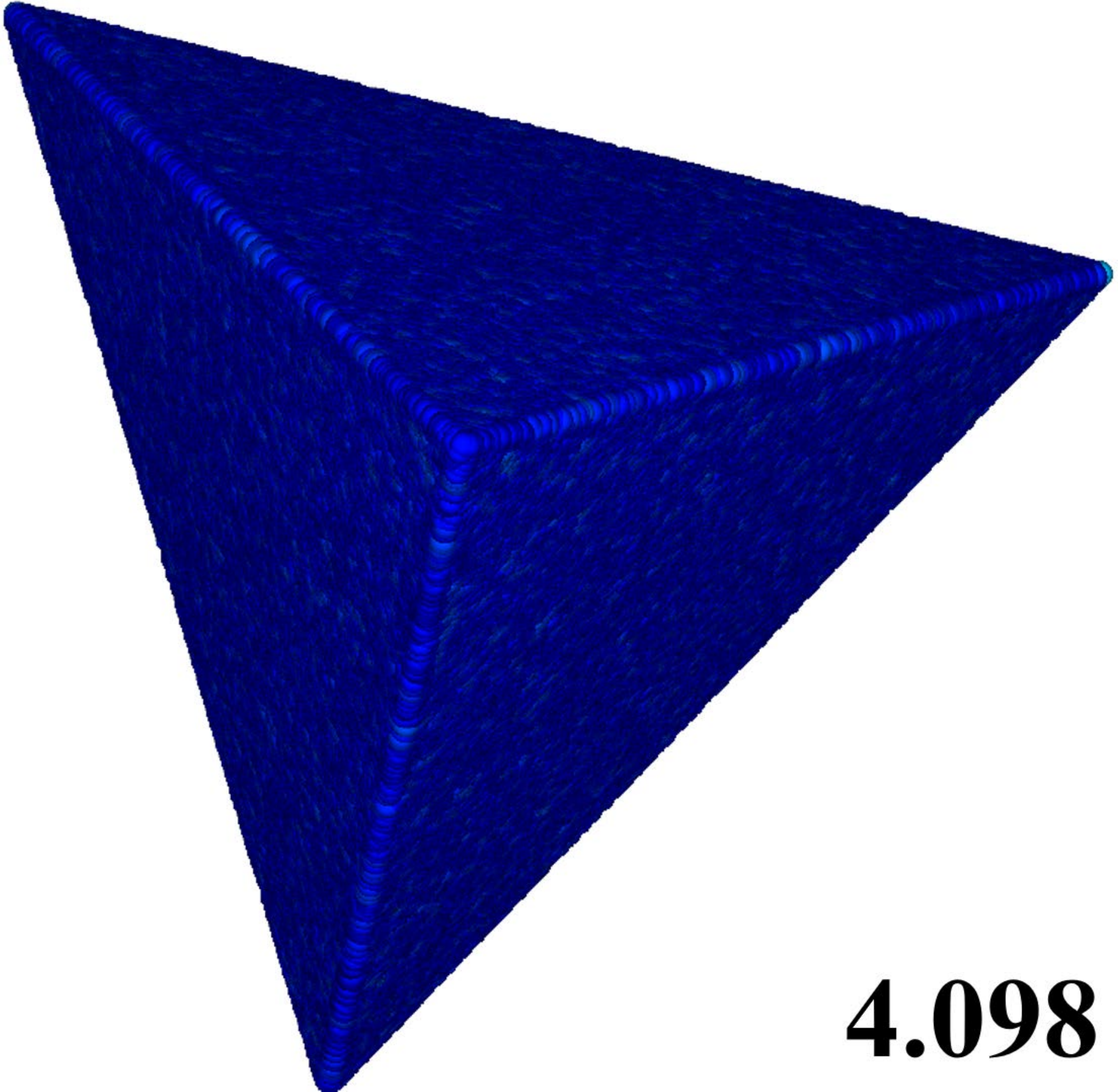}
        \end{minipage}
    }
    \subfigure
    {
        \begin{minipage}[b]{0.04\textwidth} 
        \includegraphics[width=1\textwidth]{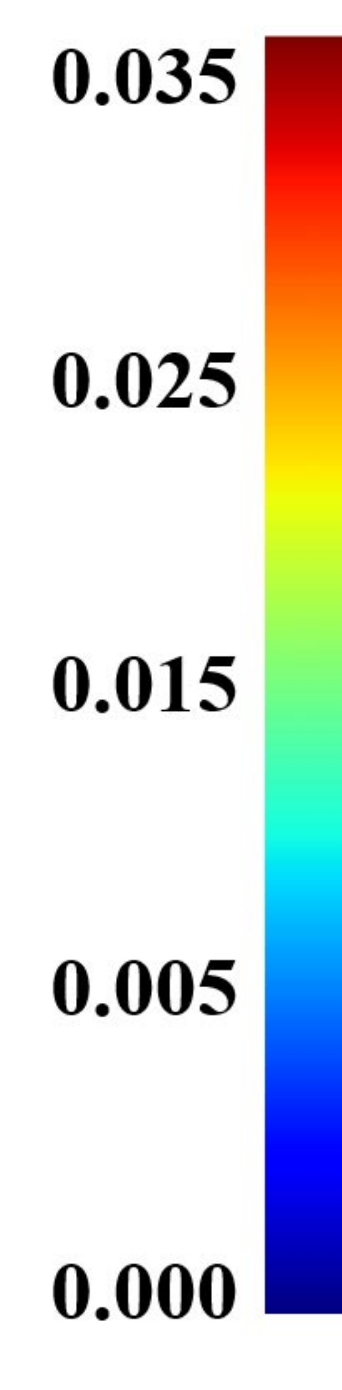}
        \end{minipage}
    }
    \caption{Quantitative comparison in mean square error (MSE). The overall errors ($\times 10^{-3}$) for different methods over four models are shown in the figure.}
    \label{fig:CPErrors}
\end{figure*}

\subsection{Evaluation Metrics}
For the sake of analysing the performance of our Pointfilter quantitatively, the evaluation metrics should be defined. The Pointfilter aims to project noisy points onto its underlying surface. It is intuitive to evaluate the distance errors by averaging the distances between a point in the ground truth and its closest points in the filtered point cloud \cite{Lu2018TVCG}. The distance error can be defined as 
\begin{equation}\label{eq:averagedistance}
\begin{aligned}
D(\mathbf{p}_{i}) = \frac{1}{M}\sum_{\bar{\mathbf{p}}_{j} \in \mathbf{NN}(\mathbf{p}_{i})} \|\mathbf{p}_{i} - \bar{\mathbf{p}}_{j}\|_{2}^{2},
\end{aligned}
\end{equation}
where $\mathbf{p}_{i}$ is the ground truth point and $\bar{\mathbf{p}}_{j}$ is one of its neighboring point in the filtered point cloud. $M = |\mathbf{NN}(\mathbf{p}_{i})|$ and $\mathbf{NN}$ represents the nearest neighbors. In our paper, we set $M$ to $10$. 
Inspired by \cite{Fan2017CVPR}, we also introduce the Chamfer distance (CD) to evaluate the error between the filtered point cloud and its corresponding ground truth (clean point cloud). CD is defined as 
\begin{equation}\label{eq:chamferDistance}
\begin{split}
C(\mathbf{P}, \bar{\mathbf{P}}) = & \frac{1}{N_{1}}\sum_{\mathbf{p}_{i} \in \mathbf{P}} \min_{\bar{\mathbf{p}}_{j} \in \bar{\mathbf{P}} } (\|\mathbf{p}_{i} - \bar{\mathbf{p}}_{j}\|_{2}^{2}) \\
& + 
 \frac{1}{N_{2}} \sum_{\mathbf{p}_{j} \in \mathbf{P}} \min_{\bar{\mathbf{p}}_{i} \in \bar{\mathbf{P}}}(\|\mathbf{p}_{j} - \bar{\mathbf{p}}_{i}\|_{2}^{2}), 
\end{split}
\end{equation}
where $N_{1}$ and $N_{2}$ represent the cardinalities of the clean point cloud $\mathbf{P}$ and the filtered point cloud $\bar{\mathbf{P}}$, respectively. The CD metric finds the nearest neighbor in the other set and sums the squared distances up. It can be viewed as an indicator function which measures the ``similarity" between two point sets. Also, it can be easily implemented in parallel. Besides the two above metrics, the point-to-surface (P2F) \cite{Yifan2019CVPR} distance against the ground truth mesh is also used in our evaluation.
\begin{figure}[htb!]
  \centering
  \subfigure[Noisy]
  {
    \begin{minipage}[b]{0.1\textwidth} 
    \includegraphics[width=1\textwidth]{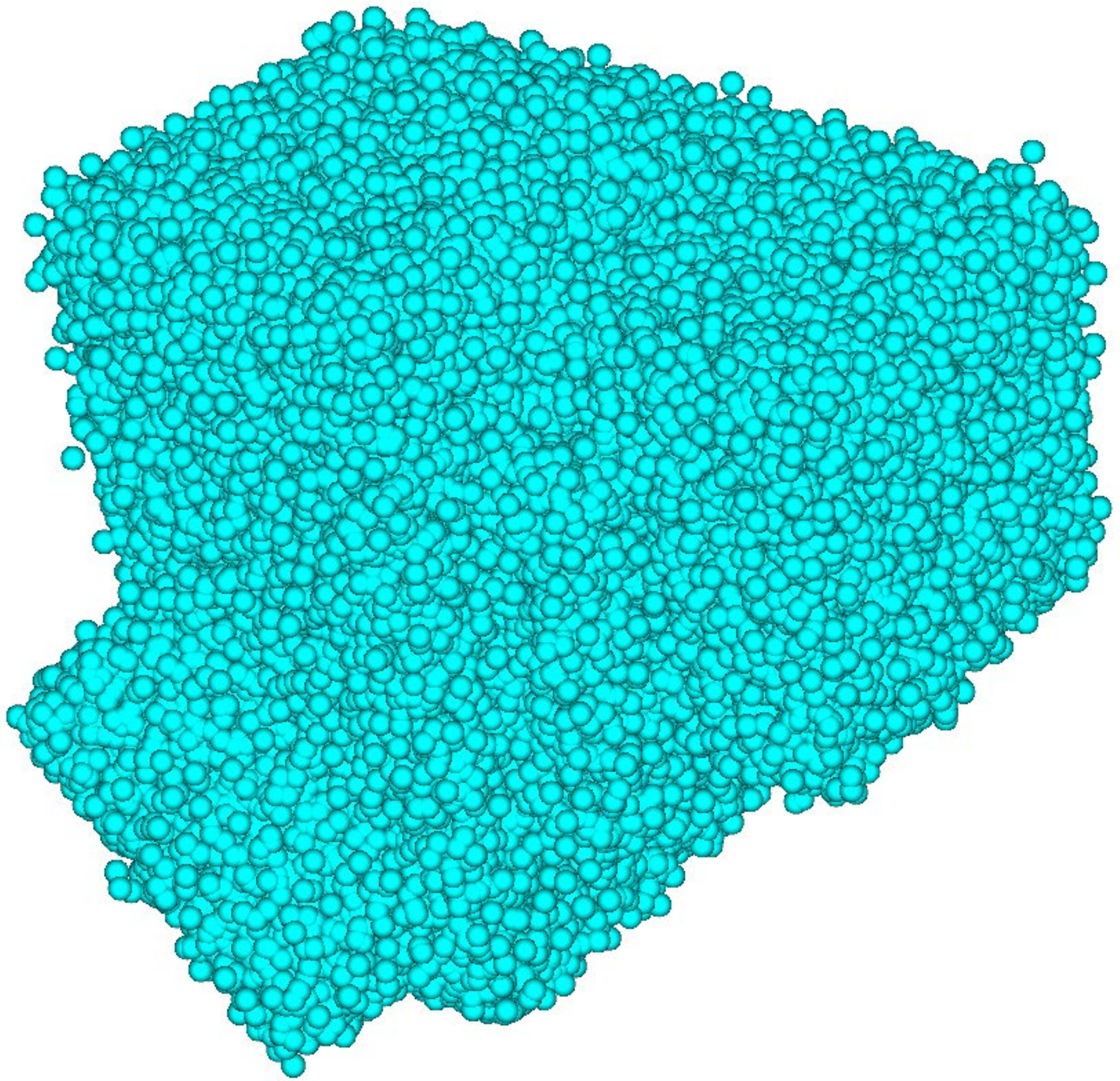}\\ 
    \includegraphics[width=1\textwidth]{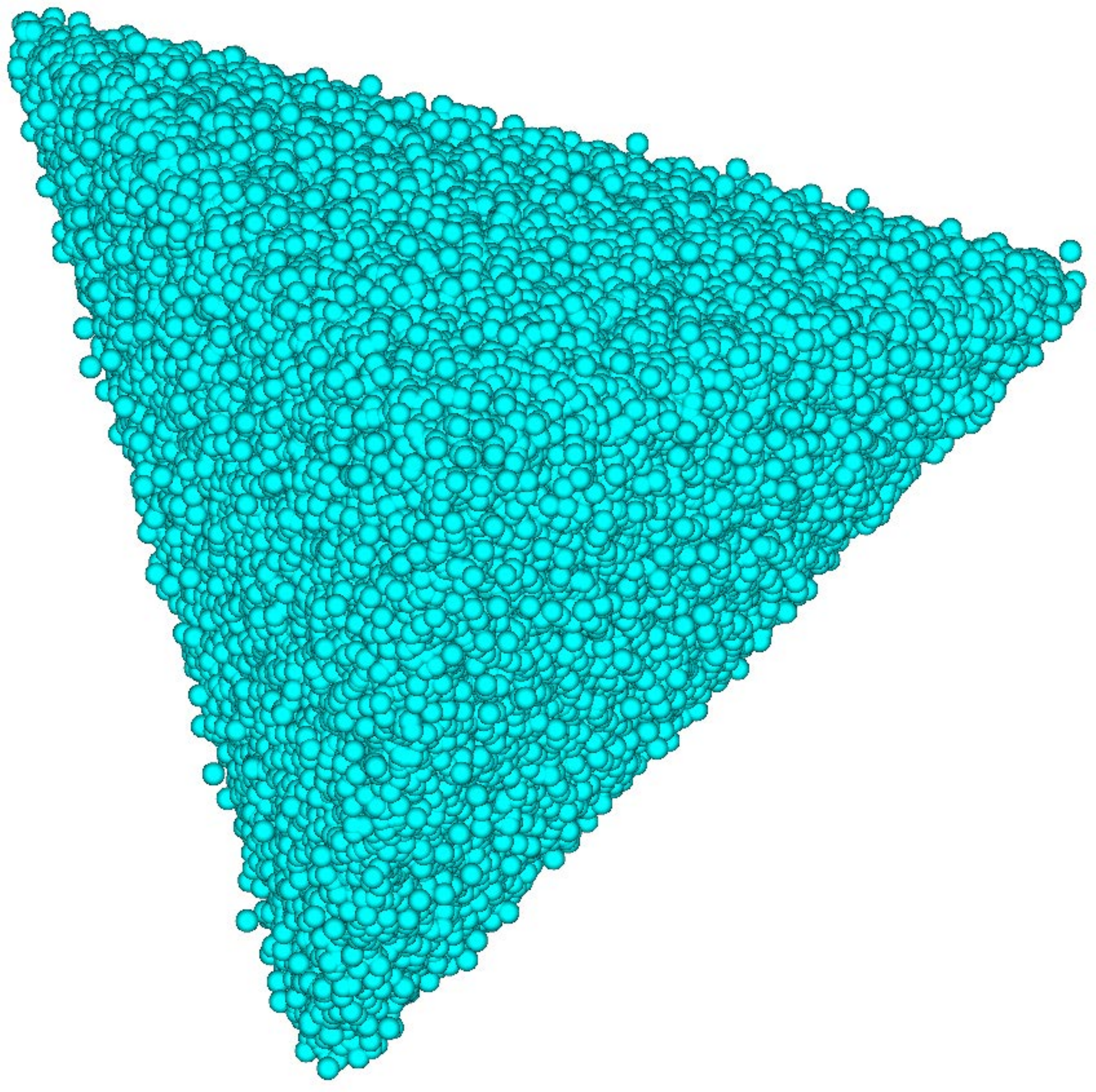}
    \end{minipage}
  }
  \subfigure[RIMLS]
  {
    \begin{minipage}[b]{0.1\textwidth} 
    \includegraphics[width=1\textwidth]{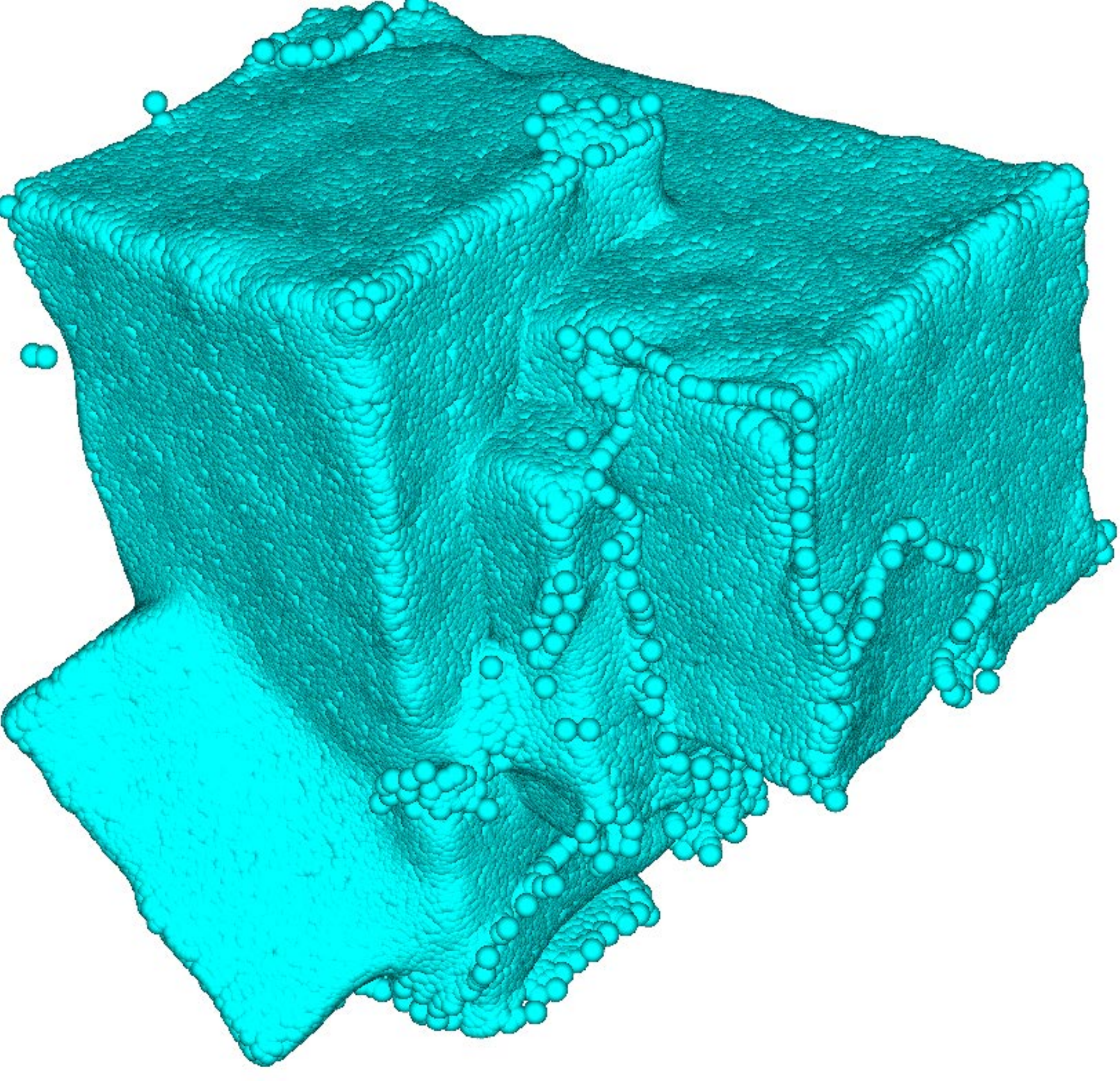}\\ 
    \includegraphics[width=1\textwidth]{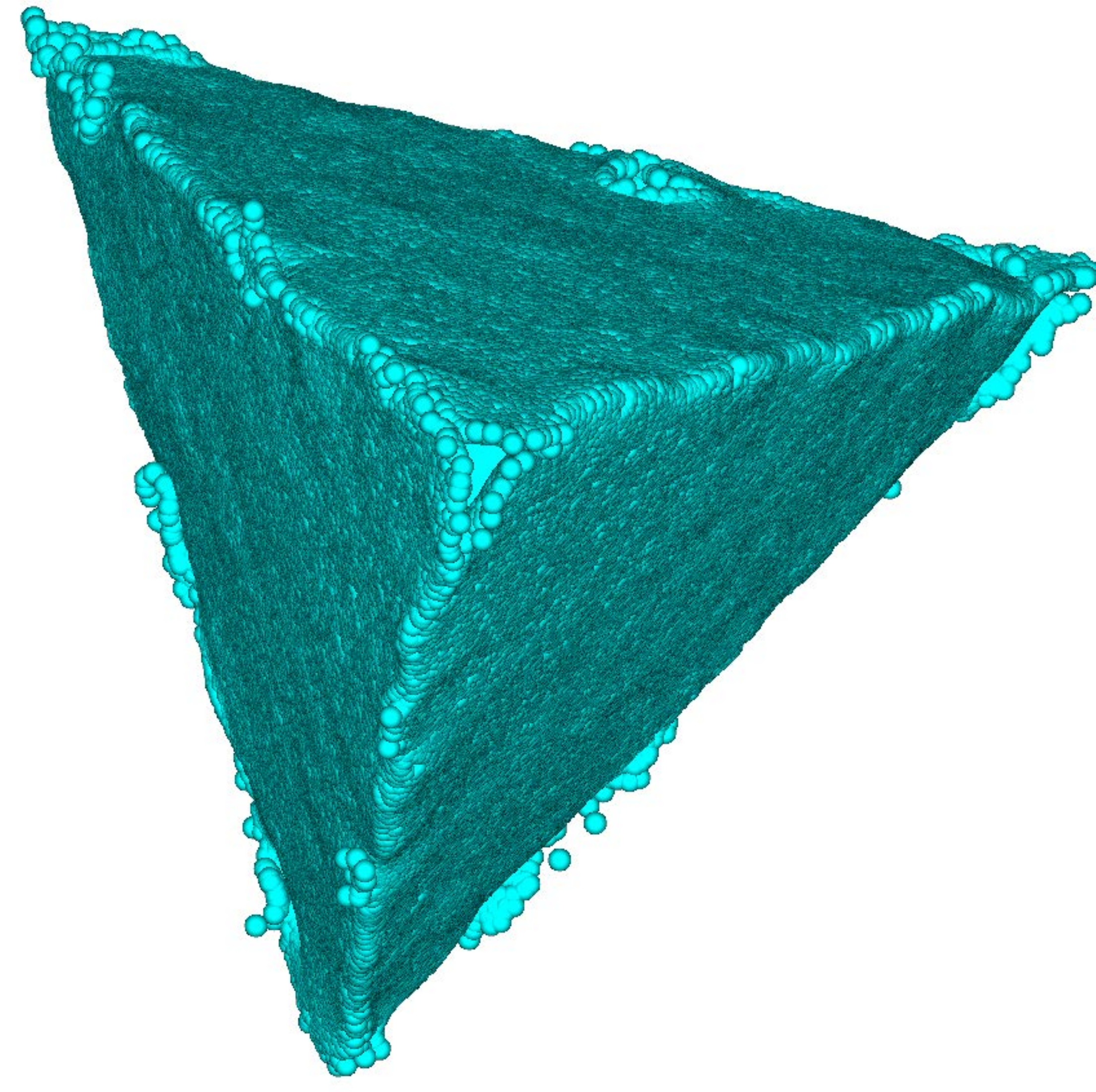}
    \end{minipage}
  }
  \subfigure[GPF]
  {
    \begin{minipage}[b]{0.1\textwidth} 
    \includegraphics[width=1\textwidth]{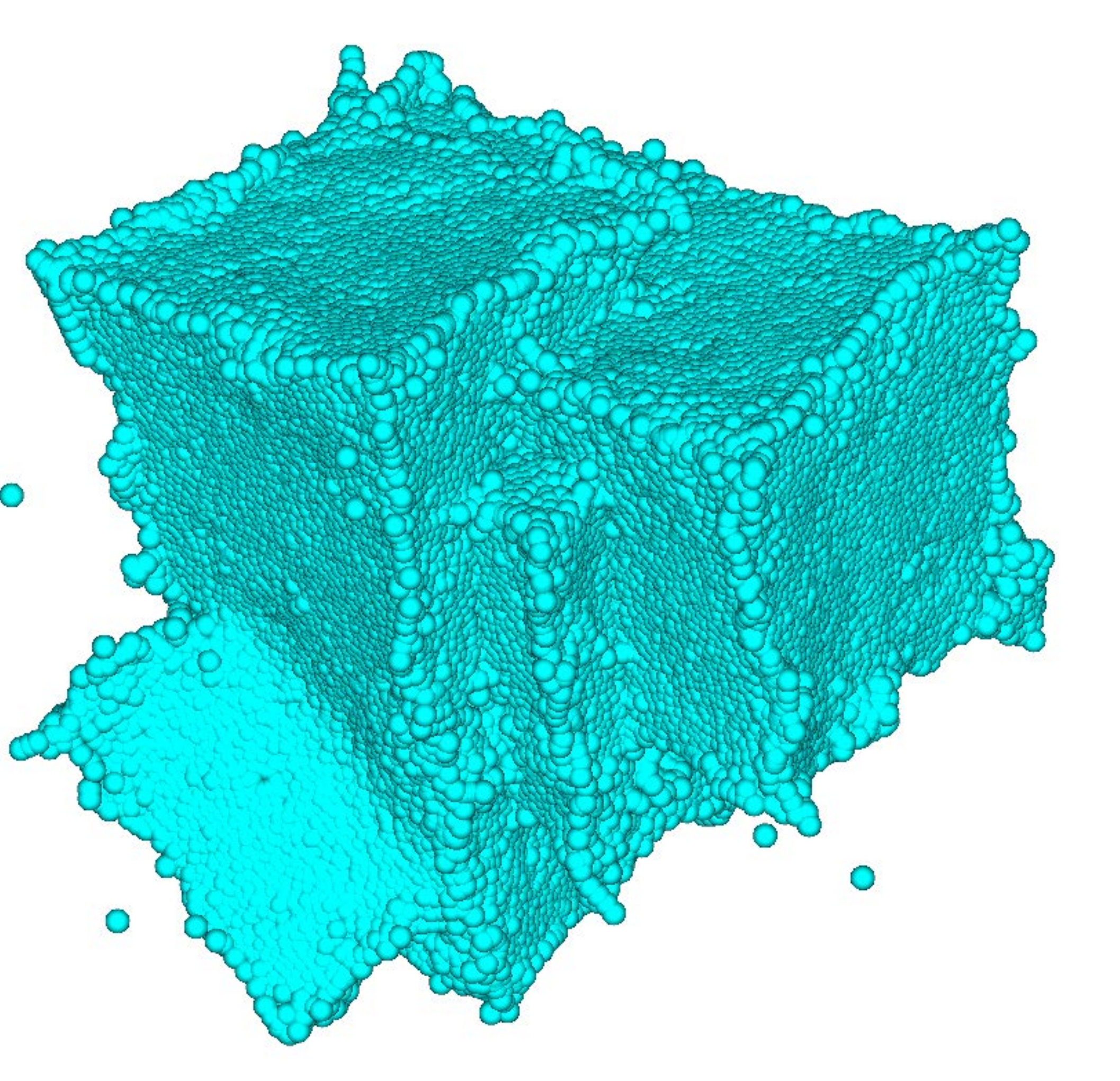}\\ 
    \includegraphics[width=1\textwidth]{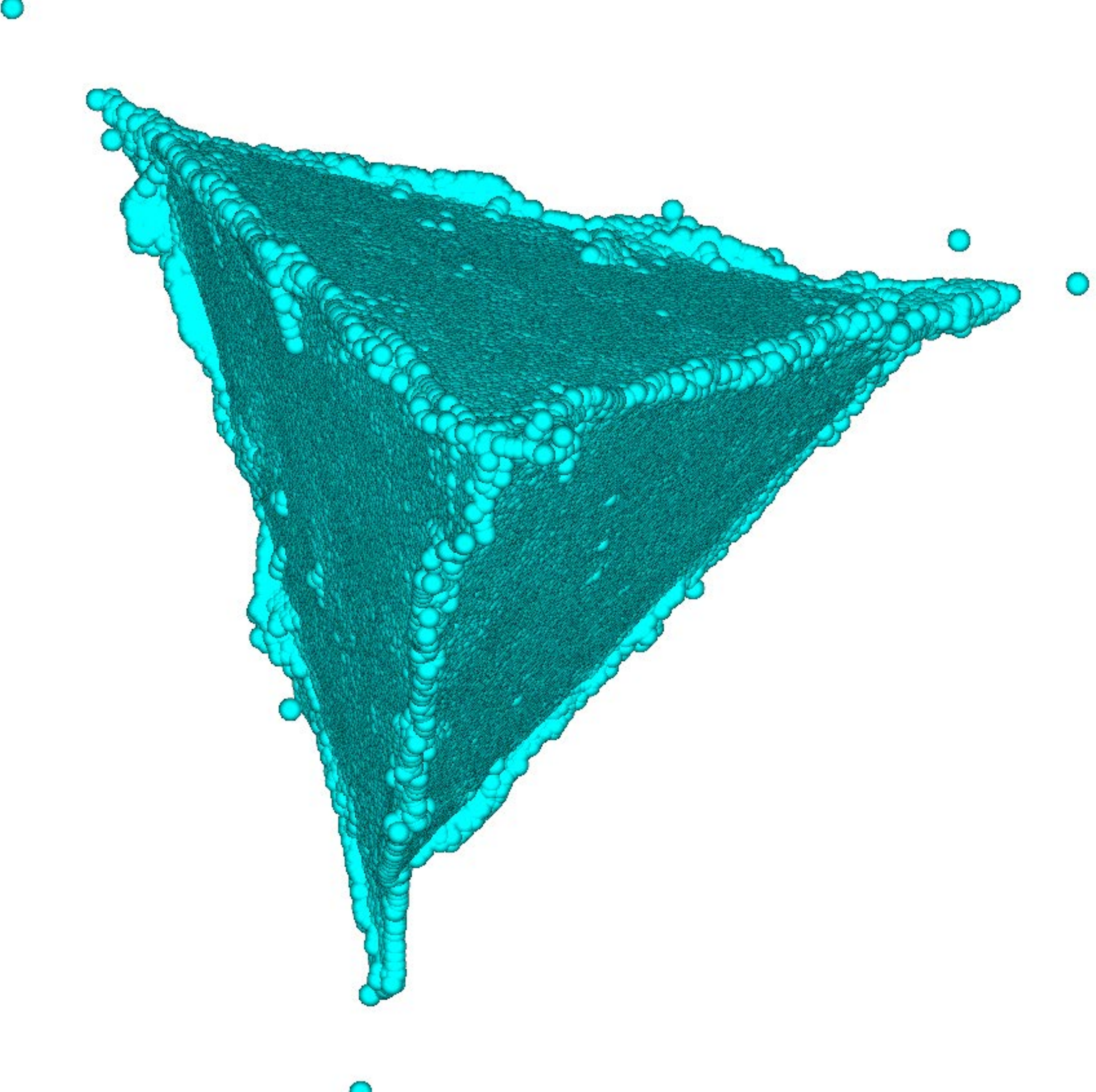}
    \end{minipage}
  }
  \subfigure[Ours]
  {
    \begin{minipage}[b]{0.1\textwidth} 
    \includegraphics[width=1\textwidth]{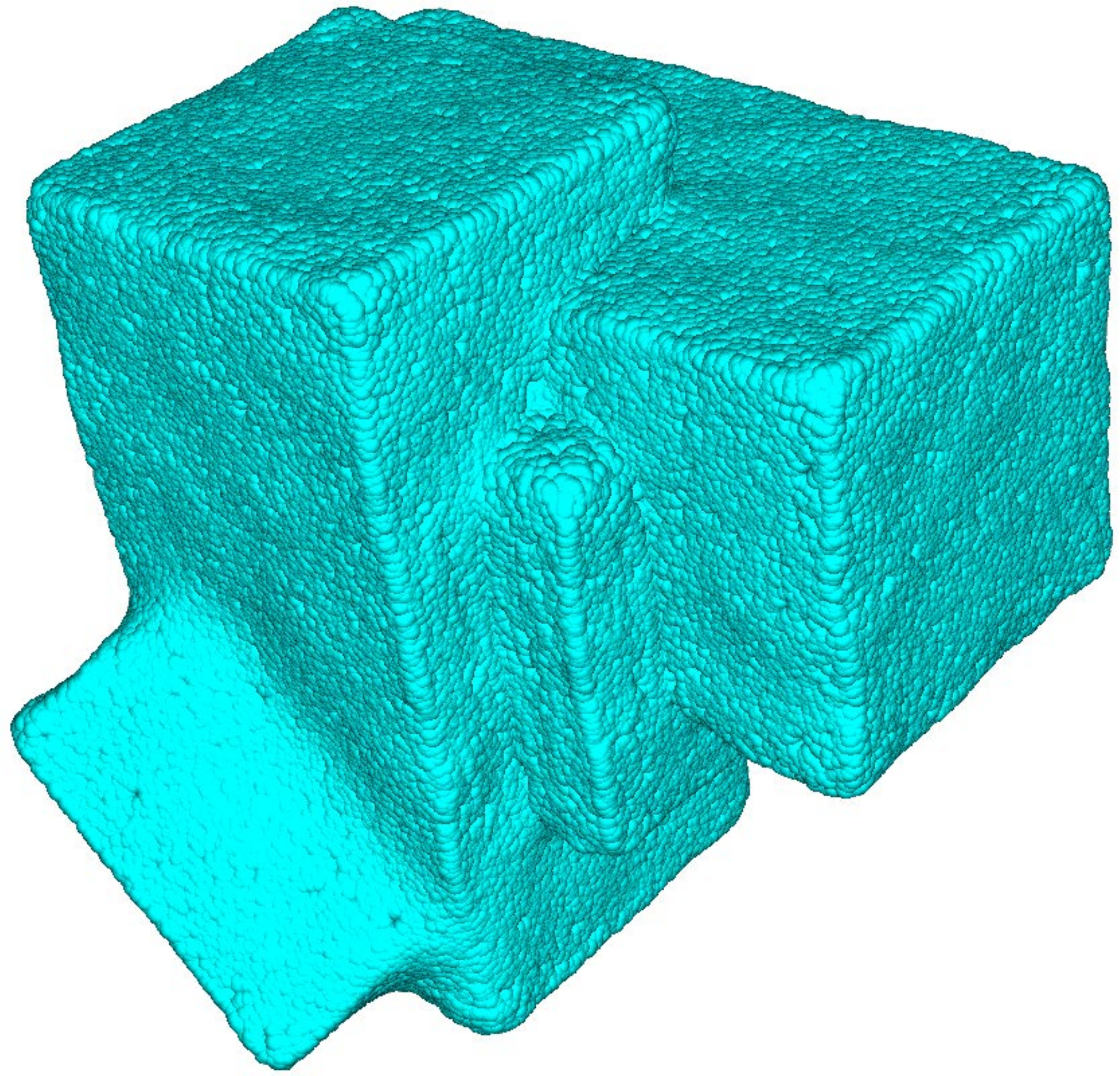}\\ 
    \includegraphics[width=1\textwidth]{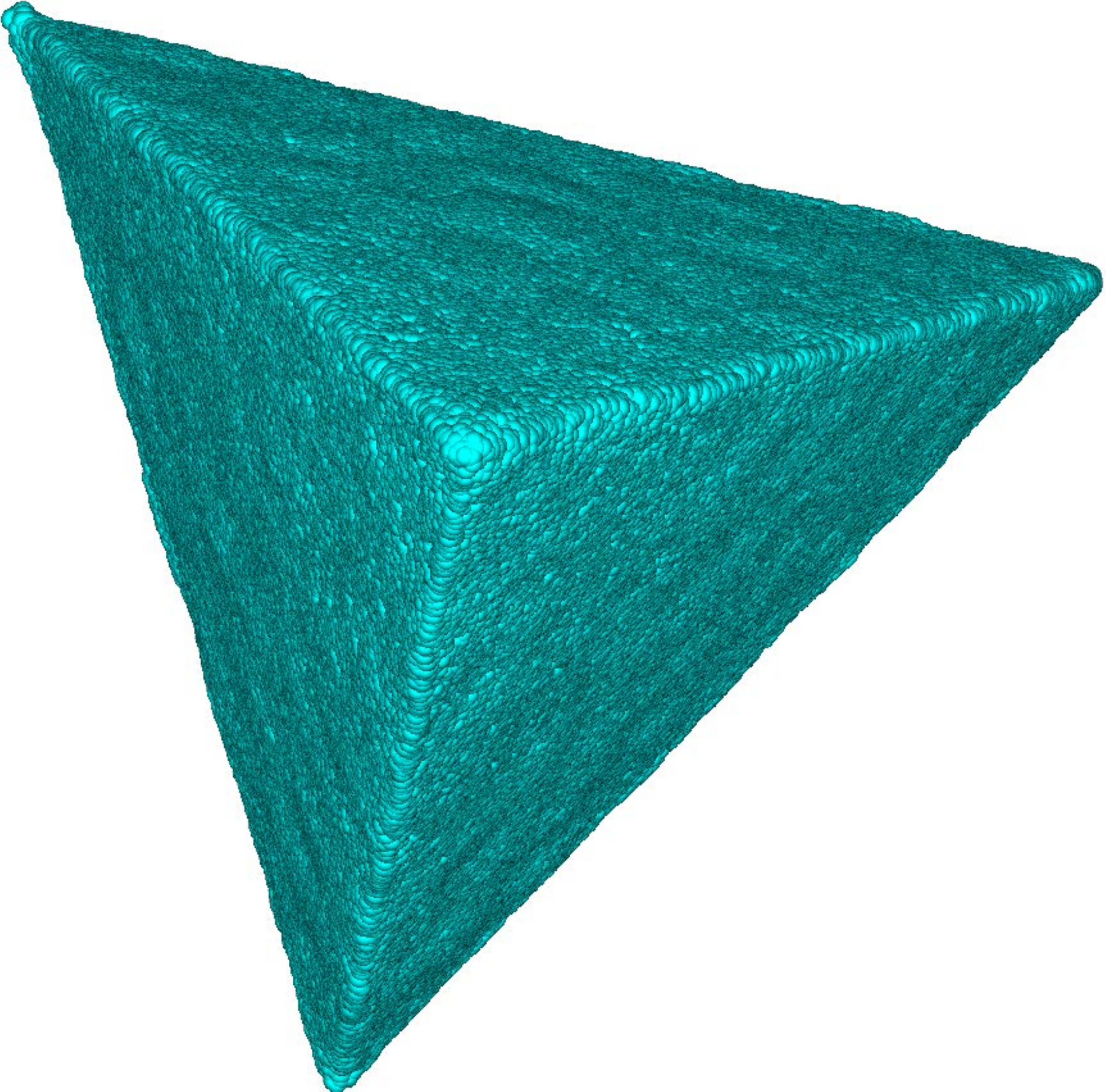}
    \end{minipage}
  }
  \caption{Point clouds with large noise (1.0\%).}
  \label{fig:normal_smoothing}
\end{figure}
\begin{figure*}[htb!]
  \centering
  \begin{tabular}{lllll}
  \tabincell{c}{\subfigure[Noisy]
  {
    \includegraphics[width=0.08\textwidth]{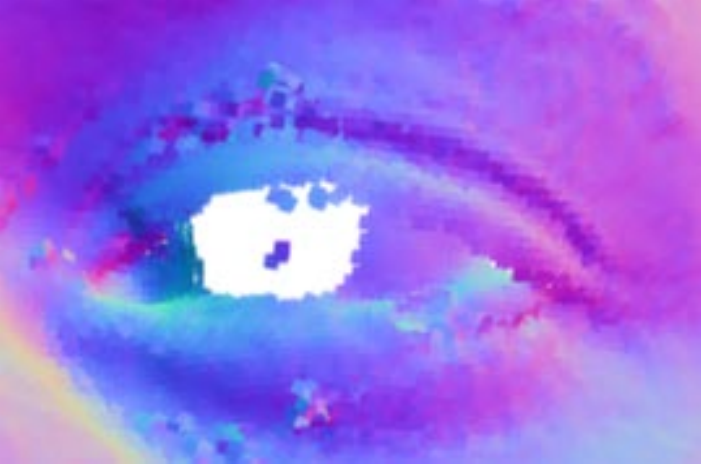}
    \includegraphics[width=0.08\textwidth]{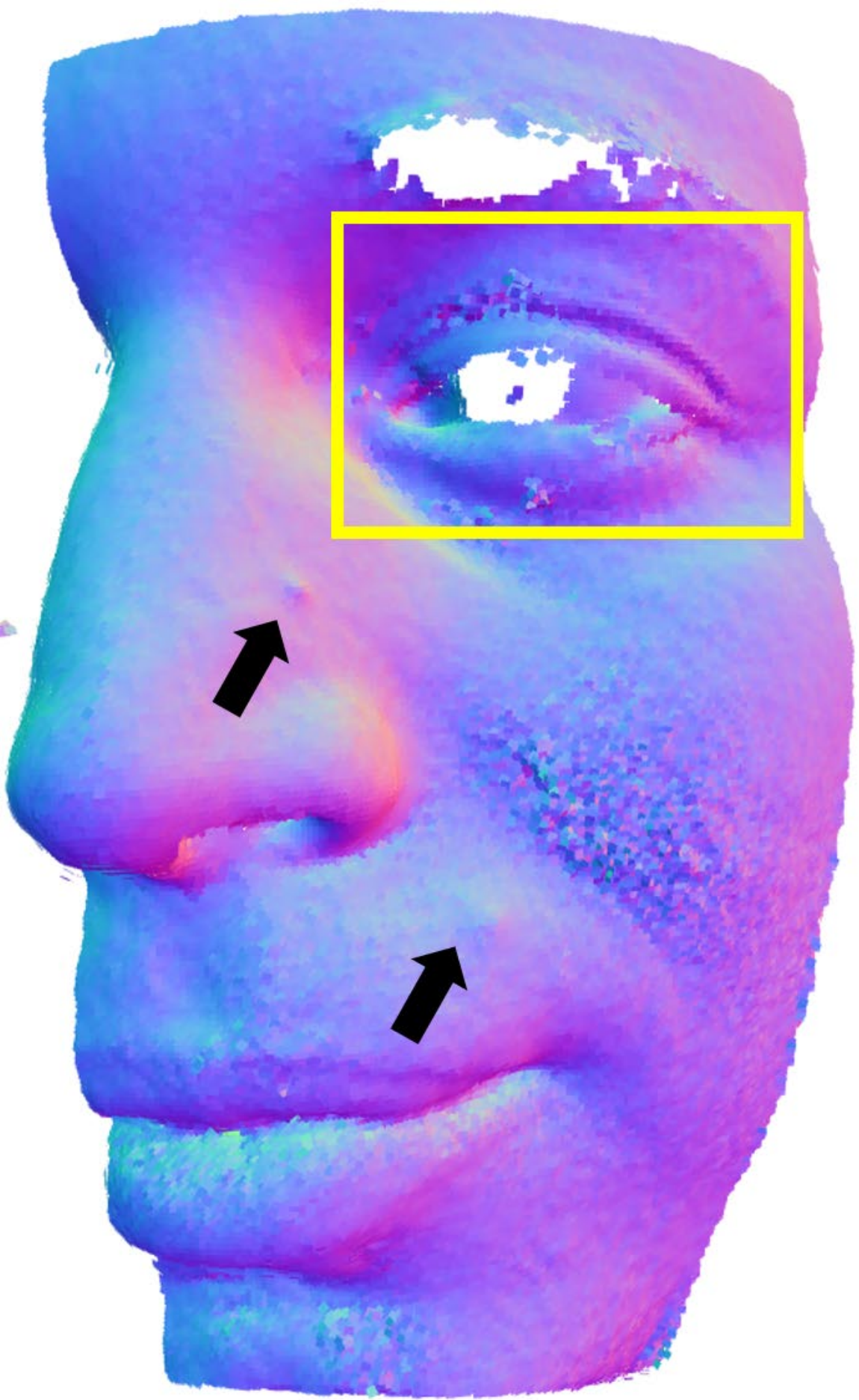}
  }} & \tabincell{c}{\subfigure[RIMLS]
  {
    \includegraphics[width=0.08\textwidth]{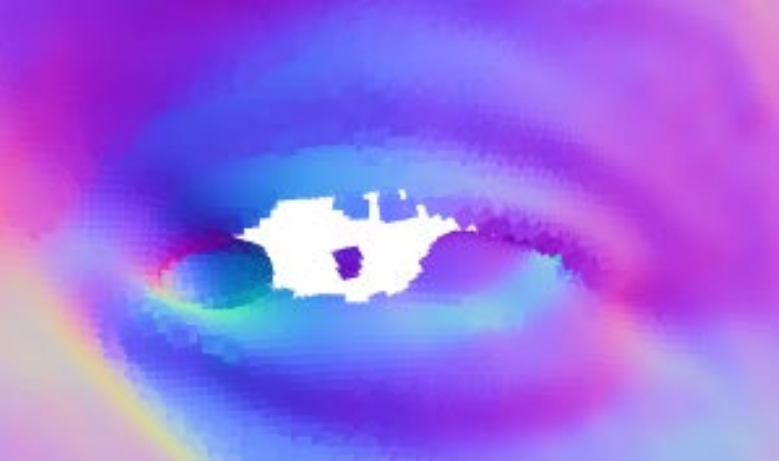}
    \includegraphics[width=0.08\textwidth]{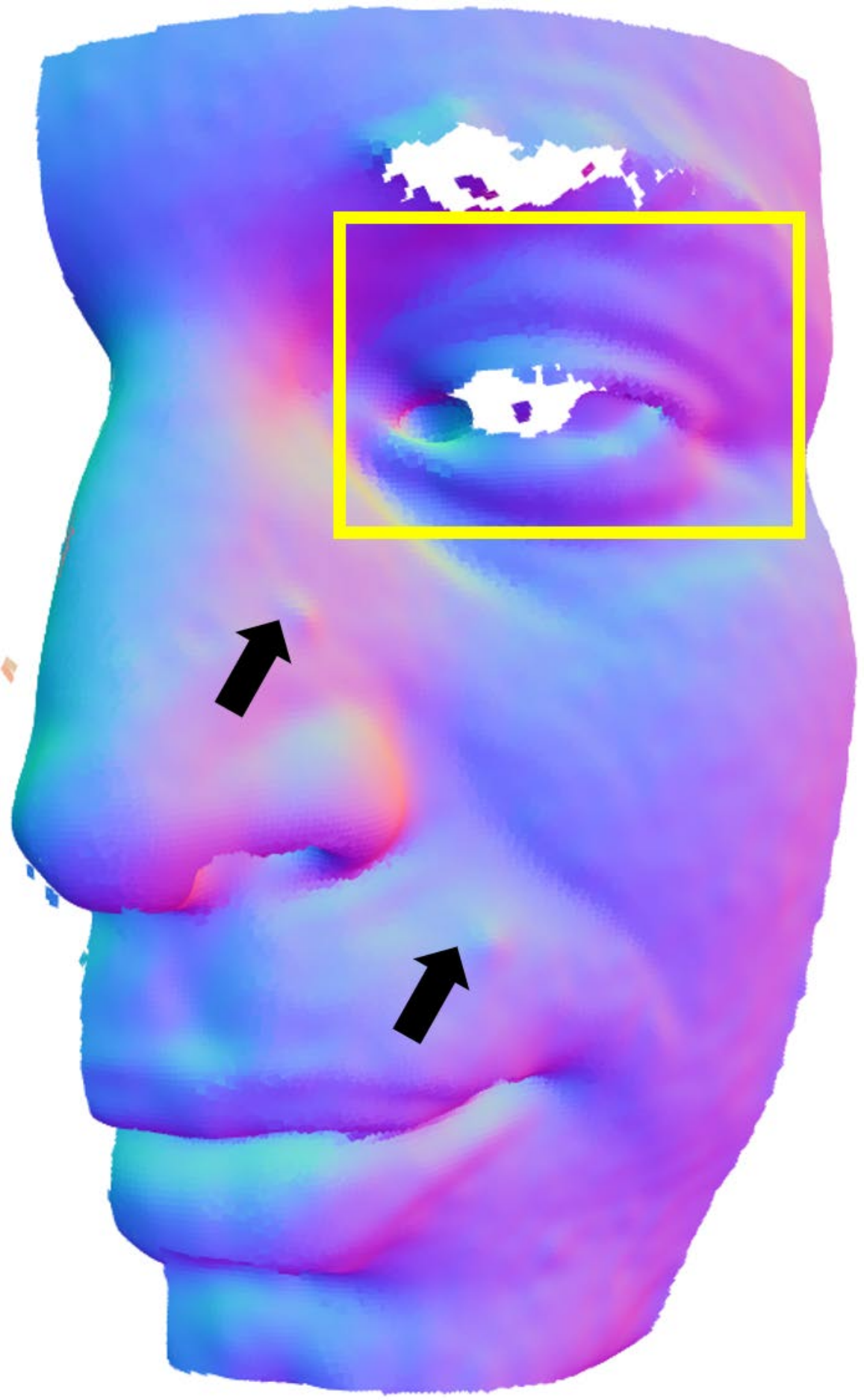}
  }\\ \subfigure[GPF]
  {
    \includegraphics[width=0.08\textwidth]{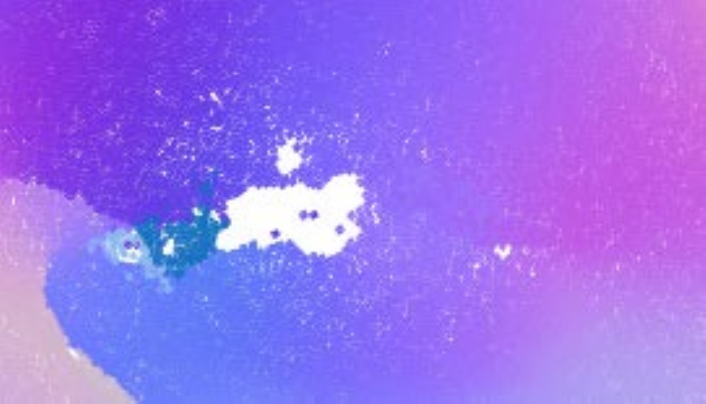}
    \includegraphics[width=0.08\textwidth]{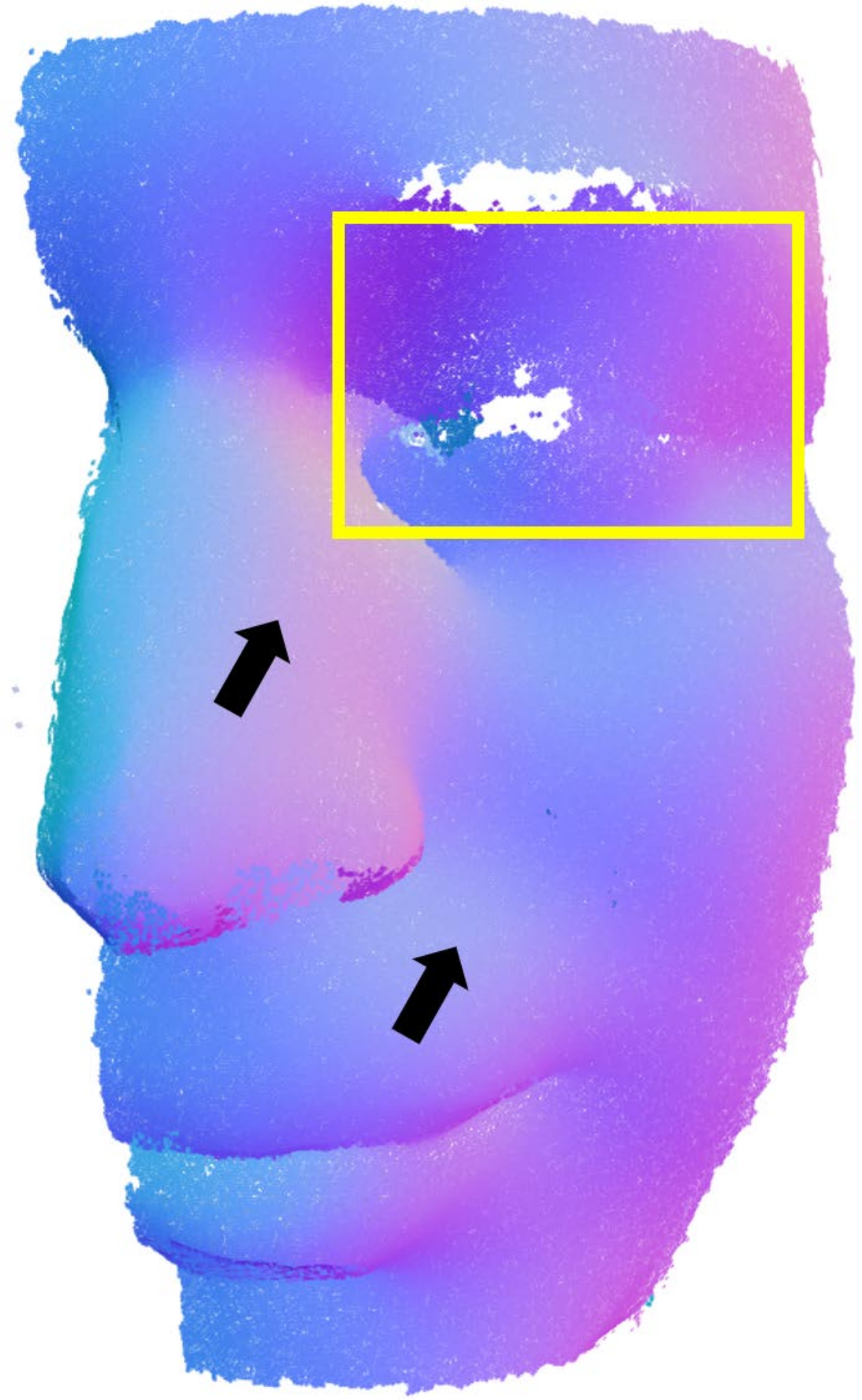}
  }} & \tabincell{c}{\subfigure[WLOP]
  {
    \includegraphics[width=0.08\textwidth]{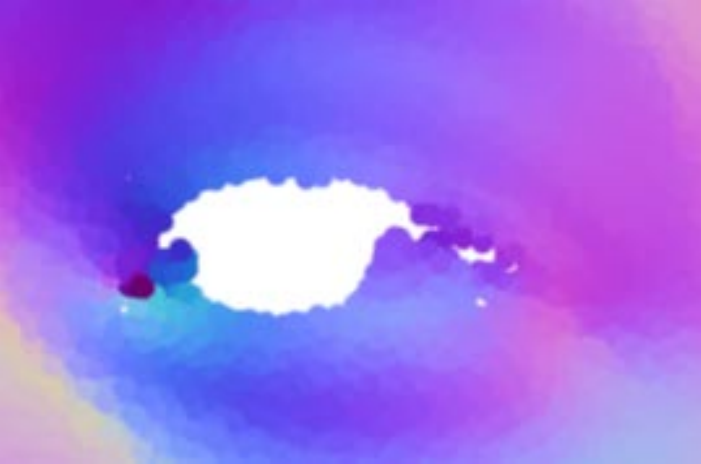}
    \includegraphics[width=0.08\textwidth]{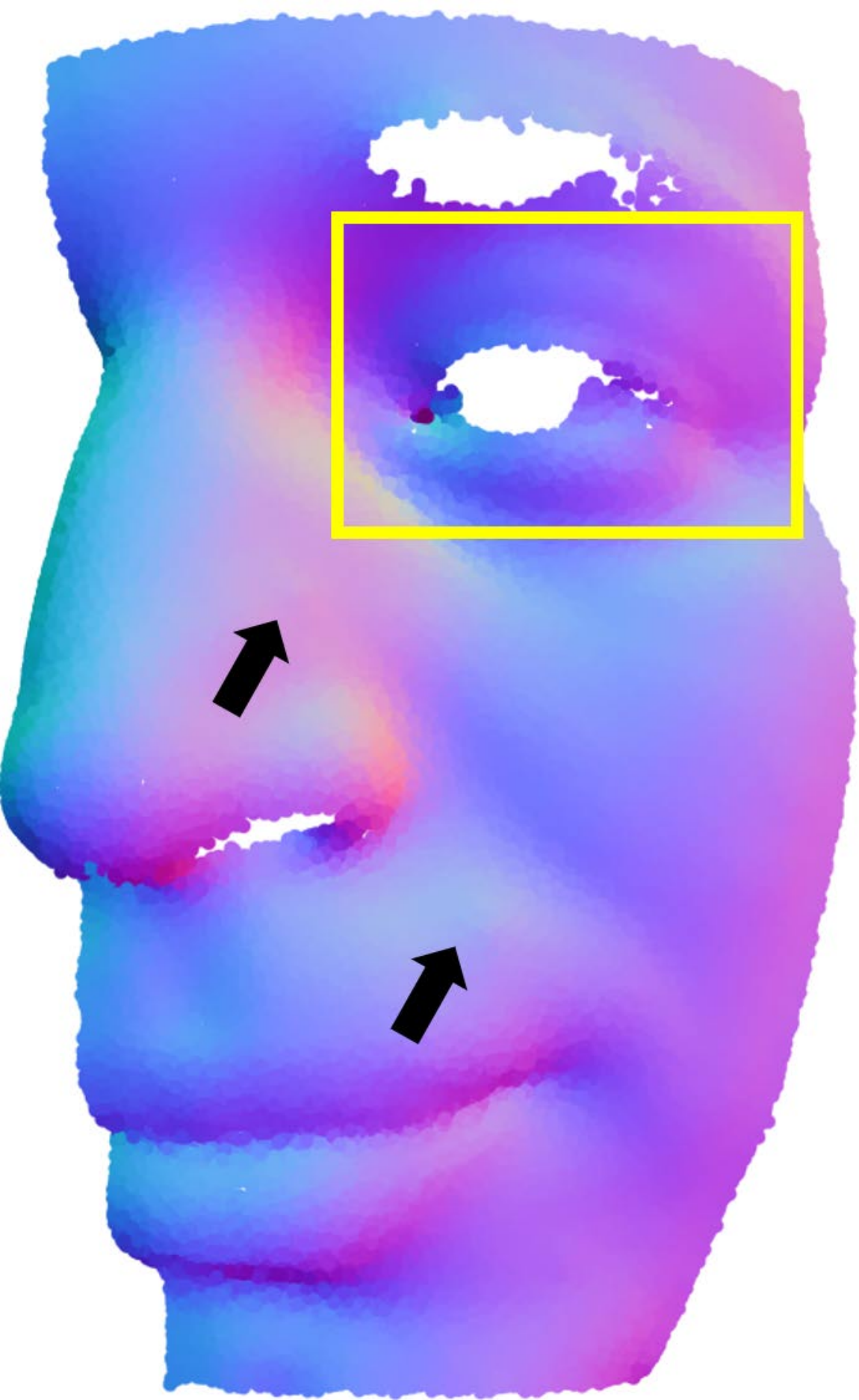}
  } \\ \subfigure[CLOP]
  {
    \includegraphics[width=0.08\textwidth]{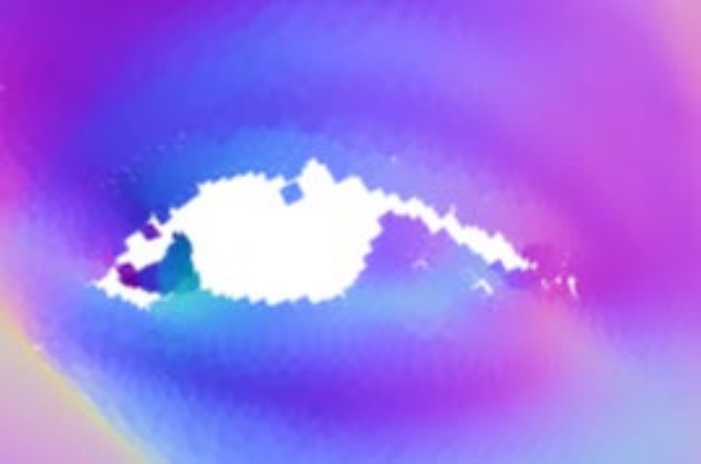}
    \includegraphics[width=0.08\textwidth]{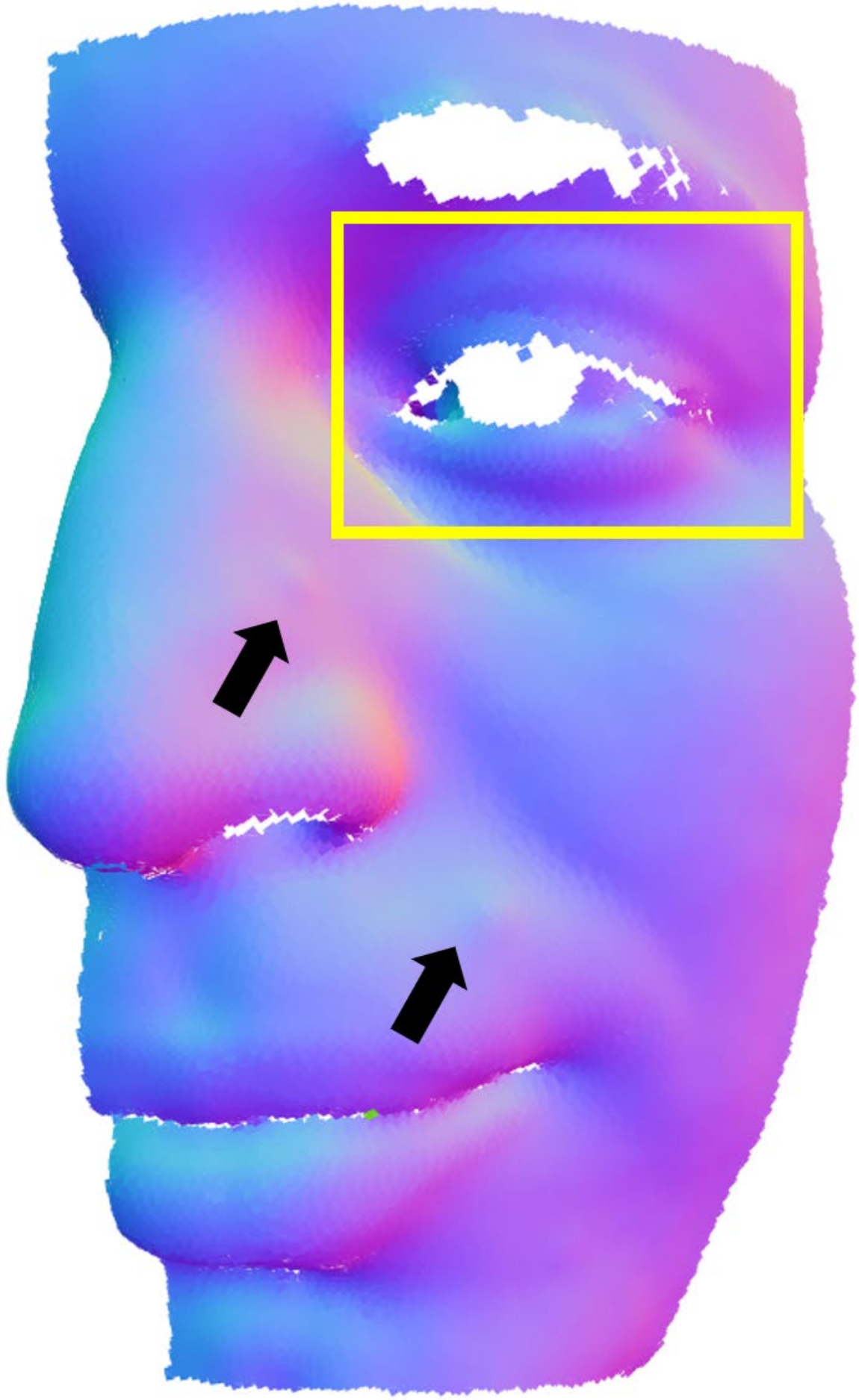}
  }} & \tabincell{c}{\subfigure[EC-Net]
  {
    \includegraphics[width=0.08\textwidth]{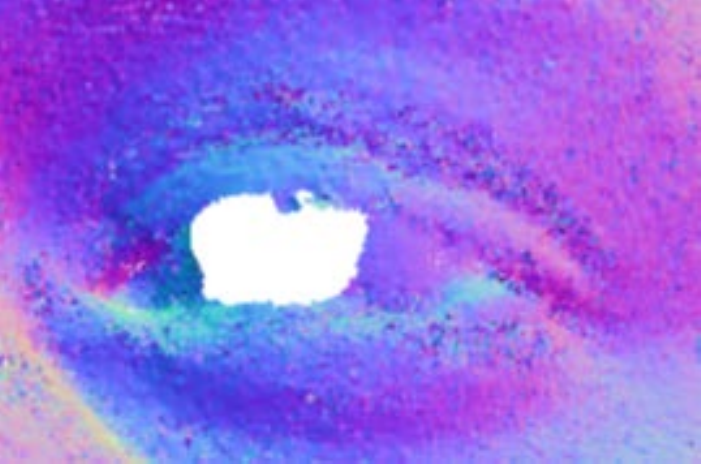}
    \includegraphics[width=0.08\textwidth]{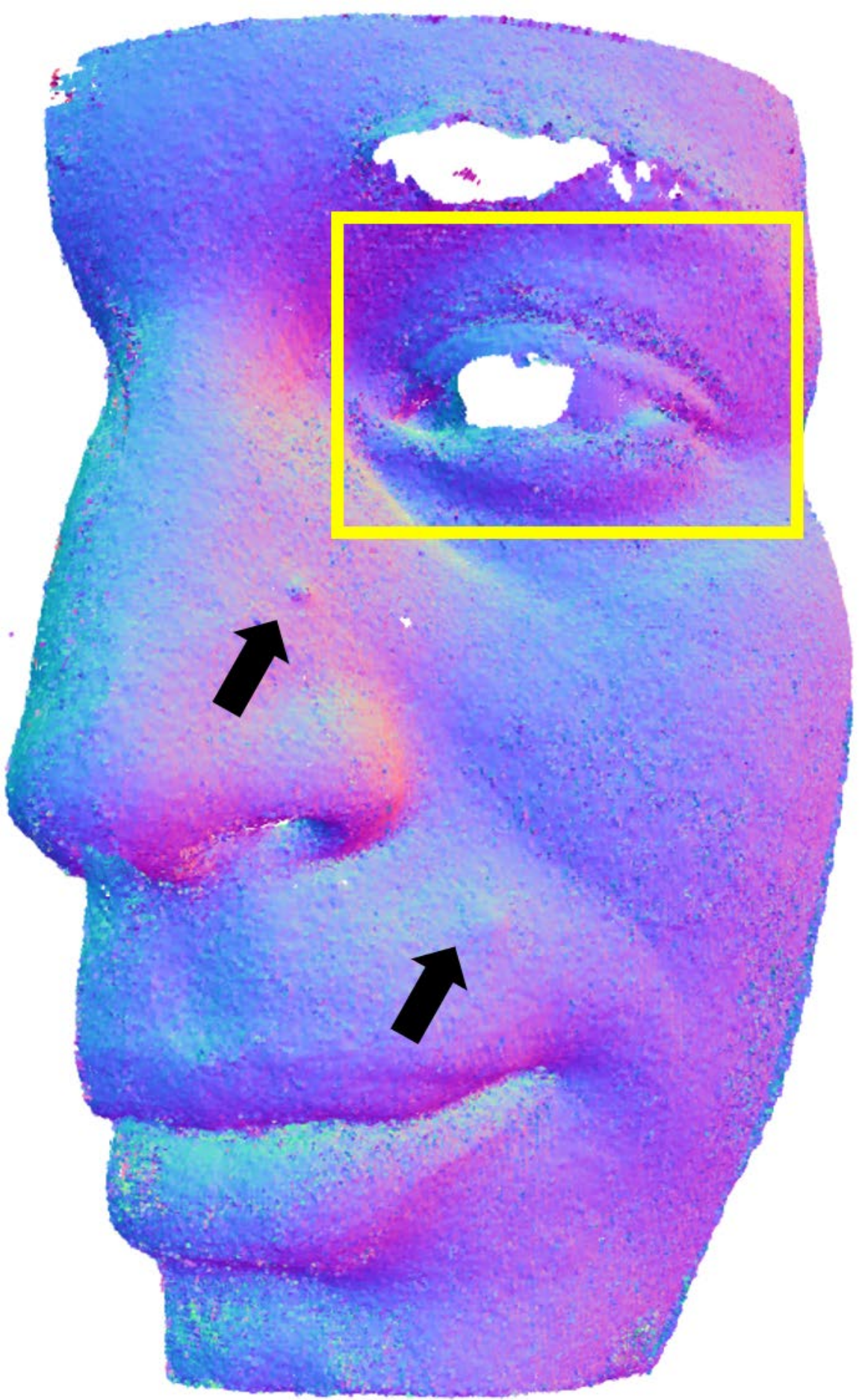}
  } \\ \subfigure[PCN]
  {
    \includegraphics[width=0.08\textwidth]{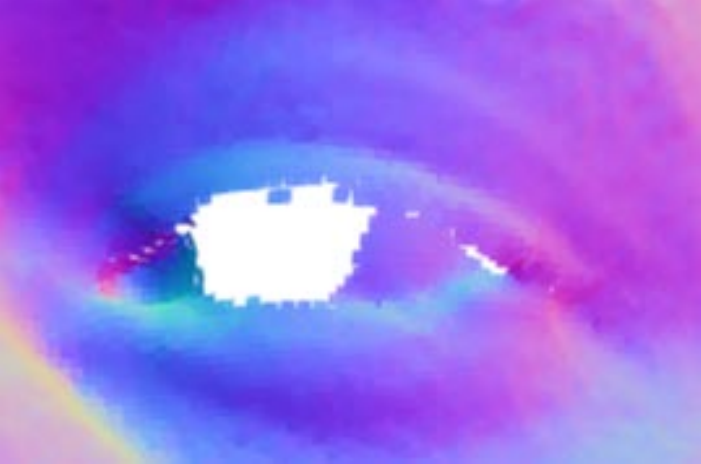}
    \includegraphics[width=0.08\textwidth]{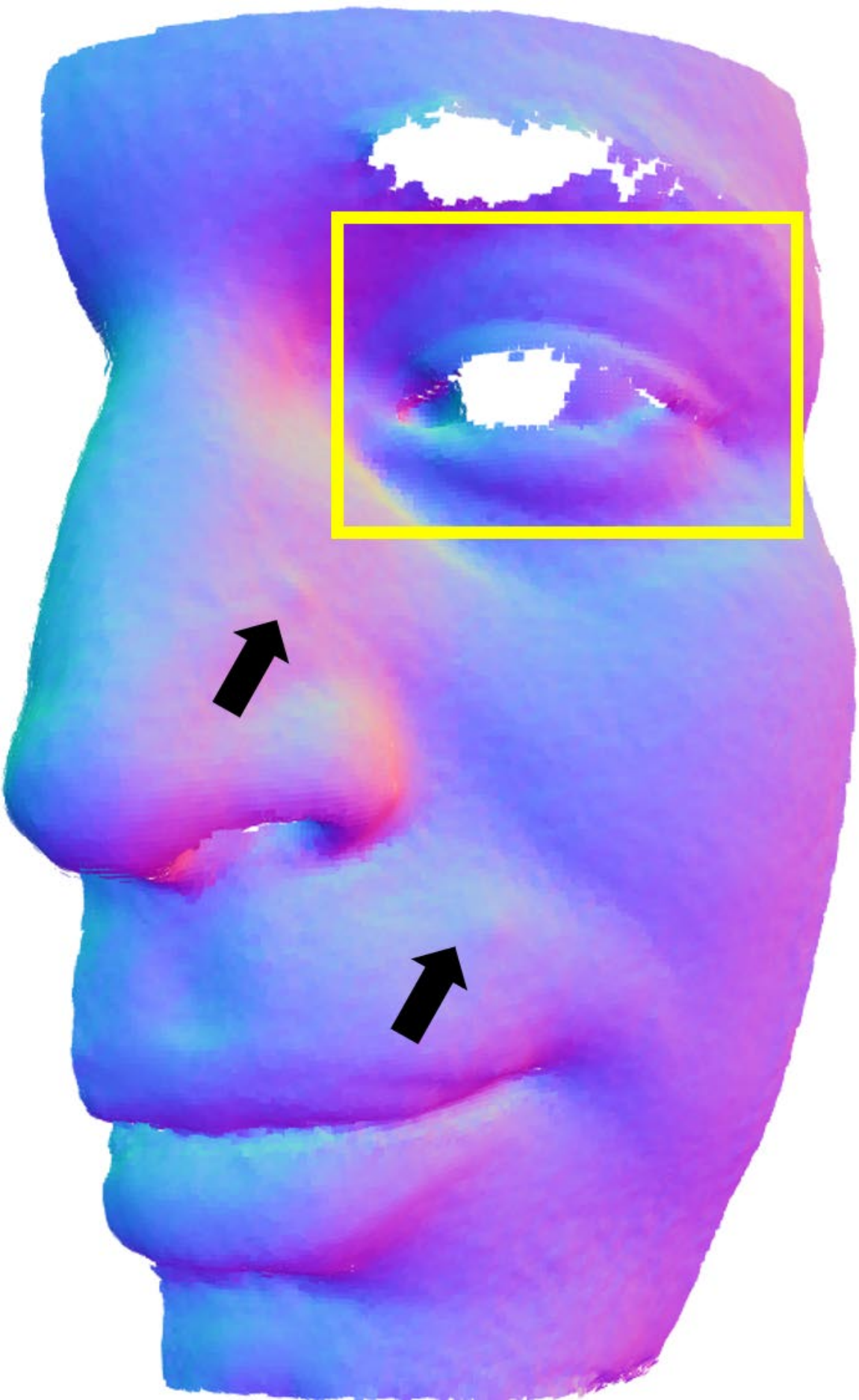}
  }} & \tabincell{c}{\subfigure[TD]
  {
    \includegraphics[width=0.08\textwidth]{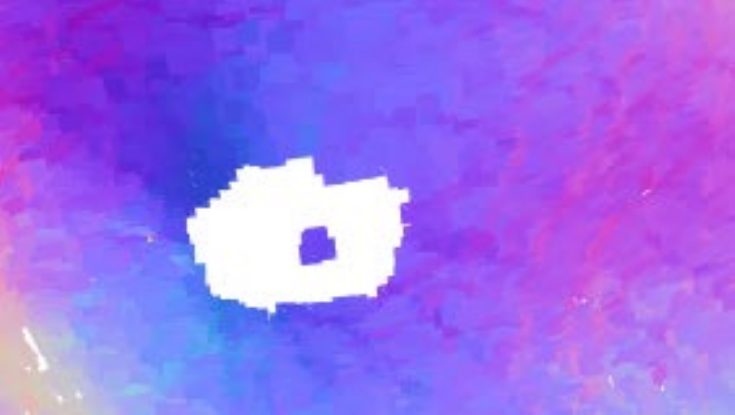}
    \includegraphics[width=0.08\textwidth]{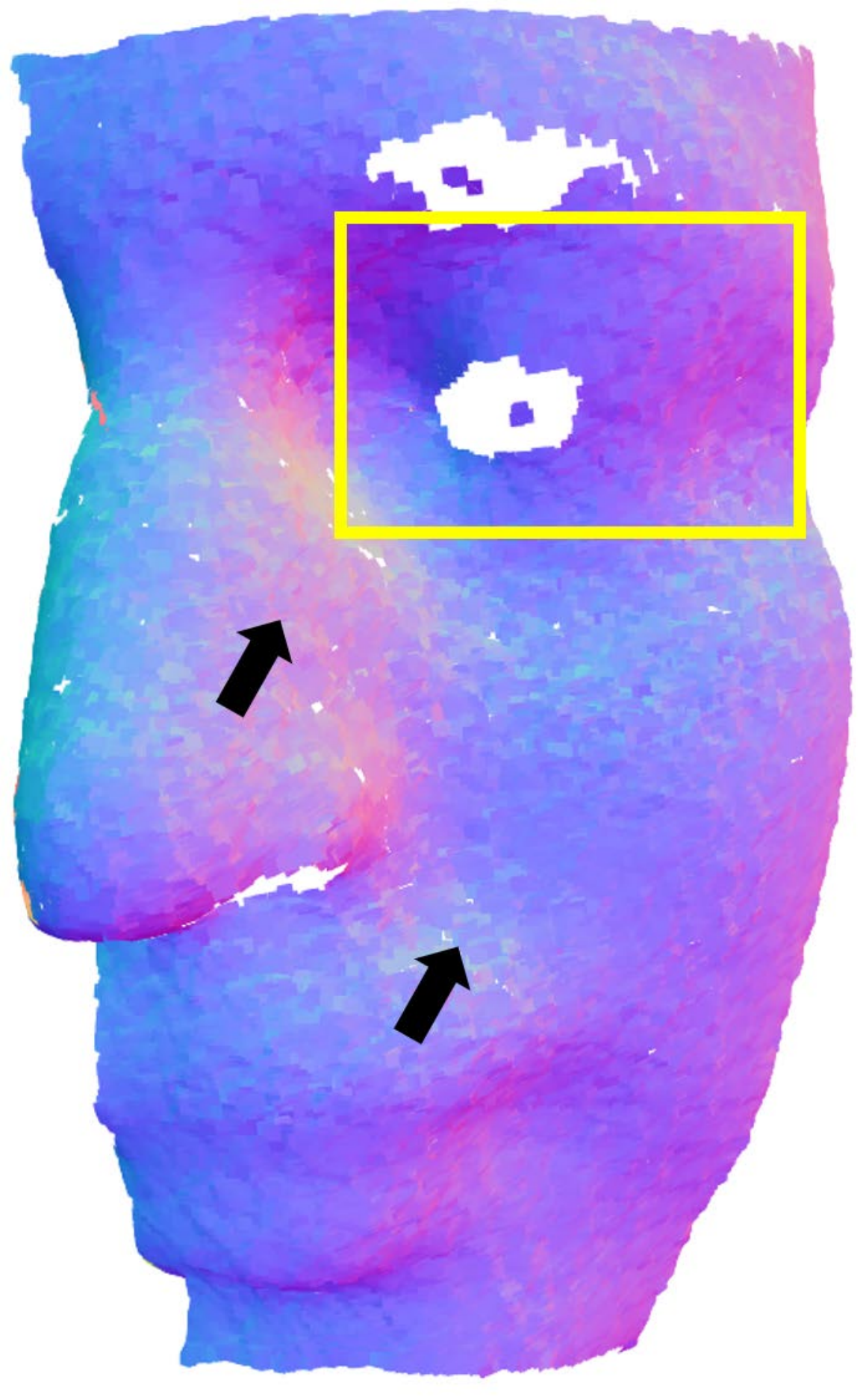}
  } \\ \subfigure[Ours]
  {
    \includegraphics[width=0.08\textwidth]{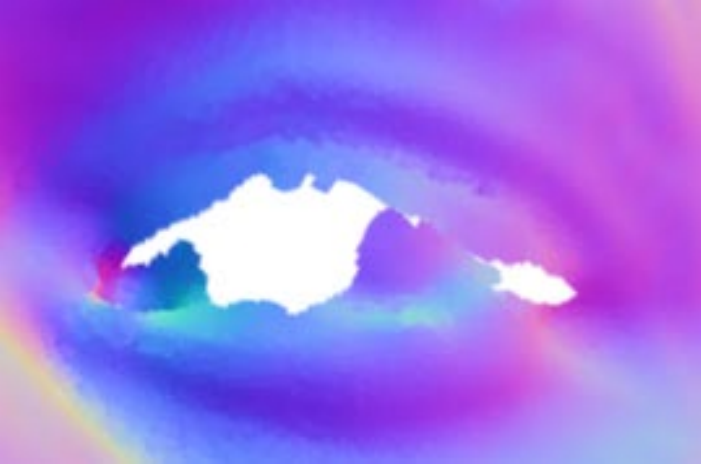}
    \includegraphics[width=0.08\textwidth]{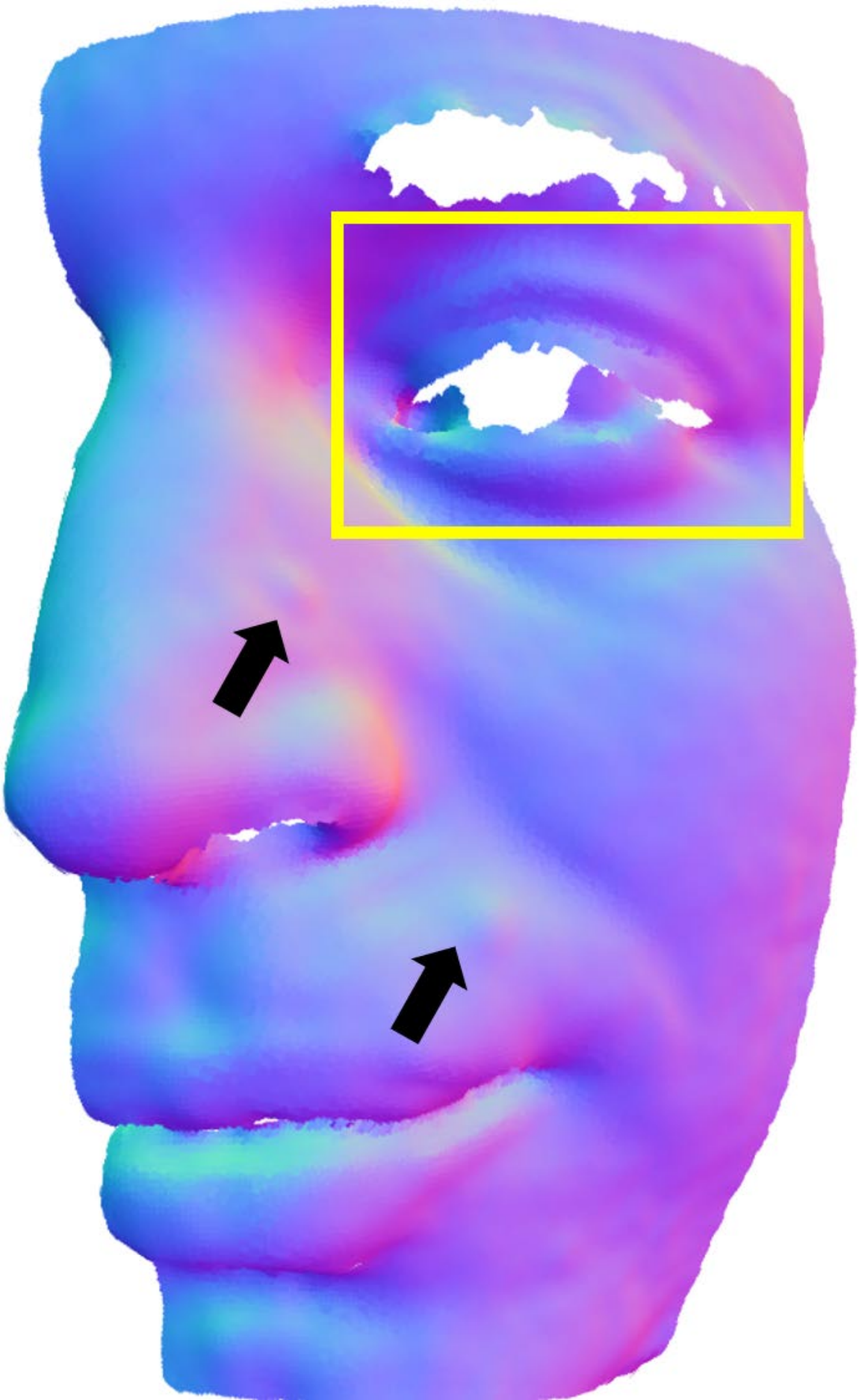}
  }}
  \end{tabular}
  \caption{Results on the raw Face model. We highlight the artifacts using arrows and show close-up views for better visualization.}
  \label{fig:rawfacefiltering}
\end{figure*}
\begin{figure*}[htb!]
    \subfigure[Noisy]
    {
        \begin{minipage}[b]{0.095\textwidth} 
        \includegraphics[width=1\textwidth ]{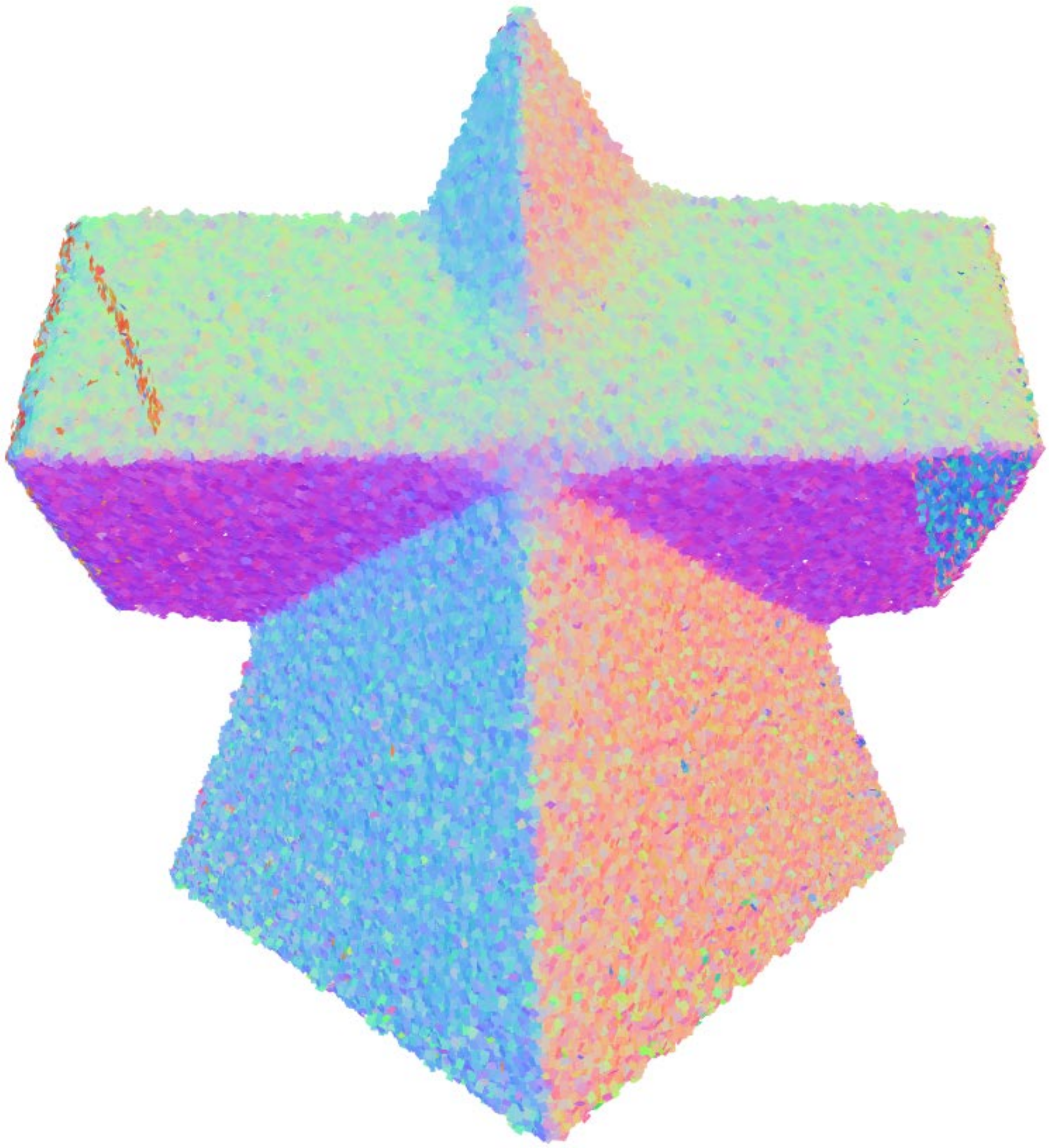}\\
        \includegraphics[width=1\textwidth ]{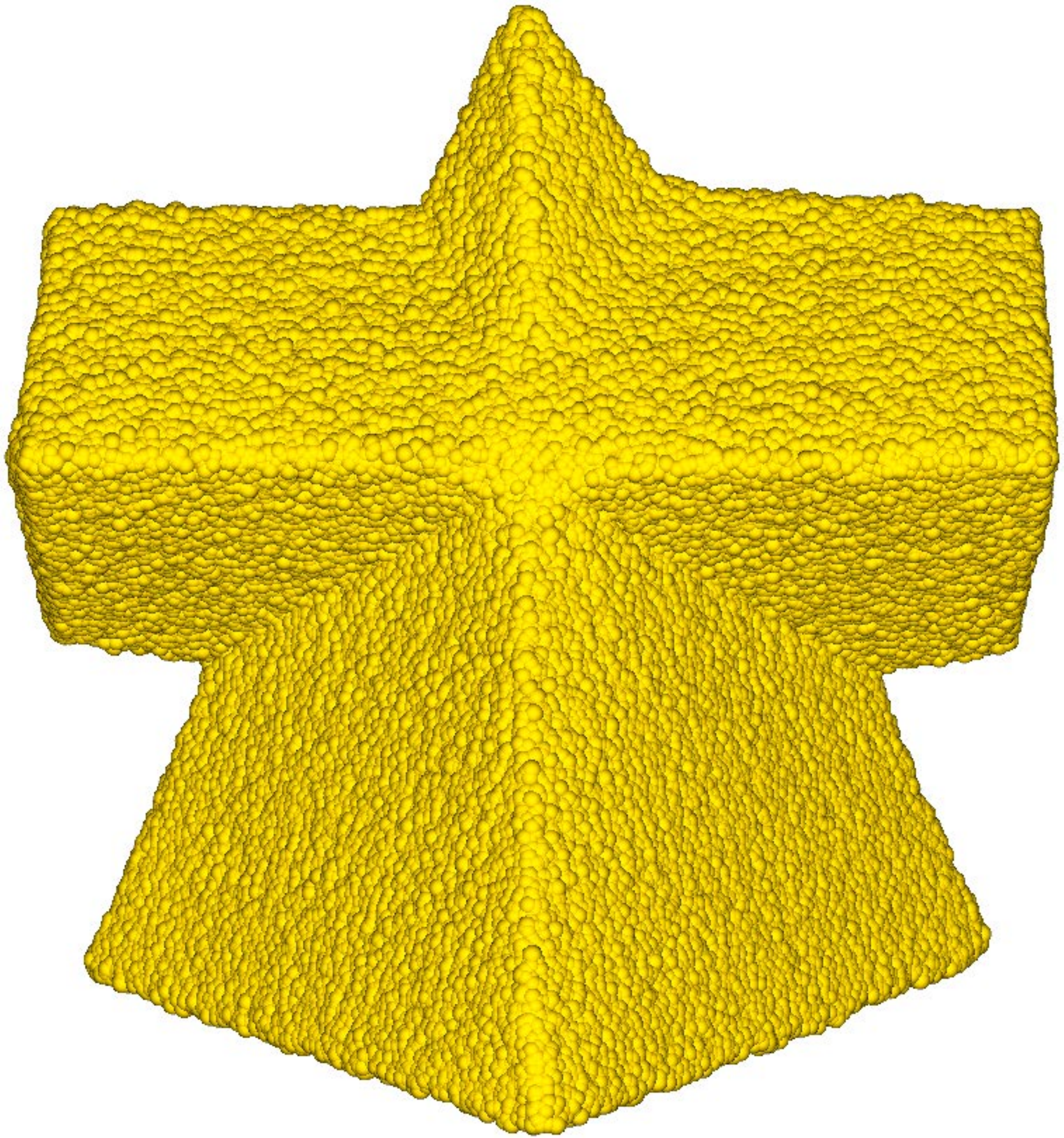}
        \end{minipage}
    }
    \subfigure[RIMLS]
    {
        \begin{minipage}[b]{0.095\textwidth} 
        \includegraphics[width=1\textwidth ]{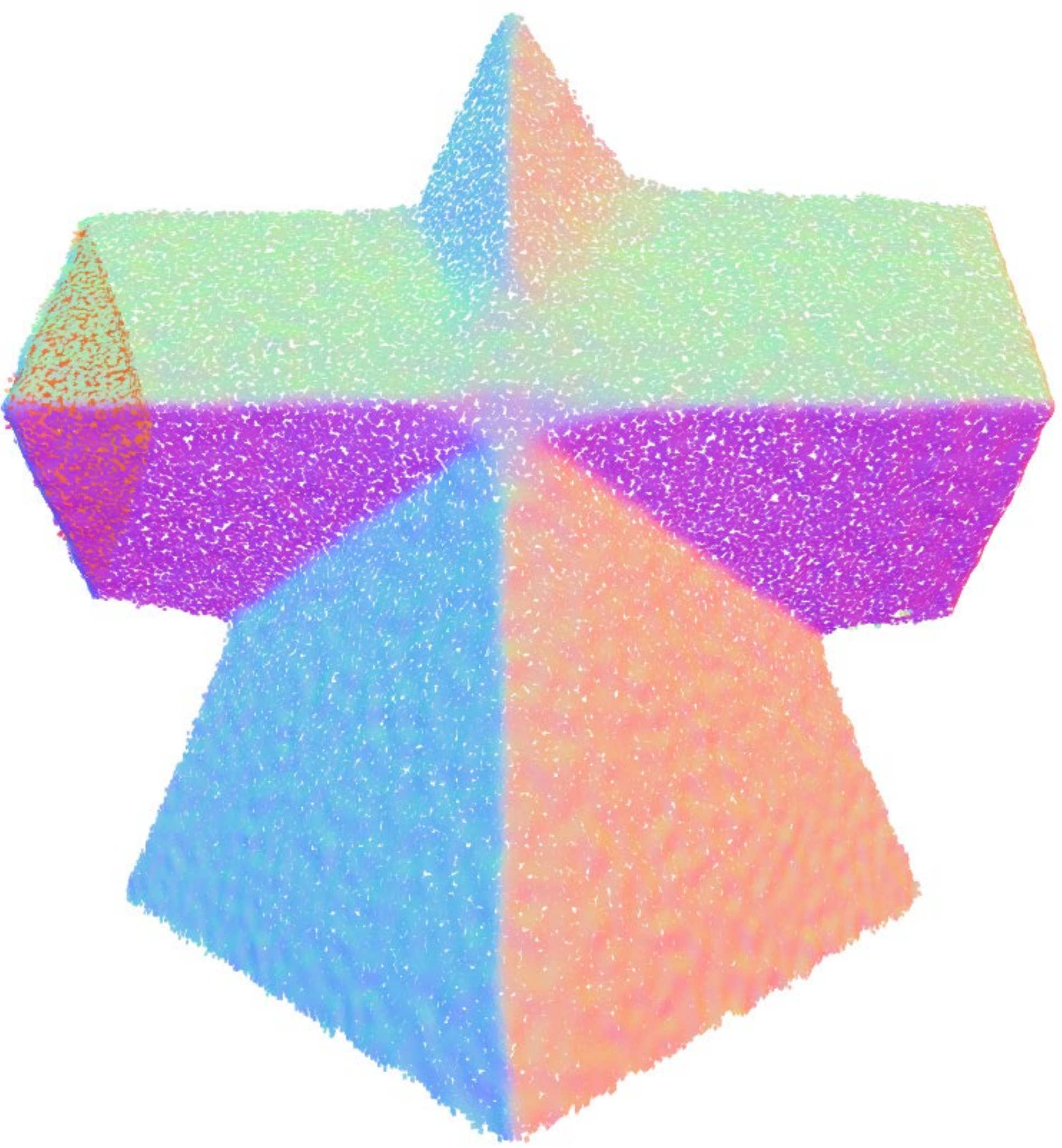}\\
        \includegraphics[width=1\textwidth ]{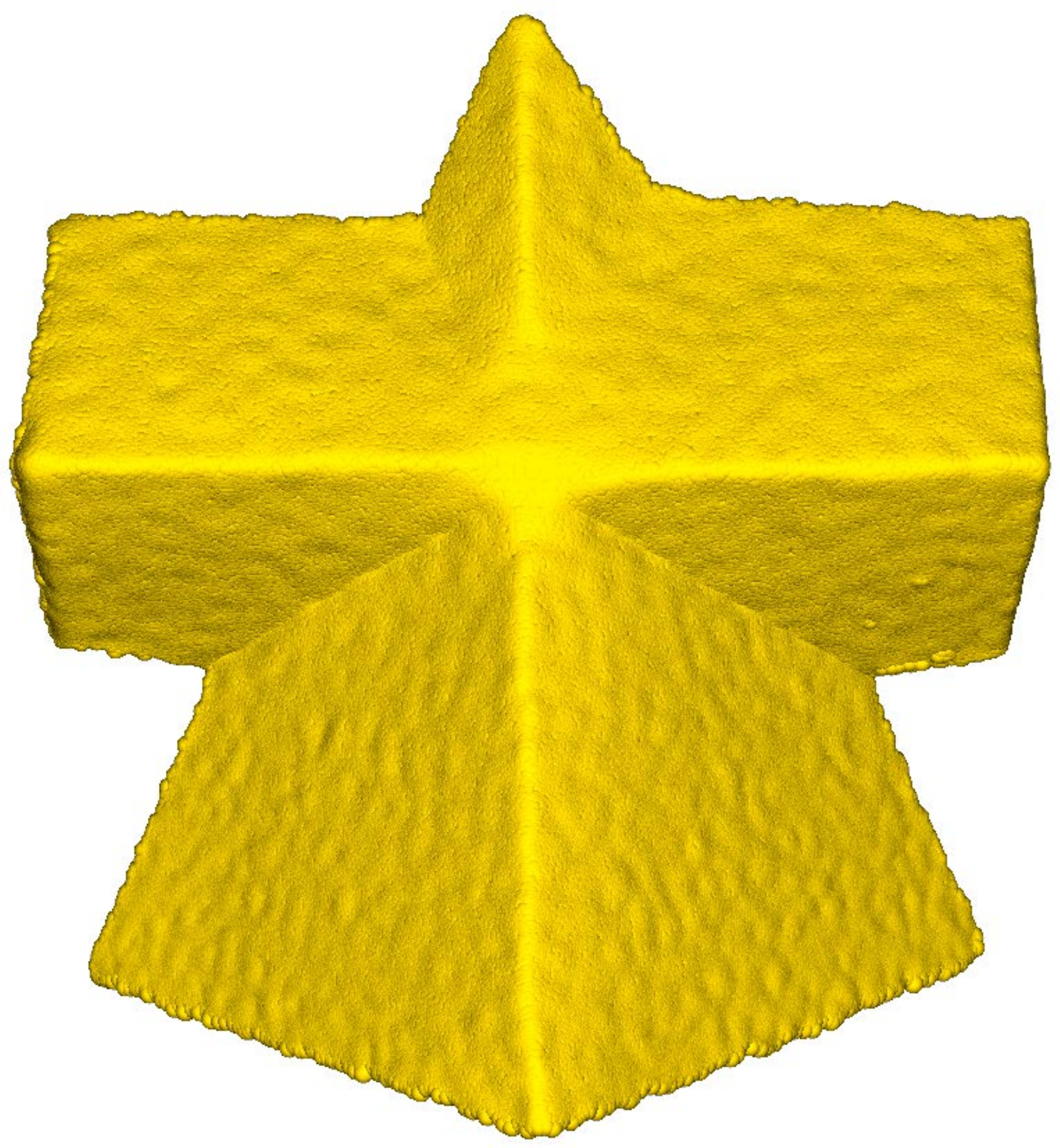}
        \end{minipage}
    }
    \subfigure[GPF]
    {
        \begin{minipage}[b]{0.095\textwidth} 
        \includegraphics[width=1\textwidth ]{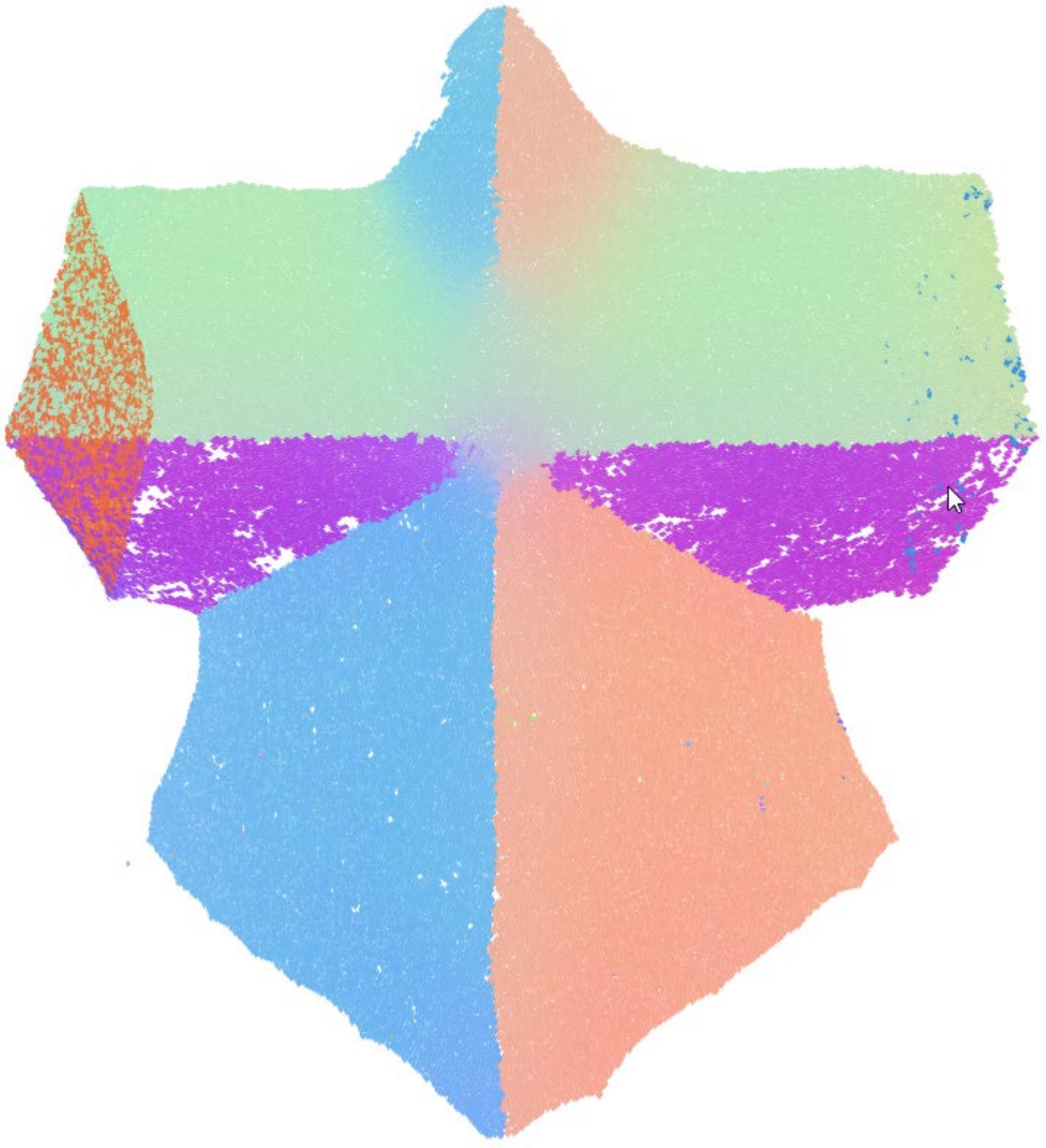}\\
        \includegraphics[width=1\textwidth ]{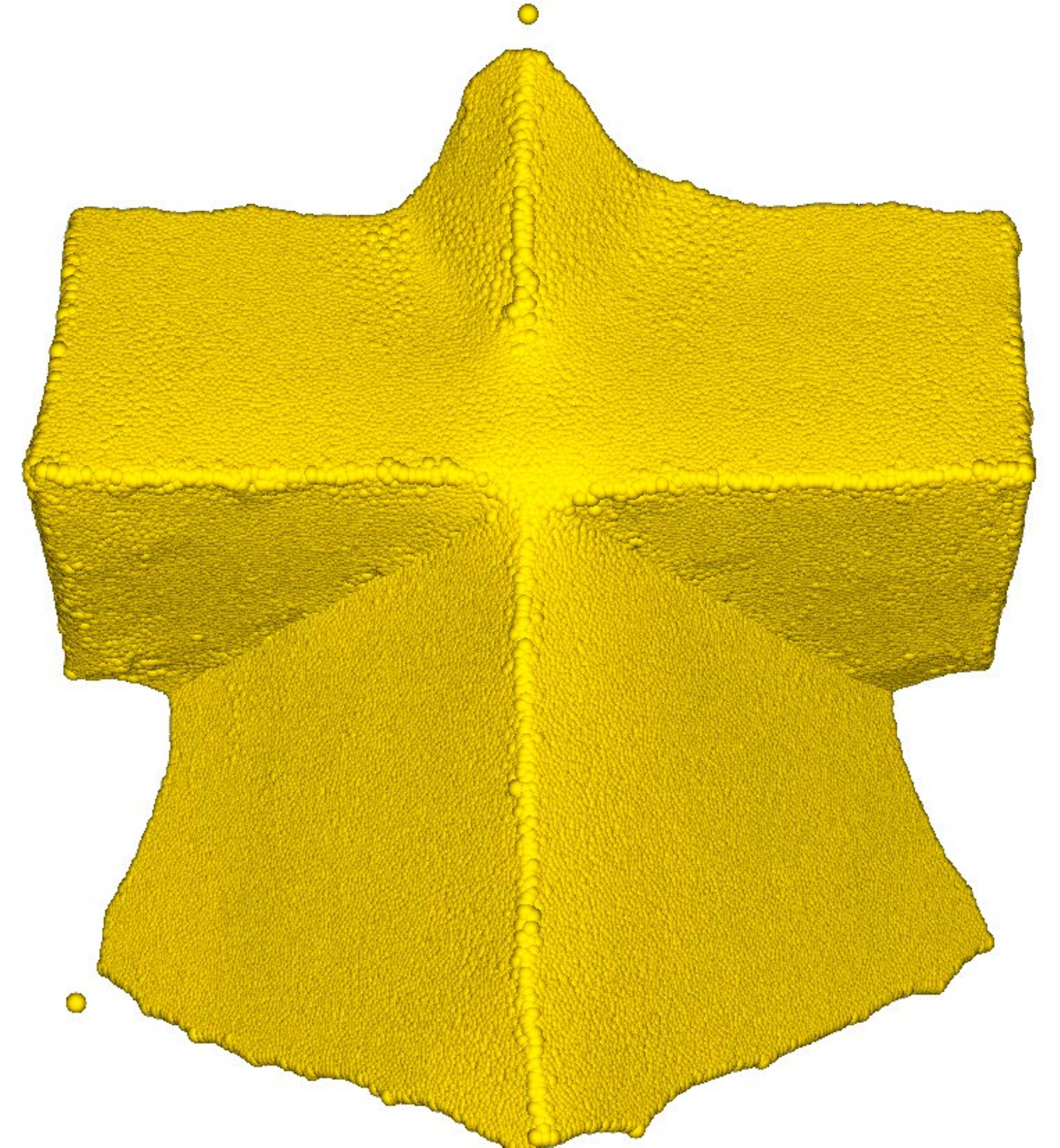}
        \end{minipage}
    }
    \subfigure[WLOP]
    {
        \begin{minipage}[b]{0.095\textwidth} 
        \includegraphics[width=1\textwidth ]{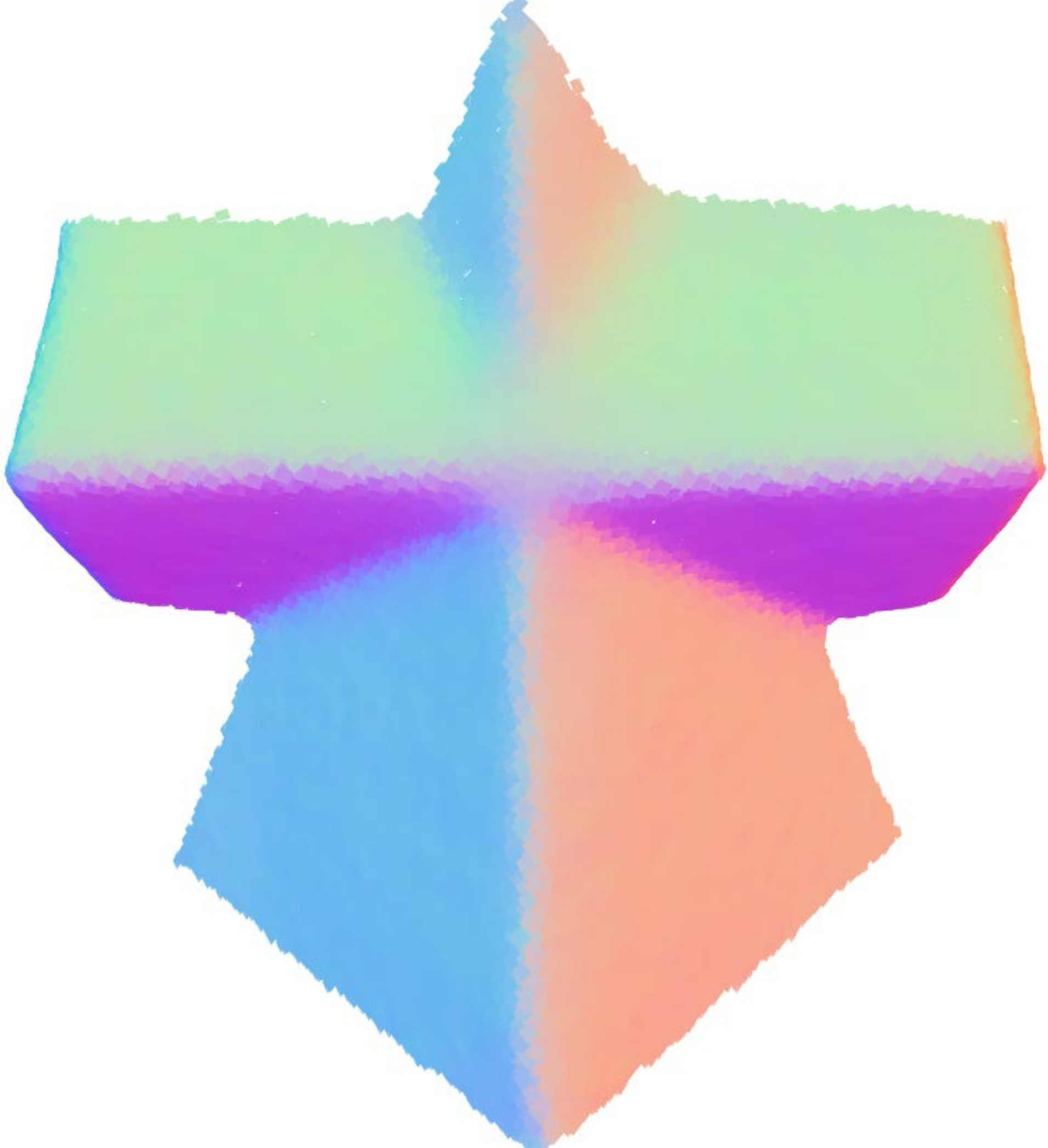}\\
        \includegraphics[width=1\textwidth ]{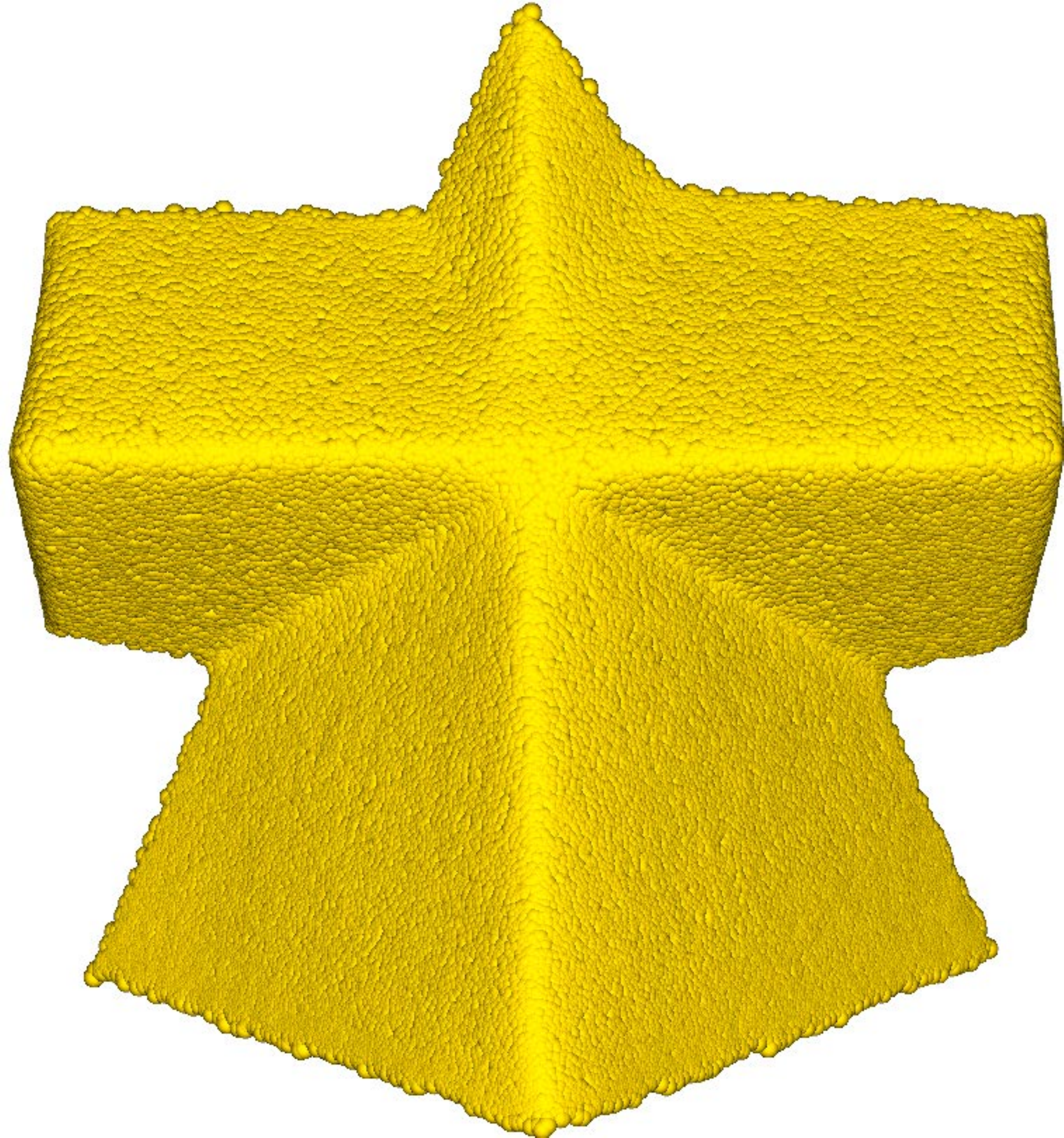}
        \end{minipage}
    }
    \subfigure[CLOP]
    {
        \begin{minipage}[b]{0.095\textwidth} 
        \includegraphics[width=1\textwidth ]{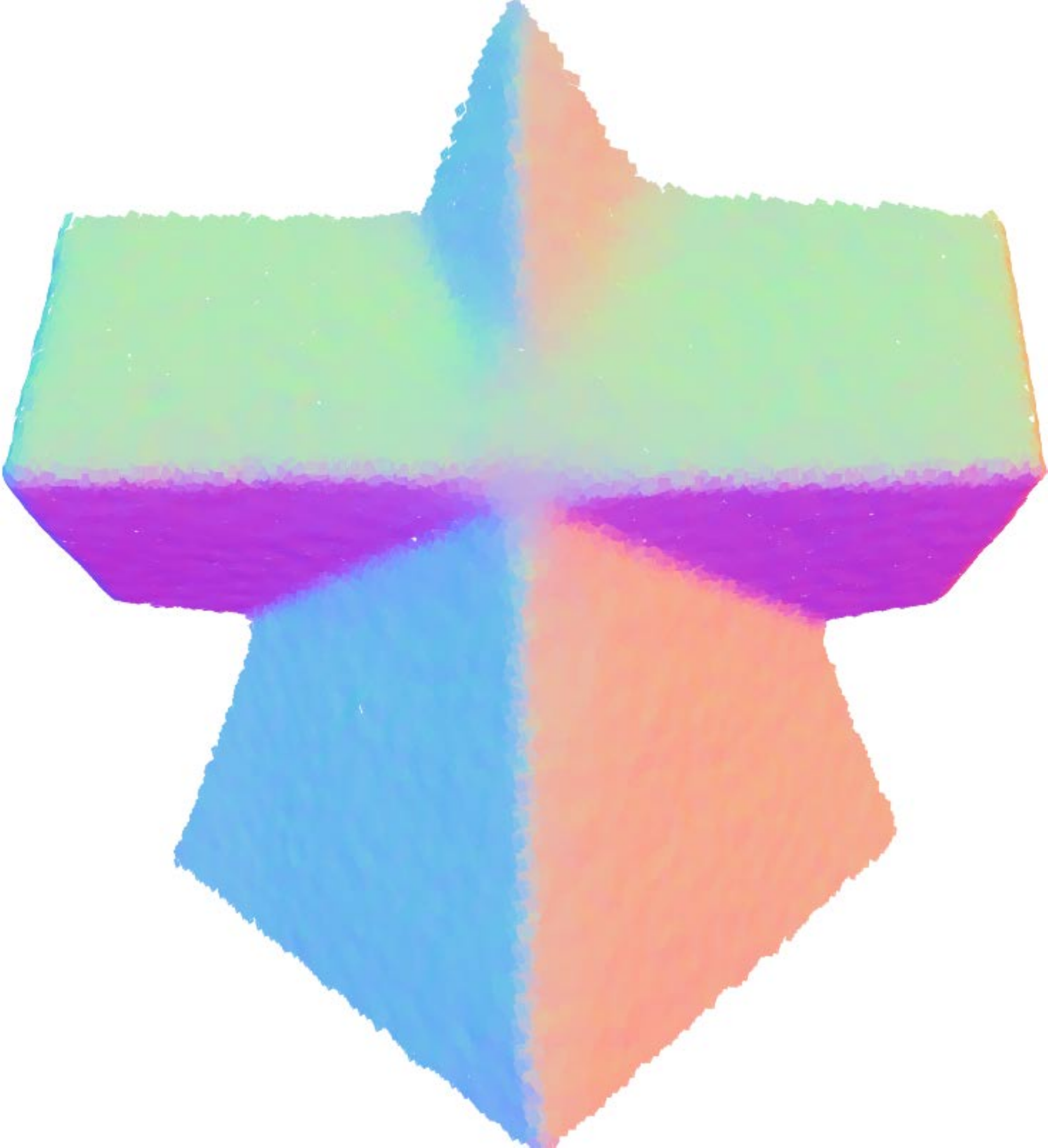}\\
        \includegraphics[width=1\textwidth ]{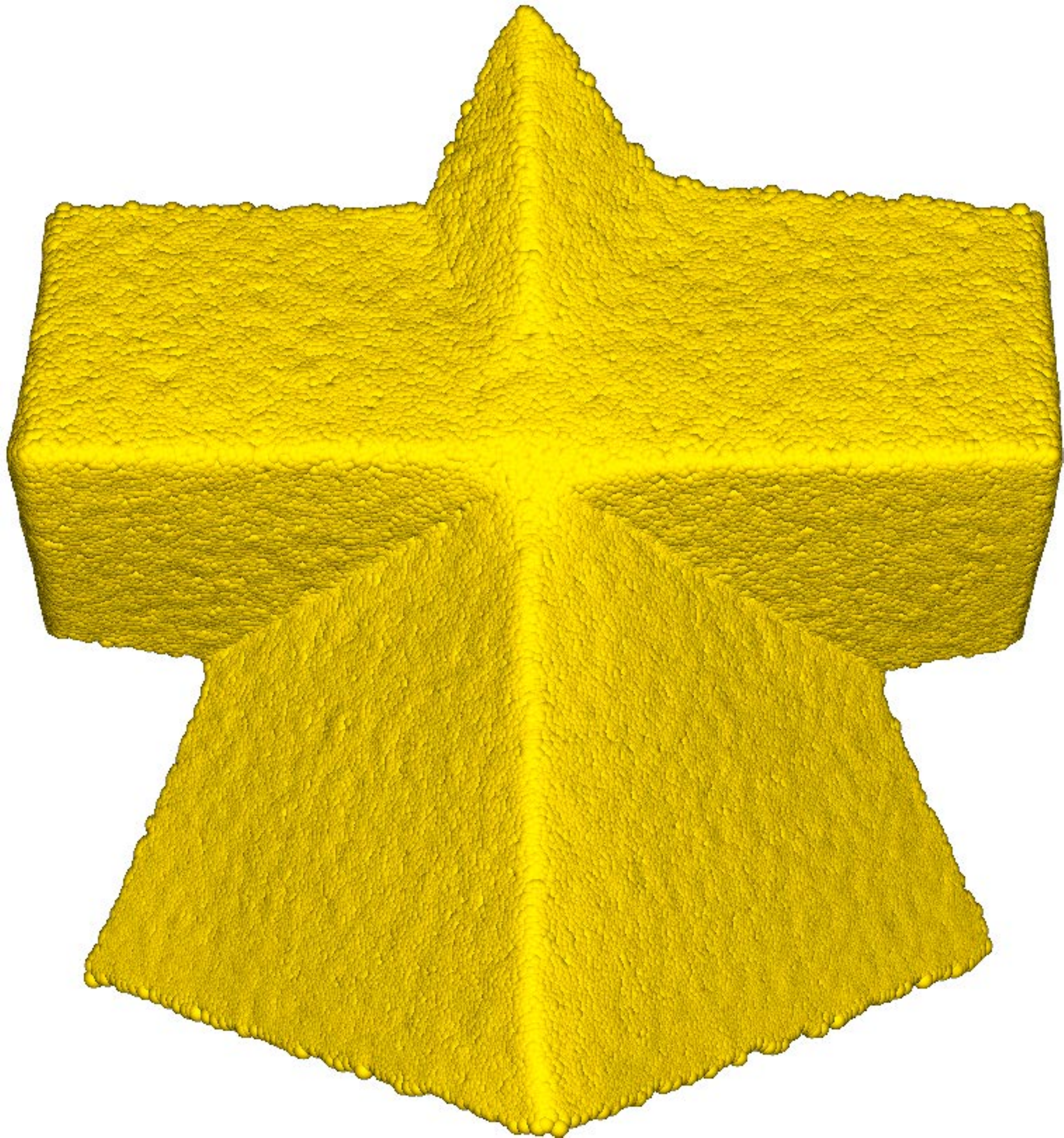}
        \end{minipage}
    }
    \subfigure[EC-Net]
    {
        \begin{minipage}[b]{0.095\textwidth} 
        \includegraphics[width=1\textwidth ]{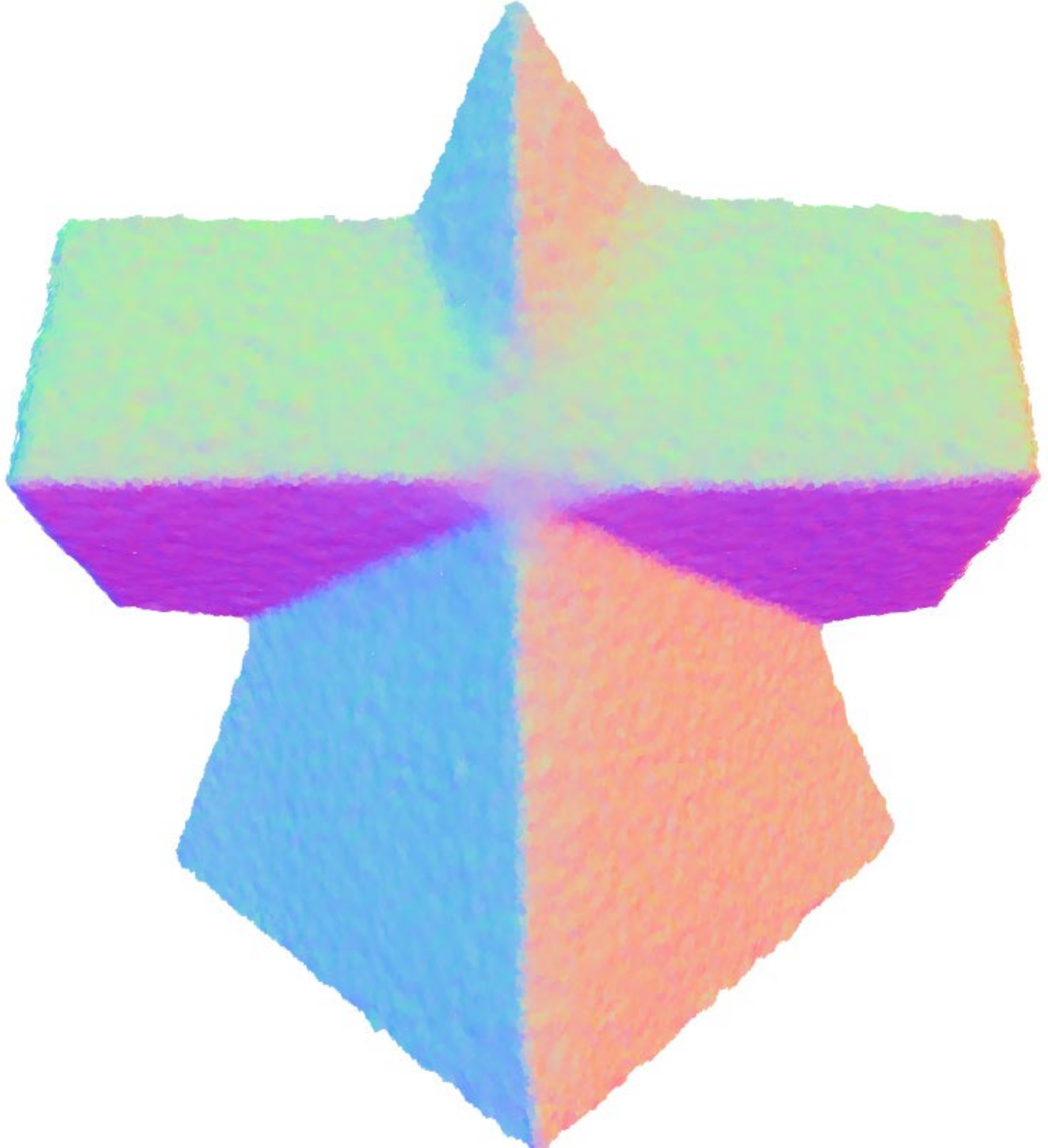}\\
        \includegraphics[width=1\textwidth ]{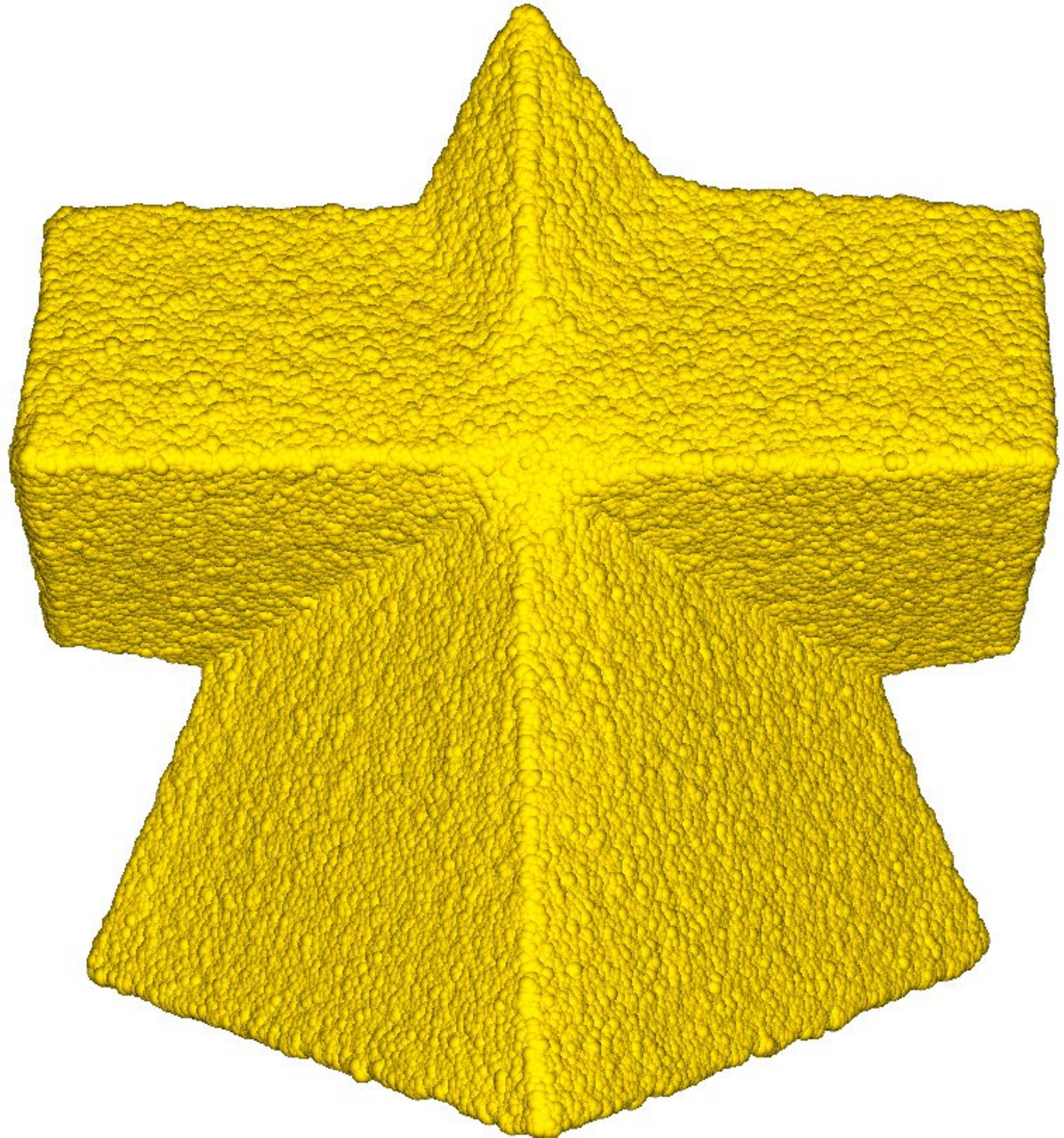}
        \end{minipage}
    }
    \subfigure[PCN]
    {
        \begin{minipage}[b]{0.095\textwidth} 
        \includegraphics[width=1\textwidth ]{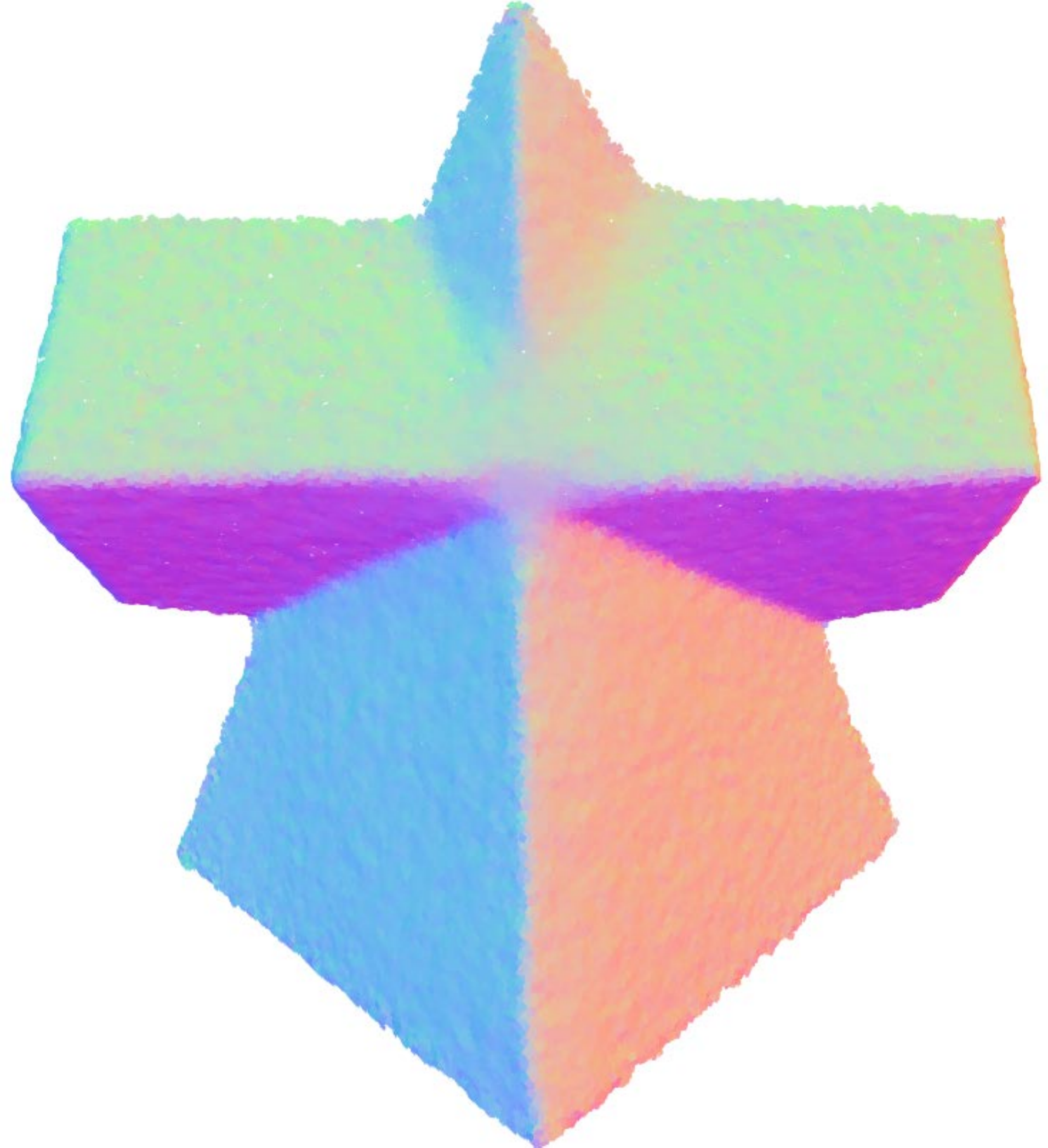}\\
        \includegraphics[width=1\textwidth ]{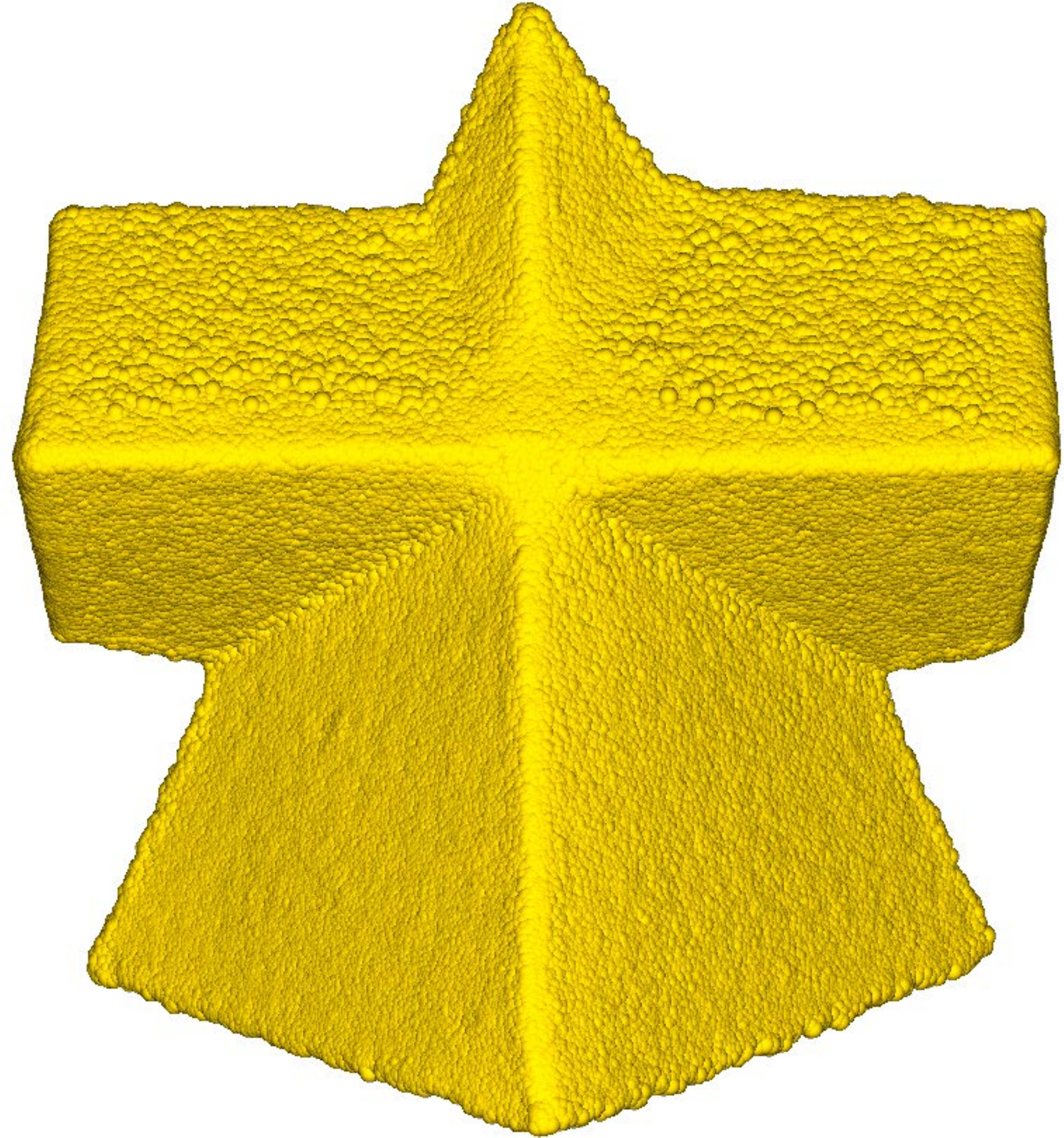}
        \end{minipage}
    }
    \subfigure[TD]
    {
        \begin{minipage}[b]{0.095\textwidth} 
        \includegraphics[width=1\textwidth ]{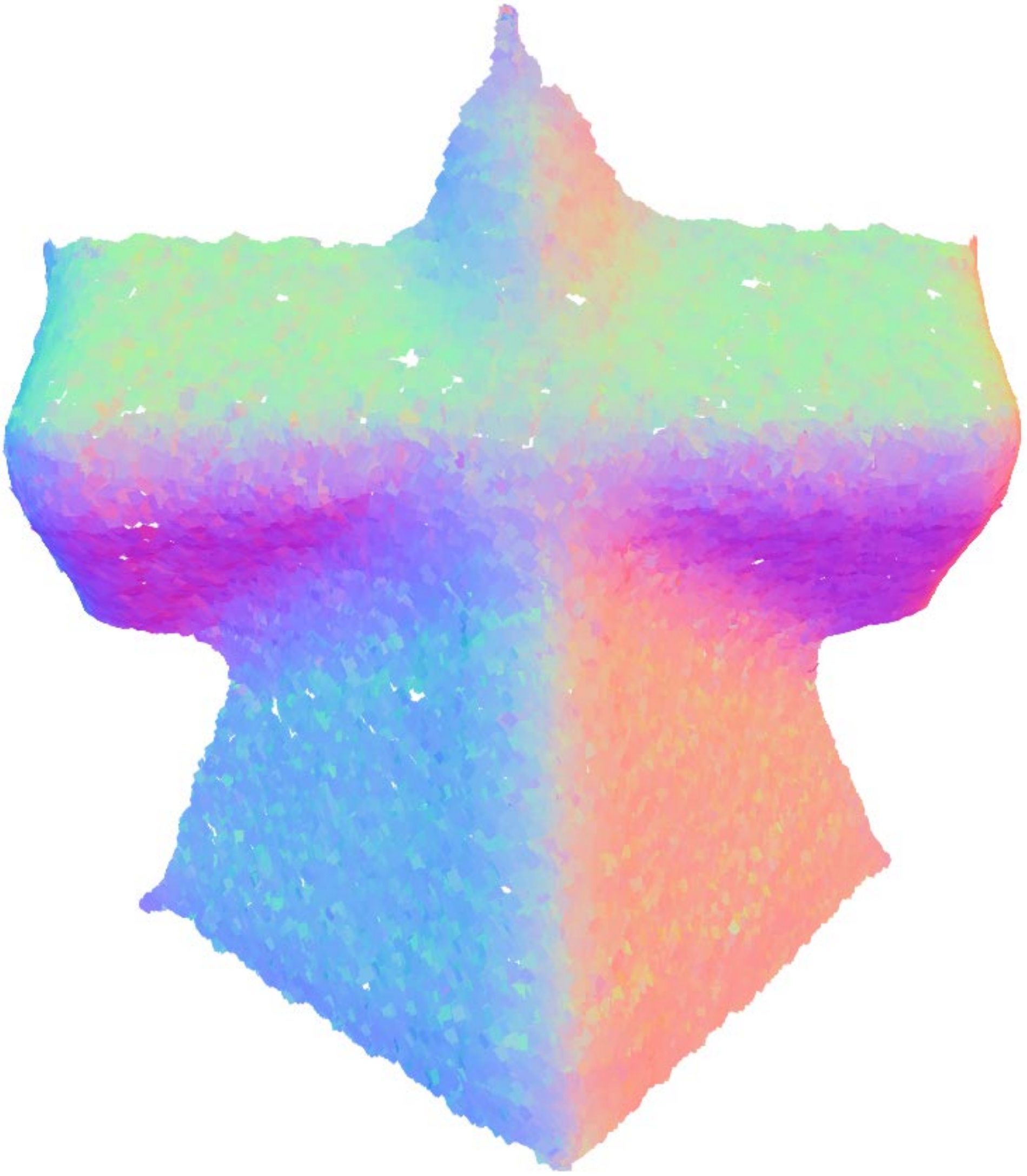}\\
        \includegraphics[width=1\textwidth ]{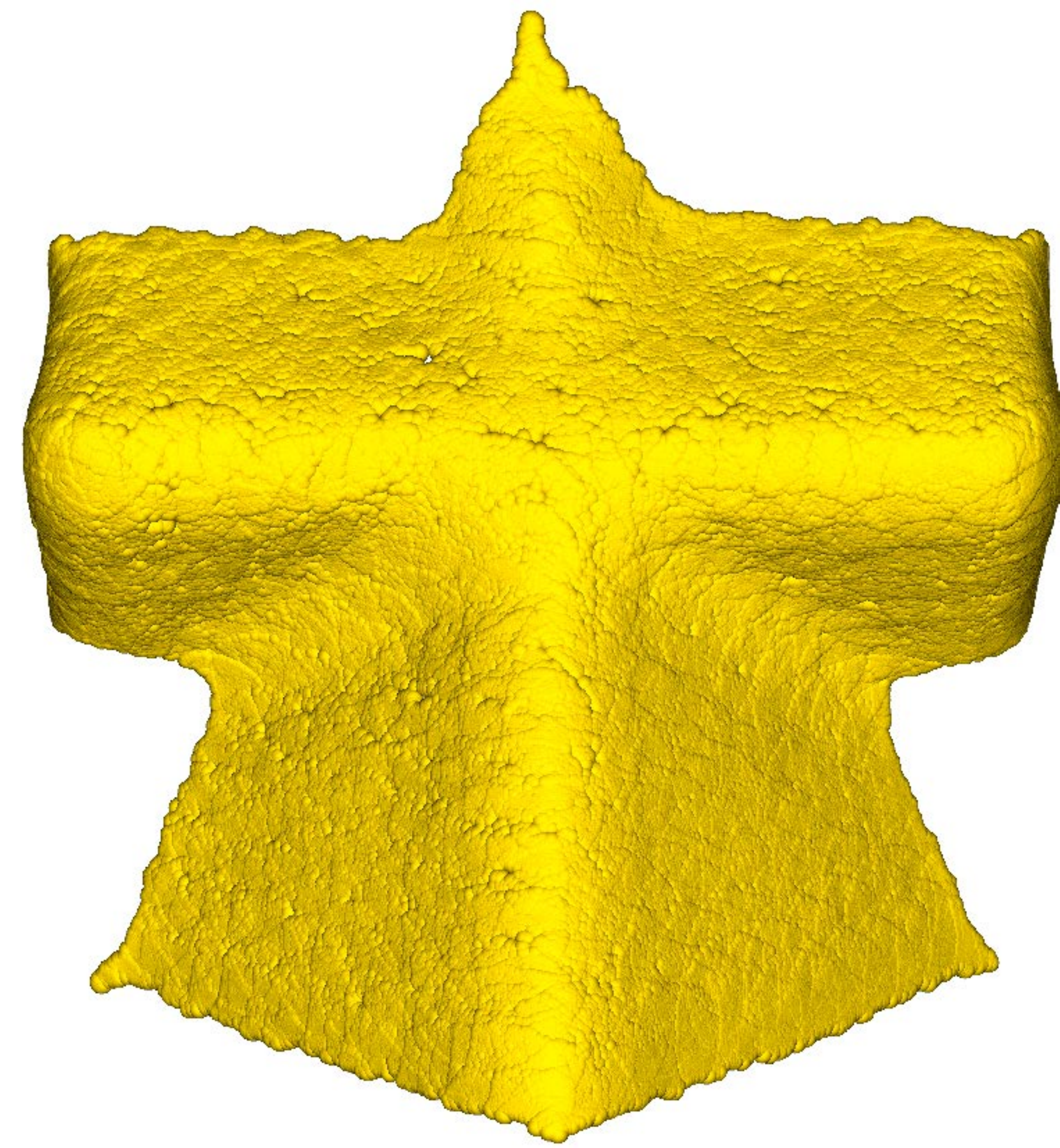}
        \end{minipage}
    }
    \subfigure[Ours]
    {
        \begin{minipage}[b]{0.095\textwidth} 
        \includegraphics[width=1\textwidth ]{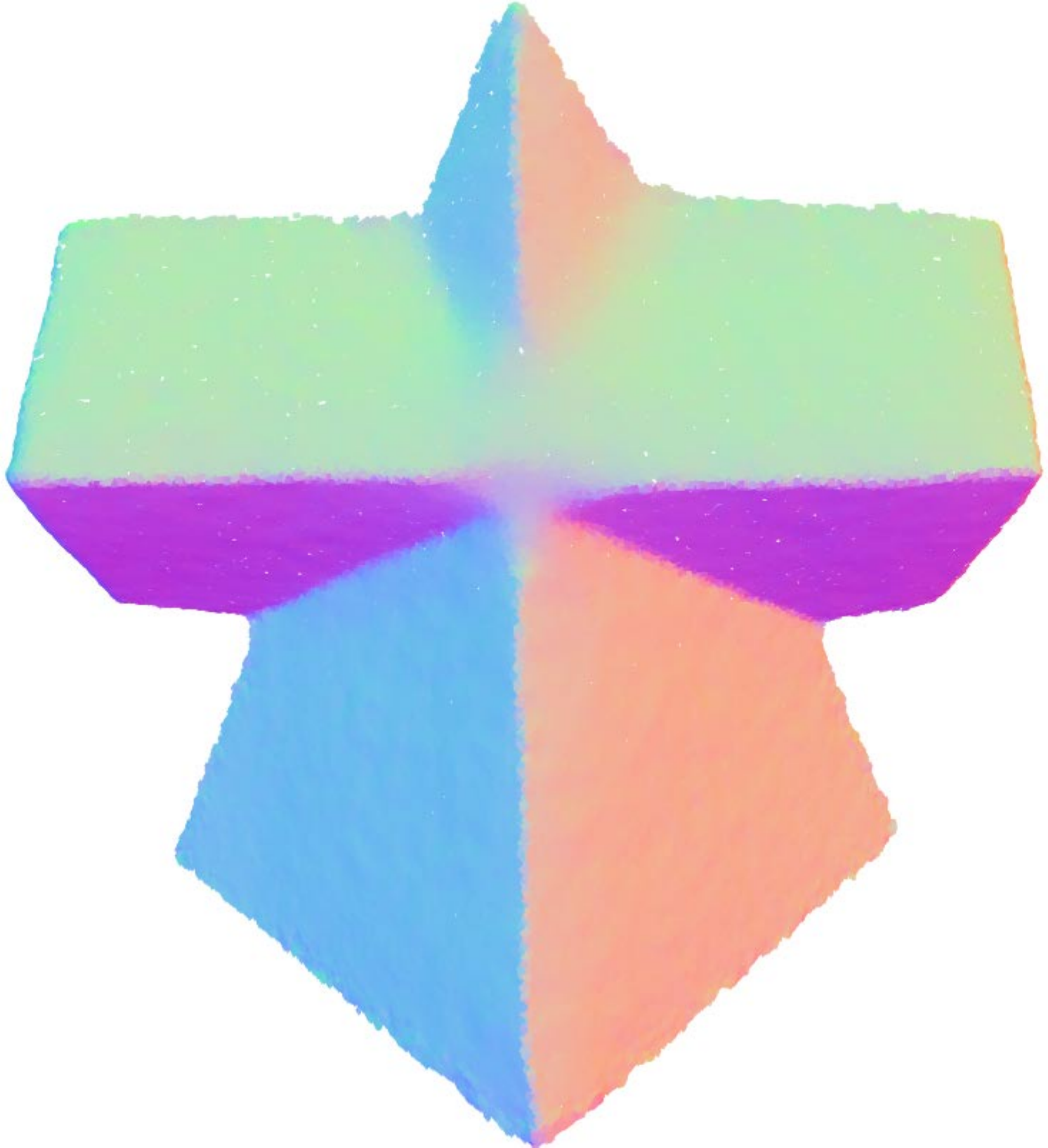}\\
        \includegraphics[width=1\textwidth ]{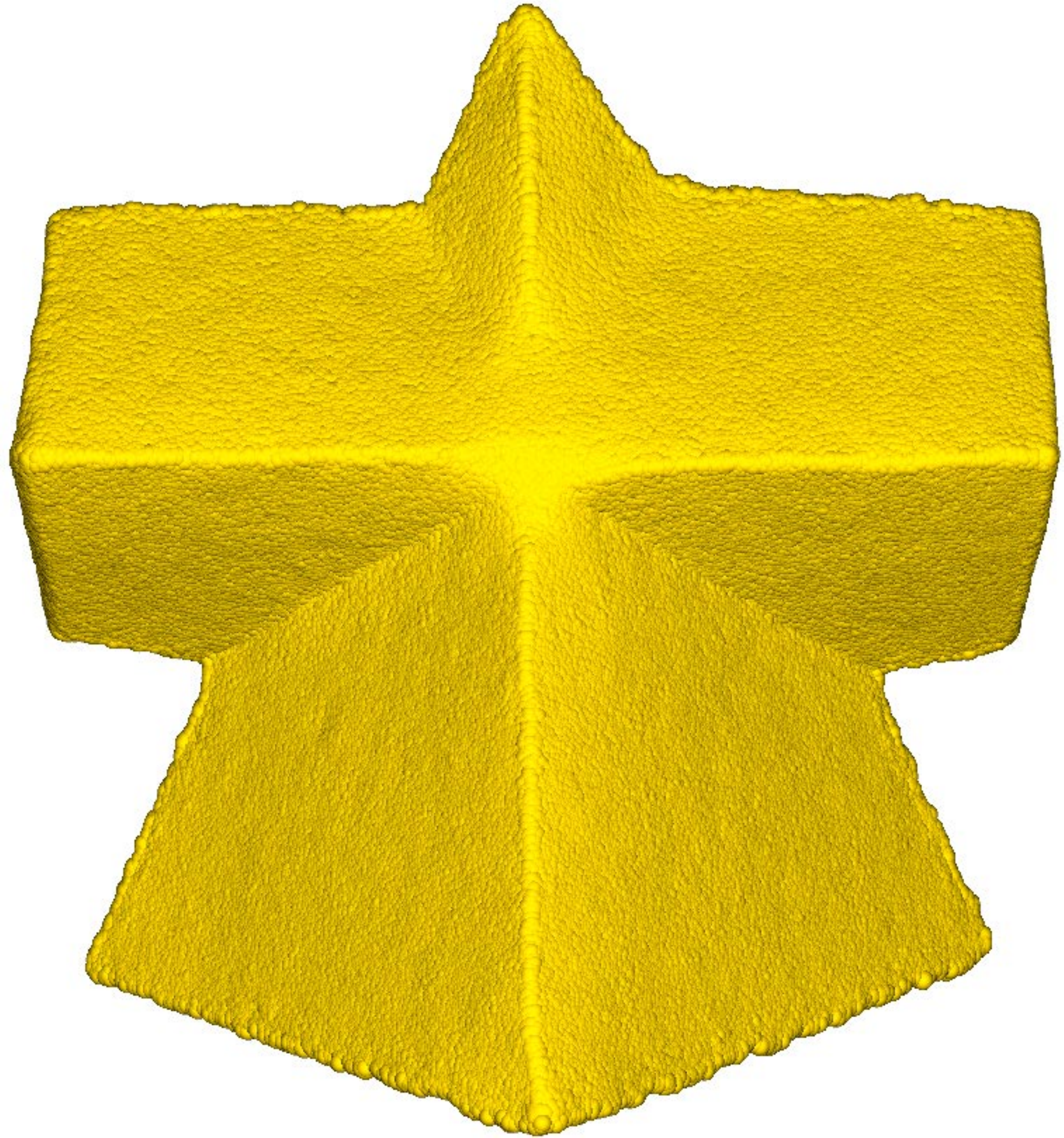}
        \end{minipage}
    }
    \caption{Filtered results of the raw Pyramid point cloud. }
    \label{fig:pyramid_sharp_feature}
\end{figure*}
\begin{figure*}[htb!]
    \subfigure[Noisy]
    {
        \begin{minipage}[b]{0.095\textwidth} 
        \includegraphics[width=1\textwidth]{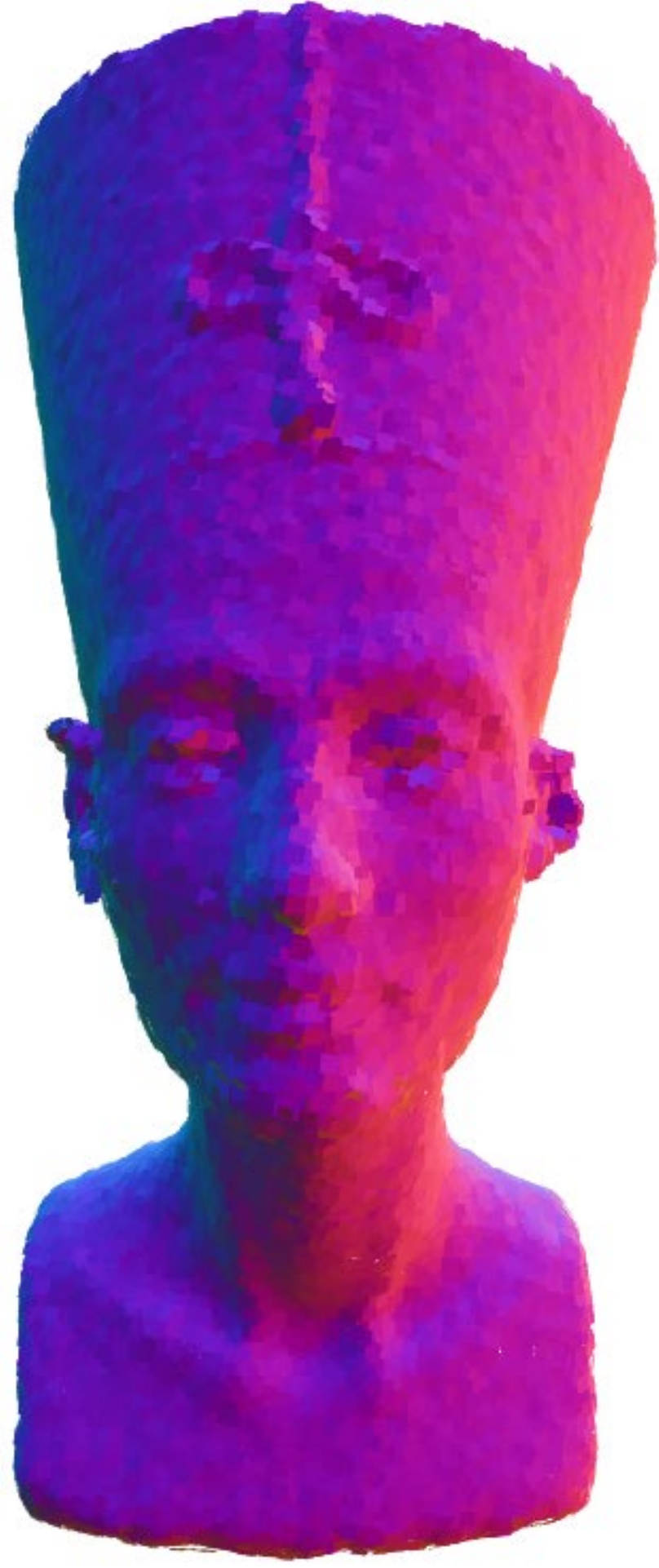}\\
        \includegraphics[width=1\textwidth]{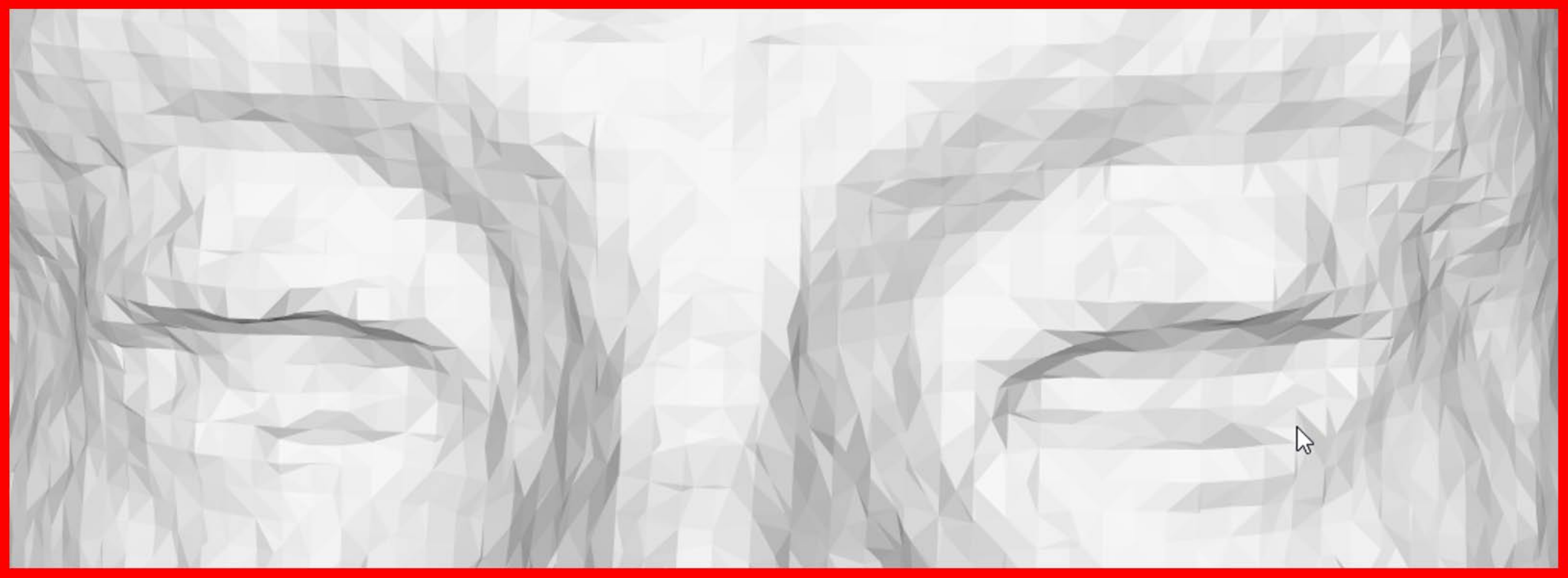}\\
        \includegraphics[width=1\textwidth]{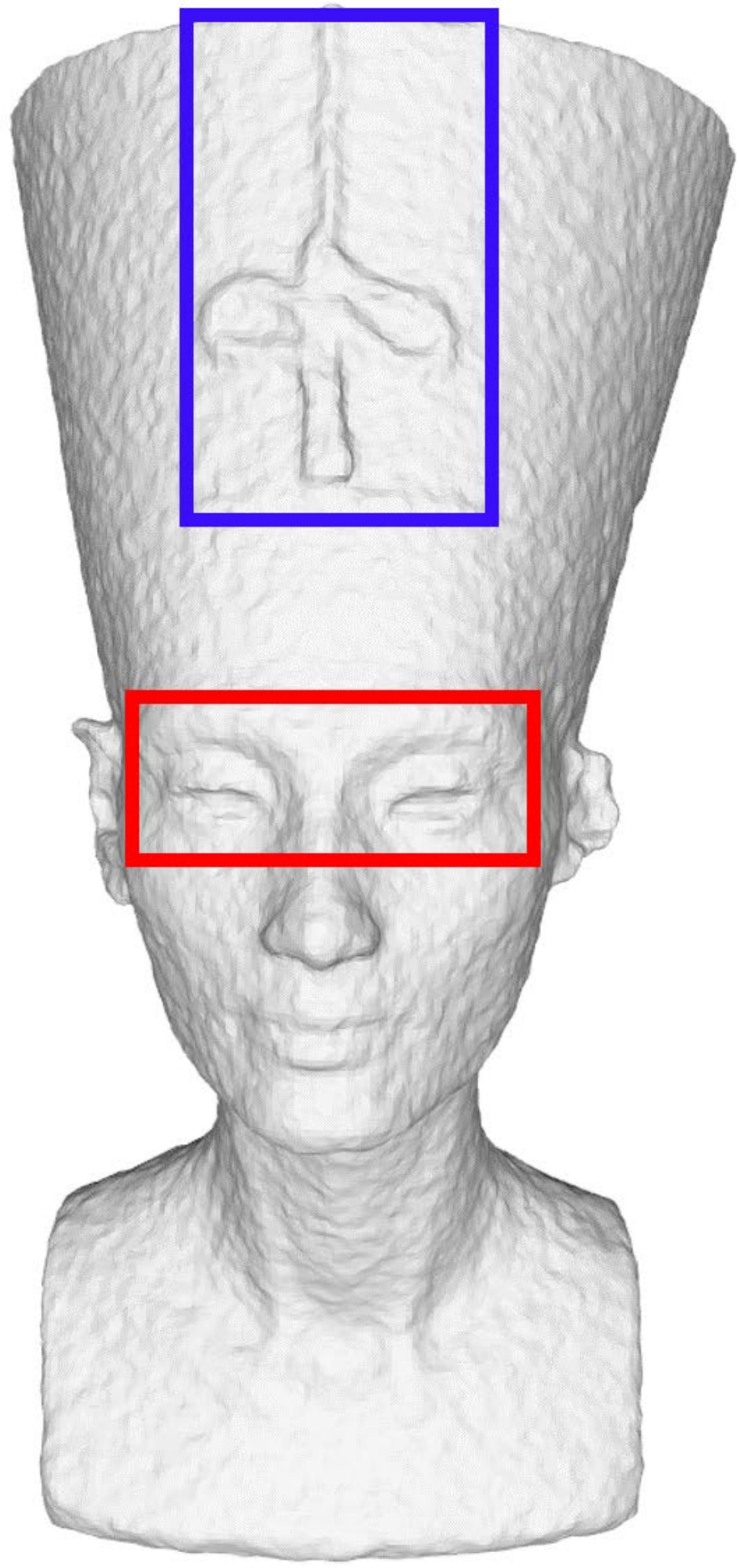}
        \end{minipage}
    }
    \subfigure[RIMLS]
    {
      \begin{minipage}[b]{0.095\textwidth}
        \includegraphics[width=1\textwidth]{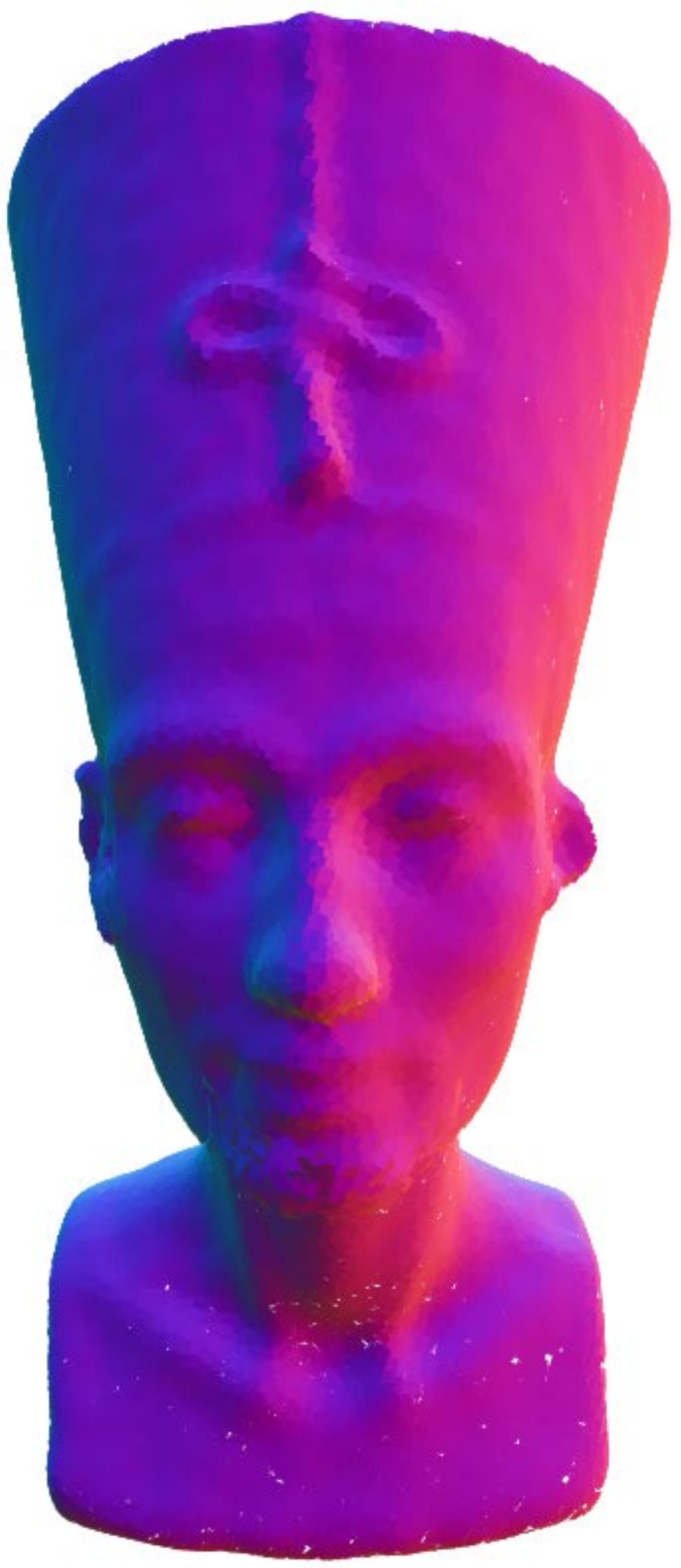}\\
        \includegraphics[width=1\textwidth]{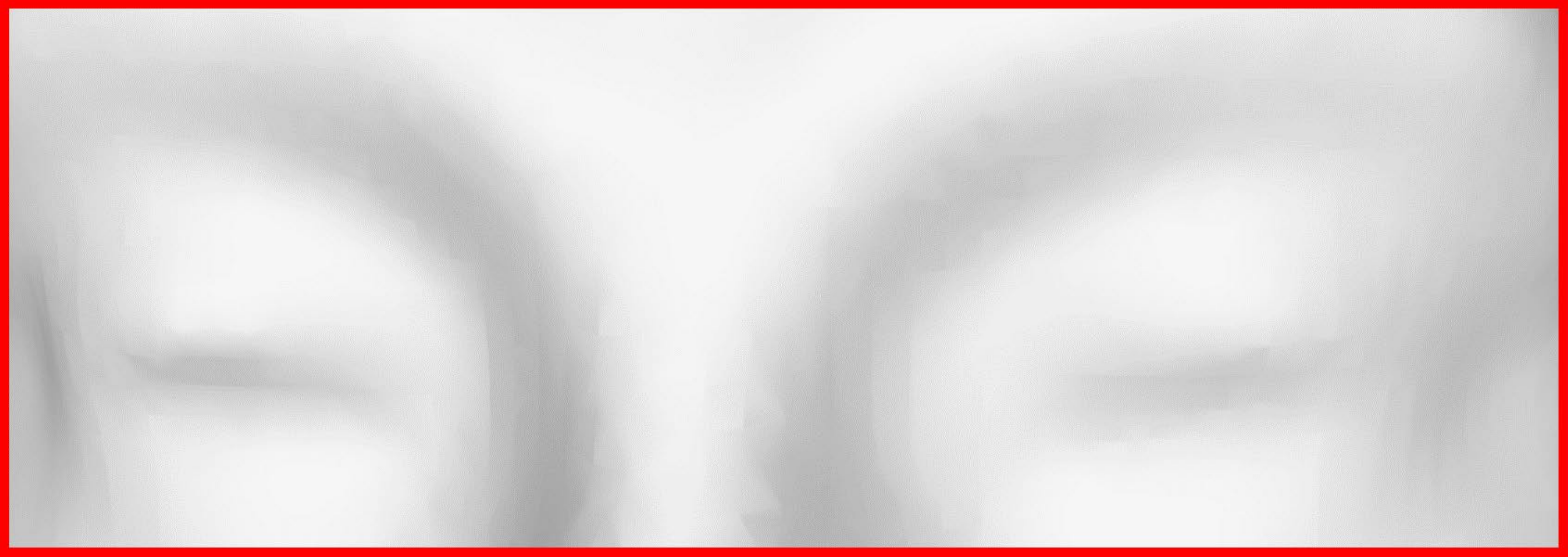}\\
        \includegraphics[width=1\textwidth]{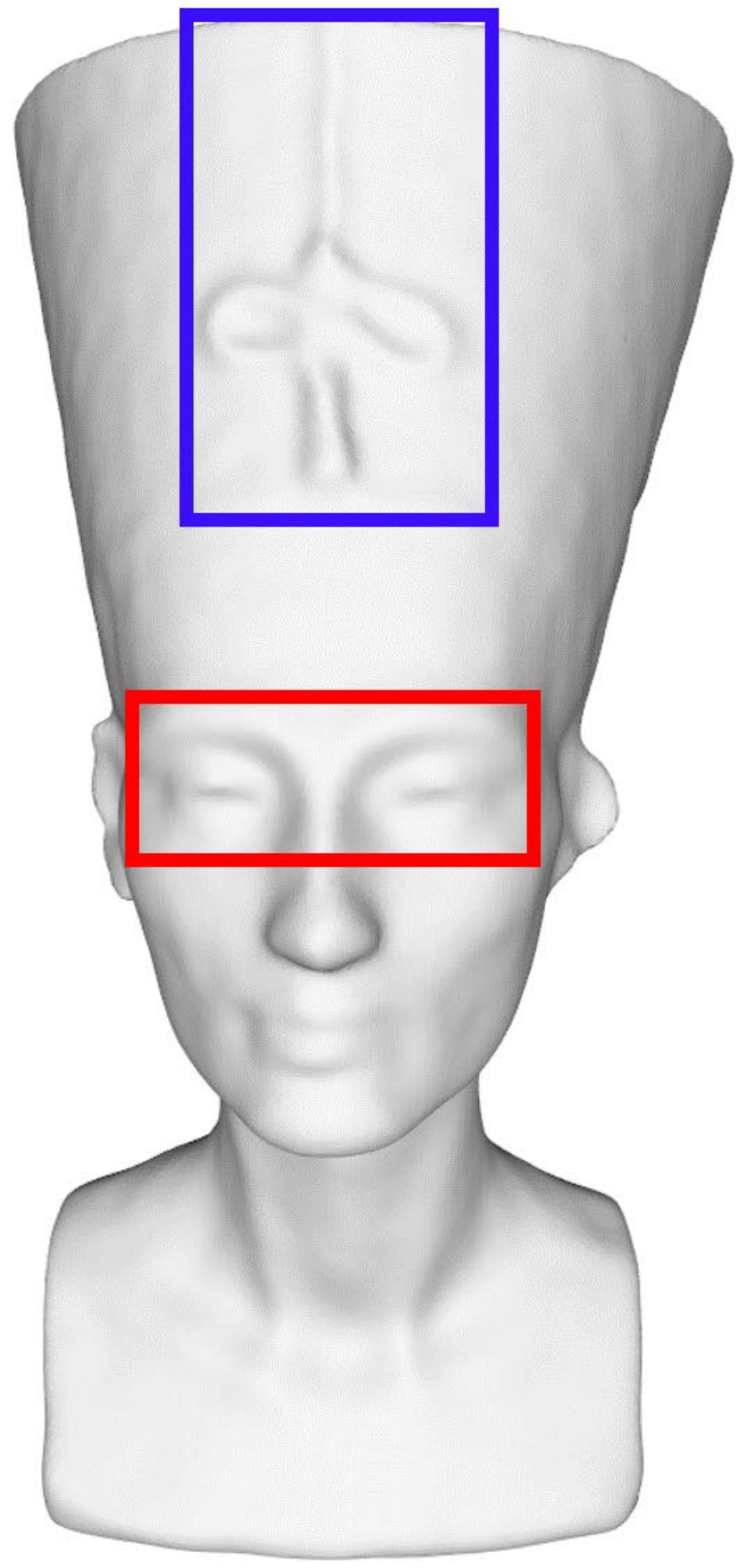}
        \end{minipage}
    }
     \subfigure[GPF]
    {
      \begin{minipage}[b]{0.095\textwidth}
        \includegraphics[width=1\textwidth]{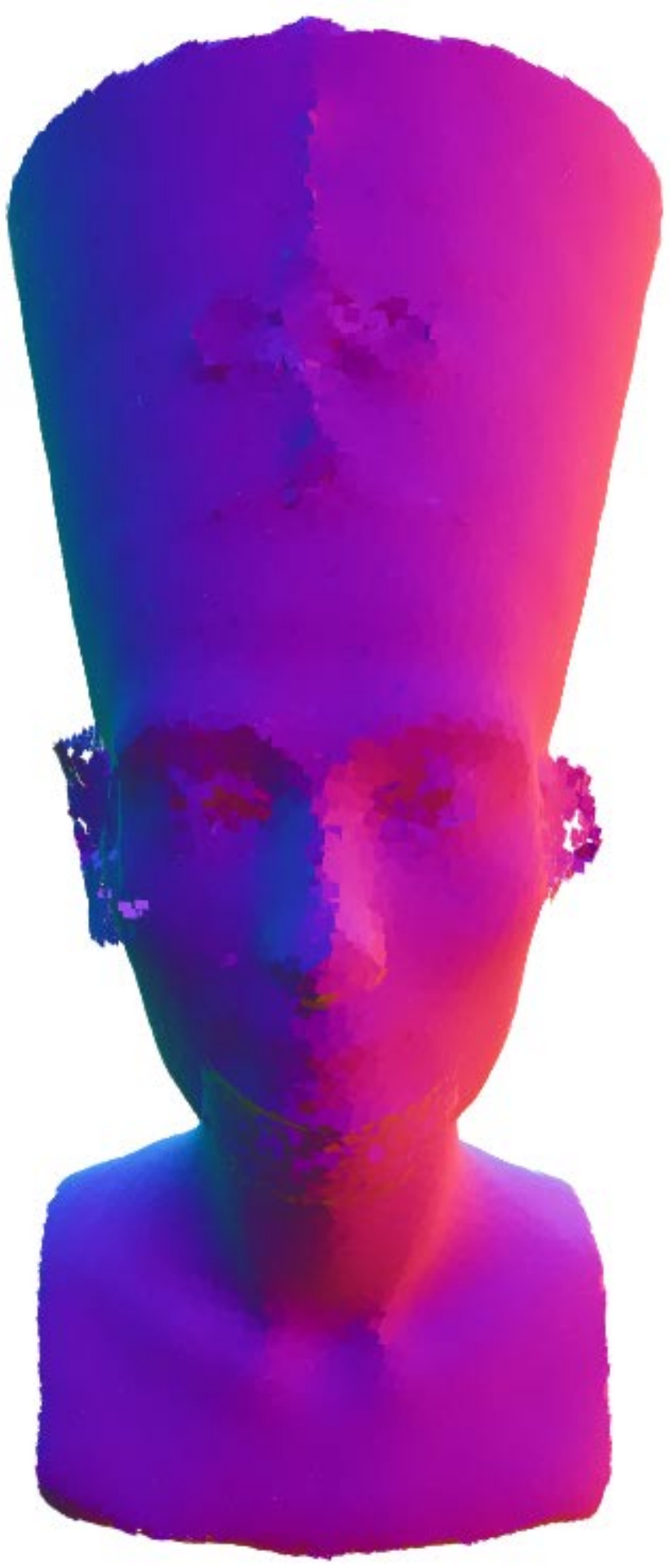}\\
        \includegraphics[width=1\textwidth]{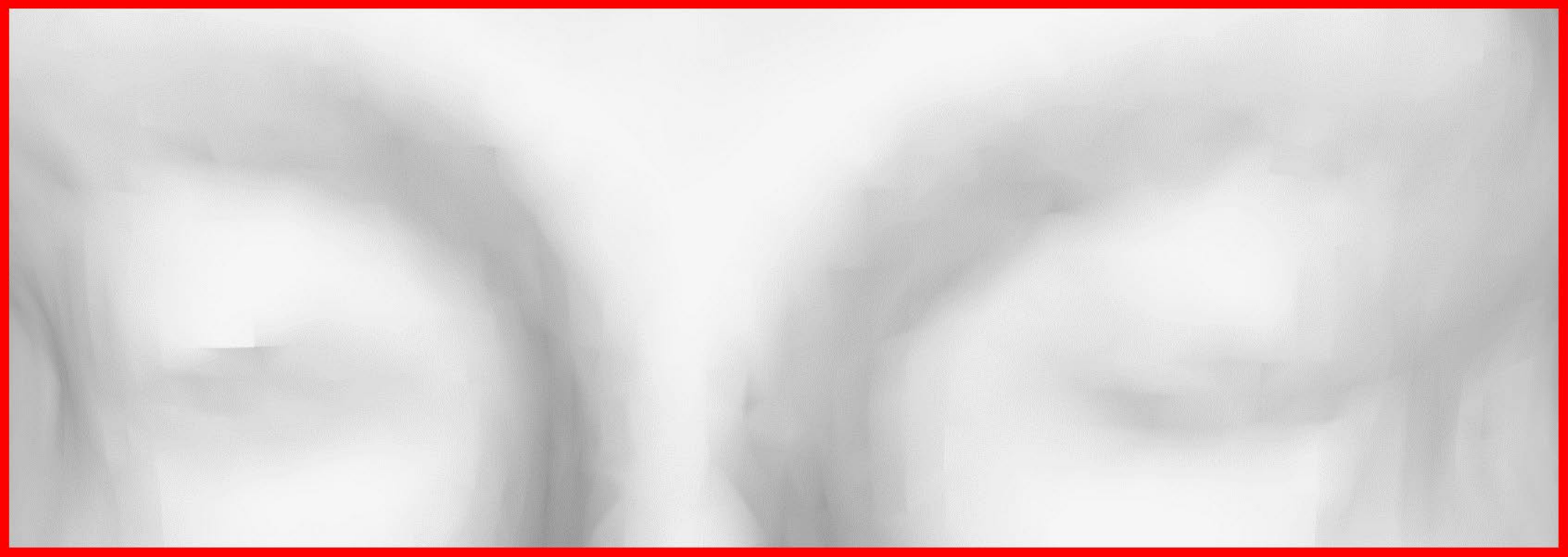}\\
        \includegraphics[width=1\textwidth]{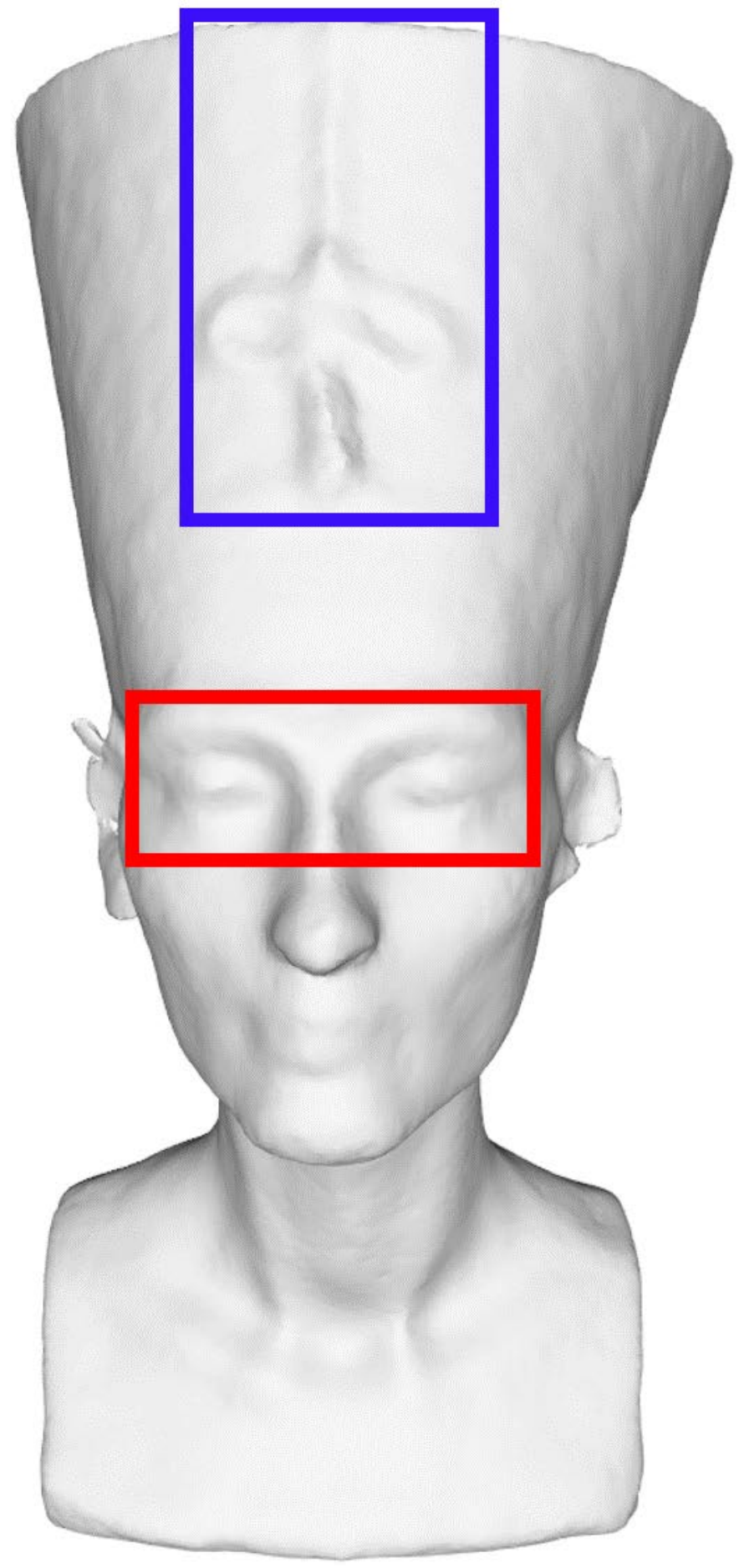}
        \end{minipage}
    }
    \subfigure[WLOP]
    {
      \begin{minipage}[b]{0.095\textwidth}
        \includegraphics[width=1\textwidth]{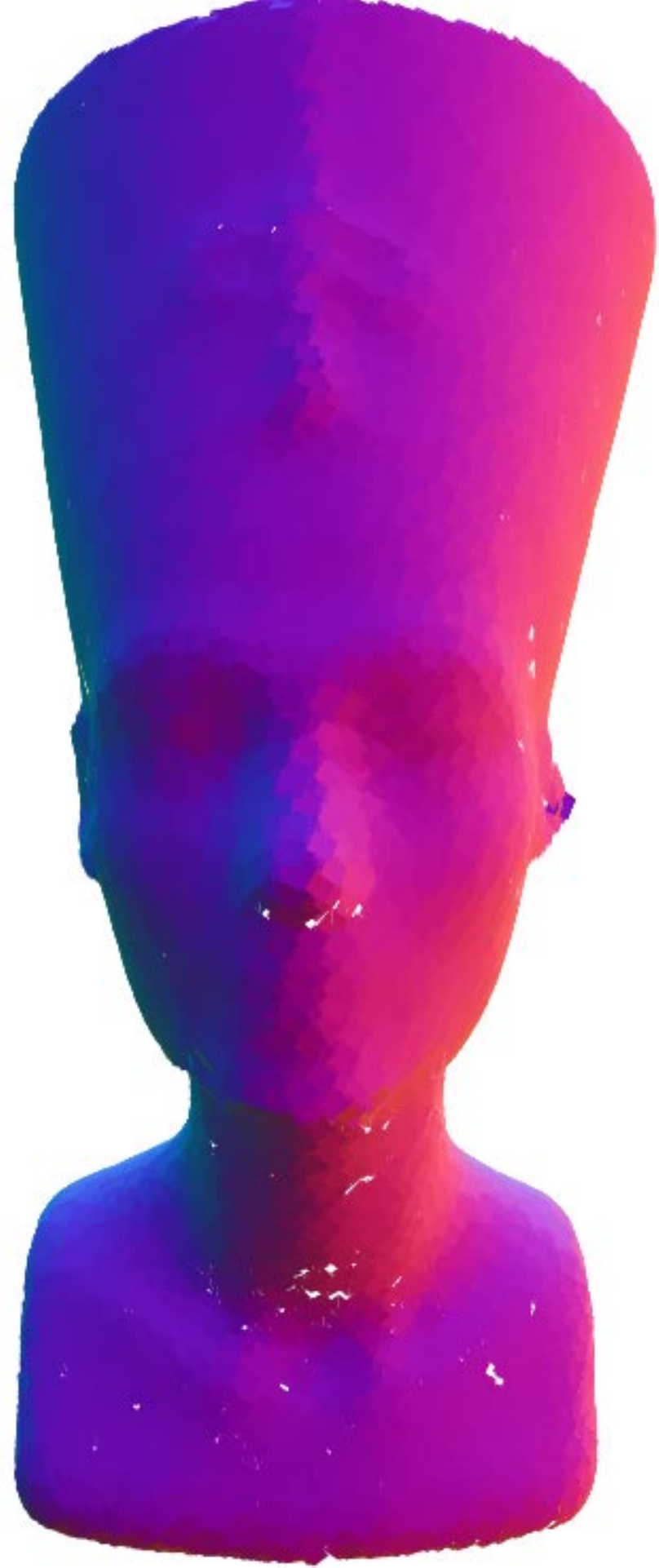}\\
        \includegraphics[width=1\textwidth]{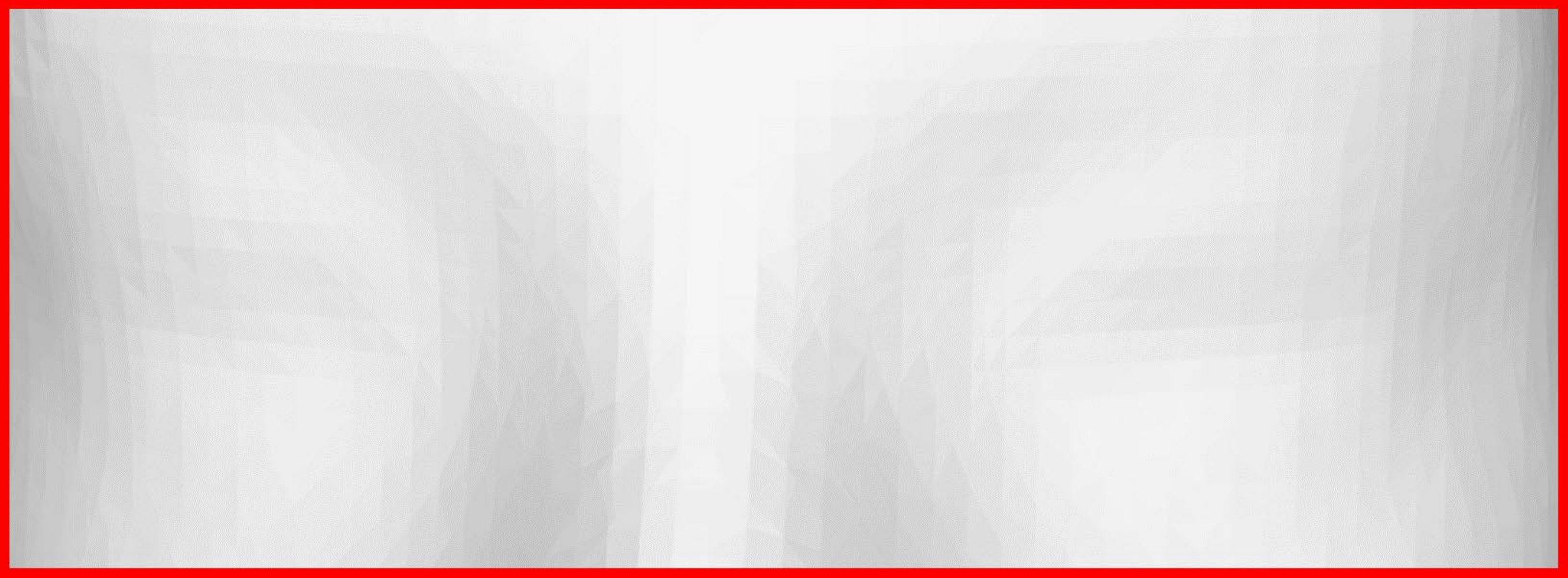}\\
        \includegraphics[width=1\textwidth]{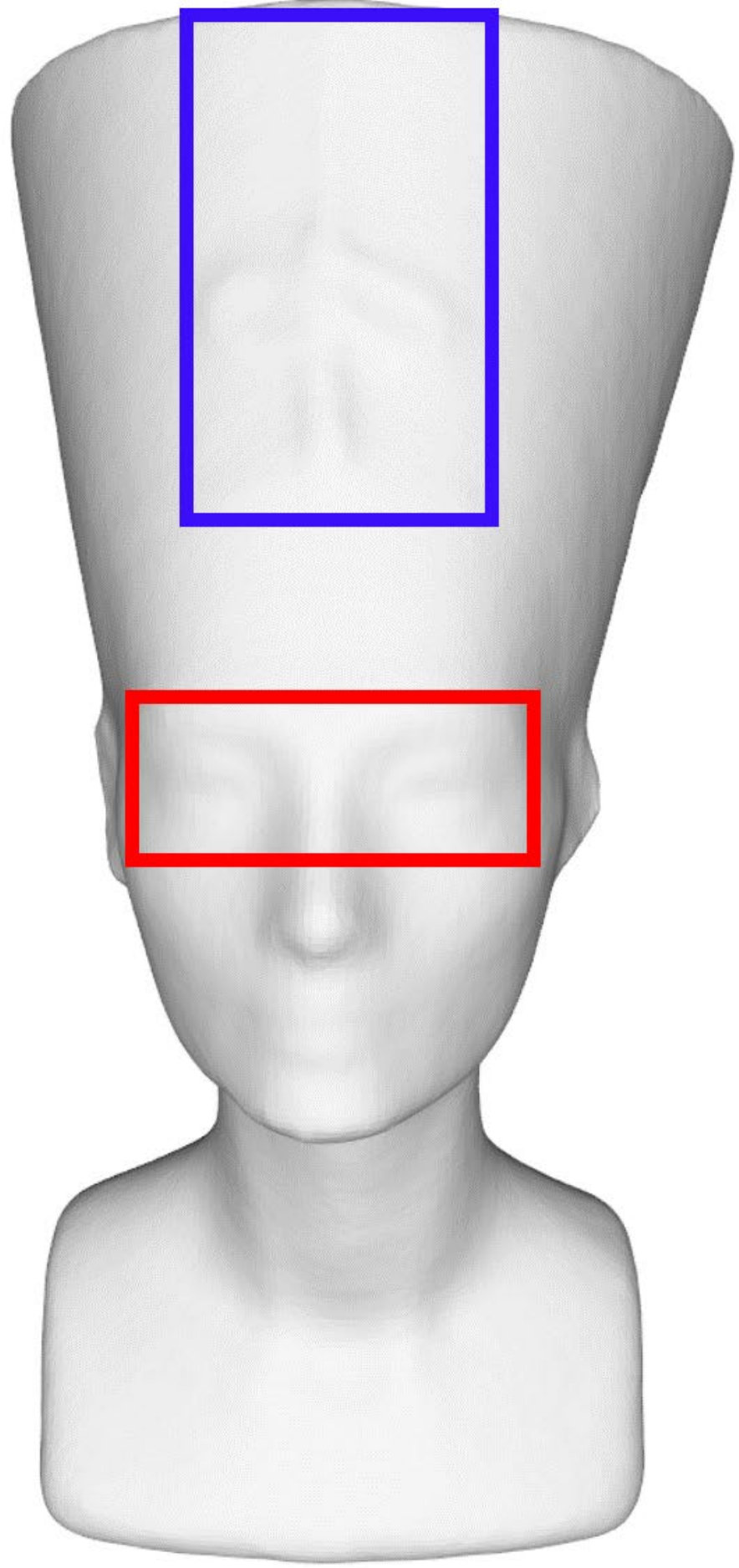}
        \end{minipage}
    }
    \subfigure[CLOP]
    {
      \begin{minipage}[b]{0.095\textwidth}
        \includegraphics[width=1\textwidth]{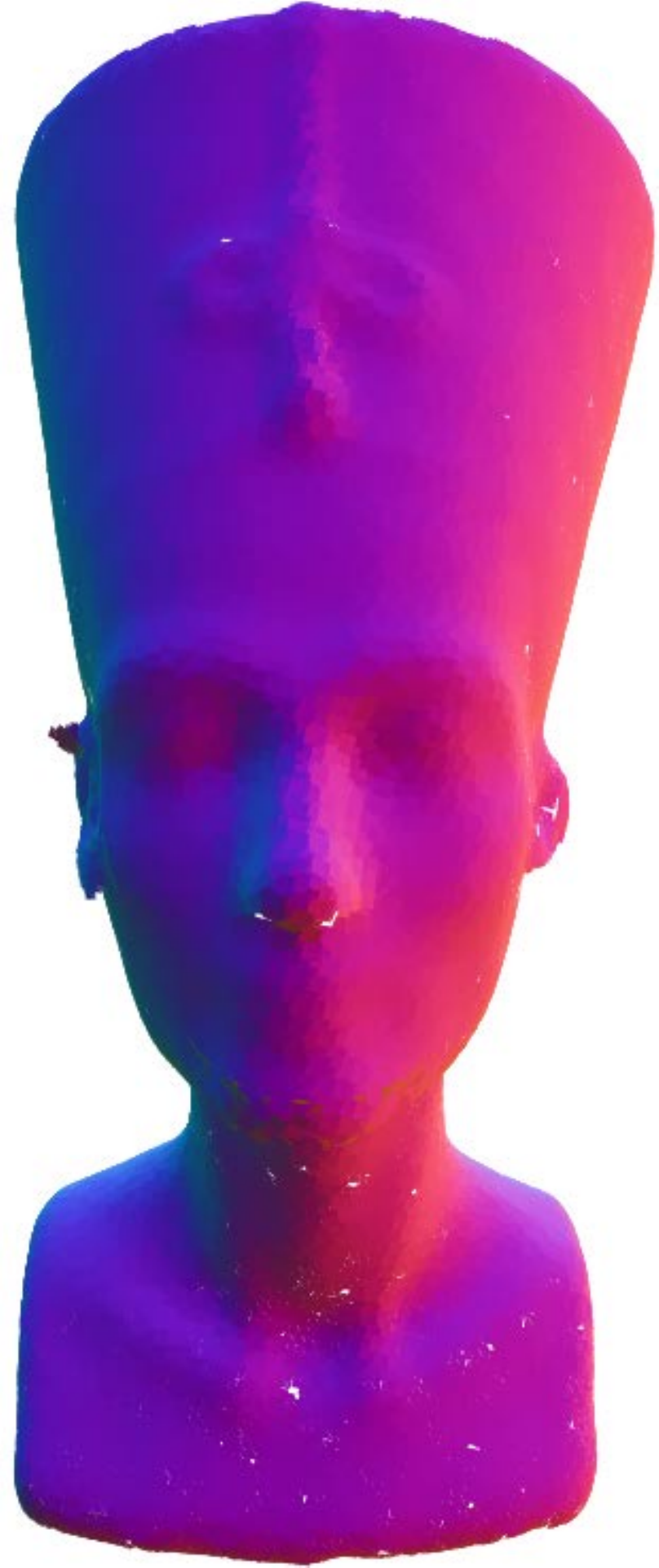}\\
        \includegraphics[width=1\textwidth]{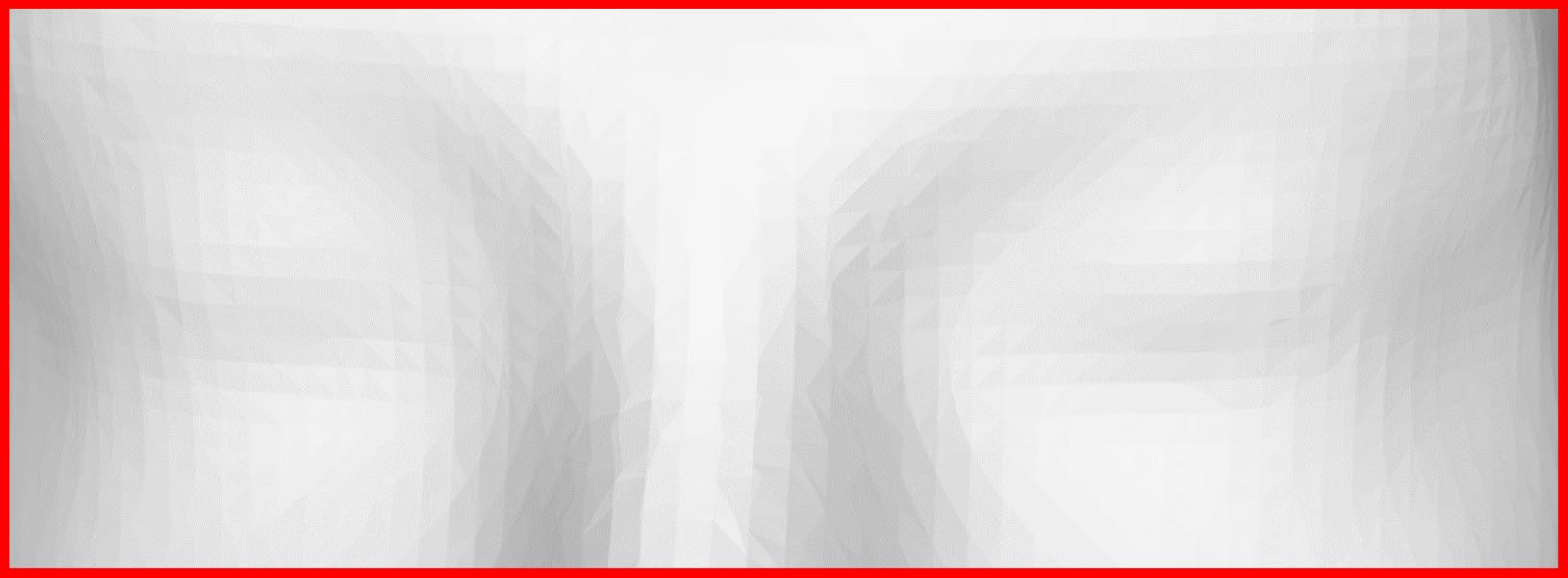}\\
        \includegraphics[width=1\textwidth]{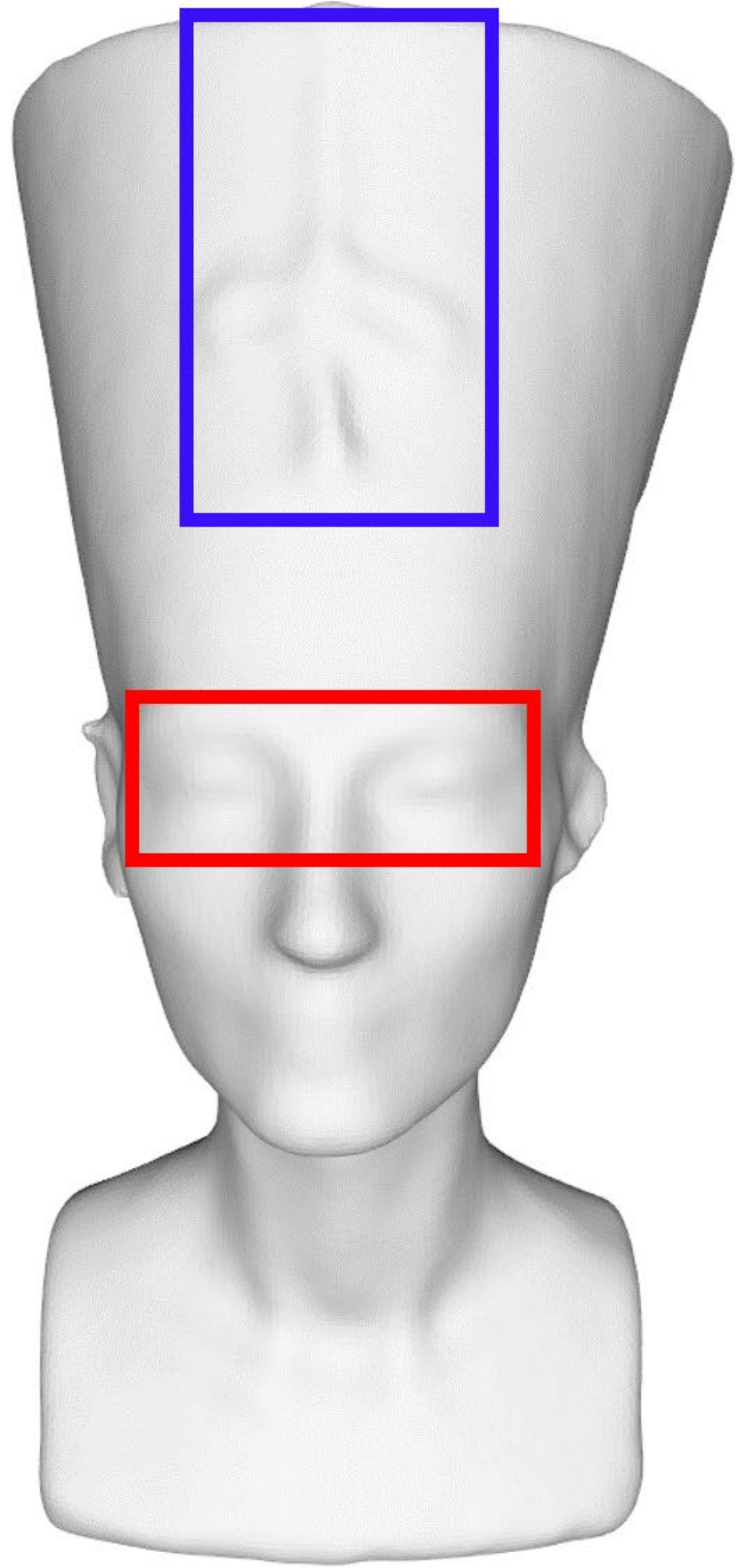}
        \end{minipage}
    }
    \subfigure[EC-Net]
    {
       \begin{minipage}[b]{0.095\textwidth}
        \includegraphics[width=1\textwidth]{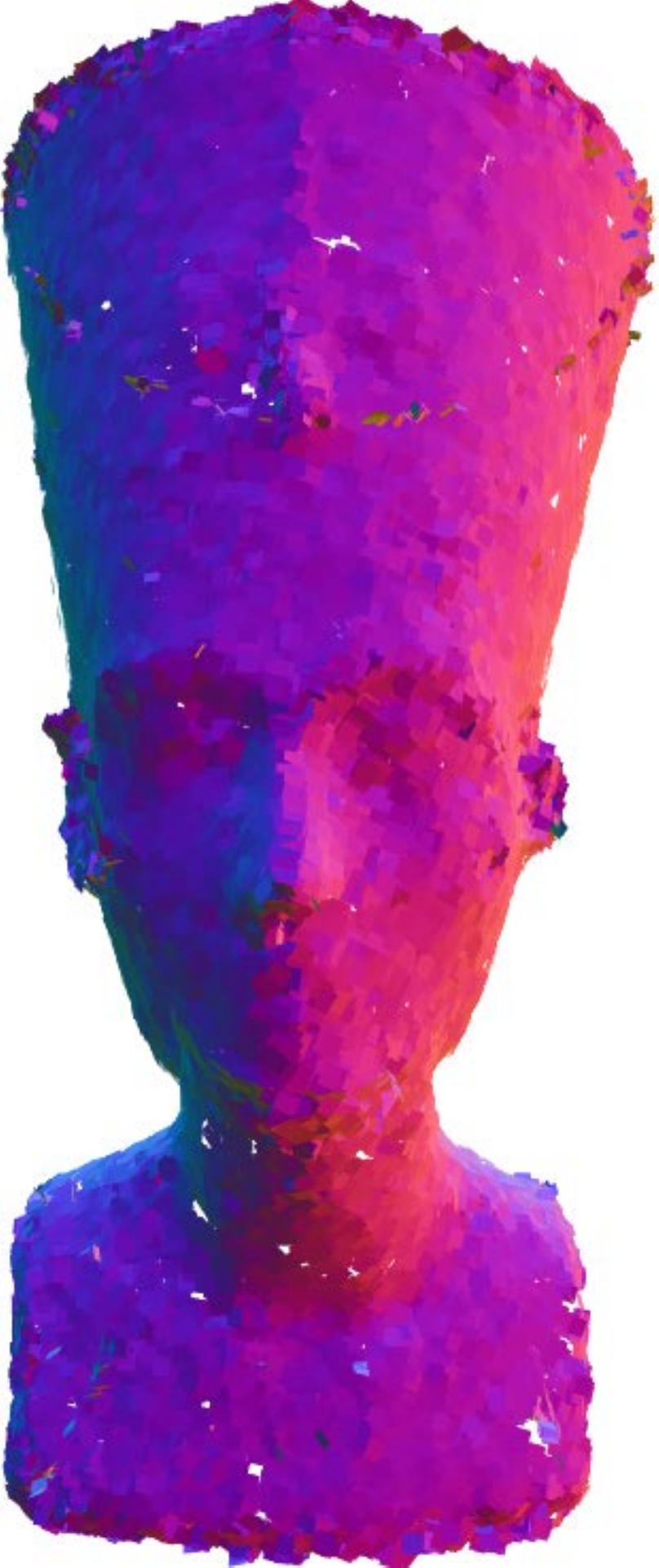}\\
        \includegraphics[width=1\textwidth]{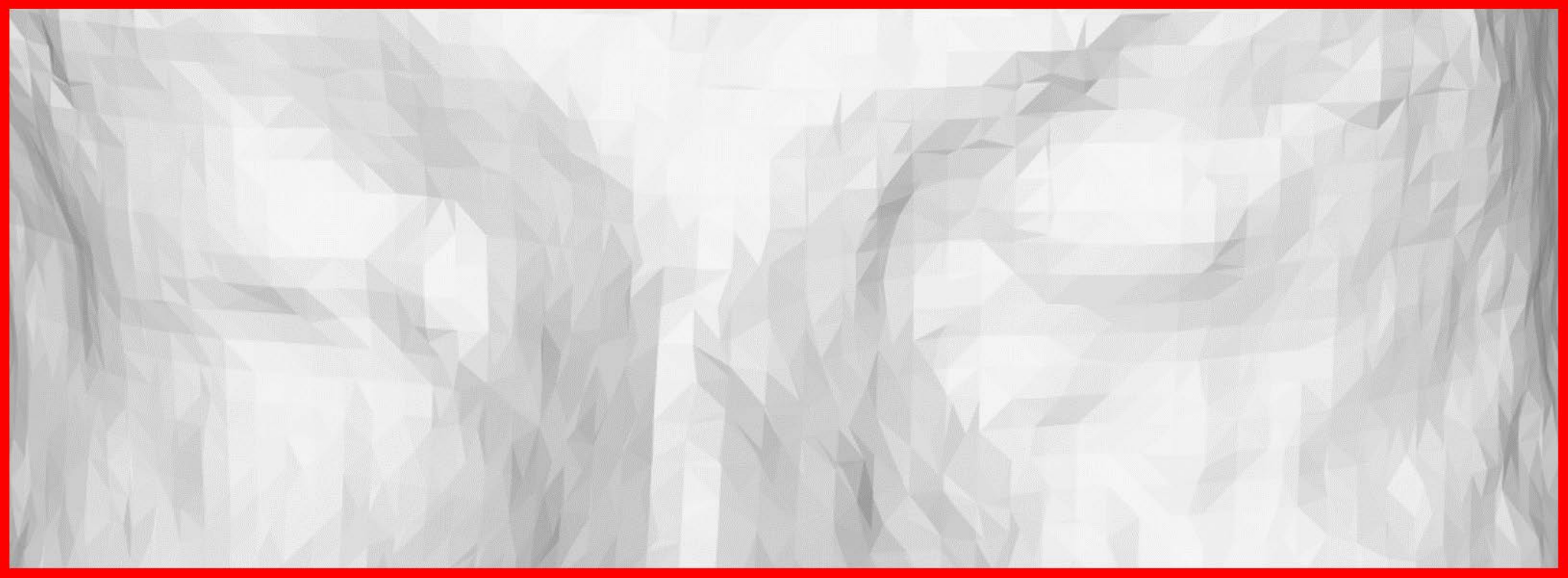}\\
        \includegraphics[width=1\textwidth]{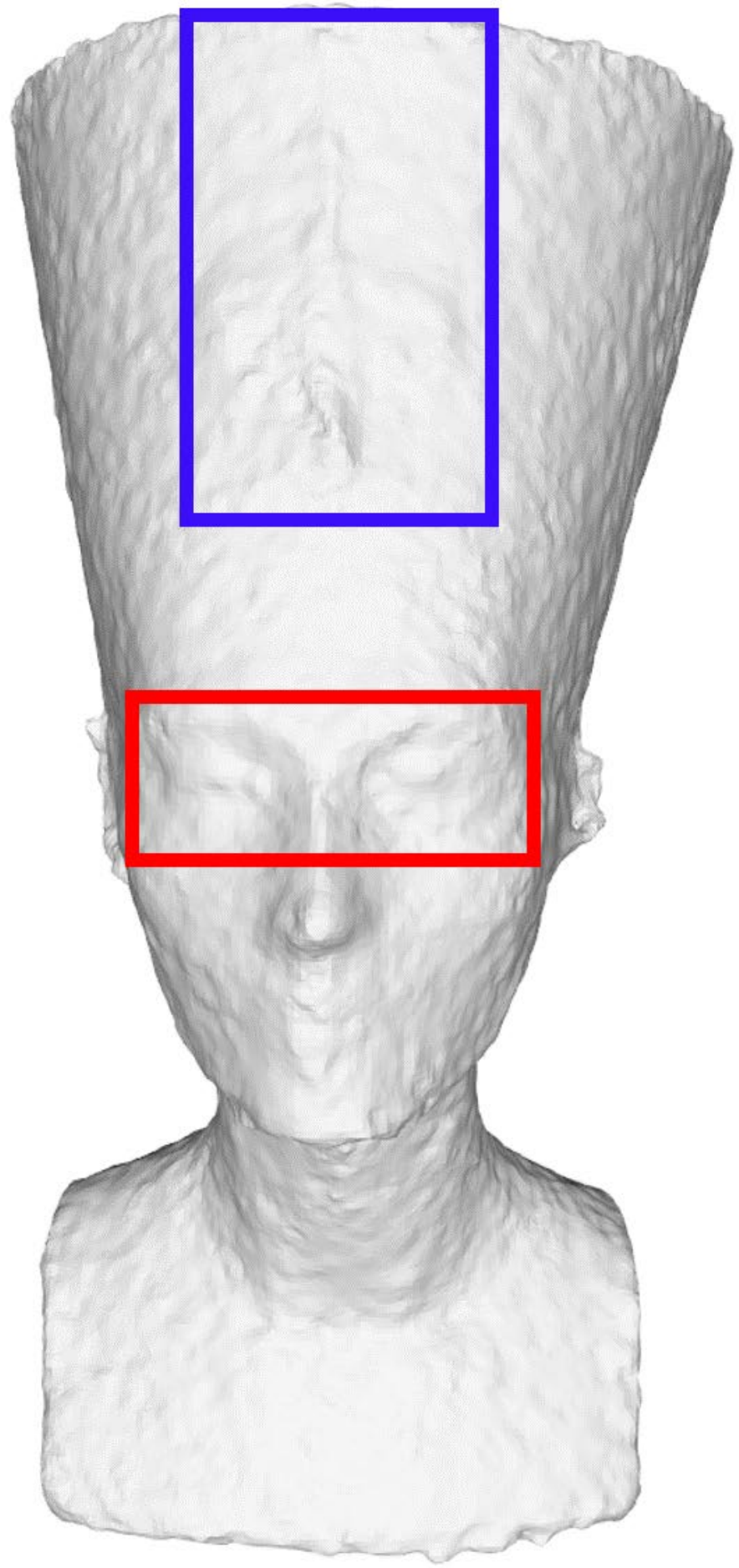}
        \end{minipage}
    }
    \subfigure[PCN]
    {
       \begin{minipage}[b]{0.095\textwidth}
        \includegraphics[width=1\textwidth]{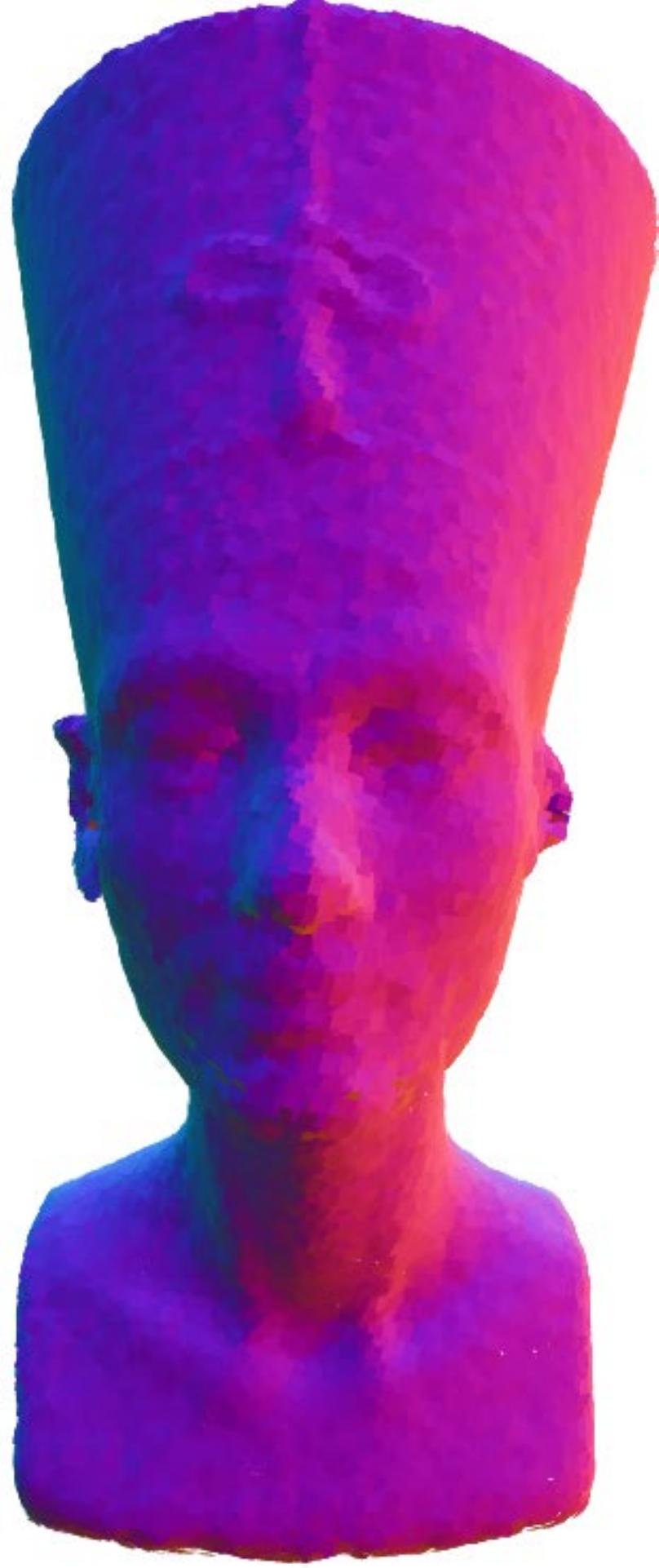}\\
        \includegraphics[width=1\textwidth]{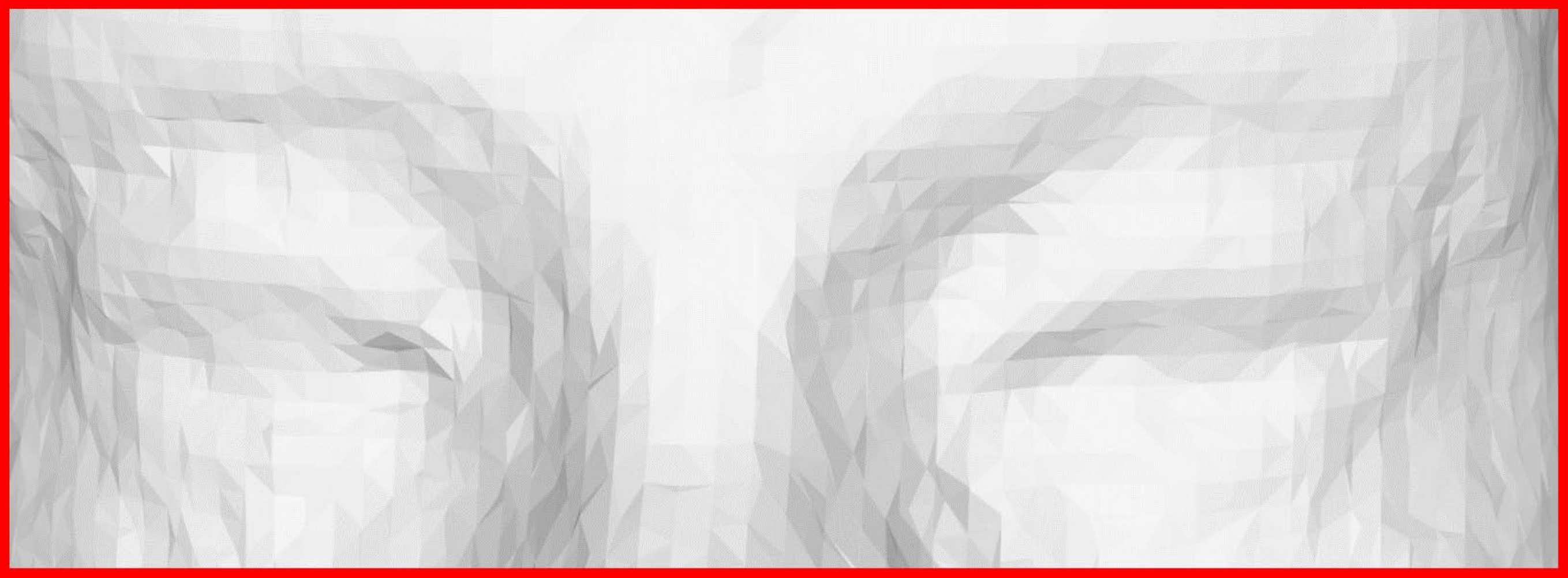}\\
        \includegraphics[width=1\textwidth]{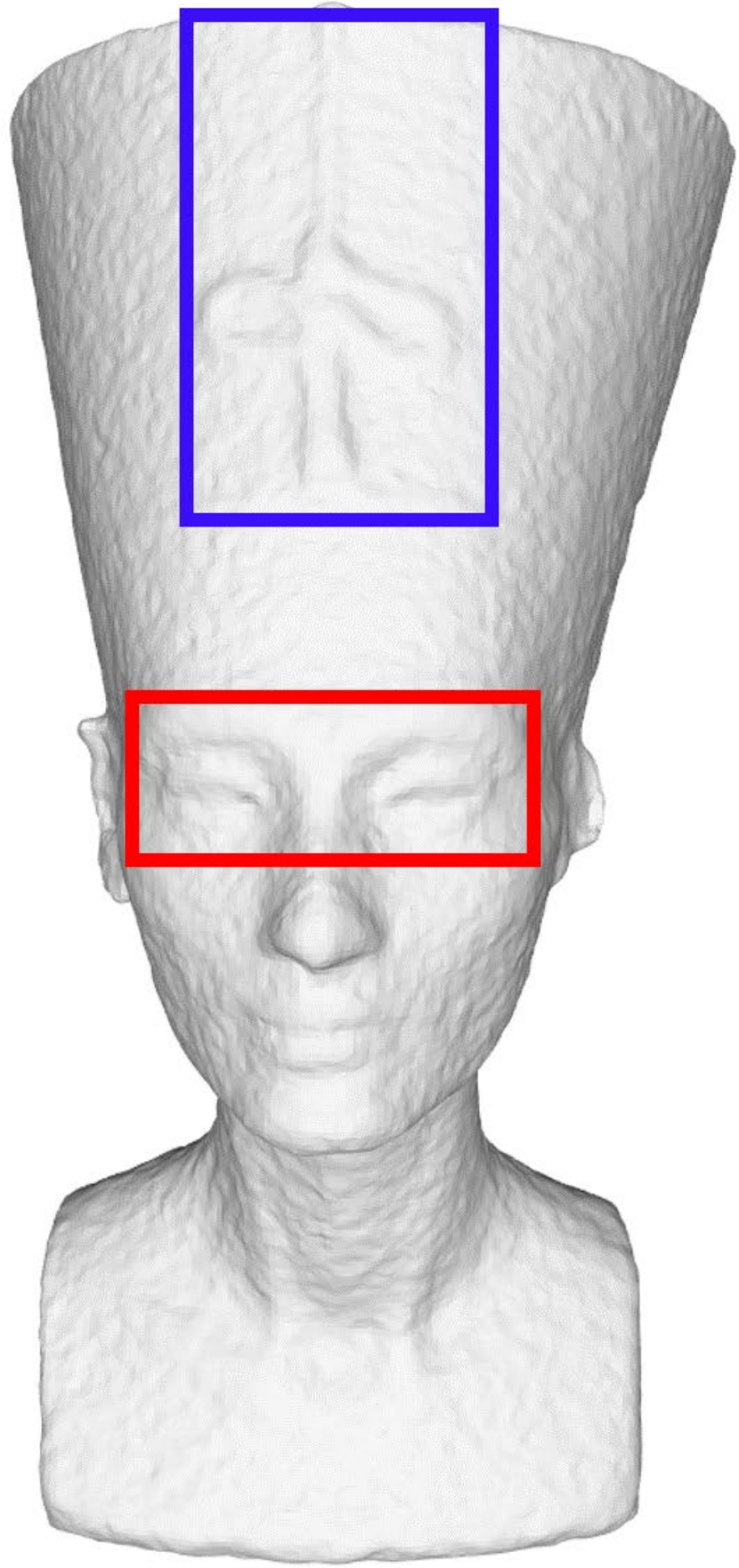}
        \end{minipage}
    }
    \subfigure[TD]
    {
        \begin{minipage}[b]{0.095\textwidth}
        \includegraphics[width=1\textwidth]{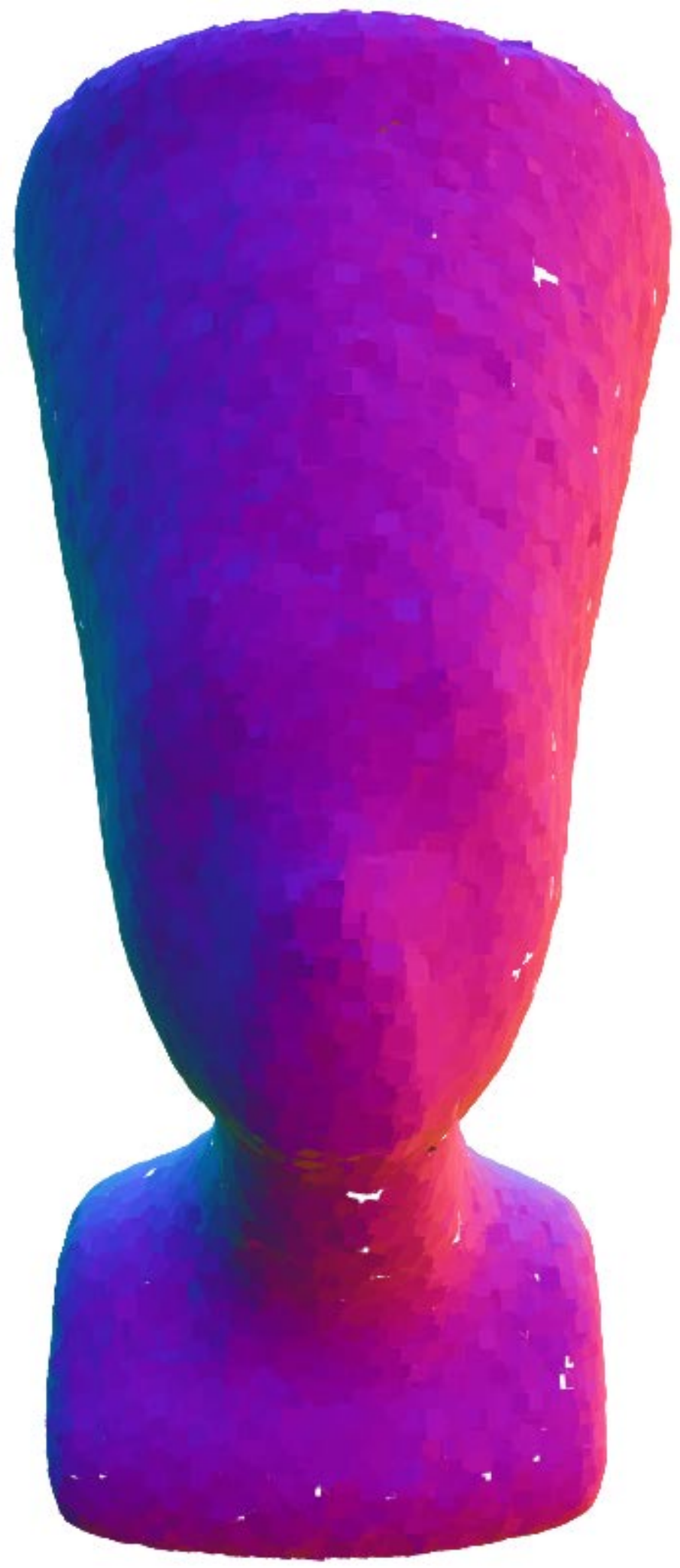}\\
        \includegraphics[width=1\textwidth]{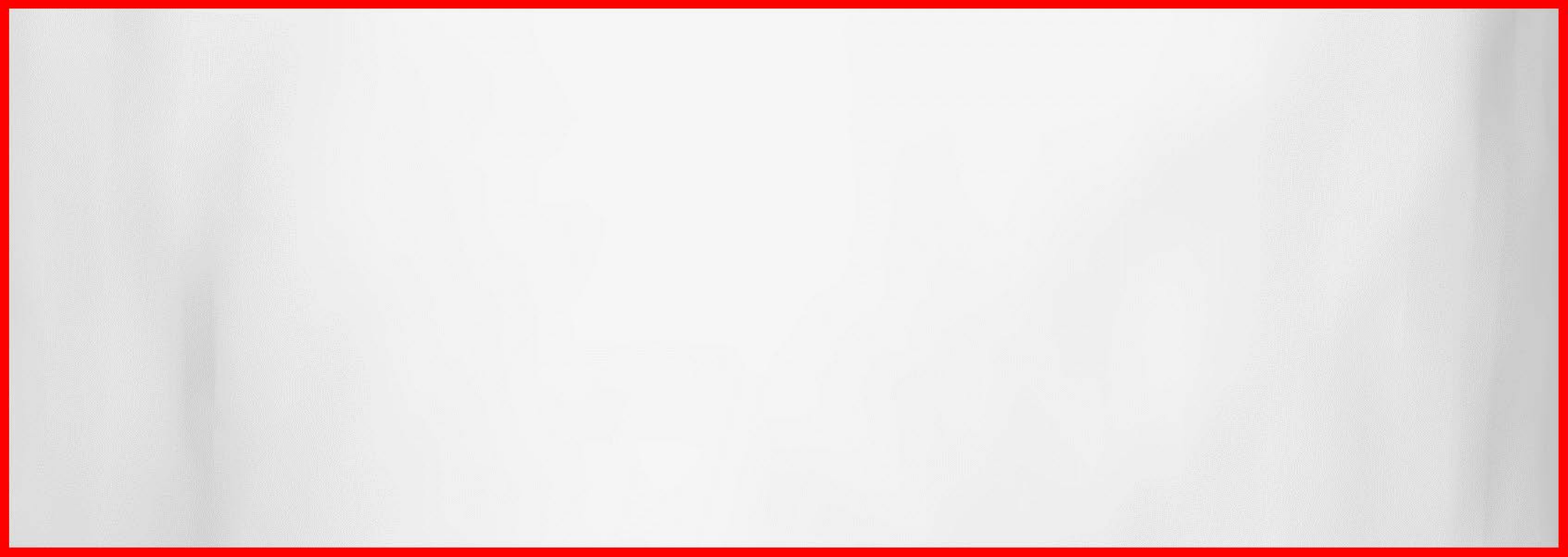}\\
        \includegraphics[width=1\textwidth]{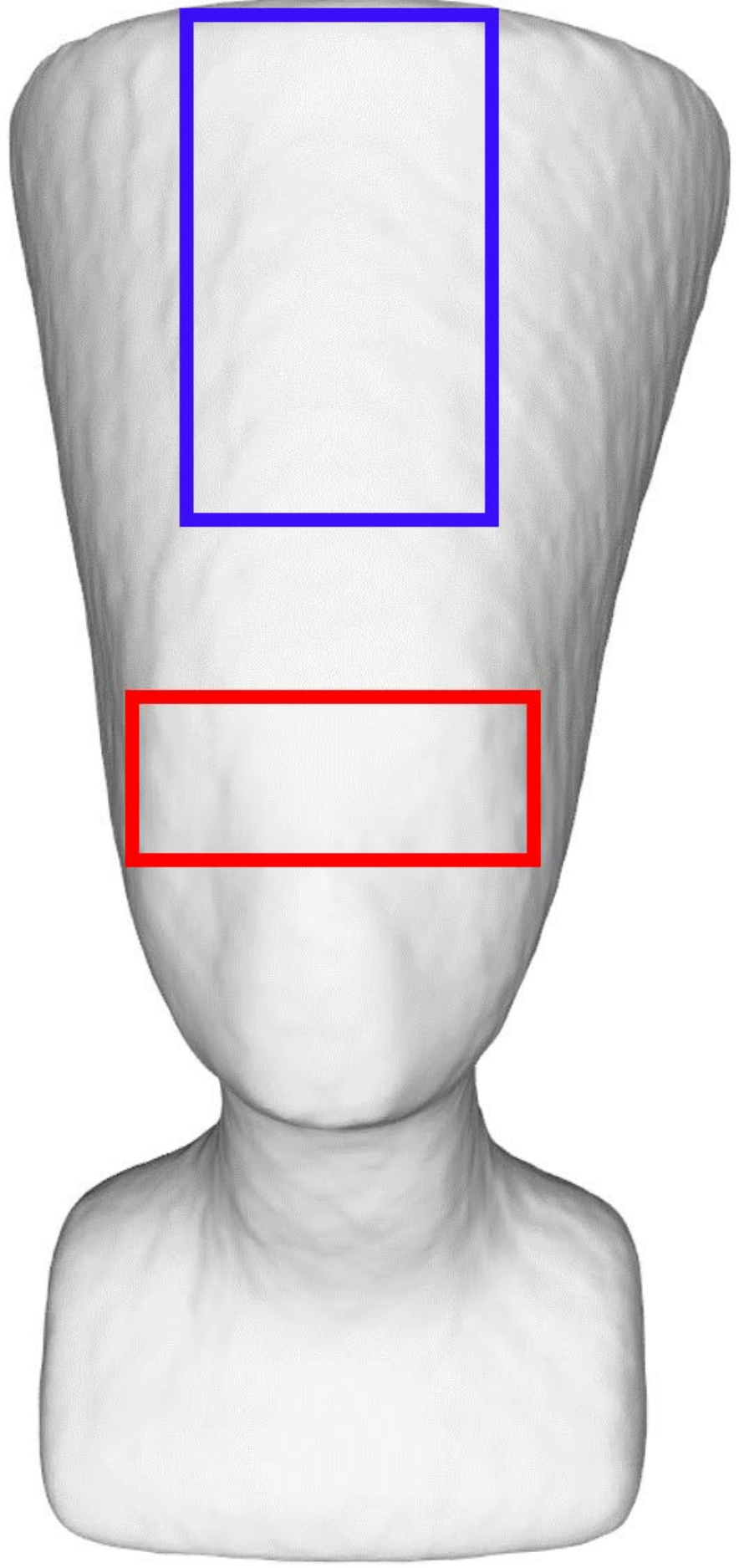}
        \end{minipage}
    }
    \subfigure[Ours]
    {
       \begin{minipage}[b]{0.095\textwidth}
        \includegraphics[width=1\textwidth]{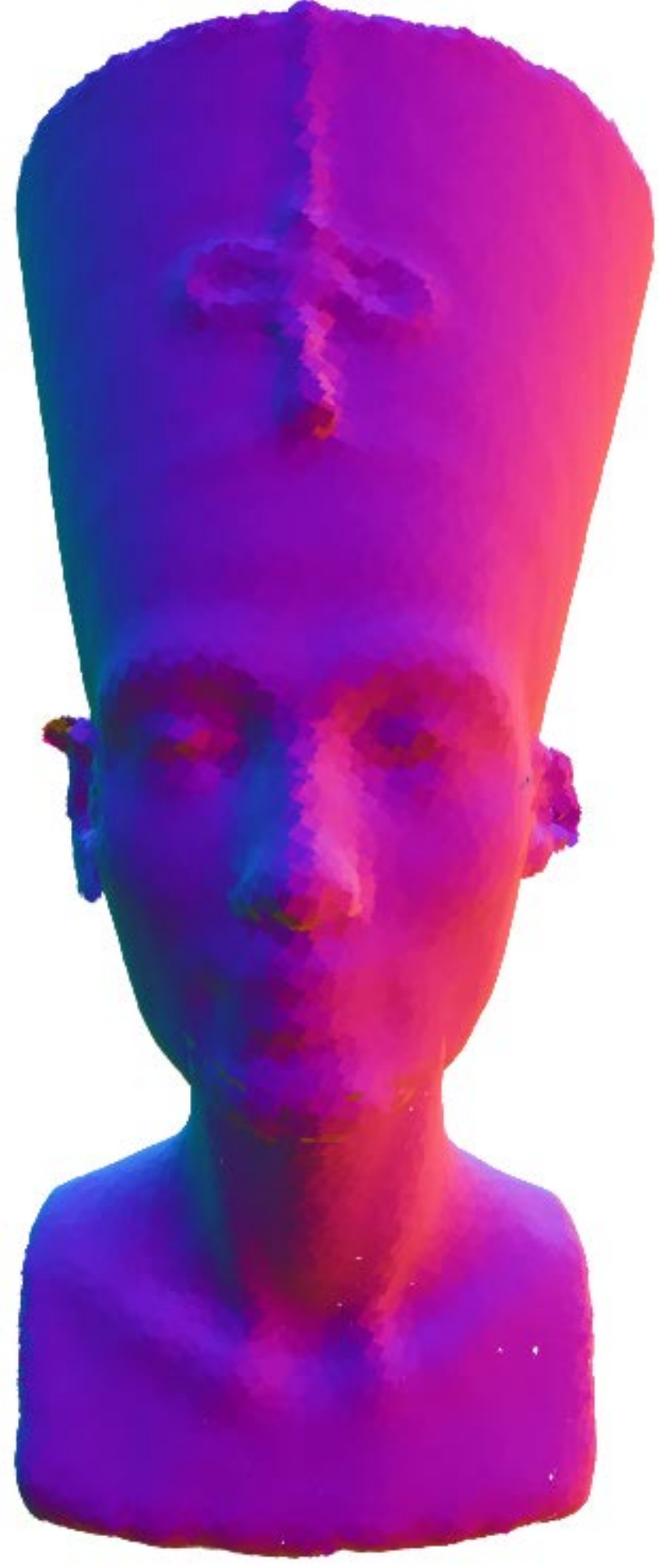}\\
        \includegraphics[width=1\textwidth]{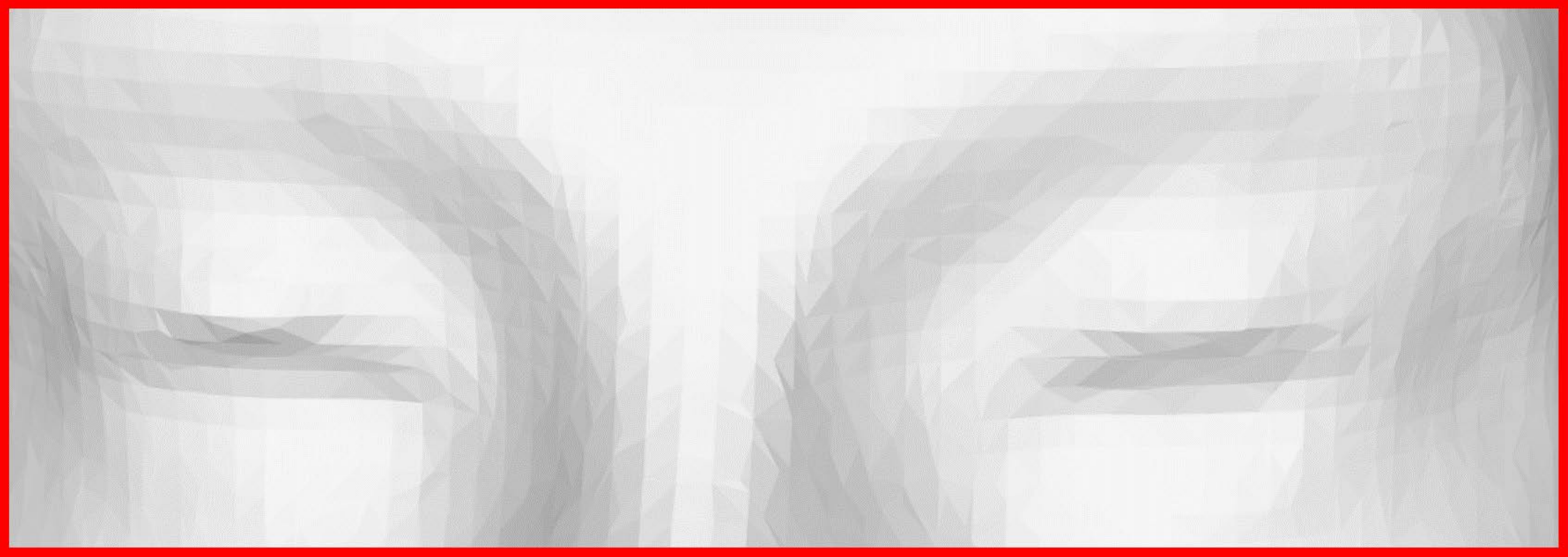}\\
        \includegraphics[width=1\textwidth]{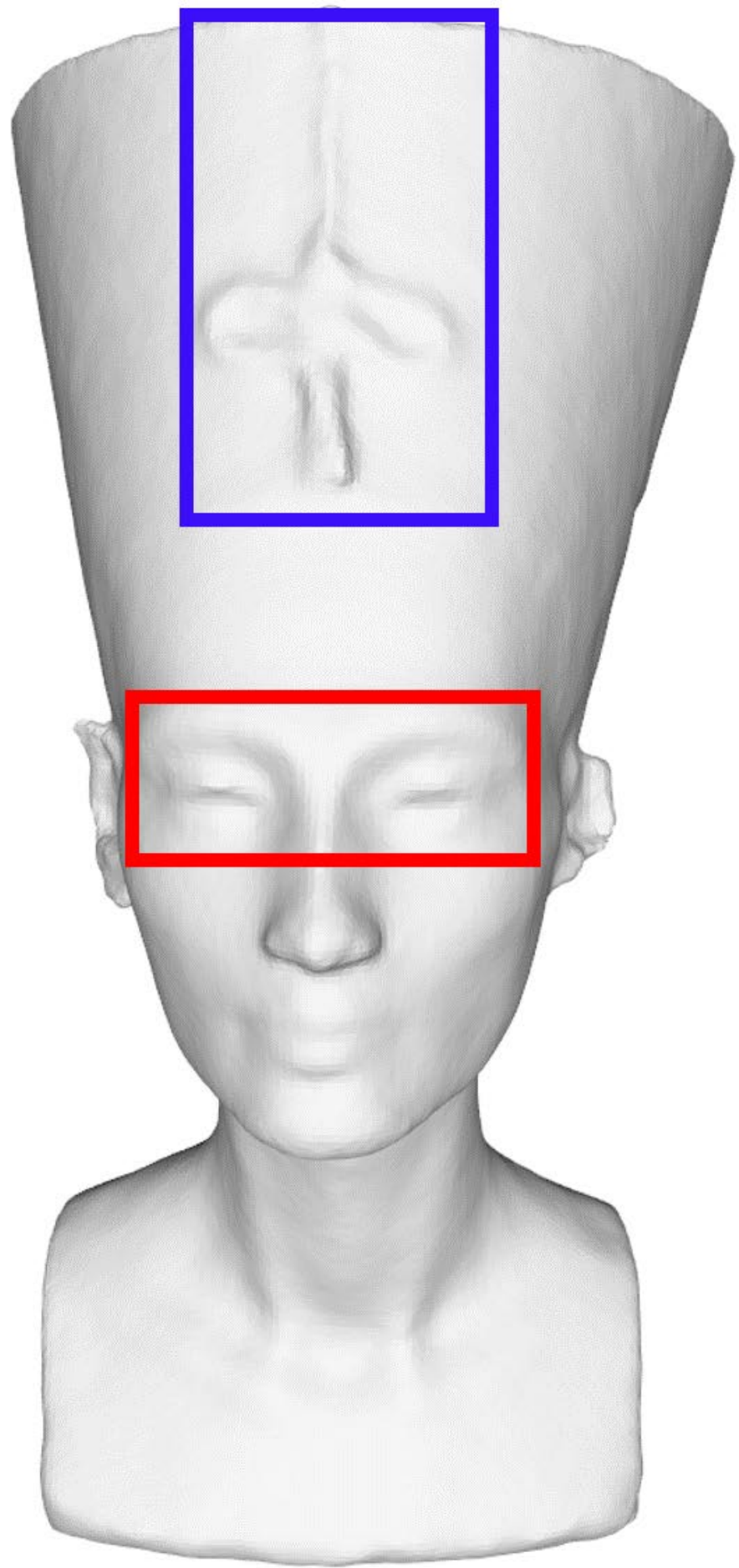}
        \end{minipage}
    }
    \caption{Results on the raw Nefertiti model. The corresponding Poisson reconstructed meshes are shown in the bottom.}
    \label{fig:Nefertiti}
\end{figure*}

\subsection{Visual Comparisons}
\label{sec:visualcomparison}
\begin{figure}[htb!]
    \subfigure[Noisy]
    {
        \begin{minipage}[b]{0.1\textwidth} 
        \includegraphics[width=1\textwidth]{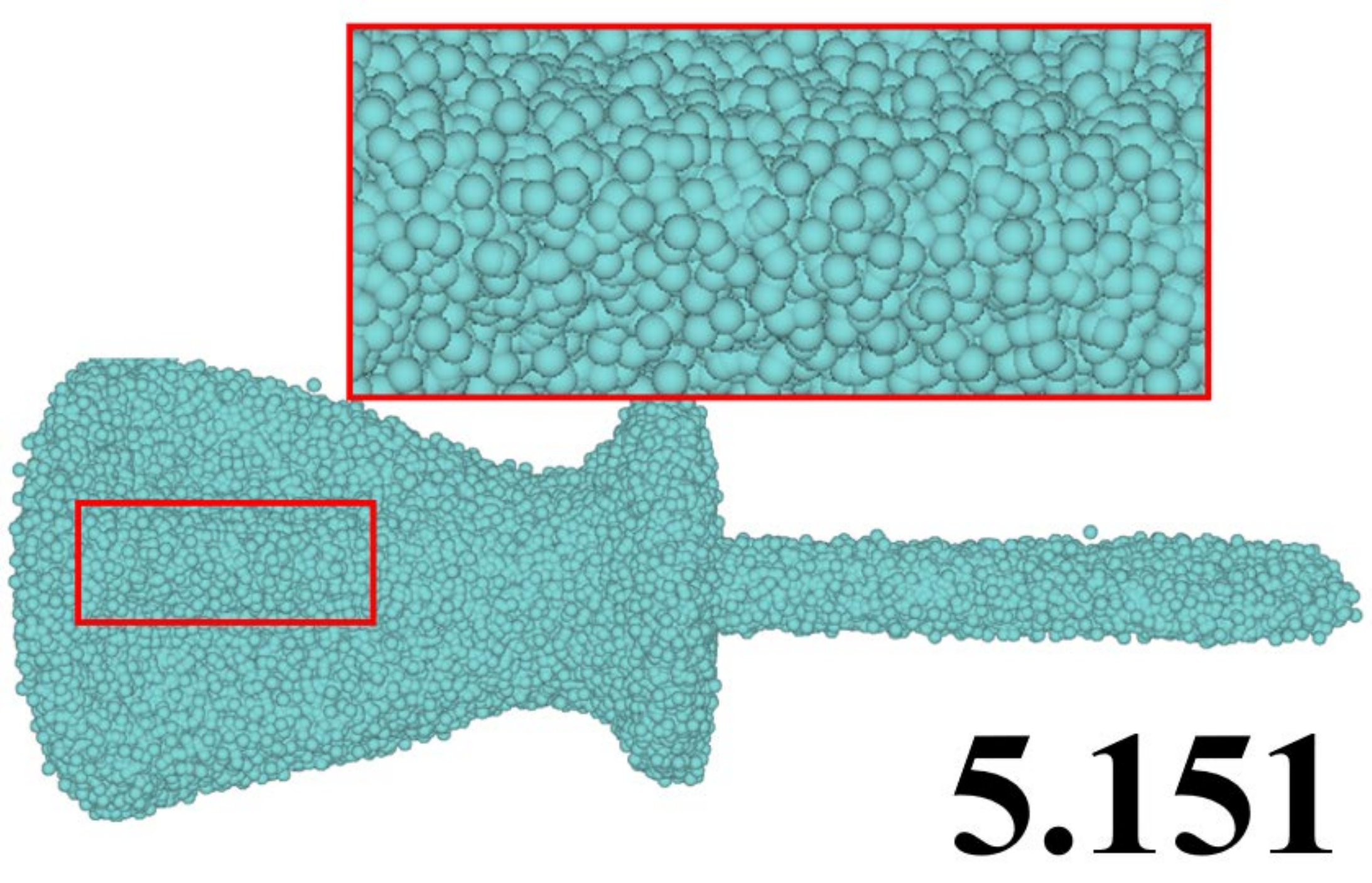}\\ 
        
        \includegraphics[width=1\textwidth]{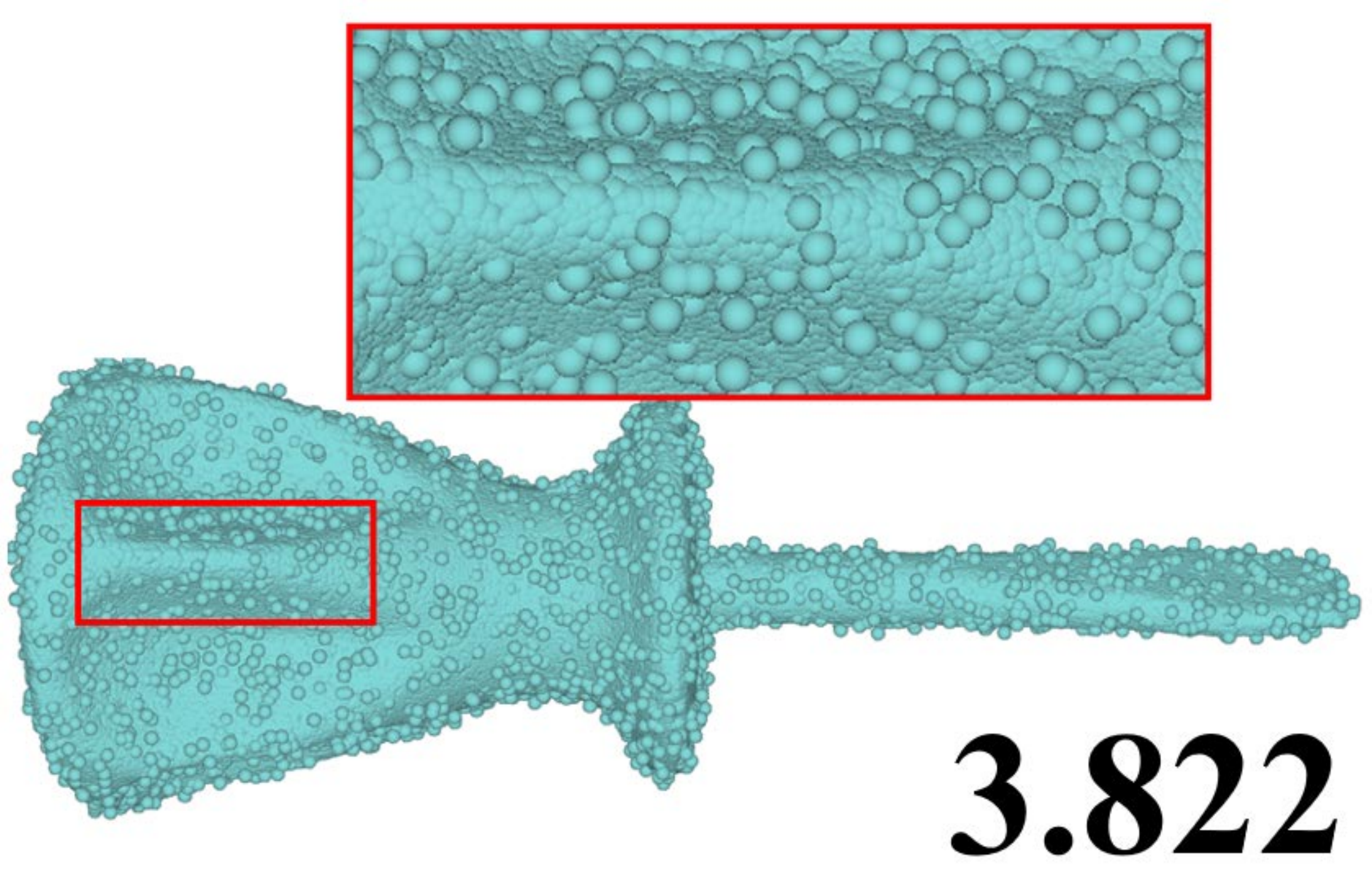}\\
        
        \includegraphics[width=1\textwidth]{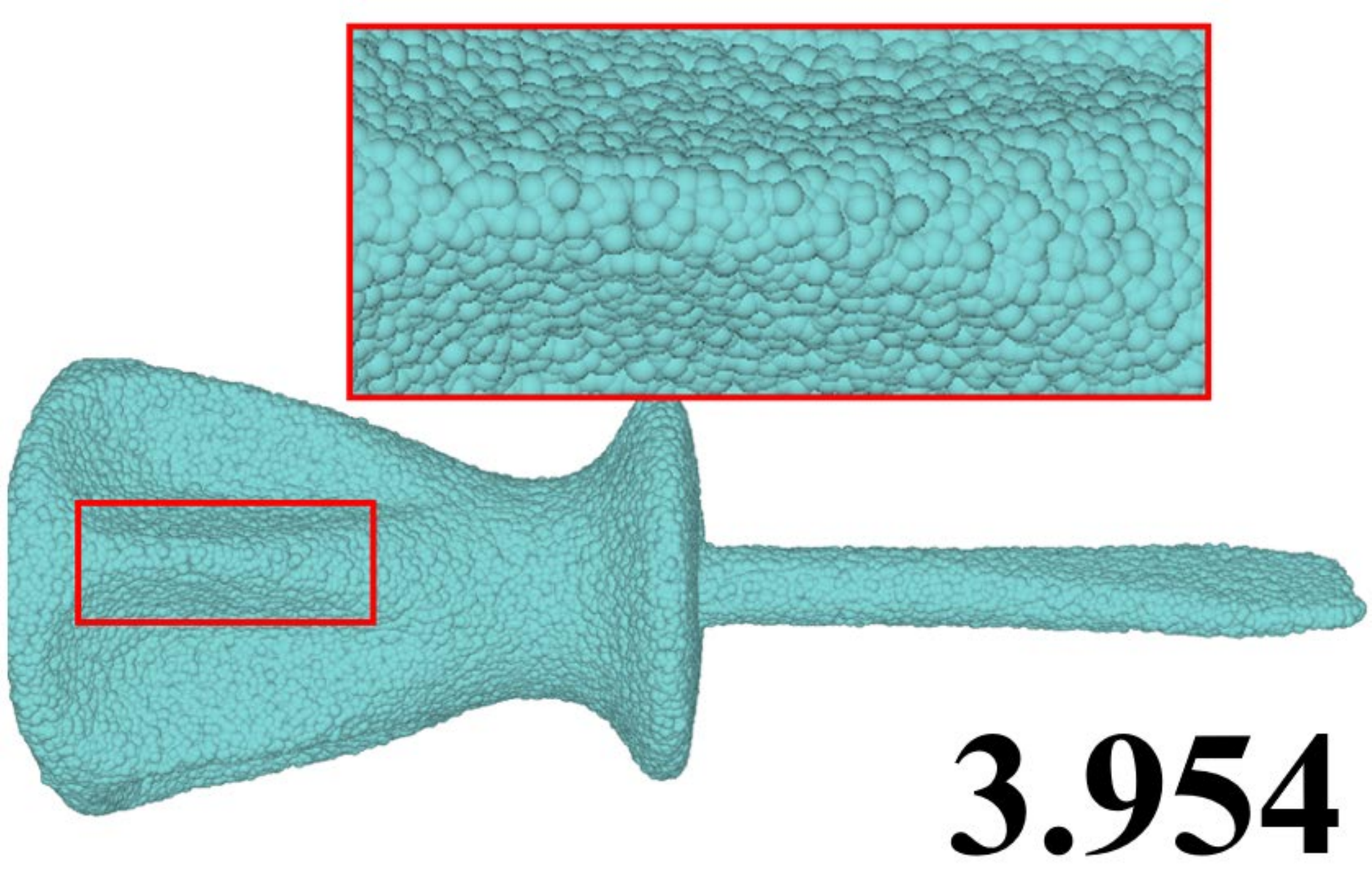}
        \end{minipage}
    }
    \subfigure[PCN]
    {
        \begin{minipage}[b]{0.1\textwidth} 
        \includegraphics[width=1\textwidth]{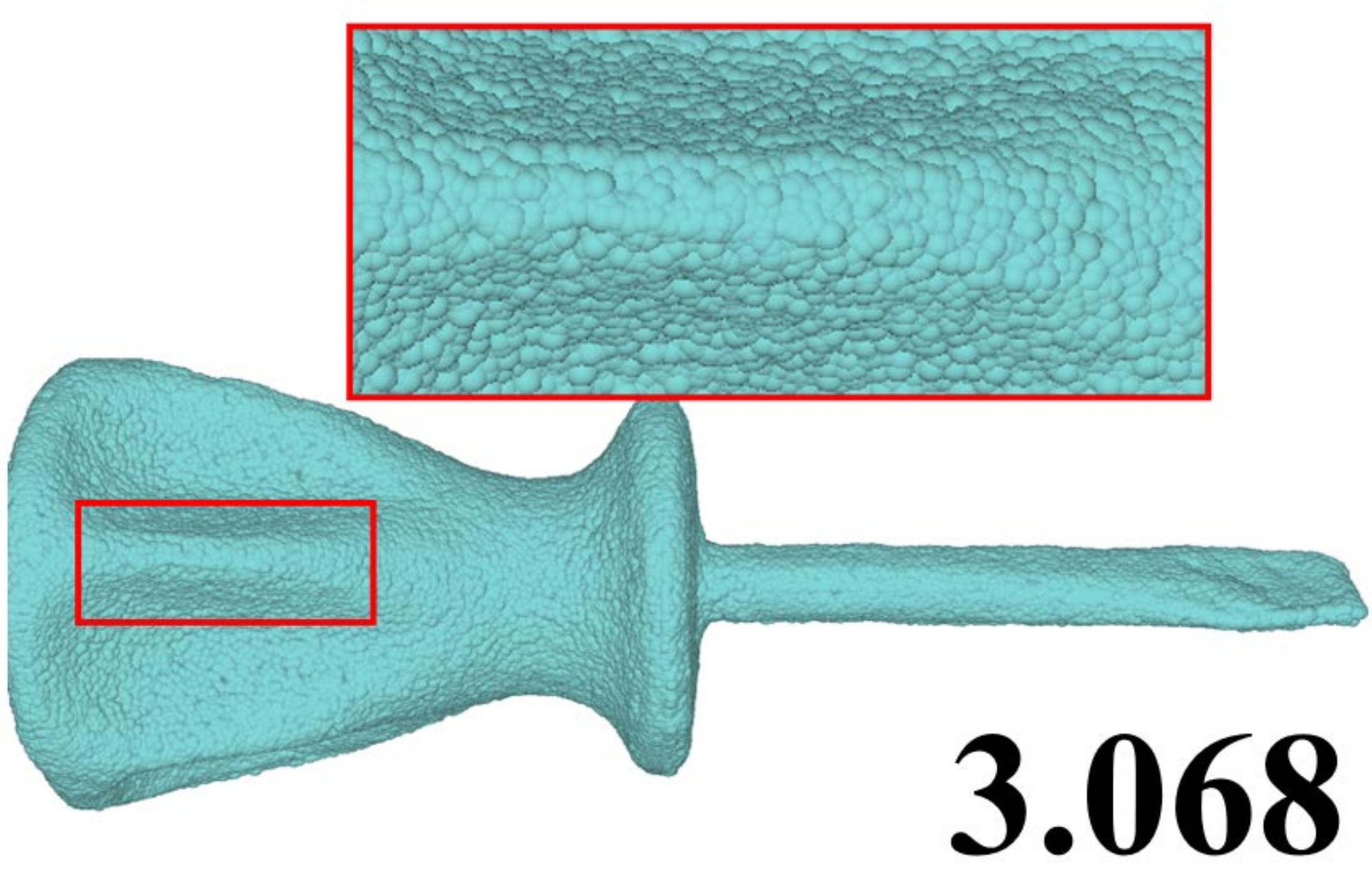}\\
        
        \includegraphics[width=1\textwidth]{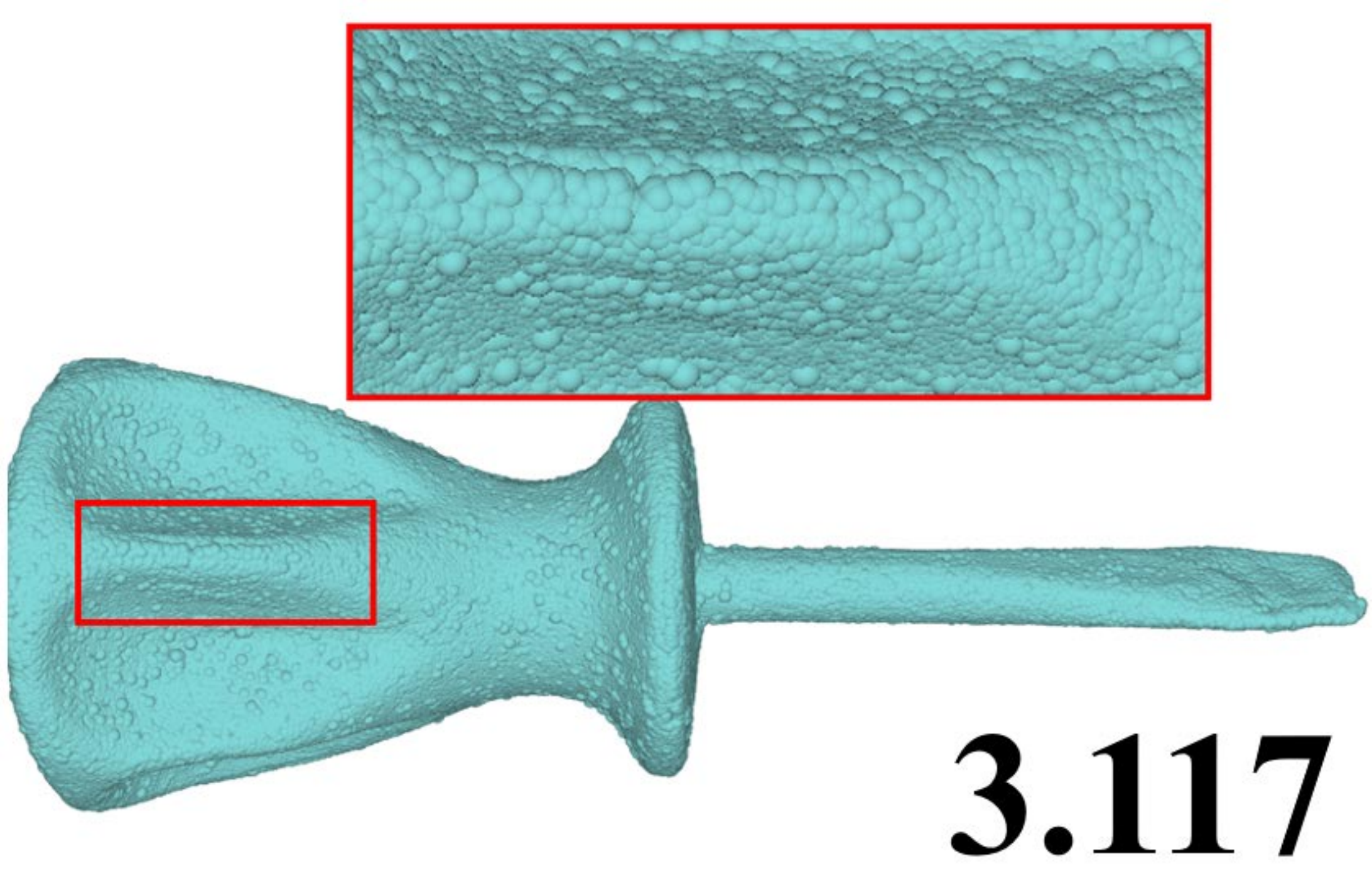}\\
        
        \includegraphics[width=1\textwidth]{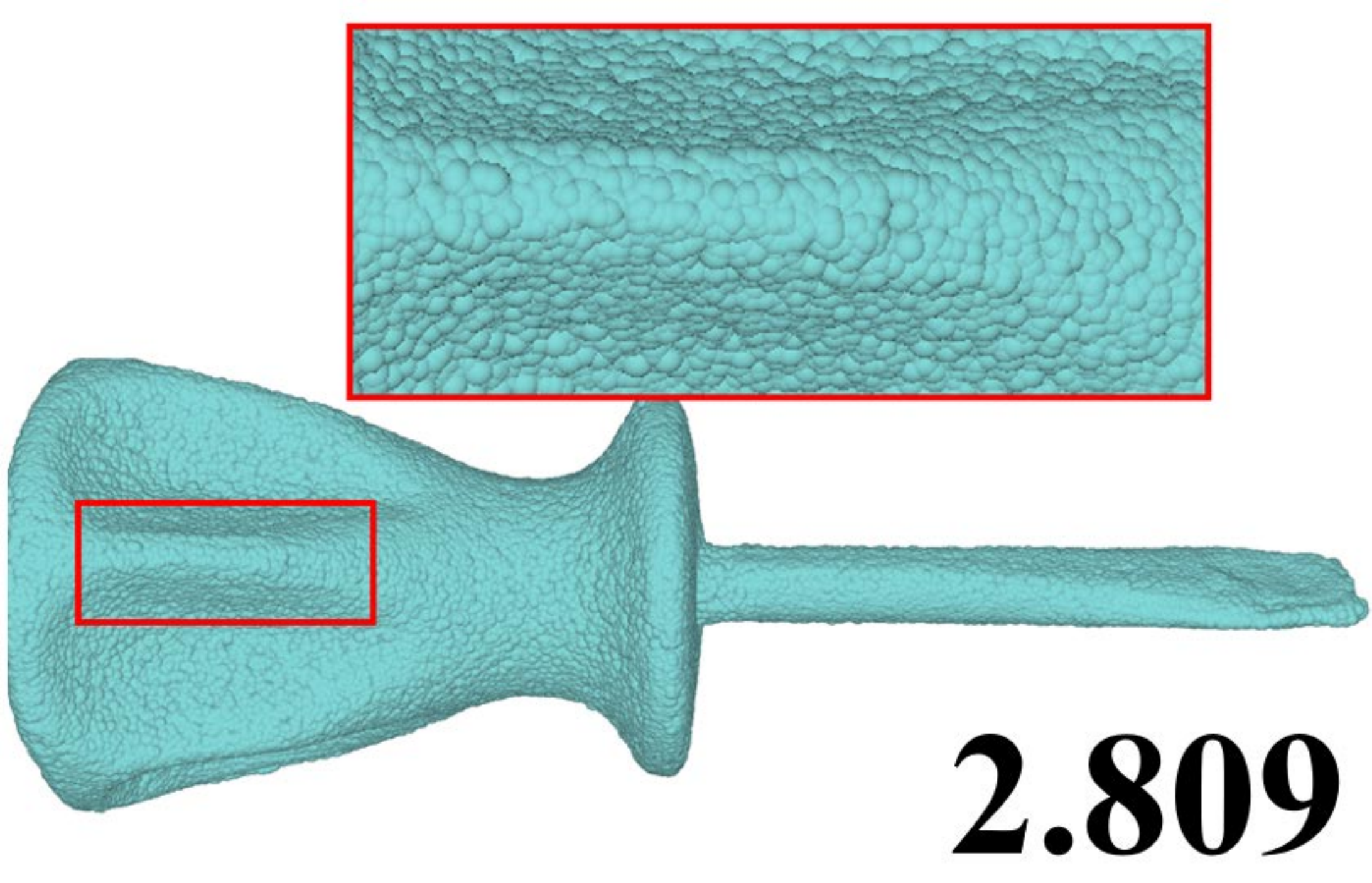}
        \end{minipage}
    }
    \subfigure[TD]
    {
       \begin{minipage}[b]{0.1\textwidth} 
        \includegraphics[width=1\textwidth]{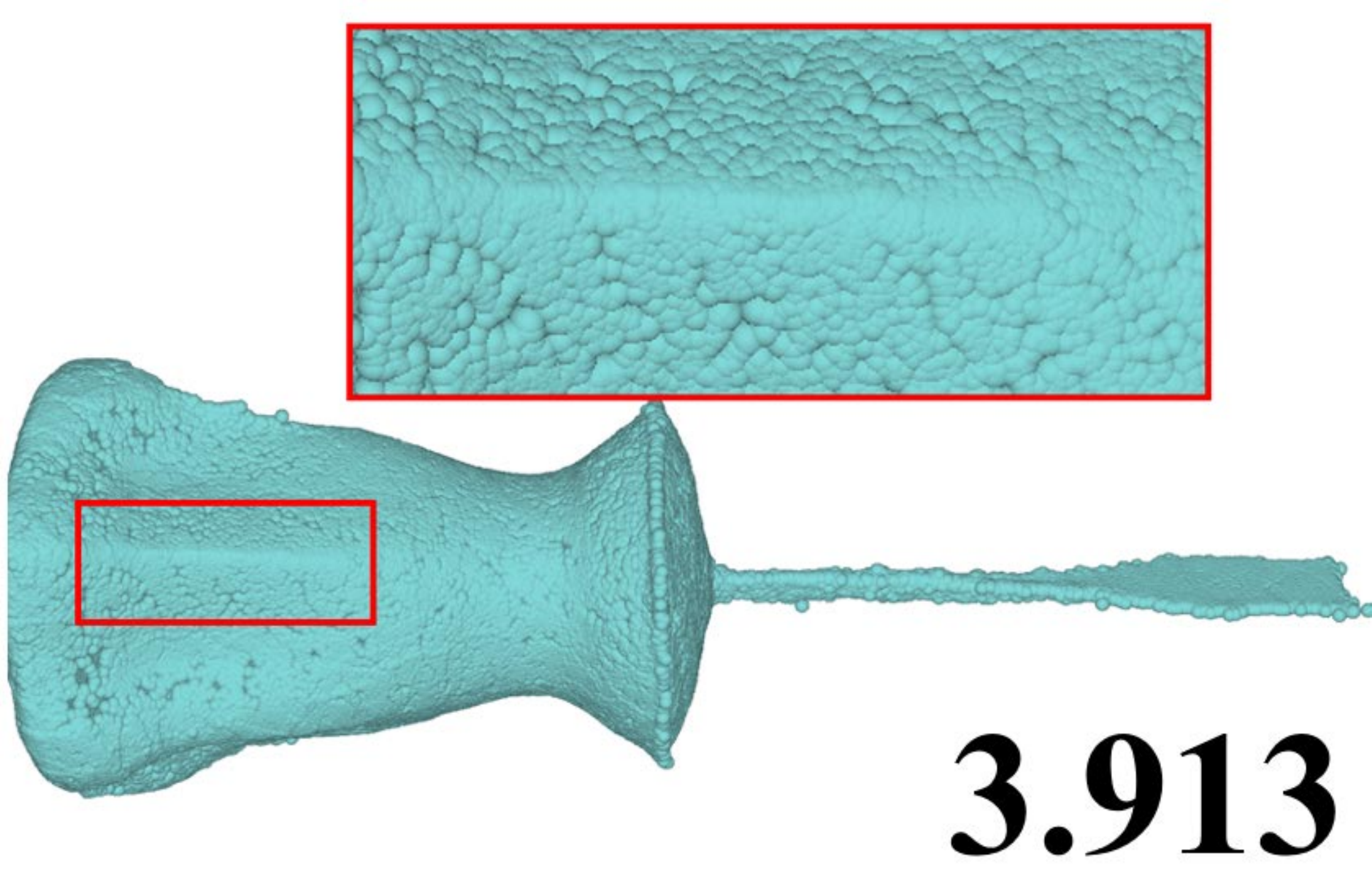}\\
        
        \includegraphics[width=1\textwidth]{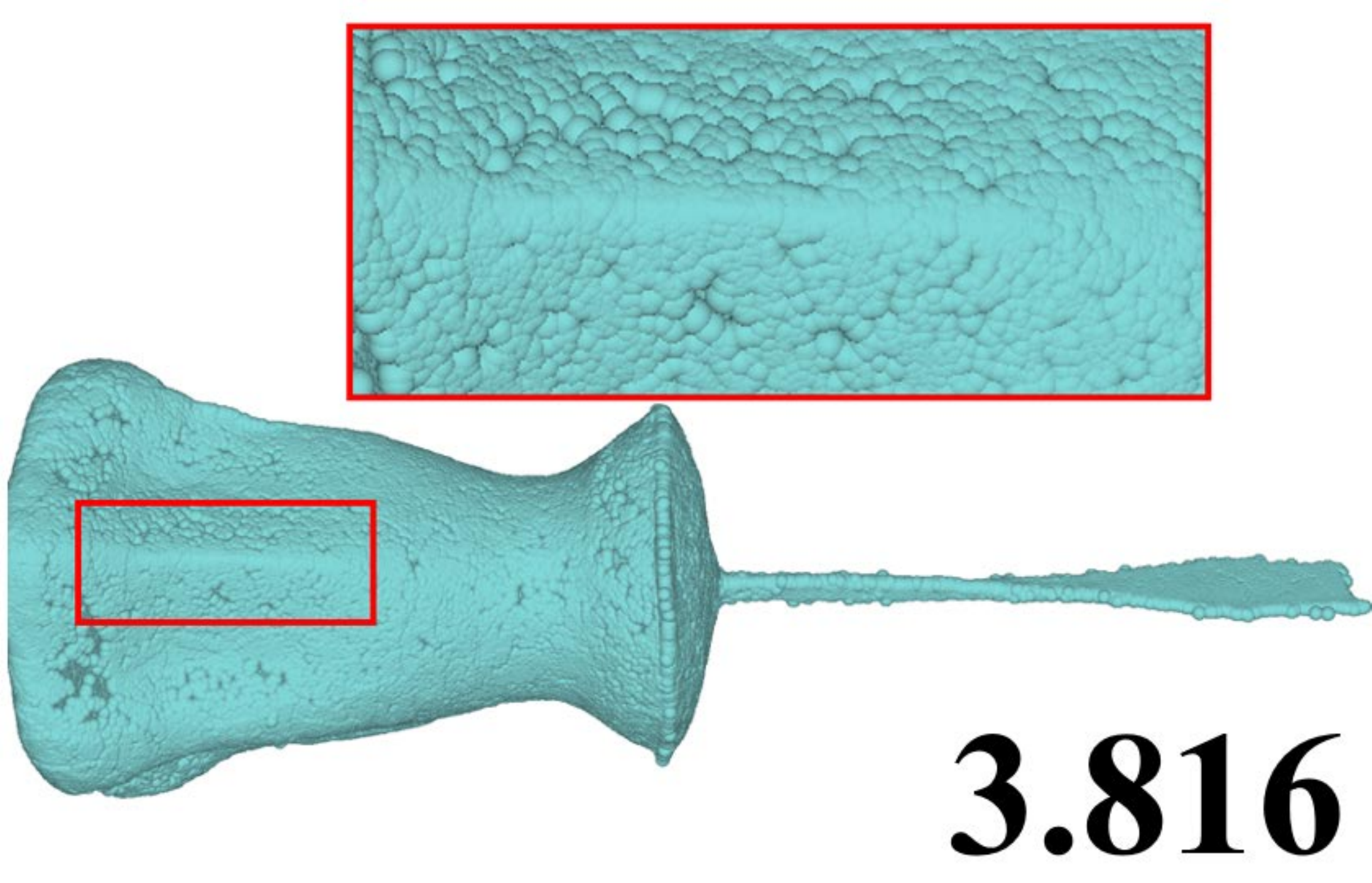}\\
        
        \includegraphics[width=1\textwidth]{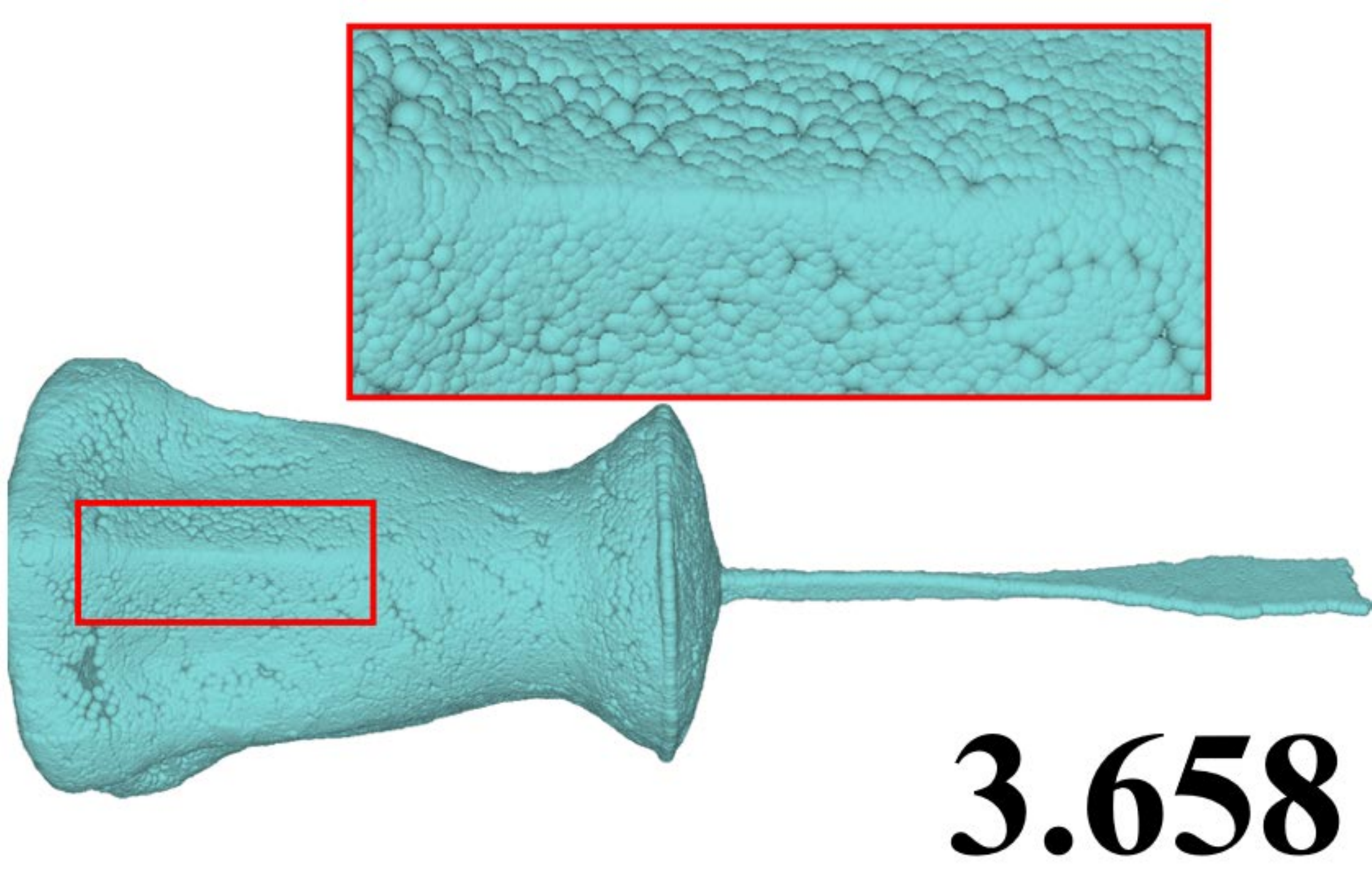}
        \end{minipage}
    }
    \subfigure[Ours]
    {
       \begin{minipage}[b]{0.1\textwidth} 
        \includegraphics[width=1\textwidth]{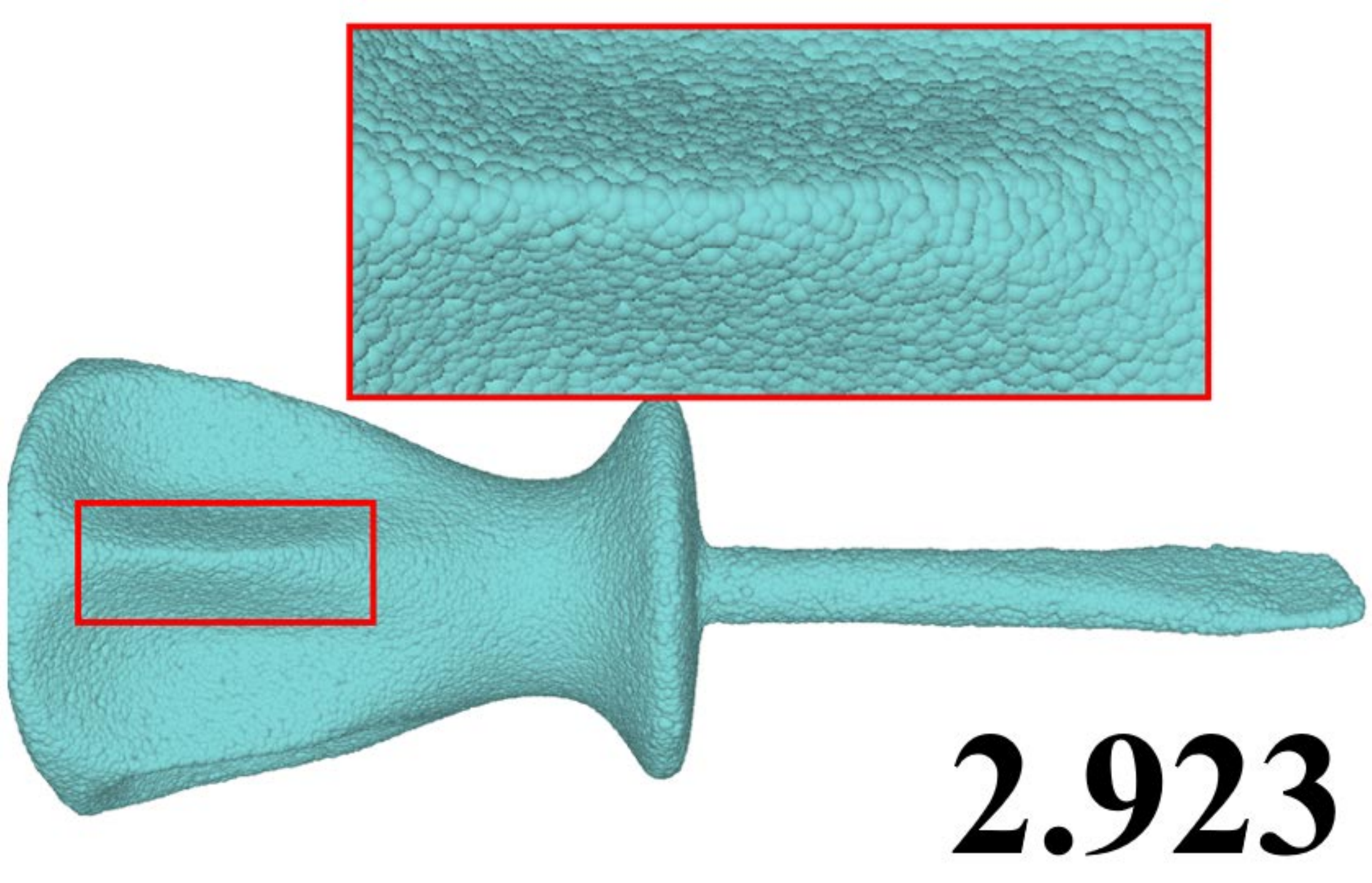}\\
        
        \includegraphics[width=1\textwidth]{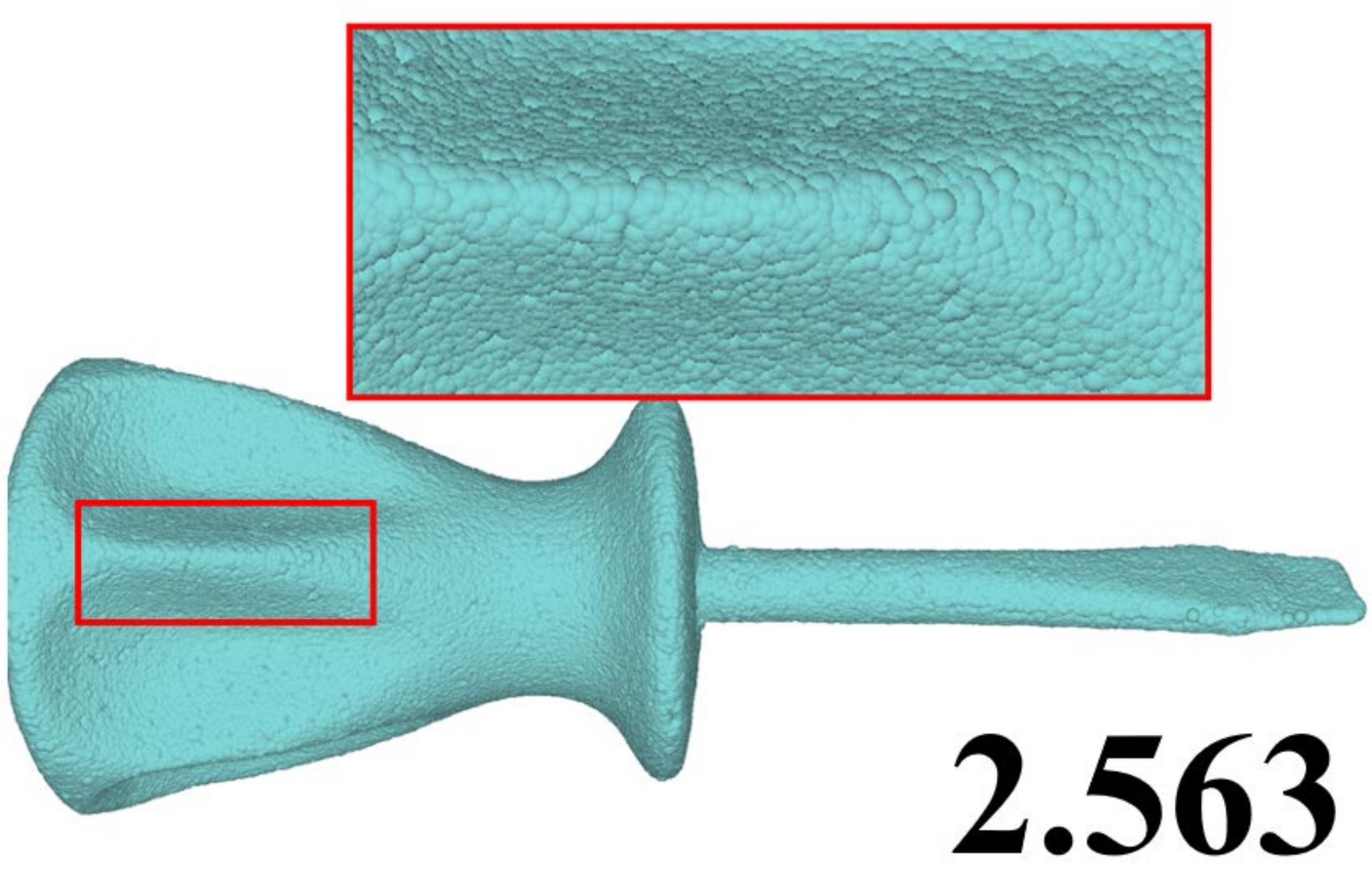}\\
        
        \includegraphics[width=1\textwidth]{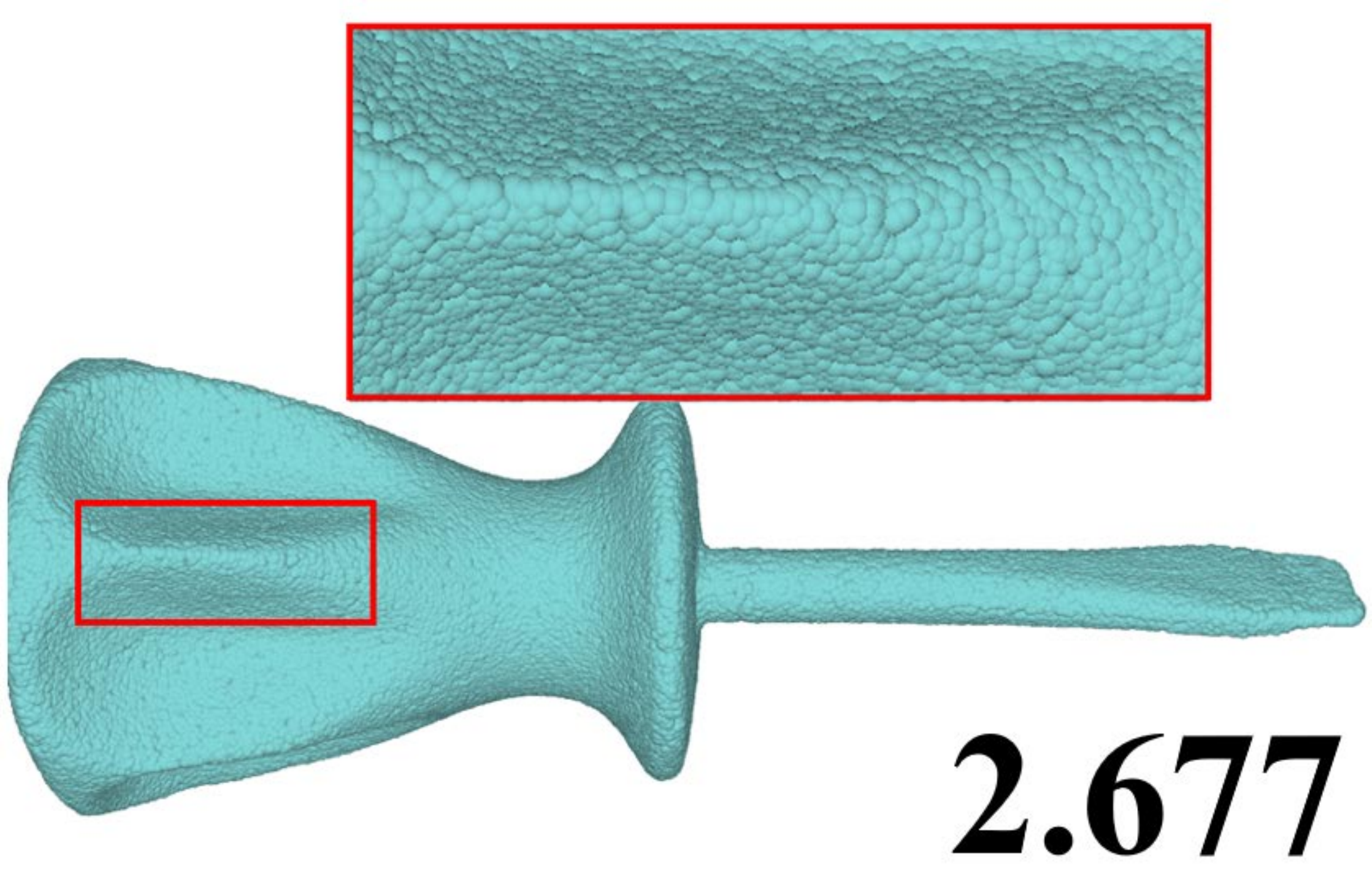}
        \end{minipage}
    }
    \caption{Point cloud filtering for three different types of synthetic noise (from top to bottom: Gaussian, impulsive, uniform). See the close-up views for differences. The overall MSEs ($\times 10^{-3}$) for different methods are shown in the figure.}
    \label{fig:multi_type_noise}
\end{figure}

\textbf{Point clouds with synthetic noise.} The synthetic noise level is estimated by the diagonal length of the bounding box. For example, $0.5\%$ noise denotes $0.5\%$ of the diagonal length. As shown in Fig. \ref{fig:cadvisualcomparison}, we test four CAD models (Boxunion, Cube, Fandisk and Tetrahedron) with $0.5\%$ noise. Compared with the state-of-the-art point cloud filtering techniques, we observe that results by our Pointfilter generates visually better results, in terms of noise removal and features preservation. Note that RIMLS and GPF can also preserve sharp features to some extent; however, they depend greatly on the capability of normal filters which become less robust when meeting large noise. Compared to RIMLS and GPF, we elegantly detour the normal filtering issue since our framework requires the easily obtained ground-truth normals for training only. As shown in Fig. \ref{fig:normal_smoothing}, RIMLS and GPF produce less desired results when handling $1.0\%$ noise, for example, obvious gaps in sharp edges (\ref{fig:normal_smoothing} (b)) and striking outliers (\ref{fig:normal_smoothing} (c)). By contrast, our Pointfilter is robust in preserving sharp features. Since the outliers exist in RIMLS and GPF results, their filtered point clouds involve shrinkage to some extent.  Despite that WLOP and CLOP are good at generating smooth results, they still fail to retain sharp features. Regarding EC-Net, it generates less pleasant results, in terms of removing noise. In addition to CAD models, we also test some non-CAD models corrupted with synthetic noise. As shown in Fig. \ref{fig:facesurfacereconstruction}, our proposed Pointfilter can also output visually decent results while preserving sharp features. For relatively smooth models shown in Fig. \ref{fig:surfacereconstruction}, our Pointfilter provides competitive filtering results, without any trial-and-error parameter tuning. Moreover, since WLOP and CLOP rely on the support radius to generate desirable results, the results of their methods usually have slight shrinking in narrow cylinder area (Fig. \ref{fig:surfacereconstruction} (b) and \ref{fig:surfacereconstruction} (c)). Besides Gaussian noise, we also test our Pointfilter on other types of synthetic noise including Impulsive and Uniform noise. It should be noted that all compared models in Fig. \ref{fig:multi_type_noise} are  trained under Gaussian noise only. From Fig. \ref{fig:multi_type_noise}, we can see that the results generated by our Pointfilter are better than other methods. Although TotalDenoising outputs smoothing results, it fails to preserve sharp features due to the non-consideration of sharp features information. In contrast, our Pointfilter has a better generalization capability in dealing with different types of noise.

\begin{figure}[htb!]
    \subfigure[Noisy]
    {
        \begin{minipage}[b]{0.08\textwidth} 
        \includegraphics[width=1\textwidth]{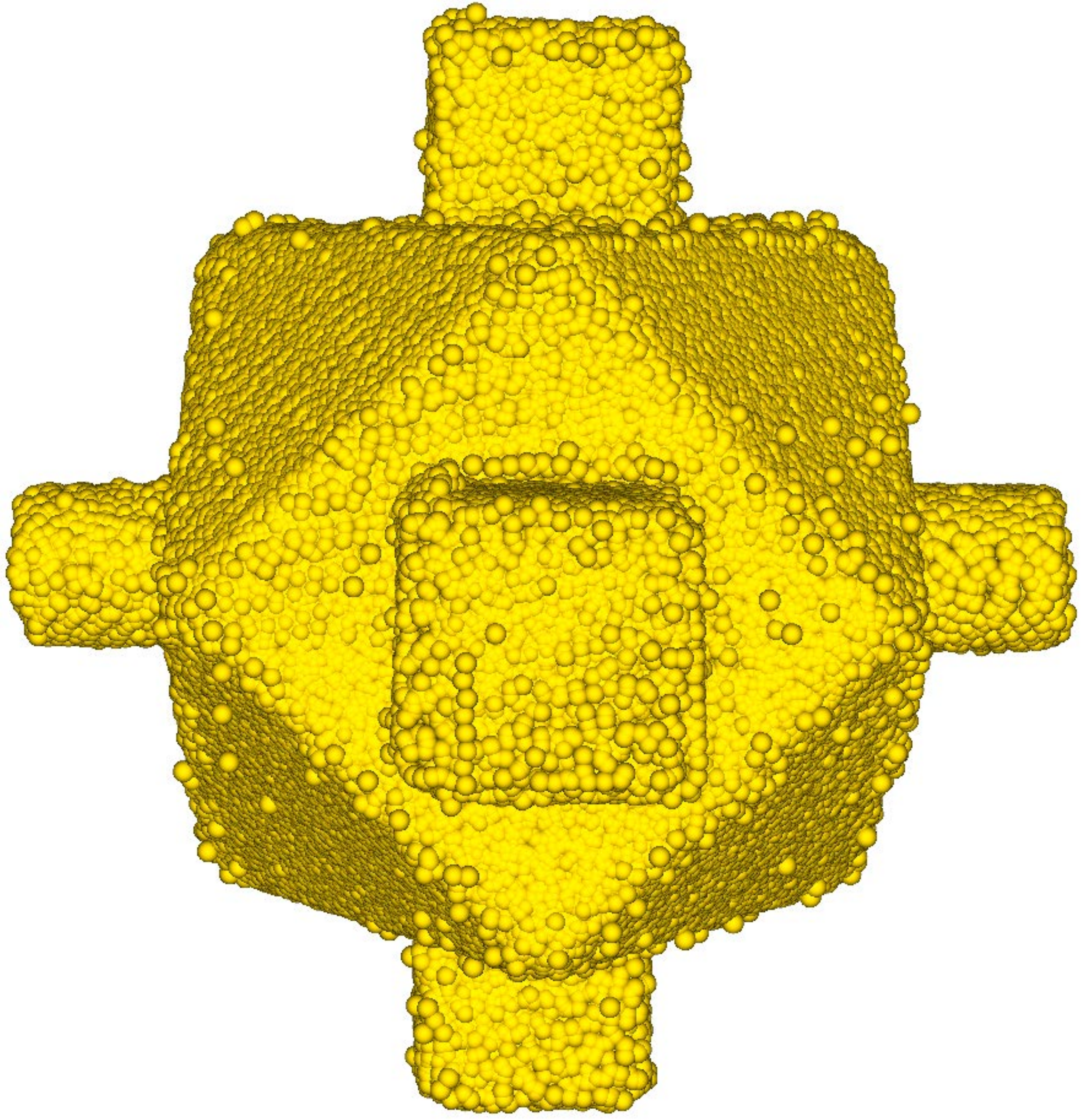}\\
        \includegraphics[width=1\textwidth]{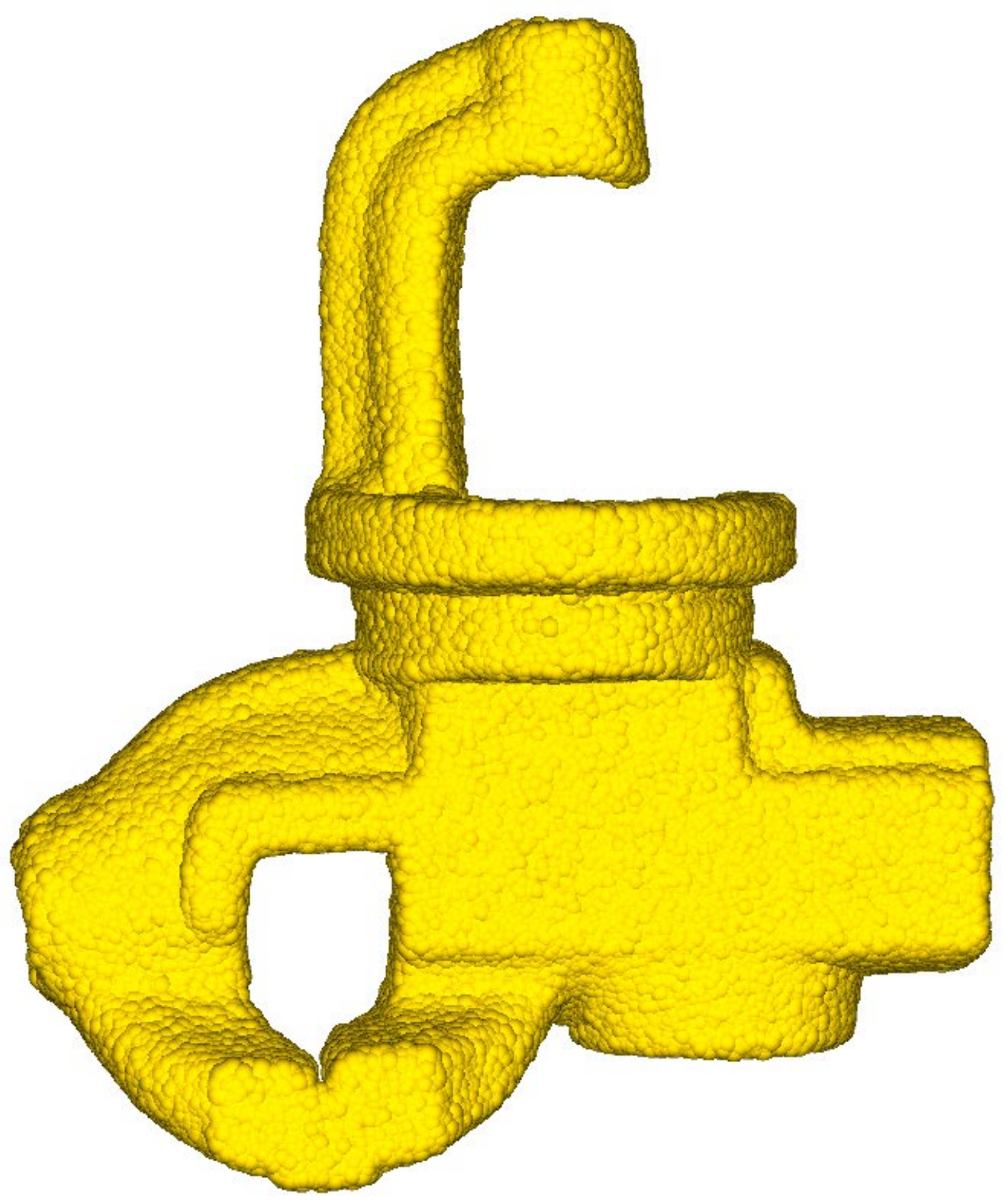}
        \end{minipage}
    }
    \subfigure[EC-Net]
    {
        \begin{minipage}[b]{0.08\textwidth} 
        \includegraphics[width=1\textwidth]{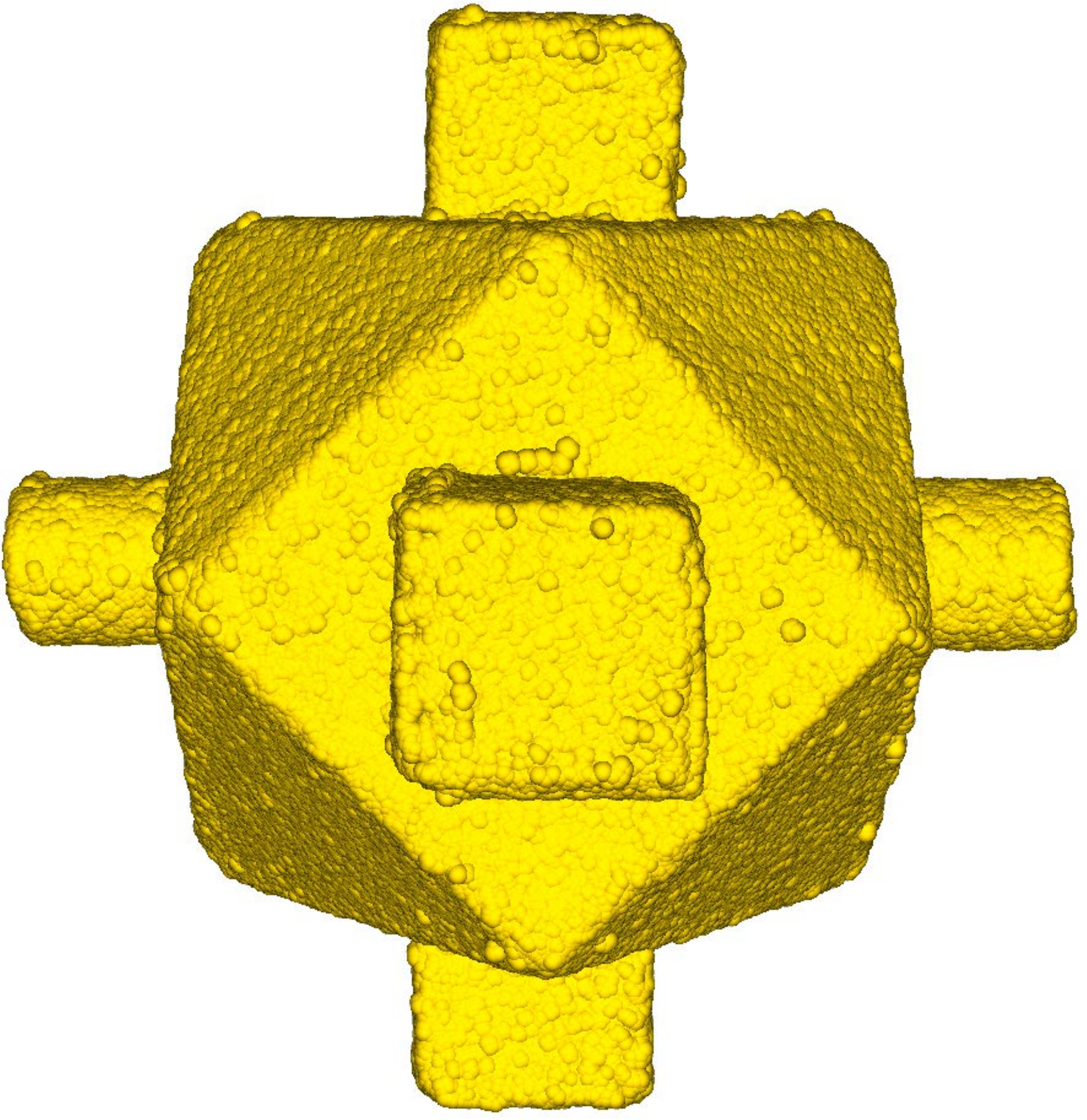}\\
        \includegraphics[width=1\textwidth]{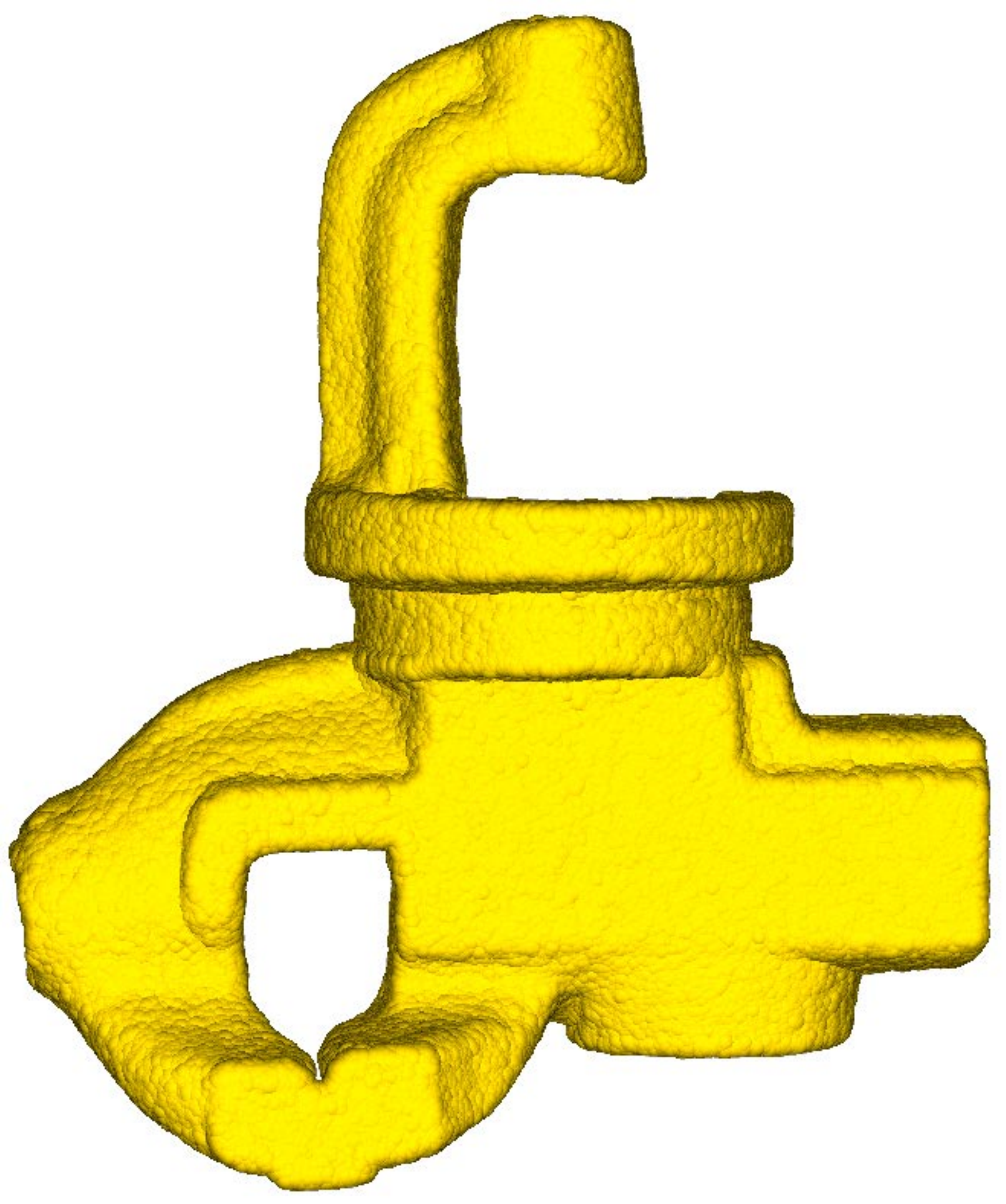}
        \end{minipage}
    }
    \subfigure[PCN]
    {
        \begin{minipage}[b]{0.08\textwidth} 
        \includegraphics[width=1\textwidth]{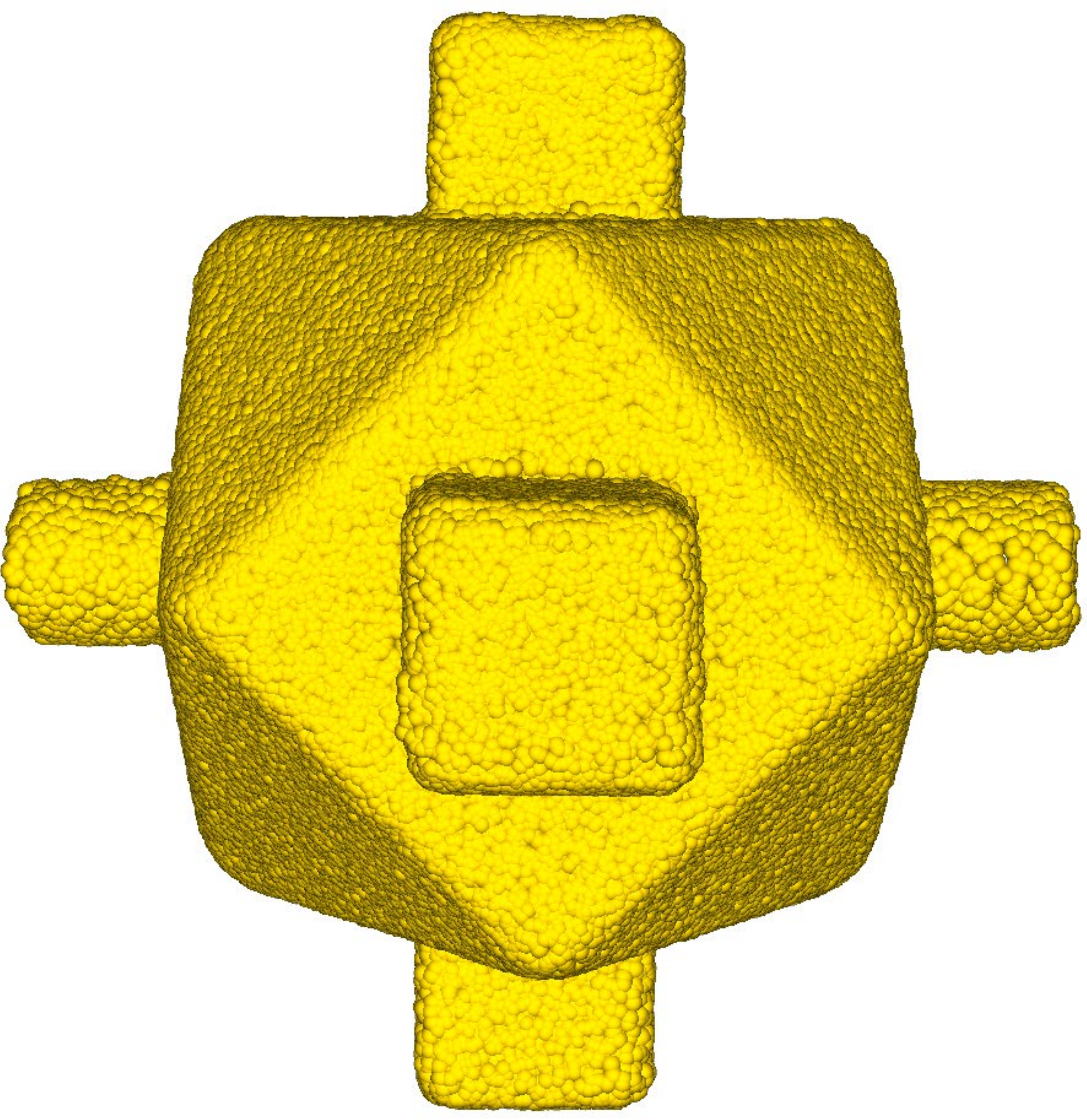}\\
        \includegraphics[width=1\textwidth]{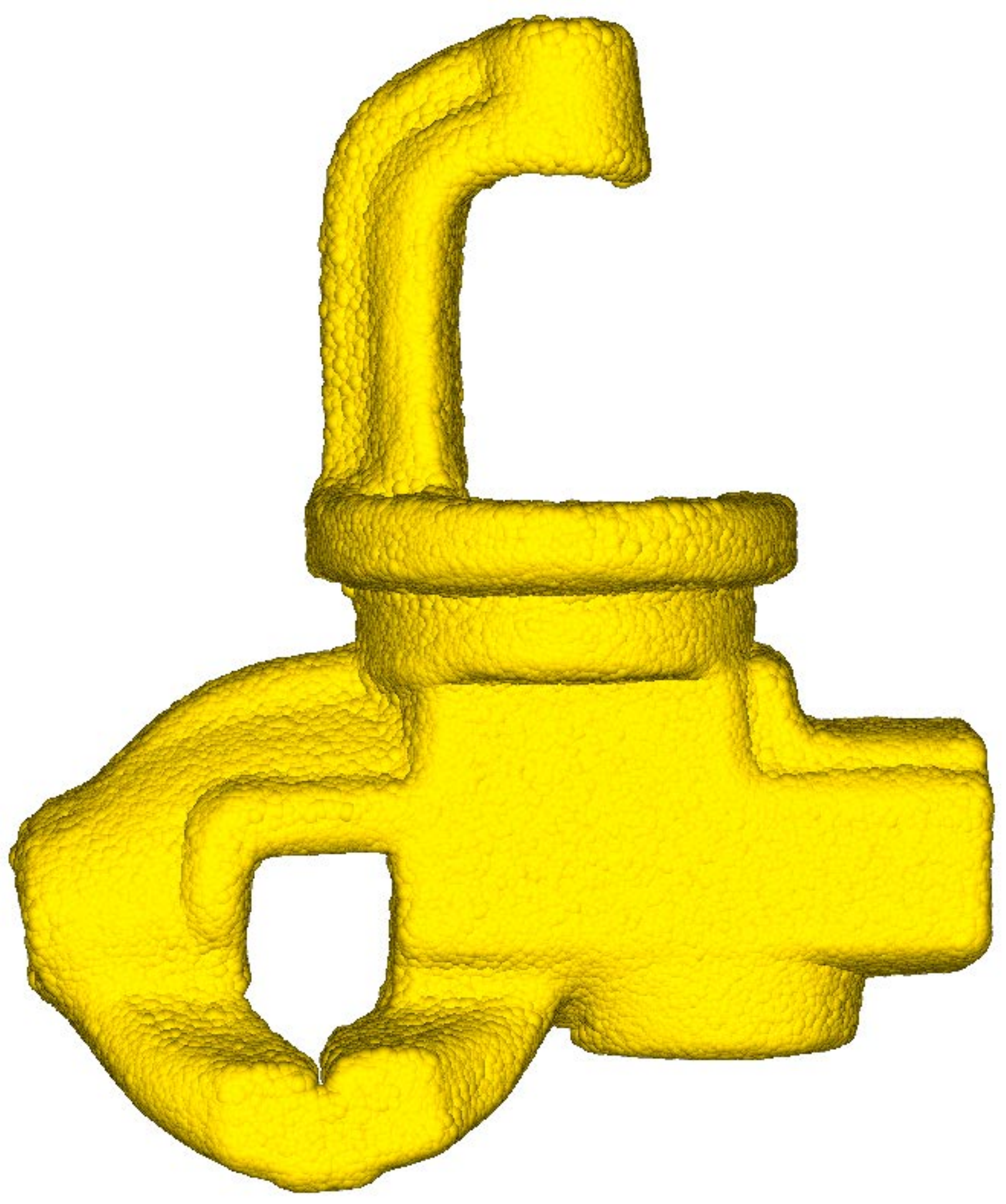}
        \end{minipage}
    }
    \subfigure[TD]
    {
        \begin{minipage}[b]{0.08\textwidth} 
        \includegraphics[width=1\textwidth]{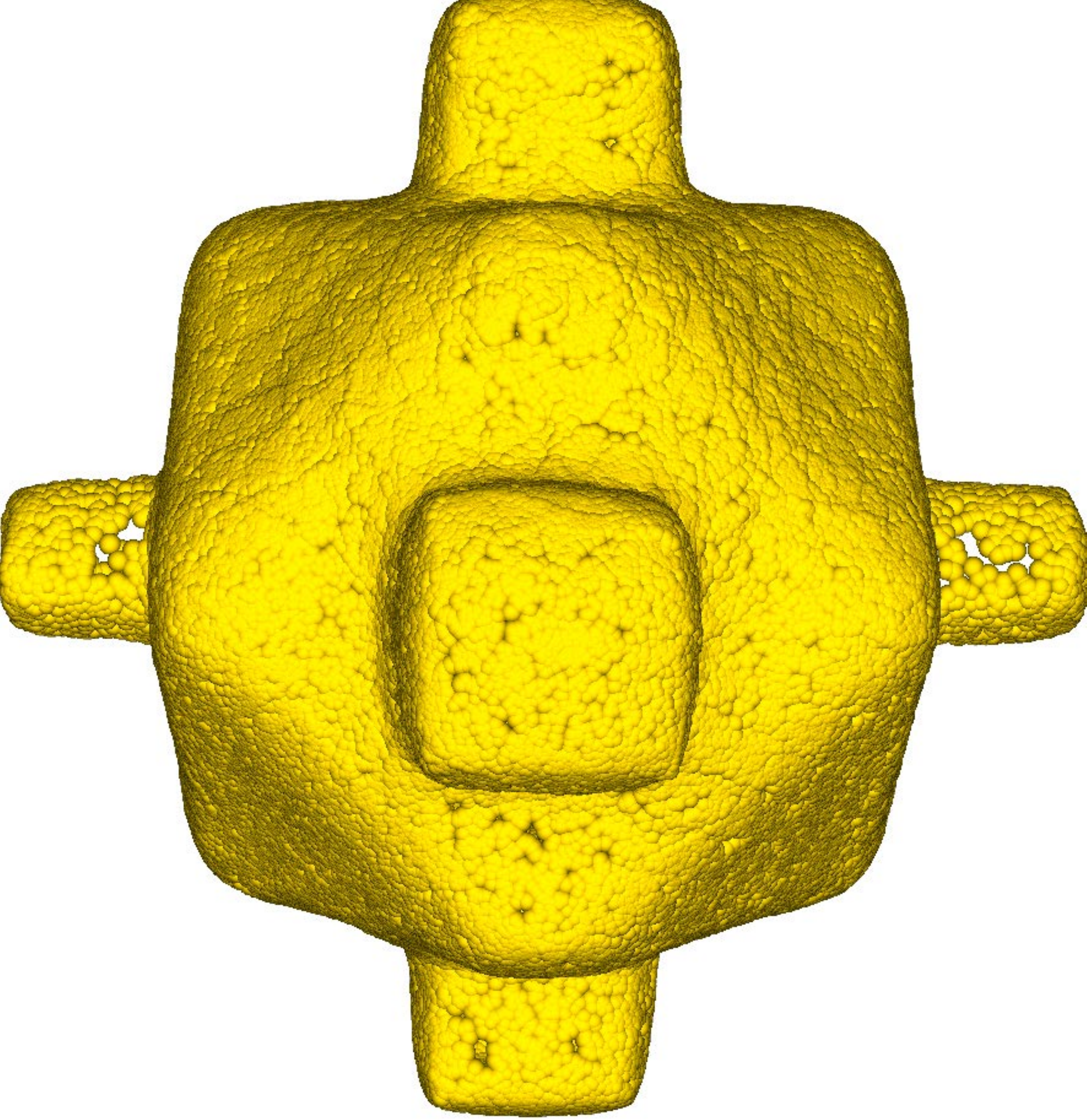}\\
        \includegraphics[width=1\textwidth]{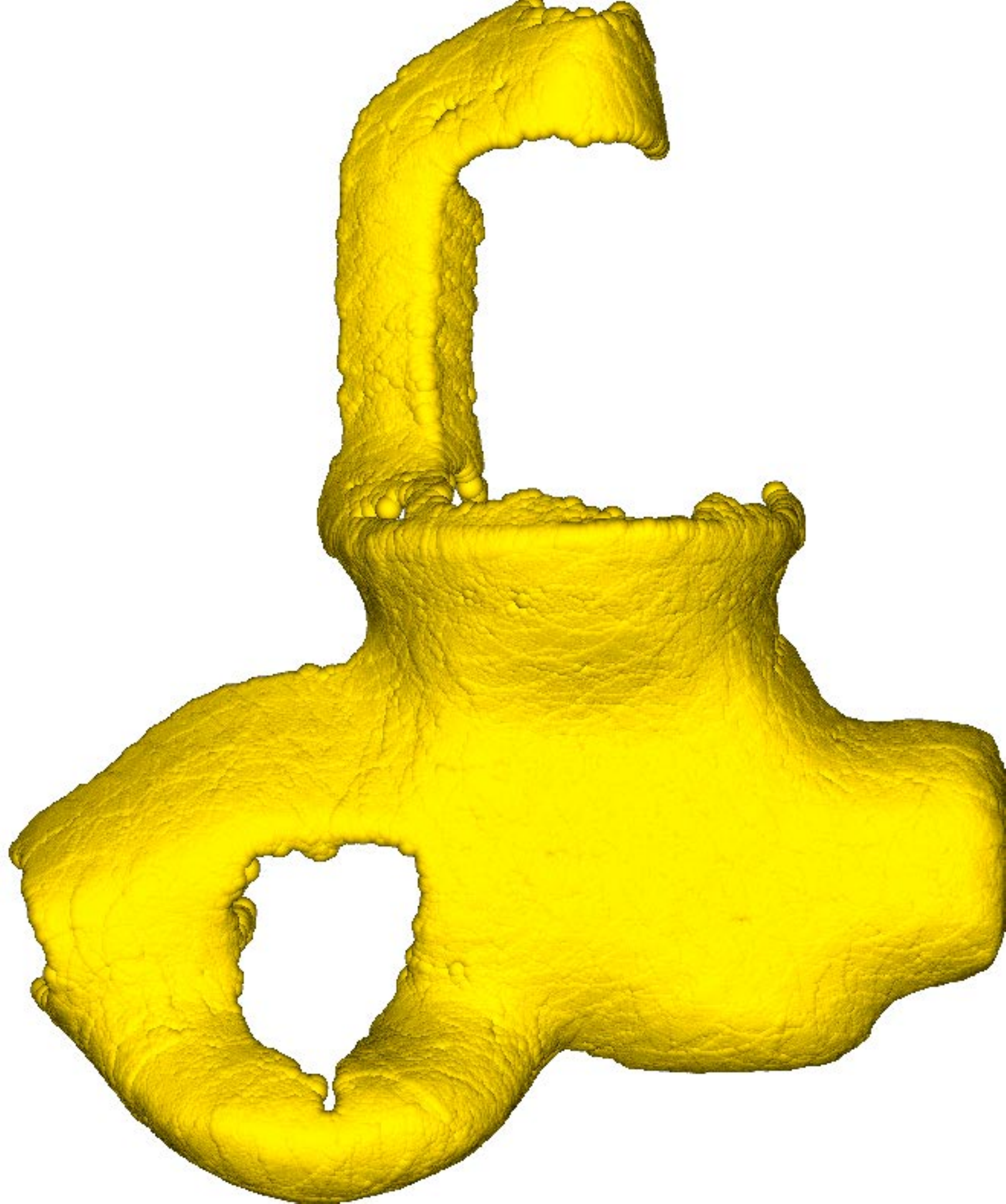}
        \end{minipage}
    }
    \subfigure[Ours]
    {
        \begin{minipage}[b]{0.08\textwidth} 
        \includegraphics[width=1\textwidth]{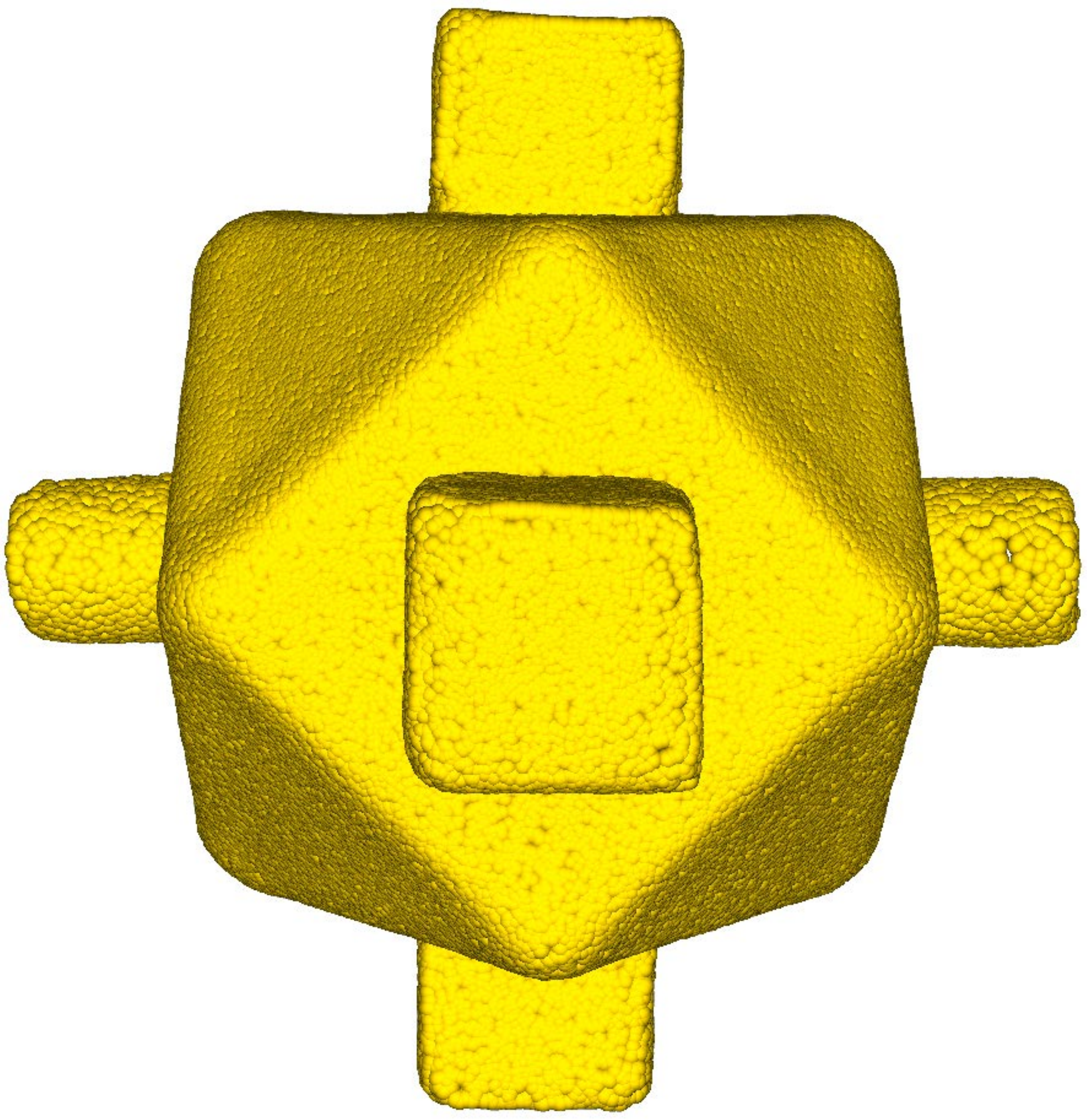}\\
        \includegraphics[width=1\textwidth]{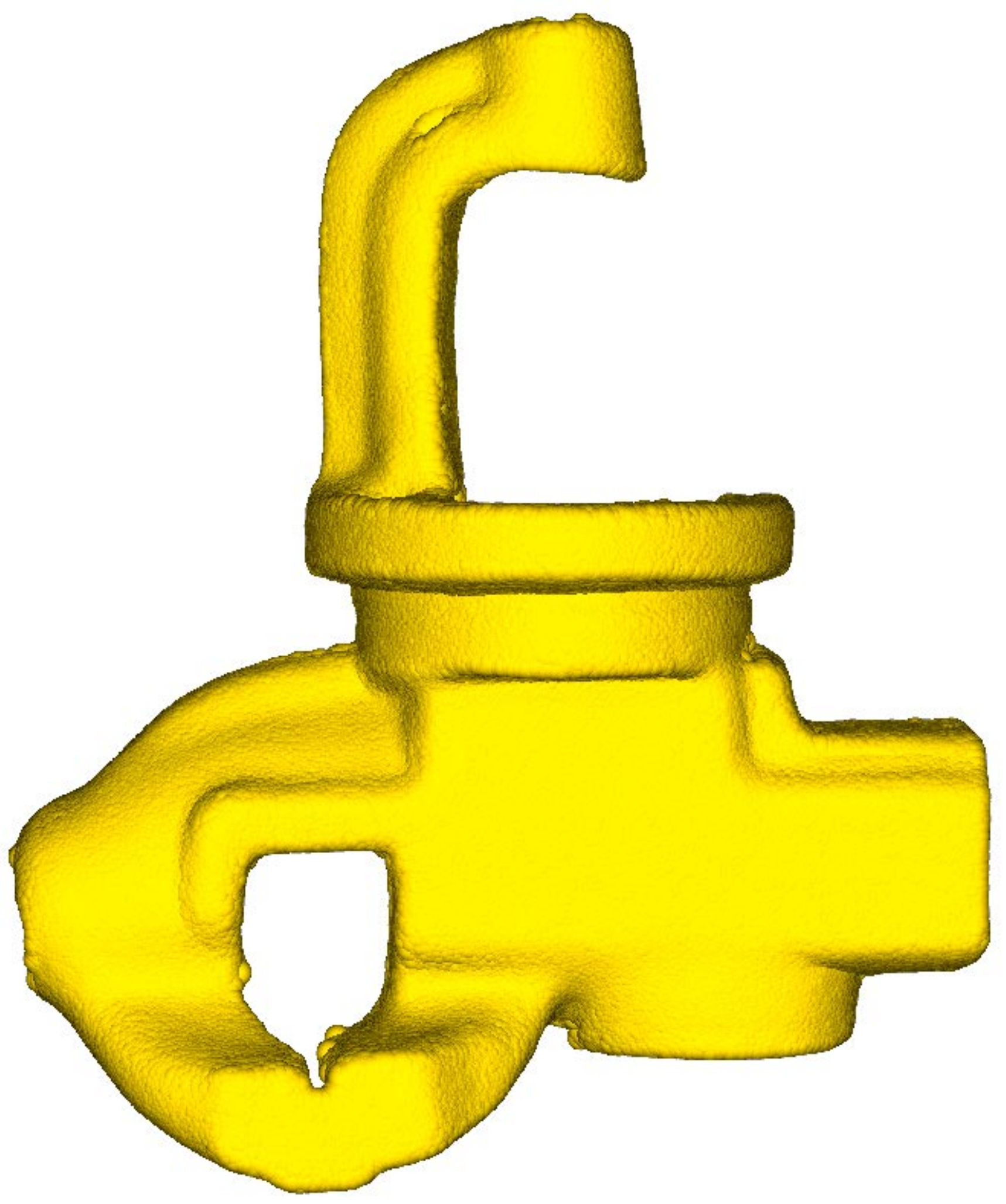}
        \end{minipage}
    }
    \caption{Filtered results of point clouds with raw noise.}
    \label{fig:real_scaned_noise_armadillo_iron}
\end{figure}

\begin{figure*}[htb!]
\subfigure[Noisy input]
    {
        \begin{minipage}[b]{0.18\textwidth} 
        \includegraphics[width=1\textwidth]{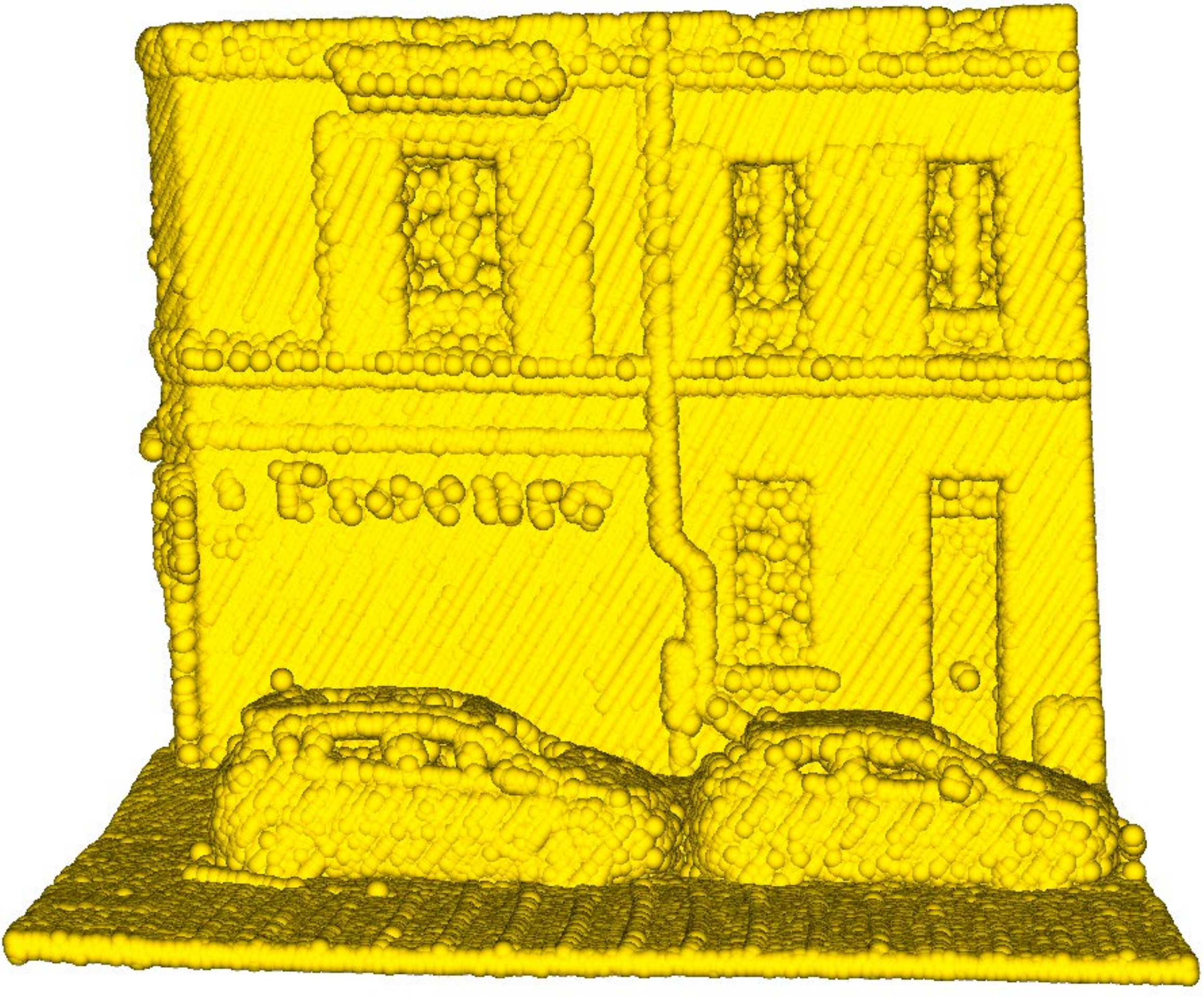}\\
        \includegraphics[width=1\textwidth]{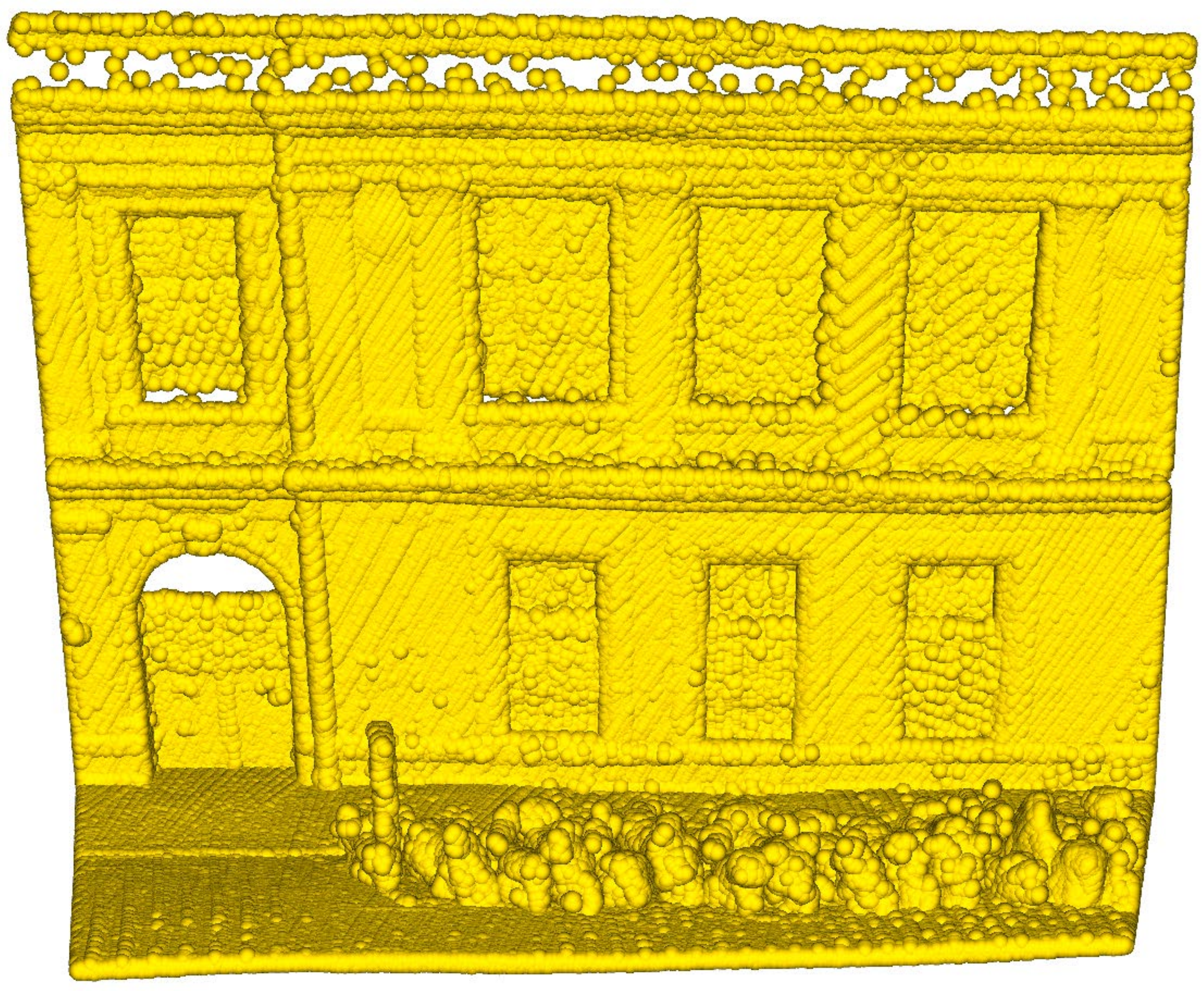}
        \end{minipage}
    }
    \subfigure[EC-Net]
    {
        \begin{minipage}[b]{0.18\textwidth} 
        \includegraphics[width=1\textwidth]{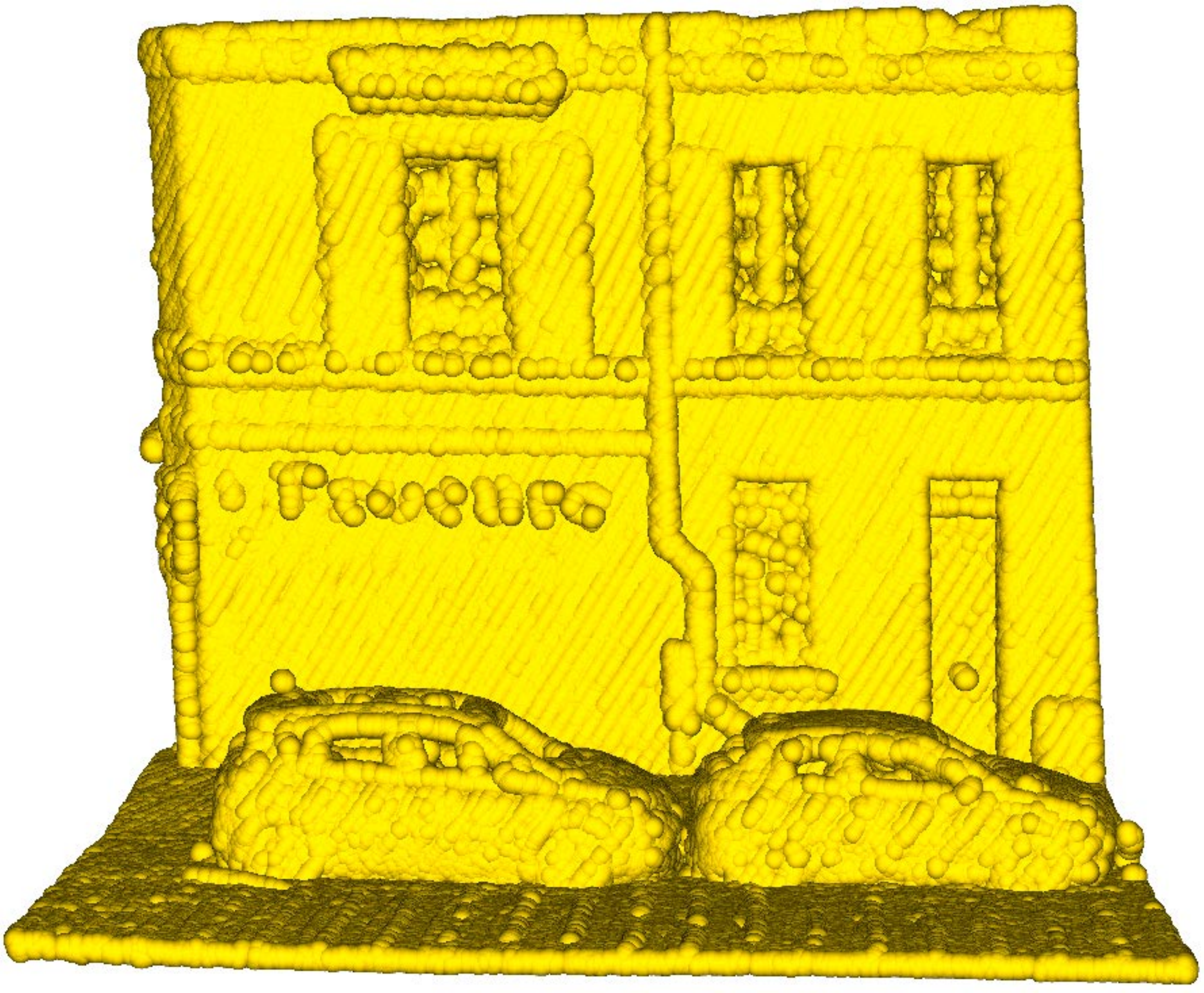}\\
        \includegraphics[width=1\textwidth]{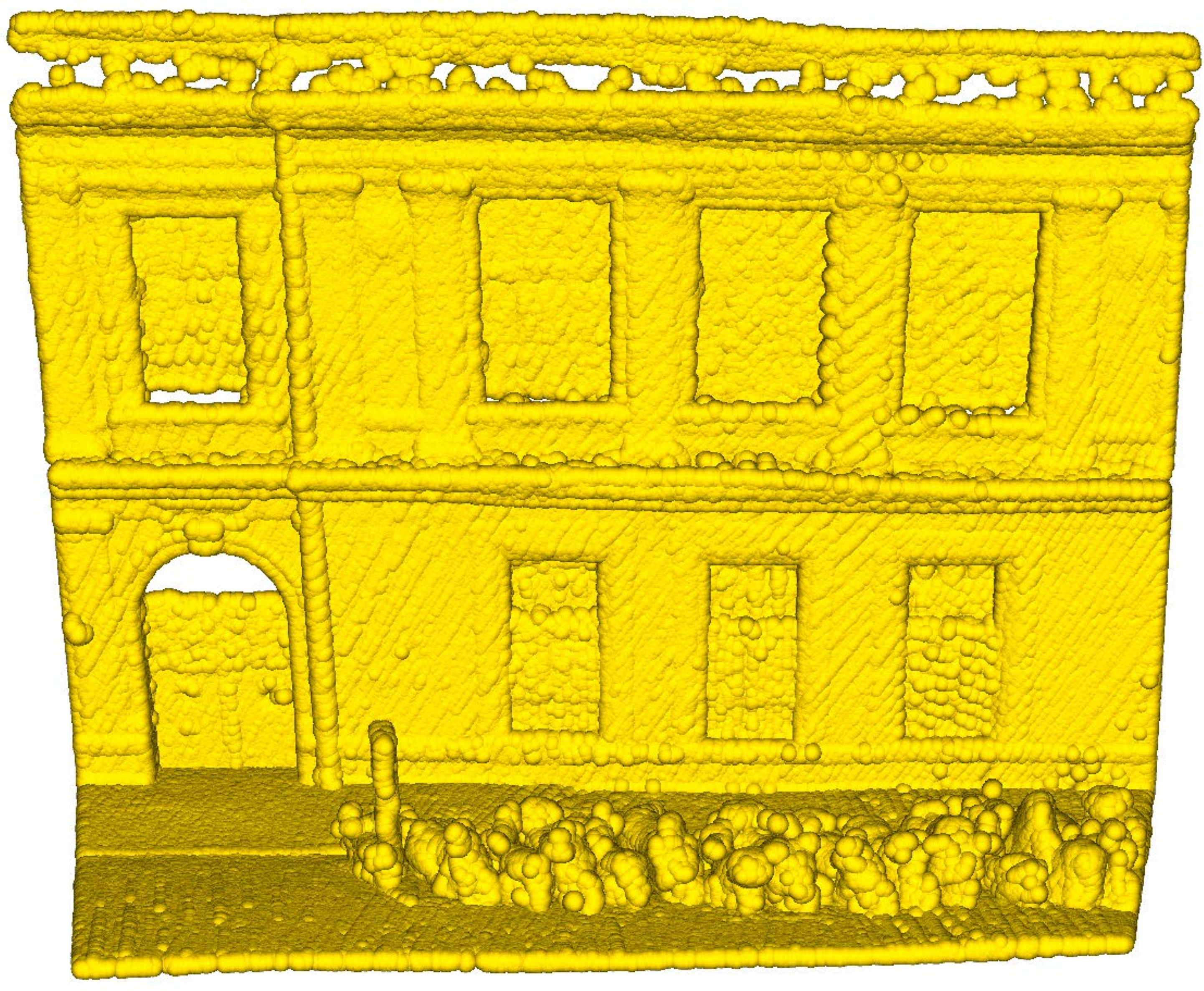}
        \end{minipage}
    }
    \subfigure[PCN]
    {
        \begin{minipage}[b]{0.18\textwidth} 
        \includegraphics[width=1\textwidth]{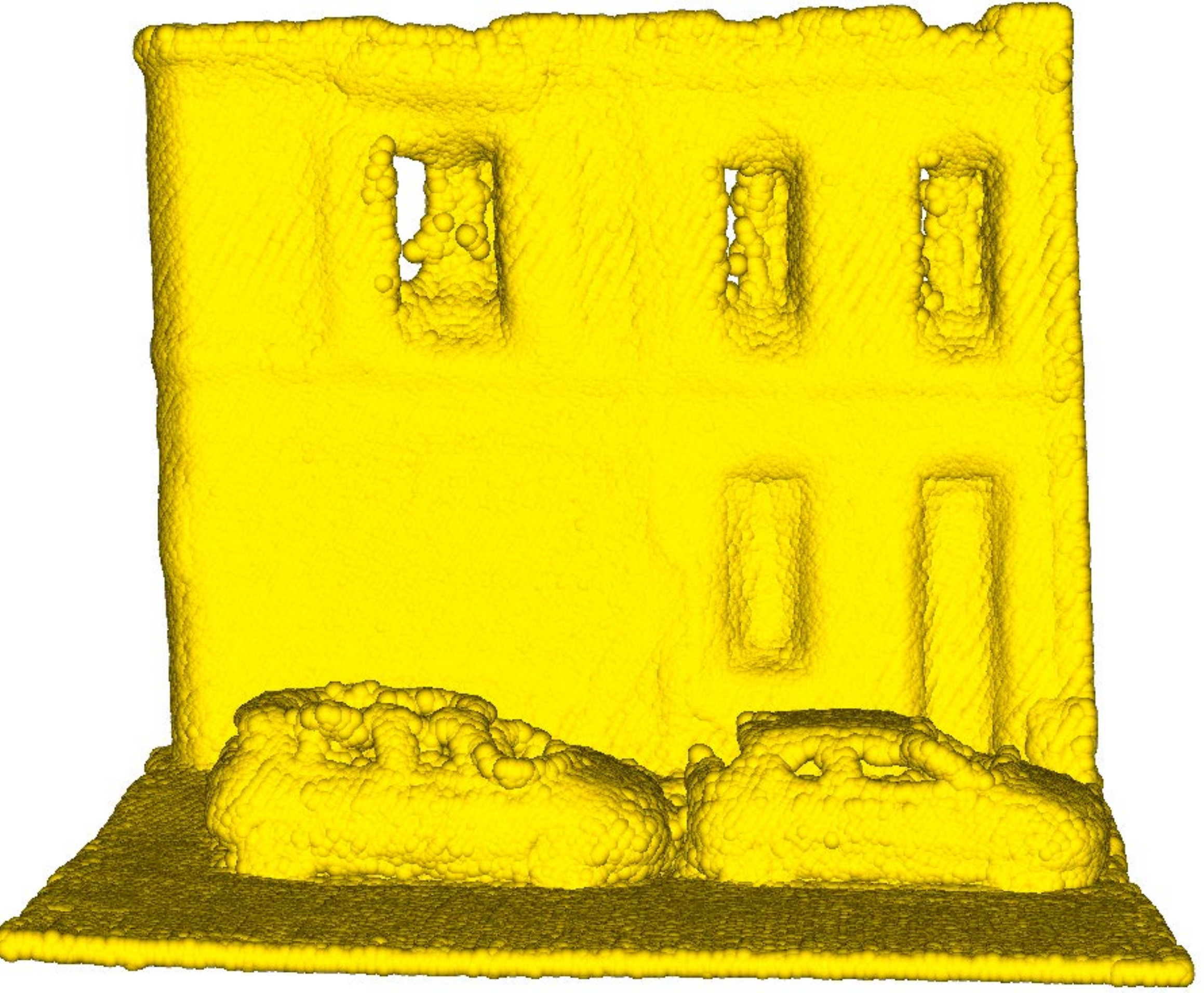}\\
        \includegraphics[width=1\textwidth]{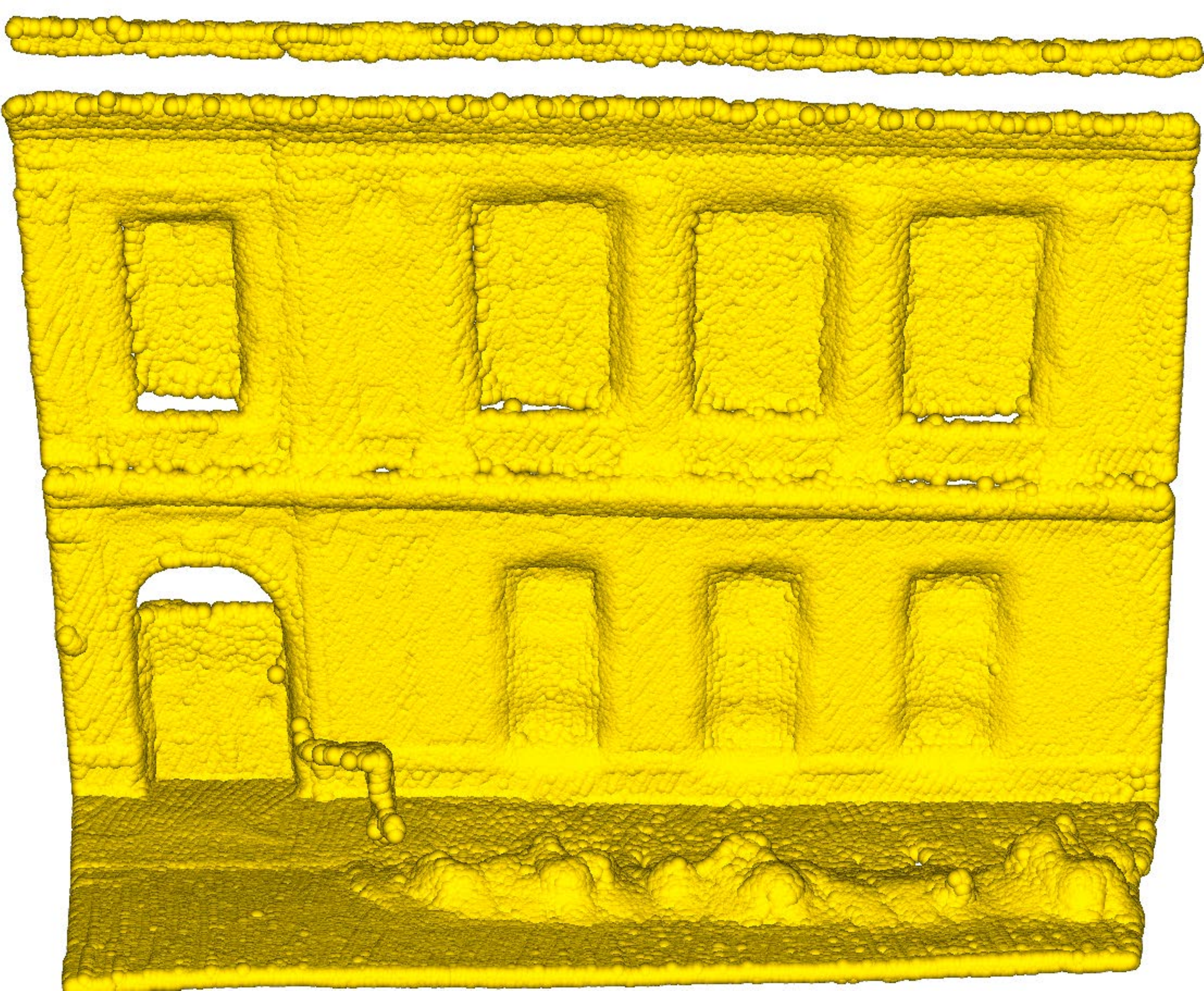}
        \end{minipage}
    }
    \subfigure[TD]
    {
        \begin{minipage}[b]{0.18\textwidth} 
        \includegraphics[width=1\textwidth]{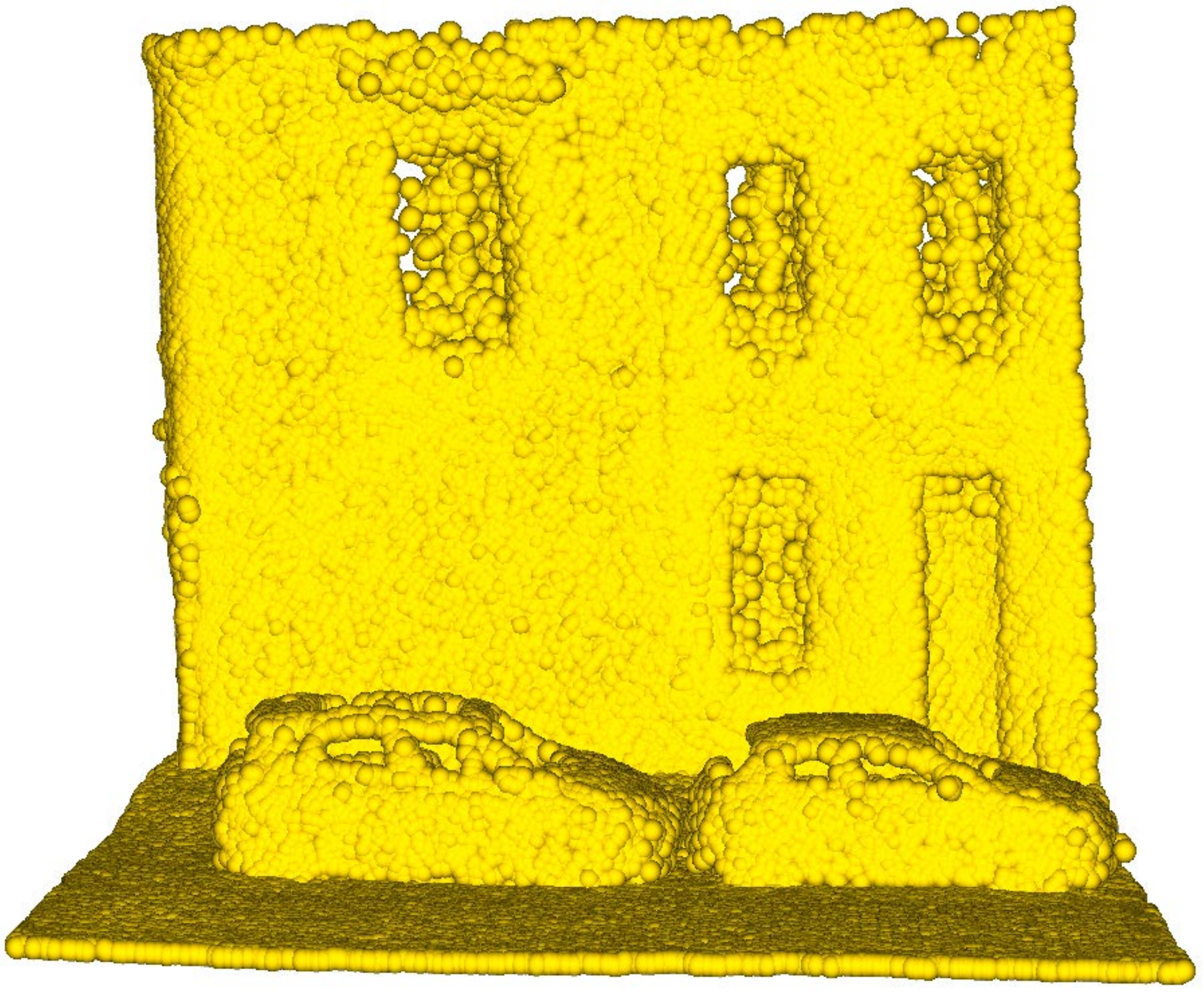}\\
        \includegraphics[width=1\textwidth]{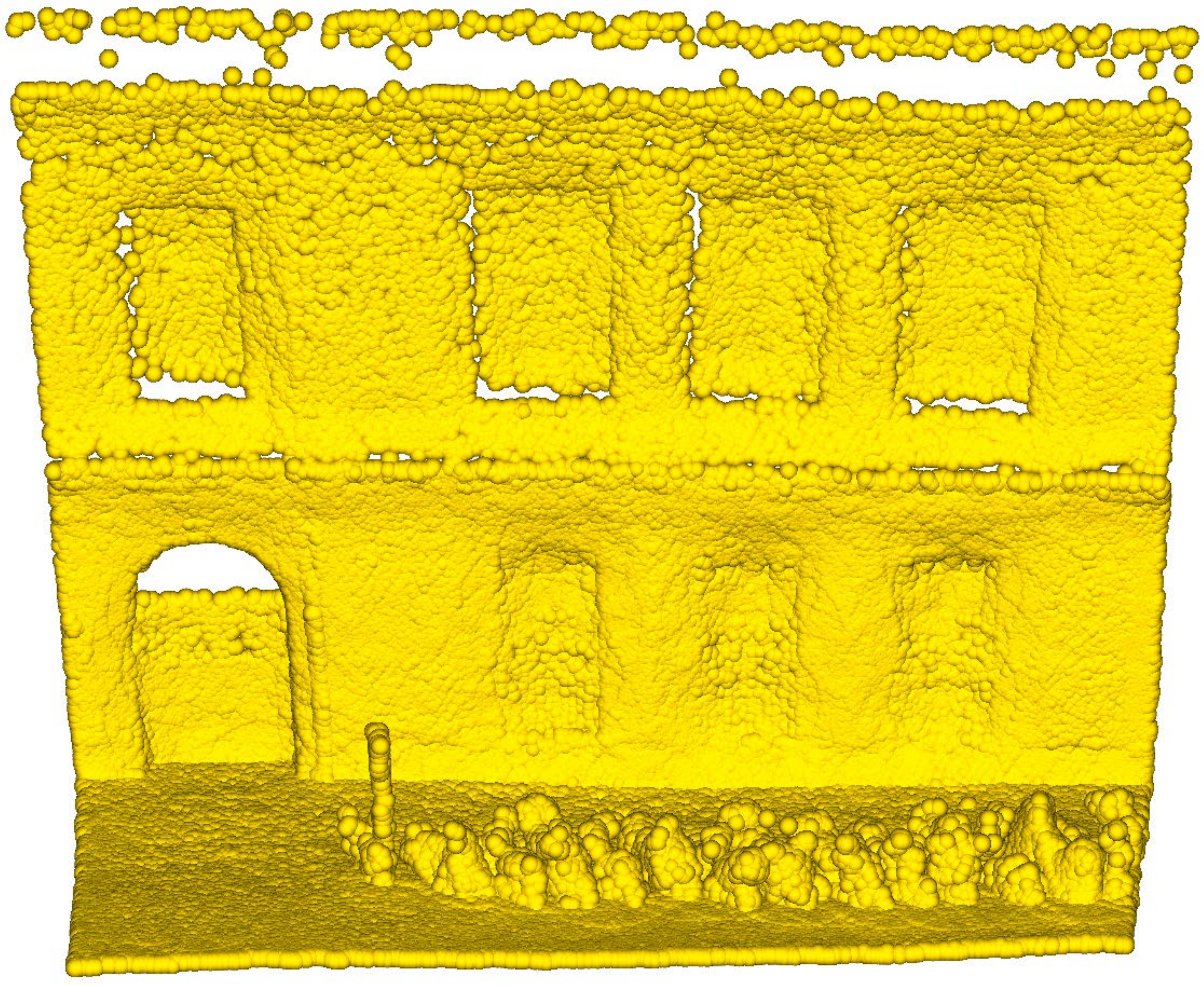}
        \end{minipage}
    }
    \subfigure[Ours]
    {
        \begin{minipage}[b]{0.18\textwidth} 
        \includegraphics[width=1\textwidth]{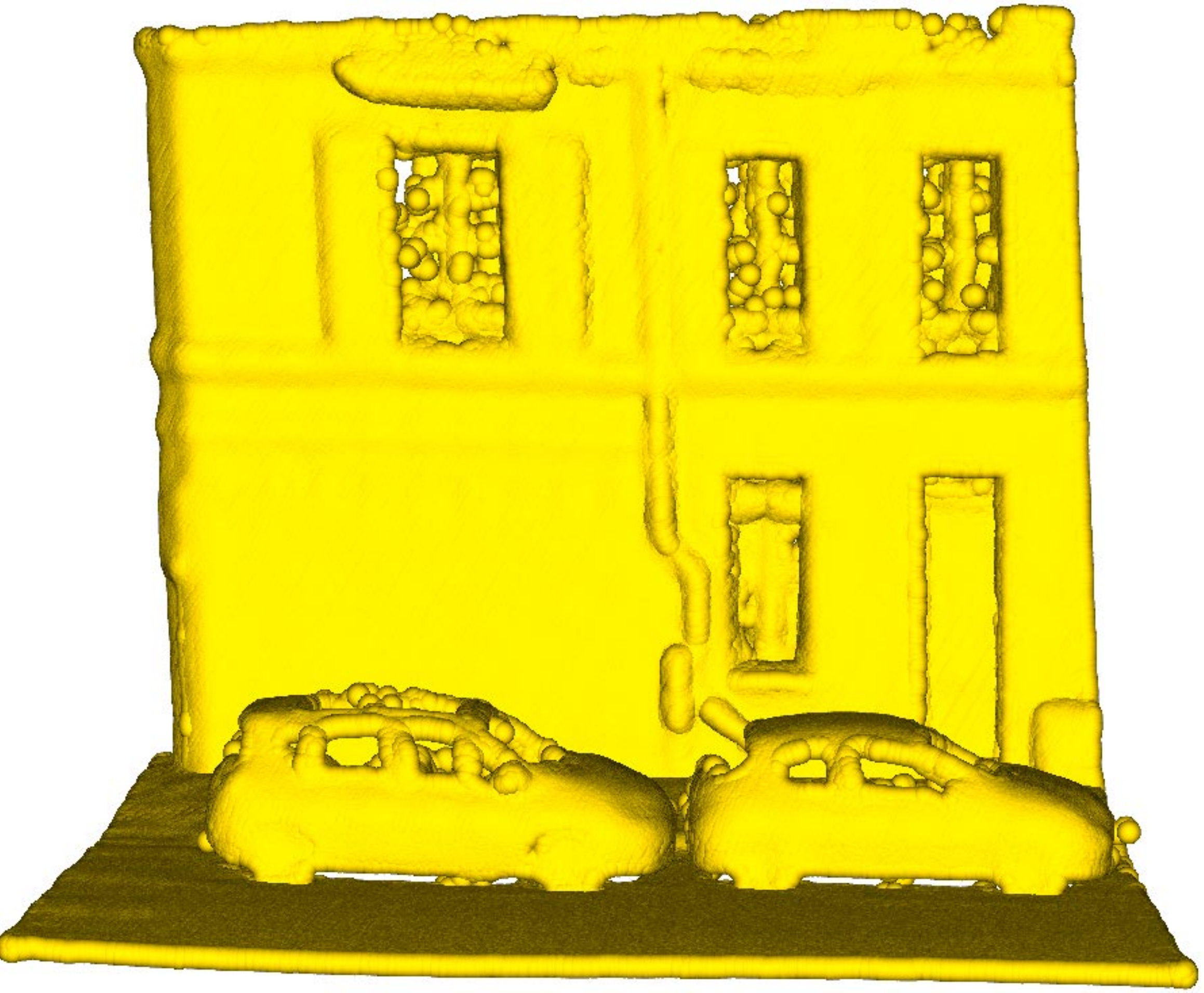}\\
        \includegraphics[width=1\textwidth]{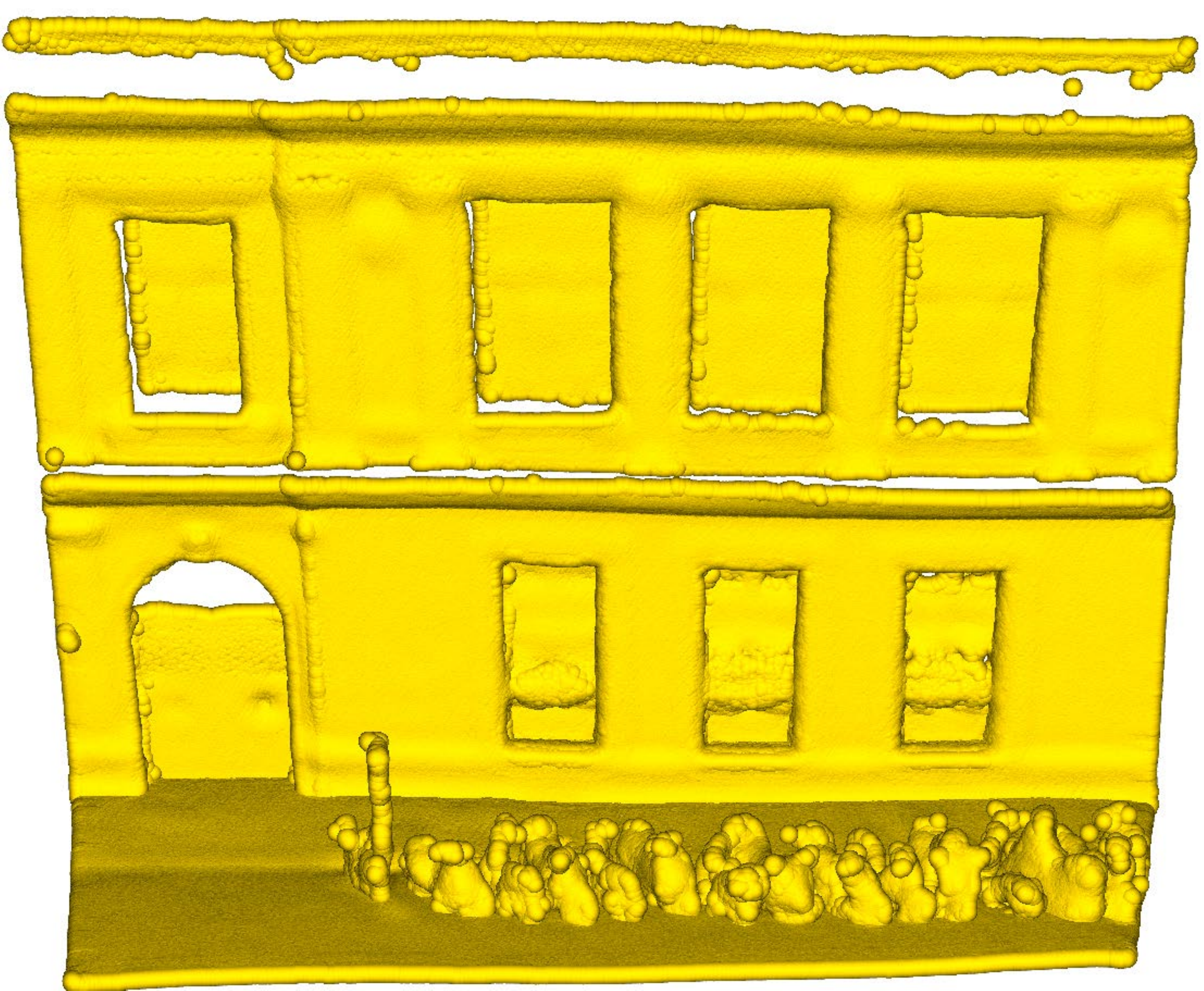}
        \end{minipage}
    }
    \caption{Results on two scanned models (top $600K$ and bottom $1000K$) from the Paris-rue-Madame Database \cite{Serna2014ICPRAM}.}
    \label{fig:Scene_1}
\end{figure*}

\begin{figure}[htb]
    \centering
    \subfigure[Noisy]
    {
        \begin{minipage}[b]{0.08\textwidth} 
        \includegraphics[width=1\textwidth]{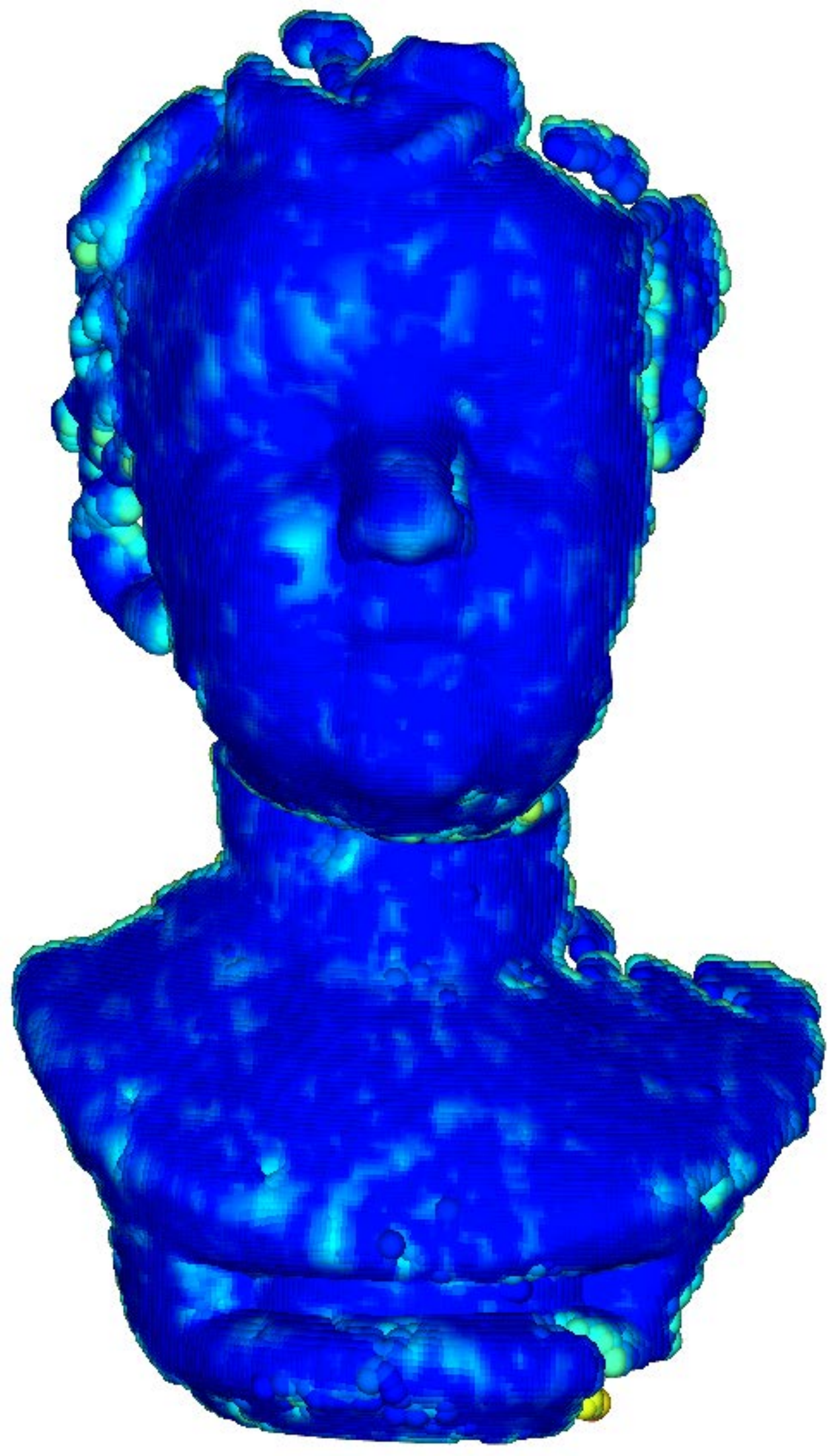}
        \includegraphics[width=1\textwidth]{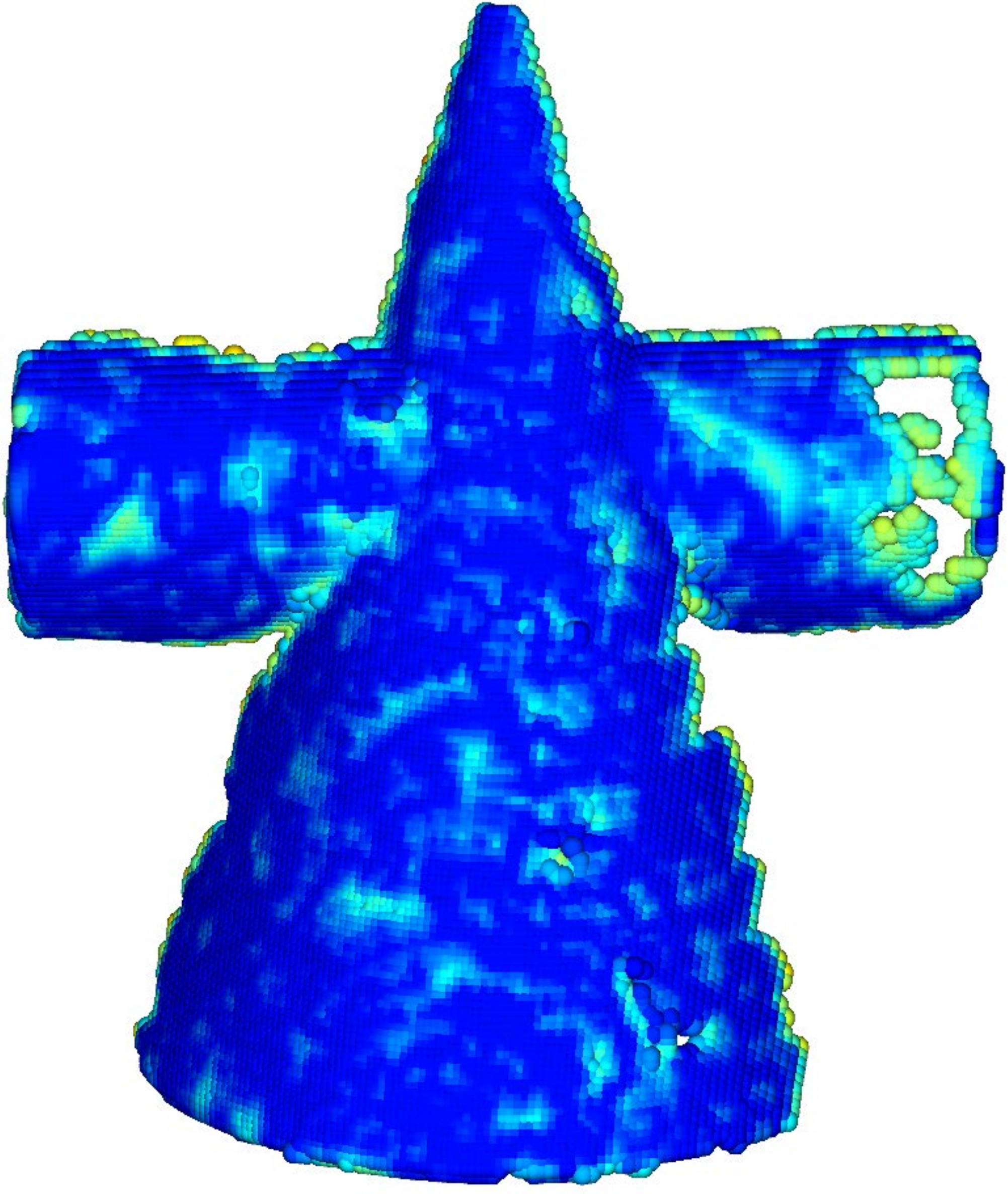}
        \includegraphics[width=1\textwidth]{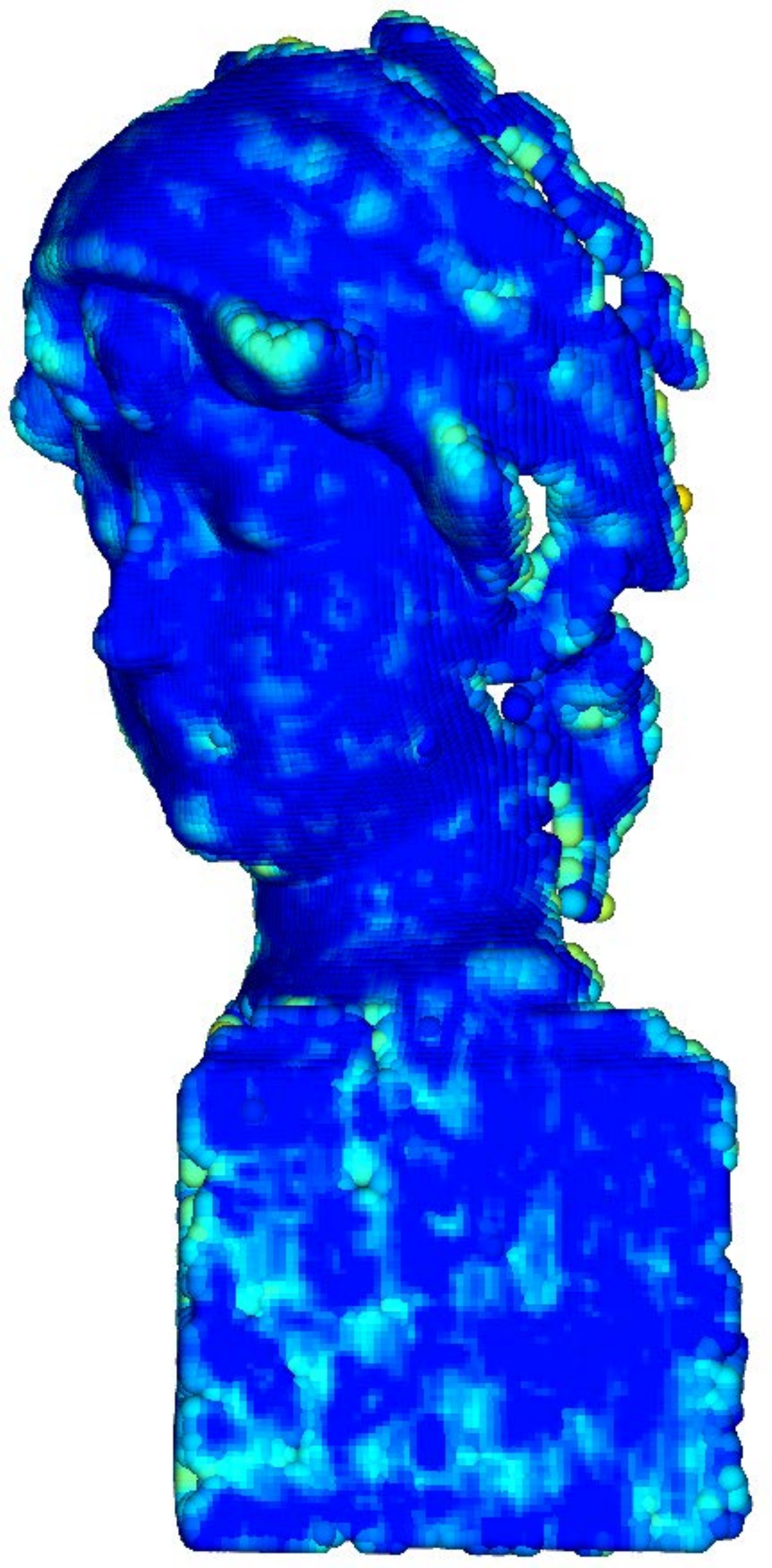}
        \end{minipage}
    }
    \subfigure[PCN-R]
    {
        \begin{minipage}[b]{0.08\textwidth} 
        \includegraphics[width=1\textwidth]{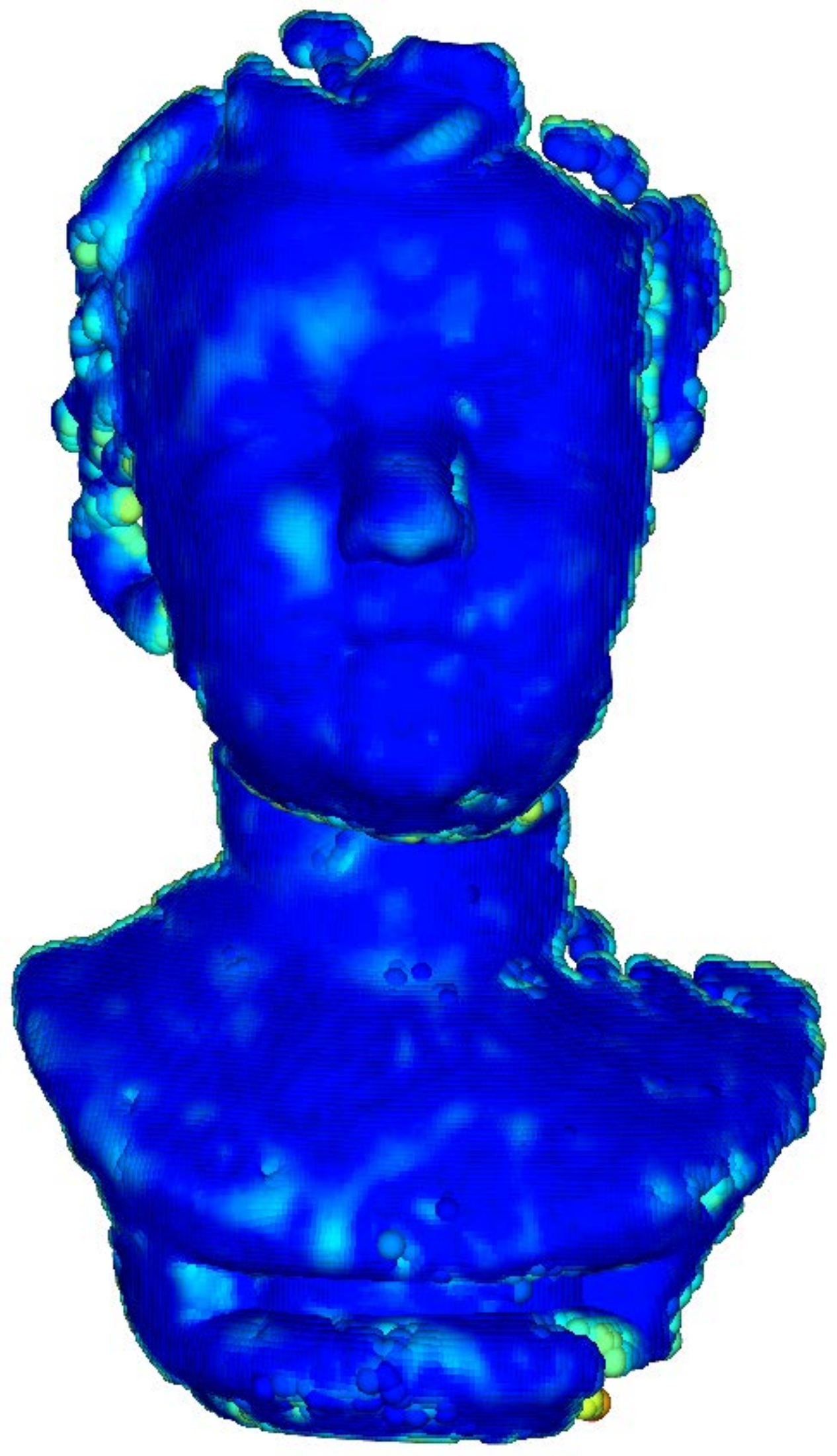}
        \includegraphics[width=1\textwidth]{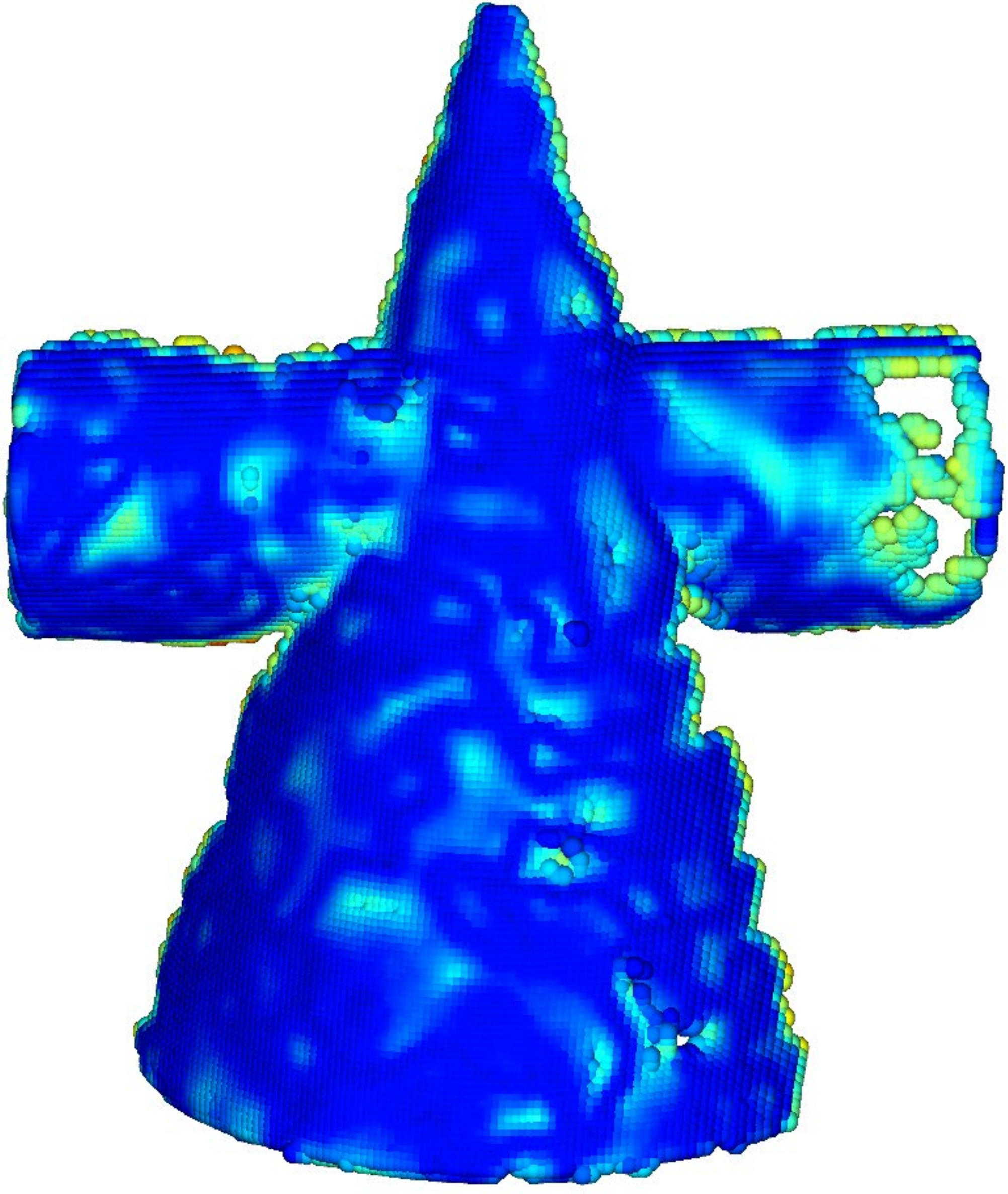}
        \includegraphics[width=1\textwidth]{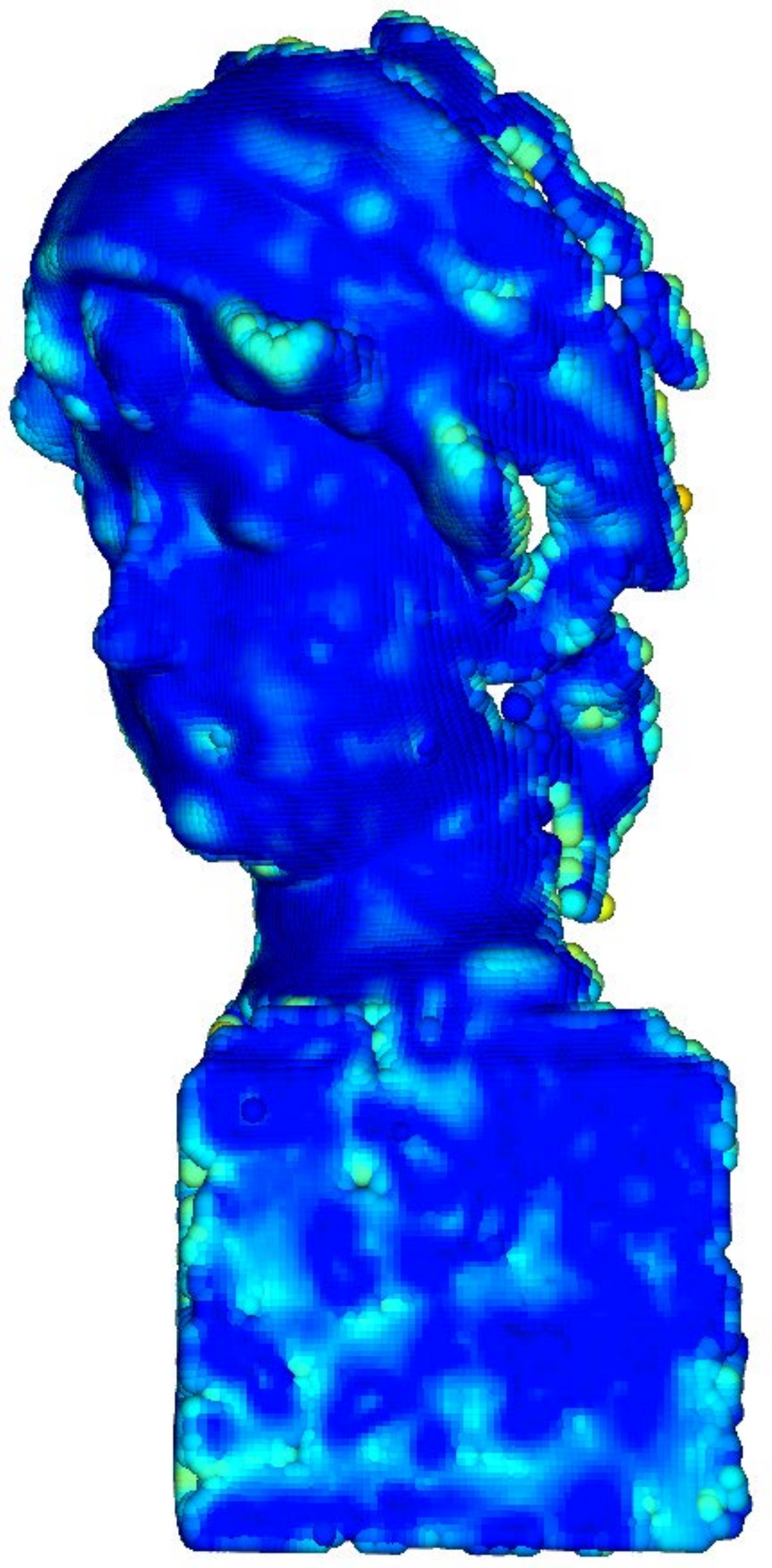}
        \end{minipage}
    }
    \subfigure[Ours]
    {
        \begin{minipage}[b]{0.08\textwidth} 
        \includegraphics[width=1\textwidth]{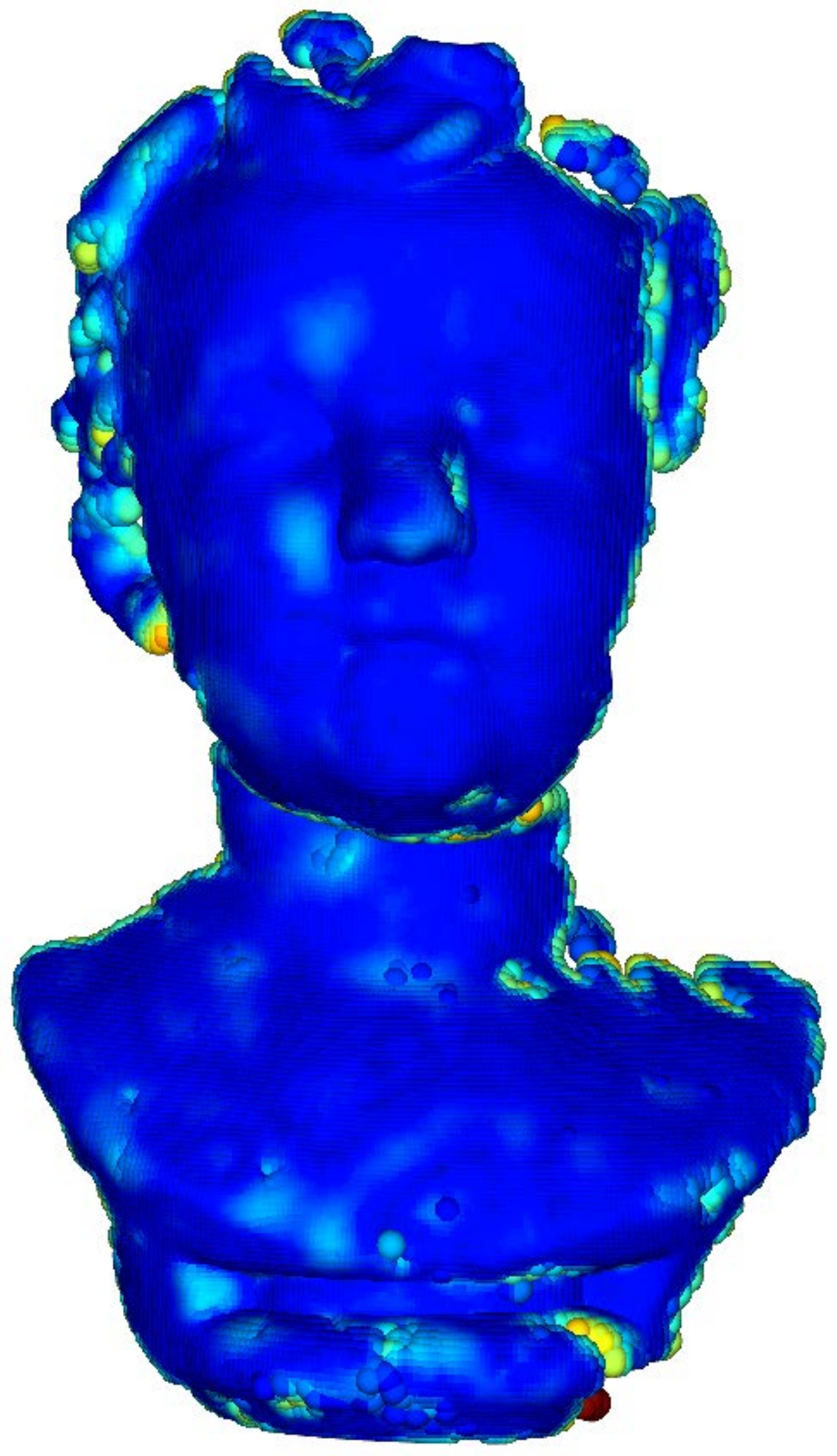}
        \includegraphics[width=1\textwidth]{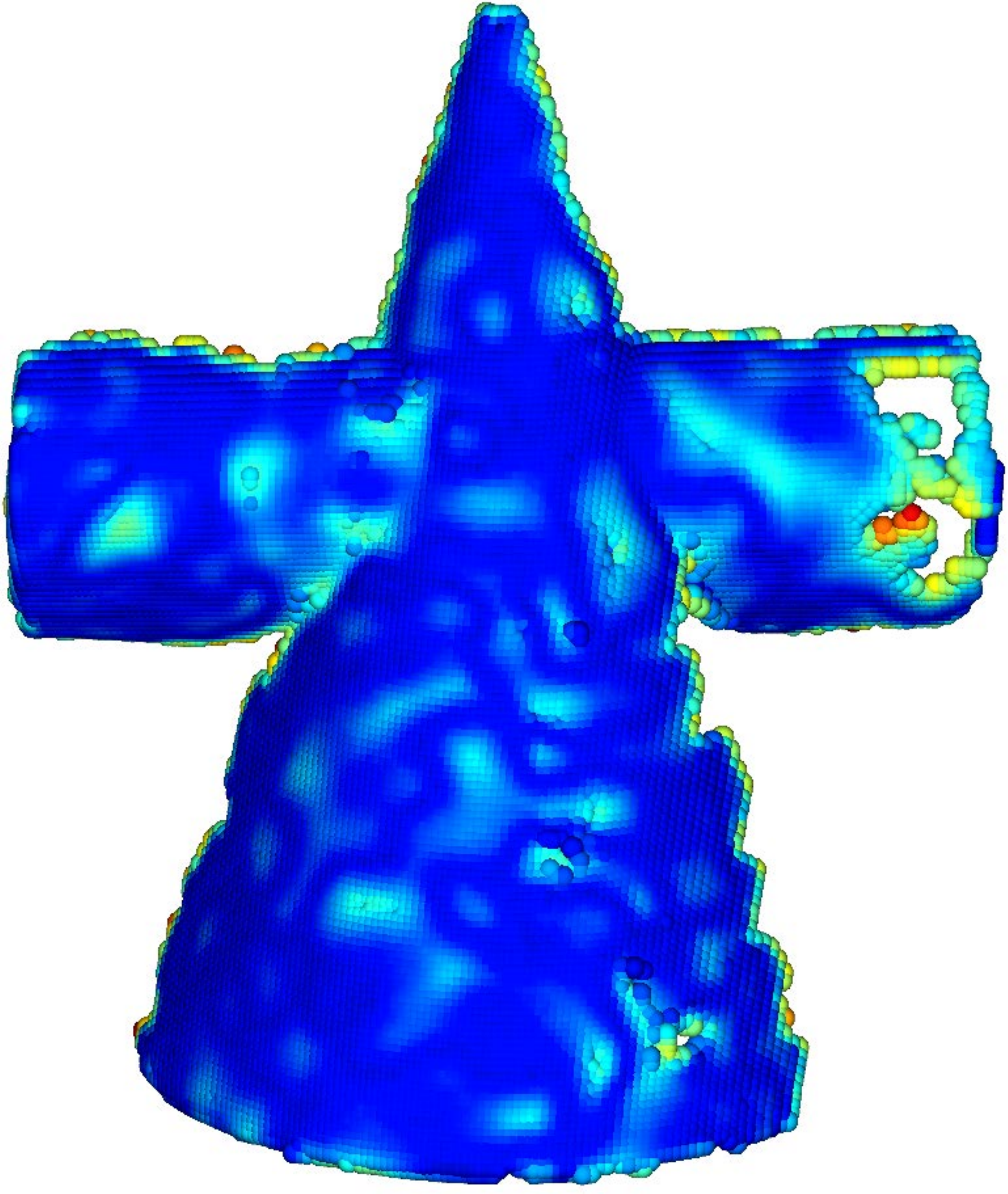}
        \includegraphics[width=1\textwidth]{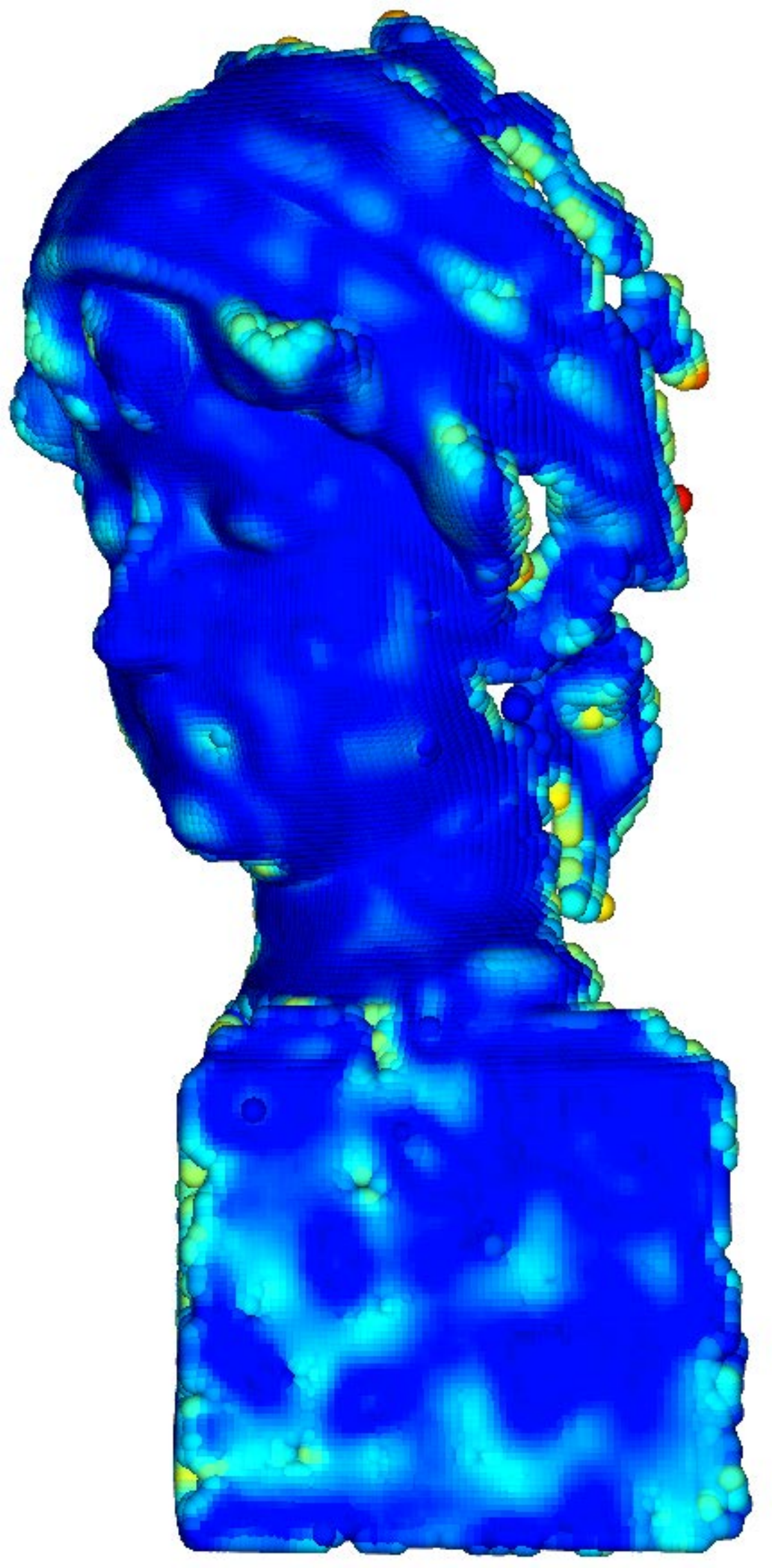}
        \end{minipage}
    }
    \subfigure[Ours-R]
    {
        \begin{minipage}[b]{0.08\textwidth} 
        \includegraphics[width=1\textwidth]{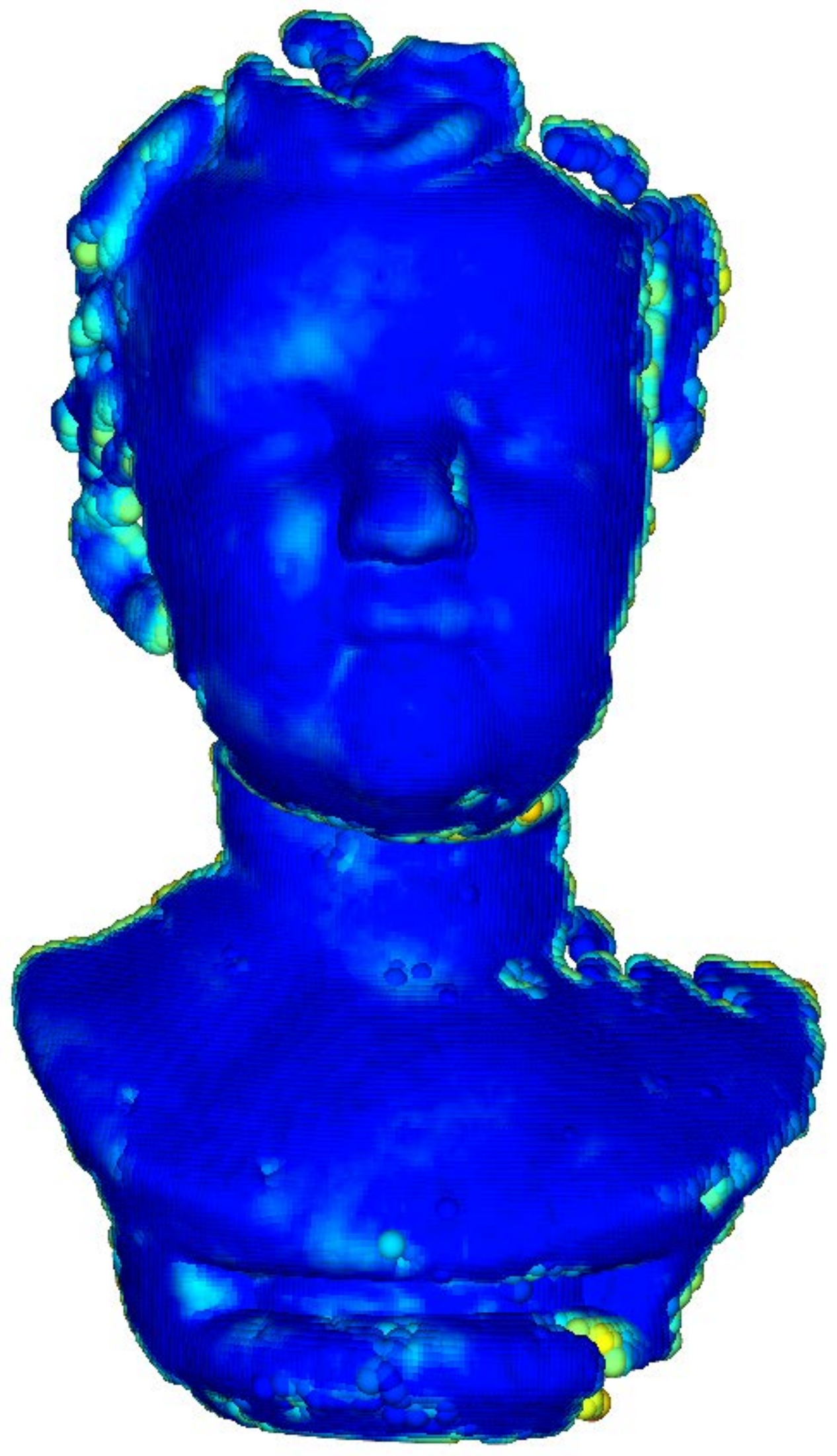}
        \includegraphics[width=1\textwidth]{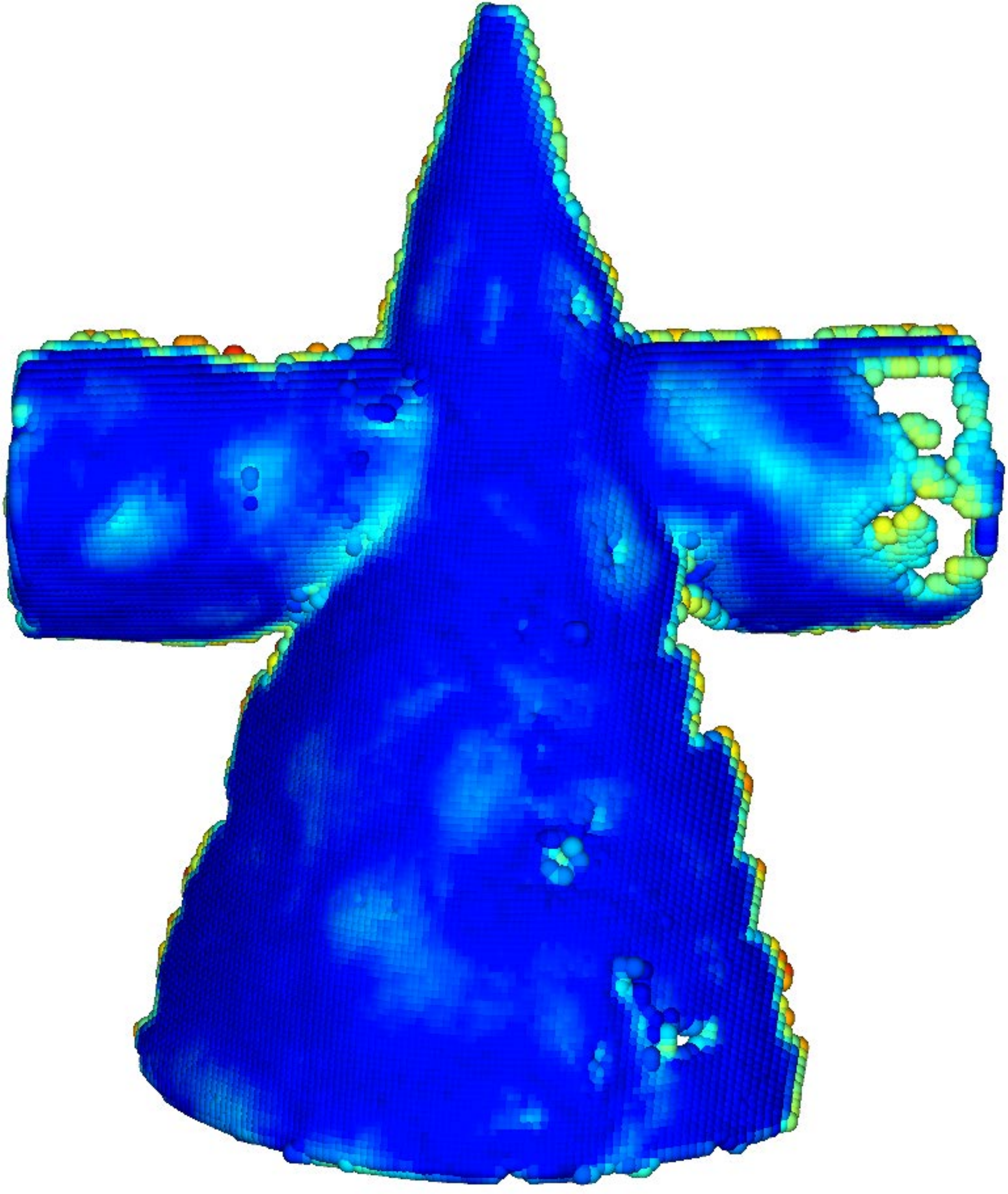}
        \includegraphics[width=1\textwidth]{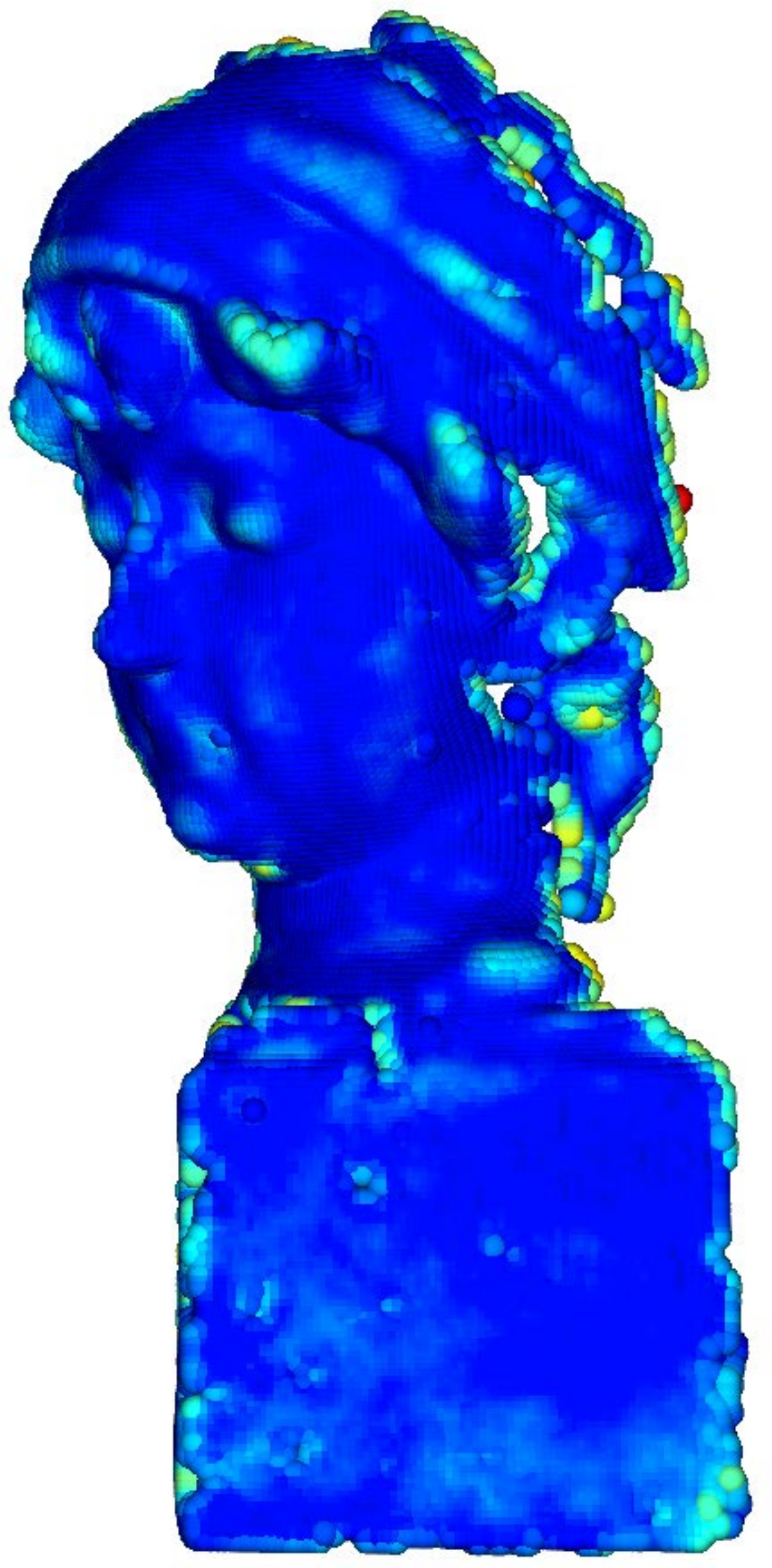}
        \end{minipage}
    }
     \subfigure
    {
        \begin{minipage}[b]{0.04\textwidth} 
        \includegraphics[width=1\textwidth]{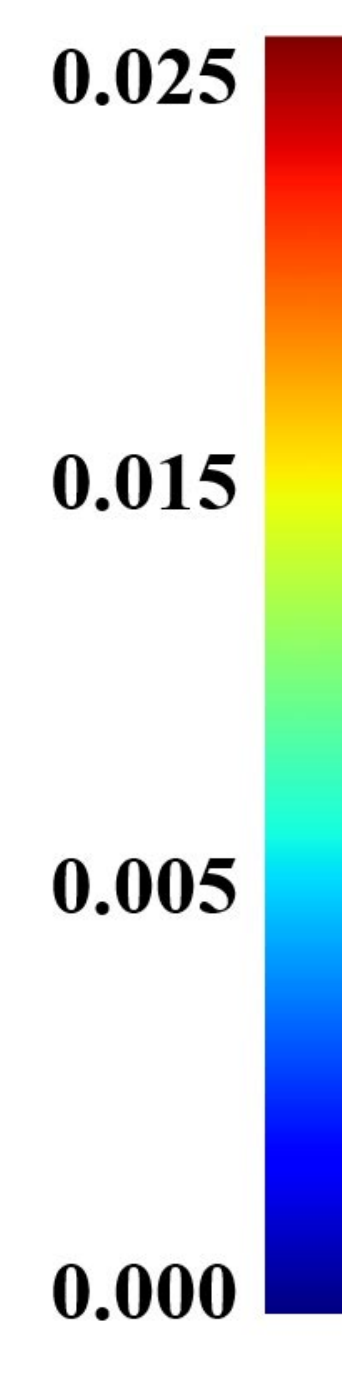}
        \end{minipage}
    }
    \caption{Visual Comparisons with PCN on the KinectV1 and KinectV2 datasets.}
    \label{fig:real_scan_visual}
\end{figure}

\noindent\textbf{Point clouds with raw noise.} We also evaluate our Pointfilter on raw scanned point clouds corrupted with raw noise. Since the ground truth models of these raw scanned point sets are not available, we demonstrate the visual comparisons with other methods, as suggested by previous techniques \cite{Lu2018TVCG}. \textit{Notice that we do not re-train our Pointfilter for the type of raw noise (except Fig. \ref{fig:real_scan_visual} and Table \ref{tab:real_scan_errors}).}  From Fig. \ref{fig:rawfacefiltering}, we see that our results are better than the state-of-the-art techniques. Besides noise removal, Pointfilter is capable of retaining sharp features which are marked by yellow box and black arrows in Fig. \ref{fig:rawfacefiltering}. Fig. \ref{fig:pyramid_sharp_feature} shows a virtually scanned point cloud model. Compared with other filtering methods, Pointfilter still produces higher quality results, in terms of preserving sharp edges. Fig. \ref{fig:Nefertiti} shows that our approach induces a better enhancement on the surface reconstruction quality, in terms of preserving sharp features. Fig. \ref{fig:real_scan_visual} shows  Pointfilter results on KinectV1 dataset \cite{wang2016mesh}. For fair comparisons, we re-train both PCN and our Pointfilter according to the training/test set configuration in this dataset. To further evaluate Pointfilter, we also compare the re-trained version and the original version trained by the synthetic noise. As shown in Fig. \ref{fig:real_scan_visual}, our re-trained version of Pointfilter (Fig. \ref{fig:real_scan_visual} (d)) is better than the retrained PCN, and produces better results than the originally trained Pointfilter. In addition, we also test our Pointfilter on scene-level models from the Paris-rue-Madame Database \cite{Serna2014ICPRAM} (see Fig. \ref{fig:Scene_1}). As we can see from these figures, our Pointfilter is still able to produce better results. More filtering results for raw noise are demonstrated in Fig. \ref{fig:real_scaned_noise_armadillo_iron}.

\begin{figure}[htb]
    \centering
    \subfigure[Noisy input]{\includegraphics[width=0.1\textwidth]{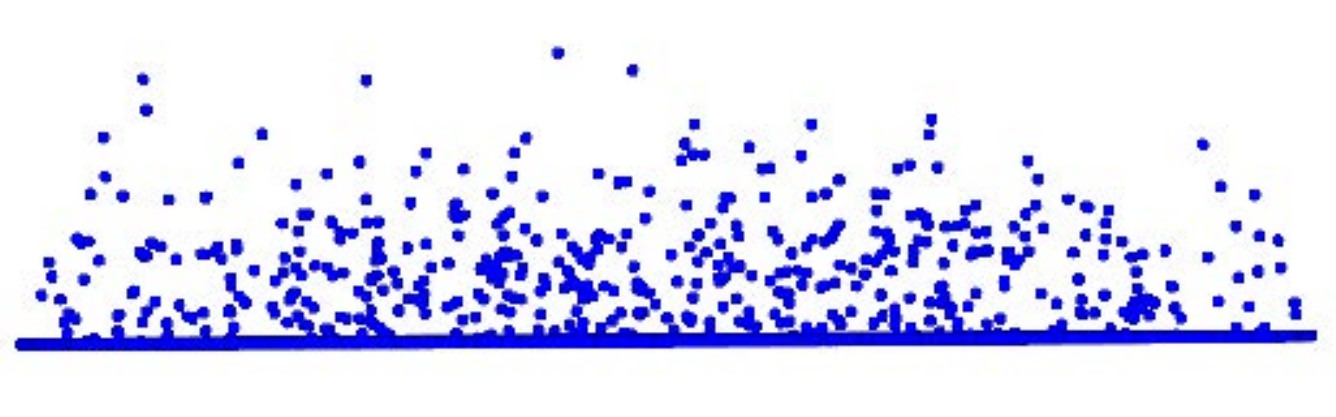}}
    \hfill
    \subfigure[WLOP]{\includegraphics[width=0.1\textwidth]{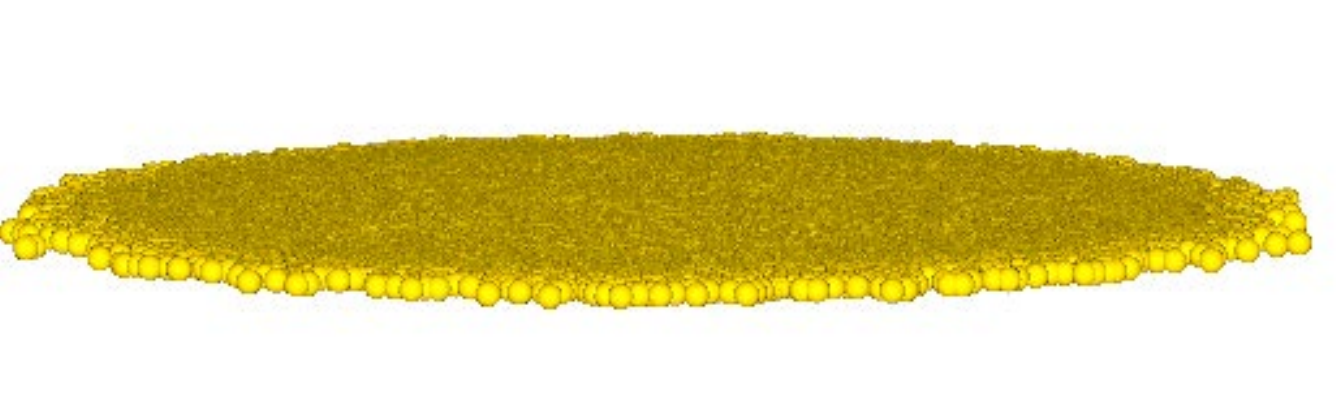}}
    \hfill
    \subfigure[CLOP]{\includegraphics[width=0.1\textwidth]{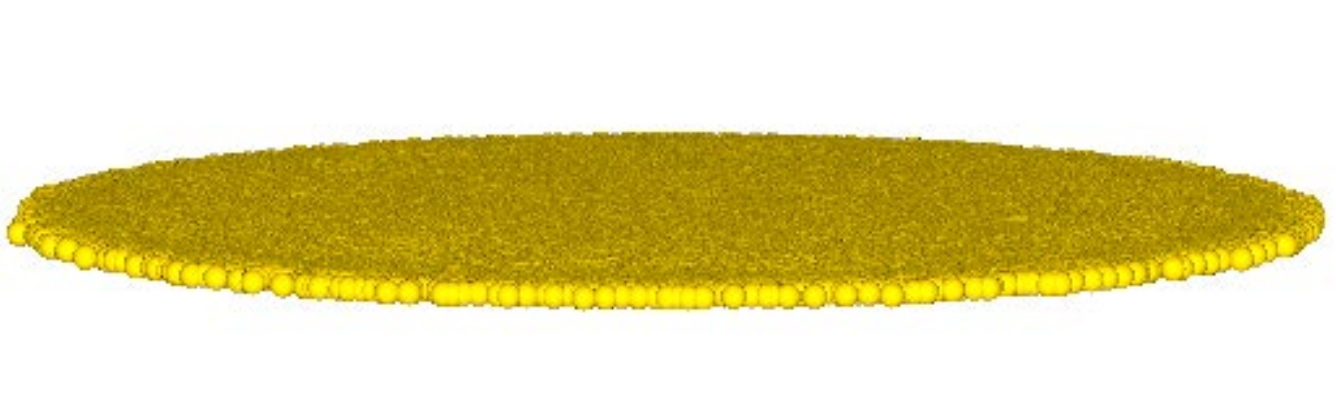}}
    \hfill
    \subfigure[Ours]{\includegraphics[width=0.1\textwidth]{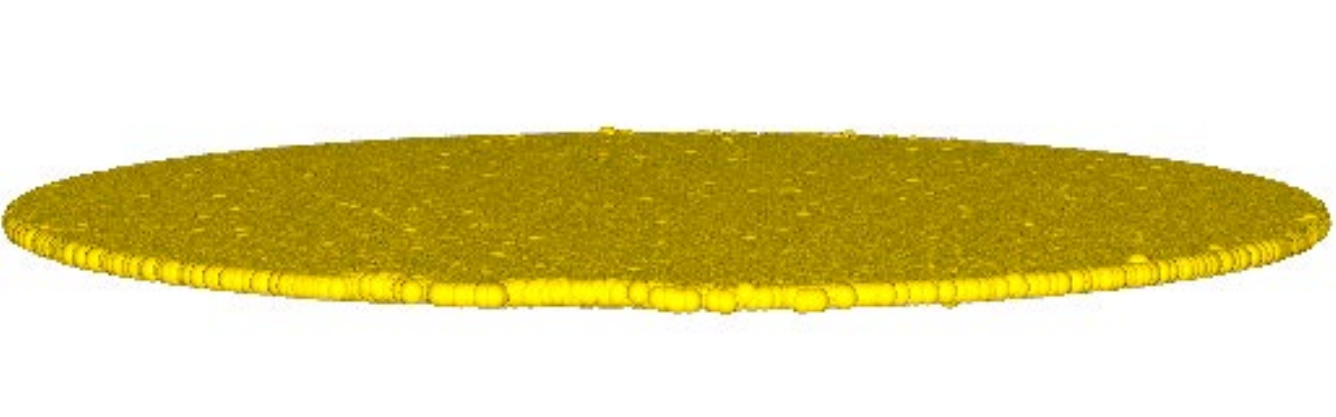}}
    \caption{Pointfilter can produce comparable results to the LOP-based methods (e.g., WLOP and CLOP) in presence of noticeable outliers. Big radii are employed here for these methods.}
    \label{fig:strongoutliers}
\end{figure}

\noindent\textbf{Point clouds with noticeable outliers.} Although Pointfilter is not particularly designed for removing outliers, we observe it produces competitive results in point clouds with noticeable outliers. Following~\cite{Lu2018TVCG}, we also conduct an experiment by comparing our method with the LOP-based methods (WLOP \& CLOP) which use $L_1$-norm and are robust to outliers. This experiment demonstrates the capability of our Pointfiler in dealing with noticeable outliers. As Fig. \ref{fig:strongoutliers} shows, Pointfilter can generate a comparable result to WLOP and CLOP. 

\begin{table}[htb!]
    \centering
    \caption{Average errors of all filtered point clouds over our test synthetic models ($15$ models with $0.5\%$ Gaussian noise), in terms of Chamfer distance (CD), mean square error (MSE) and mean point-to-surface distance (P2F).} \label{table:cadmeasurement}
    \begin{tabular}{|l|ccc|} 
    \hline
    \diagbox[width=8em]{\textbf{Methods}}{\textbf{Metrics}} & CD  ($10^{-5}$) & MSE ($10^{-3}$) & P2F ($10^{-3}$)\\
    \hline
    Noisy         & 3.655 & 5.869 & 4.058\\
    RIMLS         & 1.102 & 4.082 & 1.549\\
    GPF           & 2.273 & 4.791 & 2.446\\
    WLOP          & 1.888 & 4.222 & 1.726\\
    CLOP          & 1.421 & 4.059 & 1.439\\
    PCN           & 0.942 & 3.981 & 1.351\\
    TD            & 1.483 & 4.353 & 1.878\\
    Ours          & \textbf{0.833} & \textbf{3.884} & \textbf{1.147}\\  
    \hline
    \end{tabular}
\end{table}

\subsection{Quantitative Comparisons}
\label{sec:quantitativecomparison}

\textbf{Errors.} We calculate the above metrics of the methods on some point clouds. These point sets are achieved by adding synthetic noise to the ground truth. To depict the distance errors, we calculate the mean square error (MSE) for each ground-truth point via Eq. \eqref{eq:averagedistance} and visualize the results in Fig. \ref{fig:CPErrors}. We observe that Pointfiler generates comparable or better results, especially for sharp features. To comprehensively evaluate our Pointfiler, we also calculate the overall errors over the 15 synthetic models in the test set, in terms of the Chamfer distance (CD), mean square error (MSE) and point-to-surface distance (P2F). As illustrated in Table \ref{table:cadmeasurement}, our method averagely achieves the lowest errors.

Despite that the results of RIMLS are comparable to ours, it requires quality normals (we used bilateral smoothing \cite{Huang2013TOG} for RIMLS) and trial-and-error parameter tuning to obtain satisfactory results. Moreover, such parameter tuning is tedious, time-consuming, and especially difficult for users who do not have any background knowledge. In contrast, our method is automatic and easy to use, and is generally the most accurate one among all the compared approaches.
We also evaluate our Pointfilter on the KinectV1 and KinectV2 datasets \cite{wang2016mesh}. Table \ref{tab:real_scan_errors} shows our re-trained Pointfilter produces lower errors (Chamfer distance) in the smoothing results than the re-trained PCN.

\noindent\textbf{Runtime.} Because surface reconstruction is an application over point clouds, we only calculate the runtime of each point set filtering method. In particular, optimized-based methods (RIMLS, GPF, WLOP and CLOP) involve multiple steps and require trial-and-error efforts to tune parameters to produce decent visual results, which means these methods generally require a much longer ``runtime'' in practice. Thus, we only consider learning-based methods (EC-Net, PCN and Ours), in terms of time consumption in the test stage. Table \ref{tab:runtime} summaries the runtime of each learning-based method on some point clouds using the same configuration. Table \ref{tab:runtime} sees that EC-Net is the fastest method among the learning-based methods, as it is an upsampling method and only requires a few patches evenly distributed on the noisy input in the test phase. For fair comparisons, we only consider the runtime of the noise removal module in PCN, because extra time consumption would be introduced for the outliers removal module. In spite of this, PCN is still the slowest one. Our approach ranks the second in speed, and we suspect that it is due to the point-wise manner. 

\begin{table}[htb!]
    \centering
    \caption{Runtime performance (in seconds) for three learning-based methods in the test stage. All examples were run on the same computer configurations (Section .\ref{sec:networktraining}).}\label{tab:runtime}
    \begin{tabular}{|l|cccc|} 
    \hline
    \diagbox[width=8em]{\textbf{Models}}{\textbf{Methods}} & EC-Net & PCN & TD & \textbf{Ours}\\
    \hline
    Cube          & 27.73 & 360.64 & 96.43  &62.34\\
    Fandisk       & 26.92 & 369.67 & 96.52  &62.45\\
    Boxunion      & 26.15 & 365.09 & 95.28  &65.41\\
    Tetrahedron   & 28.64 & 326.11 & 97.88  &63.21\\
    Horse         & 26.91 & 365.62 & 98.21  &63.55\\
    Face-Yo       & 27.27 & 362.26 & 97.68  &63.08\\
    Fertility-tri & 27.43 & 370.21 & 98.36  &63.52\\
    Face-Raw      & 22.63 & 306.39 & 85.17  &55.71\\
    Pyramid-Raw   & 44.74 & 618.98 & 163.91 & 105.85\\
    Nefertiti-Raw & 27.00 & 353.44 & 96.00  &62.38\\
    \hline
    \end{tabular}
\end{table}

\begin{table}[htbp]
  \centering
  \caption{Comparison with PCN on KinectV1 and KinectV2 datasets \cite{wang2016mesh}. The Chamfer distance ($10^{-5}$) is used here.} \label{tab:real_scan_errors}
  \begin{tabular}{l|cc}
  \hline
  \textbf{Methods} & KinectV1 & KinectV2\\
  \hline
  Noisy                    & 2.467 & 3.569 \\
  PCN                      & 2.078 & 3.188 \\
  Ours                      & 2.356 & 3.411 \\
  \textbf{Ours (re-trained)}& \textbf{1.884} & \textbf{3.149} \\
  \hline
  \end{tabular}
\end{table}%

\begin{figure}[htb]
    \centering
    \includegraphics[width=0.45\textwidth]{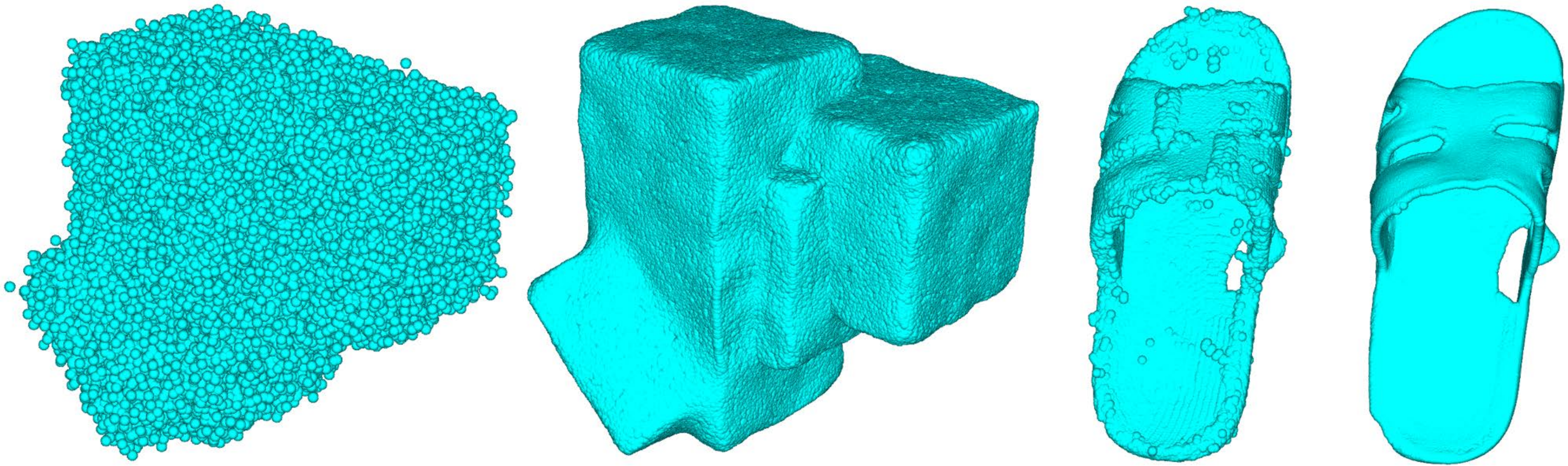}
    \makebox[0.8in]{(a)}
    \makebox[0.8in]{(b)}
    \makebox[0.8in]{(c)}
    \makebox[0.8in]{(d)}
    \caption{Failure cases: excessive noise ((a)(b)) and large holes ((c)(d)).}
    \label{fig:failure_cases}
\end{figure}

\subsection{Robustness} 
\begin{figure}[htb!]
  \centering
   \includegraphics[width=0.45\textwidth]{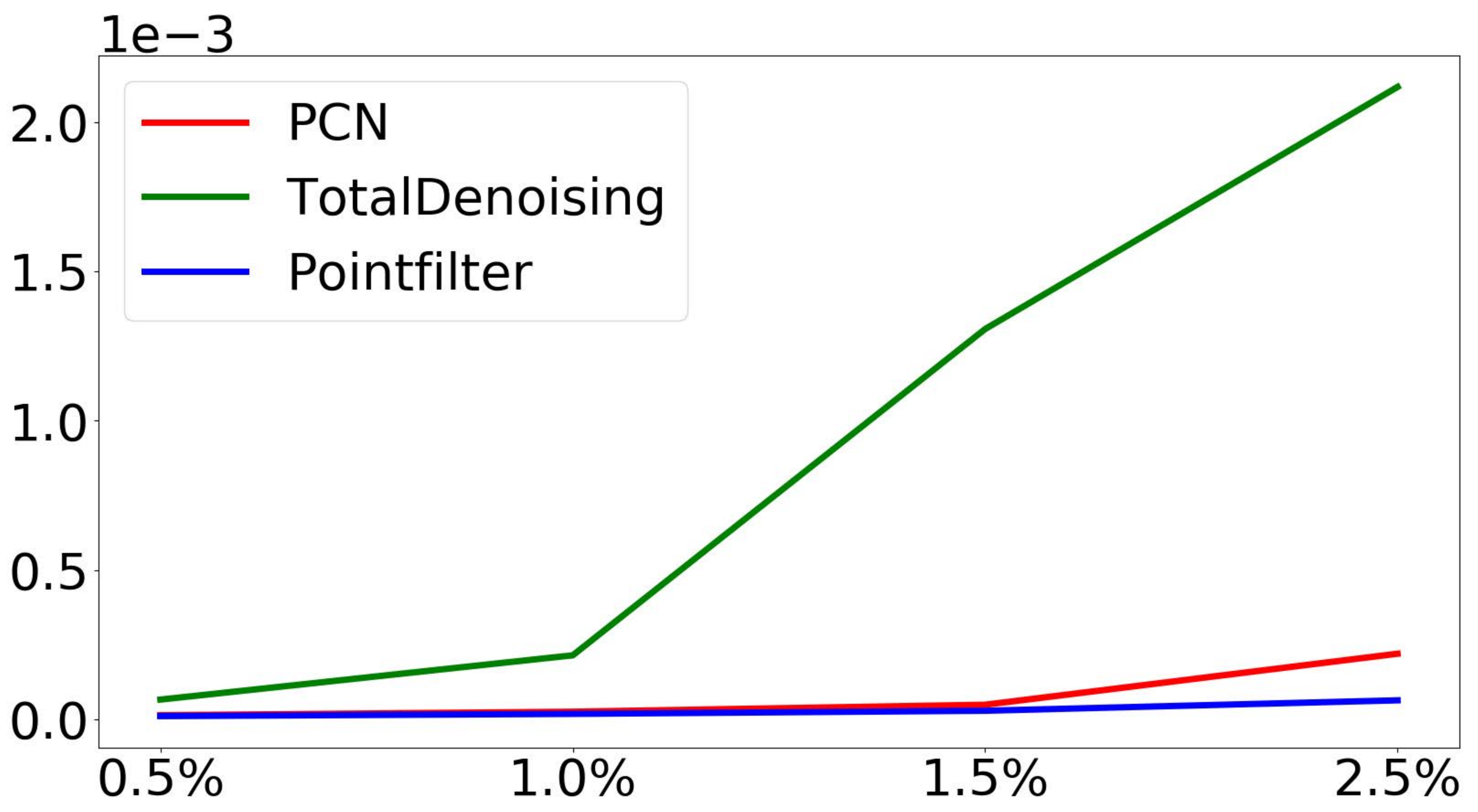}
  \caption{Average CD errors ($10^{-5}$) of filtered point clouds in the test set ($15$ synthetic models), in terms of different noise levels.}
  \label{fig:robutness_mln}
\end{figure}

\textbf{Different levels of noise.} From Fig. \ref{fig:robutness_mln}, we can observe that our Pointfilter achieves superior performance, especially for heavy noise like $2.5\%$. As such, our Pointfilter is more robust than other compared methods in handling increasing levels of noise.

\begin{figure*}[htb]
    \centering
    \subfigure[10K]
    {
        \includegraphics[width=0.22\textwidth]{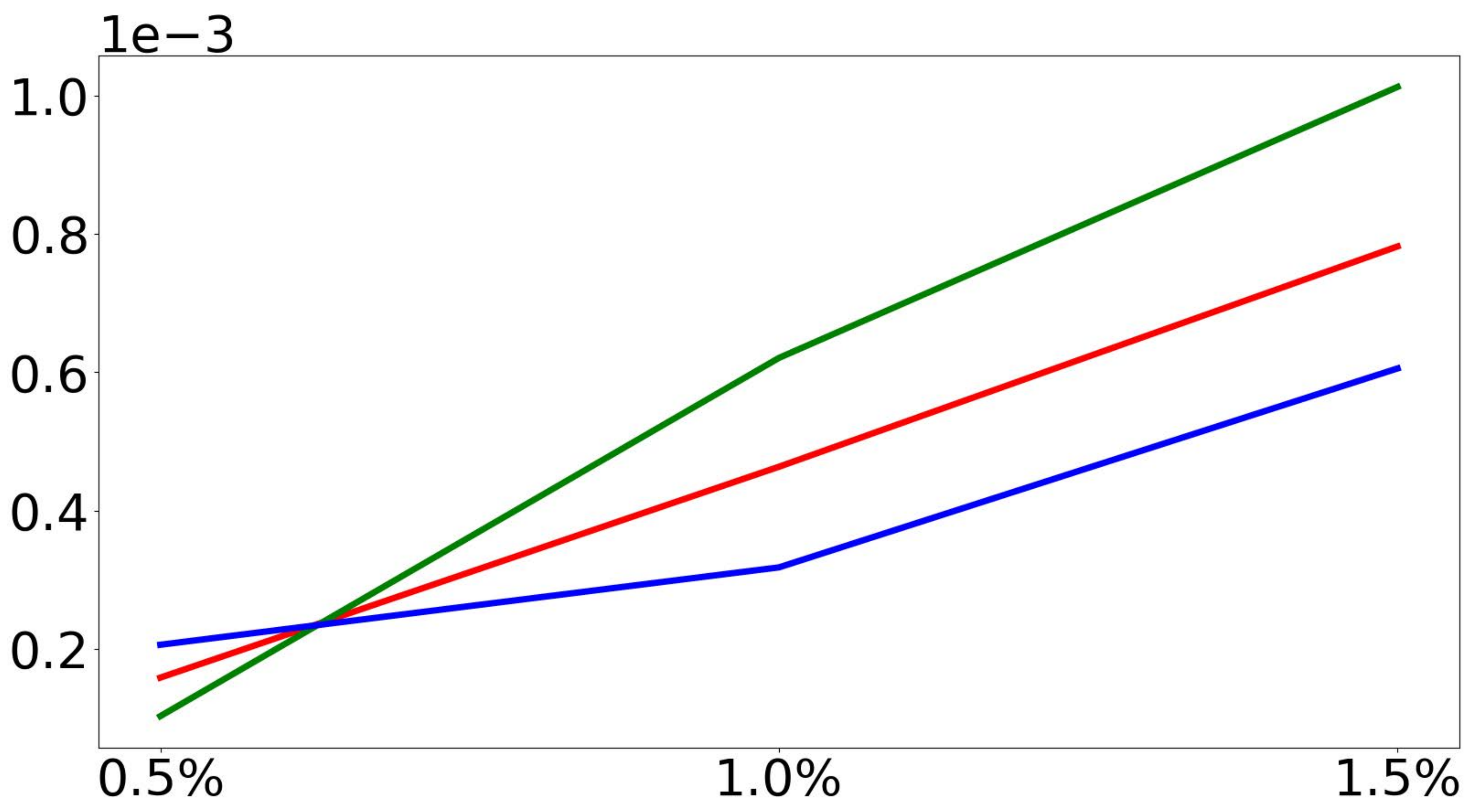}
    }
    \subfigure[30K]
    {
        \includegraphics[width=0.22\textwidth]{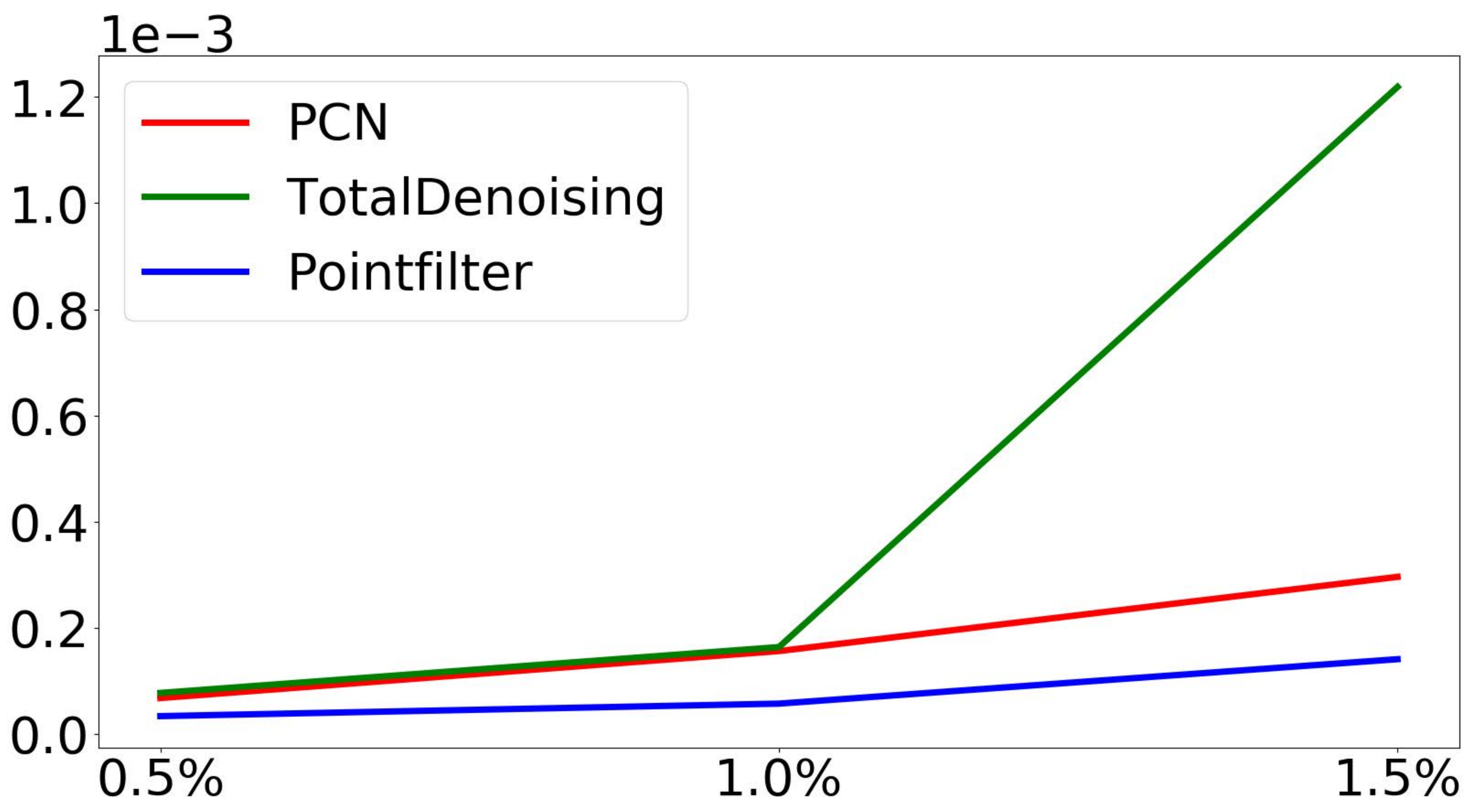}
    }
    \subfigure[50K]
    {
        \includegraphics[width=0.22\textwidth]{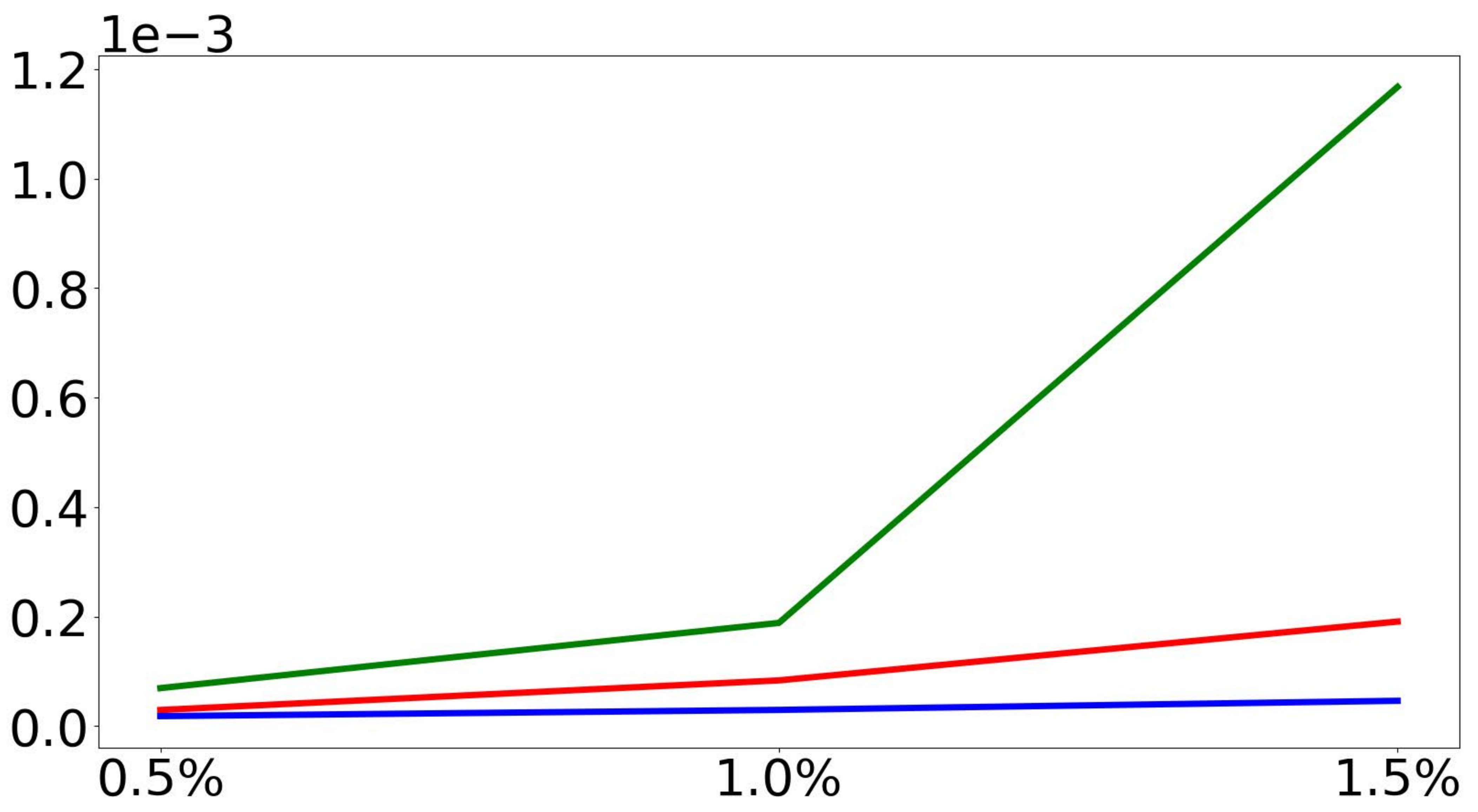}
    }
    \subfigure[80K]
    {
        \includegraphics[width=0.22\textwidth]{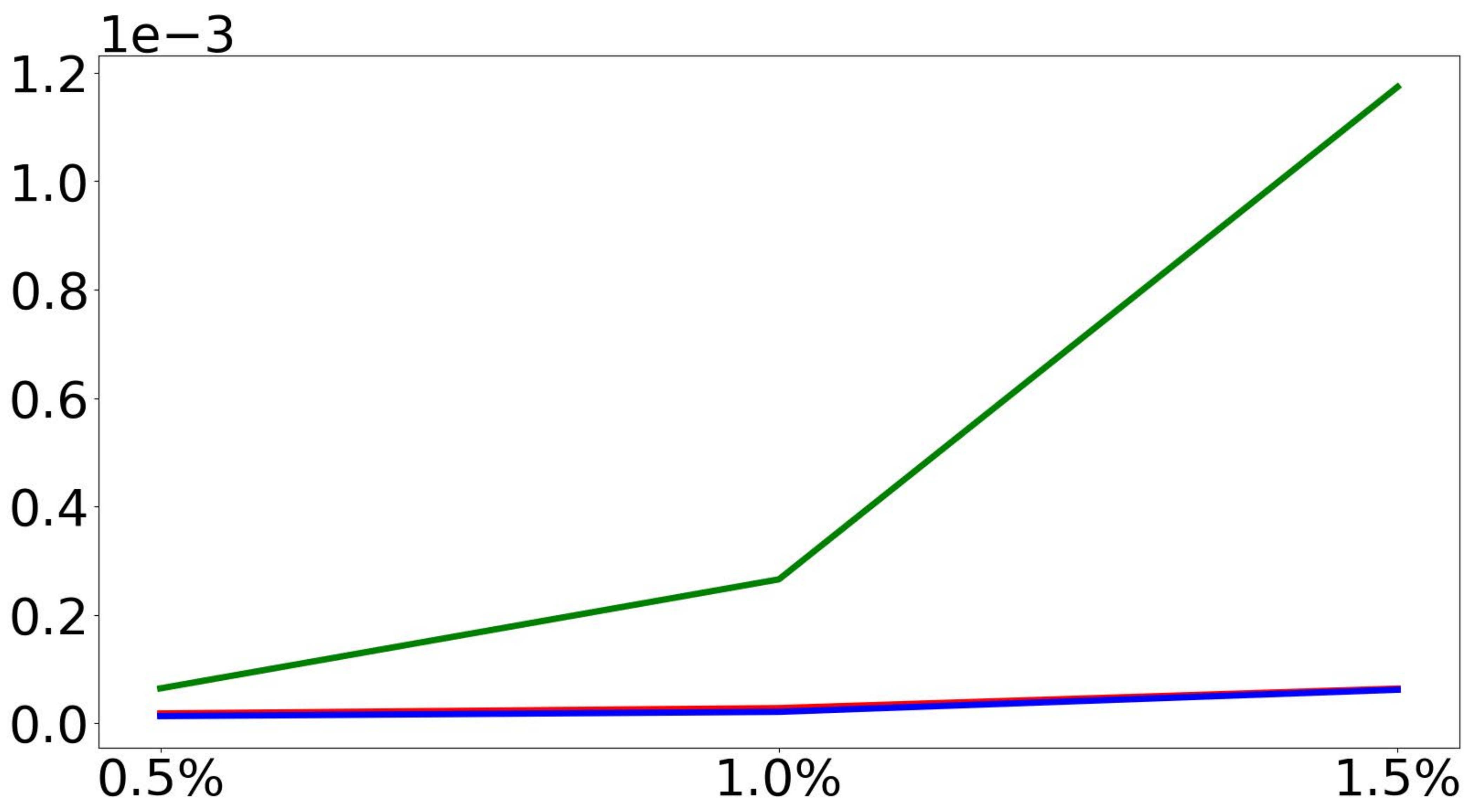}
    }
    \caption{Average CD errors ($10^{-5}$) of filtered point clouds in the test set ($15$ synthetic models), in terms of different numbers of points.}
    \label{fig:robutness_mrp}
\end{figure*}

\noindent\textbf{Different numbers of points.} We implement experiments over the test dataset, containing different numbers of points ($10K$, $30K$, $50K$ and $80K$). Fig. \ref{fig:robutness_mrp} obviously shows that the performance of our Pointfilter is better than other competitors (except the case of $10K$ points). We suspect that the relatively sparse neighboring points of a patch (comparing to training dataset) are insufficient to depict the local structure, which leads to projecting noisy points onto a misleading underlying surface and less desired results. This is often involved in patch based methods, such as PCN and Pointfilter.

\begin{figure}[htb!]
    \subfigure[Noisy]
    {
        \begin{minipage}[b]{0.1\textwidth} 
        \includegraphics[width=1\textwidth]{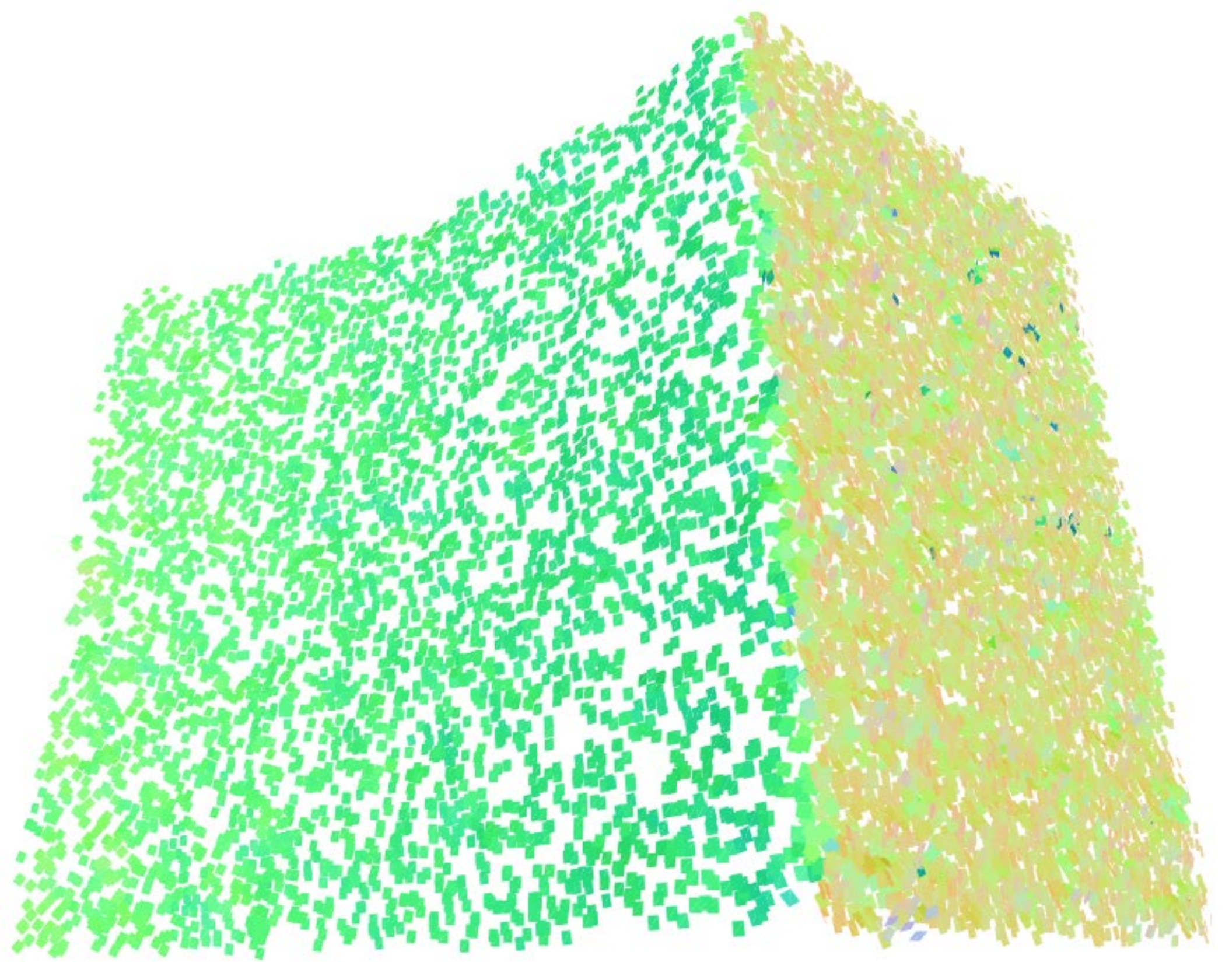}\\
        \includegraphics[width=1\textwidth]{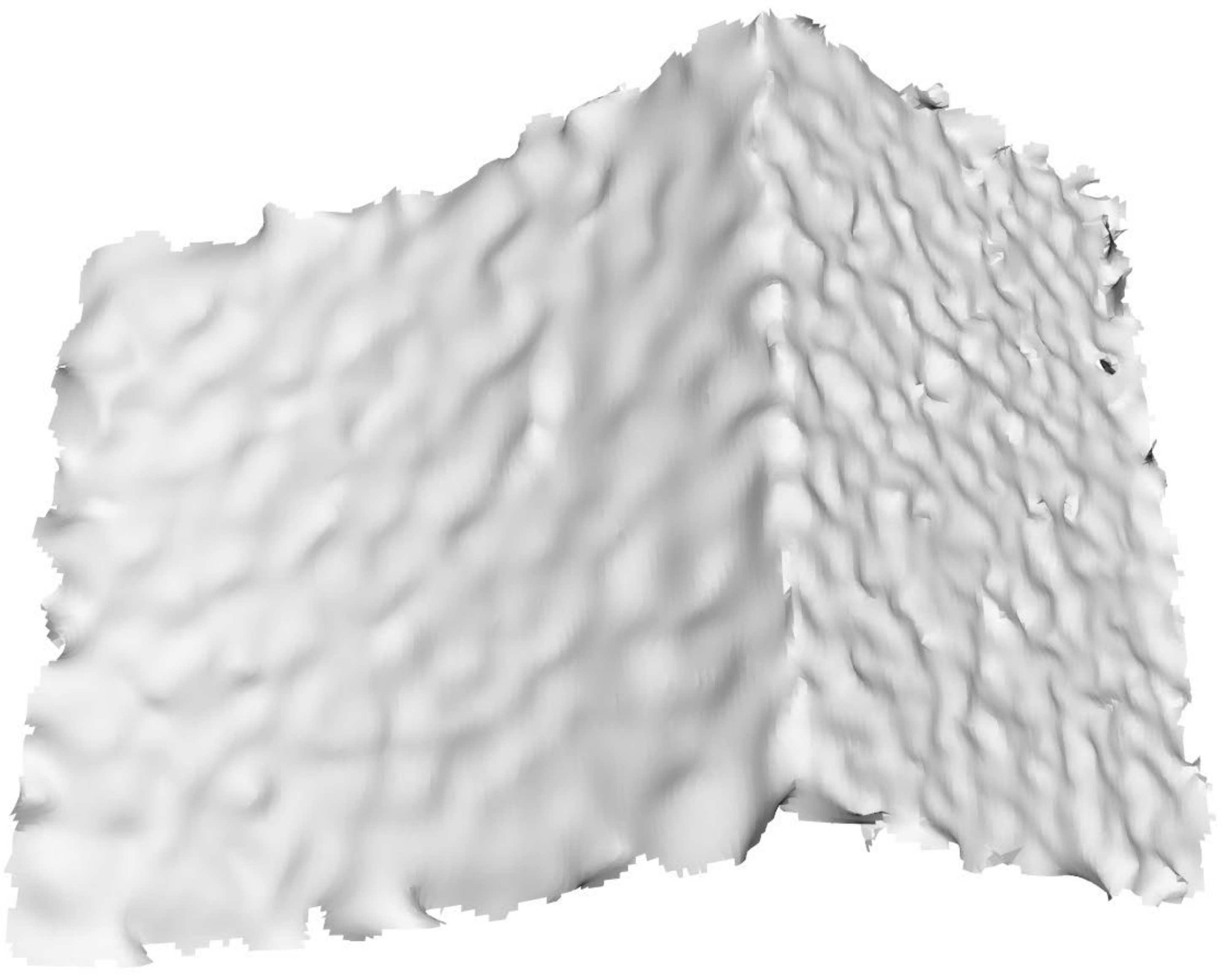}
        \end{minipage}
    }
    \subfigure[PCN]
    {
        \begin{minipage}[b]{0.1\textwidth} 
        \includegraphics[width=1\textwidth]{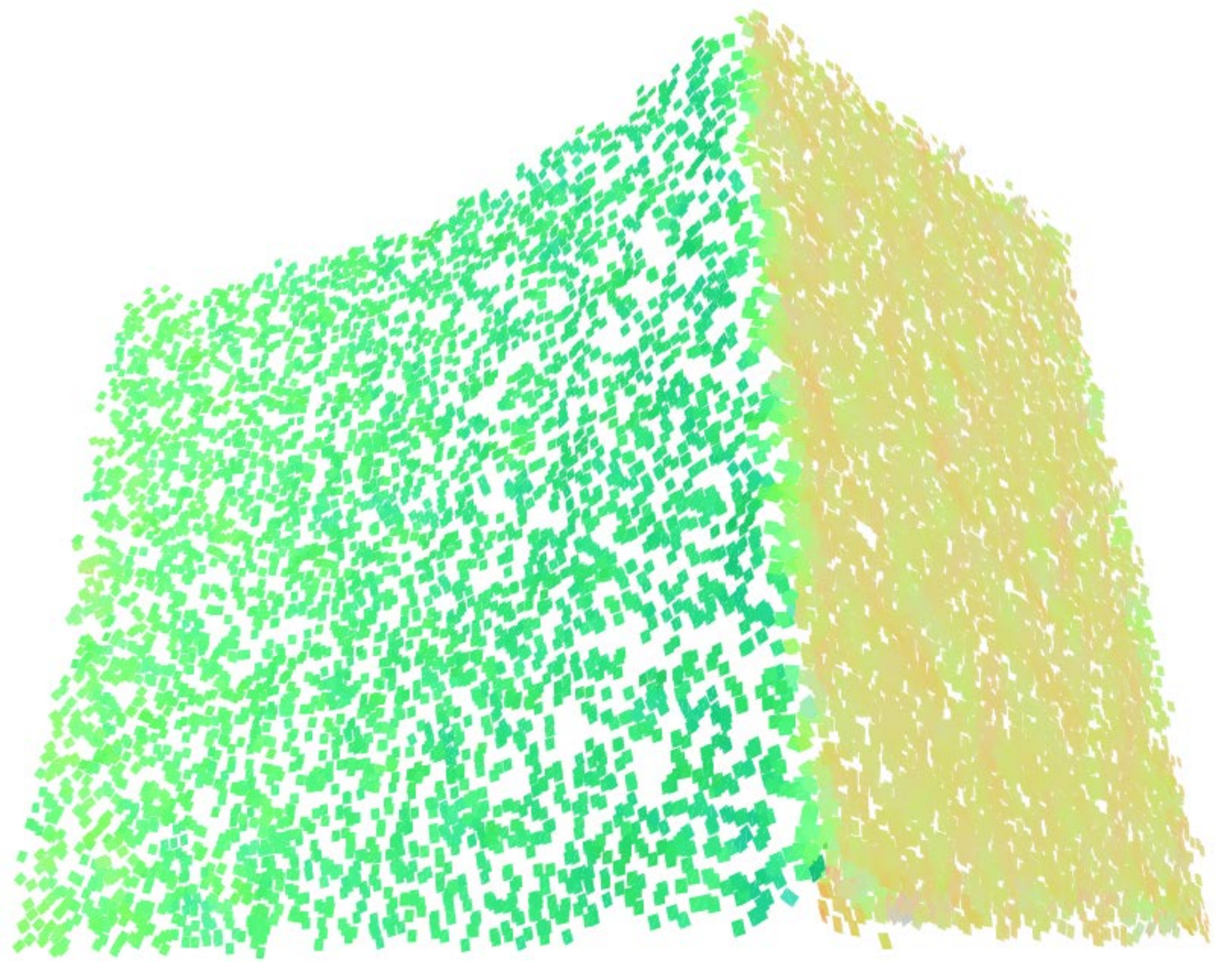}\\
        \includegraphics[width=1\textwidth]{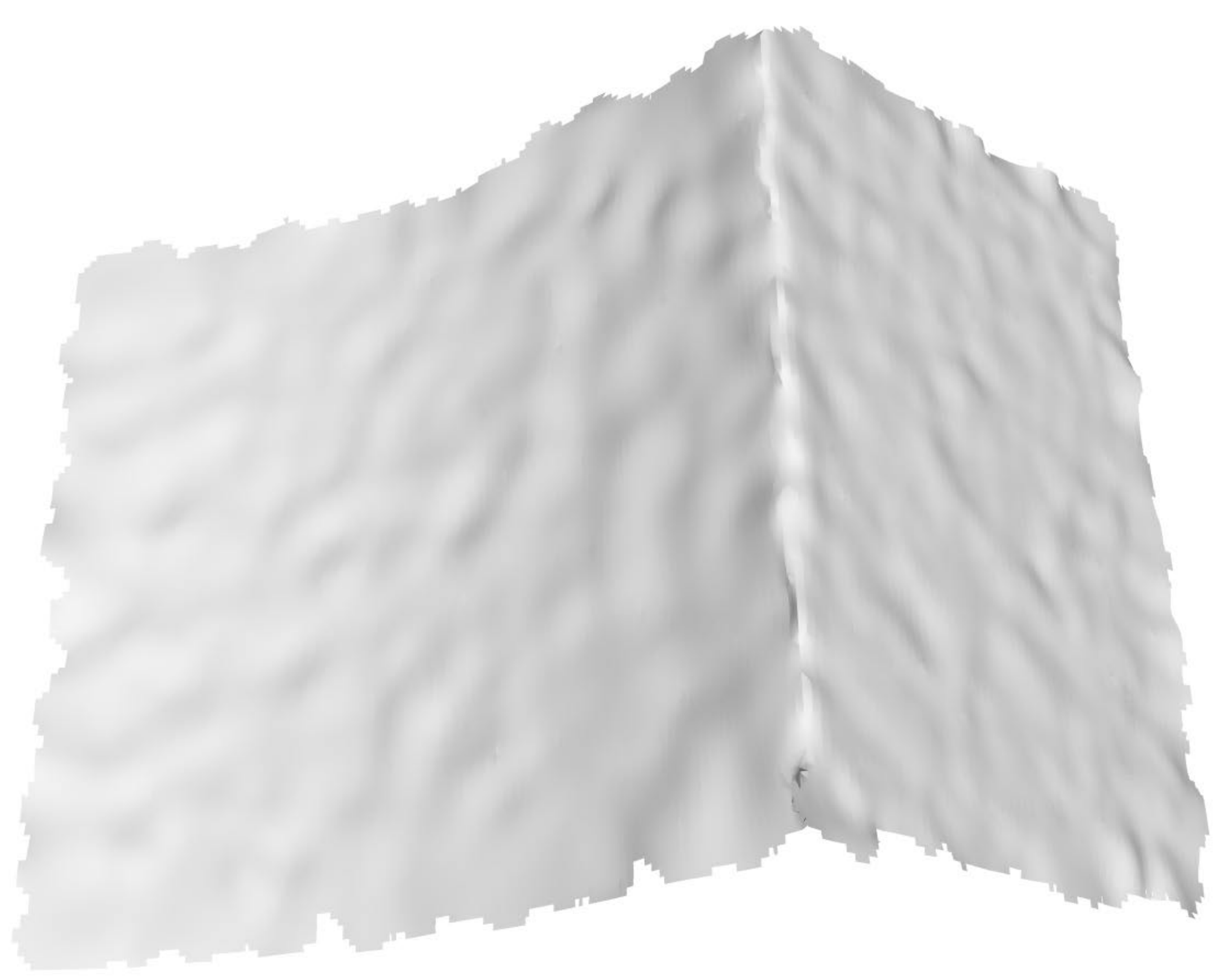}
        \end{minipage}
    }
    \subfigure[TD]
    {
        \begin{minipage}[b]{0.1\textwidth} 
        \includegraphics[width=1\textwidth]{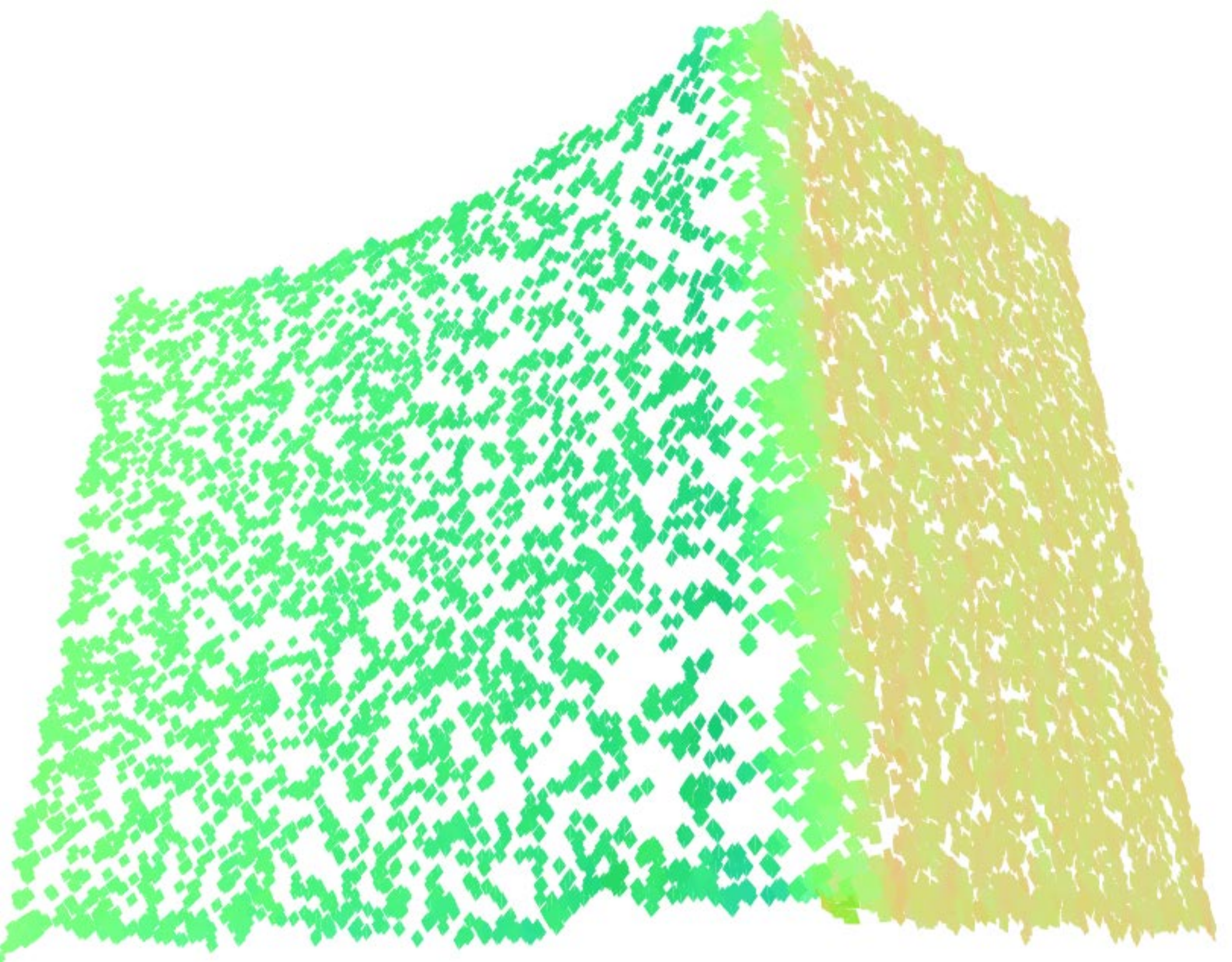}\\
        \includegraphics[width=1\textwidth]{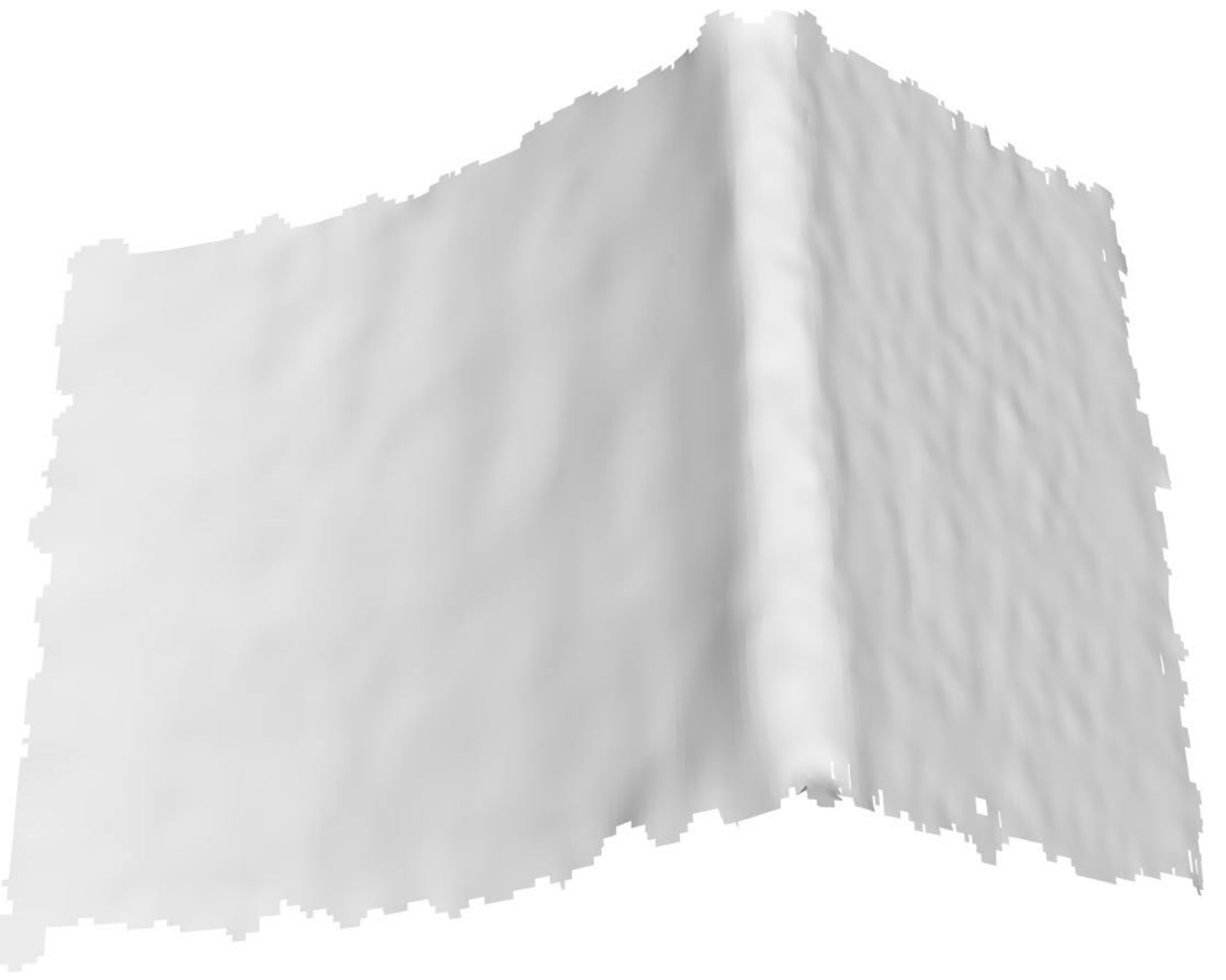}
        \end{minipage}
    }
    \subfigure[Ours]
    {
        \begin{minipage}[b]{0.1\textwidth} 
        \includegraphics[width=1\textwidth]{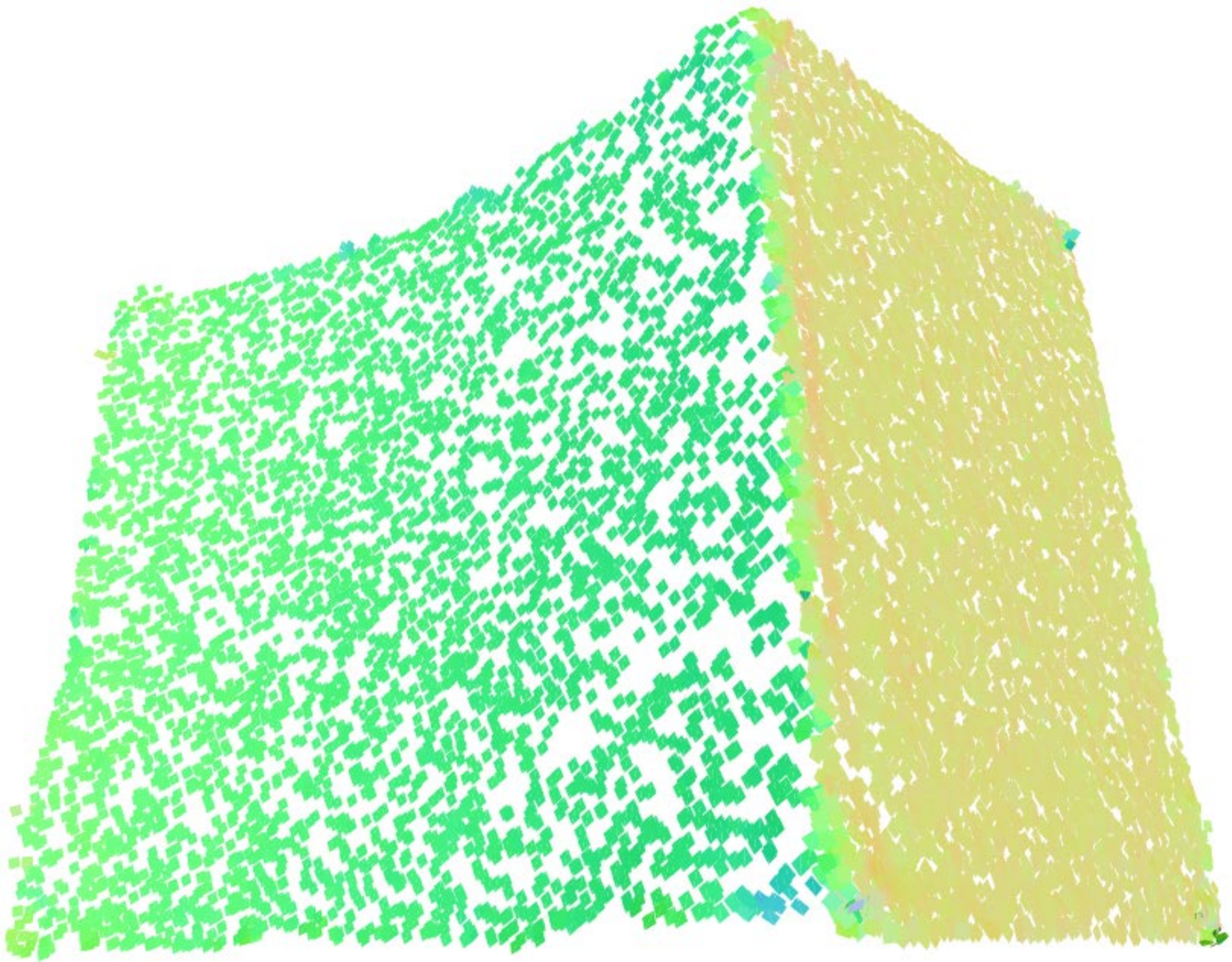}\\
        \includegraphics[width=1\textwidth]{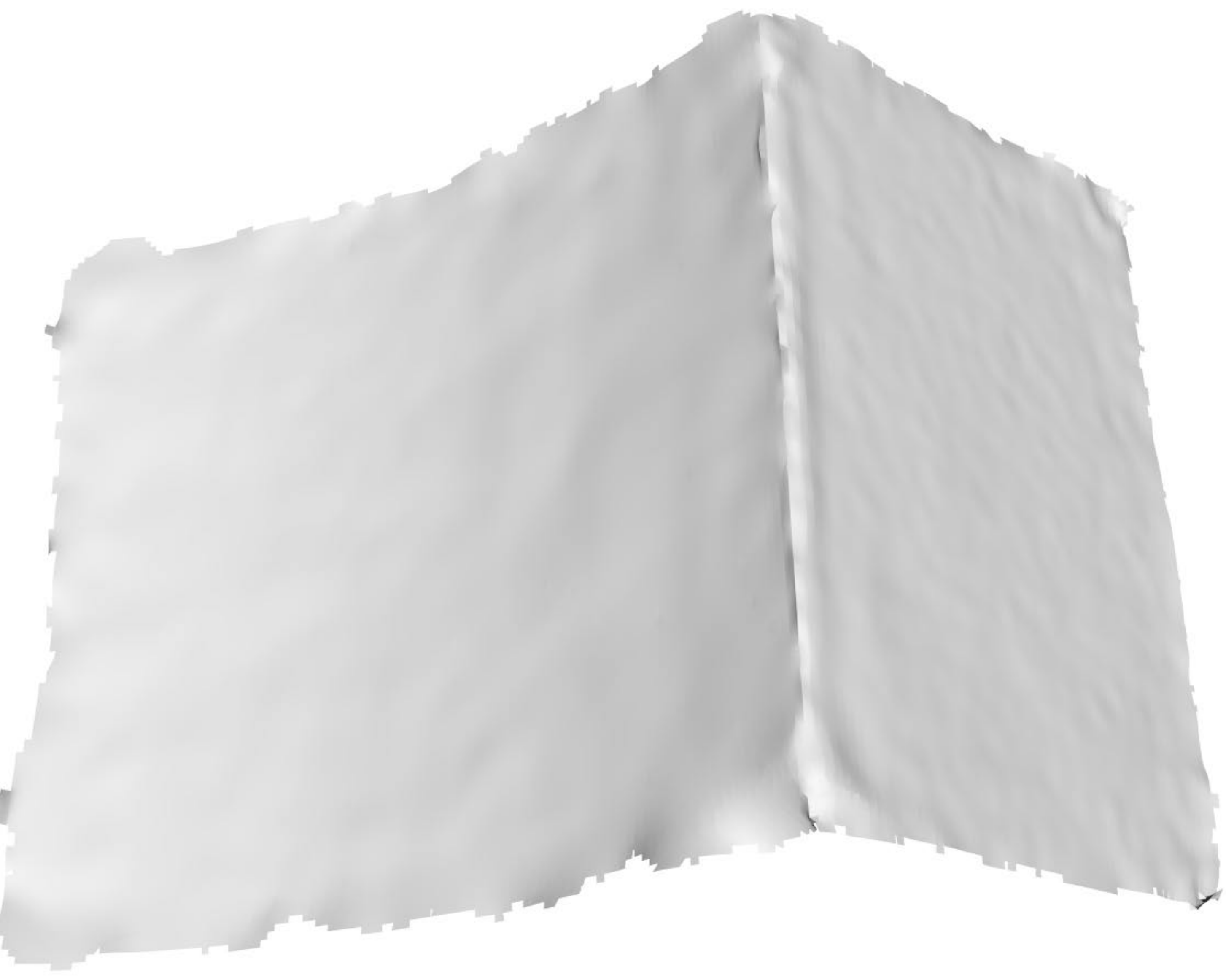}
        \end{minipage}
    }
    \caption{Non-uniform point cloud.}
    \label{fig:non_uniform_distribution}
\end{figure}

\noindent\textbf{Non-uniform point distribution.}  We also test our Pointfilter under non-uniform point distribution. We synthesize a model which consists of a curved surface and a plane surface. As analyzed above, patch-based methods tend to generate less desired results for less dense points. As shown in Fig. \ref{fig:non_uniform_distribution}, we can see that our Pointfilter can achieve a better result on the dense part, though it outperforms other compared methods.

\subsection{Ablation Study}
\label{sec:ablationstudy}
\begin{figure}[htb]
    \centering
    \subfigure[Input]
    {
        \includegraphics[width=0.14\textwidth]{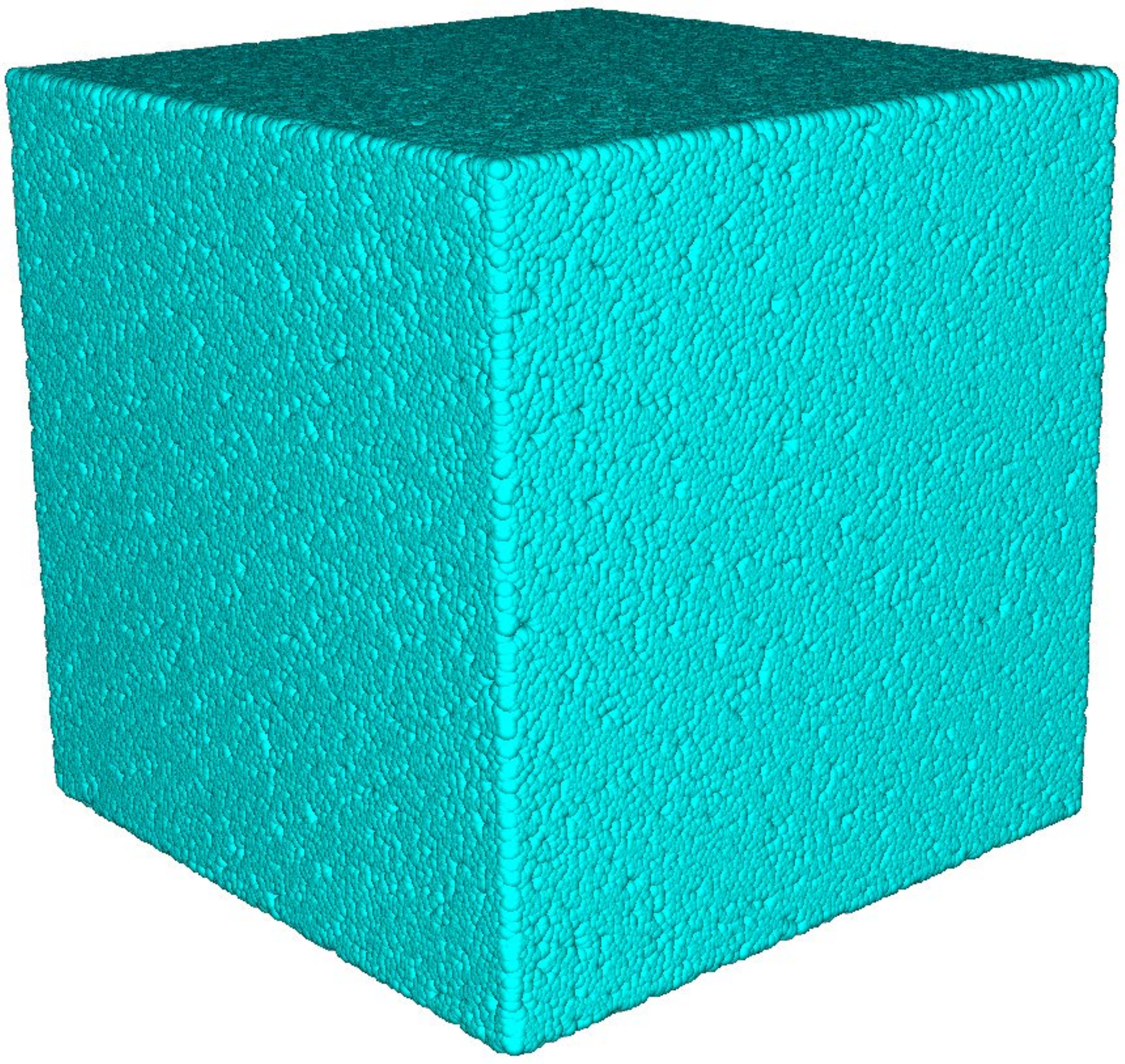}
    }
    \hfill
    \subfigure[Without repulsion]
    {
        \includegraphics[width=0.14\textwidth]{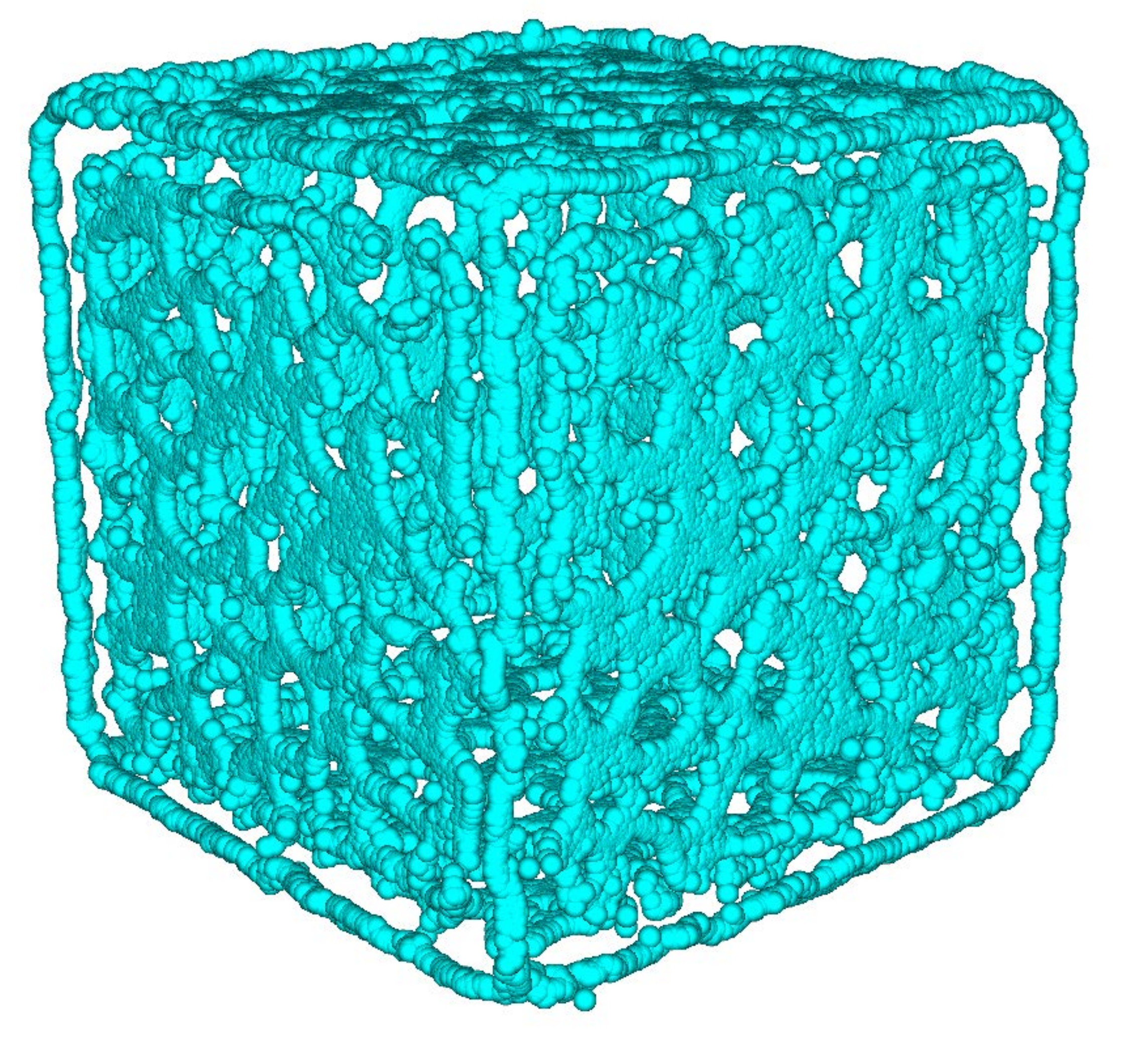}
    }
    \hfill
    \subfigure[With repulsion]
    {
        \includegraphics[width=0.14\textwidth]{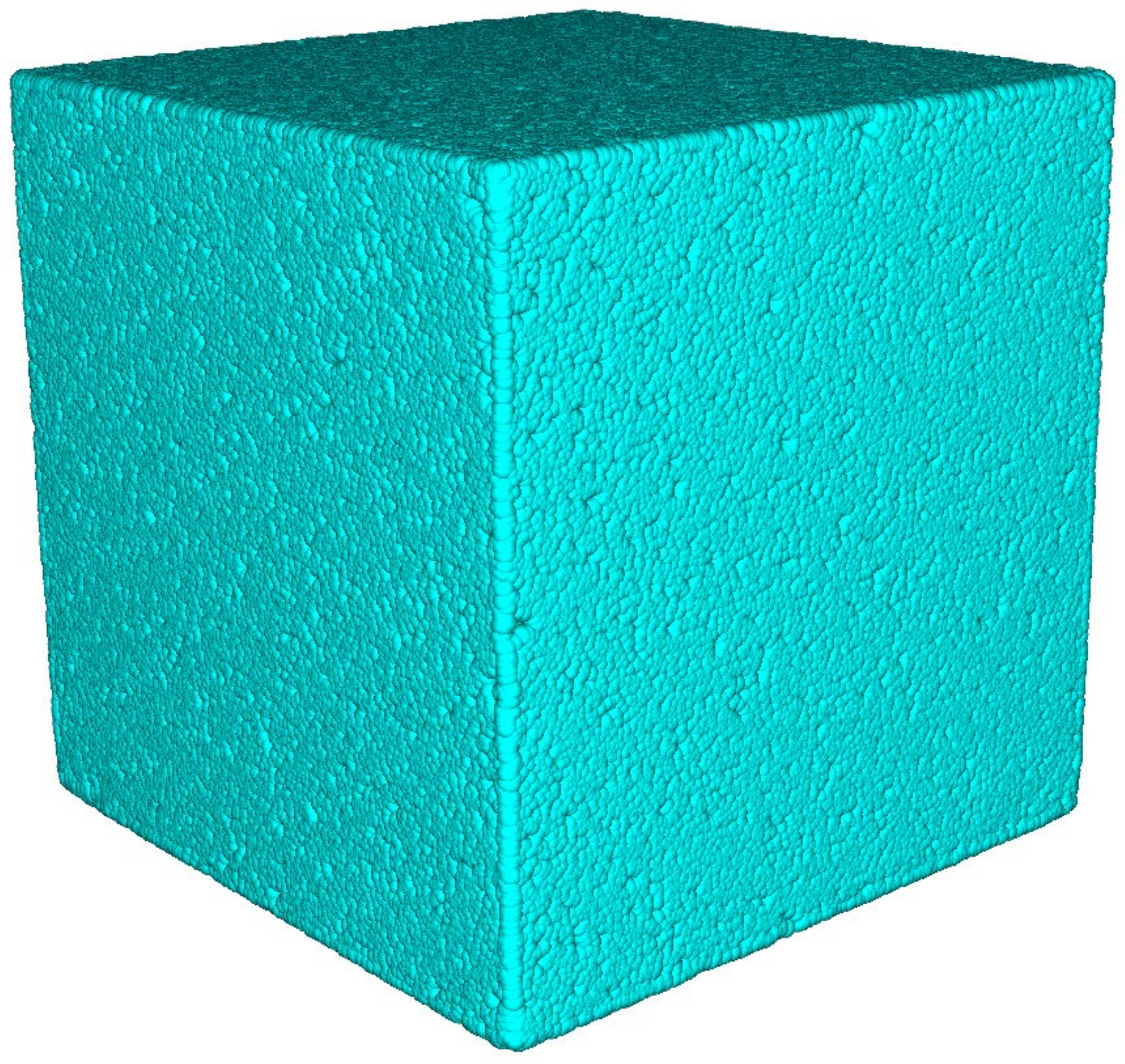}
    }
    \caption{Results with and without the repulsion term ($L_{rep}$).}
    \label{fig:w/wo_repulsion}
\end{figure}

\begin{figure}[htb]
    \centering
    \includegraphics[width=0.45\textwidth]{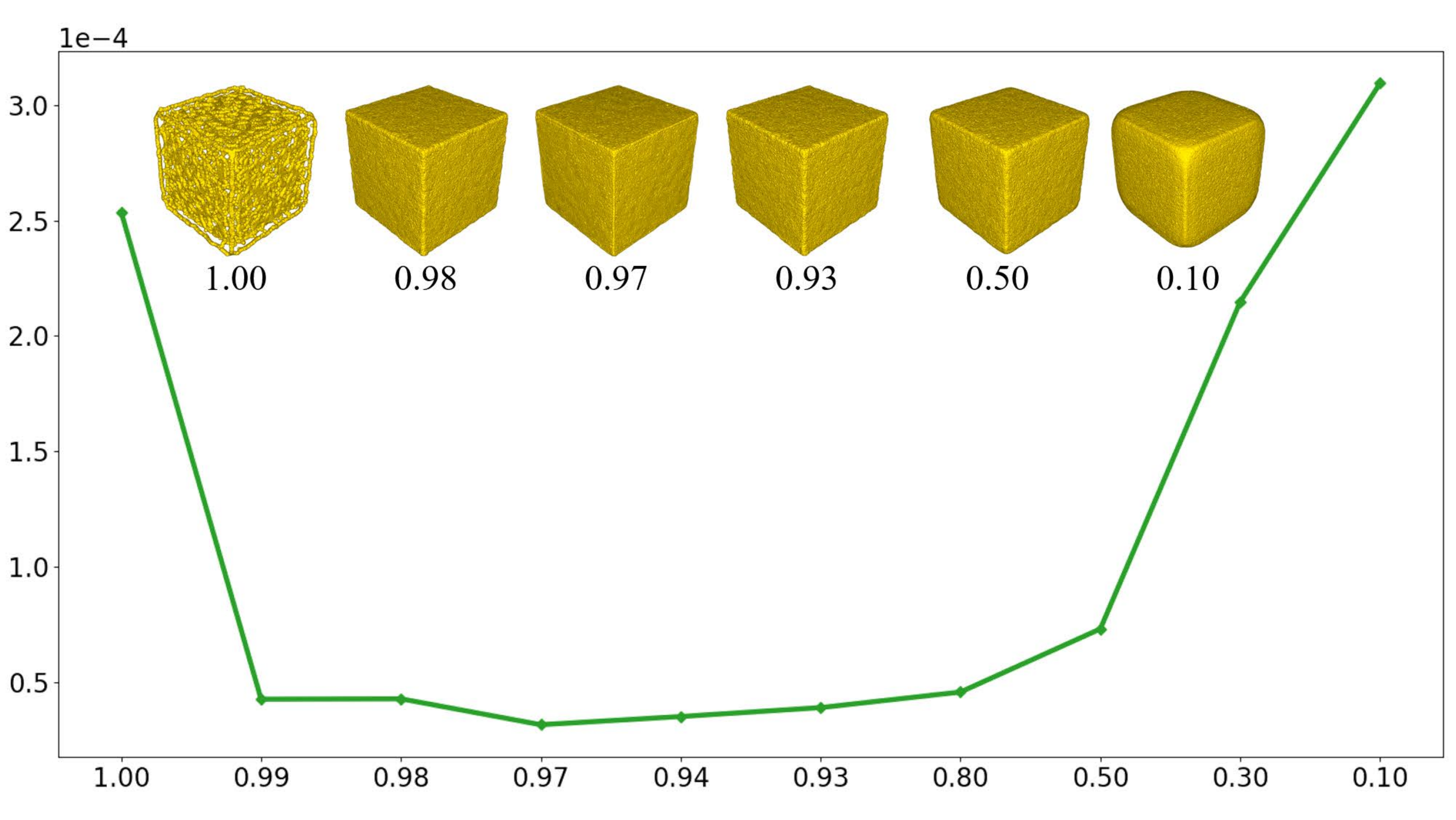}
    \caption{Average CD errors of filtered point clouds on the test set ($15$ synthetic models), in terms of different trad-off parameter $\eta$ in Eq. \eqref{eq:Initial_Loss}. }
    \label{fig:differ_repulsion_paramter}
\end{figure}

\textbf{With/without the repulsion term.} We found that points would aggregate together without the repulsion term. The repulsion term is introduced (Eq. \eqref{eq:Initial_Loss}) to mitigate this issue. As shown in Fig. \ref{fig:w/wo_repulsion}, our repulsion term plays a significant role in maintaining a relatively uniform distribution for points. In addition, we also evaluate the influence of the trade-off parameter $\eta$ in Eq. \eqref{eq:Initial_Loss}. In Fig. \ref{fig:differ_repulsion_paramter}, we empirically found that $\eta$ with $0.97$ achieves a superior performance to other choices, based on the statistics over all models in the test set. Some selected visual results with different repulsion parameters are shown in Fig. \ref{fig:differ_repulsion_paramter}. It is obvious that a small $\eta$ cannot preserve share features, especially for corner regions.

\noindent\textbf{Alternative loss.} For fairness, we train three loss functions ($L_{2}$, $L_{proj}^{a}$ (Eq. \eqref{eq:Initial_Loss})  and $L_{proj}^{b}$ (Eq. \eqref{eq:lossfunction})) separately under the same training configuration. Table \ref{tab:alternative_loss} manifests that our Pointfilter with $L_{proj}^{b}$ has the best performance. Thus, the proposed loss function is crucial and useful for boosting the performance in filtering noisy point clouds.

\begin{table}[htbp]
  \centering
  \caption{Average performance of Pointfilter with alternative loss functions on our test dataset ($15$ synthetic models). The Chamfer distances ($10^{-5}$) under different levels of noise are evaluated.} \label{tab:alternative_loss}
  \begin{tabular}{l|llll}
  \hline
  \textbf{Losses} & $0.5\%$ & $1.0\%$ & $1.5\%$ & $2.5\%$\\
  \hline
  $L_{proj}^{a}$ & 2.434 & 4.488 & 7.337 & 13.766 \\
  $L_2$          & 1.843 & 2.371 & 5.171 & 9.053  \\
  $L_{proj}^{b}$ & 1.196 & 1.951 & 3.044 & 6.414 \\ 
  \hline
  \end{tabular}
\end{table}

\noindent\textbf{Alternative backbones.} In this work, we adopt PointNet \cite{Qi2017CVPR}, which is simple and easy to use, as the backbone in the encoder part (see Figure. \ref{fig:overview}). It is necessary to test other advanced backbones such as DGCNN \cite{Wang2019TOG} and PointCNN \cite{yangyan2018arXiv}. For fair purpose, we only replace corresponding layers in the encoder and all results are measured by the Chamfer distance. As shown in Table \ref{tab:alternative_backbone}, other backbones do not show the noticeable performance gain over PointNet. For variants with DGCNN and PointCNN, we suspect that the neighboring information in the presence of noise may pose negative impact on the two backbones which depend on finding neighboring information.

\begin{table}[htbp]
  \centering
  \caption{Average performance of Pointfilter with alternative backbones on our test dataset ($15$ synthetic models). The Chamfer distances ($10^{-5}$) under different levels of noise are evaluated.} \label{tab:alternative_backbone}
  \begin{tabular}{l|llll}
  \hline
  \textbf{Backbones} & $0.5\%$ & $1.0\%$ & $1.5\%$ & $2.5\%$\\
  \hline
  PointCNN  & 2.446 & 3.592 & 4.839 & 8.767\\ 
  DGCNN     & 2.441 & 3.342 & 4.682 & 8.423 \\
  PointNet  & 1.196 & 1.951 & 3.044 & 6.414 \\
  \hline
  \end{tabular}
\end{table}

\section{Conclusion}
\label{sec:discussionconclusion}
In this paper, we proposed a Pointfilter framework for feature-preserving point cloud filtering. Our architecture can be easily trained. Given an input noisy point cloud, our method can automatically infer the involved displacement vectors and further the filtered point cloud with preserved sharp features. Extensive experiments and comparisons showed that our method outperforms the state-of-the-art point set filtering techniques (or comparable to optimization based methods like RIMLS which need quality normals and trial-and-error parameter tuning), in terms of both visual quality and evaluation errors. Our approach is automatic and also achieves impressive performance on test time. Compared to PCN \cite{Rakotosaona2019CGF}, it should be noted that our Pointfilter is not designed for large-scale outliers removing, and our Pointfilter and PCN are thus complementary in terms of sharp features preservation and heavy outliers removal.

Our method involves a few limitations. First, our Pointfilter becomes hard to retain the sharp features when handling excessive noise (see Fig. \ref{fig:failure_cases} left). Secondly, our method fails to handle significant holes in point clouds (see Fig. \ref{fig:failure_cases} right). In future, we would like to incorporate global shape information into our framework to help guide point cloud filtering.


\bibliographystyle{ACM-Reference-Format}
\bibliography{tvcg-bibliography.bib} 

\end{document}